\documentclass[
	fontsize=10pt, % Base font size
	DIV=calc,
	twoside=true, % Use different layouts for even and odd pages (in particular, if twoside=true, the margin column will be always on the outside)
	numbers=noenddot, % Comment to output dots after chapter numbers; the most common values for this option are: enddot, noenddot and auto (see the KOMAScript documentation for an in-depth explanation)
]{kaobook}

\ifxetexorluatex
	\usepackage{polyglossia}
	\setmainlanguage{english}
\else
	\usepackage[english]{babel} % Load characters and hyphenation
\fi
\usepackage[english=british]{csquotes}	% English quotes

\usepackage{setspace}
\usepackage{blindtext}

\usepackage{gensymb}

\usepackage[backend=bibtex]{kaobiblio}
\usepackage{kaorefs}

\makeindex[columns=3, title=Alphabetical Index, intoc] % Make LaTeX produce the files required to compile the index

\makenomenclature % Make LaTeX produce the files required to compile the nomenclature
\usepackage{slashed}
\usepackage{amsfonts}
\usepackage{upgreek}
\usepackage{bigints}
\usepackage{booktabs}
\usepackage{hyperref}
\usepackage{physics}
\usepackage{amsmath}
\usepackage{slashed}
\usepackage{array}
\usepackage{booktabs}
\usepackage{pdfpages}
\DeclareOldFontCommand{\rm}{\normalfont\rmfamily}{\mathrm}

\renewcommand*{\multicitedelim}{\addcomma\nobreakspace}

\emergencystretch=\maxdimen
\DeclareUnicodeCharacter{202F}{\textcolor{red}{FIX ME!!!!}}
\begin{document}
\includepdf[fitpaper]{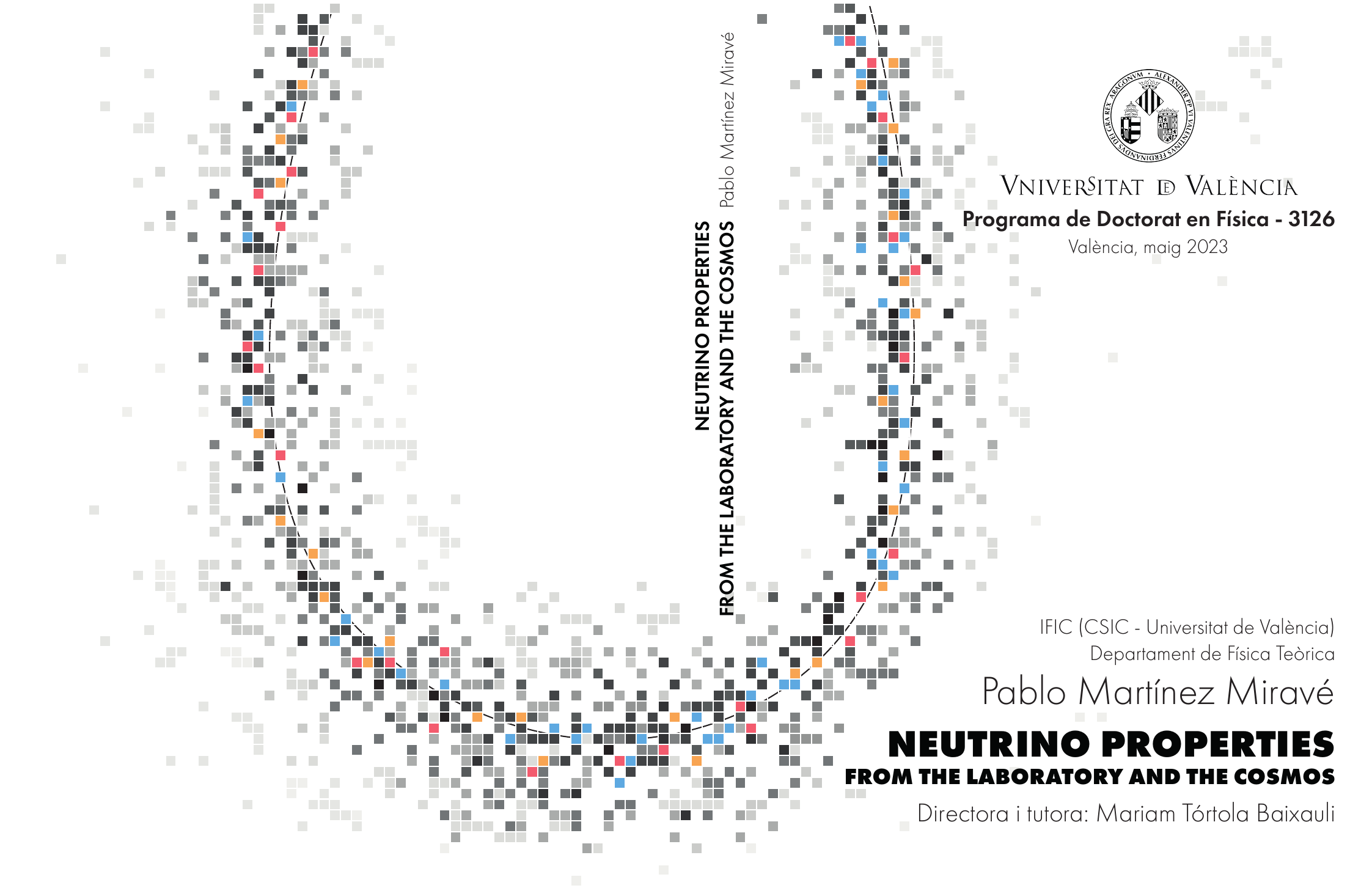}
%----------------------------------------------------------------------------------------
%	BOOK INFORMATION
%----------------------------------------------------------------------------------------

\titlehead{Tesi Doctoral}

\subject{ \includegraphics[width=9cm,height=3cm]{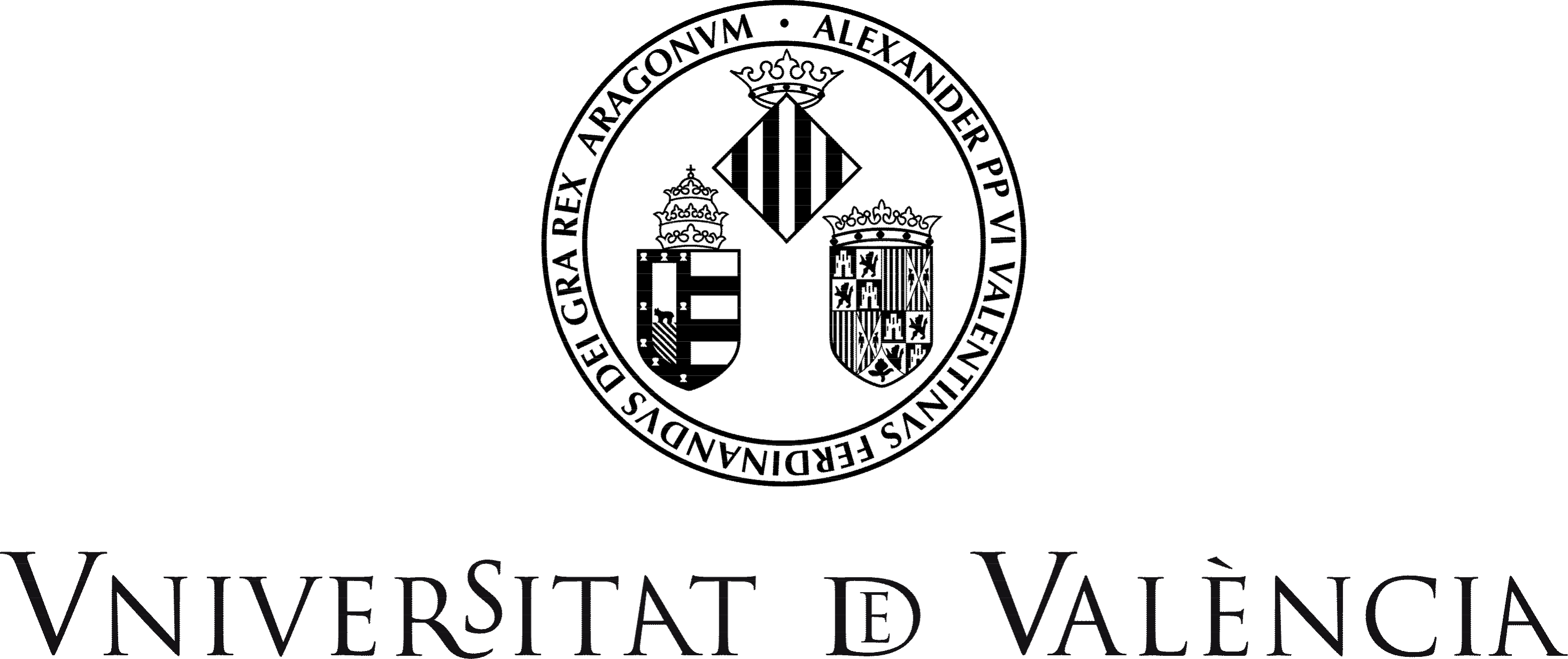}\\ Programa de Doctorat en Física - 3126}

\title[Neutrino properties from the laboratory and the cosmos]{Neutrino properties \\ from the laboratory and the cosmos}
\author[Pablo Martínez Miravé]{\textbf{ Pablo Martínez Miravé}}
\publishers{Directora i tutora: María Amparo Tórtola Baixauli}

\date{{\large IFIC (CSIC- Universitat de València) \\ Departament de Física Teòrica\\}\vspace{0.5cm} València, maig 2023}

%----------------------------------------------------------------------------------------

\frontmatter % Denotes the start of the pre-document content, uses roman numerals

\dedication{ \small \flushleft
As you set out for Ithaka \\
hope your road is a long one, \\
full of adventure, full of discovery.\\
Laistrygonians, Cyclops,\\
angry Poseidon—don’t be afraid of them:\\
you’ll never find things like that on your way\\
as long as you keep your thoughts raised high,\\
as long as a rare excitement\\
stirs your spirit and your body.\\
Laistrygonians, Cyclops,\\
wild Poseidon—you won’t encounter them\\
unless you bring them along inside your soul,\\
unless your soul sets them up in front of you.\\
\medskip
Hope your road is a long one.\\
May there be many summer mornings when,\\
with what pleasure, what joy,\\
you enter harbors you’re seeing for the first time;\\
may you stop at Phoenician trading stations\\
to buy fine things,\\
mother of pearl and coral, amber and ebony,\\
sensual perfume of every kind—\\
as many sensual perfumes as you can;\\
and may you visit many Egyptian cities\\
to learn and go on learning from their scholars.\\
\medskip
Keep Ithaka always in your mind.\\
Arriving there is what you’re destined for.\\
But don’t hurry the journey at all.\\
Better if it lasts for years,\\
so you’re old by the time you reach the island,\\
wealthy with all you’ve gained on the way,\\
not expecting Ithaka to make you rich.\\
\medskip
Ithaka gave you the marvelous journey.\\
Without her you wouldn't have set out.\\
She has nothing left to give you now.\\
\medskip
And if you find her poor, Ithaka won’t have fooled you.\\
Wise as you will have become, so full of experience,\\
you’ll have understood by then what these Ithakas mean.\\
\medskip
	\flushright -- C. P. Cavafy
}

%----------------------------------------------------------------------------------------
%	OUTPUT TITLE PAGE AND PREVIOUS
%----------------------------------------------------------------------------------------

% Note that \maketitle outputs the pages before here

\maketitle
\vspace*{5cm}
María Amparo Tórtola Baixauli, \\
Profesora Titular de la Universitat de València

CERTIFICA

Que la presente memoria \textit{"Neutrino properties from the laboratory and the cosmos"} ha sido realizada bajo su dirección en el Instituto de Física
Corpuscular, centro mixto de la Universitat de València y del Consejo Superior de Investigaciones Científicas, por Pablo Martínez Miravé y constituye su Tesis para optar al grado de Doctor en Física.

Y para que así conste, en cumplimiento de la legislación vigente, presenta en
el Departament de Física Teòrica de la Universitat de València la referida Tesis
Doctoral, y firma el presente certificado.\\

València, mayo 2023

\vspace*{0.8cm}
María Amparo Tórtola Baixauli

%----------------------------------------------------------------------------------------
%	PREFACE
%----------------------------------------------------------------------------------------
\chapter*{Acknowledgements}
\begingroup
\flushright \textit{Per curt que siga el camí \\ qui xafa fort deixa emprempta.}\\
\endgroup
\vspace{2cm}

\textit{First things first}, we say, and that is why this is the first part of the thesis that I started writing. It is also the most important to me and thus, it is also that last one that I have proof-read. The reason is that, amongst all, I feel deeply thankful for all these years, for the experiences, for what I learned about science and for my personal growing too. Thus, I wanted to dedicate a piece of this thesis to the people that were with me at some step of this long journey.

En primer lugar, quiero dar las gracias a mis padres, a mi hermana Teresa y a mi hermano Javier, por ser un pilar fundamental en mi vida y un apoyo constante, aunque eso de \textit{el olor de los neutrinos} os ha tenido un poco confusos todo este tiempo. Gracias también a ti, Clara, por acompañarme y compartir conmigo todos estos años. Me siento infinitamente afortunado de tener esta familia. \texttt{GRACIAS} por todo.

Tot el treball recollit en aquesta tesi i tot el meu aprenentatge no hauria sigut possible sense la supervisió i el cuidat de la meua directora de tesi. Gràcies, Mariam, per ensenyar-me amb paciència i per deixar-me espai per a creixer. Gràcies per totes les discussions, per traure temps per escoltar-me i també per ensenyar-me la importància de la dimensió humana en la ciència. Quiero darte también las gracias a ti, Gabriela, por creer en mí y porque, desde la carrera, siempre me has sabido contagiar la ilusión y motivación por la física. Por último, gracias, Sergio, por abrirme los ojos al mundo de la academia, por orientarme y por compartir tus conocimientos. Esta tesis no existiría sin vosotros tres.

I want to thank the people from the Astroparticle and High-Energy Physics group. You have been a scientific family during this time, the people with whom I discovered and discussed science. Pablo III, tú has estado ahí siempre en estos cuatro años y quiero darte las gracias por aguantarme y por no dejar que te estresase demasiado. A Pablo IV, Jorge y Mario agradeceros el tiempo compartido como compañeros de despacho y también a Rebeca, Omar, Antonio, Víctor y Gonzalo por todos los buenos ratos que he pasado con vosotros. I also want to thank you, Julio, Stefano, Ivania, and Valentina, for your friendship and for sharing your wisdom. I can not forget to mention you, amigo Dimitris, for your support when I was starting this journey. Thank you all. It was a great pleasure to share these years with you.

No puedo olvidarme de la gente que ha sido un apoyo para mi en este difícil último año de doctorado. Gracias, Álvaro, por esos coffee breaks al sol que tanta calma me daban y gracias, Juan, por mantenerme siempre animado. Gracias, Carlos, por hacerme sentir como en casa en mis días por la facultad. Gràcies, Neus, per compartir els dubtes i lespreocupacions del doctorat. Y grazie, Daniele, por tu apoyo constante y tu ayuda ante mi indecisión.

A vosotros, Jose y Maite, también quiero daros las gracias por seguir ahí después de tantos años. Es una suerte contaros entre mis amigos y que me hayáis acompañado en este GR. 

During these years, I have also had the chance to travel and get to know many fascinating people and made good friends. I want to thank you, Pedro, for welcoming at Fermilab and for all the valuable lessons you taught me. I also want to thank you, Evgeny, for you role as a mentor both during and after my time at the Max Planck Institute for Nucelar Physics. Likewise, I want to thank you, Manibrata, Tim, Ting, Nele, Hannes, Florian, Quim... for being so welcoming and nice to me during my stay there. I want to thank the people at Instituto Técnico Superior for making me feel at home and in particular, thank you, Filipe, for your friendship, your wisdom and the nice time we shared. Finally, I want to thank the Astroparticle community at the Niels Bohr Institute and particularly, Irene, for hosting me and encouraging me to keep on learning new things everyday.

Finally, I want to thank the memebers of the jury for the time and dedication to reading this manuscript. Besides that, I also acknowledge the financial support from the Ministerio de CIencia, Innovación y Universidades (MICIU) via the \textit{ayudas para la formación de profesorado universitario} with reference FPU18/04571.

\begingroup
\flushright \Large \texttt{THANK YOU. GRÀCIES. GRACIAS.}\\
\endgroup

%\index{preface}
\chapter*{Preface}
\addcontentsline{toc}{chapter}{Preface} % Add the preface to the table of contents as a chapter

Since its birth in 1930, the field of neutrino physics has been surrounded by more questions than answers. Hypothesised as an attempt to ensure energy conservation, it took two decades for neutrinos to be first detected. Later on, the observation of flavour oscillations manifested that the Standard Model of particle physics is an incomplete theory. At present, the study of neutrinos from different sources and at different energies using state-of-the-art technology is entering an era of precision. Nonetheless, the number of questions has kept on growing. \textit{What is the mechanism behind neutrino masses?} \textit{What other properties arise from this mass mechanism?} \textit{Do neutrinos interact with matter or among themselves in a way that differs from the Standard Model prediction?} \textit{Are neutrinos connected to the dark matter or the dark energy content of the universe?} Besides those theoretical questions, some experimental challenges exist too. Some of them are related to the need for a better understanding of neutrino fluxes, interaction cross sections, characterisation of backgrounds and the technical difficulties arising from the increasing size of neutrino detectors. All these factors manifest that the lessons yet to be learned from neutrinos in the near --- and not-so-near --- future are numberless.

This thesis narrates my journey along the blurred border between the knowns and unknowns in neutrino physics. It is structured in three main parts containing original research regarding neutrino properties.

The first part of the thesis starts with an introduction to cosmology, emphasising the role of neutrinos and dark matter in several epochs of the universe. It is followed by an introduction to neutrino mass models and some of their frequent phenomenological consequences. The following chapter is then devoted to the determination of neutrino masses and mixing, focusing on flavour oscillations in terrestrial experiments. 

Other properties such as magnetic moments, non-standard interactions and a non-unitary three-neutrino mixing are generally predicted in many extensions of the Standard Model accounting for neutrino masses. The second part of this thesis presents several studies of these properties from solar neutrinos and in medium-baseline reactor neutrino experiments. In addition, we discuss how neutrinos can test fundamental symmetries like CPT. We also analyse the impact of some of these neutrino properties in cosmological observables.

Motivated by the feebly interacting nature of neutrinos and dark matter, the third part of this manuscript explores possible connections between the two of them, concentrating on two specific candidates: primordial black holes and ultralight scalars. Finally, I provide a concluding analysis of the contents of the thesis and project my hopes and expectations for prospects in the field. A summary of the thesis --- i un \textit{Resum de la tesi} --- serve as the closing for this manuscript.

Special effort has been made to contextualise the research presented in each chapter. In particular, original material such as plots and summary tables are included in this manuscript, aiming to provide an introductory overview of the topics here discussed and to serve as a reference for their current status. When suitable, a critical discussion of the results is also provided.  

To me, this thesis is the logbook of the marvellous journey Ithaka has given me so far. It is with great pleasure that I share it with you all.
\chapter*{List of publications}
\addcontentsline{toc}{chapter}{List of publications} 
This doctoral thesis is based on the following scientific publications, ordered according to the content of the manuscript:\\

\begin{enumerate}[label=\textbf{\arabic*}.] 
\item \textbf{2020 global reassessment of the neutrino oscillation picture}\\
P. F. de Salas, D. V. Forero, S. Gariazzo, \textit{P. Martínez-Miravé}, O. Mena, C. A. Ternes, M. Tórtola, J. W. F. Valle\\
\href{https://doi.org/10.1007/JHEP02(2021)071}{JHEP 02 (2021) 071} • e-Print: \href{https://arxiv.org/abs/2006.11237}{2006.11237} [hep-ph] \\

\item \textbf{Solar $\mathbf{\overline{\nu}_e}$ flux: Revisiting bounds on neutrino magnetic moments and solar \mbox{magnetic} field}\\
E. Akhmedov, \textit{P. Martínez-Miravé}\\
\href{https://doi.org/10.1007/JHEP10(2022)144}{JHEP 10 (2022) 144} • e-Print: \href{https://arxiv.org/abs/2207.04516}{2207.04516} [hep-ph]\\

\item \textbf{Nonstandard interactions from the future neutrino solar sector}\\
\textit{P. Martínez-Miravé}, S. Molina Sedgwick, M. Tórtola\\
\href{https://doi.org/10.1103/PhysRevD.105.035004}{Phys.Rev.D 105 (2022) 3, 035004}• e-Print: \href{https://arxiv.org/abs/2111.03031}{2111.03031} [hep-ph]\\

\item \textbf{Neutrino CPT violation in the solar sector}\\
G. Barenboim, \textit{ P. Martínez-Miravé}, C. A. Ternes, M. Tórtola\\
\href{https://doi.org/10.1103/PhysRevD.108.035039}{Phys.Rev.D 108 (2023) 3, 035039}
e-Print: \href{https://arxiv.org/abs/2305.06384}{2305.06384} [hep-ph] \\

\item \textbf{Cosmological radiation density with non-standard neutrino-electron \mbox{interactions}} \\
P. F. de Salas, S. Gariazzo, \textit{P. Martínez-Miravé}, S. Pastor, M. Tórtola \\
\href{https://doi.org/10.1016/j.physletb.2021.136508}{Phys.Lett.B 820 (2021) 136508} • e-Print: \href{https://arxiv.org/abs/2105.08168}{2105.08168} [hep-ph] \\

\item \textbf{Non-unitary three-neutrino mixing in the early Universe}\\
S. Gariazzo, \textit{P. Martínez-Miravé}, O. Mena, S. Pastor, M. Tórtola\\
\href{https://doi.org/10.1088/1475-7516/2023/03/046}{JCAP 03 (2023) 046} • e-Print: \href{https://arxiv.org/abs/2211.10522}{2211.10522} [hep-ph]\\

\item \textbf{Signatures of primordial black hole dark matter at DUNE and THEIA} \\
V. De Romeri, \textit{P. Martínez-Miravé}, M. Tórtola \\
\href{https://doi.org/10.1088/1475-7516/2021/10/051}{JCAP 10 (2021) 051} • e-Print: \href{https://arxiv.org/abs/2106.05013}{2106.05013} [hep-ph] \\

\item \textbf{Signatures of ultralight dark matter in neutrino oscillation experiments}\\
A. Dev, P. A. N. Machado, \textit{P. Martínez-Miravé}\\
\href{https://doi.org/10.1007/JHEP01(2021)094}{JHEP 01 (2021) 094} • e-Print: \href{https://arxiv.org/abs/2007.03590}{2007.03590} [hep-ph] \\

\item \textbf{Cosmology-friendly time-varying neutrino masses through the sterile portal}\\
G.-Y. Huang, M. Lindner, \textit{P. Martínez-Miravé}, M. Sen \\
\href{https://doi.org/10.1103/PhysRevD.106.033004}{Phys.Rev.D 106 (2022) 3, 033004} • e-Print: \href{https://arxiv.org/abs/2205.08431}{2205.08431} [hep-ph]
\end{enumerate}

%%%%%%%%%%%%%%%%%%%%%%%%%%%%%%%%%%%%%%%%%
Other scientific publications and reviews not included in this doctoral thesis

\begin{enumerate}[label=\textbf{\arabic*}.]
\item \textbf{Sterile neutrinos with altered dispersion relations revisited} \\
G. Barenboim, \textit{P. Martínez-Miravé}, C. A. Ternes, M. Tórtola \\
\href{https://doi.org/10.1007/JHEP03(2020)070}{JHEP 03 (2020) 070} • e-Print: \href{https://arxiv.org/abs/1911.02329}{1911.02329} [hep-ph]\\

%\item \textbf{Cosmological radiation density with non-standard neutrino-electron \mbox{interactions}}\\
%\textit{P. Martínez-Miravé}\\
%J.Phys.Conf.Ser. 2156 (2021) 1, 012011 • TAUP2021 • e-Print: 2110.09988 [hep-ph]\\
%
%\item \textbf{Cosmological radiation density and neutrino NSI with electrons}\\
%\textbf{P. Martínez-Miravé}\\
%PoS CORFU2021 (2022) 332 • Contribution to: CORFU2021, 332\\
%
%\item  \textbf{Neutrino physics}\\
%\textbf{P. Martínez-Miravé}, Kristjan Müürsepp, M.Tórtola (lecturer) \\
% Lecture notes
%PoS CORFU2021 (2022) 321 • Contribution to: CORFU2021, 321\\

\item \textbf{Quantum gravity phenomenology at the dawn of the multi-messenger era - A review} • A. Addazi et al.\\
\href{https://doi.org/10.1016/j.ppnp.2022.103948}{Prog.Part.Nucl.Phys. 125 (2022) 103948} • e-Print: \href{https://arxiv.org/abs/2111.05659}{2111.05659} [hep-ph] \\

\item \textbf{Synergy between cosmological and laboratory searches in neutrino physics: a white paper} • K. Abazajian et al.  \\
Contribution to 2021 Snowmass Summer Study • e-Print: \href{https://arxiv.org/abs/2203.07377}{2203.07377} [hep-ph]\\
\end{enumerate}

%----------------------------------------------------------------------------------------
%	TABLE OF CONTENTS & LIST OF FIGURES/TABLES
%----------------------------------------------------------------------------------------

\begingroup % Local scope for the following commands

% Define the style for the TOC, LOF, and LOT
\setstretch{1} % Uncomment to modify line spacing in the ToC
\setlength{\textheight}{240\hscale} % Manually adjust the height of the ToC pages

% Turn on compatibility mode for the etoc package
\etocstandarddisplaystyle % "toc display" as if etoc was not loaded
\etocstandardlines % "toc lines" as if etoc was not loaded

\addtocontents{toc}{\protect\setcounter{tocdepth}{-1}}
\tableofcontents % Output the table of contents
\addtocontents{toc}{\protect\setcounter{tocdepth}{3}}

%\addcontentsline{toc}{chapter}{\listfigurename}
%\listoffigures % Output the list of figures

% Comment both of the following lines to have the LOF and the LOT on different pages
%\let\cleardoublepage\bigskip
%\let\clearpage\bigskip
%\addcontentsline{toc}{chapter}{\listtablename}
%\listoftables % Output the list of tables

\endgroup

%----------------------------------------------------------------------------------------
%	MAIN BODY
%----------------------------------------------------------------------------------------

\mainmatter % Denotes the start of the main document content, resets page numbering and uses arabic numbers
\setchapterstyle{kao} % Choose the default chapter heading style

%%%%%%%%%%%%%%%%%%%%%%%%%%%%

\pagelayout{wide} % No margins
\addpart{Fundamental cosmology and the physics of massive neutrinos}
\pagelayout{margin} % Restore margins
\raggedbottom
\chapter{Introductory cosmology with a taste of neutrinos and dark matter}
\labch{ch2-intro-cosmo}

Our current understanding of the universe and its evolution is captured on a theoretical model referred to as $\Lambda$CDM. It is based on three pillars: (i) the universe is expanding, (ii) it is homogeneous and isotropic on large scales, and (iii) there existed an initial inflationary era which can explain the initial conditions. 

This chapter serves as an introduction to the standard cosmological model and some concepts that are of relevance for the study of neutrino properties and dark matter --- a form of feebly-interacting or non-interacting matter. In Section~\ref{sec:ch1-metric}, we review the description of such an expanding universe. Section~\ref{sec:ch1-thermo} introduces the role the different species play in its evolution and Section~\ref{sec:ch1-story} provides a quick overview of the different stages of the evolution of the universe, emphasising the role of neutrinos in it. Finally, the pieces of evidence of the existence of dark matter are outlined in Section~\ref{sec:ch1-dark}, along with a review of some interesting dark matter candidates.

%%%%%%%%%%%%%%%%%%%%%%%%%%%%%%%%%%%%%%%%%%
%%%%%%%%%%%%%%%%%%%%%%%%%%%%%%%%%%%%%%%%%%
\section{Living in an FLRW universe}
\label{sec:ch1-metric}
%%%%%%%%%%%%%%%%%%%%%%%%%%%%%%%%%%%%%%%%%%
%%%%%%%%%%%%%%%%%%%%%%%%%%%%%%%%%%%%%%%%%%
The most general metric tensor describing a homogeneous and isotropic spacetime is the  Friedmann-Lemaître-Robertson-Walker metric,
\begin{equation}
\text{d}s^2 = g_{\mu\nu} \text{d}x^\mu \text{d}x^\nu = \text{d}t^2 - a^2(t)\left(\frac{\text{d}r^2}{1-kr^2} + r^2\text{d}\theta^2 + r^2\sin^2\theta \text{d}\phi^2 \right)\, ,
\end{equation}
where $t$ is the proper time, the set $(r, \, \theta, \, \phi)$ are comoving variables and $a(t)$ is the scale factor. Besides that, $k$ refers to the curvature of the 3-space, which can (i) be flat, $k=0$, (ii) have positive curvature, $k = +1$, or (iii) negative curvature $k = -1$.

We are interested in studying the role of photons, electrons, positrons, neutrinos and other species in the evolution of the Universe. To do so, we describe them via the energy-momentum tensor of a perfect fluid,
\begin{align}
T^{\mu}\,_{\nu} = (\rho + P)u^{\mu}u_{\nu} + P \delta^{\mu}_{\nu}\, ,
\end{align}
where $\rho$ is the energy density, $P$ is the pressure of the fluid, and we consider a fluid with zero velocity in the comoving frame so that $u^\mu = (1,\, 0, \, 0, \, 0)$. In General Relativity, conservation of the energy-stress tensor corresponds to requiring that its covariant derivative is zero, i.e.
\begin{align}
\nabla_\mu T^\mu \, _\nu = \partial_\mu T ^{\mu}\, _{\nu}+\Gamma^\sigma_{\sigma \mu} T ^{\mu}\, _{\nu} - \Gamma ^{\sigma}_{\mu \nu}T^{\mu}_{ \, \sigma} = 0\, ,
\end{align}
where we have introduced the Christoffel symbols
\begin{align}
\Gamma^\sigma_{\alpha\beta} = \frac{1}{2}g^{\sigma \rho}(\partial_{\alpha} g_{\beta \rho} + \partial_{\beta}g_{\alpha\rho} - \partial_{\rho}g_{\alpha\beta})\, .
\end{align}
The time component of this equation gives the evolution of the energy density,
\begin{align}
\dot{\rho} + 3 H (\rho + P) = 0\, ,
\label{eq:ch1-rhodot}
\end{align}
where a dot denotes the derivative with respect to proper time, $t$, and we have introduced the Hubble rate $H = \dot{a}/a$. Rewriting Equation \ref{eq:ch1-rhodot}  in terms of the equation of state of the fluid, $\omega$, defined as the ratio between energy density and pressure, i.e. $P = \omega \rho$, one finds that
\begin{align}
\dot{\rho} + 3 H (1 + \omega)\rho = 0\, ,
\label{eq:ch1-rhodot-2}
\end{align}
For a constant equation of state, one can solve Equation \ref{eq:ch1-rhodot-2} as
\begin{align}
\rho \propto a ^{-3(1 + \omega)} \, \longrightarrow \, \begin{cases}
      \rho \propto a^{-4} & \omega = 0 \quad \text{(non-relativistic matter)}\\
      \rho \propto a^{-3} & \omega = 1/3 \quad \text{(relativistic matter)}\\
      \rho = \text{const.} & \omega = -1 
    \end{cases}  
\end{align}
to find the evolution of relativistic and non-relativistic matter as the universe expands. Notice also that the energy density of a fluid with negative pressure, such that $P= -\rho$, remains constant.

Einstein's equations capture the interplay between the geometry of spacetime and its physical content, namely
\begin{align}
G_{\mu\nu} \equiv R_{\mu\nu} - \frac{1}{2}R g_{\mu \nu} = 8 \pi G T_{\mu\nu} + \Lambda g_{\mu\nu}\, .
\label{eq:ch1-einstein}
\end{align}
In this expression, we have separated the contributions of the Ricci tensor, $R_{\mu\nu}$ and the Ricci scalar, $R$, to Einstein's tensor, $G_{\mu\nu}$.
The Ricci tensor is defined as 
\begin{align}
R_{\mu\nu} = \partial_{\sigma}\Gamma^{\sigma}_{\mu\nu} -\partial_\nu \Gamma^\sigma_{\mu\sigma} + \Gamma^{\sigma}_{\sigma \rho}\Gamma ^{\rho}_{\mu\nu} - \Gamma ^{\rho}_{\mu\sigma}\Gamma^{\sigma}_{\nu\rho }\, ,
\end{align}
as a function of the Ricci scalar is $R = g^{\alpha\beta}R_{\alpha\beta}$.

In natural units, the gravitational constant $G$ defines an energy scale, which we will refer to as the Planck scale --- or Planck mass --- namely \mbox{$m_{Pl}  = G^{-1/2} = 1.22 \times 10^{19}$ GeV.} In Equation \ref{eq:ch1-einstein}, we have also introduced the energy-momentum tensor of the content of the universe, $T_{\mu\nu}$, and the contribution from Einstein's cosmological constant, $\Lambda$.

From Einstein's equations, one can recover the Friedmann equations,
\begin{align}
H^2 = \left(\frac{\dot{a}}{a}\right) ^2 = \frac{8\pi G}{3}\rho - \frac{k}{a^2}
\label{eq:ch1-Friedmann-1}
\end{align}
and
\begin{align}
\frac{\ddot{a}}{a} = - \frac{4 \pi G}{3} (\rho + 3 P)\, ,
\end{align}
which again relate the content of the universe and its expansion rate. 

Let us define the critical density at present as the current energy density under the assumption that the universe is flat, $k=0$,
\begin{align}
\rho_{\text{crit.},0} = \frac{3H^2_0}{8\pi G}\, ,
\end{align}
where $H_0$ is the value of the Hubble rate at present. Then, one can define the dimensionless energy density of each of the species today as
\begin{align}
\Omega_{i,0} \equiv \frac{\rho_i}{\rho_{\text{crit.},0}}
\end{align}
and rewrite Friedmann equation \ref{eq:ch1-Friedmann-1} as
\begin{align}
H^2(a) = H^2_0\sqrt{\Omega_{r,0}\left(\frac{a_0}{a}\right)^4 + \Omega_{m,0}\left(\frac{a_0}{a}\right)^3 + \Omega_{k,0}\left(\frac{a_0}{a}\right)^2+ \Omega_{\Lambda,0}}\, .
\end{align}
In these expressions, the subscript 0 indicates the current value of the dimensionless energy densities of radiation --- $\Omega_{r,0}$ --- and matter --- $\Omega_{m,0}$, which have equation of state $\omega = 1/3$ and $\omega = 0$ respectively. We have also recast the contribution of curvature and Einstein's cosmological constants into $\Omega_{k,0}$ and $\Omega_{\Lambda,0}$ respectively. In particular, for curvature, the dimensionless energy density at present reads
\begin{align}
\Omega_{k,0} = - \frac{k}{H^2_0}\, .
\end{align}

For a universe dominated by only one of the species, the scale factor evolves according to
\begin{align}
a(t) \propto \begin{cases} t ^{\frac{2}{3(1+\omega_i)}} & \omega_i \neq -1
\\ e^{\widetilde{H}t} & \omega_i = -1  \end{cases} \, .
\end{align}
Here, $\widetilde{H}$ is the value of the Hubble rate at the time the cosmological constant dominates the expansion of the universe and all the other contributions are negligible.

According to which is the constituent that contributes the most to the energy density budget of the universe, three different eras can be identified: \mbox{(i) radiation} domination, (ii) matter domination and (iii) dark energy era. This is illustrated in \reffig{fig:ch1-universe}. The expansion in the early universe was dominated by radiation, i.e. by the contribution of neutrinos and photons. Then, an epoch of matter domination happened, where cold dark matter and baryons where the main species contribution to the evolution of the universe. Since recent times --- given the timescales of the evolution of the universe --- the expansion of the universe has accelerated, as dictated by the cosmological constant, $\Lambda$.
\begin{figure}
\includegraphics[width = 0.95\textwidth]{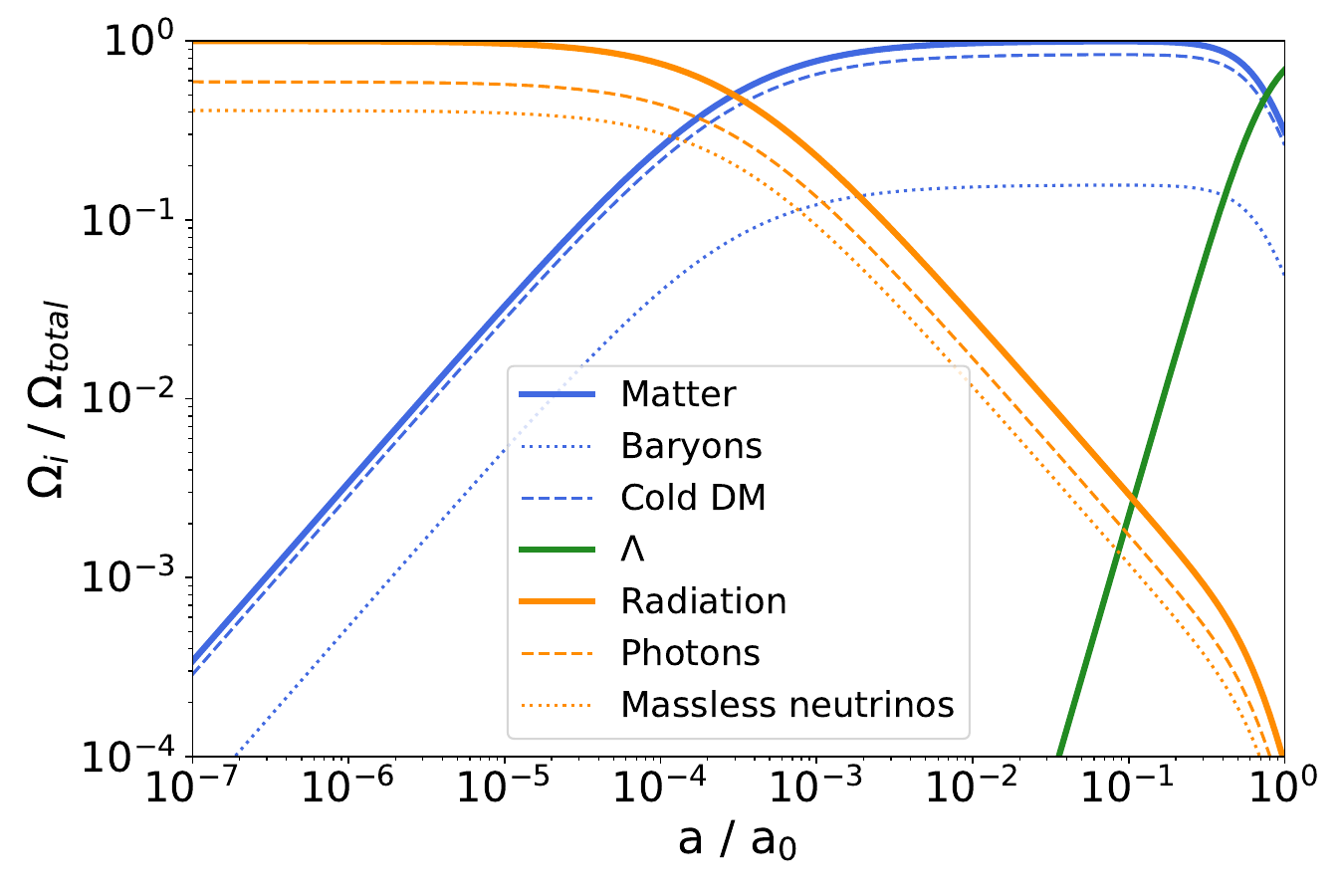}
\caption{Evolution of the fraction of the energy density in the form of matter ---cold dark matter and baryonic matter --- radiation  --- photons and massless neutrinos --- and a cosmological constant, $\Lambda$, as a function of the scale factor, $a$, normalised to its value at the present epoch, $a_0$. The evolution was computed employing \texttt{CLASS} \cite{Blas:2011rf}. \labfig{fig:ch1-universe}}
\end{figure}

So far, we have only discussed the evolution of the universe and its content at the background level and considering that it is homogeneous and isotropic. For a review on how to account for small departures from these hypotheses using perturbation theory, see \cite{Ma:1995ey,Dodelson:2003ft,Baumann:2022mni}.

%%%%%%%%%%%%%%%%%%%%%%%%%%%%%%%%%%%%%%%%%%
%%%%%%%%%%%%%%%%%%%%%%%%%%%%%%%%%%%%%%%%%%
\section{Thermodynamics in the expanding universe} 
\label{sec:ch1-thermo}
%%%%%%%%%%%%%%%%%%%%%%%%%%%%%%%%%%%%%%%%%%
%%%%%%%%%%%%%%%%%%%%%%%%%%%%%%%%%%%%%%%%%%
The energy density of the different constituents of the universe, as well as other thermodynamical quantities, depends on the distribution function of the different species, $f(t, \textbf{x}, \textbf{p})$, where a bold font indicates that the quantities are 3-vectors. Assuming an isotropic and homogeneous universe, as follows from the \textit{cosmological principle}, distribution functions also inherit these characteristics and read
\begin{align}
f (p) = \frac{1}{e^{\frac{E(p)-\mu}{T}}\pm 1}\, .
\end{align}
where $p = |\textbf{p}|$. Here, the plus sign (+) corresponds to fermions --- i.e. Fermi-Dirac distribution --- and the minus sign (-) is for bosons --- i.e. Bose-Einstein distribution.
In this expression, $E(p) = \sqrt{{p}^2 + m^2}$ is the energy of a particle of mass $m$, $\mu$ is the chemical potential and $T$ is the temperature. We have also assumed thermal equilibrium all across the universe so that distribution functions do not show any explicit time dependence when expressed in terms of the right variables. 

The number density, energy density, pressure and entropy of each constituent are
\begin{align}
n_i &= \frac{g_i}{(2\pi)^3}\int f(p) 4\pi p^2 \text{d}p\, ,\\
\rho_i &= \frac{g_i}{(2 \pi)^3} \int f(p) E(p) 4\pi p^2\text{d}p \, , \\
P_i &= \frac{g_i}{(2\pi)^3} \int f(p)\frac{p^2}{3E(p)}4\pi p^2\text{d}p\, , \\
s_i &= \frac{\rho_i + P_i}{T}\, ,
\end{align}
respectively, where $g_i$ accounts for the degrees of freedom of the species $i$. Table \ref{table:ch1-thermo} contains a summary of the value of each thermodynamical quantity for ultrarelativistic bosons and fermions ($p \gg m$) and for non-relativistic ones. Notice that in the case of non-relativistic species, no difference between fermions and bosons is found. This is a consequence of the fact that in the limit $T \ll m$ both Fermi-Dirac and Bose-Einstein distributions recover the Maxwell-Boltzmann distribution.

\begin{table*}
\renewcommand{\arraystretch}{1.8}
\centering
\begin{tabular}{lcccc}
\toprule[0.25ex]
  & $n$ & $\rho$ & $P$  & $s$  \\ 
\midrule
Ultrarelativistic boson & $g_i\frac{\zeta{3}}{\pi^2} T^3$ & $g_i \frac{\pi^2}{30}T^4$ & $\frac{1}{3}\rho$ & $\frac{4\rho}{3 T}$  \\[2.4mm]
Ultrarelativistic fermion & $\frac{3}{4}g_i\frac{\zeta{3}}{\pi^2} T^3$ & $\frac{7}{8} g_i \frac{\pi^2}{30}T^4$ & $\frac{1}{3}\rho$ &  $\frac{4\rho}{3 T}$\\[2.4mm]
Non-relativistic species & $g_i\left(\frac{mT}{2\pi}\right)^{3/2} e^{-m/T}$ & $m n$ & $n T \ll \rho$ & $\frac{\rho}{T}$ \\
\bottomrule[0.25ex]
\end{tabular}
\caption{%
Thermodynamical quantities for ultrarelativistic and \mbox{non-relativistic} bosons and fermions with mass $m$ and $g_i$ degrees of freedom and where $\zeta(x)$ denotes the Riemann zeta function. }
\label{table:ch1-thermo}
\end{table*}

During radiation domination, the main contribution to the evolution of the universe comes from the energy density of ultrarelativistic species, which can be parametrised as
\begin{align}
\rho_r = g_* (T) \frac{\pi^2}{30}T^4\, ,
\end{align}
where $g_*(T)$ is the effective number of degrees of freedom,
\begin{align}
g_*(T) = \sum_{i, \, \text{bosons}}g_i\left(\frac{T_i}{T}\right) ^4 + \frac{7}{8}g_j\sum_{j, \, \text{fermions}}\left(\frac{T_j}{T}\right)^4\, .
\label{eq:ch1-edof}
\end{align}
Notice that different species are not necessarily described by the same temperature if they are not in equilibrium one with the other. A similar definition can be made for the radiation entropy,
\begin{align}
s_r = g_s(T)\frac{4\pi^2}{90}T^3 \, ,
\end{align}
where the entropy effective relativistic degrees of freedom $g_s (T)$ are given by
\begin{align}
g_s (T) = \sum_{i, \, \text{bosons}}g_i\left(\frac{T_i}{T}\right) ^3 + \frac{7}{8}\sum_{j, \, \text{fermions}}g_j\left(\frac{T_j}{T}\right)^3\, .
\end{align}
Notice that if $T=T_i$ for all species $i$, then the effective number of degrees of freedom and the entropy effective relativistic degrees of freedom coincide. These concepts will prove useful when studying the process of neutrino decoupling in \refch{ch7-nsicosmo} and \refch{ch8-nucosmo}. 

%%%%%%%%%%%%%%%%%%%%%%%%%%%%%%%%%%%%%%%%%%
%%%%%%%%%%%%%%%%%%%%%%%%%%%%%%%%%%%%%%%%%%
\section{A brief story of the universe}
\label{sec:ch1-story}
%%%%%%%%%%%%%%%%%%%%%%%%%%%%%%%%%%%%%%%%%%
%%%%%%%%%%%%%%%%%%%%%%%%%%%%%%%%%%%%%%%%%%

%%%%%%%%%%%%%%%%%%%%%%%%%%%%%%%%%%
\subsection{Inflation}
%%%%%%%%%%%%%%%%%%%%%%%%%%%%%%%%%%
There are questions that are not properly addressed by the standard cosmological model as we have introduced it so far, mainly why is the universe so isotropic, homogeneous and flat as suggested by observations. These are the so-called \textit{horizon} and \textit{flatness problems}.

A process of accelerated expansion in the very early stages of the universe can elegantly address these issues. This process is known as inflation \cite{Linde:1990flp}. It is generally assumed that this inflationary process was triggered by a scalar field, the \textit{inflaton}. If coupled to Standard Model particles, it could have decayed into them, populating the different species in a process called \textit{reheating}~\cite{Dolgov:1982th,Abbott:1982hn}.

%%%%%%%%%%%%%%%%%%%%%%%%%%%%%%%%%%
\subsection{Neutrino decoupling and electron-positron annihilation}
%%%%%%%%%%%%%%%%%%%%%%%%%%%%%%%%%%
Some time after inflation and reheating, the baryon asymmetry of the universe originated and the electroweak and QCD phase transitions took place~\cite{Kolb:1990vq}. During this radiation-dominated era, neutrinos also decoupled from the rest of the cosmic plasma at temperatures around $\sim$ 1 MeV. After that, they started free streaming. Durings these epochs and due to their relativistic behaviour, neutrinos contributed to the energy budget of the universe as radiation. Nonetheless, as the universe expanded and cooled down, two neutrino families --- at least --- became non-relativistic. In the present era, those relic neutrinos form the cosmic neutrino background (C$\nu$B).

After neutrino decoupling, when the temperature of the plasma dropped below the mass of the electron, $T\sim m_e$, electron-positron production from a pair of photons stopped being possible and only electron-positron annihilations could take place. This resulted in an entropy injection and the corresponding heating of the photon bath. 

%%%%%%%%%%%%%%%%%%%%%%%%%%%%%%%%%%
\subsection{Big Bang Nucleosynthesis}
%%%%%%%%%%%%%%%%%%%%%%%%%%%%%%%%%%
Likewise, the synthesis of primordial elements also happened during radiation domination. At some point, when $T\sim 0.8$ MeV the processes keeping neutrons and protons in chemical equilibrium,
\begin{align}
n\, \longleftrightarrow \, p \, + e^-\,+ \, \bar{\nu}_e \nonumber \\
\nu_e\, + n \, \longleftrightarrow \, p \, + \, e^- \nonumber \\
e^+ \, + \, n \longleftrightarrow\, p \,+ \, \bar{\nu}_e \, ,
\end{align}
stopped being efficient and the neutron-to-proton ratio froze --- except for a small decrease due to neutron decay. Then, neutrons were bounded off in light nuclei, including deuterium. Elements heavier than $^4$He and $^7$Li were produced in very small quantities and quickly decayed. Once primordial nucleosynthesis of light elements had occurred, the universe was left mainly with hydrogen and $^4$He, with a mass fraction \cite{Iocco:2008va}
\begin{align}
Y_p \sim \frac{4n_{^4\text{He}}}{n_H + 4n_{^4\text{He}}} \sim 0.25\, .
\end{align}
The yields of other isotopes such as $^2$H, $^3$He and $^7$Li are known to be much smaller.

Since primordial nucleosynthesis happened right after neutrino decoupling, modifications in the process of decoupling can alter Big Bang Nucleosynthesis (BBN) predictions. For instance, depending on how neutrinos contribute to the radiation density of the universe and, therefore, to its expansion, the abundance of light elements predicted is altered. Additionally, differences in the momentum distributions can also modify the neutron-to-proton ratio. Hence, BBN can also provide interesting limits on neutrino physics~\cite{Iocco:2008va,Lesgourgues:2013sjj,Froustey:2019owm}.

%%%%%%%%%%%%%%%%%%%%%%%%%%%%%%%%%%
\subsection{Cosmic microwave background}
%%%%%%%%%%%%%%%%%%%%%%%%%%%%%%%%%%
As the universe kept on expanding and cooling down, the contribution of matter and radiation energy densities equalled at the so-called \textit{matter-radiation equality}. After that, the universe entered an epoch of matter domination. When temperatures reached $\sim$ 0.3 eV, free electrons got bind to ionised nuclei. This process is known as \textit{recombination}. From that moment, photons did not undergo  Compton scattering any more and they started free-streaming. Those emitted photons followed an almost perfect black body isotropic and homogeneous distribution and formed the cosmic microwave background (CMB).

The temperature of the CMB, its anisotropies and polarisation have been measured with accuracy and encode very relevant information on the story of the universe. For instance, two-point correlations for the anisotropies in temperature maps are studied in terms of Legendre polynomials $P_l$,
\begin{align}
4\pi \biggl< \frac{\delta T (\hat{n})}{\overline{T} } \frac{\delta T (\hat{n}')}{\overline{T} }\biggr> = \sum _{l = 0}(2l +1)\, \mathcal{C}^{TT}_l \, P_l(\hat{n}\cdot \hat{n}') \, ,
\end{align} 
where $l$ are the multipoles in the expansion, $\mathcal{C}^{TT}_{l}$ are the corresponding coefficients, $\overline{T}$ is the average temperature of the CMB and $\delta T(\hat{n})$ denotes the difference between the temperature in a given direction $\hat{n}$ and the average value $\overline{T}$. Then, the expression in brakets denotes the average product of two temperature contrasts measured in a fixed relative orientation on the sky. In the last decade, data from the Planck satellite on CMB temperature and polarisation anisotropies~\cite{Planck:2013pxb,Planck:2015fie,Planck:2018nkj} has incredibly boosted our understanding of the universe and allowed the determination of the parameters in $\Lambda$CDM cosmological model with unprecedented accuracy.

%%%%%%%%%%%%%%%%%%%%%%%%%%%%%%%%%%
\subsection{Structure formation}
%%%%%%%%%%%%%%%%%%%%%%%%%%%%%%%%%%
During matter domination, the growth of energy density perturbations in the universe led to the formation of large-scale structures. This process can be studied from surveys measuring objects at different redshifts, like Euclid~\cite{Euclid:2019clj}. Additional information can be derived from the study of baryon acoustic oscillations (BAO), resulting from the interplay between the electromagnetic pressure and the attraction towards gravitational wells originating from overdensities in the photon-baryon plasma. Besides that, the study of small-scale structures is known to provide information on the nature of dark matter. For instance, a notably large component of \textit{hot} or \textit{warm} dark matter --- i.e. relativistic or close to relativistic dark matter --- would have prevented the formation of small-scale structures. From the non-observation of such erasing of structures, one can infer some relevant properties of dark matter and neutrinos, as will be discussed in several Sections. 

\section{Dark matter}
\label{sec:ch1-dark}
%%%%%%%%%%%%%%%%%%%%%%%%%%%%%%%%%%%%%%%%%%
%%%%%%%%%%%%%%%%%%%%%%%%%%%%%%%%%%%%%%%%%%
\subsection{Evidence}
At present, there is solid evidence that our universe contains a form of matter that does not interact with light or other particles --- or that does it very feebly --- and consequently, it is referred to as \textit{dark matter}. Some of the first hints of its existence go back to the 1930s and 1970s, when Fritz Zwicky and Vera C. Rubin found a discrepancy between observations of the Coma cluster and the rotational curve of galaxies and the prediction from Newtonian dynamics \cite{Zwicky:1933gu,Rubin:1978kmz}. For instance, in the case of Vera C. Rubin, she found that the stars and celestial bodies in the outer regions of galaxies were orbiting with velocities that were too large for those bodies to be gravitationally bounded. The prediction from Newtonian mechanics is clear: given the distribution of mass observed in a galaxy, the rotational velocity of objects in the outer regions of the galaxy must decrease with the distance. However, the rotation curves as determined in \cite{Rubin:1978kmz} showed a flattening --- i.e. rotational velocities remained constant with distance in the outskirts of the studied galaxies. Assuming that galaxies are surrounded by a dark matter halo, extending beyond the region where visible --- baryonic --- matter sits, would explain these observations.

Another clear piece of evidence supporting the hypothesis of dark matter comes from gravitational lensing. It is known that massive objects bend light as it travels from the source to the observer, leading to distortions in the image of the source. The more massive the lensing object is, the stronger the effect is. Hence, gravitational lensing can be used to map the mass distribution of the object acting as a lens. The Bullet Cluster is a very illustrative example of how gravitational lensing can point towards the existence of dark matter \cite{Clowe:2006eq}. The system of interest consists of two clusters colliding, being the smallest one of the two the one called the Bullet Cluster. The centre of mass and mass distribution of the system was inferred using gravitational lensing and found not to match with the centre of mass and mass distribution determined via electromagnetic radiation. Nonetheless, the existence of a dark matter halo component in both clusters reconciles both observations.

A third piece of evidence comes from the cosmic microwave background itself. Baryonic matter and dark matter do not evolve in the same way, and neither do their background perturbations. The reason is that baryonic matter gets ionised and interacts with radiation. Conversely, dark matter does not. Yet it affects the CMB through its contribution to gravitational potentials and lensing. 

From the comparison between observations and simulations of the structures formed in the universe, it is known that not only more matter beyond baryonic matter is present in the universe, but also that this dark matter component has to be --- mainly --- \textit{cold}, i.e. non-relativistic.

Alternative hypotheses to dark matter exist. Modified Newtonian Dynamics (MOND) \cite{Milgrom:1983ca} and extensions which correspond to modifications of General Relativity are amongst the most well-known. However, so far none of them has proven capable of explaining all the different mismatches in observations that support the dark matter hypothesis. The reason is that evidence comes from so many independent approaches that, even if these theories can explain individual phenomena, it is very difficult to address all of the evidence simultaneously without invoking the existence of dark matter.

\subsection{A plethora of dark matter candidates}
The landscape of dark matter candidates is very broad and spans more than 40 orders of magnitude in the mass of the candidates. With the hope that its existence can be proved through methods beyond the study of their gravitational impact, possible portals to the Standard Model are being explored both theoretically and experimentally. As a result, there is a vast worldwide program of dark matter searches addressing a gigantic number of experimental signatures in a wide parameter space. 

\begin{figure*}
\includegraphics[width = 0.72\paperwidth]{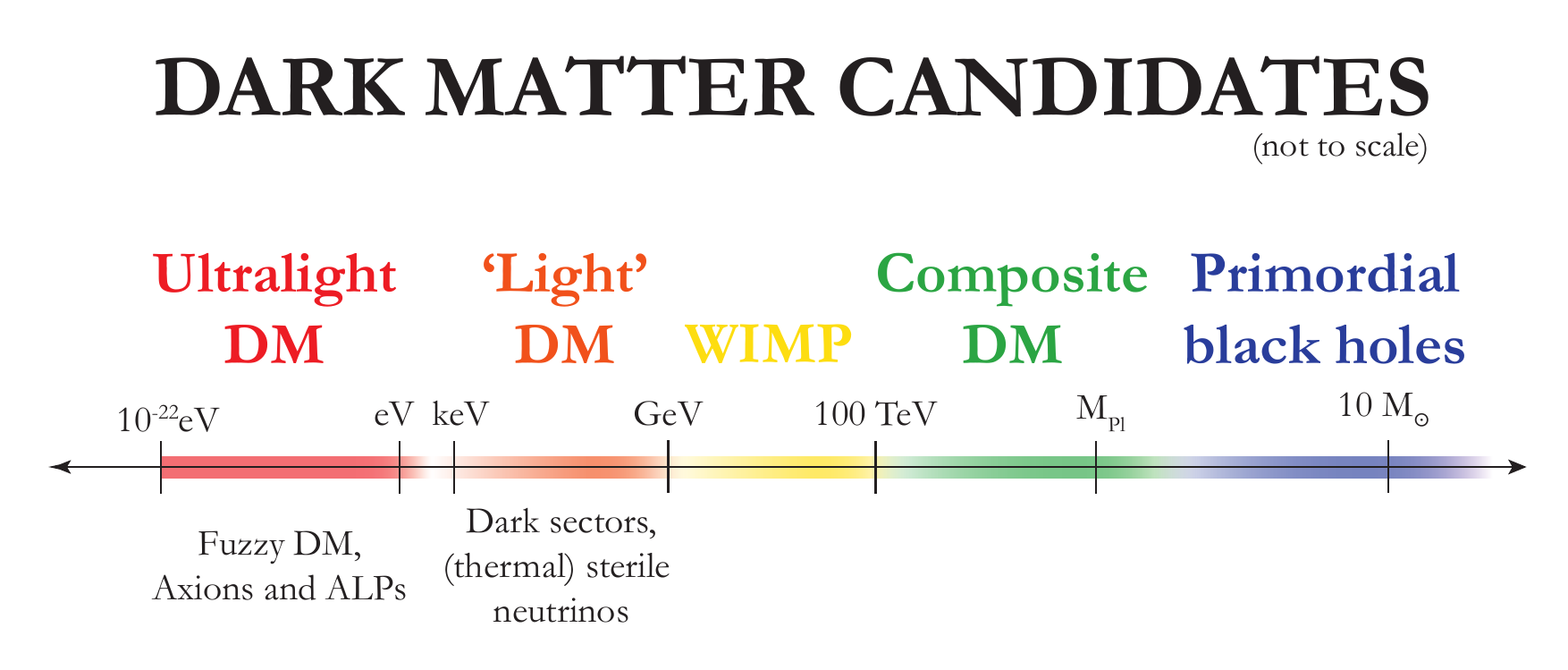}
\caption{Graphical representation -not to scale- of the different dark matter candidates according to their typical mass. Adapted from \cite{Lin:2019uvt,Ferreira:2020fam}. \labfig{fig:ch1-dm} }
\end{figure*}

Among the many dark matter candidates, one finds, for instance:

\begin{itemize}
\item \textbf{Ultralight dark matter fields.} Ultralight dark matter fields have gained recent attention due to their capability to solve some small-scale discrepancies between observations in simulations. These are discussed in more detail in~\refch{ch10-uldm}. For a fermionic candidate, there exists a stringent lower limit in the mass \cite{Tremaine:1979we} whereas the same limit for bosonic candidates is much less stringent. This opened the possibility of studying ultralight scalars~\cite{Hui:2016ltb},~vectors\cite{Fabbrichesi:2020wbt}, and spin-2 particles~\cite{Marzola:2017lbt} as constituents of the dark matter of the universe.

\item \textbf{Axions and ALPs.} Axions were proposed as a dynamical solution to the anomalous strong CP problem \cite{Peccei:1977hh,Weinberg:1977ma,Wilczek:1977pj,Crewther:1979pi}. The existence of other pseudo-scalar bosons with low masses, often referred to as \mbox{Axion-like} particles (ALPs) \cite{Marsh:2017hbv}, is a common prediction in many different contexts like string theory. A variety of experiments are searching for signs of their existence, mainly exploiting the axion-photon coupling, for example in haloscopes, \textit{light-shinning-through-a-wall} experiments and from stellar cooling \cite{Galanti:2022ijh}. 

\item \textbf{Sterile neutrinos.} Standard Model neutrinos, due to their small masses, do not act as \textit{cold} dark matter. However, heavier neutral leptons or sterile neutrinos could play that role if their masses are in the keV range \cite{Boyarsky:2018tvu}. Despite being unstable, their lifetime could be longer than the age of the universe. Moreover, they can be produced in the early universe via the Dodelson-Widrow mechanism \cite{Dodelson:1993je} although extensions and alternatives exist. The simplest scenarios in which keV sterile neutrinos compose the totality of the dark matter in the universe are currently tightly constrained, yet not completely ruled out.

\item \textbf{WIMPs and FIMPs.} Weakly interacting massive particles (WIMPs) were considered, for a long time, ideal dark matter candidates. The reason behind this belief is the fact that the right dark matter abundance of the universe could be achieved through the thermal production ---  via freeze-out --- of a weakly interacting particle with a mass near the electroweak scale, i.e. $\sim 100$ GeV \cite{Feng:2022rxt}. At present, the WIMP paradigm faces constraints from collider, direct and indirect searches. Feebly interacting massive particles \cite{Steigman:1984ac,Lanfranchi:2020crw} is the term used for particle candidates that extend the WIMP paradigm. In general, they are predicted in high-energy theories and interact feebly with Standard Model content.

\item \textbf{Primordial black holes}. Dark matter consisting of primordial black holes (PBHs) is an exciting possibility due to the large number of phenomenological consequences they would have. Their formation can result from the collapse of large primordial fluctuations in the early universe, the collapse of topological defects or phase transitions. There exist constraints on their mass and the fraction of dark matter they could compose from the study of microlensing, the non-observation of evaporation products, and from CMB, among others \cite{Green:2020jor}. 

\end{itemize}

\reffig{fig:ch1-dm} provides a graphical overview of the typical mass scale of some dark matter candidates. Some benchmark masses are given for reference although one should bear in mind that none of these are strict since the allowed mass spectrum of the different candidates is strongly model-dependent.

In our quest for the nature of dark matter, little is known beyond what is observed from its gravitational interactions. The third part of this thesis projects our hope to shed some light on this subject based on two different ways in which neutrinos could be connected to dark matter.

\chapter{Neutrino masses and phenomenological implications}
%\addcontentsline{toc}{chapter}{}
\labch{ch2-masses}
In 2015, Takaaki Kajita and Arthur B. McDonald were awarded the Nobel Prize in Physics \textit{for the discovery of neutrino oscillations, which shows that neutrinos have mass}. This discovery proves that there is physics beyond the Standard Model and possibly some new physics connected to the origin of neutrino masses too. 

The goal of this Chapter is to introduce the phenomenology that is explored in this thesis and contextualise it regarding its relation with neutrino mass models. Firstly, Section \ref{sec:ch2-mass-models} gives a quick overview of some well-known mechanisms and approaches to explain neutrino masses. Next, Section \ref{sec:ch2-intro-osc} describes the phenomenon of neutrino oscillations. Section \ref{sec:ch2-intro-mass-scale} provides an introduction to different probes of the absolute neutrino mass scale and finally, we overview other phenomenological implications that can result from different neutrino mass models in Section \ref{sec:ch2-intro-pheno}

\section{The zoo of neutrino mass models}
\label{sec:ch2-mass-models}

Once the phenomenon of flavour oscillation was observed --- implying that at least two of the neutrino masses were non-vanishing --- and established as the answer to the solar puzzle, the question that arose was: \textit{what is the origin of neutrino masses?} As an attempt to answer this question, theoretical particle physicists have proposed a plethora of neutrino mass models. In this section, we provide a lightning review of some benchmark --- minimal --- extensions of the Standard Model (SM) accounting for non-zero neutrino masses. The aim is to highlight some of the common phenomenological consequences of these mass models.

Using left-handed (LH) and right-handed (RH) neutrino ($\nu$) fields, and without taking into account the Standard Model symmetries, one can write three types of mass terms: a LH Majorana mass term
\begin{align}
\mathcal{L}^L_M = -\frac{1}{2}\sum_{\alpha, \beta = e, \mu,\tau} \overline{\nu^C_{\alpha L}}\left(\mathbf{M}_L\right)_{\alpha \beta }\nu_{\beta L} + \text{h.c.}\, ,
\end{align}
a RH Majorana mass term
\begin{align}
\mathcal{L}^R_M = -\frac{1}{2}\sum_{i,j = 1,...,n_R } \overline{\nu^C_{i R}}\left(\mathbf{M}_R\right)_{i,j}\nu_{j R} + \text{h.c.}\, ,
\end{align}
and a Dirac mass term
\begin{align}
\mathcal{L}_D = - \sum_{\alpha, i} \overline{\nu_{iR}}\left(\mathbf{M}_D\right)_{i\alpha}\nu_{\alpha L}+ \text{h.c.} \, ,
\end{align}
where the subscripts $L$ and $R$ denote the left-handed and right-handed chiral projects of the fields. The most general Lagrangian including both Majorana and Dirac mass terms leads to Majorana neutrinos.

If one were to write an effective field theory using Standard Model fields and respecting symmetries as
\begin{align}
\mathcal{L}_{\text{eff}} = \mathcal{L}_{\text{SM}} + \delta \mathcal{L}_{\text{D=5}} + \delta \mathcal{L}_{\text{D=6}} +... \, ,
\end{align}
then it is known that the lowest order effective operator leading to neutrinos masses and compatible with Standard Model symmetries is the so-called Weinberg operator~\cite{Weinberg:1979sa}, 
\begin{align}
\delta \mathcal{L}_{\text{D=5}} = \frac{1}{2} \sum_{\alpha, \beta}\kappa_{\alpha \beta}\left(\overline{l^C_{\alpha L}} \tilde{h}^* \right)\left(\tilde{h}^\dagger l_{\beta L}\right) + \text{h.c.}\, ,
\end{align}
where $l_{\alpha L}$ are the Standard Model lepton doublets, $h$ is the Higgs doublet, and $\tilde{h} \equiv i\sigma_2 h^*$. After electroweak symmetry breaking, this operator gives rise to an effective Majorana mass term,
\begin{align}
\mathcal{L}_{\text{Maj. mass}} = -\frac{1}{2}\sum_{\alpha, \beta}\overline{\nu^C _{\alpha L}}\left(\mathcal{M}\right)_{\alpha\beta}\nu_{\beta L} + \text{h.c} =  -\frac{1}{2}\sum_k m_k\overline{\nu^C _{k L}}\nu_{k L} + \text{h.c}\, ,
\end{align}
where $\left(\mathcal{M}\right)_{\alpha\beta} = - \kappa_{\alpha\beta}v^2/2 $, being $v/\sqrt{2}$ the Higgs vacuum expectation value (vev). Many neutrino mass models try to find which kind of physics beyond the Standard Model gives rise to the dimension-5 operator here presented, which is known to generate neutrino masses.

\subsection{Seesaw mechanisms}
There are three different tree-level realisations of the effective operator: the so-called seesaw mechanisms. The type-I seesaw mechanism~\cite{Minkowski:1977sc,Gell-Mann:1979vob,Yanagida:1979as,Schechter:1980gr,Magg:1980ut,Cheng:1980qt,Mohapatra:1980yp} considers the inclusion of a number $n_R$ of right-handed fermionic singlets, $\nu_R$. Then, the Lagrangian  reads
\begin{align}
\mathcal{L} = \mathcal{L}_{\text{SM}} + i \overline{\nu_{R}}\slashed{\partial}\nu_{R} - \left[ \mathbf{Y}^\dagger_\nu \overline{l_L} \, \tilde{h} \, \nu_{R} -\frac{1}{2}\mathbf{M}_R \overline{\nu^C_{R}} \nu_{R} + \text{h.c.}\,\right]\, ,
\end{align}
and then, for very large $M_R$ --- i.e. larger than the Dirac mass term --- neutrino masses are
\begin{align}
\mathcal{M}_\nu = \frac{v^2}{2}\mathbf{Y}^T_\nu\, \mathbf{M}^{-1}_R \ \mathbf{Y}_\nu \sim \frac{y_\nu v^2}{M}\, .
\end{align}
Here, $M$ denotes the approximate mass scale of the eigenvalues of $M_R$, which approximately corresponds to the mass scale of the right-handed fermionic singlets $\nu_R$, and $y_\nu$ denotes the order of magnitude of the Yukawa couplings in $\mathbf{Y}_\nu$. Notice that, since $\mathbf{M}_R$ is not protected by any symmetry, it can be very large. In that case, if  $M \gg v$, neutrino masses are naturally small.

The type-III seesaw model~\cite{Foot:1988aq} introduces a number $n_\Sigma$ of fermionic triplets $\Sigma_{Rj}$, which in a 2$\times$2 representation read
\begin{align}
\Sigma_{Rj} = \begin{pmatrix}
\Sigma^0_{Rj}/\sqrt{2} & \Sigma^+_{Rj} \\ \Sigma^-_{Rj} & -\Sigma^0_{Rj}/\sqrt{2} 
\end{pmatrix}\, ,
\end{align}
with $j = 1,\, ..., \, n_\Sigma$. Then, the Lagrangian reads
\begin{align}
\mathcal{L} = \mathcal{L}_{\text{SM}} &+\text{Tr}\left[\overline{\Sigma}_{R}\slashed{D}\Sigma_{R}\right] \nonumber \\ &+ \bigg\{ \mathbf{Y}_\Sigma \tilde{h}^\dagger \, \overline{\Sigma}_{R} \, l_{L} - \frac{1}{2}\text{Tr}\left[\overline{\mathbf{M}_\Sigma \Sigma}_{R}\Sigma^C_{R}\right] + \text{h.c.}\bigg\}
\end{align}
and it leads to neutrino masses,
\begin{align}
\mathcal{M}_\nu = - \frac{v^2}{2}\mathbf{Y}^T_\Sigma \mathbf{M}^{-1}_\Sigma \mathbf{Y}_\Sigma\,.
\end{align}
As in the type-I seesaw model, the smallness of neutrino masses in the type-III seesaw model is a result of the different scale between the Higgs vev and the Majorana masses of the heavy neutral fermionic triplets introduced.

A third realisation, the type-II seesaw model~\cite{Schechter:1980gr}, includes $n_\Delta$ scalar triplets with hypercharge Y = 2,
\begin{align}
\Delta_a = \begin{pmatrix}
\Delta^+_{a}/\sqrt{2} & -\Delta^{++}_{a} \\ \Delta^0_{a} & -\Delta ^+_{a}/\sqrt{2} 
\end{pmatrix}\, ,
\end{align}
where $ a = 1, ..., n_\Delta$. In this model, the Lagrangian is
\begin{align}
\mathcal{L} = \mathcal{L}_{\text{SM}} &+ \text{Tr}\left[\left(D_\mu \Delta_a\right)^\dagger \left(D^\mu \Delta_a\right)\right] \nonumber\\&- \left[ \mathbf{Y}_{\Delta_a} \overline{l^C_L} \,  \Delta_a \, l_{L} + \text{h.c.}\right] - V_\Delta\, .
\end{align}
Here, $V_\Delta$ denotes the scalar potential, which is not explicitly shown in this text. If the components $\Delta^0_a$ acquire a vev, $\langle \Delta^0_a \rangle = u_a e^{i\theta_a}$, the neutrino mass matrix is
\begin{align}
\mathbf{M}_\nu = 2 \sum_{a= 1}^{n_\Delta}\mathbf{Y_{\Delta_a}}u_a e^{i\theta_a}\, .
\end{align}
Actually, only one scalar triplet is needed to account for neutrino masses. In that case, the smallness of the masses depends on two scales: the mass of the scalar triplet and a parameter from the scalar potential, often denoted by $\mu$, which softly breaks lepton number,
\begin{align}
V \supset \mu \tilde{h}^T\Delta \tilde{h}\, .
\end{align}
This term can be naturally small in the t'Hooft sense~\cite{tHooft:1979rat} and then, since neutrino masses are proportional to $\mu/M^2_\Delta$, it allows to lower the mass scale of the scalar triplet. 

\begin{figure*}
\includegraphics[width = 0.68\paperwidth]{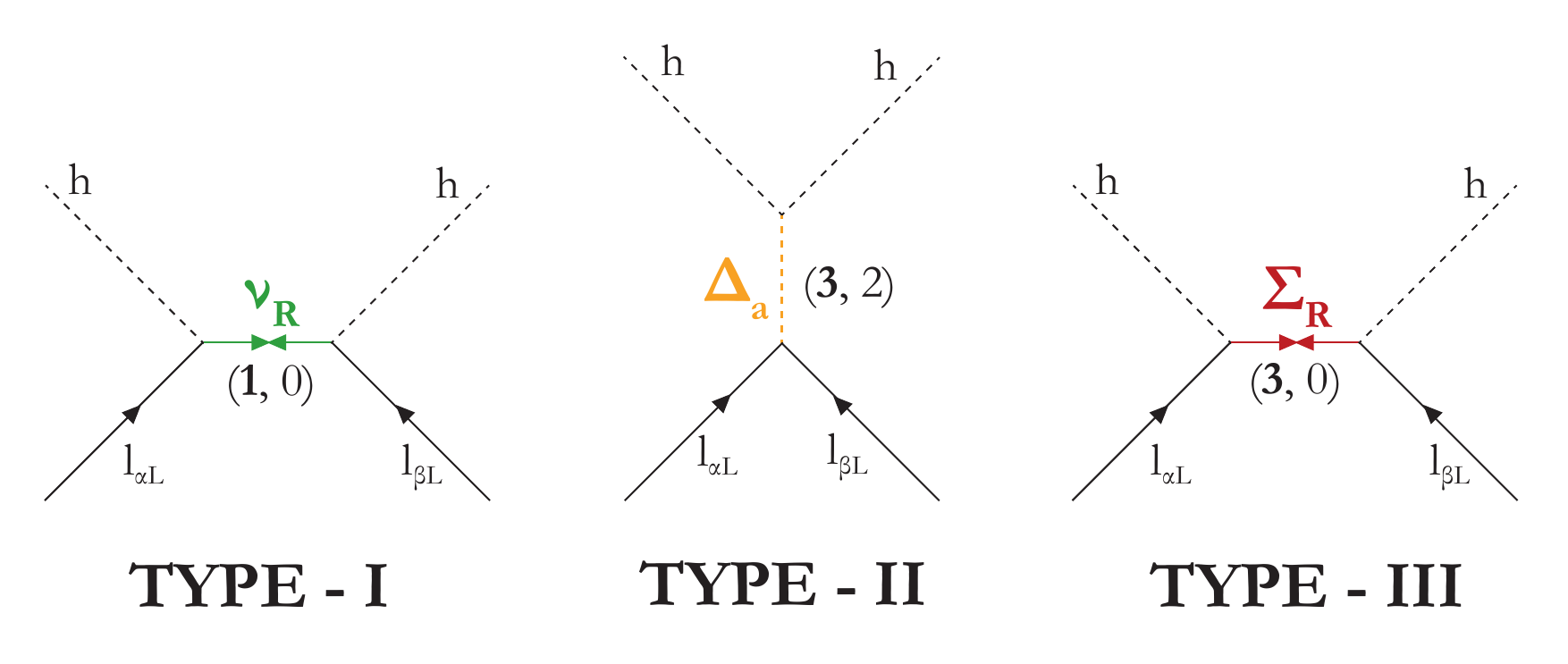}
\caption{Diagrams responsible for Majorana masses in seesaw mechanisms of type I, type II and type III. \labfig{fig:ch2-seesaws}}
\end{figure*}
These three seesaw mechanisms are summarised in \reffig{fig:ch2-seesaws}
There exist low-scale seesaw mechanisms too. For instance, the \mbox{linear~\cite{Akhmedov:1995ip,Malinsky:2005bi}} and inverse seesaw models~\cite{Mohapatra:1980yp,Wyler:1982dd,Mohapatra:1986bd,Gonzalez-Garcia:1988okv} extend the Standard Model with two sets of fermion singlets and explicit terms breaking lepton number. The success of these models relies on their capability to generate small neutrino masses without imputing tiny Dirac masses and with a low new physics scale, at the expense of introducing small lepton number violation scales.

If neutrinos were Dirac fermions, analogous mechanisms to generate neutrino masses can also be invoked. Such realisations can naturally arise in scenarios in which additional symmetries exist and prevent the existence of Majorana mass terms for the fermion singlets~\cite{Ma:2016mwh}.

\subsection{Radiative mass models}
Loop diagrams provide corrections to the tree-level neutrino mass models presented above. However, in radiative neutrino mass models, neutrino masses do not occur at tree level but at a certain loop order instead. Among radiative models, one finds for instance the Zee model~\cite{Zee:1980ai,Wolfenstein:1980sy} and the Zee-Babu model~\cite{Cheng:1980qt,Babu:2002uu}. The phenomenology of radiative models is very rich and includes non-standard interactions, charged-lepton flavour violating decays, instability issues in the scalar potential, and anomalous g-2 of charged fermions, among others. In addition, other radiative neutrino mass models can provide a viable and stable dark matter candidate. That is the case of the \textit{scotogenic} model~\cite{Ma:2006km,Escribano:2020iqq}. For a recent review on radiative neutrino mass models and the associated phenomenology, see~\cite{Cai:2017jrq}.

\subsection{Flavoured mass models}
There exists a large family of models that tries to address simultaneously the origin of neutrino masses and the flavour puzzle~\cite{Feruglio:2019ybq}. The latter has two main aspects: the existence of three replicated fermion families 
%\pmm{- or equivalently, the accidental $U(3)^4$ global symmetry of the Standard Model Lagrangian -}
and the large differences between the mass scales and mixing patterns in the quark and lepton sectors. 

Flavour models consider flavour symmetries which act linearly or non-linearly on the fields and commute with the Poincaré and gauge transformations. Different types of flavour groups $G_f$ can be considered: $G_f$ can be a Lie or a discrete group, it can be abelian or non-abelian, and the symmetry can be global or gauged. The general approach to follow is to assume that a high-scale theory obeys some flavour symmetry given by $G_f$ and then consider that this symmetry is explicitly or spontaneously broken. Depending on the breaking pattern, one can predict the mixing of quarks and leptons and extract relations between fermion masses. Among common choices for flavour symmetries --- see~\cite{Meloni:2017cig} for a recent review --- one finds discrete symmetries like $G_f = S_4$, which can generate tri-bi-maximal and bimaximal textures for the lepton mixing matrix, or the modular symmetry $G_f = A_4$, which can be motivated by the compactification of orbifolds~\cite{deAnda:2021jzc}.

\section{Neutrino oscillations}
\label{sec:ch2-intro-osc}
\subsection{Neutrino oscillations in vacuum}
Neutrinos are produced and detected as flavour eigenstates, which we will denote by $\ket{\nu_\alpha}$ with $\alpha = \lbrace e, \, \mu, \, \tau \rbrace$. Such flavour eigenstates can be expressed as a combination of the eigenstates of the free Hamiltonian, which we denote by $\ket{\nu_i}$ with $i = \lbrace 1,\, 2, \, 3 \rbrace$ and satisfy that
\begin{align}
H_0\ket{\nu_i} = E_i \ket{\nu_i} \quad \text{and} \quad E_i = \sqrt{\mathbf{p}^2 + m^2_i}\, \text{.\footnotemark}
\end{align}
\footnotetext{Remember that, following the notation already introduced in the previous Chapter, vectors are denoted by bold characters.}
The relation between both sets of eigenstates is given by
\begin{align}
\ket{\nu_\alpha} = \sum_{i } U^*_{\alpha i}\ket{\nu_i} \, ,
\end{align}
where we have introduced the lepton mixing matrix $U$. The evolution of the flavour eigenstate is obtained by solving the Schrödinger equation
\begin{align}
i \frac{\text{d}}{\text{d}t}\ket{\nu_i} = H_0 \ket{\nu_i}\, .
\end{align}
Then, the evolution of a flavour eigenstate is
\begin{align}
\ket{\nu_\alpha (t)} = \sum_{i} U_{\alpha i}^* e ^{-i E_i t}\ket{\nu_i} =  \sum_{\beta }\sum_{i } U_{\alpha i}^* U_{\beta i}e ^{-i E_i t}\ket{\nu_\beta} \, .
\end{align}

The flavour transition amplitude reads
\begin{align}
\mathcal{A}_{\nu_\alpha \rightarrow \nu_\beta}(t)\equiv \varphi_{\alpha \beta} = \bra{\nu_\beta}\ket{\nu_\alpha(t)} = \sum_i U^*_{\alpha i} U_{\beta i} e^{-i E_i t}\, ,
\end{align}
and the flavour transition probability is
\begin{align}
P_{\alpha\beta}(t) = |\mathcal{A}_{\nu_\alpha \rightarrow \nu_\beta}(t)|^2 =\sum_{i,j} U^*_{\alpha i}U_{\beta i}U_{\alpha j}U^*_{\beta j}e^{-i (E_i - E_j)t} \, .
\end{align}

Assuming that neutrinos are ultrarelativistic particles, the energy can be expanded as
\begin{align}
E_i = \sqrt{\mathbf{p}^2 + m^2_i} \simeq |\mathbf{p}| + \frac{m_i^2}{2 |\mathbf{p}|} \, .
\end{align}
Defining the mass splitting as $\Delta m^2_{ij} = m^2_i - m^2_j$, one can rewrite the oscillation probability as
\begin{align}
P_{\alpha\beta}(t)  =\sum_{i,j} U^*_{\alpha i}U_{\beta i}U_{\alpha j}U^*_{\beta j}\,\text{exp}\left(-i \frac{\Delta m^2_{ij} L}{2E}\right) \, .
\label{eq:ch2-osc-prob}
\end{align}
Notice that in the expression above, we have replaced the neutrino \mbox{time-of-flight,} $t$, with the baseline, $L$. This is a well-justified approximation for ultrarelativistic neutrinos. 

Actually, a more precise derivation should be performed considering neutrino wave-packets in quantum mechanics~\cite{Giunti:1991ca,Giunti:1991sx} or, even more generally, a quantum field theory description~\cite{Akhmedov:2010ms}. It can be shown that as long as coherence conditions in production detection and propagation are satisfied, the result here derived in terms of equal-momentum plane waves --- see Equation \ref{eq:ch2-osc-prob} --- is valid~\cite{Akhmedov:2019iyt}. 

The oscillation probability in Equation \ref{eq:ch2-osc-prob} can be rewritten as
\begin{align}
P_{\alpha\beta}(t)  & = \sum_j |U_{\alpha j}|^2 |U_{\beta j}|^2 + \sum_{j\neq i}U^*_{\alpha i}U_{\beta i}U_{\alpha j}U^*_{\beta j}\, \text{exp}\left(-i \frac{\Delta m^2_{ij} L}{2E}\right) \nonumber \\
& = \sum_j |U_{\alpha j}|^2 |U_{\beta j}|^2 + 2 \text{Re}\left[\sum_{j > i}U^*_{\alpha i}U_{\beta i}U_{\alpha j}U^*_{\beta j}\,\text{exp}\left(-i \frac{\Delta m^2_{ij} L}{2E}\right) \right] \nonumber \\
& = \sum_j |U_{\alpha j}|^2 |U_{\beta j}|^2 + 2 \sum_{j > i}\text{Re}\left[U^*_{\alpha i}U_{\beta i}U_{\alpha j}U^*_{\beta j}\right] \cos \left(\frac{\Delta m^2_{ij} L}{2E}\right) \nonumber \\ & - 2 \sum_{j > i}\text{Im}\left[U^*_{\alpha i}U_{\beta i}U_{\alpha j}U^*_{\beta j}\right] \sin \left(\frac{\Delta m^2_{ij} L}{2E}\right) \, .
\end{align}

Using the fact that 
\begin{align}
\sum_j |U_{\alpha j}|^2 |U_{\beta j}|^2 = \delta_{\alpha\beta} - 2 \sum_{j> i}\text{Re}\left[U^*_{\alpha i}U_{\beta i}U_{\alpha j}U^*_{\beta j}\right]\, ,
\end{align}
one arrives at the well-known expression
\begin{align}
P_{\alpha\beta}(t)  = \delta_{\alpha \beta} &- 4 \sum_{j > i} \text{Re}\left[U^*_{\alpha i}U_{\beta i}U_{\alpha j}U^*_{\beta j}\right]\sin^2 \left(\frac{\Delta m^2_{ij}L}{4E}\right) \nonumber \\ &+2 \sum_{j > i} \text{Im}\left[U^*_{\alpha i}U_{\beta i}U_{\alpha j}U^*_{\beta j}\right]\sin \left(\frac{\Delta m^2_{ij}L}{2E}\right)\, .
\end{align}

For Dirac neutrinos, the lepton mixing matrix, $U$, can be parametrised in terms of $n(n-1)/2$ angles and $(n-1)(n-2)/2$ phases. Nonetheless, for Majorana neutrinos, $n-1$ additional phases are required. The parametrisation of the lepton mixing matrix adopted in the different chapters of this thesis is specified when required.

Notice that, for the mixing of three active neutrinos, the phenomenon of oscillation can be parametrised in terms of 3 mixing angles and one CP phase --- independently of the Dirac or Majorana nature of neutrinos~\cite{Rodejohann:2011vc} --- and two mass splitting, $\Delta m^2_{21}$ and $\Delta m^2_{31}$. The first of the mass splittings, $\Delta m^2_{21}$ is positive by definition. In contrast, the sign of $\Delta m^2_{31}$ has to be determined experimentally. So far, our ignorance in this regard means that two possible patterns or orderings need to be considered. We refer to the mass ordering as normal (NO) when $\Delta m^2_{31} > 0$ --- and hence the lightest neutrino is the one the largest fraction of the electron neutrino flavour state -- or inverted ordering (IO), when $\Delta m^2_{31} < 0 $. \reffig{fig:ch2-ordering} shows the two possible scenarios for the mass hierarchy and the composition of the mass eigenstates in terms of the flavour eigenstates given our knowledge of the oscillation parameters and the uncertainties in the picture.

\begin{figure}
\includegraphics[width = \textwidth]{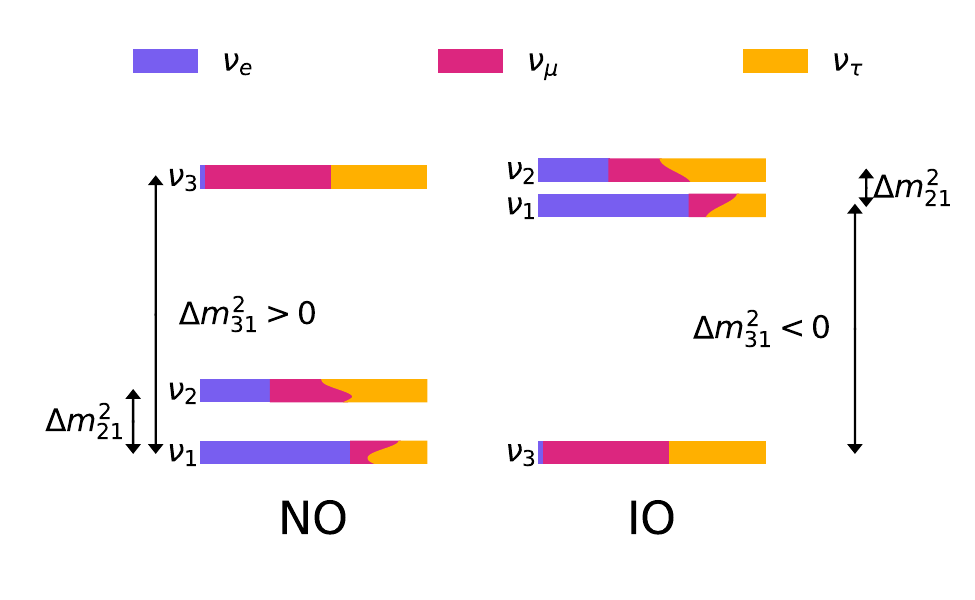}
\caption{Representations of the two possible orderings of the neutrino mass eigenstates: normal and inverted --- denoted by NO and IO respectively. The flavour composition of the three mass eigenstates is shown, including the dependence of $\delta_{\text{CP}}$. The uncertainties in the flavour content of each mass eigenstate are illustrated in the form of a varying composition. \labfig{fig:ch2-ordering}}
\end{figure}

Finally, it is possible to define the following quantities~\cite{Jarlskog:1985ht,Jarlskog:1985cw,Greenberg:1985mr,Dunietz:1985uy,Wu:1985ea}
\begin{align}
J_{\alpha\beta i j} = \text{Im}\left[U^*_{\alpha i}U_{\beta i}U_{\alpha j}U^*_{\beta j}\right] = J\sum_\gamma \epsilon_{\alpha\beta\gamma}\sum_l \epsilon_{ijl}\, ,
\end{align}
where $\epsilon_{ijl}$ and $\epsilon_{\alpha\beta\gamma}$ are Levi-Civita symbols. The quantity $J$ is known as the Jarlskog invariant. These quantities are invariant under any rephasing of the fields,
\begin{align}
\nu_\alpha \, \longrightarrow \, e^{i\phi_\alpha}\nu_\alpha \quad \text{and}\quad \nu_j \, \longrightarrow \, e^{i\phi_j} \nu_j\, .
\end{align}
For $\alpha\neq\beta$ and $i\neq j$, these invariants are equal except for an overall sign and measure the amount of CP violation in the leptonic mixing matrix. This parameter $J$ is only non-zero if all three mixing angles are different from zero and it can have a maximum value $J_{\text{max}} = 1/(6\sqrt{3})$.

\subsection{Neutrino oscillations in matter}
The evolution of neutrinos in a non-polarised electrically neutral medium is altered by two effective potentials arising from charged-current (CC) and neutral-current (NC) coherent elastic forward scattering, namely
\begin{align}
V_{\text{matt}} \ket{\nu_\alpha} = (V_{CC} + V_{NC})\ket{\nu_\alpha}\, .
\end{align}
The tree-level contribution to these potentials is given by the diagrams in \reffig{fig:ch2-matter}. Since ordinary matter is composed of electrons, neutrons and protons, the above expression simplifies to
\begin{align}
V_{\text{matt}} \ket{\nu_\alpha} =\sqrt{2} G_{\text{F}} \left( N_e \delta_{\alpha e} - \frac{1}{2}N_n \right) \ket{\nu_\alpha} \equiv \left( V_e \delta_{\alpha e} - V_n\right) \ket{\nu_\alpha}\, ,
\end{align}
which depends on the electron and neutron number densities, $N_e$ and $N_n$ respectively, and where we have defined
\begin{align}
V_e = \sqrt{2}G_{\text{F}} N_e \quad \quad \text{and} \quad \quad V_n = - \frac{G_\text{F}}{\sqrt{2}} N_n \, ,
\label{eq:ch2-matterpot}
\end{align}
as they will be used in \refch{ch4-magnetic}. Notice that the neutral-current contribution of protons and electrons is of opposite sign and hence, they cancel out in electrically neutral media.
Then, the evolution of a given flavour eigenstate is given by
\begin{align}
i\frac{\text{d}}{\text{d}t}\ket{\nu_\alpha} = (H_0 + V_{\text{matt}}) \ket{\nu_\alpha}\,.
\label{eq:ch2-hamiltonian-std}
\end{align}
Using CP properties, the evolution for antineutrinos is obtained by replacing $V_e \, \rightarrow \, -V_e$, $V_n \, \rightarrow \, -V_n$ and $U \, \rightarrow \, U^*$. Notice that the terms related to forward scattering on neutrons are proportional to the identity and hence, they can be rephased. However, they are included here for completeness and they will play a role in~\refch{ch4-magnetic}. 

\begin{figure}
\includegraphics[width = 0.8\textwidth]{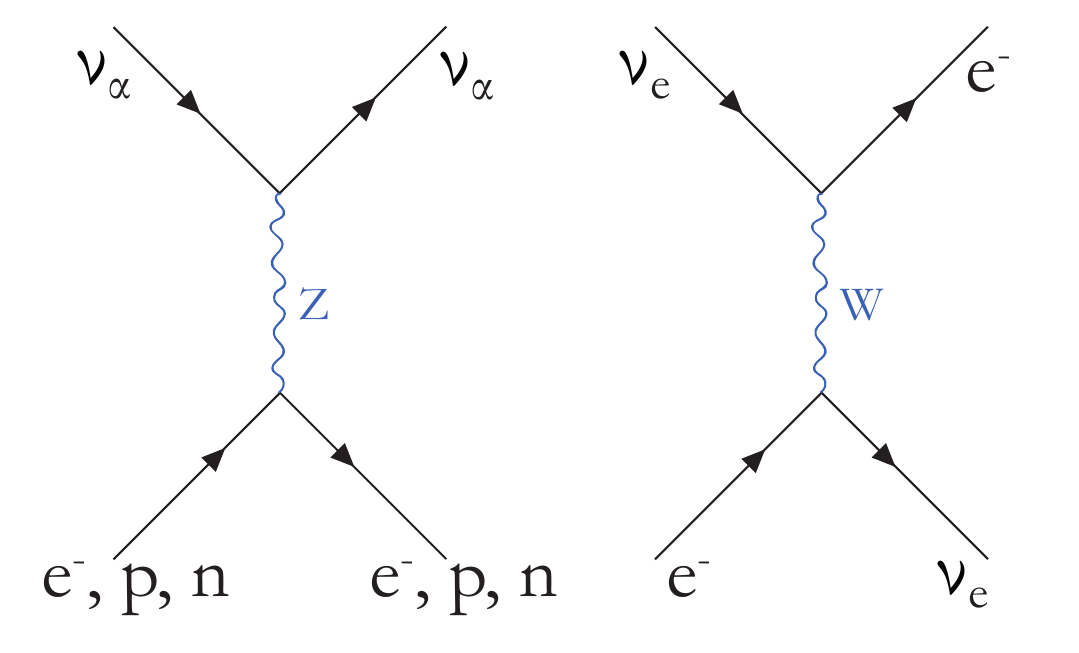}
\caption{Standard Model diagrams corresponding to the neutral-current --- in the left --- and charged-current interactions --- in the right --- between neutrinos and matter fields, giving rise to the effective potentials in Equation~\ref{eq:ch2-matterpot}. \labfig{fig:ch2-matter}}
\end{figure}

\section{Direct and indirect experimental probes of the absolute neutrino mass scale}
\label{sec:ch2-intro-mass-scale}
\subsection{Beta-decay experiments}
The measurement of beta decay --- also denoted as $\beta$ decay throughout the text --- provides a direct and reliable measurement of the absolute neutrino mass scale. The reason is that, from energy conservation arguments, part of the energy released in a beta decay must go to the electron neutrino --- or antineutrino --- in the final state. Then, the existence of a non-zero neutrino mass produces a shift in the maximum energy of this positron --- or electron. The differential rate of tritium beta decay,
\begin{align}
^3\text{H} \, \rightarrow ^3\text{He}\, + \,  e^-\, + \, \bar{\nu}_e \, ,
\end{align}
when expressed as a function of the electron kinetic energy, $K_e$, is proportional to~\cite{Long:2014zva}
\begin{align}
\frac{\text{d}\Gamma_\beta}{\text{d}K_e}  \propto \sum^3_{i = 1}|U_{ei}|^2 \sqrt{y \left(y + \frac{2m_im_{^3\text{He}}}{m_{^3 \text{H}}}\right)} \left[y + \frac{m_i}{m_{^3\text{H}}}(m_{^3\text{He}} + m_i )\right]  \, ,
\end{align} 
where $y = K_{\text{end}} - K_e$. Here, $K_{\text{end}}$ denotes the kinetic energy of the detected positron --- or electron --- at the end-point of the spectrum.

Let us define 
\begin{align}
Q_\beta = m_{3\text{H}} - m_{^3\text{He}} - m_e\, .
\end{align}
and let $m_{\text{lightest}}$ be the mass of the lightest neutrino.
Then, massive neutrinos shift the end-point spectrum from $K_{\text{end},0}= Q_\beta$ to $K_{\text{end}}= Q_\beta - m_{\text{lightest}}$. Kinks at the decay spectrum are also expected for the remaining neutrino masses --- i.e. when $K_{e} = Q_{\beta} - m_i$. Given the energy resolution at current experiments, these kinks can not be resolved and the decay rate can be expressed as \cite{Giunti:2007ry}
\begin{align}
\frac{\text{d}\Gamma_\beta}{\text{d}K_e} \propto  \sqrt{(Q_\beta - K_e) - m^2_\beta}(Q_\beta - K_e) \, ,
\end{align} 
where we have introduced the \textit{effective neutrino mass in $\beta$ decays},
\begin{align}
m^2_\beta = \sum^3_{i= 1} |U_{ei}|^2m^2_i.
\end{align}
Notice that the relation between the effective mass parameter $m^2_\beta$, the lightest neutrino mass and the mass splittings depends on the mass hierarchy, namely.
\begin{align}
m^2_\beta = \begin{cases} m_{\text{lightest}} + |U_{e2}|^2 \Delta m^2_{21} + |U_{e3}|^2 \Delta m^2_{31} & \text{(NO)} \\  m_{\text{lightest}}+ |U_{e2}|^2\Delta m^2_{21} + (1- |U_{e3}|^2) |\Delta m^2_{31}| & (\text{IO}) \end{cases} \, ,
\end{align}
Then, for normal ordering (NO), $m_{\text{lightest}} = m_1$, the lower limit on the effective mass parameter reads
\begin{align}
m^2_\beta > |U_{e2}|^2 \Delta m^2_{21} + |U_{e3}|^2 \Delta m^2_{31}\, ,
\end{align}
whereas for inverted ordering (IO), $m_{\text{lightest}} = m_3$, and the lower limit is 
\begin{align}
m^2_\beta > |U_{e2}|^2\Delta m^2_{21} + (1- |U_{e3}|^2) |\Delta m^2_{31}|\, .
\end{align} 
Thus, the determination of the mass splitting from oscillation data sets a sensitivity goal for beta decay experiments.

Note also that other limits on neutrino masses exist from kinematic measurements, in particular from pion and tau decays, but those are not competitive since they are of order $m_{\text{lightest}} \sim \mathcal{O} (0.1-10)$ MeV~\cite{Giunti:2007ry}.

\subsection{Neutrinoless double-beta decay}
\begin{figure*}
\includegraphics[width = 0.76\paperwidth]{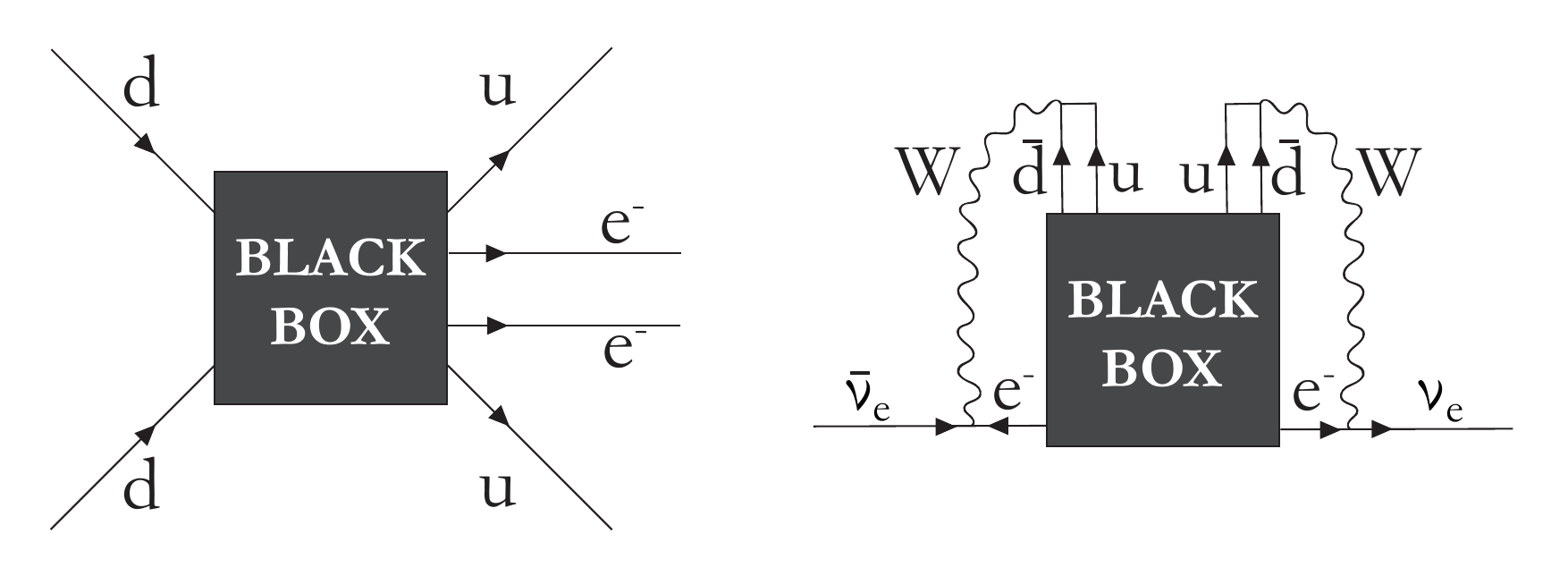}
\caption{Left sketch shows the diagram responsible for $0\nu\beta\beta$ decay. The right sketch illustrates the correspondence between the $0\nu\beta\beta$ decay process and the origin of Majorana neutrino masses. This serves as a graphical illustration of the Black Box theorem~\cite{Schechter:1981bd}. \labfig{fig:ch2-blackbox}}
\end{figure*}
Neutrinoless double-beta decay --- or $0\nu\beta\beta$ decay --- is a hypothetical process that would violate lepton number by two units. The Black Box theorem~\cite{Schechter:1981bd} states that the diagram responsible for $0\nu\beta\beta$ necessarily generates a contribution to the Majorana mass of the electron neutrino through radiative corrections at some order of perturbation theory, even if there is no tree level Majorana neutrino mass term, as illustrated in \reffig{fig:ch2-blackbox}. Note that this does not necessarily mean that the main contribution to the $0\nu\beta\beta$ decay is coming from the exchange of Majorana massive neutrinos. 

Neutrinoless double beta decay experiments aim to measure the lifetime of the decay, $T^{0\nu}_{1/2} (\mathcal{N})$. If mediated by the exchange of light Majorana neutrinos, it is given by
\begin{align}
\frac{1}{T^{0\nu}_{1/2} (\mathcal{N})} = G^N_{0\nu} |\mathcal{M}^{0\nu}_N|^2 \left(\frac{m_{\beta\beta}}{m_e}\right)^2 \, ,
\end{align}
where $G^N_{0\nu}$ is the phase-space factor, $\mathcal{M}^{0\nu}_N$ is nuclear matrix element (NME), $m_e$ is the electron mass and we have introduced the effective mass parameter, $m_{\beta\beta}$, namely
\begin{align}
m_{\beta\beta} = \sum_i U^2_{ei}m_i \, .
\end{align}

Note that determining the nuclear matrix elements of $0\nu\beta\beta$ decay is highly non-trivial. They are not identical to the matrix elements for $2\nu\beta\beta$ decay --- which can be experimentally measured. Hence, they have to be determined directly from nuclear theory. Among the different methods used to calculate these nuclear matrix elements, one finds \textit{Quasiparticle Random Phase Approximation (QRPA)}~\cite{Mustonen:2013zu,Hyvarinen:2015bda,Simkovic:2018hiq,Fang:2018tui,Terasaki:2020ndc}, \textit{Nuclear Shell Model (NSM) calcualtions}~\cite{Horoi:2015tkc,Menendez:2017fdf,Coraggio:2020hwx,Coraggio:2022vgy}, the \textit{Energy-Density Functional theory (EDF)}~\cite{Rodriguez:2010mn,LopezVaquero:2013yji,Song:2017ktj}, and the \textit{Interacting Boson Model (IBM)}~\cite{Barea:2015kwa,Deppisch:2020ztt}. This uncertainty in nuclear matrix elements would potentially limit the possibility to determine with accuracy the absolute neutrino mass scale from the measurement of $0\nu\beta\beta$~\cite{Pompa:2023jxc}.

Finally, note that, as a function of the lightest neutrino mass, the effective mass parameters for normal ordering reads
\begin{align}
&m_{\beta\beta} =  U^2_{e1} m_{\text{lightest}} \nonumber \\ 
&\hspace{2cm}+ U^2_{e2}\sqrt{\Delta m^2 _{21} + m^2_{\text{lightest}}} 
+ U^2_{e3}\sqrt{\Delta m^2_{31} + m^2_{\text{lightest}}}\, ,
\end{align}
whereas for inverted ordering
\begin{align}
&m_{\beta\beta} =  U^2_{e1}\sqrt{-\Delta m^2_{31} - m^2_{\text{lightest}}} \nonumber \\ &\hspace{2cm} + U^2_{e2}\sqrt{-\Delta m^2_{31} + \Delta m^2_{21} + m_{\text{lightest}}} + U^2_{e3} m_{\text{lightest}}\, .
\end{align}
The process is sensitive to Majorana phases of the lepton mixing matrix to the point that, for normal ordering, $m_{\beta\beta}$ could be zero, even if the individual masses are not~\cite{Giunti:2007ry}.

\subsection{Cosmological observables}

Neutrinos have an impact on the cosmological evolution of the universe in different ways~\cite{Lesgourgues:2013sjj}. Being relativistic in the early universe, they act as a radiation component. During this epoch, neutrinos decouple from the thermal plasma and start to propagate freely until today. Because of the dilution of energy during the expansion of the universe, neutrinos are expected to have a temperature of the order of $10^{-4}$ eV today. This means that at least two over three mass eigenstates are non-relativistic at present times, thus contributing to the matter energy density. The particular behaviour, with a non-relativistic transition occurring at some point, leaves an imprint on several observables.

There are several ways in which massive neutrinos may leave their signature in the cosmic microwave background (CMB). The reason is that they shift the time of matter-radiation equality and the radiation energy density of the universe at late times. In the first place, massive neutrinos enhance the Hubble expansion. This leads to a lower angular diameter distance to the last-scattering surface and a consequent shift of the CMB peaks to smaller multipoles --- larger angular scales. Two additional signatures arise due to the Integrated Sachs-Wolfe (ISW) effect~\cite{Sachs:1967er,Rees:1968zza,1990ApJ355L5M,Sugiyama:1994ed},  an anisotropy in the CMB caused by the uneven redshift of CMB photons due to the evolution of the large-scale gravitational potential. The first one is related to the early ISW effect and how an earlier \mbox{matter-radiation} equality suppresses the temperature anisotropies, especially in the multipole range 20 < $l$ <500 --- mainly the first peak. The second one is related to the late ISW effect and the fact that massive neutrinos delay the epoch of dark energy domination and decrease the temperature anisotropies in the range of multipoles 2 < $l$ < 50~\cite{Lesgourgues:2006nd}.

Unfortunately, the degeneracies between the parameters in the model can easily compensate for these effects, especially in extended models --- i.e. beyond $\Lambda$CDM. For instance, the shift in the acoustic peaks induced by massive neutrinos is highly degenerate with the value of $H_0$. Regarding the effects on the low multipoles induced by the late ISW effect, they happen in a region in which cosmic variance does not allow for accurate measurements of the CMB anisotropies.

Since neutrinos are hot --- relativistic --- relics, their free-streaming results in a suppression of the growth of matter fluctuations at small scales and impacts the formation of structures~\cite{Bond:1980ha}. Information on the evolution of large-scale structures in the universe can be extracted from baryon acoustic oscillation, Lyman-$\alpha$ forest, galaxy clusters, and galaxy surveys and hence, they provide constraints on the sum of neutrino masses~\cite{Lesgourgues:2006nd,Gariazzo:2018pei}.

Neutrino free-streaming properties also manifest indirectly in the lensing of the CMB resulting from the distortion of photon propagation due to the inhomogeneities in matter along the line of sight. As a consequence of lensing, there is an effective remapping of the CMB fluctuations which gives an indirect measurement of the integrated mass distribution back to the last scattering surface~\cite{Kaplinghat:2003bh,Lesgourgues:2006nd}. Thus, this is a different manifestation of the same effect: the suppression of small-scale fluctuations due to neutrino free-streaming.

The impact of neutrinos in cosmology is generally studied in terms of two parameters: $N_\text{eff}$ and $\sum m_\nu$. The former is the so-called \textit{effective number of neutrinos} and is related to the contribution of neutrinos to the radiation energy of the universe. Further discussion will follow in \refch{ch7-nsicosmo} and \refch{ch8-nucosmo}. The latter parameter is the sum of neutrino masses. For normal ordering, it reads
\begin{align}
\sum m_\nu = m_{\rm lightest} + \sqrt{\Delta m^2_{21}-m^2_\text{lightest}}+ \sqrt{\Delta m^2_{31}-m^2_\text{lightest}}\, ,
\end{align} 
while for inverted ordering, it is given by
\begin{align}
\sum m_\nu &= \sqrt{-\Delta m^2_{31}-m^2_\text{lightest}}\nonumber \\ & \hspace{1cm}+ \sqrt{-\Delta m^2_{31}+\Delta m^2_{21}-m^2_\text{lightest}} +m_{\rm lightest} \, .
\end{align} 

It is known that neutrinos with masses larger than $\mathcal{O}(10^{-4}$eV) would be non-relativistic at present. For non-relativistic neutrinos at present, their energy density would be $\rho_{\nu_i} = m_i n^0_{\nu, i}$ where $n^0_{\nu_i}$ is the current neutrino number density for the mass eigenstate $i$. Then, if all neutrinos were non-relativistic, their contribution to the energy budget of the universe would be, in terms of the dimensionless energy density,
\begin{align}
\Omega_\nu h^2= \frac{\sum m_\nu}{93.14 \text{ eV}}\, .
\end{align}
Here, we have also introduced the reduced Hubble parameter $h = H_0/ (100 \text{km s}^{-1} \text{Mpc}^{-1})$.

\section{Some phenomenological consequences of neutrino mass models}
\label{sec:ch2-intro-pheno}
To explain the origin of neutrino masses, the content of the Standard Model has to be extended --- as it was discussed in Section \ref{sec:ch2-mass-models}. In many cases, the inclusion of new symmetries is also relevant to make the mass model compatible with other observables, stable from quantum corrections or as an attempt to increase the capacity of the model to make predictions and hence, its testability.

As a result of non-zero neutrino masses, a plethora of other neutrino properties arise. For instance, massive neutrinos are known to have \mbox{non-zero} electromagnetic properties~\cite{Marciano:1977wx,Lee:1977tib,Fujikawa:1980yx,Shrock:1982sc,Petcov:1976ff,Bilenky:1987ty,Giunti:2014ixa}. The magnitude of such electromagnetic properties strongly depends on the underlying realisation of neutrino masses.

Similarly, many mass models are known to give rise to non-standard neutrino interactions (NSIs) with other Standard Model fermions~\cite{Wolfenstein:1977ue,Abada:2007ux,Gavela:2008ra,Antusch:2008tz,Farzan:2017xzy}. In particular, radiative mass models predict relatively sizeable new neutrino interactions~\cite{Babu:2019mfe}. Nevertheless, note that due to SU(2) symmetry of the Standard Model, NSIs are often accompanied by lepton-flavour violating interactions, which are strongly constrained experimentally~\cite{Davidson:2019iqh}. \footnote{These constraints, however, do not apply if NSIs are a consequence of the existence of light mediators~\cite{Farzan:2015doa,Farzan:2015hkd}.}

Heavy neutral leptons~\cite{Abdullahi:2022jlv} are also a common ingredient in extensions of the Standard Model accounting for neutrinos masses, such as type-I and type-III seesaw mechanisms. If they were heavy enough so that they are not kinematically accessible in experiments and if their admixture with the SM states were non-zero, their presence would manifest as a non-unitary three-neutrino mixing matrix~\cite{Antusch:2006vwa,Abada:2007ux,Forero:2021azc}.

Since neutrinos are massive, the two heavier ones are also unstable and can decay. These decays can be radiative~\cite{Giunti:2014ixa} or involving other species such as Majorons~\cite{Gelmini:1980re}. If neutrinos' lifetime is larger than the age of the universe, they are effectively stable. Nonetheless, the predicted lifetime is strongly dependent on the underlying model and hence, worth testing experimentally.

The smallness of their masses and the fact that they are fundamental particles also make neutrinos ideal to probe some fundamental aspects of our understanding of nature. For instance, the fact that particles and antiparticles have the same mass and lifetime --- if unstable --- is a consequence of CPT invariance in local relativistic quantum field theories. The determination of the mass splittings $\Delta m^2_{ij}$ and mixing angles independently both for neutrinos and antineutrinos provides a valuable and fundamental test of CPT invariance~\cite{Tortola:2020ncu}.

Other exotic scenarios are also of interest in the neutrino phenomenology community. Some examples are the study of neutrino decoherence~\cite{Akhmedov:2019iyt}, altered dispersion relations~\cite{Barenboim:2019hso}, the existence of light sterile neutrinos \cite{Giunti:2019aiy} --- not necessarily related to the mass mechanism --- or the existence of large extra dimensions~\cite{Arkani-Hamed:1998wuz,Dienes:1998sb,Dvali:1999cn,Barbieri:2000mg}, among others. 

%\setchapterpreamble[u]{\margintoc\chapter{Global analysis of neutrino data}
%\addcontentsline{toc}{chapter}{Global analysis of neutrino data}
\chapter{Global analysis of neutrino data} 
\labch{ch3-fit}

This chapter is devoted to the current status of the determination of neutrino masses and mixing from a global fit to all neutrino data, following~\cite{deSalas:2020pgw}. Section \ref{sec:ch3-data} describes the different datasets included in the analysis and Section \ref{sec:ch3-results} highlights the contribution of each one to the global fit. From this combined analysis, we extract information on the mass splittings $\Delta m^2_{21}$ and $\Delta m^2_{31}$, the angles and the Dirac phase characterising neutrino mixing. For the lepton mixing matrix, we adopt the following parametrisation,
\begin{align}
&U = \nonumber \\ &\hspace{0.1cm}\begin{pmatrix}
1 & 0 & 0 \\ 0 & c_{23} & s_{23} \\ 0 & -s_{23} & c_{23}
\end{pmatrix} \begin{pmatrix}
c_{13} & 0 & s_{13} e ^{-i \delta_{\text{CP}}} \\ 0 & 1 & 0 \\ -s_{13}e^{i\delta_{\text{CP}}} & 0 & c_{13}  
\end{pmatrix}\begin{pmatrix}
c_{12} & s_{12} & 0 \\ -s_{12} & c_{12} & 0 \\ 0 & 0 & 1
\end{pmatrix}\, ,
\label{eqn:lmmatrix}
\end{align}
where we use the short-hand notation $c_{ij} = \cos \theta_{ij}$ and $s_{ij} = \sin \theta_{ij}$. We do not include the two additional Majorana phases since they do not have any impact on flavour oscillations.

In Section \ref{sec:ch3-bayes}, we introduce a Bayesian approach to the data analysis. We do so to include information from the $\beta$-decay experiment KATRIN, \mbox{$0\nu\beta\beta$-decay} searches and cosmology. We address the determination of the absolute neutrino mass scale and the mass ordering too. Finally, in Section \ref{sec:ch3-summary-fit}, we summarise the results from this global fit.

%%%%%%%%%%%%%%%%%%%%%%%%%%%
%%%%%%%%%%%%%%%%%%%%%%%%%%%
\section{Neutrino data from oscillation experiments \label{sec:ch3-data}}
%%%%%%%%%%%%%%%%%%%%%%%%%%%
%%%%%%%%%%%%%%%%%%%%%%%%%%%
\subsection{Solar neutrino experiments and KamLAND}
%%%%%%%%%%%%%%%%%%%%%%%%%%%

\textbf{Solar neutrino experiments}

\begin{figure*}[t!]
\centering
\includegraphics[width = 0.75\paperwidth]{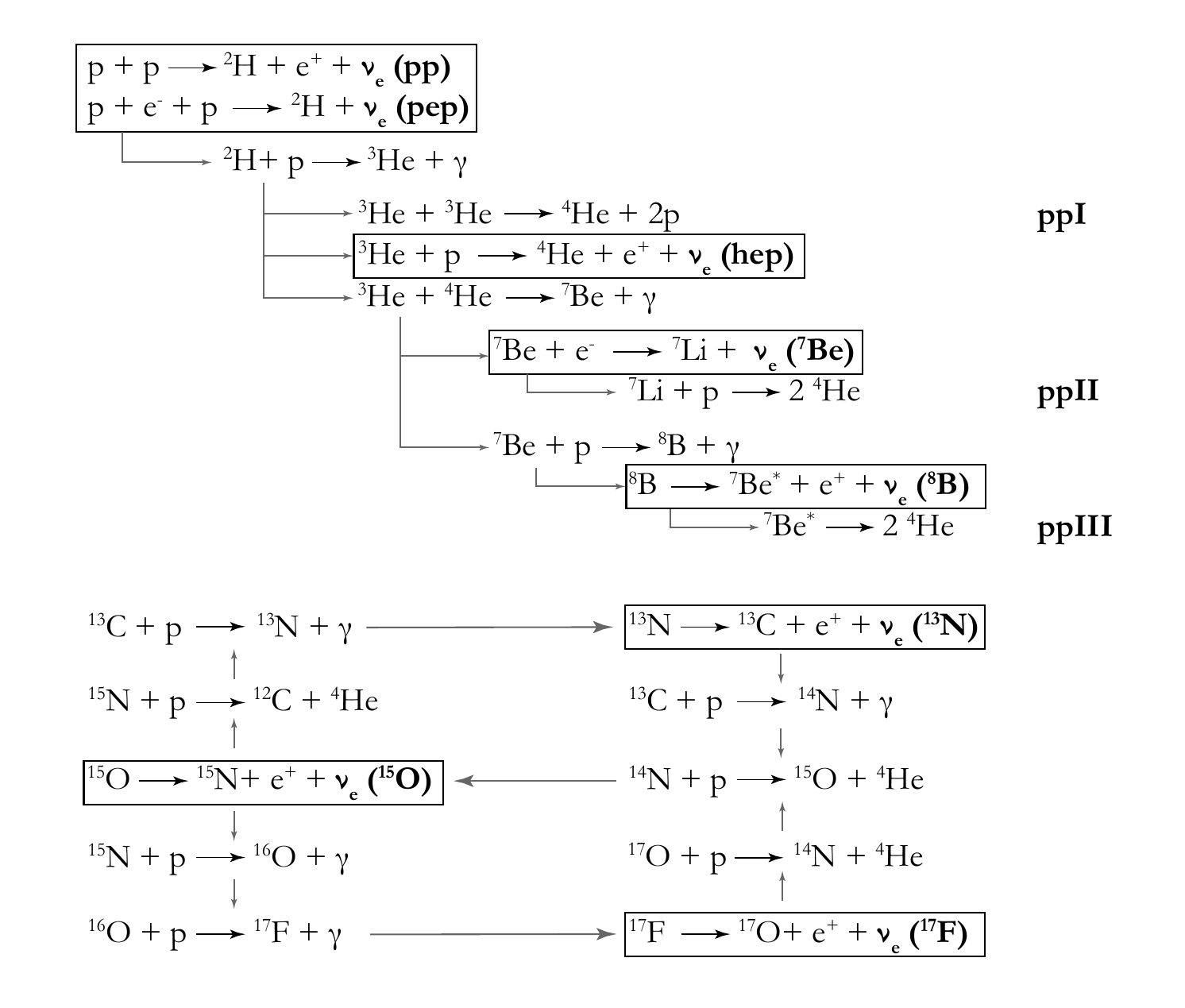}
\caption{Thermonuclear reactions responsible for the flux of solar neutrinos. The upper diagram illustrates the so-called $pp$-chains and the lower diagram shows the CNO cycle.\labfig{fig:ch3-solarchains}}
\end{figure*}

\begin{figure*}
\centering
\includegraphics{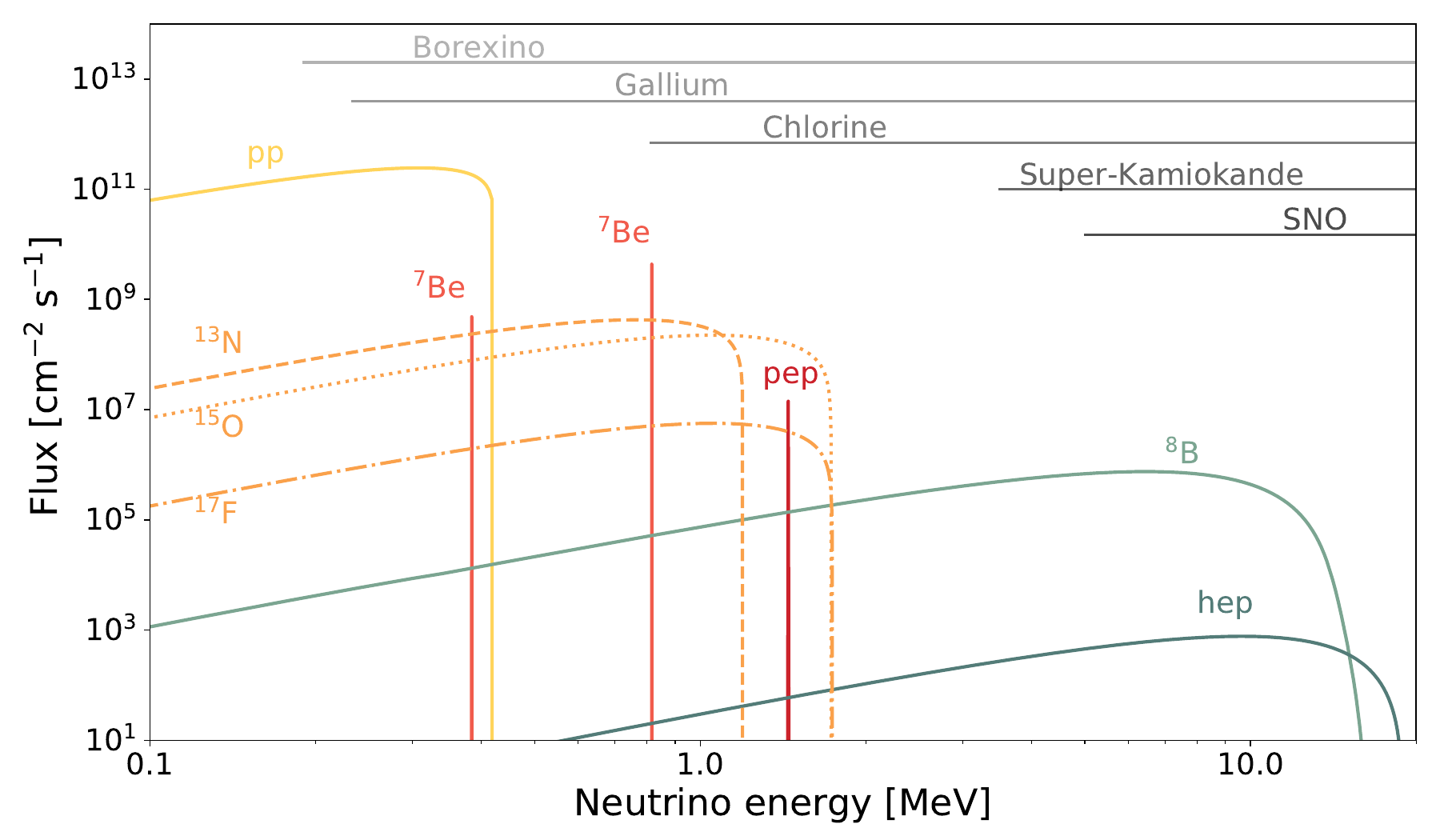}
\caption{Flux of solar neutrinos as a function of neutrino energy. The different contributions from the different thermonuclear reactions are indicated as from~\cite{Bahcall:1987jc,Bahcall:1994cf,Bahcall:1996qv,Bahcall:1997eg,Vinyoles:2016djt}. On top, the range of energy accessible at different experiments is indicated. \labfig{fig:ch3-solarflux}}
\end{figure*}
Solar neutrinos are produced in the thermonuclear reactions powering the core of the Sun. Among those, one can find the \textit{pp}-chain reactions --- giving rise to $pp$, $pep$, $^7$Be, $^8$B and $hep$ neutrinos --- and the CNO cycle --- responsible for the flux of $^{13}$N $^{15}$O and $^{17}$N neutrinos. \reffig{fig:ch3-solarchains} shows these processes diagrammatically. The spectrum resulting from all these reactions is shown in \reffig{fig:ch3-solarflux}.

Several detection techniques have been developed and optimised for the different neutrino energies involved. For instance, the first studies of solar neutrinos were based on neutrino capture reactions,
\begin{align}
\nu_e \, + \, (\text{A, Z})\, \rightarrow \, e^- \, + \, (\text{A, Z+1})\, ,
\end{align}
in counting experiments. Due to the small cross-sections and in order to prevent cosmic-ray interactions from mimicking the signal, experiments were conducted deep underground using tons of target material, (A, Z). After a certain exposure time, the number of isotopes \mbox{(A, Z+1)} produced by neutrino captures would be counted. This is the case of the Homestake experiment, which studied neutrino capture in Chlorine, 
\begin{align}
\nu_e \, + \, ^{37}\text{Cl} \,\rightarrow \, ^{37}\text{Ar} + \, e^{-} \, .
\end{align}
GALLEX/GNO and SAGE performed similar studies using neutrino capture in Gallium,
\begin{align}
\nu_e \, + \, ^{71}\text{Ga} \rightarrow \, ^{71}\text{Ge} + e^{-}\, .
\end{align} Our analysis includes the total rate measurements from the three radiochemical experiments: Homestake~\cite{Cleveland:1998nv}, SAGE~\cite{SAGE:2009eeu}, and GALLEX/GNO~\cite{Kaether:2010ag}.

We also use low-energy spectral information from Borexino~\cite{Bellini:2011rx,Borexino:2013zhu}. This liquid-scintillator detector measures low-energy solar neutrinos and it is sensitive to the three flavours via elastic neutrino scattering on electrons,
\begin{align}
\nu_\alpha \,  + \, e^{-} \, \rightarrow \, \nu_\alpha \, + \, e^{-}\, .
\end{align} Note, however, that this is not a flavour-blind process, since the cross-section of this process is larger for $\nu_e$ than for $\nu_\mu$ and $\nu_\tau$.

The water Cherenkov neutrino telescope Super-Kamiokande also relies on this channel --- elastic neutrino scattering on electrons --- to detect neutrinos. However, unlike Borexino, it identifies the charged leptons in the final state via the Cherenkov light produced as they travel across the water tank. In the global fit in~\cite{deSalas:2020pgw}, we include the day-night spectral information from the first four runs of the detector, Super-Kamiokande I-IV~\cite{Super-Kamiokande:2005wtt,Super-Kamiokande:2008ecj,Super-Kamiokande:2010tar,Nakano:2016uws}.

Finally, the high-energy end of the spectrum is also measured by the Sudbury Neutrino Observatory (SNO)~\cite{SNO:2011hxd}. This experiment consists of a tank filled with heavy water, D$_2$O --- that is water containing only deuterium instead of $^1 _1$H. It is sensitive to solar neutrinos via elastic scattering on electrons, neutrino captures in deuterium, D,
\begin{align}
\nu_e \,+\, \text{D}\, \rightarrow\, p\, +\,  p \, + \, e^{-}
\end{align}
and the neutral current reaction \begin{align}
\nu_\alpha \, + \, \text{D} \, \rightarrow\, n\, +\, p \,+\, \nu_\alpha \, ,
\end{align} where $p$ and $n$ denote the protons and neutrons in the final state. The latter reaction was particularly relevant historically since it provided a \mbox{flavour-blind} measurement of the solar neutrino flux and helped solve the solar neutrino puzzle.

The main neutrino energies accessible at Gallium-based, Chlorine-based, liquid-scintillators (Borexino), heavy-water tank (SNO), and water Cherenkov detectors (Super-Kamiokande) are shown in~\reffig{fig:ch3-solarflux}.

We present the results of our combined analysis of solar neutrino data in \reffig{fig:ch3-solarbf2020}. This analysis considers the low-metallicity Standard Solar Model AGSS09~\cite{Vinyoles:2016djt}. Nonetheless, note that the measurement of CNO neutrinos from Borexino results in a preference for high-metallicity composition~\cite{BOREXINO:2022abl}.\footnote{This will also be of relevance for future solar neutrino observatories since the predicted flux of $^8$B and $hep$ neutrinos from high-metallicity solar models is larger than for low-metallicity ones.} Note also that solar neutrinos are mainly sensitive to $\theta_{12}$ and $\Delta m^2_{21}$. A more detailed discussion on the sensitivity of solar neutrino experiments to the oscillation parameters can be found in \refch{ch5-nsisolar}.

\textbf{KamLAND}

The long-baseline reactor neutrino experiment KamLAND (Kamioka Liquid Scintillator Antineutrino Detector) measured the flux of antineutrinos from 56 different reactors located at an average distance of 180 km from the detector. Like solar neutrino experiments, it was sensitive to the oscillation parameters $\theta_{12}$ and $\Delta m^2_{21}$~\cite{KamLAND:2008dgz,KamLAND:2010fvi,KamLAND:2013rgu}.\footnote{This statement is true under the assumption that CPT invariance holds.} In our global fit, we include the data from their analysis in~\cite{KamLAND:2010fvi}. 

Reactor electron antineutrinos are measured through inverse beta decay (IBD),
\begin{align}
\bar{\nu}_e \, + \, p \longrightarrow \, n\, + \, e^+\, ,
\end{align}
 and the neutrino energy is estimated to be $E_{\nu} \simeq E_{\text{prompt}} + 0.8 \, \text{MeV}$, where $E_{\text{prompt}}$ is the prompt energy of the positron emitted. Following \cite{Gando:2010aa}, the energy resolution is assumed to be
\begin{align}
\frac{\sigma(E)}{E} = \frac{0.064}{\sqrt{E\text{[MeV]}}}\, .
\end{align}
and additional information on the average isotope composition in the reactors is incorporated into the analysis.

Our frequentist approach relies on the definition and minimisation of a $\chi^2$-test, which relates the number of events observed at the experiments and the number of events predicted for a given value of the parameters involved. The latter is computed using the software \texttt{GLoBES} (General Long Baseline Experiment Simulator)~\cite{Huber:2004ka,Huber:2007ji}, after having adapted it for the experimental configuration of the KamLAND experiment. For the analysis, one has to simulate the expected flux from each of the 20 reactor sites, taking into account the number of cores, their thermal power and their distance to the detector. This information is summarised in \reftab{ch3-KamLAND}.

\begin{table*}[t!]
\renewcommand*{\arraystretch}{1.2}
\centering
\begin{tabular}{lccc}
\toprule[0.25ex]
         Site & Distance [km] & Number of cores & Thermal power [GW]  \\ \midrule
         Kashiwazaki& 160 &  7 & 24.3 \\
         Ohi & 179 & 4 & 13.7 \\
         Takahama & 191 & 4 & 10.2 \\
         Hamoka & 214 & 4 & 10.6 \\
         Tsuruga & 138 & 2 & 4.5 \\
         Shiga & 88 & 1 & 1.6 \\
         Mihama & 146 & 3 & 4.9 \\
         Fukushima-I & 349 & 6 & 14.2 \\
         Fukushima-II & 345 & 4 & 13.2 \\
         Tokai-II & 294 & 1 & 3.3 \\
         Simane & 401 & 2 & 3.8 \\
         Onagawa & 431 & 2 & 6.5 \\
         Ikata & 561 & 3 & 6.0 \\
         Genkai & 755 & 4 & 10.1 \\
         Sendai & 830 & 2 & 5.3 \\
         Tomari & 783 & 2 & 3.3 \\\midrule
         Ulchin & 712 & 4 & 11.5 \\
         Yonggwang & 986 & 4 & 17.4 \\
         Kori & 735 & 6 & 9.2 \\
         Wolsong & 709 & 4 & 8.2 \\
         \bottomrule[0.25ex]
\end{tabular}
    \caption{Summary of reactor complexes in Japan and Korea contributing to the electron antineutrino flux detected in KamLAND, combining information from~\cite{KamLANDtable1,KamLANDtable2}.}
    \label{tab:ch3-KamLAND}
\end{table*}  

In the first place, one has to normalise the total rate to the number of events expected in the absence of oscillations from reactor antineutrinos, which is given by the KamLAND Collaboration. Then, we also include the backgrounds provided in the data release together with several free --- pull --- parameters. In this way, all 76 bins were used for the analysis, including those with contributions from geoneutrinos.\footnote{Note that we do not model the flux of geoneutrinos. Instead, we take the expected event rate per energy bin as given by the collaboration.}

Our $\chi^2$-test,
\begin{equation}
    \chi^2 = \sum_{i=1}^{76} \frac{\left(N_i^{\text{obs}} - N_i^{\text{teo}}\right)^2}{N_i^{\text{obs}}} + \sum_j^{N_{\text{pulls}}}\left(\frac{\eta_j}{\sigma_j}\right)^2 + \frac{\left(N_{\text{total}}^{obs} - N_{\text{total}}^{teo}\right)^2}{N_{\text{total}}^{obs}}
\label{eqn:ch3-chi2KamLAND}
\end{equation}    
depends on the observed and predicted number of events in the $i$-th energy bin, $N_i^{\text{obs}}$ and $N_i^{\text{teo}}$ respectively. It includes several pull parameters, denoted by $\eta_j$ that account for systematics on the total power of each reactor site ($\sigma_p= 2$\%), an overall normalisation ($\sigma_n =2$\%),  changes in fuel composition ($\sigma_c = 2$\%), systematics on the energy resolution ($\sigma_r = 1.5$\%), systematics on the normalisation of the geoneutrino backgrounds ($\sigma_b =  15 - 50$\%) and spectral systematic errors ($\sigma _s = 3$\%). The third term in Equation~\ref{eqn:ch3-chi2KamLAND} accounts for the information on the total rate of events measured so that the spectral and total rates are fitted. No uncorrelated errors are considered and we assume that the number of events follows a Gaussian distribution. Hence, since we only consider the statistical errors, $\sigma_i^2 = N_i^{\text{obs}}$.

\textbf{Solar sector}

In \reffig{fig:ch3-solarbf2020}, we show the result of our analysis and compare it with the one from solar experiments. Whereas KamLAND provides an accurate determination of the mass splitting, solar neutrino oscillation data gives a better measurement of the mixing angle $\theta_{12}$. Actually, KamLAND is mainly sensitive to $\sin^2 2\theta_{12}$ and only shows a slight preference for the first octant ($\theta_{12} < \pi/4$) due to matter effects. The allowed value in the second octant is excluded once KamLAND is combined with solar neutrino data. The reason is that adiabatic flavour conversion in the Sun is sensitive to $\sin^2\theta_{12}$ and hence, it breaks the octant degeneracy.  This is a clear example of the potential of combining complementary datasets in global analyses.

The plane in parameter space determined by the solar mixing angle $\theta_{12}$ and the solar mass splitting $\Delta m^2_{21}$ is often referred to as the solar sector. Nevertheless, both solar neutrino experiments and long-baseline reactor experiments have a mild dependence on the mixing angle $\theta_{13}$ which is further exploited in global analyses. In the results, here presented we marginalised over $\theta_{13}$ without taking into account the constraints from reactor experiments discussed in Subsection~\ref{subsec:ch3-reactor}.

%\begin{figure}
%\centering
%\includegraphics[width = 0.65\textwidth]{sol_gf2020}
%\caption{Allowed regions at 90\% and 99\% confidence level --- using solid and dashed lines --- in the $\theta_{21}$ - $\Delta m^2_{21}$ for KamLAND, and solar experiments ---in blue and black respectively. Filled contours correspond to the same allowed regions for the combined analysis. Results are shown for two degrees of freedom, after marginalising over $\theta_{13}$, without including information from short-baseline reactor experiments \labfig{fig:ch3-solarbf2020}.}
%\end{figure}

\begin{figure}
\centering
\includegraphics[width = 0.61\textwidth]{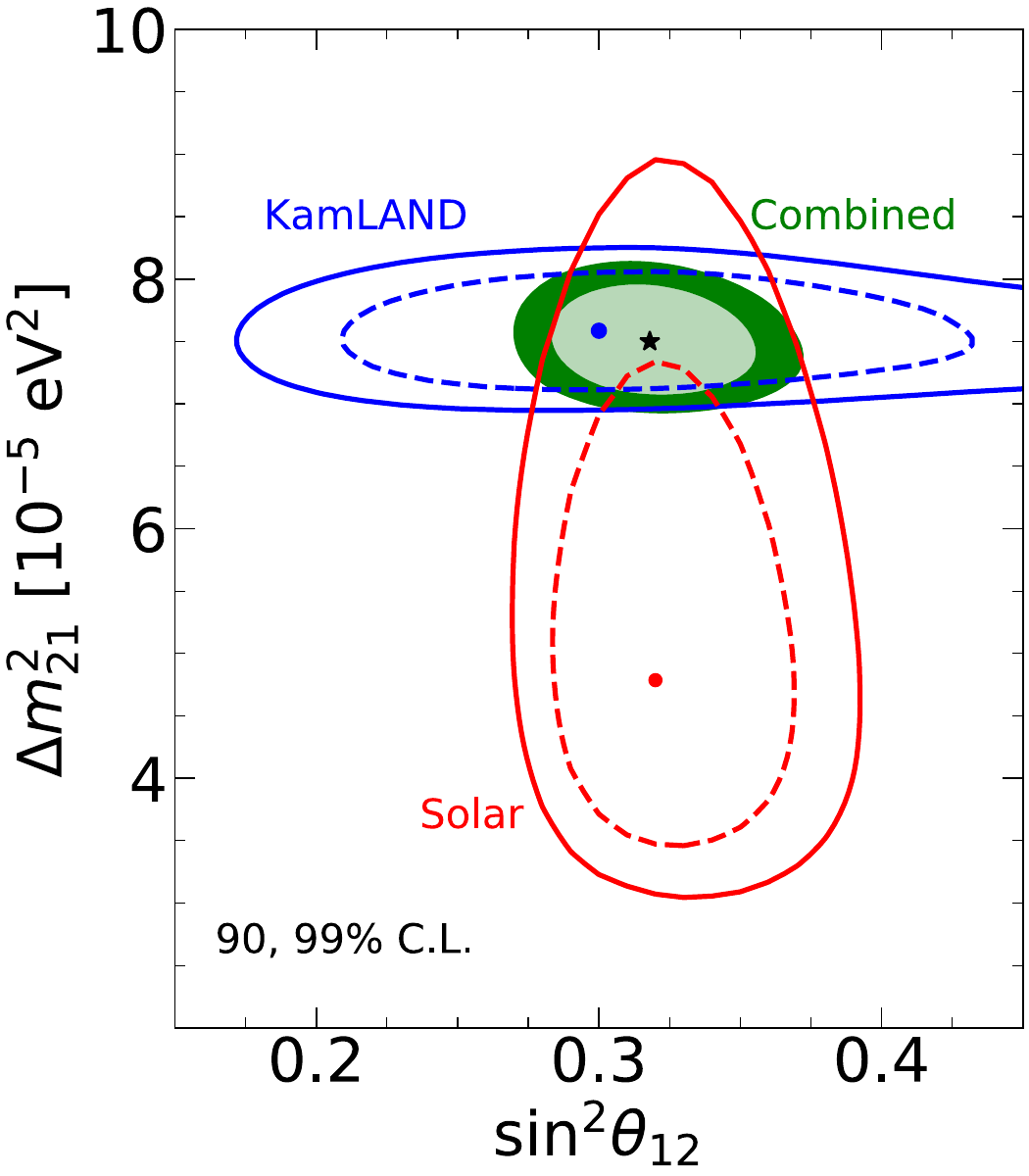}
\caption{Allowed regions at 90\% and 99\% confidence level --- using solid and dashed lines --- in the $\theta_{21}$ - $\Delta m^2_{21}$ for KamLAND, and solar experiments ---in blue and red respectively. Filled contours correspond to the same allowed regions for the combined analysis. Results are shown for two degrees of freedom, after marginalising over $\theta_{13}$, without including information from short-baseline reactor experiments \labfig{fig:ch3-solarbf2020}.}
\end{figure}

Notice that solar experiments find a best-fit value of $\Delta m^2_{21} = 4.8 \times 10 ^{-5} \text{eV}^2$, which is excluded with a high confidence level by KamLAND. However, there is considerable overlap in the preferred regions at the 99\% confidence level.

The final results from Super-Kamiokande IV were presented in~\cite{yasuhiro_nakajima_2020_4134680}. Using 2970 days of data, the Collaboration reported a larger ratio between the data and the prediction in the absence of oscillations and a larger day-night asymmetry. Both facts result in a preference for a larger $\Delta m^2_{21}$ than in previous analyses. Hence, a better agreement in the determination of the solar mass splitting is expected once this dataset is incorporated into the global fit. This is a matter of relevance for the consolidation of the three-neutrino oscillation picture. However, in practice, the measurement of $\Delta m^2_{21}$ is dominated by KamLAND and will be only mildly affected by this result.

%%%%%%%%%%%%%%%%%%%%%%%%%%%
\subsection{Short-baseline reactor experiments}
\label{subsec:ch3-reactor}
%%%%%%%%%%%%%%%%%%%%%%%%%%%
In short-baseline reactor experiments like RENO~\cite{RENO:2018dro}, Daya Bay~\cite{DayaBay:2018yms} and Double Chooz~\cite{DoubleChooz:2019qbj}, detectors are placed close to the nuclear power plants. Such experiments are sensitive to the reactor mixing angle $\theta_{13}$ and the absolute value of the mass splitting~\cite{Nunokawa:2005nx}
 \begin{align}
\Delta m^2_{ee} = \cos^2\theta_{12} \Delta m^2_{31} + \sin^2\theta_{32}\Delta m^2_{32}\, ,
\end{align}
through the study of electron antineutrino disappearance. 

In the analysis presented here, we include only the latest data available from inverse beta decay followed by neutron capture in Gadolinium from Daya Bay and RENO. We comment on the implications of other measurements from these experiments as well as from Double Chooz~\cite{DoubleChooz:2011ymz,DoubleChooz:2012gmf,DoubleChooz:2014kuw,DoubleChooz:2019qbj} at the end of the subsection.

\textbf{RENO}

The Reactor Experiment for Neutrino Oscillation (RENO) is located at the Hanbit Nuclear Power Plant in South Korea. Electron antineutrinos are produced by six reactors with thermal powers of 2.6 GW$_{\text{th}}$ or 2.8 GW$_{\text{th}}$, which are equally separated along a 3 km line. Two detectors of 16 tons each are positioned at 0.295 km and 1.383 km from the centre of the line along which the reactors are placed. In~\cite{jonghee_yoo_2020_4123573}, the RENO Collaboration reported their results after 2900 days of data taking, resulting in a value of the reactor mixing angle and mass splitting of 
\begin{align}
&\sin^22\theta_{13} = 0.0892 \pm 0.0063 \quad \text{and} \nonumber\\
&|\Delta m^2_{ee}| = (2.74\pm 0.12)\times 10^{-3}\text{eV}^2\, ,
\end{align} 
respectively. Our analysis takes into account the events in the near and far detectors after background subtraction. As in~\cite{RENO:2016ujo}, we also assume Gaussian energy smearing with an energy resolution
\begin{align}
\frac{\sigma(E)}{E} = \frac{0.07}{\sqrt{E \text{[MeV]}}}\, .
\end{align}
We define the following $\chi^2$ function,
\begin{equation}
    \chi^2 = \sum_i^{N_{\text{bins}}} \frac{\left(\mathcal{O}^{F/N}_i - T^{F/N}_i\right)^2}{\sigma_i^2} + \sum_j^{N_{\text{pulls}}} \left(\frac{\xi_j}{\sigma_{\xi_j}}\right)\, ,
\end{equation}
which depends on the observed and predicted far-over-near ratios, $\mathcal{O}^{F/N}_i$ and $T^{F/N}_i$, respectively. 

A total of six pull parameters, $\xi_j$, were considered in order to account for systematic uncertainties, including a difference in efficiency between both detectors ($\sigma_n = 0.22$\%), the fission fractions ($\sigma_f = 0.7$\%) and the energy scale on each detector ($\sigma_e = 0.15$\%). The $\chi ^2 $ function depends on $\sigma_i = \sigma(F/N)_i$,
\begin{equation}
    \sigma(F/N)_i ^2 = \left(\frac{\sigma(F)_i}{N_i}\right)^2 + \left(\frac{F_i\sigma(N)_i}{N_i^2}\right)^2\, ,
\end{equation}
where the statistical and systematic error in both near $\sigma(N)_i$ and far $\sigma(F)_i$ detectors in the $i$-th energy bin is given as part of the data release.

\textbf{Daya Bay}

The Daya Bay reactor neutrino experiment studies the disappearance of electron antineutrinos produced by six reactor cores at the power plants of Daya Bay and Ling Ao. It consists of two near experimental halls --- referred to as EH1 and EH2 --- with two detectors each, and four detectors at the far experimental hall --- or EH3. The distance from the cores to the different experimental halls ranges from 0.3 to 1.3 km in the near experimental halls and from 1.5 to 1.9 km in the far one. After analysing the data collected during 1958 days, the collaboration reported the measurement of the oscillation parameters~\cite{DayaBay:2018yms}
\begin{align}
&\sin^2 2\theta_{13} = 0.856 \pm 0.0029 \quad \text{and} \nonumber\\ 
&|\Delta m^2_{ee}| = (2.522^{+0.068}_{0.07})\times 10^{-3} \text{eV}^2 
\end{align}

In our analysis, we consider the ratio events between EH3 and EH1 and also between EH2 and EH1 after background subtraction. The information regarding the fusion fractions and thermal power of the reactors was determined from specific studies (see Table 9 in~\cite{DayaBay:2016ssb} and Table I in~\cite{DayaBay:2016ggj}, respectively). The collaboration performed three different analyses using a covariance matrix, nuisance parameters and a hybrid approach, finding consistent results with the three methods. We chose to use nuisance parameters in our analysis, with the following $\chi^2$ definition:
\begin{align}
    \chi^2 &= \sum_i^{N_{bins}} \frac{\left(\mathcal{O}^{\text{ EH3/EH1}}_i - T^{\text{ EH3/EH1}}_i\right)^2}{\sigma_i^2} \nonumber \\ & \hspace{2cm}+ \sum_i^{N_{bins}} \frac{\left(\mathcal{O}^{\text{ EH2/EH1}}_i - T^{\text{ EH2/EH1}}_i\right)^2}{\sigma_i^2}
   + \sum_j^{N_{\text{pulls}}} \left(\frac{\xi_j}{\sigma_{\xi_j}}\right)\, ,
\end{align}

where $\mathcal{O}^{\text{ EH3/EH1}}_i$ and $\mathcal{O}^{\text{ EH2/EH1}}_i$ are the observed far-over-near ratios and $T^{\text{ EH3/EH1}}_i$ and $T^{\text{ EH2/EH1}}_i$ are the predicted ones. We account for the uncertainties from the fission fractions and the thermal power at each of the nuclear reactors, $\sigma_{\text{frac}} = 0.1\%$ and $\sigma_{r} = 0.2\%$ respectively. We also account for different efficiencies and running times of each detector and for other sources of uncertainties, such as shifts in the energy scale ($\sigma_{\text{scale}}= 0.6\%$).

\textbf{Reactor plane}

Given the smallness of the solar mass splitting $\Delta m^2_{21}$ determined by KamLAND, the sensitivity of reactor experiments to the parameters of the solar sector $\theta_{21}$ and $\Delta m^2_{21}$ is only marginal~\cite{Hernandez-Cabezudo:2019qko}. In \reffig{fig:ch3-reac_2020} we present our results in the $\sin^2\theta_{13}$ - $|\Delta m^2_{31}|$ plane, for the two different mass orderings and after fixing the values of $\theta_{12}$ and $\Delta m^2_{21}$ to their best-fit value from solar neutrino experiments and KamLAND.

The data included in this analysis corresponds only to the inverse beta decay events followed by neutron capture in Gadolinium. However, Daya Bay and RENO have also presented analyses of part of the data corresponding to neutron capture in hydrogen~\cite{RENO:2019otc,DayaBay:2016ziq,Kohn:2021kmh}. These results were not included in the global fit in~\cite{deSalas:2020pgw} since, given their precision, they would barely help determine the oscillation parameters more precisely. Nonetheless, they provide a very valuable cross-check of our understanding of flavour oscillations and detection strategies. 

\begin{figure*}[t!]
\includegraphics{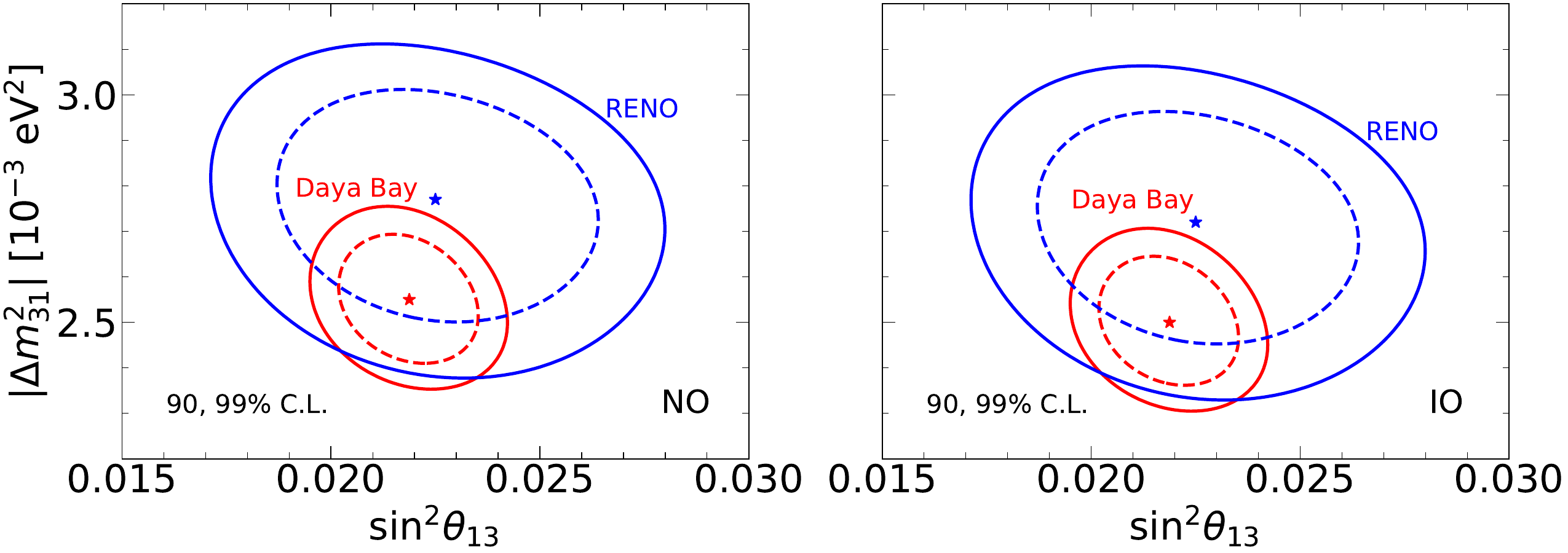}
\caption{Allowed regions at 90\% and 99\% confidence levels --- using dashed and solid lines respectively ---- in the $\sin^2 \theta_{13}$ - $|\Delta m^2_{31}|$ plane for 2 degrees of freedom for RENO and Daya Bay --- with the values of $\theta_{12}$ and $\Delta m^2_{21}$ fixed. Best fit points are indicated with a star. The left and right panels correspond to normal (NO) and inverted ordering (IO) respectively. \labfig{fig:ch3-reac_2020}}
\end{figure*}

In addition to the experiments mentioned above, the Double Chooz Experiment also measured reactor antineutrinos in France. It was the first experiment to observe the disappearance of electron antineutrinos from neutrino~\cite{DoubleChooz:2011ymz}, pointing towards a non-zero value \mbox{of $\theta_{13}$.} Recently, the Double Chooz Collaboration also performed the first measurement of $\theta_{13}$ via total neutron capture~\cite{DoubleChooz:2019qbj}, i.e. considering neutron capture in all the targets --- Gadolinium, Hydrogen and Carbon. Again, and for analogous reasons, these results are not part of the analysis in~\cite{deSalas:2020pgw}.

%%%%%%%%%%%%%%%%%%%%%%%%%%%
\subsection{Atmospheric neutrino experiments}
%%%%%%%%%%%%%%%%%%%%%%%%%%%
Cosmic rays hitting the atmosphere produce a cascade of particles including multiple kaons and pions. Atmospheric neutrinos are produced as secondary particles from the decay of these mesons. The energy of these muons and electron neutrinos --- and antineutrinos --- ranges from a few MeV to approximately $10^9$ GeV. However, flavour oscillations studies focus on neutrino energies between 100 MeV and 100 GeV. Atmospheric neutrino experiments sensitivity strongly relies on the observation of muon neutrino disappearance, hence being sensitive to $\sin^2\theta_{23}$ and $\Delta m^2_{31}$. Note, however, that the observation of electron neutrino appearance gives neutrino telescopes a partial sensitivity to the Dirac CP phase, $\delta_{\text{CP}}$, and to the reactor mixing angle $\theta_{13}$. In the global fit~\cite{deSalas:2020pgw}, we include data from IceCube DeepCore~\cite{IceCube:2017lak,IceCube:2019dqi} and Super-Kamiokande~\cite{Super-Kamiokande:2017yvm,Super-Kamiokande:2019gzr}.

The analysis of the Super-Kamiokande atmospheric data can not be precisely reproduced outside the collaboration due to its complexity. Consequently, in our global fit we include directly the $\chi^2$ tables made available by the collaboration.\footnote{\href{http://www-sk.icrr.u-tokyo.ac.jp/sk/publications/data/sk.atm.data.release.tar.gz}{http://www-sk.icrr.u-tokyo.ac.jp/sk/publications/data/sk.atm.data.release.tar.gz}} Although updated results have been presented --- see for instance~\cite{linyan_wan_2022_6694761} --- those are preliminary and the corresponding $\chi^2$ tables that would update the analysis have not been included in our global fit yet. 

Regarding the IceCube Collaboration, the latest data release corresponds to three years of data taking and includes both track-like and shower-like events. Two different sets of selection cuts are applied, defining the so-called Sample A and Sample B~\cite{IceCube:data}. We performed our analysis using Sample A and we have included several different systematics related either to the flux or the detector. For instance, we take into account the uncertainties in the optical efficiencies, neutrino scattering and absorption on ice, uncertainties on the ratio of neutrino to antineutrino and on the ratio of muon to electron antineutrinos or the overall normalisation, among others.

\reffig{fig:ch3-atm_2020} shows the result of our analysis of IceCube DeepCore data and the one from the Super-Kamiokande Collaboration. There is good agreement between both experiments although one can see that \mbox{Super-Kamiokande} provides a better measurement of the mixing angle, whereas IceCube DeepCore gives a better determination of the atmospheric mass splitting.

%\begin{figure*}
%\includegraphics{atm_gf2020}
%\caption{Allowed regions at 90\% and 99\% confidence levels  --- using dashed and solid lines respectively --- in the $\sin^2 \theta_{23}$ - $|\Delta m^2_{31}|$ plane for 2 degrees of freedom for Super-Kamiokande (SK) and IceCube DeepCore. Best fit points are indicated with a star. The left and right panels correspond to normal (NO) and inverted ordering (IO) respectively. \labfig{fig:ch3-atm_2020}}
%\end{figure*}

\begin{figure*}
\includegraphics{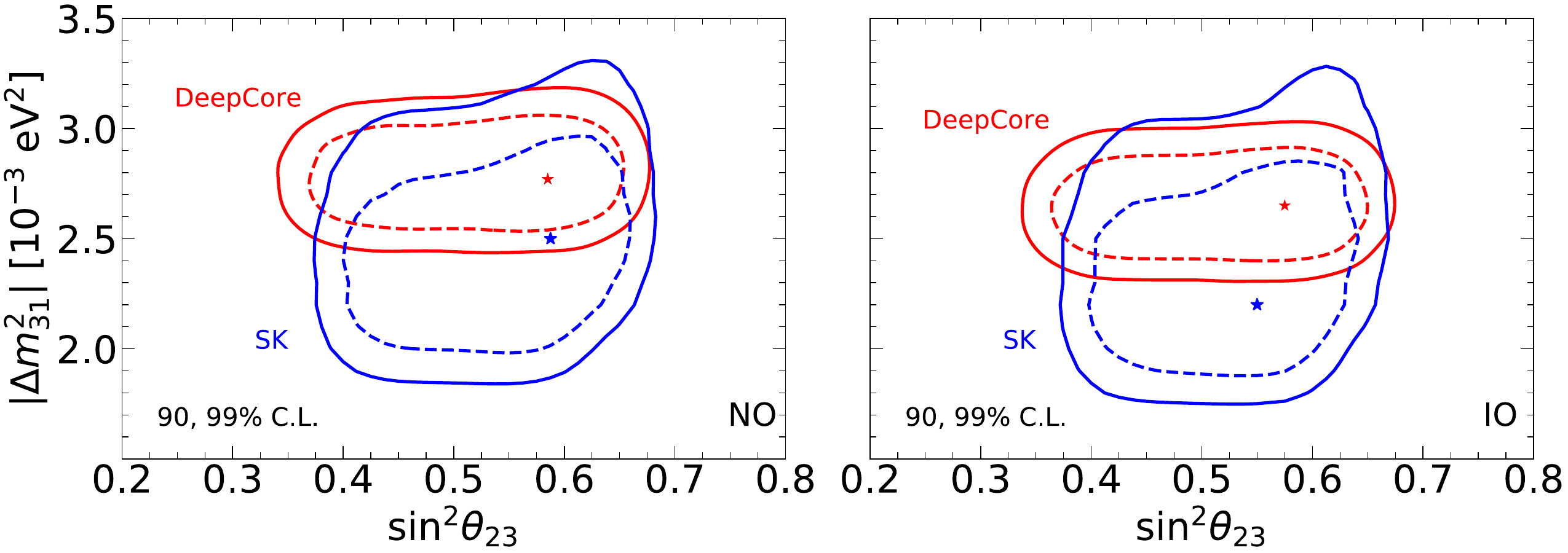}
\caption{Allowed regions at 90\% and 99\% confidence levels  --- using dashed and solid lines respectively --- in the $\sin^2 \theta_{23}$ - $|\Delta m^2_{31}|$ plane for two degrees of freedom for Super-Kamiokande (SK) and IceCube DeepCore. Best fit points are indicated with a star. The left and right panels correspond to normal (NO) and inverted ordering (IO) respectively. \labfig{fig:ch3-atm_2020}}
\end{figure*}

%%%%%%%%%%%%%%%%%%%%%%%%%%%
\subsection{Long-baseline accelerator experiments}
%%%%%%%%%%%%%%%%%%%%%%%%%%%
Long-baseline accelerator experiment study flavour oscillations using a human-made beam of neutrinos originating at particle accelerators. Protons are impinged into a target, producing mesons --- mainly kaons and pions --- which are focused using magnetic horns and allowed to decay. The polarisation of the focusing horns is used to separate mesons from antimesons. The result after they decay is a relatively pure beam of neutrinos or antineutrinos. Nevertheless, small contamination of so-called \textit{wrong-sign} neutrinos is always present. Long-baseline oscillation experiments generally consist of at least one near detector, which allows the characterisation of the initial flux, and a far detector, which measures the oscillated flux. The study of muon neutrino disappearance and electron neutrino appearance makes long-baseline experiments sensitive to $\theta_{23}$, $\theta_{13}$, $\Delta m^2_{31}$ and $\delta_{\text{CP}}$.

\begin{figure*}
\includegraphics{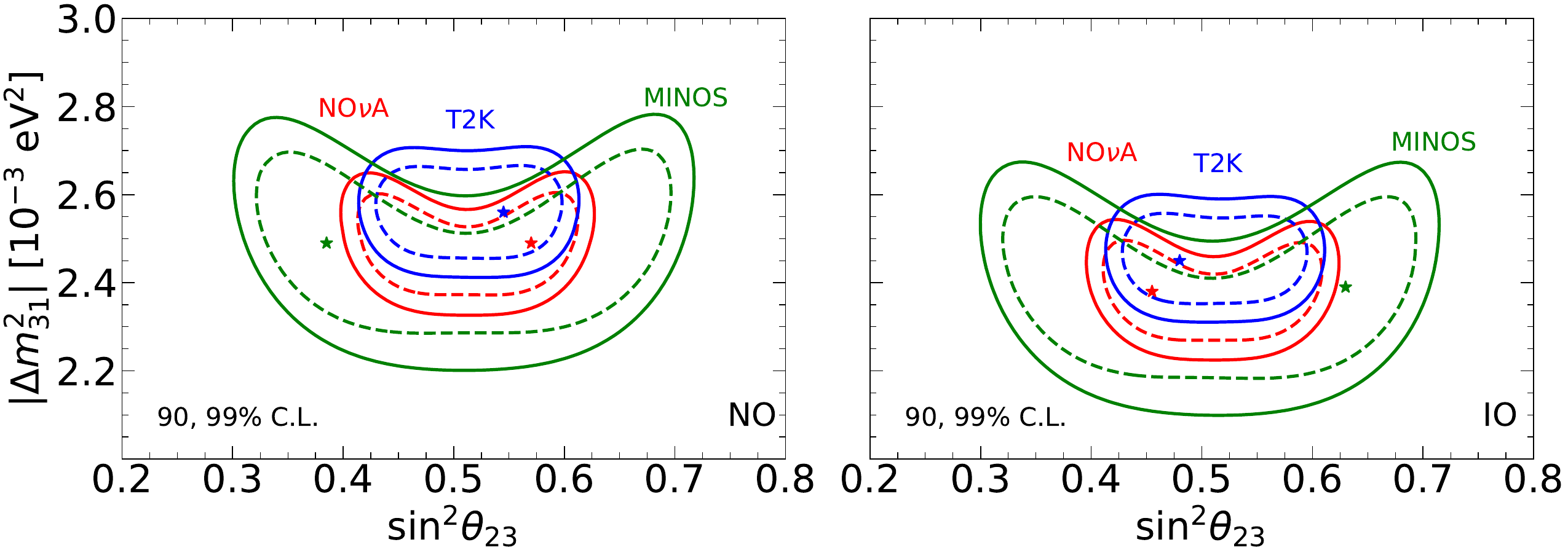}
\includegraphics{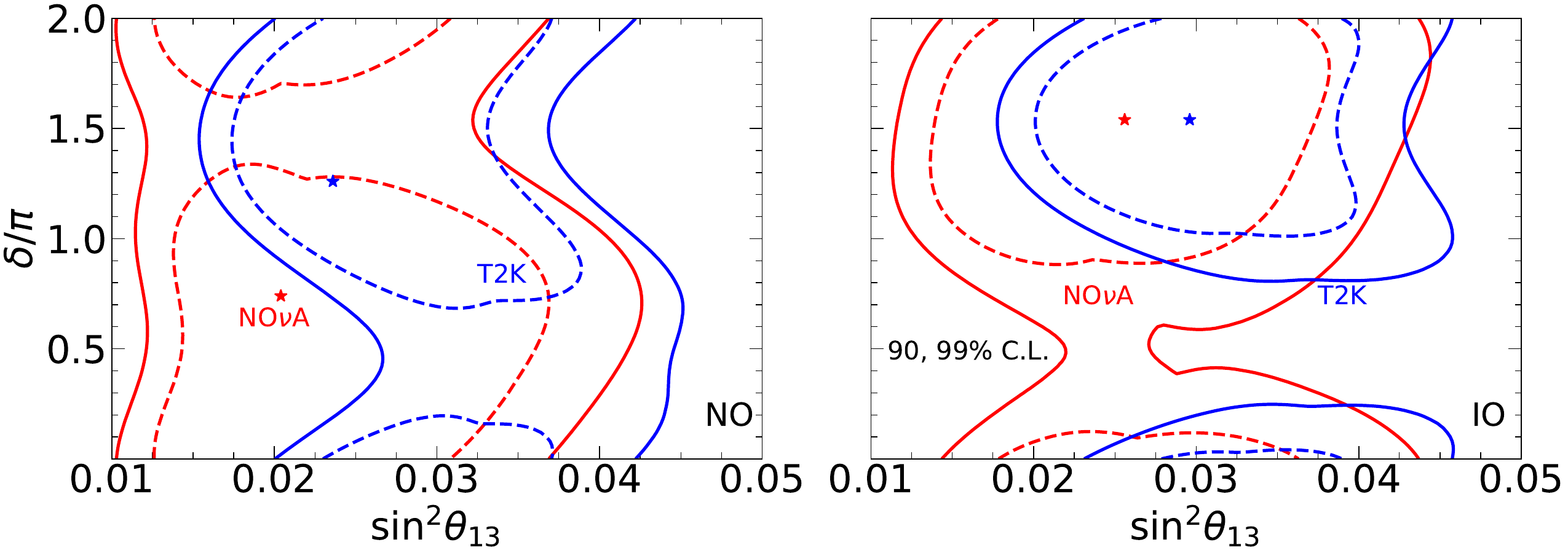}
\caption{Upper panels show the allowed regions at 90\% and 99\% confidence level for 2 degrees of freedom in the $\sin^2\theta_{23}$ - $|\Delta m^2_{31}|$ plane for NOvA, T2K and MINOS --- in red, blue, and green respectively. The lower panel depicts the allowed regions at the same confidence levels in the $\sin^2\theta_{13}$ - $\delta_{\text{CP}}$ plane for NOvA and T2K. The left panels and right panels correspond to normal (NO) and inverted ordering (IO) respectively. \labfig{fig:ch3-lbl_2020}}
\end{figure*}

In our analysis, we include the results from the following experiments: NOvA~\cite{alex_himmel_2020_4142045}, T2K~\cite{patrick_dunne_2020_4154355}, MINOS~\cite{MINOS:2014rjg} and K2K~\cite{K2K:2006yov}. In the upper panels of \reffig{fig:ch3-lbl_2020} we show the results from our analysis in the $\sin^2 \theta_{23}$ - $|\Delta m^2_{31}|$  plane for NOvA, T2K and MINOS. The lower panels show the results for the $\delta_{\text{CP}}$ - $\sin^2\theta_{13}$ plane. MINOS is not sensitive to the latter pair of parameters. Similarly, although K2K is included in the analysis, its sensitivity has been overcome by more recent experiments.

%%%%%%%%%%%%%%%%%%%%%%%%%%%
%%%%%%%%%%%%%%%%%%%%%%%%%%%
\section{Global fit to neutrino oscillation data \label{sec:ch3-results}}
%%%%%%%%%%%%%%%%%%%%%%%%%%%
%%%%%%%%%%%%%%%%%%%%%%%%%%%
\subsection[The knowns: $\theta_{12}, \, \theta_{13}, \, \Delta m^2_{21}$ and $|\Delta m^2_{31}|$]{The knowns: $\mathbf{\theta_{12}, \, \theta_{13}, \, \Delta m^2_{21}}$ and $\mathbf{|\Delta m^2_{31}|}$}
\label{subsec:knowns}
%%%%%%%%%%%%%%%%%%%%%%%%%%%
At present, our knowledge of the parameters of the solar sector, $\theta_{12}$ and $\Delta m^2_{21}$, comes solely from the combination of KamLAND and solar neutrino experiments. Both datasets also show a marginal dependence on the reactor mixing angle $\theta_{13}$. Hence, after combining all datasets in the fit, the measurement of the solar parameters improves slightly since the degeneracy between $\theta_{12}$ and $\theta_{13}$ is lifted.
The reactor mixing angle $\theta_{13}$ is also well-measured and dominated by reactor experiments. Including the latest dataset from Daya Bay~\cite{DayaBay:2022orm} will further improve the precision of this measurement. In contrast, the absolute value of the atmospheric mass splitting $|\Delta m^2_{31}|$ receives contributions from reactor and long-baseline experiments, and to a minor extent, from the measurement of atmospheric neutrinos. 

Shortly, the medium-baseline reactor experiment JUNO will significantly improve the measurement of the solar parameters, $\theta_{12}$ and $\Delta m^2_{21}$, while also providing a good measurement of $\theta_{13}$ and $|\Delta m^2_{31}|$, together with a direct determination of the mass ordering~\cite{JUNO:2022mxj}.

%%%%%%%%%%%%%%%%%%%%%%%%%%%
\subsection{The unknown octant of $\mathbf{\theta_{23}}$}
%%%%%%%%%%%%%%%%%%%%%%%%%%%
Long-baseline accelerator and atmospheric neutrino experiments are mainly sensitive to the disappearance of muon neutrinos --- and antineutrinos. Hence, they measure $\sin^2 2\theta_{23}$ accurately. In the case of atmospheric neutrinos, the octant degeneracy --- i.e. whether $\theta_{23} > \pi/4$ or $\theta_{23} < \pi/4$ --- is slightly lifted due to the matter effects experienced by neutrinos while travelling across the Earth. Besides that, the observation of electron neutrino appearance provides a direct measurement of $\sin^2 \theta_{23}$. As a consequence, this channel breaks the octant degeneracy too. There is an additional degeneracy between the atmospheric and reactor mixing angles, since a smaller value of $\sin^2 \theta_{23}$ can be compensated by a larger value of $\sin^2\theta_{13}$. Global fits exploit the fact that reactor experiments determine $\theta_{13}$ very precisely in a way that is independent of $\theta_{23}$. This fact results in the aforementioned degeneracy being further lifted. 

%\begin{figure*}
%\includegraphics{sq23_sq13_2020}
%\caption{Allowed regions at 90\% and 99\% confidence level regions for 2 degrees of freedom in the $\sin^2\theta_{23}$ - $\sin^2\theta_{13}$ plane. Results for long-baseline (LBL) alone, with atmospheric data (LBL+ATM) and with reactor data (LBL+REAC) are shown in black, blue, and red, respectively. The allowed regions from the global fit correspond to the coloured regions. The left and right panels correspond to normal (NO) and inverted ordering (IO) respectively. \labfig{fig:ch3-sq23_sq13}}
%\end{figure*}

\begin{figure*}
\includegraphics{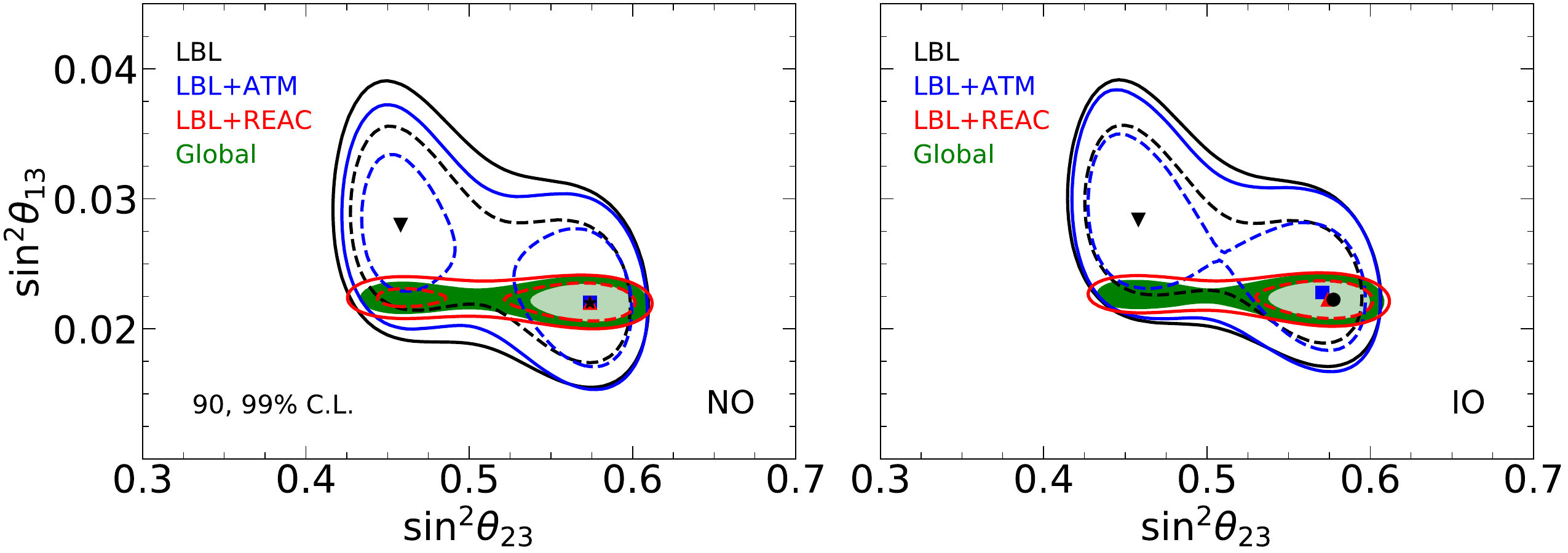}
\caption{Allowed regions at 90\% and 99\% confidence level regions for two degrees of freedom in the $\sin^2\theta_{23}$ - $\sin^2\theta_{13}$ plane. Results for long-baseline (LBL) alone, with atmospheric data (LBL+ATM) and with reactor data (LBL+REAC) are shown in black, blue, and red, respectively. The allowed regions from the global fit correspond to the filled regions. Left and right panels correspond to normal (NO) and inverted ordering (IO) respectively. \labfig{fig:ch3-sq23_sq13}}
\end{figure*}

In \reffig{fig:ch3-sq23_sq13}, we show the confidence regions in the plane $\sin^2\theta_{23} - \sin^2\theta_{13}$ from long-baseline accelerator experiments individually, combined with atmospheric and reactor data, and from the global fit of~\cite{deSalas:2020pgw}. One can see the role of atmospheric and reactor data in moving the preference to the upper octant and breaking the degeneracy between the atmospheric and reactor mixing angles. Nevertheless, recent data from Super-Kamiokande atmospheric might shift the preference back to the upper octant~\cite{linyan_wan_2022_6694761}. The exact impact of this dataset is yet to be analysed.

%%%%%%%%%%%%%%%%%%%%%%%%%%%
\subsection[The CP phase $\delta_{\text{CP}}$]{The CP phase, $\mathbf{\delta_{\text{CP}}}$}
%%%%%%%%%%%%%%%%%%%%%%%%%%%
The measurement of the CP-violating phase relies on the fact that, in the presence of CP violation, i.e. for $\delta_{\text{CP}} \neq \lbrace 0, \pi\rbrace$, it induces opposite shifts in the electron neutrino appearance channel. Consequently, the joint analysis of $\nu_\mu \rightarrow \nu_e$ and $\bar{\nu}_\mu \rightarrow \bar{\nu}_e$ channels provides a measurement of this parameter. That is the origin of the sensitivity to $\delta_{\text{CP}}$ reported by the long-baseline accelerator experiments NOvA and T2K, and from Super-Kamiokande atmospheric data. In addition, the inclusion of reactor data can help improve the determination of $\delta_{\text{CP}}$, since it helps to break the degeneracy between $\theta_{13}$ and $\delta_{\text{CP}}$ in this channel.

\begin{figure*}
\includegraphics{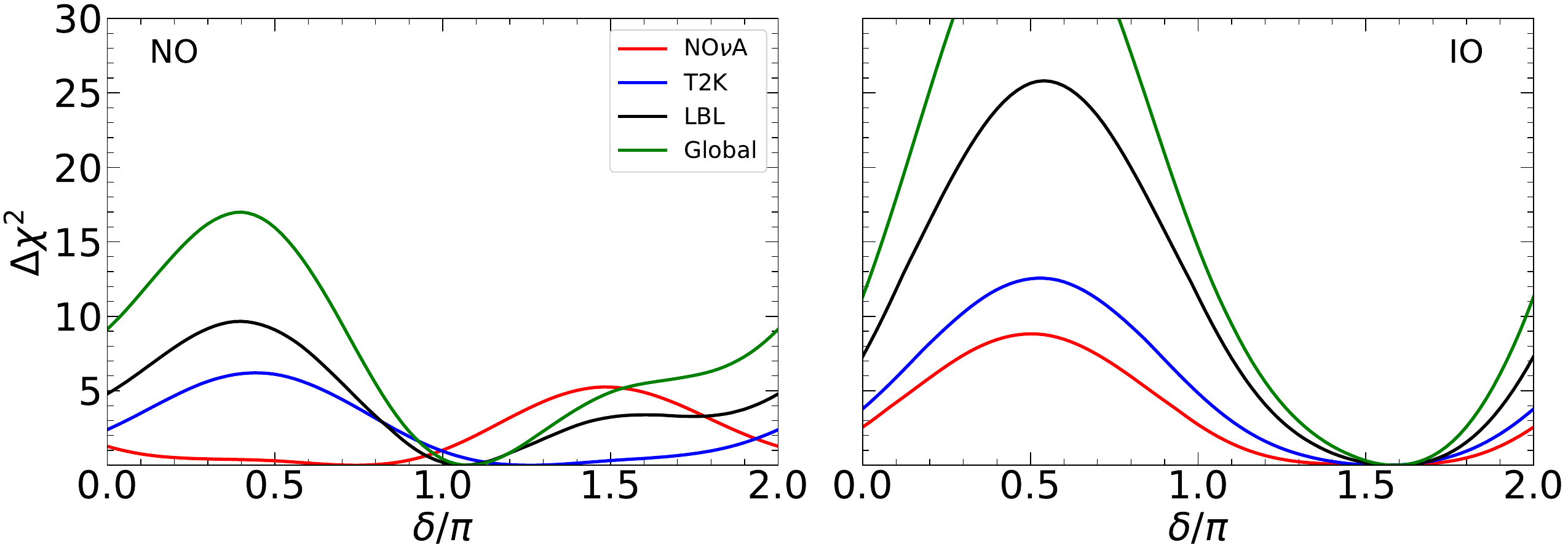}
\caption{$\Delta \chi^2 $ profiles for $\delta_{\text{CP}}$  from the analysis of NOvA, T2K, all long-baseline experiments (LBL) and from the global fit~\cite{deSalas:2020pgw}. The left panel and right panel correspond to normal (NO) and inverted ordering (IO) respectively. \labfig{fig:ch3-deltacp}}
\end{figure*}

The status of this open question is summarised in \reffig{fig:ch3-deltacp}. For normal ordering, there is a mismatch between the preferred values for T2K and NOvA. The best fit value of the combination is found for $\delta_{\text{CP}}$ = 1.08$\pi$, whereas the CP-conserving value $\delta_{\text{CP}}=0$ is disfavoured at $\Delta \chi^2 \approx 9.1$ and $\delta_{\text{CP}} = \pi$ is allowed with $\Delta \chi ^2 \approx 0.4$. For inverted ordering, both long-baseline experiments agree and the best-fit value of the global fit is found to be close to maximal CP-violation, particularly for $\delta_{\text{CP}} = 1.58\pi$. The CP-conserving values $\delta_{\text{CP}} = 0$ and $\delta_{\text{CP}} = \pi$ are excluded at $\sim 3.4\sigma$ and $\sim 3.8\sigma$, respectively.
%with $\Delta \chi^2 \approx 11.3$ and $\Delta \chi^2 \approx 14.6$, respectively.

%%%%%%%%%%%%%%%%%%%%%%%%%%%
\subsection{The neutrino mass ordering}
%%%%%%%%%%%%%%%%%%%%%%%%%%%
The last unknown in the three-neutrino oscillation picture is the mass ordering. Individually, long-baseline analyses are marginally sensitive to the ordering due to the subleading role played by matter effects. However, a combined fit of all available accelerator data shows a preference for inverted ordering due to the mismatch in the preferred value of $\delta_{\text{CP}}$ between T2K and NOvA in normal ordering. When combining accelerator and reactor data, the preference shifts towards normal ordering. That is a result of the mismatch between the preferred value of the atmospheric mass splitting $|\Delta m^2_{31}|$ between both sets of experiments. In parallel, from matter effects experienced by atmospheric neutrinos, both IceCube DeepCore and Super-Kamiokande prefer normal ordering. Again, after combining with reactor data, the small tension in the value of the mass splitting preferred by atmospheric and reactor neutrino experiments enhances the preference for normal ordering.

The quantitative status of this matter in~\cite{deSalas:2020pgw} was the following. T2K and NOvA separately prefer NO with $\Delta \chi^2 \approx 0.4$ Nonetheless, the combination of all accelerator data shows a preference for inverted ordering of 2.4 units of $\Delta \chi^2$, and after combining with reactor data, the preference is shifted back to normal ordering with $\Delta \chi^2\approx 1.4$. Regarding atmospheric neutrinos, \mbox{Super-Kamiokande} and IceCube DeepCore prefer NO with $\Delta \chi^2 \approx 3.5$ and $\Delta \chi^2 \approx 1.0$, respectively. The combination of all datasets from reactor, atmospheric and long-baseline experiments results in a preference for NO of $\Delta \chi^2 \approx 6.4$, which corresponds to a significance of 2.5$\sigma$. However, this preference for normal ordering is still far from conclusive and could rapidly change once ongoing analyses by IceCube DeepCore, Super-Kamiokande atmospheric data, T2K and NOvA are released. Likewise, the medium baseline reactor experiment JUNO aims to provide a direct measurement of the ordering from the precise measurement of the oscillated spectrum~\cite{JUNO:2015zny}.

%%%%%%%%%%%%%%%%%%%%%%%%%%%
%%%%%%%%%%%%%%%%%%%%%%%%%%%
\section{Bayesian approach to neutrino data analysis \label{sec:ch3-bayes}}
%%%%%%%%%%%%%%%%%%%%%%%%%%%
%%%%%%%%%%%%%%%%%%%%%%%%%%%
\subsection{Bayesian analysis}
%%%%%%%%%%%%%%%%%%%%%%%%%%%
For our Bayesian analysis of neutrino oscillation data, we convert the \mbox{$\chi^2$-functions} described in the previous sections into a likelihood, as follows
\begin{align}
\ln \mathcal{L} = - \frac{\chi^2}{2}\, ,
\end{align}
and consider flat priors on the oscillation parameters. We use \texttt{MontePython}~\cite{Audren:2012wb,Brinckmann:2018cvx} to compute the likelihoods and run the Markov Chain Monte Carlo (MCMC) simulations. We also rely on \texttt{MontePython} to post-process the MCMC outputs and obtain the marginalised posteriors and credible intervals. 

To quantify the preference of one neutrino mass ordering over the other one, we compute the Bayes factor $B_{\text{NO, IO}} =  \mathcal{Z}_{\text{NO}}/\mathcal{Z}_{\text{IO}}$. We calculate the marginalised likelihood or model evidence, $\mathcal{Z}$, for NO and IO, by means of \texttt{MCEvidence}~\cite{Heavens:2017afc}. In this way, we derive that neutrino oscillation data alone give now 
\begin{align} 
\text{ln}B_{\text{NO, IO}} = 3.01 \pm 0.04\, .
\end{align} 
For a Gaussian variable, this value of the Bayes factor corresponds to a $2.0 \sigma$ probability.

%%%%%%%%%%%%%%%%%%%%%%%%%%%
\subsection{Absolute neutrino mass scale and mass ordering}
%%%%%%%%%%%%%%%%%%%%%%%%%%%

Neutrino oscillations are only sensitive to the mass splittings and not to the absolute neutrino mass scale. Thus, weighting neutrinos requires the input of additional datasets. In this subsection, we describe the contribution of \mbox{$\beta$-decay} experiments, $0\nu\beta\beta$-decay searches and cosmological probes to this purpose. We also discuss whether they help determine the mass ordering. Besides those probes, there exist other proposals to determine neutrino masses using supernova neutrinos~\cite{Pompa:2022cxc} or from the study of neutrino capture on a $\beta$-decaying target at an experiment like PTOLEMY~\cite{PTOLEMY:2019hkd}. These will not be discussed in the manuscript.

In this subsection, when discussing neutrino mass bounds, we impose a logarithmic prior in the mass of the lightest neutrino, in the range \mbox{[$10^{-3}$, 10] eV}~\cite{Gariazzo:2018pei,DeSalas:2018rby}. Conversely, when computing limits on $\sum m_\nu$, $m_\beta$ and $m_{\beta\beta}$, we consider a linear prior on the range [0, 10] eV. Such parametrisations and prior choices prevent artificial preferences for normal ordering~\cite{Gariazzo:2018pei,Gariazzo:2022ahe}.

\newpage
\textbf{The endpoint of the $\mathbf{\beta}$-decay spectrum.}

Currently, the KATRIN experiment holds the strongest limit on the effective electron antineutrino mass: $m_\beta < 0.8$ eV at 90\% C.L.~\cite{KATRIN:2021uub}. This bound applies irrespectively of whether neutrinos are Dirac or Majorana particles. In our analysis, we take into account the results from KATRIN's first campaign~\cite{KATRIN:2019yun} --- which determined $m_\beta$ < 1.1 eV at 90\% C.L. --- through the approximated analytical likelihood proposed in~\cite{Huang:2019tdh}:
\begin{align}
\mathcal{L}_{\text{KATRIN}} \propto \frac{1}{\sqrt{2\pi} \sigma} \exp \left[-\frac{1}{2}\left(\frac{m^2_\beta - \mu}{\sigma}\right)^2\right] \text{Erfc}\left(-\frac{\alpha}{\sqrt{2}}\frac{m^2_\beta - \mu}{\sigma}\right)\, ,
\end{align}
where Erfc is the complementary error function, $\sigma$ = 1.506 eV$^2$, $\mu$ = 0.0162 eV$^2$, $\alpha$ = 2.005 and $m_\beta$ is in units of eV. We did not include data from previous experiments such as MAINZ ~\cite{Kraus:2004zw} and TROITSK~\cite{Aseev:2012zz}, since their constraints are much weaker than those from KATRIN.

\textbf{Neutrinoless double beta decay.}

Various experiments have set lower limits on the neutrinoless double-beta decay half-life --- denoted by $T^{0\nu}_{1/2}(\mathcal{N})$ --- using different isotopes, among which  $^{76}$Ge, $^{130}$Te and $^{136}$Xe. 

\begin{table}
\renewcommand*{\arraystretch}{1.2}
\centering
\begin{tabular}{cccc}
\toprule[0.25ex]
 Experiment & 90\% C.L. limit & Isotope & Reference \\
\midrule
GERDA & $T^{0\nu}_{1/2}(\mathcal{N})> 9 \times 10^{25}$ yr & $^{76}$Ge &~\cite{GERDA:2019ivs} \\
CUORE & $T^{0\nu}_{1/2}(\mathcal{N})> 3.2 \times 10^{25}$ yr & $^{130}$Te &~\cite{CUORE:2019yfd} \\
KamLAND-Zen & $T^{0\nu}_{1/2}(\mathcal{N})> 1.07 \times 10^{26}$ yr yr & $^{136}$Xe &~\cite{KamLAND-Zen:2016pfg} \\
\bottomrule[0.25ex]
\end{tabular}
\caption{\label{tab:ch3-0nubb}Strongest limits on the neutrinoless double-beta decay half-life at 90\% C.L. for different isotopes. }
\end{table}

We consider the bounds from the experiments in \reftab{ch3-0nubb} using approximate analytical expressions for the likelihoods, mainly
\begin{align}
&-\text{ln}\,\mathcal{L}_\text{GERDA} \propto -5.5 + 26.7 (T^{0\nu}_{1/2})
^{-1} + 38.4 (T^{0\nu}_{1/2})^{-2}\, , \\
&-\text{ln}\,\mathcal{L}_\text{CUORE} \propto 4.02 + 10.5 (T^{0\nu}_{1/2})
^{-1} + 8.6 (T^{0\nu}_{1/2})^{-2}\, , \\
&-\text{ln}\,\mathcal{L}_\text{KamLAND-Zen} \propto 9.71 (T^{0\nu}_{1/2})
^{-1} + 28.1 (T^{0\nu}_{1/2})^{-2} \, .
\end{align}
The expression for KamLAND-ZEN was proposed in~\cite{Caldwell:2017mqu}, whereas those for CUORE and GERDA have been obtained using the approach proposed in~\cite{Caldwell:2017mqu} and information from~\cite{CUORE:2019yfd} and~\cite{GERDA:2019ivs}, respectively.

In the analyses including constraints from $0\nu\beta\beta$-decays, we marginalise over the two Majorana phases. In addition, we account for the theoretical uncertainties in the nuclear matrix elements by varying them in the 1$\sigma$ ranges proposed in~\cite{Vergados:2016hso}, namely,
\begin{align}
\mathcal{M}^{^{76}\text{Ge}}_{0\nu} \in [3.35 , 5.75]\, , \\
\mathcal{M}^{^{130}\text{Te}}_{0\nu} \in [1.75 , 5.09]\, , \\
\mathcal{M}^{^{136}\text{Xe}}_{0\nu} \in [1.49 , 3.69]\, .
\end{align}

\textbf{Cosmological probes.}

The combination of the measurement of the cosmic microwave background (CMB) and baryon acoustic oscillations (BAO) set limits on the sum of neutrino masses, $\sum m_\nu$. However, owing to the anticorrelation between the sum of the neutrino masses and the Hubble parameter, it is relevant to consider constraints on the latter quantity too.

We consider the most recent CMB observations by Planck~\cite{Planck:2018nkj,Planck:2018vyg}. They include the temperature and polarisation spectra~\cite{Planck:2019nip}, determined in a wide range of multipoles, as well as the measurements of the lensing potential~\cite{Planck:2018lbu}. Regarding BAO observations, we include information from the 6dF~\cite{Beutler:2011hx}, SDSS DR7 Main Galaxy Sample (MGS)~\cite{Ross:2014qpa} and BOSS DR12~\cite{BOSS:2016wmc} galaxy redshift surveys. Measurements at redshift z = 0.45 constrain the Hubble parameter H(z)~\cite{Moresco:2016mzx}, and hence, the expansion history of the universe. We account for bounds from observations of Type Ia supernovae through the Pantheon sample~\cite{Pan-STARRS1:2017jku}. In this manuscript, we denote the analyses including the aforementioned datasets by \textit{COSMO}. In some cases, we include the recent local determination of the Hubble parameter, $H_0$ = 74.03 $\pm$ 1.42 km/s/Mpc from~\cite{Riess:2019cxk}. This will illustrate the impact of the $\sum m_\nu \, - \, H_0$ degeneracy.

We calculate the predicted cosmological observables using the Boltzmann solver code \texttt{CLASS}~\cite{Lesgourgues:2013bra,Blas:2011rf,Lesgourgues:2011re}. Our fiducial cosmological model is a minimal extension of $\Lambda$CDM, described by the baryon and cold dark matter densities $\Omega_b h^2$ and $\Omega_{\text{cdm}}h^2$, the angular size of the sound horizon at last-scattering $\theta_s$, the optical depth to reionisation $\tau$, the amplitude and tilt of the primordial scalar power spectrum $A_s$ and $n_s$, and the mass of the lightest neutrino, $m_{\text{lightest}}$.

\textbf{Results on the absolute mass scale and mass ordering.}

In \reffig{fig:ch3-mass_params} we report the allowed regions at 1$\sigma$ and 2$\sigma$ for the parameters $m_{\text{lightest}}$, $m_\beta$, $m_{\beta\beta}$ and $\sum m_\nu$. The results correspond to the analysis of the likelihoods derived from our oscillation analysis together with those from the \textit{COSMO} dataset --- referred to as OSC+COSMO. Those constraints are much stronger than the ones including solely information from terrestrial experiments, i.e. oscillation experiments, $\beta$-decay measurements and \mbox{$0\nu\beta\beta$-decay} searches.

\begin{table}
\renewcommand*{\arraystretch}{1.2}
\centering
\begin{tabular}{ccc}
\toprule[0.25ex]
data set & lnB$_{NO,IO}$ & N$\sigma$ \\
\midrule
OSC & 3.01 $\pm$ 0.04 & 2.00 \\
OSC + $\beta$ decay & 3.22 $\pm$ 0.03 & 2.07 \\
OSC + $0\nu\beta\beta$ & 3.46 $\pm$ 0.25 & 2.17 \\
OSC + COSMO &4.90 $\pm$ 0.50 & 2.68 \\
OSC + COSMO + $H_0$ & 4.98 $\pm$ 0.34 & 2.70 \\
\bottomrule[0.25ex]
\end{tabular}
\caption{\label{tab:ch3-NONI} Bayes factors and significance in terms of standard errors of normal versus inverted mass ordering for the different data combinations.}
\end{table}

\begin{figure*}[t]
\includegraphics{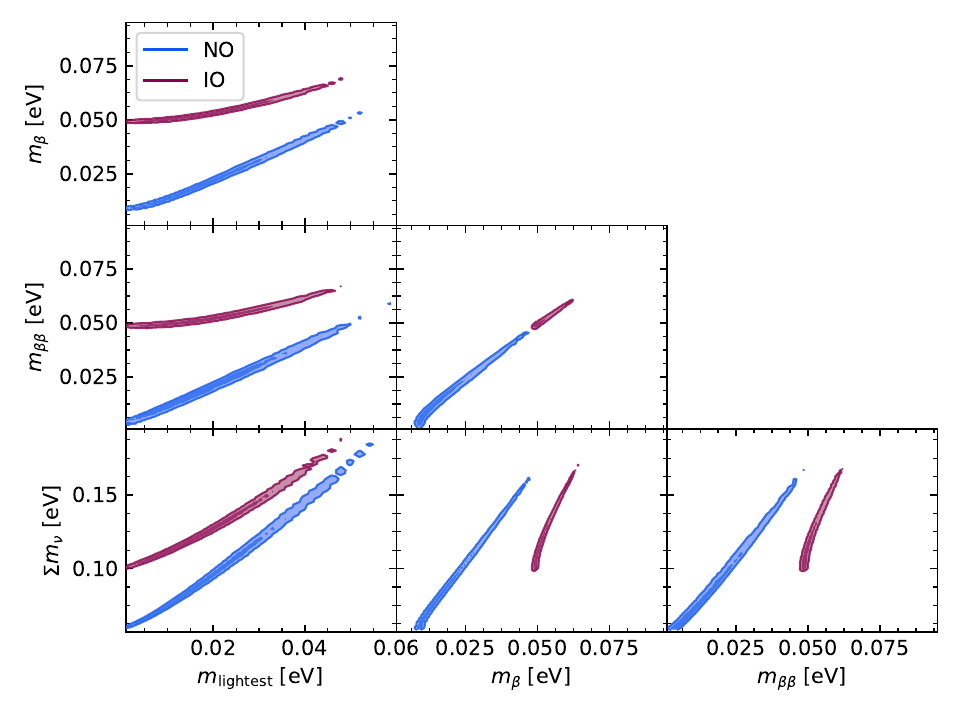}
\caption{Marginalised allowed regions at 1$\sigma$ and 2$\sigma$ --- darker and lighter areas respectively --- for $m_{\text{lightest}}$, $m_\beta$, $m_{\beta\beta}$and $\sum m_{\nu}$ --- for  the data combination we denote as OSC+COSMO. Blue and purple regions correspond to normal (NO) and inverted ordering (IO) respectively.\labfig{fig:ch3-mass_params}}
\end{figure*}

Finally, we present how the absolute mass measurement affects the preference for normal ordering reported from oscillation data only. We report the Bayes factor for different datasets in \reftab{ch3-NONI}. One can see that $\beta$-decay limits barely help to discriminate the ordering. For Majorana neutrinos, $0\nu\beta\beta$ experiments shift the preference for normal ordering from 2.00$\sigma$ --- obtained from considering oscillations only --- to 2.17$\sigma$. Besides, the significance increases even more when the constraints on the neutrino mass from cosmology are taken into account. In such case, when considering the OSC+COSMO dataset, we obtain a preference of  $\approx 2.68 \sigma$. This result does not vary significantly if a prior on the Hubble parameter is also included. For normal ordering, the minimum value of the sum of neutrino masses is $\sum m_\nu \approx 0.06$ eV, whereas, for inverted ordering, the minimum is $\sum m_\nu \approx 0.1$ eV. Since the available parameter-space volume is larger in the former case, when including the COSMO dataset, the preference for NO is enhanced.

\section{Summary of the global fit to neutrino data \label{sec:ch3-summary-fit}}
The determination of the parameters in the three-neutrino oscillation picture has become increasingly accurate in the last few years. In light of the data available in 2020, four of the parameters are well-measured. Those are the solar mixing angle $\theta_{12}$, the solar mass splitting $\Delta m^2_{21}$, the reactor mixing angle $\theta_{13}$ and the absolute value of atmospheric mass splitting $|\Delta m^2_{31}|$. However, there are still a few open questions, mainly the octant of $\theta_{23}$, the value of the CP phase $ \delta_{\text{CP}}$ and the mass ordering --- the sign of $\Delta m^2_{31}$. The best-fit values and confidence intervals for these parameters are summarised in \reftab{ch3-sum-2020} and the one-dimensional $\chi^2$-profiles depicted in \reffig{fig:ch3-sum_2020}. In addition, we have shown how measurements of the beta-decay endpoint, cosmological probes and, in the case of Majorana neutrinos, searches for neutrinoless double-beta decay, can shed light on the neutrino mass ordering and absolute mass scale.
\begin{table}
\renewcommand*{\arraystretch}{1.2}
\centering
\begin{tabular}{lccc}
\toprule[0.25ex]
parameter & best fit $\pm$ $1\sigma$ & \hphantom{x} 2$\sigma$ range \hphantom{x} & \hphantom{x} 3$\sigma$ range \hphantom{x}
\\ \midrule 
$\Delta m^2_{21} [10^{-5}$eV$^2$]  &  $7.50^{+0.22}_{-0.20}$  &  7.12--7.93  &  6.94--8.14  \\[2.4mm]
$|\Delta m^2_{31}| [10^{-3}$eV$^2$] (NO)  &  $2.55^{+0.02}_{-0.03}$  &  2.49--2.60  &  2.47--2.63  \\
$|\Delta m^2_{31}| [10^{-3}$eV$^2$] (IO)  &  $2.45^{+0.02}_{-0.03}$  &  2.39--2.50  &  2.37--2.53  \\[2.4mm]
$\sin^2\theta_{12} / 10^{-1}$         &  $3.18\pm0.16$  &  2.86--3.52  &  2.71--3.69  \\[2.4mm]

$\sin^2\theta_{23} / 10^{-1}$       (NO)  &  $5.74\pm0.14$  &  5.41--5.99  &  4.34--6.10  \\
$\sin^2\theta_{23} / 10^{-1}$       (IO)  &  $5.78^{+0.10}_{-0.17}$  &  5.41--5.98  &  4.33--6.08  \\[2.4mm]

$\sin^2\theta_{13} / 10^{-2}$       (NO)  &  $2.200^{+0.069}_{-0.062}$  &  2.069--2.337  &  2.000--2.405  \\
$\sin^2\theta_{13} / 10^{-2}$       (IO)  &  $2.225^{+0.064}_{-0.070}$  &  2.086--2.356  &  2.018--2.424  \\[2.4mm]

$\delta/\pi$                        (NO)  &  $1.08^{+0.13}_{-0.12}$  &  0.84--1.42  &  0.71--1.99  \\
$\delta/\pi$                        (IO)  &  $1.58^{+0.15}_{-0.16}$  &  1.26--1.85  &  1.11--1.96  \\
\bottomrule[0.25ex]
\end{tabular}
\caption{
Summary of neutrino oscillation parameters from the global analysis. The 1$\sigma$, 2$\sigma$ and 3$\sigma$ ranges for inverted ordering are obtained with respect to the local minimum for this neutrino mass ordering.}
\label{tab:ch3-sum-2020}
\end{table}

\begin{figure*}
\includegraphics{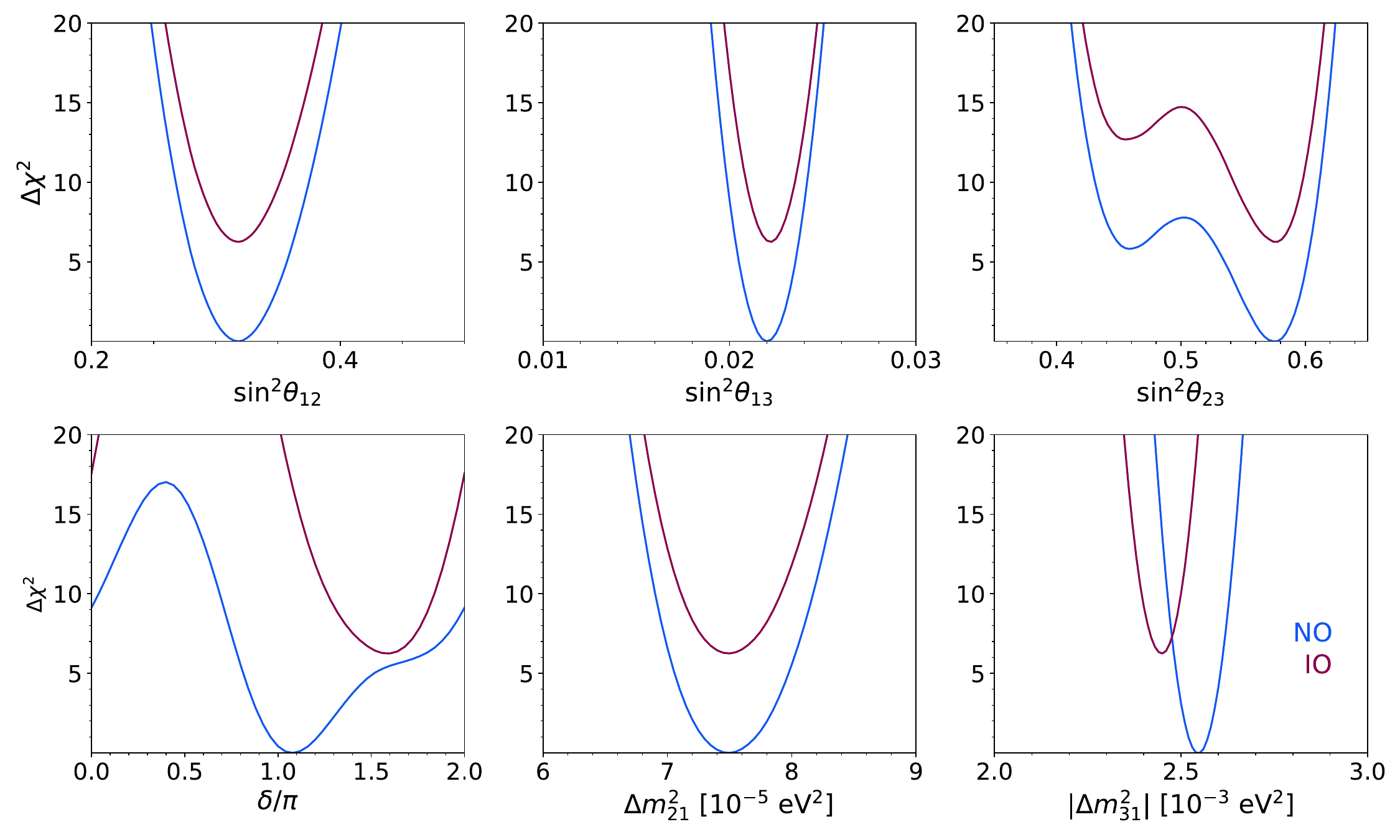}
\caption{One-dimensional $\Delta \chi^2$ profiles for the six oscillation parameters for normal (NO) and inverted (IO) ordering, in blue and purple, respectively.\labfig{fig:ch3-sum_2020}}
\end{figure*}

Precision measurements of the oscillation parameters and the determination of the neutrino mass scale serve as a guide for theoretical particle physics. For instance, extensions of the Standard Model often aim to explain simultaneously the origin neutrino mass and the differences between the mixing in the leptonic and the quark sectors.  Besides that, the question of CP violation in the neutrino sector can be related to the origin of the matter-antimatter asymmetry. Hence, an accurate measurement of these parameters can falsify model-building proposals.

Regarding neutrino phenomenology, in light of the precision being achieved in current experiments and the expectations for next-generation ones, it is possible to search for BSM physics in the neutrino sector. Deviations from the three-neutrino picture or tensions between experimental results could be interpreted as a signature of new physics.

Finally, on the experimental side, improving the existing results is pushing the development of detector technologies, like Liquid Argon time projector chambers (LAr TPCs), and novel analysis techniques, including those based on Machine Learning, among others~\cite{Psihas:2020pby}. Likewise, the measurement of Coherent Elastic Neutrino-Nucelus Scattering (CE$\nu$NS) has opened a new window for the study of neutrinos~\cite{COHERENT:2017ipa}. For instance, current and future detectors based on this process would be sensitive to the solar neutrino flux~\cite{XENON:2020gfr,Aalbers:2022dzr}. However, being a flavour-blind process at tree level, it is not that suitable for the study of oscillations.

In the years to come, substantial improvement is expected from next-generation experiments, including JUNO~\cite{JUNO:2021vlw}, DUNE~\cite{DUNE:2020ypp}, Hyper-Kamiokande~\cite{Hyper-Kamiokande:2018ofw}, KM3NET/ORCA~\cite{KM3Net:2016zxf}, the IceCube Upgrade~\cite{IceCube-Gen2:2020qha} and ESS$\nu$SB~\cite{ESSnuSB:2021azq}, among others. They have been conceived as multi-purpose experiments with the capability to reach unprecedented sensitivity and simultaneously explore neutrinos from different sources and at different energy ranges.

\pagelayout{wide} % No margins
\addpart{Neutrino properties beyond masses and mixing}
\pagelayout{margin} % Restore margins
\chapter{Neutrino magnetic moments and spin-flavour precession in the Sun}
%\addcontentsline{toc}{chapter}{Neutrino magnetic moments and spin-flavour precession} 
\labch{ch4-magnetic}
Since neutrinos are electrically neutral, they do not couple directly to electromagnetic fields. Then, their electromagnetic properties originate through radiative corrections. For instance, in the Standard Model, neutrinos have a non-zero charge radius generated by quantum loop effects. Technically, that is the only non-zero electromagnetic form factor massless left-handed Weyl neutrino can have. Other non-zero neutrino electromagnetic properties are a common prediction in many extensions of the Standard Model accounting for neutrino masses. The focus of this chapter is the study of neutrino magnetic dipole moments through the process of spin-flavour precession (SFP). If neutrinos are Majorana particles, the interaction of their flavour off-diagonal magnetic moments with the solar magnetic field can result in the conversion of a fraction of left-handed $\nu_{e}$ produced in the Sun into right-handed antineutrinos $\bar{\nu}_{\mu}$ and $\bar{\nu}_{\tau}$. Although at present it is firmly established that the observed deficit of solar $\nu_e$ is due to the MSW effect, SFP could still be present at a subdominant level. 
The combined action of neutrino SFP and flavour conversions would then produce a small but potentially observable flux of solar electron antineutrinos $\bar{\nu}_e$ at the Earth --- see for instance~\cite{Akhmedov:2002mf,Guzzo:2012rf} and references therein. The detection of such a flux would therefore be a clear signature of both non-zero magnetic moment and the Majorana nature of neutrinos.

In the first place, \refsec{ch4-magneticmom} introduces neutrino electromagnetic properties in the one-photon approximation and, in particular, neutrino magnetic dipole moments. Then, we present an analytical calculation of the SFP for solar neutrinos in \refsec{ch4-analytical} and check the validity of our results in \refsec{ch4-magn_results}. Finally, we conclude with some discussion of the results in \refsec{ch4-discussion}.

%%%%%%%%%%%%%%%%%%%%%%%%%%%%%%%%%%%%%%%%%%%%
%%%%%%%%%%%%%%%%%%%%%%%%%%%%%%%%%%%%%%%%%%%%
\section{Neutrino magnetic moments}
\labsec{ch4-magneticmom}
%%%%%%%%%%%%%%%%%%%%%%%%%%%%%%%%%%%%%%%%%%%%
%%%%%%%%%%%%%%%%%%%%%%%%%%%%%%%%%%%%%%%%%%%%

\begin{figure}
\vspace{-40pt}
\includegraphics[width=0.68\textwidth]{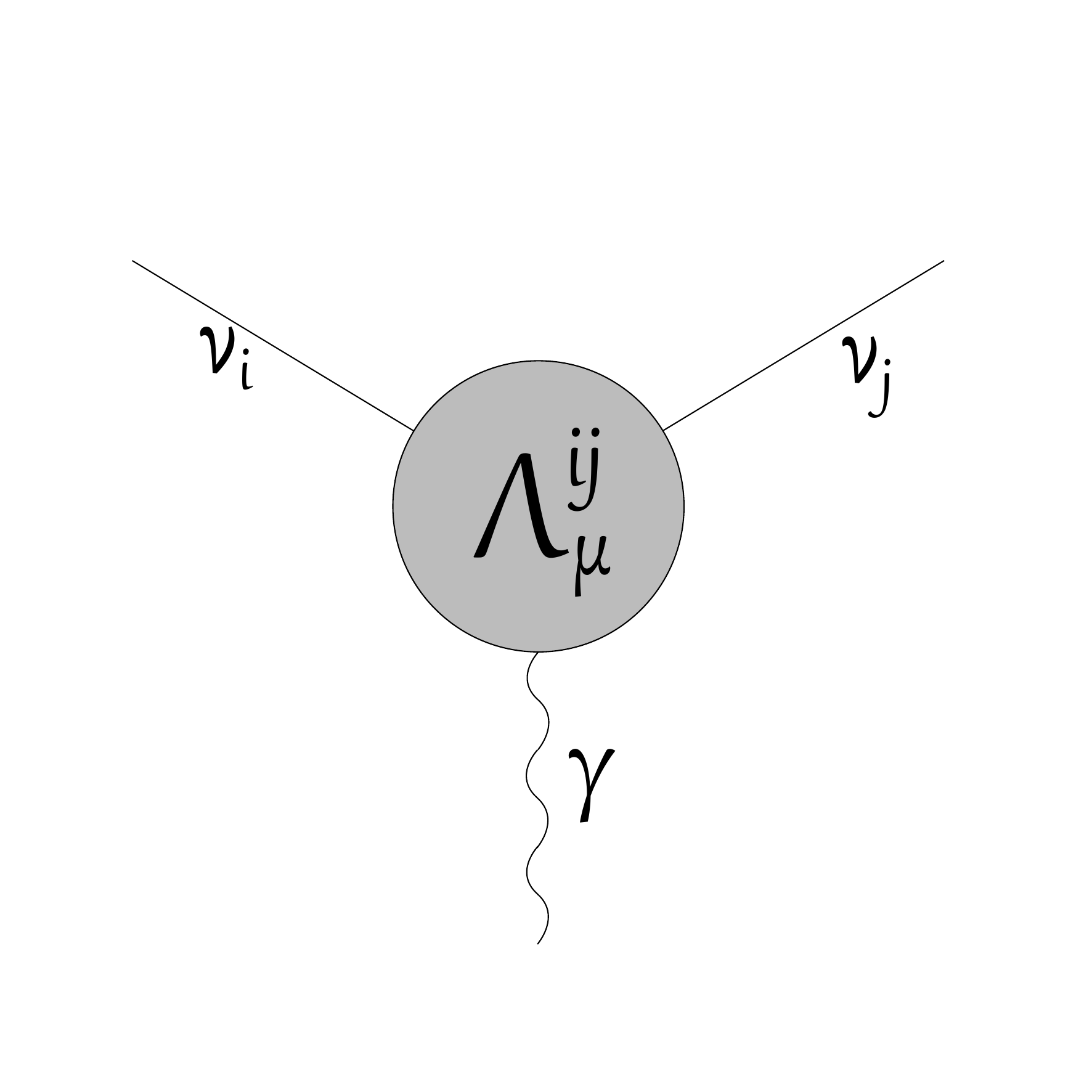}
\vspace{-30pt}
\caption{Effective electromagnetic vertex accounting for neutrino electromagnetic interactions in the one-photon approximation. \labfig{emvertex}}
\end{figure}

In the one-photon approximation, electromagnetic interactions of the neutrino fields can be described by the effective electromagnetic vertex~\cite{Giunti:2014ixa}
\begin{align}
\Lambda^{ij}_{\mu}(q) = (\gamma_\mu - q_\mu\slashed{q}/q^2)\left[\mathsf{f}^{ij}_{Q}(q^2) + \mathsf{f}^{ij}_{A}(q^2) q^2\gamma_5\right] \nonumber\\ -i \sigma_{\mu\nu}q^\nu[\mathsf{f}^{ij}_M(q^2) + i \mathsf{f}^{ij}_E(q^2)\gamma_5]\, ,
\end{align}
as shown in  \reffig{emvertex}. It depends solely on the momentum transferred to the photon, $q$, and on the electromagnetic form factors $\mathsf{f}^{ij}_{x}$, with $x = Q, A, M, E$. The two superindices $i$ and $j$ account for the two neutrino mass states involved in the interaction. Electromagnetic form factors with $i=j$ are often referred to as \textit{diagonal}, whereas those with $i\neq j$ are known as \textit{transition} form factors. For an interaction with a real photon, $q^2=0$, we define the neutrino charge $\mathsf{q}_{ij}  = \mathsf{f}^{ij}_Q (0)$, the magnetic and electric dipole moment --- denoted by $\mathsf{\mu} = \mathsf{f}^{ij}_M(0)$ and $\mathsf{\epsilon} = \mathsf{f}^{ij}_E(0)$ respectively --- and the anapole moment $\mathsf{a} = \mathsf{f}^{ij}_A(0)$. \footnote{Note that the matrices of electromagnetic form factors are Hermitian. In the case of Majorana neutrinos, charge, magnetic and electric form factor matrices are antisymmetric and the anapole form factor matrix is symmetric.}

From this moment on, the discussion will focus on neutrino magnetic dipole moments. Neutrino electromagnetic dipole moment interactions flip chirality since
\begin{align}
&\bar{\nu}_i \sigma^{\mu\nu} \nu_j = \bar{\nu}_{iL} \sigma^{\mu\nu} \nu_{jR} +  \bar{\nu}_{iR} \sigma^{\mu\nu} \nu_{jL}  \, , \nonumber \\
&i\bar{\nu}_i \sigma^{\mu\nu} \gamma_5\nu_j = i\bar{\nu}_{iL} \sigma^{\mu\nu} \nu_{jR}  -  i\bar{\nu}_{iR} \sigma^{\mu\nu} \nu_{jL} \, .
\end{align}
Notice therefore that massless neutrinos have zero electromagnetic dipole moments.

Generically, neutrino magnetic moments are expected to be proportional to neutrino masses. For instance, in the minimal extension of the Standard Model with right-handed singlet fermions, neutrinos are Dirac fermions and their magnetic moments are~\cite{Shrock:1982sc}
\begin{align}
&\mu^D_{ij} \simeq  \frac{3eG_{\text{F}}}{16\sqrt{2}\pi^2}(m_i + m_j)\left(\delta_{ij} - \frac{1}{2}\sum_{l = e, \, \mu,\,\tau} U^*_{lj}U_{li} \frac{m^2_l}{m^2_W}\right) \, .
\end{align}
In the expression above, $G_{\text{F}}$ denotes the Fermi constant and $e$ is the electron charge. The result depends on the mass of each neutrino, $m_i$, and the ratio between the charged-lepton masses, $m_l$, and the W-boson mass, $m_W$. This ratio is  $5\times10^{-4}$ for the tau lepton, which means that transition magnetic moments are around three to four orders of magnitude smaller than the diagonal ones~\cite{Giunti:2014ixa},
\begin{align}
\mu^D_{kk} \sim 10^{-20}\mu_B \left(\frac{m_k}{0.1\text{eV}}\right)\, ,
\end{align}
where $\mu_B$ denotes the Bohr magneton.

In the simplest extensions of the Standard Model leading to Majorana neutrinos --- i.e. in the seesaw models presented in \refch{ch2-masses} --- and neglecting the model dependent diagrams arising from differences in the scalar potential, the magnetic moments read~\cite{Shrock:1982sc}
\begin{align}
&\mu^M_{ij} \simeq - i\frac{3eG_{\text{F}}}{32\sqrt{2}\pi^2}(m_i +m_j)\sum_{l = e,\, \mu, \, \tau}\text{Im}[U^*_{li}U_{lj}]\frac{m^2_l}{m^2_W} \, .
\end{align}
In this case, only transition magnetic moments are allowed, and the largest expected values are $\mathcal{O}(10^{-23}\mu_B)$~\cite{Giunti:2014ixa}.

In several extensions of the Standard Model, neutrino magnetic moments are enhanced and reach values much closer to the existing experimental limits. For instance, in left-right symmetric models, the right-handed neutrino couples to a $W_R$ gauge boson --- which also mixes with the Standard Model $W$ boson. In most minimal versions of these models, the magnetic moment for Dirac neutrinos becomes proportional to the charged-lepton masses, instead of being proportional to neutrino masses \cite{Shrock:1982sc,Czakon:1998rf,Fukugita:2003en}. This leads to a significant enhancement of its value. However, it is also proportional to the mixing between the $W$ and $W_R$ bosons. Existing constraints on the parameter space of the minimal models imply that magnetic moments are at most $\mathcal{O}(10^{-16}\mu_B)$~\cite{Giunti:2014ixa}.  
 
A different family of models is the one featuring new light particles ---  with masses below $\sim$1 GeV --- and a fractional charge --- as large as possible given the existing constraints from millicharged particles. In these scenarios, the expected neutrino magnetic moments would be smaller than $\mathcal{O}(10^{-15}\mu_B)$~\cite{Lindner:2017uvt}. Besides that, in the context of Supersymmetry, it is also possible to achieve relatively large neutrino magnetic moments~\cite{Babu:1989px}. As an example, values of the order $\mathcal{O}(10^{-14}\mu_B)$ can be reached in the Minimal Supersymmetric Standard Model extended with one vector-like lepton generation~\cite{Aboubrahim:2013yfa}.

Typically, theoretical models that predict large magnetic moments also predict large neutrino masses --- incompatible with existing data --- unless some fine tuning is invoked. A solution to this issue would be to suppress the mass-to-magnetic-moment ratio by some symmetry. Along these lines, one can propose a horizontal SU(2)$_\nu$ symmetry that transforms $\nu$ into $\nu^C$ --- the left-handed antiparticle of the right-handed neutrino~\cite{Voloshin:1987qy,Babu:2020ivd}. If this symmetry was exact, the neutrino mass term would be forbidden but not the magnetic moment one. If this symmetry only holds approximately, then the ratio between masses and magnetic moments could still be small. In this way, magnetic moments of $\mathcal{O}(10^{-11}\mu_B)$ would be consistent with current limits on neutrino masses~\cite{Brdar:2020quo,Babu:2020ivd}. A different approach consists in suppressing the neutrino mass diagram without suppressing the magnetic moment one. As an example, this can be achieved with the inclusion of three Higgs doublets and a charged scalar that is a singlet of SU(2)$_L$~\cite{Barr:1990um}. In this case, spin conservation can be used to suppress neutrino masses with respect to magnetic moments. This spin symmetry mechanism allows reaching values of the magnetic moment of $\mathcal{O}(10^{-12}\mu_B)$~\cite{Barr:1990um,Babu:1992vq,Lindner:2017uvt,Babu:2020ivd}.

\begin{figure*}
\includegraphics[width = 0.72\paperwidth]{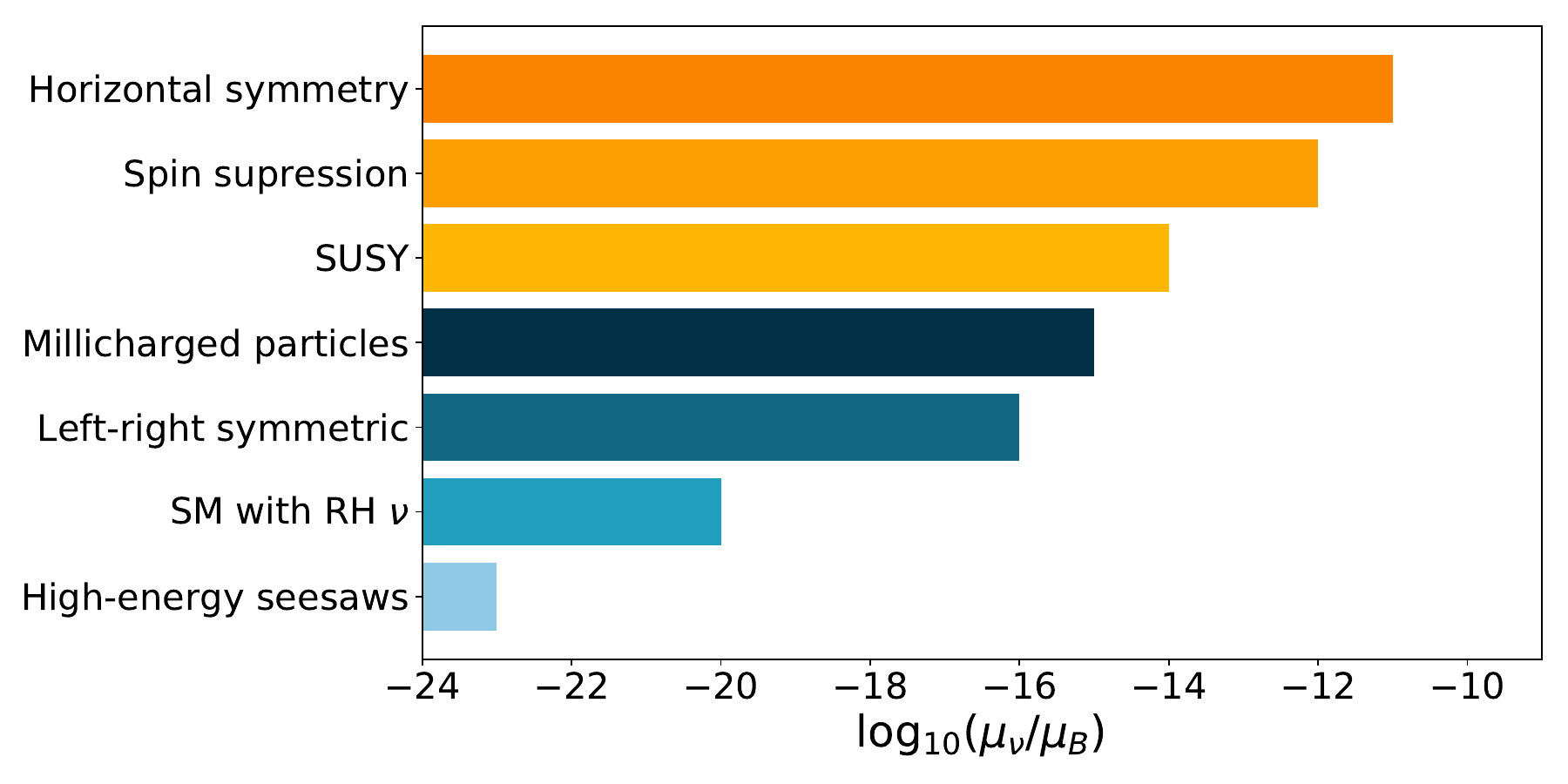}
\caption{Typical values of the neutrino magnetic moment predicted in different extensions of the Standard Model, as commented in the main text. \labfig{fig:ch4-munulimits}}
\end{figure*}

\reffig{fig:ch4-munulimits} presents a graphical summary of the typical magnitude of the magnetic moments predicted in different theoretical models, as in~\cite{Jana:2022xru}. In Appendix \ref{app1-magnetic}, an updated compilation of limits on neutrino magnetic moments is presented, for comparison with the predictions here discussed. It also serves to contextualise the results derived in the following sections from spin-flavour precession.

%%%%%%%%%%%%%%%%%%%%%%%%%%%%%%%%%%%%%%%%%%%
%%%%%%%%%%%%%%%%%%%%%%%%%%%%%%%%%%%%%%%%%%%%

%%%%%%%%%%%%%%%%%%%%%%%%%%%%%%%%%%%%%%%%%%%%
%%%%%%%%%%%%%%%%%%%%%%%%%%%%%%%%%%%%%%%%%%%%
\section{Spin-flavour precession in the Sun}
\labsec{ch4-analytical}
%%%%%%%%%%%%%%%%%%%%%%%%%%%%%%%%%%%%%%%%%%%%
%%%%%%%%%%%%%%%%%%%%%%%%%%%%%%%%%%%%%%%%%%%%
From this section on, this chapter is restricted to the discussion of \mbox{spin-flavour} precession for Majorana neutrinos. The reason is that, in the case of Dirac neutrinos, the signature expected from SPF is a reduction of the overall flux of solar neutrinos due to the conversion of left-handed neutrino fields into the inert right-handed component. Given the current limits on neutrino magnetic moments and the uncertainties in the flux prediction, such an effect would be practically unobservable. However, for Majorana neutrinos, spin-flavour precession, together with flavour conversion, would result in a flux of antineutrinos detectable at neutrino observatories searching for a flux of $\bar{\nu}_e$ of astrophysical origin.

The evolution of neutrinos in the Sun, in the absence of interactions with the magnetic field, is given by Equation \ref{eq:ch2-hamiltonian-std}. However, in the current scenario, it is more convenient to define a new basis as follows
\begin{align}
\nu_{fl L} = O_{23} \Gamma_\delta O_{13} \nu'_L \, ,
\label{eq:ch4-primed-basis}
\end{align}
where $\nu_{flL} = (\nu_{eL},\, \nu_{\mu L} ,\, \nu_{\tau L})^T$ is the three-vector of amplitudes in flavour basis and $\nu'_{L} = (\nu'_{eL},\, \nu'_{\mu L} ,\, \nu'_{\tau L})^T$ is the three-vector in the so-called \textit{primed basis}. Here $O_{ij}$ are the orthogonal matrices of rotation with the angle $\theta_{ij}$ in the $i-j$ plane and $\Gamma_{\delta} = \text{diag}(1,\, 1,\, e^{i\delta_{\text{CP}}} )$. With this notation, the lepton flavour mixing matrix is expressed as
\begin{align}
U = O_{23}\Gamma_\delta O_{13} \Gamma^\dagger_{\delta} O_{12}\, .
\end{align}
In the case of Majorana neutrinos, the leptonic mixing matrix depends on two additional phases, $U_M = U\cdot K$, where $K = \text{diag}(1, e^{i\lambda_2}, e^{i\lambda_3} )$. 

In this new basis, the evolution is given by
\begin{align}
&i\frac{\text{d}}{\text{d}t}\begin{pmatrix}\nu'_{eL} \\ \nu'_{\mu L} \\ \nu'_{\tau L}\end{pmatrix} = H \begin{pmatrix}\nu'_{eL} \\ \nu'_{\mu L} \\ \nu'_{\tau L}\end{pmatrix} \, ,
\label{eq:ch4-vac_evol}
\end{align}
with 
\begin{align}
H = \begin{pmatrix}2 \delta s^2_{12} + c^2_{13}V_e + V_n & 2 \delta s_{12}c_{12} & s_{13}c_{13}V_e \\ 2 \delta s_{12}c_{12} & 2\delta c^2_{12} + V_n & 0 \\ c_{13}s_{13}V_e & 0 & 2\Delta + s^2_{13}V_e + V_n \end{pmatrix} \, ,
\label{eq:ch4-vac_evol-2}
\end{align}
where we have introduced the short-hand notation $c_{ij} \equiv \cos\theta_{ij}$ and  \mbox{$s_{ij} \equiv \sin  \theta_{ij}$}. Remember that the evolution for antineutrinos is computed directly taking into account that they are CP conjugate fields. This is equivalent to changing $V_e \rightarrow -V_e$, $V_n \rightarrow -V_n$ and $\delta_{\text{CP}} \rightarrow -\delta_{\text{CP}}$. We will denote the Hamiltonian dictating antineutrino evolution by $\bar{H}$.

Once one includes the effect of neutrino interactions with external magnetic fields due to non-zero magnetic moments, the evolution equation as a function of the position $r$ in the trajectory reads~\cite{Lim:1987tk,Akhmedov:1989df,Akhmedov:1993sh}
\begin{equation}
    i \frac{\text{d}}{\text{d}r} \begin{pmatrix}
    \nu'_{L} \\ \bar{\nu}'_{ R}
    \end{pmatrix} = \begin{pmatrix}
    H & \mathcal{B} \\ \mathcal{B}^ {\dagger} & \bar{H}
    \end{pmatrix}\begin{pmatrix}
    \nu'_{L} \\ \bar{\nu}'_{R}
    \end{pmatrix}\, ,
\label{eq:ch4-evol}
\end{equation}
where $\bar{\nu}_R'=(\bar{\nu}'_{eR}\;\bar{\nu}'_{\mu R}\;
\bar{\nu}'_{\tau R})^T$ is the vector of the right-handed antineutrino amplitudes in the \textit{primed basis}. The submatrices $H$ and $\bar{H}$ are defined in Equation \ref{eq:ch4-vac_evol-2} and in the subsequent description. The matrix $\mathcal{B}$, describing Majorana neutrino interactions with the external magnetic field, can be written as 
\begin{align}
    \mathcal{B} = & \begin{pmatrix}
    \mathcal{B}_{e'e'} & \mathcal{B}_{e'\mu'} & \mathcal{B}_{e'\tau'}\\
    \mathcal{B}_{\mu'e'} & \mathcal{B}_{\mu'\mu'} & \mathcal{B}_{\mu' \tau'} \\
    \mathcal{B}_{\tau'e'} & \mathcal{B}_{\tau'\mu'} & \mathcal{B}_{\tau'\tau'}
    \end{pmatrix} \nonumber \\ = &\begin{pmatrix}
    0 & \mu_{e'\mu'} & \mu_{e'\tau'}\\
    -\mu_{e'\mu'} & 0 & \mu_{\mu' \tau'} \\
    - \mu_{e'\tau'} & -\mu_{\mu'\tau'} & 0
    \end{pmatrix} B_{\perp} (r) e^{i\phi(r)} 
%\cdot B(r) 
\equiv \upmu' \cdot 
B_{\perp} (r) e^{i\phi(r)} 
\label{eq:ch4-B} \, . 
\end{align}
Here $\upmu'$ denotes the matrix of transition magnetic moments in the primed basis. For simplicity, we have introduced the short-hand notation $\mu_{\alpha' \beta'}$ for the elements in the matrix $\upmu'$. The factor $B_{\perp}(r)e^{i\phi(r)}$ describes the external magnetic field in the plane transverse to the neutrino momentum, where $B_{\perp}(r)>0$ is the field strength and the azimuthal angle $\phi(r)$ defines the direction of the magnetic field in this plane. The contribution of the longitudinal component of the magnetic field is inversely proportional to the  Lorenz factor of the neutrino and hence, it is negligible in all situations of practical interest.

The magnetic moments in the \textit{primed basis} can be related to the magnetic moments in the mass basis, $\upmu_m$, which are of a more fundamental nature, namely
\begin{equation}
    \upmu'=\Gamma_\delta O_{12} K^* \upmu_{m} K^* O^T_{12} \Gamma_\delta \,.
\end{equation}
Then, one finds the following relations
\begin{align}
    \mu_{e'\mu'} & = \mu_{12}e^ {-i\lambda_{2}}\,, 
\label{eq:ch4-emu}
\\
    \mu_{e'\tau'} & = \left(\mu_{13}c_{12} + \mu_{23}s_{12}
e^{-i \lambda_{2} }\right) e^{-i(\lambda_{3}-\delta_{\text{CP}})}\,,
\label{eq:ch4-etau}\\
    \mu_{\mu' \tau'} & = \left(\mu_{23}c_{12}e^{-i\lambda_{2}} - 
\mu_{13}s_{12}\right)e^{-i(\lambda_{3}-\delta_{\text{CP}})} \, ,
\label{eq:ch4-mutau}
\end{align}
which depend on the mixing angles and the CP-violating phases previously defined.

Taking into account that the diagonal elements of the matrix $\mathcal{B}$ vanish --- see Equation \ref{eq:ch4-B} --- and $H_{\mu' \tau'} = H_{\tau' \mu'} = \bar{H}_{\mu'\tau'} = \bar{H}_{\tau'\mu'} $ are also zero, the Equation \ref{eq:ch4-evol} can be rewritten more explicitly, namely
\begin{subequations}
\begin{align}
i\frac{\text{d}}{\text{d}r}\nu'_{eL} = H_{e'e'}\nu'_{eL} + H_{e'\mu'}\nu'_{\mu L} &+ 
H_{e'\tau'}\nu'_{\tau L}\nonumber \\ &+ \mathcal{B}_{e'\mu'}\bar{\nu}'_{\mu R}   		+ \mathcal{B}_{e'\tau'}\bar{\nu}'_{\tau R} \, ,
\\
i\frac{\text{d}}{\text{d}r}\nu'_{\mu L} = H_{\mu'e'}\nu'_{eL} + H_{\mu'\mu'}\nu'_{\mu L}  &+ \mathcal{B}_{\mu'e'}\bar{\nu}'_{e R} + \mathcal{B}_{\mu'\tau'}\bar{\nu}'_{\tau R} \, ,
\\
i\frac{\text{d}}{\text{d}r}\nu'_{\tau L} = H_{\tau'e'}\nu'_{eL} + H_{\tau'\tau'}\nu'_{\tau' L} &+ \mathcal{B}_{\tau'e'}\bar{\nu}'_{eR} + \mathcal{B}_{\tau'\mu'}\bar{\nu}'_{\mu R}\, , 
\\
i\frac{\text{d}}{\text{d}r}\bar{\nu}'_{e R} = \bar{H}_{e'e'}\bar{\nu}'_{eR} + \bar{H}_{e'\mu'}\bar{\nu}'_{\mu R} &+\bar{H}_{e'\tau'}\bar{\nu}'_{\tau R}  \nonumber\\&- \mathcal{B}^*_{e'\mu'} \nu'_{\mu L}  - \mathcal{B}^*_{e'\tau '}\nu'_{\tau L}\, , 
\\  
i\frac{\text{d}}{\text{d}r}\bar{\nu}'_{\mu R} = \bar{H}_{\mu'e'}\bar{\nu}'_{eR} +\bar{H}_{\mu'\mu'}\bar{\nu}'_{\mu R}  &- \mathcal{B}^*_{\mu'e'} \nu'_{e L} - \mathcal{B}^*_{\mu'\tau '}\nu'_{\tau L}\, ,
\\
i\frac{\text{d}}{\text{d}r}\bar{\nu}'_{\tau R} = \bar{H}_{\tau'e'}\bar{\nu}'_{eR} +\bar{H}_{\tau'\tau'}\bar{\nu}'_{\tau R} &-\mathcal{B}^*_{\tau'e'} 
\nu'_{e L} - \mathcal{B}^*_{\tau'\mu '}\nu'_{\mu L} \, . 
\end{align}
\label{eq:ch4-evol3}
\end{subequations}

%%%%%%%%%%%%%%%%%%%%%%%%%%%%%%%%%%%%%%%%%%%%%%%%%%%%%%%%%%%%%
\subsection{Analytical solution to the evolution of the amplitudes}
%%%%%%%%%%%%%%%%%%%%%%%%%%%%%%%%%%%%%%%%%%%%%%%%%%%%%%%%%%%%%
Since Majorana neutrinos can only have non-zero transition magnetic moments, direct conversion of $\nu_{eL}$ produced in the Sun into $\bar{\nu}_{eR}$  is not possible. Nonetheless, the transition can still occur in two steps, i.e.  
\begin{subequations}
\begin{align}
    \nu_{eL} \overset{\text{osc.}}{\longrightarrow} \nu_{\mu L} 
\overset{\text{SFP}}{\longrightarrow} \bar{\nu}_{eR}\,,
%\indent \text{and} \indent
\label{eq:ch4-trans1}
\\
    \nu_{eL} \overset{\text{SFP}}{\longrightarrow} \bar{\nu}_{\mu R} 
    \overset{\text{osc.}}{\longrightarrow} \bar{\nu}_{eR} \, ,
\label{eq:ch4-trans2}
\end{align}
\end{subequations}
and similarly for transitions involving $\nu_{\tau L}$ and $\bar\nu_{\tau R}$  as intermediate states. In vacuum, the amplitudes of the processes in (\ref{eq:ch4-trans1}) and (\ref{eq:ch4-trans2}) are of opposite signs and cancel each other exactly at first order. In the Sun, due to matter effects, the cancellation is not exact but conversions remain strongly suppressed. The same argument applies to the transitions between states in the \textit{primed basis}, $\nu_{eL}' \to \bar{\nu}_{eR}'$.\footnote{Note that in the presence of a non-zero twist in the magnetic field --- i.e. if the azimuthal angle $\phi$ in the parametrisation of the magnetic field changes $\text{d}\phi/\text{d}r \neq 0$ --- the cancellation is also not exact~\cite{Akhmedov:1993sh}.}

Nonetheless, if flavour conversions and spin-flavour precession took place in spatially separated regions, the transitions $\nu_{eL}\to \bar{\nu}_{eR}$ trough the processes in (\ref{eq:ch4-trans1}) and (\ref{eq:ch4-trans2}) --- and the analogous ones with $\nu_{\tau L}$ and $\bar\nu_{\tau R}$ as intermediate states ---- would not be suppressed. 
Magnetic fields outside the Sun are very weak and thus, the only possibility left is that SPF occurs inside the Sun and subsequently, flavour conversions take place between the Sun and the Earth.  Hence, in order to calculate the flux of solar electron antineutrinos, one first needs to know the flux of antineutrinos at the surface of the Sun.

To approach this problem analytically we proceed to make the following approximations in the set of Equations \ref{eq:ch4-evol3}. In the first place, and owing to the arguments presented above, one can neglect $\nu'_{eL} \to \bar{\nu}'_{eR}$ inside the Sun. Consequently, we skip the evolution equation for electron antineutrinos and we disregard the terms proportional to the $\bar{\nu}'_{eR}$ amplitude in the remaining equations. Secondly, we omit from the evolution equations the terms related to $H_{e'\tau'} = H_{\tau' e'}$ and $\bar{H}_{e'\tau'} = \bar{H}_{\tau' e'}$ since they are considerably smaller than the diagonal elements in the Hamiltonian $H_{\mu'\mu'}$, $H_{\tau'\tau'}$, $\bar{H}_{\mu'\mu'}$ and $\bar{H}_{\tau'\tau'}$. Physically, it can be interpreted as neglecting transitions between $\nu'_e$ and $\nu'_\tau$, which are strongly suppressed by the smallness of $\theta_{13}$~\cite{deSalas:2017kay}. Lastly, given the agreement between solar data and the three-neutrino oscillation picture, SFP can only be a subdominant effect. It is therefore justified to work at first order in perturbation theory in $\mathcal{B}$. To do so, we will obtain the evolution of the $\nu'_{eL}$, $\nu'_{\mu L}$ and $\nu'_{\tau L}$ at zeroth order. Those amplitudes appear in the evolution equations of $\bar{\nu}'_{\mu R}$ and $\bar{\nu}'_{\tau R}$ multiplied by elements of $\mathcal{B}$, which means that they act as sources for the appearance of both muon and tau antineutrinos.

With these considerations, the system of equations can be simplified to
\begin{subequations}
\begin{align}
i\frac{d}{dr}\nu'_{eL} &= \left(2\delta s^2_{12} + c^2_{13}V_e +V_n\right) \nu'_{eL} + 2\delta s_{12}c_{12}\nu'_{\mu L}\,,
\label{eq:ch4-nuel}
\\
i\frac{d}{dr}\nu'_{\mu L} &= 2\delta s_{12}c_{12} \nu'_{eL} + \left(2\delta c^2_{12}+V_n\right)\nu'_{\mu L}\,,
\label{eq:ch4-numul}
\\
i\frac{d}{dr}\nu'_{\tau L} &= \left(2\Delta + s^2_{13}V_e + V_n\right)\nu'_{\tau  L}\,,
\label{eq:ch4-nutaul}
\\
i\frac{d}{dr}\bar{\nu}'_{\mu R} &= \left(2\delta c^2_{12} -V_n\right)\bar{\nu}'_{\mu R} \nonumber \\ &\hspace{1cm} + \mu^*_{e'\bar{\mu}'} B_{\perp}e^{-i\phi}\nu'_{eL} - 
\mu^*_{\mu' \bar{\tau}'}B_{\perp}e^{-i\phi}\nu'_{\tau L}\,,
\label{eq:ch4-numur}
\\
i\frac{d}{dr}\bar{\nu}'_{\tau R} &= \left(2\Delta -s^2_{13}V_e - V_n\right)\bar{\nu}'_{\tau R} \nonumber\\ & \hspace{1cm}+ \mu^ *_{e'\bar{\tau}'}B_{\perp}e^{-i\phi} \nu'_{eL} + \mu^*_{\mu' \bar{\tau}'}B_{\perp}e^{-i\phi}\nu'_{\mu L}\,.
\label{eq:ch4-nutaur}
\end{align}
\end{subequations}
Notice that with these approximations, the evolution of the amplitudes $\nu'_{e L}$ and $\nu'_{\mu L}$ decouples from the rest of the system and reduces to solving the standard picture of solar neutrinos.

Let us denote the coordinate of the neutrino production point as $r_0$ and define
\begin{equation}
    E_{1,2} \equiv \delta + c^2_{13}\frac{V_e}{2} +V_n \mp \sqrt{\left(\delta
\cos2\theta_{12}- c^2_{13}V_e/2\right)^ 2 + \delta^ 2\sin^2 2\theta_{12}}\,,
\label{eq:ch4-E12}
\end{equation}
and the effective mixing angle in matter, $\Tilde{\theta}$, from the relation
\begin{equation}
    \cos 2\Tilde{\theta}(r)=\frac{\cos2\theta_{12} - c^2_{13}V_e/2\delta}
{\sqrt{\left(\cos2\theta_{12}- c^2_{13}V_e/2\delta\right)^ 2 
+ \sin^2 2\theta_{12}}}\,.
\label{eq:ch4-tildetheta}
\end{equation}
Then, using the adiabatic approximation, one obtains
\begin{align}
&\nu'_{eL}(r) = c_{13}\left[\cos \Tilde{\theta}(r_0)\cos \Tilde{\theta}(r) 
e^{-i\int_{r_0}^ r E_1 dr'} \right.\nonumber \\ &\hspace{3cm}\left.+ \sin \Tilde{\theta}(r_0)\sin
\Tilde{\theta}(r) e^{-i\int_{r_0}^ r E_2 dr'}\right]\, ,\\
&\nu'_{\mu L}(r) = c_{13}\left[-\cos \Tilde{\theta}(r_0)\sin \Tilde{\theta}(r) e^{-i\int_{r_0}^ r E_1 dr'} \right. \nonumber \\ &\hspace{3cm}\left.+ \sin \Tilde{\theta}(r_0)\cos \Tilde{\theta}(r) e^{-i\int_{r_0}^ r E_2 dr'}\right] \,,
\end{align}
Here, we have also imposed the initial condition that neutrinos are produced as $\nu_{eL}$, which in the \textit{primed basis} translates to $\nu'_{eL} (r_0) = c_{13}$, $\nu'_{\mu L} (r_0) = 0$  and $\nu'_{\tau_L} (r_0) = s_{13}$.

The evolution of the amplitude $\nu_{\tau L}'$ is decoupled from the rest of the system and given by
\begin{equation}
    \nu'_{\tau L}(r) = s_{13} e^{-i\int_{r_0}^ r\left(2\Delta + s^2_{13}V_e 
+ V_n\right)dr'}.
\end{equation}

Having found the evolution for $\nu_{eL}'$, $\nu_{\mu L}'$ and $\nu_{\tau L}'$ amplitudes, one can solve Equation \ref{eq:ch4-numur} and Equation \ref{eq:ch4-nutaur} to find the amplitudes of $\bar{\nu}'_{\mu R}$ and $\bar{\nu}'_{\tau R}$. At the surface of the Sun --- which we denote by $r = R_\odot$ --- they read  
\begin{align}
    \bar{\nu}'_{\mu  R} (R_{\odot}) & = \int_{r_0}^ {R_\odot} \text{d}r \, B_{\perp} (r) 
     \left[c_{13}\mu^ *_{e' \mu'} \cos \Tilde{\theta} (r_0) \cos 
     \Tilde{\theta} (r) e ^ {-i g_1 (r)} \right.\, \nonumber \\ 
   & \, + c_{13}\mu^ *_{e' \mu'} \sin \Tilde{\theta} (r_0) \sin \Tilde{\theta}(r) 
e^{-i g_2 (r)}  \left. - s_{13}\mu ^*_{\mu' \tau'} e ^ {-ig_3(r)} \right] \,,
\label{eq:ch4-numur1}
\end{align}
and
\begin{align}
\bar{\nu}'_{\tau R} (R_{\odot}) &= \int_{r_0}^ {R_\odot} \text{d}r \,B_{\perp} (r) 
c_{13}\, \times  \nonumber \\ &\hspace{0.3cm} \left[\cos \Tilde{\theta} (r_0)\left(\mu^*_{e'\tau'}\cos \Tilde{\theta} (r) 
-\mu^*_{\mu'\tau'}\sin \Tilde{\theta}(r)\right) e^{-ig_4(r)} \nonumber \right.\\ &\left. \hspace{0.3cm} + 
    \sin\Tilde{\theta} (r_0) 
\left(\mu^*_{e'\tau'}\sin \Tilde{\theta} (r) + \mu^*_{\mu'\tau'}
\cos \Tilde{\theta} (r)\right) e^{-ig_5(r)} \right]\, ,
\label{eq:ch4-nutaur1}
\end{align}
where we have defined the following functions,
\begin{subequations}
\begin{align}
&g_{1,2}(r)  \equiv  \phi + \int_{r_0} ^ r \left[E_{1,2} - \left( 2c^2_{12}
\delta - V_n \right)\right]dr'\,,
\label{eq:ch4-g1}
\\
&g_{3}(r) \equiv \phi + \int_{r_0} ^ r \left[\left(2\Delta +s^2_{13}V_e 
+V_n \right) - \left( 2c^2_{12}\delta - V_n \right) \right]dr'\,,
\label{eq:ch4-g2}
\\
&g_{4,5}(r) \equiv \phi + \int_{r_0} ^ r \left[ E_{1,2}-\left(2\Delta 
-s^2_{13}V_e -V_n \right) \right] dr'\, .
\label{eq:ch4-g3}
\end{align}
\label{eq:ch4-gs}
\end{subequations}

Besides, we have omitted the irrelevant overall phase factors from the expressions for $\bar{\nu}'_\mu$ and $\bar{\nu}'_\tau$ since, when propagating to Earth, coherence is lost and such phase factors become inconsequential. From this point on, we consistently disregard all phase factors of this sort.

In the remaining part of this section, we discuss the details of how to obtain an analytical expression for the amplitude with an extensive discussion on the role of the magnetic field twist ---  i.e. if the azimuthal angle $\phi$ in the parametrisation of the magnetic field varies along the neutrino trajectory.
 
%%%%%%%%%%%%%%%%%%%%%%%%%%%%%%%%%%%%%%%%%%%%%%%%
\textbf{Analytical expressions for the amplitudes}
%%%%%%%%%%%%%%%%%%%%%%%%%%%%%%%%%%%%%%%%%%%%%%%%

The integrals in the amplitudes $\bar{\nu}'_\mu$ and $\bar{\nu}'_\tau$, in the set of Equations~\ref{eq:ch4-numur1} and~\ref{eq:ch4-nutaur1}, are of the form
\begin{equation}
    I = \int_{a}^b f(x) e^{-ig(x)}dx\,,
\label{eq:ch4-int1}
\end{equation}
where $f(x)$ is a slowly varying function of coordinate and $|g'(x)|$ is large except possibly in the vicinity of a finite number of points in the interval~$(a,b)$. Such integrals get their main contributions from the endpoints of the integration intervals and the static points of $g(x)$ --- i.e. where $g'(x)=0$ --- if any~\cite{Erdelyi}. This is known as the \textit{stationary phase approximation}. We will first check whether any stationary phase point exists for the integrals in Equations~\ref{eq:ch4-numur1} and \ref{eq:ch4-nutaur1}. 

Let us start with the amplitude $\bar{\nu}_{\mu R}'$, from Equation \ref{eq:ch4-numur1}, which depends on the functions $g_1$, $g_2$ and $g_3$. The stationary phase conditions are 
\begin{equation}
\frac{d}{dr}g_{1,2} = 0 \indent \longrightarrow \indent
\frac{d\phi}{dr} = 2c^2_{12}\delta -V_n -E_{1,2}\,,\qquad\quad~~
\label{eq:ch4-statpoint1}
\end{equation}
\begin{equation}
\frac{d}{dr}g_{3} = 0 \indent \longrightarrow \indent
 \frac{d\phi}{dr} = 2c^2_{12}\delta -2V_n -2\Delta -s^2_{13}V_e\,. 
\label{eq:ch4-statpoint2}
\end{equation}
Using the definition of $E_{1,2}$ in Equation \ref{eq:ch4-E12}, one can reduce Equation \ref{eq:ch4-statpoint1}  to  
\begin{equation}
    \frac{\text{d}\phi}{\text{d}r} +2V_n+c^2_{13}V_e - 2\delta \cos2\theta_{12} = \frac{\delta^2 
\sin ^2 2\theta_{12}}{d\phi/dr + 2 V_n} \,, 
\end{equation}
which has a solution as long as
\begin{equation}
    1 + \sin^2 2\theta_{12} \frac{c^2_{13}V_e}{d\phi/dr +2V_n} \geq 0 \, .
\label{eq:ch4-twistcond1}
\end{equation}
This condition is satisfied if
\begin{align}
%    \frac{d \phi}{dr} >  -2V_n \quad \text{or} \quad
    \frac{d \phi}{dr} >  2|V_n| \quad \text{or} \quad
%    \frac{d \phi}{dr}  \leq  c^ 2_{13}V_e \left( \frac{1- Y_e}{c^ 2_{13}Y_e} - 
%    \sin^2 2\theta_{12}\right)\, .
    -\frac{d \phi}{dr} \geq  c^ 2_{13}V_e 
\left(\sin^2 2\theta_{12}- \frac{1- Y_e}{c^ 2_{13}Y_e}\right), 
\label{eq:ch4-twistcond2}
\end{align}
where $Y_e$ is the number of electrons per nucleon in the medium. As $Y_e$ varies between 0.67 and 0.88 in the Sun~\cite{Serenelli:2009yc}, one can see that the expression in the brackets in the second condition in Equation \ref{eq:ch4-twistcond2} is positive and of the order of 0.3 -- 0.7. Therefore, for non-twisting magnetic fields, the stationary phase condition cannot be fulfilled. In fact, it requires $|d\phi/dr|$ to be of the same order of magnitude as $V_e$ and $|V_n|$, which vary from $\sim 7\times 10 ^{-12} \text{ eV}$ near the neutrino production point to zero at the surface of the Sun, where the solar magnetic field nearly vanishes as well.  
One can see that the stationary phase condition can be fulfilled, for instance, for magnetic fields with a constant twist $|d\phi/dr| \sim 10/R_\odot \sim 3\times 10^ {-15} \text{ eV}$~\cite{Akhmedov:1993sh}.

Regarding Equation~\ref{eq:ch4-statpoint2}, the term $2\Delta$ on its right-hand side is at least an order of magnitude larger than the other terms. \footnote{Note that $2\Delta \sim 10^{-9} - 10^ {-10}$ eV.} For the solution of this equation to exist, $d\phi/dr$ should be of the same order of magnitude, which corresponds to $\sim 1-10$ rad/km. While short-scale stochastic magnetic fields in the Sun may have such rapid twists, it is unlikely that this is possible for large-scale fields relevant for SFP.

Similarly, for the integrals in Equation \ref{eq:ch4-nutaur1}, corresponding to the amplitude $\bar{\nu}_{\tau R}'$, the stationary phase points correspond to $\frac{d}{dr}g_{1,2,3} = 0$, or   
\begin{equation}
\frac{d}{dr}g_{4,5} = 0 \indent \longrightarrow \indent 
\frac{d\phi}{dr} = 2\Delta - s^2_{13}V_e -V_n -E_{1,2} = 0\,.
\label{eq:ch4-statpoint3}
\end{equation}
Again, the term $2\Delta$ on the right side is larger than the other terms and, as the case of Equation~\ref{eq:ch4-statpoint2}, the stationary phase condition is unlikely to be satisfied.

This means that no stationary phase points are expected for the integrals in Equation~\ref{eq:ch4-nutaur1} and the integral containing $g_3$ in Equation~\ref{eq:ch4-numur1}, and they should receive their main contributions from the endpoints of the integration interval. 

In the following, we will first focus on the case in which the magnetic fields in the Sun are either non-twisting or twist slowly enough so that no stationary phase points exist. 
The effects of the possible existence of stationary phase points in the scenario with fast-twisting magnetic fields will be discussed later in the text. 

%%%%%%%%%%%%%%%%%%%%%%%%%%%%%%%%%%%%%%%%%%%%%%%%%%%%%%%%%%%%%%%%%%%% 
\textbf{Non-twisting or slowly twisting magnetic fields} 
%%%%%%%%%%%%%%%%%%%%%%%%%%%%%%%%%%%%%%%%%%%%%%%%%%%%%%%%%%%%%%%%%%%% 

In the case of a non-twisting magnetic field, the integrals we are interested in get contributions only from the endpoints of the integration interval.
\begin{align}
    \int_{a}^ {b}&f(x) e^ {-ig(x)} dx = \int_{a}^ {b} \frac{f(x)}{g'(x)}g'(x) e^ {-ig(x)} 
   \nonumber \\ &= \left[i \frac{f(x)}{g'(x)} e^{-i g(x)}\right]_a ^ b -i \int_a^ b \left(\frac{f'(x)}{g'(x)} - \frac{f(x)g''(x)}{g'(x)^ 2}\right)e^{-ig(x)} dx \nonumber \\
    &= \left[\left(i \frac{f(x)}{g'(x)} + \frac{f'(x)}{g'(x)^2} - \frac{f(x)g''(x)}{g'(x)^3}\right)e^{-i g(x)}\right]_a ^ b + \mathcal{O}\left(\frac{1}{g'(x)^3}\right).
    \label{eq:ch4-appr1}
\end{align}

It follows from the definitions of the phases $g_i$ --- with $i = 1,...,5$ --- in Equations \ref{eq:ch4-gs} that in the case of interest to us the condition 
\begin{equation}
|g''(x)|^2/g'(x)^2 \ll 1
\label{eq:ch4-cond1}
\end{equation}
is satisfied for all $g_i(x)$, mainly because the matter potentials are slowly varying functions of the position. Then, the third term in the brackets in Equation~\ref{eq:ch4-appr1} is negligible when compared to the first term.   

In addition, provided that 
\begin{equation}
    \Bigg|\frac{f(x)}{f'(x)}\Bigg| \gg \frac{1}{|g'(x)|},
\label{eq:ch4-cond2}
\end{equation}
the first term dominates over the second one. The left-hand side of this inequality is essentially the characteristic distance over which $f$ varies significantly --- often referred to as the scale height of $f(x)$.
For the case under study, $f(x)$ is proportional to $B_\perp(r)$ times $\sin\tilde{\theta}(r)$ or $\cos\tilde{\theta}(r)$. The effective mixing angle $\tilde{\theta}(r)$ varies slowly inside the Sun and as a consequence, it is justified to assume adiabatic evolution for flavour conversions in the Sun. Therefore, the scale height of $f(x)$ is basically that of the solar magnetic field, and Equation \ref{eq:ch4-cond2} reads
\begin{equation}
L_B \equiv \frac{B_\perp(r)}{|B'_\perp(r)|} \gg \frac{1}{|g_i'(r)|} .
\label{eq:ch4-cond2b}
\end{equation} 
for each of the five functions $g_i(r)$ defined. 

For neutrinos propagating in the Sun, the conditions in Equation \ref{eq:ch4-cond2b} are satisfied if $L_B \gg 10^ {-4} R_\odot$. 
Solar magnetic fields with $L_B \lesssim 10^{-4} R_\odot$ are not expected to exist over large distances in the Sun and hence, the condition \ref{eq:ch4-cond2} is satisfied. A short-scale magnetic field with a larger scale height could be present but it would not lead to sizeable SFP effects.
As a result, it is justified to keep only the first term in the brackets in Equation~\ref{eq:ch4-appr1}. Moreover, since the magnetic field strength at the surface is negligibly weak, we only take into account the contribution from the production point $r =r_0$. Under these assumptions, and given that  $g'_3 \sim 2\Delta$, $g'_{4,5} 
\sim - 2\Delta$ and that
\begin{equation}
   \bigg|\frac{1}{g'_4}-\frac{1}{g'_5} \bigg|  \ll \bigg| \frac{1}{g'_{4,5}} 
\bigg|,
\end{equation}
the amplitudes $\bar{\nu}'_{\mu R}$ and $\bar{\nu}'_{\tau R}$ from Equations \ref{eq:ch4-numur1} and \ref{eq:ch4-nutaur1} yield
\begin{align}
&\bar{\nu}'_{\mu  R} (R_\odot) \nonumber \\ 
& \hspace{0.4cm}\simeq  \, B_\perp (r_0) \left[c_{13}\mu^*_{e'\mu'}\left(\frac{\cos^2\tilde{\theta}(r_0)}{g'_1(r_0)} + \frac{\sin^2\tilde{\theta}(r_0)}{g'_2(r_0)}\right) - \frac{s_{13}\mu^*_{\mu' \tau'}}{2\Delta}\right],
\label{eq:ch4-numur2}
\\
&\bar{\nu}'_{\tau R}(R_\odot) \simeq \, B_\perp (r_0)\frac{c_{13}
\mu^*_{e' \tau'}}{2\Delta} \, .
\label{eq:ch4-nutaur2}
\end{align}
Notice that setting $\theta_{13} = 0$ and neglecting $\cos\tilde{\theta}(r_0)$ compared with $\sin\tilde{\theta}(r_0)$ one recovers the expression for the amplitude of $\bar{\nu}'_{\mu R}$ found in~\cite{Akhmedov:2002mf}.

%%%%%%%%%%%%%%%%%%%%%%%%%%%%%%%%%%%%%%%%%%%%%% 
\textbf{Fast-twisting magnetic fields} 
%%%%%%%%%%%%%%%%%%%%%%%%%%%%%%%%%%%%%%%%%%%%%% 

For large twisting magnetic fields, if one of the conditions in \ref{eq:ch4-twistcond2} is satisfied, then the contribution of stationary phase points to the integrals in the amplitudes needs to be taken into account.

For a stationary phase point $x_0$, such that $g'(x_0)= 0$, and being $f(x)$ a slowly varying function of the coordinate,
\begin{align}
    I &= \int_{a}^ b f(x) e^{ig(x)} dx \simeq  f(x_0)e^{ig(x_0)}\int_a ^b \text{exp}\left[i \frac{g''(x_0)}{2}(x-x_0)^2\right] dx \nonumber \\ &=  f(x_0)e^{ig(x_0)}\int_a ^b \left[\cos \left(\frac{g''(x_0)}{2}(x-x_0)^2\right) \right. \nonumber \\ & \hspace{4cm} \left. +i\sin \left(\frac{g''(x_0)}{2}(x-x_0)^2\right)\right] dx\, ,
\end{align}
where we have expanded
\begin{equation}
    g(x)\simeq g(x_0) + \frac{1}{2}g''(x_0) (x-x_0)^2 \, .
\end{equation}
With an appropriate change of variable, $\pi t^2 = g''(x_0) (x-x_0)^2 $, and from the fact that
\begin{equation}
    \int_0 ^ {t_a} \cos \left(\frac{\pi t^2}{2}\right)\text{d}t = \frac{1}{2} \quad \text{ and } \quad \int_0 ^ {t_a} \sin \left(\frac{\pi t^2}{2}\right)\text{d}t = \frac{1}{2} \, ,
\end{equation}
when $t_a\, \longrightarrow \, \infty$, one obtains the following approximate solution for the integral
\begin{equation}
    I \simeq f(x_0)\sqrt{\frac{2\pi}{g''(x_0)}}e ^ {i\left(\frac{\pi}{4} + g(x_0)\right)}\, .
\end{equation}

Let $r_1$ and $r_2$ be such that $g'_1(r_1) = 0$ and $g'_2(r_2) =0$. Then, the contribution of the stationary phase points to the amplitude $\bar{\nu}_{\mu R}'$ is
\begin{align}
&\bar{\nu}'_{\mu R}(R_\odot) = c_{13}\mu^*_{e'\mu'}\left[\cos \tilde{\theta} (r_0) \cos \tilde{\theta} (r_1)B_\perp(r_1)\sqrt{\frac{2\pi}{g''_1(r_1)}}\right. \nonumber \\ & \hspace{1.2cm}\left.+ \sin \tilde{\theta} (r_0) \sin \tilde{\theta} (r_2)B_\perp(r_2)\sqrt{\frac{2\pi}{g''_2(r_2)}}e^ {-i (g_2(r_2) -g_1(r_1))}\right] \,  ,
\end{align}
This contribution dominates over the ones from the endpoints of the integration interval, which can therefore be neglected. Note that this is not the case for the amplitude $\bar{\nu}_{\tau R}'$, which has no stationary phase points and consequently, it is given by Equation \ref{eq:ch4-nutaur2}, regardless of the twist of the magnetic field. Nonetheless, the validity of this approximation still relies on the assumption that there is no direct $\nu_e \rightarrow \bar{\nu}_e$ in the Sun. In the presence of a large twisting magnetic field, this might not be accurate enough.

From this point on, we will limit ourselves to the study of a non-twisting or slowly twisting magnetic field. As follows from the discussion above, the contributions from stationary phase points are larger than the ones from the endpoints. Therefore, by focusing our work on the case of non-twisting magnetic fields, we will be deriving conservative upper bounds on $\mu B_\perp$.

%%%%%%%%%%%%%%%%%%%%%%%%%%%%%%%%%%%%%%%%%%%%%%%%
%%%%%%%%%%%%%%%%%%%%%%%%%%%%%%%%%%%%%%%%%%%%%%%%
\subsection{Solar electron antineutrino flux on Earth}
\label{subsec:ch4-Flux}
%%%%%%%%%%%%%%%%%%%%%%%%%%%%%%%%%%%%%%%%%%%%%%%%
%%%%%%%%%%%%%%%%%%%%%%%%%%%%%%%%%%%%%%%%%%%%%%%%
Once we know the amplitudes $ \bar{\nu}_{\mu R}'$ and $\bar{\nu}_{\tau R}'$ at the surface of the Sun, it is straightforward to compute the expected flux of electron antineutrinos that reaches the Earth. 
Since the magnetic field between the Sun and the Earth is very weak, its effects are negligible and the evolution of neutrinos on their way to Earth is simply due to flavour transformations. Moreover, due to the loss of coherence, what arrives on Earth is an incoherent sum of mass eigenstates~\cite{Dighe:1999id}, both for neutrinos and antineutrinos.
Thus, the electron antineutrino appearance probability on Earth is
\begin{equation}
  P(\nu_{eL} \rightarrow \bar{\nu}_{eR}) = |U_{e1}|^ 2 |\bar{\nu}_{1\oplus}|^2 
+|U_{e2}|^ 2 |\bar{\nu}_{2\oplus}|^ 2 + |U_{e3}|^ 2 |\bar{\nu}_{3\oplus}|^2\,,
\label{eq:ch4-P1}
\end{equation}
where $\bar{\nu}_{i\oplus}$ ($i=1,2,3$) denote the amplitudes of the antineutrino mass eigenstates reaching the Earth.

These amplitudes are related to those in the \textit{primer basis} by $\bar{\nu}'_ R = \tilde{U}\bar{\nu}_R$, with $\tilde{U} = \Gamma^\dagger_\delta O_{12}$, and where $\bar{ \nu}'_R = (\bar{\nu}'_{eR}, \bar{\nu}'_{\mu R},\bar{\nu}'_{\tau R})^T$ and $\bar{ \nu}'_R = (\bar{\nu}'_{eR}, \bar{\nu}'_{ \mu R},\bar{\nu}'_{\tau R})^T$. Therefore 
\begin{equation}
   |\bar{\nu}_{i\oplus}|^ 2 = |\Tilde{U}_{\mu ' i}|^ 2 |\bar{\nu}'_{\mu  R }
(R_\odot)|^ 2 + |\Tilde{U}_{\tau ' i}|^2 |\bar{\nu}'_{\tau  R} (R_\odot)|^2 \,,
\end{equation}
and the electron antineutrino appearance probability can be rewritten as 
\begin{align}
P (\nu_{eL} \rightarrow \bar{\nu}_{eR}) = \frac{1}{2}c^2_{13}\sin^2 
2\theta_{12} |\bar{\nu}_{\mu R}'(R_\odot)|^ 2 +s^2_{13}|\bar{\nu}_{\tau R}' 
(R_\odot)|^2.
\end{align} 
Substituting here the approximate analytical expressions for the amplitudes $\bar{\nu}_{\mu R}'$ and $\bar{\nu}_{\tau R}'$ from Equations~\ref{eq:ch4-numur2} and \ref{eq:ch4-nutaur2}, we find 
\begin{align}
&P (\nu_{eL} \rightarrow \bar{\nu}_{eR}) = s^2_{13}B^2_\perp (r_0)\Bigg( \frac{c_{13} |\mu_{e'\tau'}|}
{2\Delta}\Bigg)^ 2\nonumber \\
&\hspace{0.2cm}+ \frac{1}{2} c^2_{13}\sin^2 
2\theta_{12}B^2_\perp (r_0)\left[c^2_{13}|\mu_{e'\mu'}|^2 
\Bigg(\frac{\cos^2\tilde{\theta}(r_0)}{g'_1(r_0)} + \frac{\sin^2\tilde{\theta}
(r_0)}{g'_2(r_0)}\Bigg) ^ 2  \right. \nonumber \\  &\hspace{1.2cm}\left.  -2 \text{ Re} 
\{ \mu_{e'\mu'}^* \mu_{\mu' \tau'} \}
\left(\frac{\cos^2\tilde{\theta}(r_0)}{g'_1(r_0)} + \frac{\sin^2\tilde{\theta}
(r_0)}{g'_2(r_0)}\right) \frac{s_{13} c_{13}}{2\Delta}  \right] \, .
    \label{eq:ch4-peebar}
\end{align}
The terms containing $|\mu_{\mu'\tau'}|^2$ and $|\mu_{e' \tau '}|^2$ are expected to give very small contributions since they are proportional to $s^2_{13}/\Delta^2$. This is true unless $|\mu_{e'\mu'}|$ were anomalously small with respect to the other transition magnetic moments.

There are three main differences between this result and that derived in the 2-flavour approach in~\cite{Akhmedov:2002mf}. 
Firstly, the main term in Equation \ref{eq:ch4-peebar}, which is proportional to $|\mu_{e'\mu'}|^2$ contains an additional factor $c^4_{13}$ that was not accounted for in the 2-neutrino result. Note that, interestingly, since $|\mu_{e'\mu'}|^2$ is equal to $|\mu_{12}|^2$, to a very good approximation and if all magnetic moments are of the same order of magnitude, the electron antineutrino appearance probability is proportional to $|\mu_{12}B_\perp(r_0)|^2$.
The second difference is the cross-term contribution in Equation \ref{eq:ch4-peebar} involving $\mu_{e'\mu'}$ and $\mu_{\mu'\tau'}$, which is absent in the two-flavour approach. This additional term may result in a \mbox{non-negligible} correction to the $\bar{\nu}_{eR}$ appearance probability.
Finally, the simplifying assumptions invoked in~\cite{Akhmedov:2002mf} restricted its applicability to neutrino energies above $\sim 5 - 8$ MeV, whereas such limitation to the validity of the result is not present for Equation~\ref{eq:ch4-peebar}. These and other aspects are discussed in detail in \refsec{ch4-magn_results}.

%%%%%%%%%%%%%%%%%%%%%%%%%%%%%%%%%%%%%%%%%%%
\subsection{Numerical calculations}
%%%%%%%%%%%%%%%%%%%%%%%%%%%%%%%%%%%%%%%%%%%

As an alternative to the approximate analytical solution derived above, one could solve the complete set of six coupled differential equations --- see Equations \ref{eq:ch4-evol3} --- numerically, tracing the evolution of the amplitudes from the production point to the Earth.
We have developed a numerical code that performs this task and computes the electron antineutrino appearance probability at the surface of the Earth. The calculation averages over the production region of neutrinos in the Sun and also takes into account the profiles of electron and neutron number densities through which neutrinos propagate.

We are mainly interested in inverse beta decay as the detection channel since the strongest limits have been derived from this process, which has a threshold of $\sim$ 1.8 MeV. Thus, we will consider only the flux from $^8$B.

We considered two Standard Solar Models (SMM) with different abundances of heavy elements and therefore having different metallicities. We present the results for the high-metallicity GS98 model and the low-metallicity AGSS09 model, as from~\cite{Serenelli:2009yc}. Regarding the magnetic field inside the Sun, it is essentially unknown and one has to resort to model field profiles. Nevertheless, the arbitrariness of the modelling of this profile is partially alleviated because one expects the $\bar{\nu}_e$ appearance probability to be most sensitive to the magnetic field strength at the neutrino production point. In order to prove this statement, we implement different magnetic field profiles with approximately the same strength at $r=0.05R_\odot$ --- see section \ref{subsec:ch4-profile} below.
If not otherwise specified, in our calculations, we assume the following linearly decreasing magnetic field profile,  
\begin{equation}
B_\perp(r)=B_0(r) \equiv 52600 (1 - r/R_\odot)\,{\text{kG}}\,, 
\label{eq:ch4-b0}
\end{equation}
which takes the value 
$B_\perp (r = r_0) \simeq 5 \times 10^{7}$\,G 
at $r_0$ = 0.05R$_\odot$ and vanishes at the surface of the Sun. The magnetic field inside the Sun is also assumed to be non-twisting.

%%%%%%%%%%%%%%%%%%%%%%%%%%%%%%%%%%%%%%%%%%%%%%%%%
%%%%%%%%%%%%%%%%%%%%%%%%%%%%%%%%%%%%%%%%%%%%%%%%%
\section{Results}
\labsec{ch4-magn_results}
%%%%%%%%%%%%%%%%%%%%%%%%%%%%%%%%%%%%%%%%%%%%%%%%%
%%%%%%%%%%%%%%%%%%%%%%%%%%%%%%%%%%%%%%%%%%%%%%%%%

%%%%%%%%%%%%%%%%%%%%%%%%%%%%%%%%%%%%%%%%%%%%%%%%%%%%%%%% 
\subsection{\label{sec:compare}Comparison between analytical and numerical results}
%%%%%%%%%%%%%%%%%%%%%%%%%%%%%%%%%%%%%%%%%%%%%%%%%%%%%%%% 

Since the main contribution to the electron antineutrino appearance probability is proportional to $|\mu_{e'\mu'}|^2=|\mu_{12}|^2$, we will first set $\mu_{13} = 0$ and $\mu_{23} = 0$ --- which also implies that $\mu_{e'\tau'}$ and $\mu_{\mu ' \tau '}$ vanish.  In this way, we will compare our analytical results with the ones from the numerical calculations and the previous analytical expression derived in the 2-neutrino framework. Under these assumptions, our analytical expression in Equation \ref{eq:ch4-peebar} becomes  
\begin{align}
&P(\nu_{eL} \rightarrow \bar{\nu}_{eR}) = \nonumber \\ &\hspace{1cm}\frac{1}{2} c^4_{13}\sin^ 2 2\theta_{12}B^2_\perp (r_0)|\mu_{e'\mu'}|^2 \Bigg(\frac{\cos^2 \tilde{\theta}(r_0)}{g'_1(r_0)} + \frac{\sin^2\tilde{\theta}(r_0)}{g'_2(r_0)}\Bigg) ^ 2.
\label{eq:ch4-peebar_full}
\end{align}
We will also consider the simplified analytical expression  
\begin{align}
P (\nu_{eL} \rightarrow \bar{\nu}_{eR})_\text{simpl.}  = \frac{1}{2} c^4_{13}\sin^2 2\theta_{12}B^2_\perp (r_0)|\mu_{e'\mu'}|^2 \Bigg( \frac{\sin^2\tilde{\theta}(r_0)}{g'_2(r_0)}\Bigg) ^ 2, 
\label{eq:ch4-peebar_simple}
\end{align}
obtained from Equation \ref{eq:ch4-peebar_full} by neglecting the first term in the brackets compared to the second one. 
This approximation is similar to the one adopted in~\cite{Akhmedov:2002mf} and it  
is expected to be valid for relatively high neutrino energies, for which $\cos^2\tilde{\theta}(r_0)\ll \sin^2\tilde{\theta}(r_0)$. \footnote{Note that $|g_1'(r_0)|$ and $|g_2'(r_0)|$ differ by less than a factor of two for all considered energies.}
\begin{figure*}[t!]
\centering
\includegraphics[width = 0.375\paperwidth]{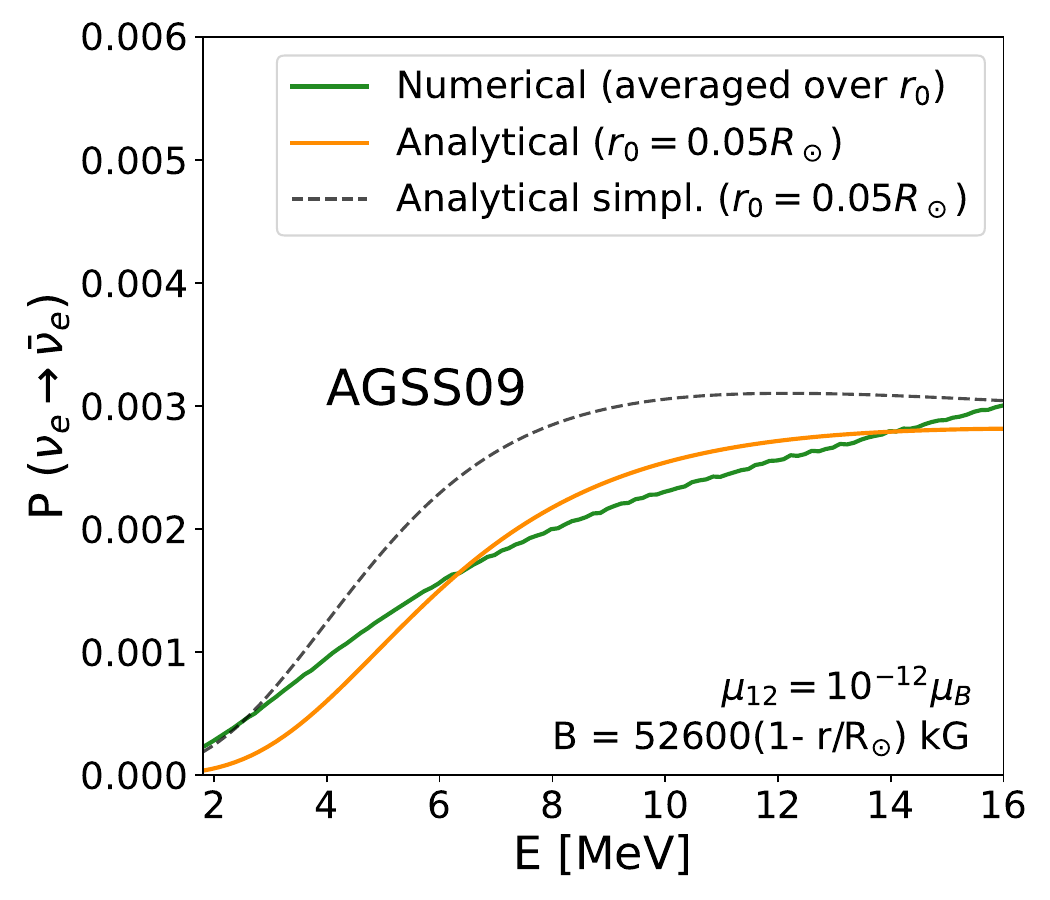}
\includegraphics[width = 0.375\paperwidth]{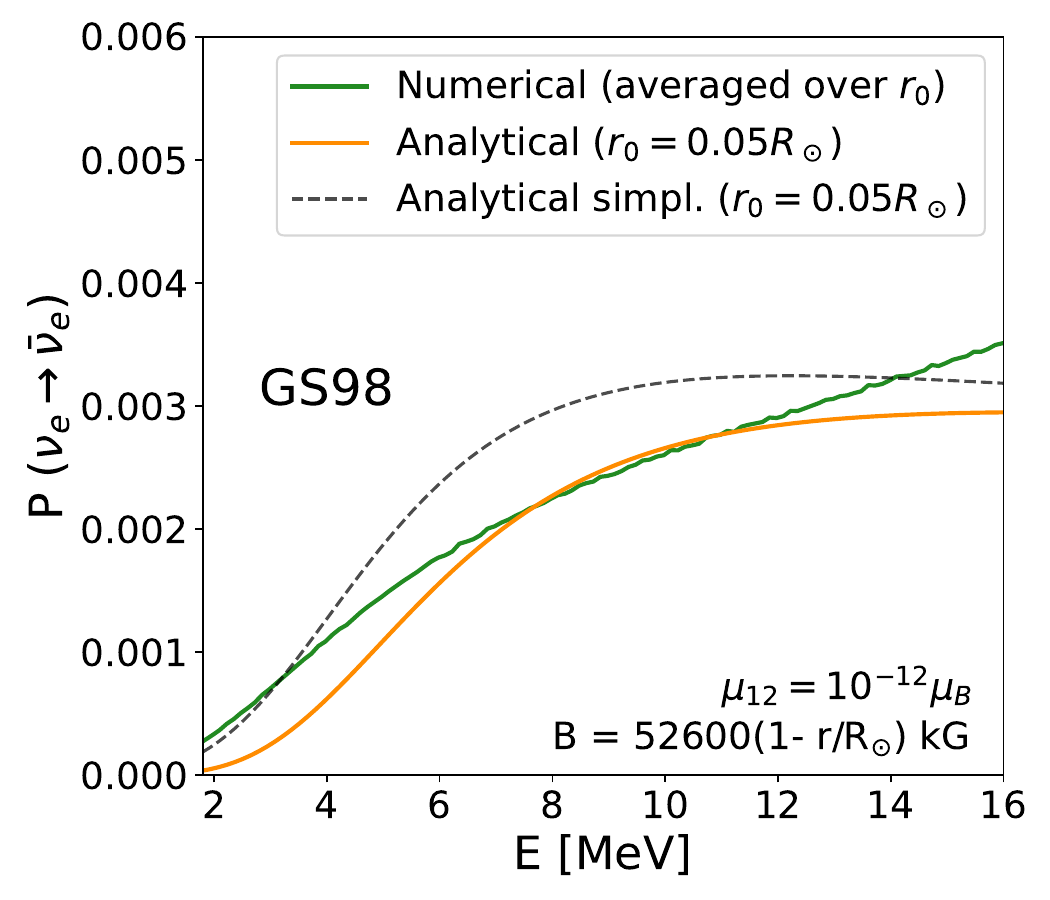}
\caption{Comparison of different calculations for $\bar{\nu}_{eR}$ appearance probability for $^8$B neutrinos. We take the values \mbox{($\sin^2\theta_{12}$, $\Delta m^2_{21}$) = (0.32, 7.5$\times 10^{-5}$ eV$^2$)} and \mbox{$\sin^2\theta_{13} = 0.022$} for the oscillation parameters~\cite{deSalas:2020pgw}. The left and right panels correspond to AGSS09 and GS98 SSM, respectively~\cite{Serenelli:2009yc}.
The green curves show numerical calculation with averaging over the neutrino production region and the orange ones are the results based on the full analytical expression in Equation~\ref{eq:ch4-peebar_full}. Grey dashed curves correspond to the simplified analytical expression in Equation~\ref{eq:ch4-peebar_simple}.
\labfig{fig:ch4-Peebar2020}}

\end{figure*}

In \reffig{fig:ch4-Peebar2020} we compare our analytical results with those found by numerical solution of the Equations \ref{eq:ch4-evol3}. The green curves show the numerical results obtained, after averaging over the neutrino production region.
The solid orange curves and grey dashed ones correspond to the analytical expressions in Equation \ref{eq:ch4-peebar_full} and Equation \ref{eq:ch4-peebar_simple}, respectively, assuming that all neutrinos are produced in the Sun at $r_0=0.05R_\odot$. 
The left and right panels show the results for AGSS09 and GS98 solar models, respectively. The figure demonstrates a good general agreement between our numerical and analytical results, especially for neutrino energies $E\gtrsim 5$ MeV. The discrepancy between the numerical and analytical results becomes larger for smaller $E$, where the $\bar{\nu}_{eR}$ appearance probability is relatively small. 

As shown in \reffig{fig:ch4-Peebar2020}, the $\bar{\nu}_{eR}$ appearance probability in Equation \ref{eq:ch4-peebar} is a relatively slowly varying function of neutrino energy for $E\gtrsim 5 - 8$ MeV, which is the relevant energy range for the experiments detecting mainly solar $^8$B neutrinos. Taking for an estimate its value at $E=12$ MeV and electron and neutron number densities at the neutrino production point $N_e \simeq 89$\,cm$^{-3}$ and $N_n \simeq 35$\,cm$^{-3}$~\cite{Serenelli:2009yc}, the electron antineutrino appearance probability can be written as 
\begin{equation}
P (\nu_{eL} \rightarrow \bar{\nu}_{eR}) \simeq 1.1\times 10^{-10} \left(
\frac{\mu_{12}B_\perp (r_0)}{10^{-12}\mu_B \cdot 10\, \text{kG}}\right)^2\,,
\label{eq:ch4-analit1}
\end{equation}
where $\mu_B$ is the electron Bohr magneton. 
An expression of this form was derived in the two-flavour approach~\cite{Akhmedov:2002mf} and has been extensively used in the literature to derive limits on magnetic moments. That previous expression differs by a factor of 1.4 with respect to the one here presented. This is partly due to 3-flavour effects and the use of updated neutrino mixing parameters and solar models and partly because of the approximation $\cos\tilde{\theta}(r_0)\ll\sin\tilde{\theta} (r_0)$ adopted in~\cite{Akhmedov:2002mf}. 

Note that Equation \ref{eq:ch4-analit1} is not suitable for experiments sensitive to $pp$, $pep$ or $^7$Be solar neutrinos, for which the electron antineutrino appearance probability is strongly suppressed. See, for instance, that at $E\sim 1$ MeV, the exact numerical result is approximately three orders of magnitude smaller than that given by Equation \ref{eq:ch4-analit1} and also exhibits a stronger energy dependence.

%%%%%%%%%%%%%%%%%%%%%%%%%%%%%%%%%%%%%%%%%%%
\subsection{Neutrino evolution inside the Sun}
%%%%%%%%%%%%%%%%%%%%%%%%%%%%%%%%%%%%%%%%%%%

It is possible to gain a better understanding of the process of antineutrino appearance by considering the evolution of the system inside the Sun as a function of distance to the centre of the Sun. We do so in a two-flavour framework --- setting $\sin^2 \theta_{13} = 0$. Since corrections are of order $\sin^2\theta_{13}$, the results derived in this way do not deviate significantly from those in the full 3-flavour approach.

In \reffig{fig:ch4-inside-bases} we show the evolution of the antineutrino appearance probabilities as a function of the distance to the centre of the Sun. Results were obtained by numerically solving the evolution equations for mass-eigenstates --- in the left panel --- and \textit{primed} states --- in the right panel. 
\begin{figure*}
\includegraphics[width = 0.76\paperwidth]{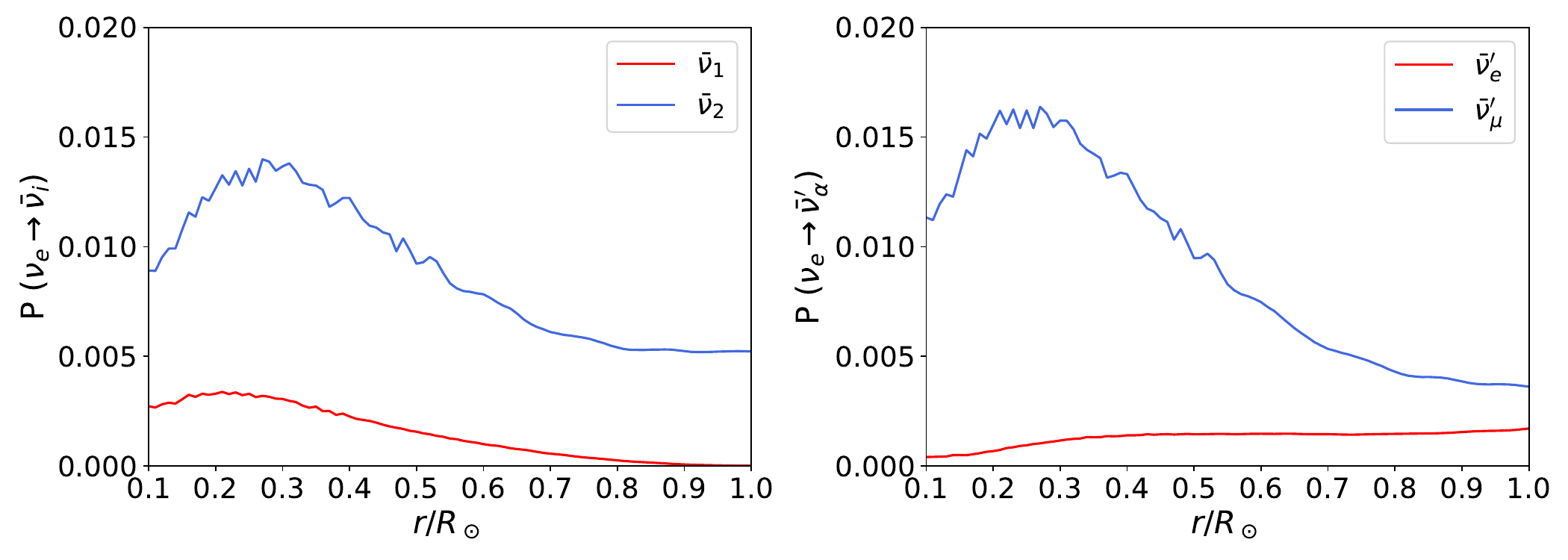}
    \caption{
Antineutrino appearance probabilities in the two-neutrino approach for the mass eigenstates and for the states in the \textit{primed basis} --- in the left and right panels respectively --- as functions of the distance to the centre of the Sun. The neutrino energy is fixed to $E=10$ MeV. Transition magnetic moments are $\mu_{12}= 10^{-12} \mu_B$, $\mu_{13} = \mu_{23}=0$ and the production point of electron neutrinos is $r_0= 0.05R_\odot$.\labfig{fig:ch4-inside-bases}}
\end{figure*}
From the right panel, one can see that considering that $\bar{\nu}_e'\sim 0$ is a reasonably good approximation inside the Sun. However, notice that it becomes less accurate when approaching the surface. 

In this two-flavour approach, spin-flavour precession converts solar $\nu_{e}$ into $\bar{\nu}_\mu'$. Close to the production point, the composition of $\bar{\nu}_\mu'$ is approximately 
\begin{align}
\bar{\nu}_\mu'\simeq - \sin \bar{\vartheta}(r_0) \bar{\nu}_{1M} + \cos \bar{\vartheta}(r_0) \bar{\nu}_{2M}\,.
\end{align}
Here, $\bar{\nu}_{1,2 M}$ are the states that diagonalise the antineutrino Hamiltonian in matter and the effective mixing angle $\bar{\vartheta}(r)$ is    
\begin{align}
\tan 2\bar{\vartheta}(r) = \frac{\sin 2 \theta_{12}}{\cos 2\theta_{12} 
+ c^2_{13}V_e(r)/2\delta}\, .
\end{align}
Given the large electron number density around the neutrino production point, for neutrino energies $E \gtrsim 2 \text{ MeV}$, the mixing is $\bar{\vartheta}(r_0)\ll 1$ and, therefore, $\bar{\nu}'_\mu \simeq \bar{\nu}_{2M}$. Since antineutrinos evolve adiabatically in the Sun and there is no level crossing, $\bar{\nu}_{2M}$ states propagate through the Sun without a noticeable transformation into $\bar{\nu}_{1M}$.
At the surface, matter density essentially vanishes. Hence, matter eigenstates become mass eigenstates there and antineutrinos emerge at the surface of the Sun as $\bar{\nu}_{2}$. This can be seen in \reffig{fig:ch4-inside-norm}, where we depict the appearance probabilities for $\bar{\nu}_1$ and $ \bar{\nu}_2$ --- in the left panel --- and for the matter eigenstates $\bar{\nu}_{1M}$ and $ \bar{\nu}_{2M}$ --- in the right panel. Notice that we have normalised the results to unit sum.  

At a distance $r=0.1R_{\odot}$ --- which is relatively close to the neutrino production point --- most of the antineutrinos are in the form of $\bar{\nu}_{2M}$ states, which are a combination of $\bar{\nu}_{1}$ and $\bar{\nu}_2$. 
At the surface of the Sun, the antineutrinos emerge as $\bar{\nu}_{2M}$ as well, which coincides there with $\bar{\nu}_2$. Again this is in agreement with the numerical results shown in the left panel of \reffig{fig:ch4-inside-bases}, where one can see that at $r=R_\odot$ we mainly find $\bar{\nu}_2$. 
Since $\bar{\nu}_2$ is a linear combination of $\bar{\nu}_e'$ and $\bar{\nu}_\mu'$  --- with weights $\sin^2\theta_{12}\simeq 1/3$ and $\cos^2\theta_{12}\simeq 2/3$ respectively --- at the surface of the Sun the appearance probability of $\bar{\nu}_\mu'$ is about twice that of $\bar{\nu}_e'$ --- as shown in the right panel of \reffig{fig:ch4-inside-bases}. 
It should be noted that, unlike for the normalised probabilities shown in \reffig{fig:ch4-inside-norm}, the sum of the antineutrino appearance probabilities presented in \reffig{fig:ch4-inside-bases} is not conserved. This results from the fact that some of the antineutrinos can precess back to neutrinos while they evolve inside the Sun. 
 
\begin{figure*}
\includegraphics[width = 0.76\paperwidth]{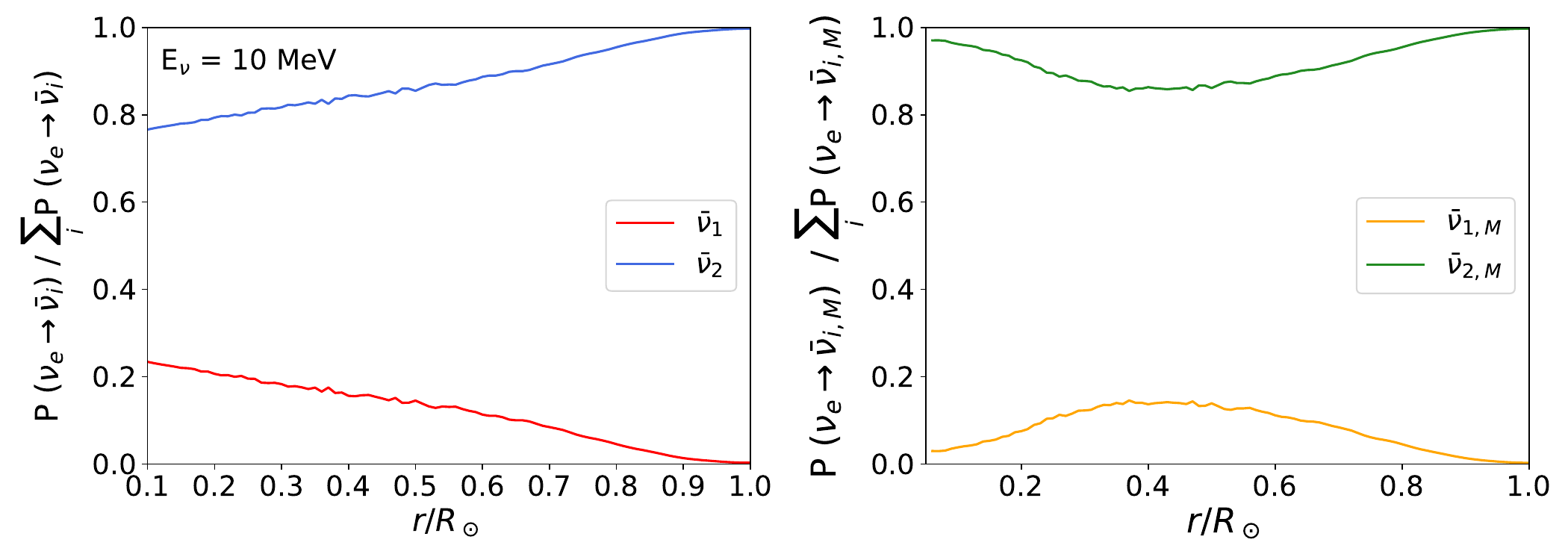}
\caption{Two-flavour evolution of antineutrino appearance probabilities inside the Sun for mass eigenstates and the matter eigenstates --- in the left and right panel respectively --- normalised to the unit total antineutrino appearance probability. The numerical values chosen for the neutrino energy, magnetic moment and the $\nu_e$ production coordinate are the same as in \reffig{fig:ch4-inside-bases}.\labfig{fig:ch4-inside-norm}}
\end{figure*}

\subsection{The roles of various transition magnetic moments and the magnetic field profile}
\label{subsec:ch4-profile}

In the above, our numerical analysis was assuming only one transition magnetic moment --- namely $\mu_{e'\mu'}=\mu_{12}$ --- to be nonzero. This was motivated by our analytical results, which showed that the contributions from $\mu_{e'\tau'}$ and $\mu_{\mu'\tau'}$ --- which are linear combinations $\mu_{13}$ and $\mu_{23}$ --- are strongly suppressed. 
This point is illustrated in \reffig{fig:ch4-comparison}, where we present the $\bar{\nu}_e$ appearance probability $P(\nu_e\to \bar{\nu}_e)$ at the Earth when one non-zero magnetic moment is allowed at a time. It demonstrates that unless $\mu_{13}$ or $\mu_{23}$ are more than one order of magnitude larger than $\mu_{12}$, the latter completely dominates the $\nu_e\to \bar{\nu}_e$ conversion.  

\begin{figure}[t]
\centering
\includegraphics[width = 0.45\paperwidth]
{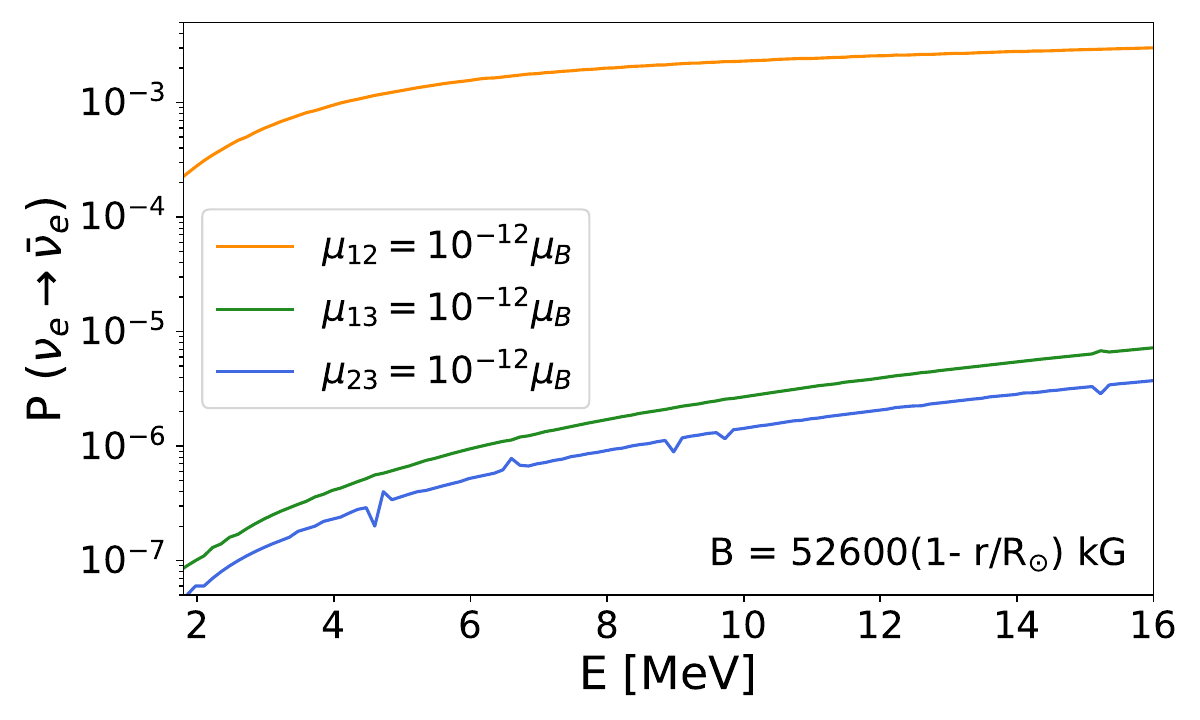}
\caption{Electron antineutrino appearance probability on the Earth as a function of neutrino energy for non-zero $\mu_{12}$, $\mu_{13}$ and $\mu_{23}$. Results are shown for the AGSS09 solar model and the evolution equations were solved numerically.\labfig{fig:ch4-comparison}}
\end{figure}

Up until this point, all the numerical results were obtained assuming that the magnetic field strength decreases linearly as in Equation~\ref{eq:ch4-b0}. We shall now study how the solar magnetic field profile affects $\nu_e\to \bar{\nu}_e$ conversion.  
For this purpose, we compare the $\bar{\nu}_e$ appearance probability for the linear profile we used above with those for two different parabolic profiles --- which we denote by $B_1(r)$ and $B_2(r)$. The three profiles are such that the strength at $R = 0.5R_\odot$ is $\sim 5\times 10^4$\,kG at $r=0.05R_\odot$ and vanishes at the surface of the Sun.
The first profile, 
    \begin{equation}
    B_1(r) = 50000 + 2632 \frac{r}{R_\odot} - 52632 \left(\frac{r}
    {R_\odot}\right)^2 \, \text{kG} \, ,
    \label{eq:ch4-b1}
    \end{equation}
is almost flat over the production region whereas the second one,  
    \begin{equation}
    B_2(r) = 55000 - 102368 \frac{r}{R_\odot} + 47368 \left(\frac{r}
    {R_\odot}\right)^2 \, \text{kG} \,,
    \label{eq:ch4-b2}
    \end{equation} 
corresponds to the magnetic field that is smaller than the linear one for $r > 0.05R_\odot$. A more detailed discussion on the possible magnitude of the magnetic field is carried out in \refsec{ch4-discussion}.

In the left panel of \reffig{fig:ch4-magn_profile} we present the magnetic field profiles we employ, whereas, in the right panel, we show the corresponding $\bar{\nu}_e$ appearance probabilities. For neutrino energies $E\lesssim 7$\,MeV all the magnetic field profiles we considered lead to $\bar{\nu}_e$ appearance probabilities that are quite similar to each other. Nevertheless, the differences between them and, therefore, the sensitivity to the magnetic field profile, increases with neutrino energy. 
The reason for this is twofold. Firstly, neutrinos are produced over extended regions. Thus, the $\nu_e\to \bar{\nu}_e$ production probability is sensitive to the magnetic profile in that region. 
Secondly, the $\bar{\nu}_e$ appearance probability depends on the effective mixing of the left-handed and right-handed neutrinos at their production point $r_0$, which is proportional to $\mu_{12}B_\perp(r_0)/(\Delta m_{21}^2/2E)$ and increases with the energy.
From \reffig{fig:ch4-magn_profile}, it follows that for energies of $E\sim 8$\,MeV, the dependence of the electron antineutrino appearance probability on the choice of the magnetic field profile is approximately 10 $-$ 15\%. 

\begin{figure*}
\centering
\includegraphics[width = 0.78\paperwidth]
{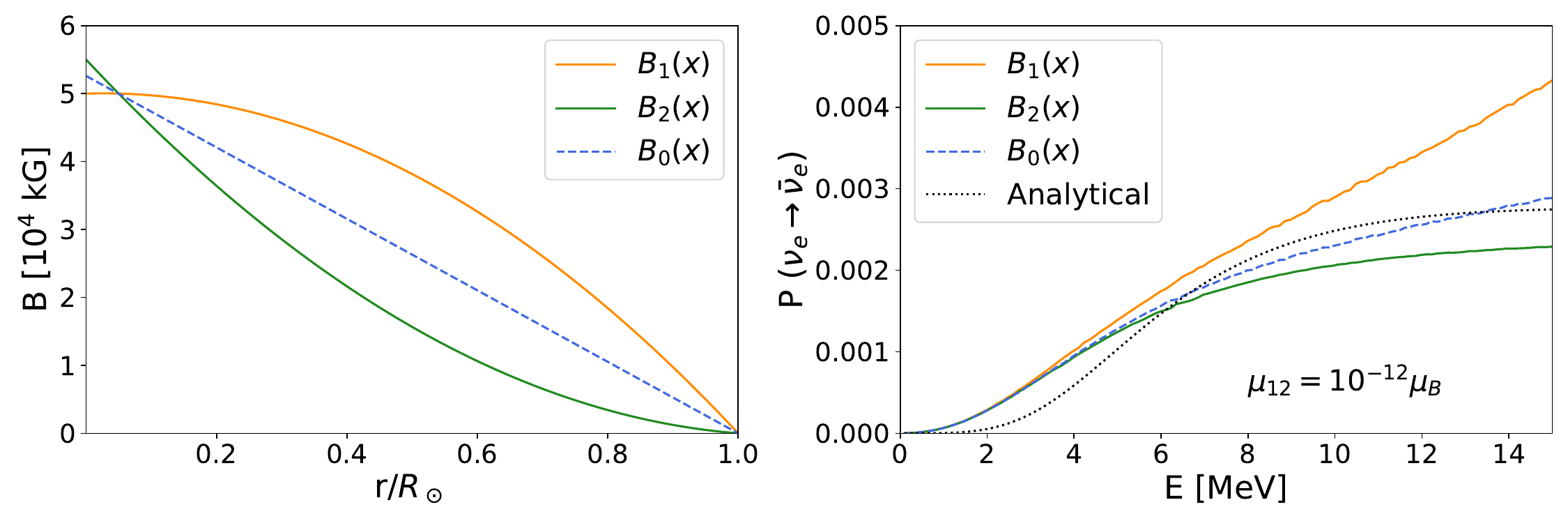}
\caption{Benchmark magnetic field profiles inside the Sun, 
as defined in Equation \ref{eq:ch4-b0}, Equation \ref{eq:ch4-b1} and Equation~\ref{eq:ch4-b2} and the corresponding electron antineutrino appearance probabilities on the Earth --- in the left and right panel respectively. The results of the analytical expression are shown by a black dotted curve. For the rest of the curves evolution equations were solved numerically.\labfig{fig:ch4-magn_profile} }
\end{figure*}

%%%%%%%%%%%%%%%%%%%%%%%%%%%%%%%%%%%%%%%%%%%%%%%%%%%%%%%%
\subsection{Average $\mathbf{\bar{\nu}_e}$ appearance probability and expected flux}
%%%%%%%%%%%%%%%%%%%%%%%%%%%%%%%%%%%%%%%%%%%%%%%%%%%%%%%%

Several experimental collaborations have reported upper bounds on electron antineutrino fluxes from astrophysical sources. These are obtained for different energy ranges as
\begin{equation}
    \Phi_\text{C.L.} = \frac{N_\text{C.L.}}{\epsilon \cdot \langle\sigma\rangle
\cdot T\cdot N_p} \, ,
\end{equation}
where $N_\text{C.L.}$ denotes the upper limit on the number of events at a given confidence level, $\epsilon$ is the average detection efficiency in the energy range considered, $\langle\sigma\rangle$ is the averaged cross-section in the same energy range, $T$ is the exposure time and $N_p$ is the number of targets in the detector.

To facilitate the use of our results in ongoing and future data analyses, we compute, for different energy bins $E \in [E_i-\Delta E/2, E_i+\Delta E/2]$, the averaged electron antineutrino appearance probability,
\begin{equation}
    \langle P_{i} \rangle = \frac{ \bigintss_{E_i - \Delta E/2}^{E_i 
+\Delta E/2} \phi(E) \sigma(E) P(E) dE}{ 
\bigintss_{E_i - \Delta E/2}^{E_i +\Delta E/2} \phi(E) \sigma(E) dE} \, ,
\label{eq:ch4-Pi}
\end{equation}
and the expected flux of electron antineutrinos,
\begin{equation}
    \langle\Phi_{i}\rangle = \frac{ \bigintss_{E_i - \Delta E/2}^{E_i +
\Delta E/2} \phi(E) \sigma(E) P(E) dE}{ \bigintss_{E_i - 
\Delta E/2}^{E_i +\Delta E/2} \sigma(E) dE} \,.
\label{eq:ch4-Phii}
\end{equation}
We assume that the detection efficiency $\epsilon$ is energy independent within each bin so that it cancels out when computing the ratios --- see Equation \ref{eq:ch4-Pi} and Equation \ref{eq:ch4-Phii}.
We also considered perfect detector energy resolution --- or at least an energy-independent one. We have checked that, for the KamLAND experiment, when including a realistic energy resolution of 6.4\%/$\sqrt{E \text{(MeV)}}$, our results deviate in less than 0.5\%. This is a consequence of $\bar{\nu}_e$ appearance probability being a rather smooth function of neutrino energy --- see the right panel of \reffig{fig:ch4-magn_profile}. 

The most restrictive limits so far have been derived using inverse beta decay as a detection channel. Hence, we focus on that process and restrict ourselves to energies above its threshold --- which is of $\sim$1.8 MeV --- where the only relevant contribution is due to $^8$B neutrinos.
We compute the $\bar{\nu}_e$ appearance probability and the expected flux numerically, using both the numerical and analytical expressions for the probabilities $P(\nu_e\to\bar{\nu}_e)$.
In \reftab{ch4-estimatesAGSS09} and \reftab{ch4-estimatesGS98} we present these probabilities and the expected $\bar{\nu}_e$ fluxes for $\mu_{12} = 10^{-12} \mu_B$ and $B_\perp(r_0) = 1$ kG. As we have shown, to a very good extent, the electron antineutrino appearance probability  --- and therefore the $\bar{\nu}_e$ flux --- is proportional to $(\mu_{12}B_\perp(r_0))^2$. Then, the values of $\langle P_i\rangle$ and $\langle \Phi_i\rangle$ for different choices of $\mu_{12}B_\perp(r_0)$ can be found simply by rescaling.

\begin{table*}[t]
\centering
\renewcommand*{\arraystretch}{1.2}
\begin{tabular}{ccccc}
\toprule[0.25ex]
 & \multicolumn{2}{c}{Numerical AGSS09} & \multicolumn{2}{c}
{Analytical AGSS09}\\[1.2mm] 
 &   & $\langle\Phi_{i}\rangle$ &  & $\langle\Phi_{i}\rangle $  \\
 E [MeV] & $\langle P_{i} \rangle$  & $[\text{cm}^{-2}  \text{s}^{-1}  \text{MeV}^{-1}]$ & $\langle P_{i}\rangle$ & $[\text{cm}^{-2}  \text{s}^{-1}  \text{MeV}^{-1}]$   \\ \midrule
1.8	- 2.8	&	1.73$\times 10^{-13}$	&
4.62$\times 10^{-8}$	&	5.08$\times 10^{-14}$	&	1.36$\times 
10^{-8}$	\\
2.8	- 3.8 &	2.95$\times 10^{-13}$	&
1.18$\times 10^{-7}$	&	1.45$\times 10^{-13}$	&	5.82$\times 
10^{-8}$	\\
3.8	- 4.8 &	4.28$\times 10^{-13}$	&
2.24$\times 10^{-7}$	&	2.97$\times 10^{-13}$	&	1.55$
\times 10^{-7}$	\\
4.8	- 5.8 &	5.52$\times 10^{-13}$	&
3.34$\times 10^{-7}$	&	4.73$\times 10^{-13}$	&	2.86$
\times 10^{-7}$	\\
5.8	- 6.8 &	6.57$\times 10^{-13}$	&
4.19$\times 10^{-7}$	&	6.38$\times 10^{-13}$	&	4.07$
\times 10^{-7}$	\\
6.8	- 7.8 &	7.45$\times 10^{-13}$	&
4.62$\times 10^{-7}$	&	7.74$\times 10^{-13}$	&	
4.80$\times 10^{-7}$	\\
 7.8	- 8.8	&	8.19$\times 10^{-13}$	&
4.55$\times 10^{-7}$	&	8.78$\times 10^{-13}$	&	
4.88$\times 10^{-7}$	\\
8.8	- 9.8 &	8.84$\times 10^{-13}$	&
4.03$\times 10^{-7}$	&	9.53$\times 10^{-13}$	&	
4.35$\times 10^{-7}$	\\
9.8  - 10.8 	&	9.38$\times 10^{-13}$	&
3.15$\times 10^{-7}$	&	1.01$\times 10^{-12}$	&	
3.38$\times 10^{-7}$	\\
10.8 - 11.8 	&	9.87$\times 10^{-13}$	&
2.11$\times 10^{-7}$	&	1.04$\times 10^{-12}$	&	
2.23$\times 10^{-7}$	\\
11.8 - 12.8 	&	1.04$\times 10^{-12}$	&
1.12$\times 10^{-7}$	&	1.07$\times 10^{-12}$	&	
1.16$\times 10^{-7}$	\\
12.8 - 13.8 &	1.08$\times 10^{-12}$	&
3.87$\times 10^{-8}$	&	1.08$\times 10^{-12}$	&	
3.89$\times 10^{-8}$	\\
13.8 - 14.8 &	1.12$\times 10^{-12}$	&
5.77$\times 10^{-9}$	&	1.09$\times 10^{-12}$	&	
5.63$\times 10^{-9}$	\\
14.8 - 15.8 &	1.16$\times 10^{-12}$	&
2.83$\times 10^{-10}$	&	1.10$\times 10^{-12}$	&	
2.68$\times 10^{-10}$	\\ \bottomrule[0.25ex]
\end{tabular}
\caption{\label{tab:ch4-estimatesAGSS09} Averaged $\bar{\nu}_e$ appearance probabilities and expected fluxes of $\bar{\nu}_e$ from the Sun for low-metallicity AGSS09 SSM. Results consider inverse beta decay as the detection channel, a value of \mbox{$\mu_{12}B_\perp(r_0) =10^{-12}\mu_B \cdot \text{kG}$,} and a magnetic field profile as in \mbox{Equation \ref{eq:ch4-b0}.} For rescaling to different values of $\mu_{12}B_\perp(r_0)$ see text.}
\end{table*}

\begin{table*}[t]
\centering
\renewcommand*{\arraystretch}{1.2}
\begin{tabular}{ccccc}
\toprule[0.25ex]
 & \multicolumn{2}{c}{Numerical GS98} & \multicolumn{2}{c}
{Analytical GS98}\\[1.2mm] 
 &   & $\langle\Phi_{i}\rangle$ &  & $\langle\Phi_{i}\rangle $  \\
 E [MeV] & $\langle P_{i} \rangle$  & $[\text{cm}^{-2}  \text{s}^{-1}  \text{MeV}^{-1}]$ & $\langle P_{i}\rangle$ & $[\text{cm}^{-2}  \text{s}^{-1}  \text{MeV}^{-1}]$   \\ \midrule
1.8	-2.8 	& 2.03$\times 10^{-13}$	 & 6.60$\times 10^{-8}$	 &  5.43$\times 10 ^{-14}$	 & 1.76$\times 10^{-8}$	\\
2.8	- 3.8	& 3.43$\times 10^{-13}$	 & 1.67$\times 10^{-7}$	 &  1.55$\times 10^{-13}$	 & 7.56$\times 10^{-8}$	\\
3.8	- 4.8	& 4.92$\times 10^{-13}$	 & 3.12$\times 10^{-7}$	 &  3.19$\times 10^{-13}$	 & 2.02$\times 10^{-7}$	\\
4.8	- 5.8	& 6.27$\times 10^{-13}$	 & 4.60$\times 10^{-7}$	 &  5.06$\times 10^{-13}$	 & 3.72$\times 10^{-7}$	\\
5.8	- 6.8	& 7.41$\times 10^{-13}$	 & 5.73$\times 10^{-7}$	 &  6.82$\times 10^{-13}$	 & 5.28$\times 10^{-7}$	\\
6.8	- 7.8	& 8.39$\times 10^{-13}$	 & 6.31$\times 10^{-7}$	 &  8.26$\times 10^{-13}$	 & 6.21$\times 10^{-7}$	\\
7.8 - 8.8	& 9.23$\times 10^{-13}$	 & 6.22$\times 10^{-7}$	 &  9.35$\times 10^{-13}$	 & 6.30$\times 10^{-7}$	\\
8.8	- 9.8	& 9.96$\times 10^{-13}$	 & 5.51$\times 10^{-7}$	 &  1.0$\times 10^{-12}$	 & 5.60$\times 10^{-7}$	\\
9.8  - 10.8	 & 1.06$\times 10^{-12}$  & 4.33$\times 10^{-7}$  &  1.07$\times 10^{-12}$	 & 4.35$\times 10^{-7}$	\\
10.8 - 11.8	& 1.12$\times 10^{-12}$	 & 2.92$\times 10^{-7}$	 &  1.10$\times 10^{-12}$	 & 2.87$\times 10^{-7}$	\\
11.8 - 12.8	& 1.18$\times 10^{-12}$	 & 1.55$\times 10^{-7}$	 &  1.13$\times 10^{-12}$	 & 1.48$\times 10^{-7}$	\\
12.8 - 13.8	& 1.24$\times 10^{-12}$	 & 5.38$\times 10^{-8}$	 &  1.14$\times 10^{-12}$	 & 4.99$\times 10^{-8}$	\\
13.8 - 14.8	& 1.29$\times 10^{-12}$	 & 8.06$\times 10^{-9}$	 &  1.15$\times 10^{-12}$	 & 7.21$\times 10^{-9}$ \\
14.8 - 15.8	& 1.34$\times 10^{-12}$	 & 3.98$\times 10^{-10}$	 &  1.16$\times 10^{-12}$	 & 3.43$\times 10^{-10}$ \\ \bottomrule[0.25ex]
    \end{tabular}
    \caption{\label{tab:ch4-estimatesGS98} Same as in \reftab{ch4-estimatesAGSS09} but for high-metallicity GS98 SSM. 
}
\end{table*}

For better illustration, \reffig{fig:ch4-pav_flux} provides a graphical comparison between the electron antineutrino $\bar{\nu}_e$ appearance probabilities and the predicted $\bar{\nu}_e$ fluxes at the Earth obtained numerically and analytically. These results correspond to the Standard Solar Model AGSS09, a magnetic field strength as in Equation \ref{eq:ch4-b0} and $\mu_{12} = 10^{-12} \mu_B$. 
One can see from the figure that there is a good accord between our numerical and analytical results for neutrino energies $E\gtrsim 6$ MeV. However, the agreement worsens towards smaller $E$. Thus, while our simple analytical results can be reliably used at relatively high neutrino energies, numerical results should preferably be used for analysing experiments sensitive to the low-energy part of the \mbox{solar $^8$B} neutrino spectrum, such as in Borexino.     

\begin{figure*}
    \centering
    \includegraphics[width = 0.78\paperwidth]{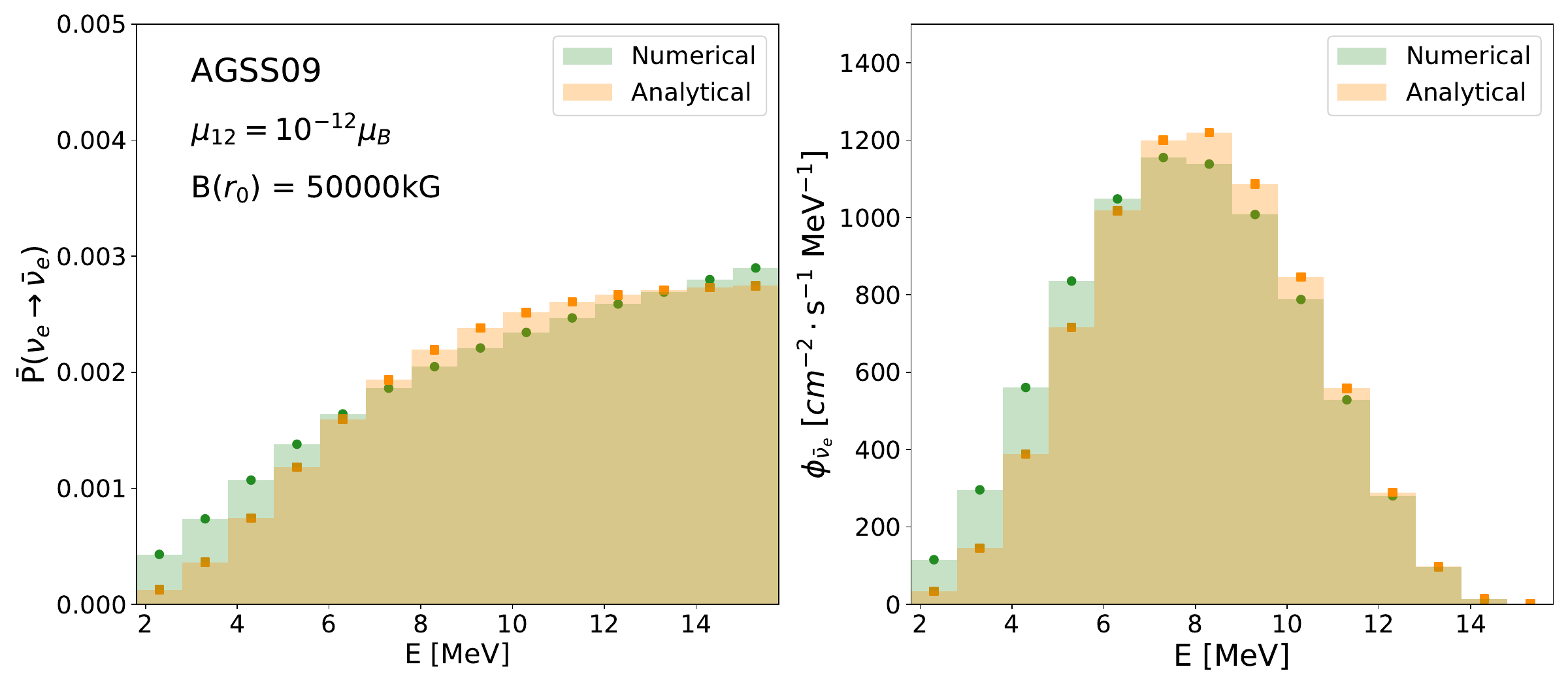}
    \caption{Comparison of the numerical and analytical results for the electron antineutrino appearance for AGSS09 SSM. Appearance probabilities and expected fluxes are shown in the left and right panels respectively, for a magnetic field as in Equation \ref{eq:ch4-b0} and a magnetic moment $\mu_{12} = 10^{-1 2} \mu_B$.\labfig{fig:ch4-pav_flux}}
\end{figure*}

%%%%%%%%%%%%%%%%%%%%%%%%%%%%%%%%%%%%%%%%%%%%%%%%%%
\subsection{Existing limits from astrophysical $\bar{\nu}_e$ fluxes revisited}
%%%%%%%%%%%%%%%%%%%%%%%%%%%%%%%%%%%%%%%%%%%%%%%%%%

In light of the new results, we revisit the existing limits on neutrino magnetic moments and solar magnetic fields coming from the upper bounds on astrophysical $\bar{\nu}_e$ fluxes and compare them with our results. 
At present, the most stringent bounds come from the KamLAND experiment~\cite{KamLAND:2021gvi}, although Borexino and Super-Kamiokande have set comparable limits~\cite{Borexino:2019wln, Super-Kamiokande:2020frs,Super-Kamiokande:2021jaq}. In all cases, the limits were derived from the inverse beta decay channel.
Historically, SNO also put constraints on astrophysical $\bar{\nu}_e$ in the MeV energy range using charged-current interactions with deuterium~\cite{SNO:2004eru}. However, these limits are not currently competitive.

We depict the model-independent limits on the $\bar{\nu}_e$ flux established by the above-mentioned experiments in \reffig{fig:ch4-limits}. For comparison, we also show our $\bar{\nu}_e$ flux prediction for solar electron antineutrinos for the AGSS09 SSM and for $\mu_{12}B(r_0) = 2.5 \times 10^{-9}\mu_B\,$kG. Notice that the experimental bounds at neutrino energies $E\sim 10$ MeV are the closest to the predicted flux. 
Notice also that the bounds are stronger at energies around 20-30 MeV. However, the flux of solar neutrinos is extremely low for energies above 16 MeV. 

\begin{figure*}[t]
    \centering
    \includegraphics[width = 0.68\paperwidth]{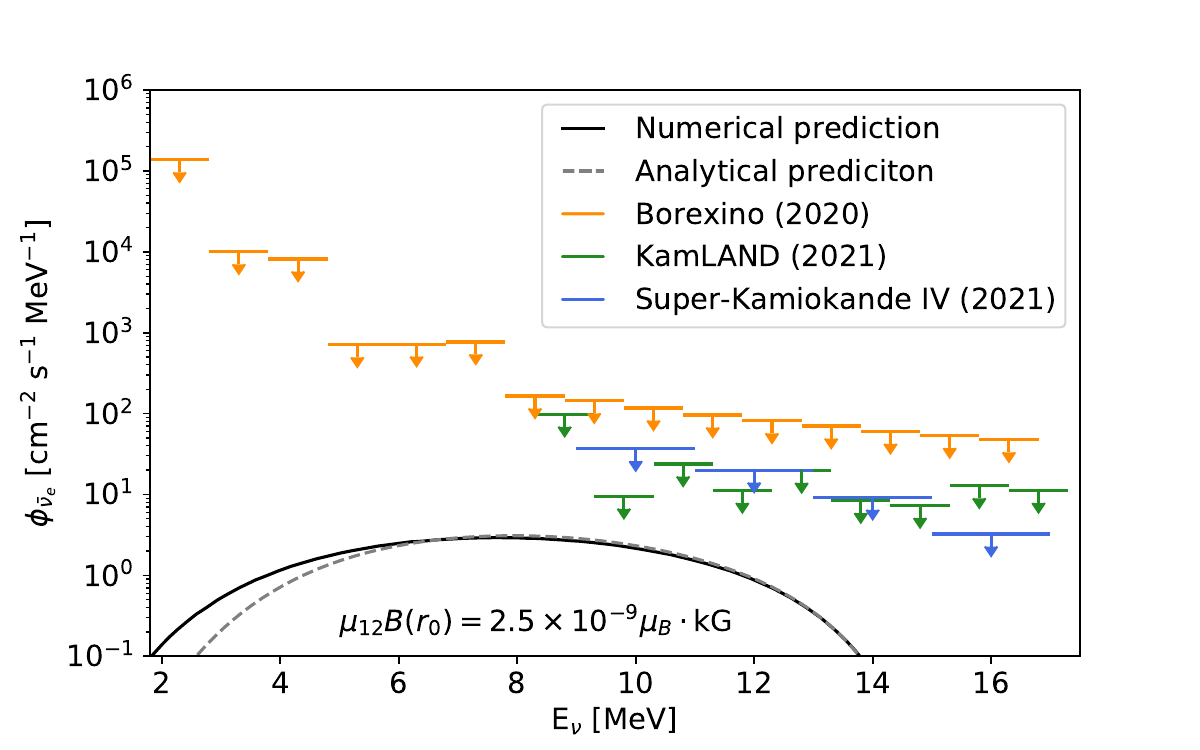}
    \caption{Model-independent limits on $\bar{\nu}_e$ flux of 
    astrophysical origin, as reported by KamLAND~\cite{KamLAND:2021gvi}, Borexino~\cite{Borexino:2019wln} and Super-Kamiokande \
\cite{Super-Kamiokande:2020frs,Super-Kamiokande:2021jaq}, together with the expected solar $\bar{\nu}_e$ flux for $\mu_{12}B_\perp(r_0)= 2.5 \times 10^{-9}\mu_B\,$kG We show the results from our numerical and analytical calculation for the AGSS09 SSM.    \labfig{fig:ch4-limits}}
\end{figure*}

The 90$\%$ C.L. upper limits on the product of the neutrino magnetic moment and the solar magnetic field strength we obtain from the KamLAND upper bound on the astrophysical $\bar{\nu}_e$ flux are
\begin{align}
&\mu_{12} B_\perp(r_0)\Big| _\text{AGSS09} < \left( 4.9-5.1\right)
\times 10^{-9} \mu_B\,\text{kG}\,, \nonumber
\\ 
&\mu_{12} B_\perp(r_0)\Big|_\text{ GS98} < \left( 4.7 -4.8\right)
\times 10^{-9} \mu_B\,\text{kG}\,.
\label{eq:ch4-KamLANDlimit}
\end{align}
where we have considered two Standard Solar Models --- AGSS09 and GS98. Here the lower numbers correspond to our analytical approximation and the higher ones, to the full numerical calculation. 
A good general accord between the results of the two approaches can be seen. These results are also consistent with the limits derived in~\cite{KamLAND:2021gvi}, $\mu B_\perp(r_0) < 4.9 \times 10^{-9} \mu_B\,\text{kG}$, using the two-neutrino prediction from~\cite{Akhmedov:2002mf}.

Similarly, one can derive the 90$\%$ C.L. limits from the Borexino results, 
\begin{align}
&\mu_{12} B_\perp(r_0)\Big|_\text{ AGSS09} < \left( 1.8-1.9\right)
\times 10^{-8} \mu_B\,\text{kG}\,, \nonumber
\\ 
&\mu_{12} B_\perp(r_0)\Big|_\text{GS98} < \left( 1.7-1.8\right)\times 
10^{-8} \mu_B\,\text{kG}\,, 
\label{eq:ch4-BorLimitOurs}
\end{align}
The previous limit from Borexino~\cite{Borexino:2019wln}, for the high-metallicity GS98 SSM, was $\mu B_\perp(r_0) < 6.9 \times 10^{-9} \mu_B \cdot \text{kG}$. Notice that there is a factor $\sim 2.6$ discrepancy between this result and our limit --- in Equation \ref{eq:ch4-BorLimitOurs}. This is presumably related to the fact that, in the Borexino analysis, the simplified energy-independent formula from~\cite{Akhmedov:2002mf} --- derived for $E\sim 5-10$ MeV --- was used for neutrinos of smaller energies, i.e. outside its range of validity. As a result, they obtained a more stringent limit.  

Regarding the Super-Kamiokande experiment, at 90\% C.L., we find the following limit :
\begin{align}
&\mu_{12} B_\perp(r_0)\Big|_\text{AGSS09} < \left( 7.1-7.3\right)
\times 10^{-9} \mu_B \, \text{kG} \nonumber
\\ 
&\mu_{12} B_\perp(r_0)\Big|_\text{GS98} < \left( 6.8-6.9\right)
\times 10^{-9} \mu_B \, \text{kG} \, .
\label{eq:ch4-SK}
\end{align}
The corresponding previous bound was $\mu B_\perp(r_0) < 1.5 \times 10^{-8} \mu_B \, \text{kG}$~\cite{Super-Kamiokande:2020frs}. This is approximately a factor 2 weaker than our limit --- in Equation \ref{eq:ch4-SK}. The difference is probably because Super-Kamiokande looked for electron antineutrinos in the energy range 9.3 to 17.3 MeV but used in their analysis the same simplified energy-independent $\bar{\nu}_e$ appearance probability that was derived in~\cite{Akhmedov:2002mf} for smaller energies.

%%%%%%%%%%%%%%%%%%%%%%%%%%%%%%%%%%%%%%%%%%%%%%%%
%%%%%%%%%%%%%%%%%%%%%%%%%%%%%%%%%%%%%%%%%%%%%%%%
\section{Discussion}
\labsec{ch4-discussion}
%%%%%%%%%%%%%%%%%%%%%%%%%%%%%%%%%%%%%%%%%%%%%%%%
%%%%%%%%%%%%%%%%%%%%%%%%%%%%%%%%%%%%%%%%%%%%%%%%

In this chapter, we have presented a study of the conversion of solar electron neutrinos into antineutrinos through the combined action of spin-flavour precession and flavour conversion. This effect is based on the assumption that neutrinos are Majorana particles. In that case, the interactions between the solar magnetic field and the non-zero transition magnetic moments of neutrinos induce a simultaneous chirality-flip and a flavour transformation giving rise to a flux of solar antineutrinos.

Our approach starts by deriving the evolution equations in the three-neutrino picture in a rotated basis --- which we referred to as the \textit{primed basis}. This basis is known to be more convenient for the description of solar neutrinos. Since SFP can only play a subleading role in the picture of solar neutrinos, we developed an approach based on perturbation theory and obtained a simple analytical expression for the appearance probability of solar $\bar{\nu}_e$ on the Earth. The possibility that the solar magnetic fields may twist was also considered. We obtain an expression that can be readily employed for the analysis and interpretation of the experimental results on searches for $\bar{\nu}_e$ fluxes of astrophysical origin.

We have checked the validity of our approximation and the accuracy of the analytical solution we obtained with the full numerical calculation that results from solving a set of coupled evolution equations for the system. We have found a good general agreement between our numerical and analytical results, especially for neutrino energies $E\gtrsim 5$ MeV. The discrepancy we find between the two approaches is larger for smaller neutrino energies. However, in that energy range, the electron antineutrino appearance probability is relatively small.

Under the assumption that the three transition magnetic moments in the mass basis are of similar magnitude, we show that the $\bar{\nu}_e$ appearance probability is --- to a good accuracy --- proportional to the product of the magnetic moment $\mu_{12}^2$ and the magnetic field at the production point $B_\perp(r_0)^2$.
The contribution of the other two transition magnetic moments, $\mu_{13}$ and $\mu_{23}$, is strongly suppressed unless they exceed $\mu_{12}$ more than one order of magnitude. We have also shown that the profile of the magnetic field in the Sun plays a minor role and the expected flux of solar electron antineutrinos is mainly dominated by the magnetic field strength at the production point. 

To facilitate future analyses from collaborations searching for astrophysical electron antineutrino fluxes, we provide the average electron antineutrino appearance probabilities and the expected fluxes on Earth, for different energy bins and two different solar models --- with high and low metallicities --- in \reftab{ch4-estimatesAGSS09} and \reftab{ch4-estimatesGS98}.   
Using these results, we have also revisited and updated the existing upper bounds on $\mu_{12}B_\perp$. The best current bound on the product of the neutrino magnetic moment and the solar magnetic field comes from the KamLAND upper limit on the astrophysical $\bar{\nu}_e$ flux, 
\begin{align} 
\mu_{12} B_{\perp} (r_0)\lesssim 5\times 10^{-9} \mu_B\,\text{kG}\,,
\end{align} 
with a mild dependence on the solar model considered.

If the magnetic field strength in the core of the Sun were known, one could use the upper bounds on $\mu_{12}B_\perp$ to derive constraints on $\mu_{12}$ for Majorana neutrinos. Unfortunately, very little --- if anything --- is known about it. There exists a very conservative upper bound $B< 10 ^9$\,G coming from the requirement that the pressure of the magnetic field in the solar core does not exceed matter pressure~\cite{Schramm:1993mv}.
For illustrative purposes, let us assume the actual value of the magnetic field strength coincides with this upper limit.
Then, the bound from KamLAND would translate into the limit $\mu_{12}<5\times 10 ^{-15}\mu_B$ for the neutrino magnetic moment. From solar oblateness and the analysis of the splitting of the solar oscillation frequencies, one can argue that the magnetic fields in the radiative zone of the Sun can not exceed a certain upper value, namely $B \lesssim 7$\,MG~\cite{Friedland:2002is}.
If one assumed --- rather arbitrarily --- that the magnetic field in the solar core was similar in magnitude,  an upper limit of $\mu_{12} < 7.1\times 10^{-13}\mu_B$ would be obtained. 
Requiring the stability of toroidal magnetic fields in the radiative zone of the Sun one can find a much more stringent limit $B\lesssim 600$\,G~\cite{Kitchatinov, Bonanno2013}. Again, if the magnetic field in the core of the Sun were of similar magnitude, one would obtain the constraint $\mu_{12}<8.3 \times 10^{-9} \mu_B$. We stress there is no \textit{a priori} reason to believe that the magnetic fields in the core of the Sun and the radiative zone are of the same order. Thus, these limits must be regarded solely as reference values.  

On the other hand, if a non-zero neutrino magnetic moment were even hinted, for instance at scattering experiments,\footnote{For a review on existing limits on neutrino magnetic moments from neutrino scattering, see~Appendix~\ref{app1-magnetic}.} non-observation of solar electron antineutrino would teach us about the magnetic field in the core of the Sun and the Majorana --- or Dirac --- nature of neutrinos.

These limits on the product of the neutrino magnetic moment and the solar magnetic field strength will be improved shortly by current and \mbox{next-generation} neutrino observatories such as Super-Kamiokande loaded with gadolinium \cite{Super-Kamiokande:2023xup}, JUNO~\cite{JUNO:2022lpc} and Hyper-Kamiokande~\cite{Hyper-Kamiokande:2018ofw}, among others. The simple analytical expression for the electron antineutrino appearance probability derived here will facilitate --- and accelerate significantly --- the analyses of forthcoming data.

\chapter{Non-standard interactions with quarks and the future solar sector}
\labch{ch5-nsisolar}
The symmetries of the Standard Model determine the strength and flavour structure of neutrino interactions. Nonetheless, many extensions of the theory include or predict modifications to this picture. Such new interactions with matter fields would modify the cross-section of processes involving neutrinos. Consequently, one can search for such hypothetical interactions in scattering experiments. Besides, since new interactions modify neutrino propagation in a medium, oscillation experiments can explore these scenarios too. Model-independent approaches in which the Lorentz structure of the interaction vertex is not specified are often referred to as Generalised Neutrino Interactions. 

In this chapter, we will limit ourselves to the discussion of vector and axial vector interactions, generally dubbed neutrino non-standard interactions (NSIs), notably with quarks. In Section~\ref{sec:ch5-intro}, we discuss several theoretical aspects of relevance for the topic of non-standard interactions. Then, in Section~\ref{sec:ch5-nsi-osc} we review the role of NSIs in neutrino oscillations. Next, we introduce the next-generation experiments Hyper-Kamiokande and JUNO and give some details on our simulations and analyses in Sections~\ref{sec:ch5-HK} and~\ref{sec:ch5-JUNO} respectively. Finally, we present our results in Section~\ref{sec:ch5-results} and we comment on the role of our assumptions in Section~\ref{sec:ch5-conclusion}.

%%%%%%%%%%%%%%%%%%%%%%%%%%%%%%%%%%%%%%%%%%%%%%%%%%
%%%%%%%%%%%%%%%%%%%%%%%%%%%%%%%%%%%%%%%%%%%%%%%%%%
\section{Neutrino non-standard interactions}
\label{sec:ch5-intro}
%%%%%%%%%%%%%%%%%%%%%%%%%%%%%%%%%%%%%%%%%%%%%%%%%%
%%%%%%%%%%%%%%%%%%%%%%%%%%%%%%%%%%%%%%%%%%%%%%%%%%

%%%%%%%%%%%%%%%%%%%%%%%%%%%%%%%%%%
\subsection{General formalism}
\label{sec:ch5-gen-form}
%%%%%%%%%%%%%%%%%%%%%%%%%%%%%%%%%%
Neutrino non-standard interactions are generally parametrised in terms of effective four-fermion operators. Particularly, for the so-called \textit{neutral-current non-standard interactions} (NC-NSIs), the effective Lagrangian is
\begin{align}
\mathcal{L}_{\text{NC-NSI}} = -2\sqrt{2}G_F \varepsilon_{\alpha\beta}^{fX}\left(\bar{\nu}_\alpha\gamma^\mu P_L\nu_\beta \right)\left(\bar{f} \gamma_\mu P_X f\right)\, ,
\label{eq:ch5-NCNSI}
\end{align}
where $G_F$ is the Fermi constant and the strength of the interaction is determined by the dimensionless parameters $\varepsilon^{fX}_{\alpha\beta}$.\footnote{Note that, whenever these parameters are equal to $\pm 1$, it means that NSIs are as sizeable as Standard Model interactions.} In the expression above, the sums over neutrino flavours $\alpha, \beta = \lbrace e , \, \mu , \, \tau \rbrace$, chirality projection \mbox{$X = \lbrace L, \, R \rbrace$} and charged fermions, $f$ is implicit. Actually, since only electrons, $u$-quarks and $d$-quarks are present in ordinary matter, studies are normally limited to NSIs with these matter fields. Similarly, one can define the following effective Lagrangian \cite{Wolfenstein:1977ue}
\begin{align}
\mathcal{L}_{\text{CC-NSI}} = -2\sqrt{2}G_F \varepsilon_{\alpha\beta}^{ff'X}\left(\bar{\nu}_\alpha\gamma^\mu P_L l_\beta \right)\left(\bar{f}' \gamma_\mu P_X f\right)\, ,
\label{eq:ch5-CCNSI}
\end{align}
with $f\neq f'$.It accounts for NSIs involving two different charged fermions, which are often referred to as \textit{charged-current non-standard interactions} \mbox{(CC-NSIs).} Similarly to the previous case, the strength of the CC-NSIs is determined by $\varepsilon^{ff'X}_{\alpha \beta}$.

Depending on the flavour of the neutrinos involved in a given process, one can differentiate between \textit{flavour-changing} NSIs --- when $\alpha \neq \beta$ --- and \textit{non-universal} NSIs --- for $\alpha = \beta$. The former ones are named in this way since they correspond to lepton flavour non-conserving interactions, whereas, for the latter, lepton universality is not preserved if the difference between two of them is non-zero --- i.e. for instance if $\varepsilon^{fX}_{\alpha\alpha} - \varepsilon^{fX}_{\beta\beta} \neq 0$.

Instead of separating the two chiral components of the new interactions, one can project the NSIs in the vector and axial-vector components. For NC-NSIs, and if one rewrites the effective Lagrangian as \cite{Wolfenstein:1977ue}
\begin{align}
\mathcal{L}_{\text{NC-NSI}} = -2\sqrt{2}G_F \left(\bar{\nu}_\alpha\gamma^\mu P_L\nu_\beta \right)\left[\varepsilon_{\alpha\beta}^{fV}\left(\bar{f} \gamma_\mu f\right) + \varepsilon_{\alpha\beta}^{fA}\left(\bar{f} \gamma_\mu \gamma_5 f\right)\right]\, ,
\end{align}
the NSI parameters read $\varepsilon^{fV}_{\alpha\beta} = \varepsilon ^{fL}_{\alpha\beta} + \varepsilon^{fR}_{\alpha\beta}$ and $\varepsilon^{fA}_{\alpha\beta} = \varepsilon ^{fR}_{\alpha\beta} - \varepsilon^{fL}_{\alpha\beta}$ respectively. This parametrisation is convenient when studying the role of NSIs in neutrino propagation, which is only altered by the vector component of the interactions. Nonetheless, whenever one is addressing how NSIs affect production and detection processes both components become relevant and there is no \textit{a priori} preference for one basis or the other.

%%%%%%%%%%%%%%%%%%%%%%%%%%%%%%%%%%%%%%%%%%%%%%%%%%%
\subsection{Ultraviolet-complete model for NSIs}
\label{sec:ch5-uv-nsi}
%%%%%%%%%%%%%%%%%%%%%%%%%%%%%%%%%%%%%%%%%%%%%%%%%%%
A major concern regarding non-standard interactions is related to the connection between the formalism previously presented and the fundamental --- renormalisable --- theory from which NSIs arise. Imposing $SU(2)_L$ gauge symmetry in Equations \ref{eq:ch5-NCNSI} and \ref{eq:ch5-CCNSI} necessarily leads to additional interactions between charged fermions. For instance, the effective Lagrangian in Equation \ref{eq:ch5-NCNSI} could result from a dimension-6 operator~\cite{Broncano:2002rw,Gavela:2008ra,Antusch:2008tz}
\begin{align}
\varepsilon^{fV}_{\alpha \beta}(\bar{l}_{\alpha L}\gamma^\sigma l_{\beta L})(\bar{l}_{f L} \gamma_\sigma l_{f L}) \, ,
\end{align}
where $l_{\alpha L}$ is the $SU(2)$ doublet. However, such an operator would also induce flavour-violating decays. For example, a non-zero $\varepsilon^{eV}_{e\mu}$, would be constrained not only from neutrino data but also from $\mu \to e \, \gamma$ searches. Using arguments of this nature, one can set stringent constraints on NSI parameters from charged-lepton flavour-violating observables. Depending on the exact ultraviolet completion, the allowed values for the NSI parameters generally turn out to be well below the sensitivity of current and next-generation neutrino experiments.

The four-fermion operators presented before could also arise from dimension-8 operators~\cite{Gavela:2008ra,Antusch:2008tz}. In that case, NSIs could be present only in the neutrino sector. See for instance, that the operator
\begin{align}
\varepsilon^{fV}_{\alpha \beta}(\bar{l}_{\alpha L}\Phi)\gamma_\mu(\Phi^\dagger l_{\beta L})(\bar{l}_{f L}\gamma^\mu l_{f L}) \, ,
\end{align}
where $\Phi$ is the Higgs doublet with hypercharge $Y = +1/2$, generates a dimension-6 operator of the form of Equation~\ref{eq:ch5-NCNSI} once the Higgs takes a vacuum expectation value \cite{Davidson:2019iqh}. An alternative to having sizeable enough NSIs, so that they could manifest in current and next-generation experiments, would be to generate the operators presented above from dimension-6 and dimension-8 effective operators simultaneously~\cite{Proceedings:2019qno}.

Large NSIs are also known to be present, for instance, in radiative neutrino mass models with additional charged scalars or leptoquarks~\cite{Babu:2019mfe}. Likewise, models with light mediators could also evade existing bounds and yet lead to observable NSIs~\cite{Farzan:2015doa,Farzan:2015hkd}. The reason is that, for such light mediators, the effective description in terms of dimension-6 operators is not valid. Nonetheless, limits from neutrino propagation in matter and CE$\nu$NS would still apply since for coherent forward scattering only the component with zero squared four-momentum transfer, $q^2=0$, would contribute. This scenario is still well-described by Equations~\ref{eq:ch5-NCNSI} and~\ref{eq:ch5-CCNSI}.

%%%%%%%%%%%%%%%%%%%%%%%%%%%%%%%%%%%%%%%%%%%%%%%%%%%%%%%%%%%%%%%%%%
%%%%%%%%%%%%%%%%%%%%%%%%%%%%%%%%%%%%%%%%%%%%%%%%%%%%%%%%%%%%%%%%%%
\section{Neutrino oscillations with NSIs}
\label{sec:ch5-nsi-osc}
%%%%%%%%%%%%%%%%%%%%%%%%%%%%%%%%%%%%%%%%%%%%%%%%%%%%%%%%%%%%%%%%%%
%%%%%%%%%%%%%%%%%%%%%%%%%%%%%%%%%%%%%%%%%%%%%%%%%%%%%%%%%%%%%%%%%%

From this moment on, we will focus on NC-NSIs, since CC-NSIs are very strongly constrained \cite{Farzan:2017xzy}. NSIs alter the picture of neutrino oscillations since they modify neutrino propagation and can also introduce new sources of CP-violation \cite{Denton:2020uda,Chatterjee:2020kkm}. In addition, when analysing neutrino data, one needs to take into account the changes that NSIs could also induce in the production and detection processes.

Regarding propagation, as it was mentioned previously, only the vector component contributes significantly to coherent forward scattering and gives rise to an effective potential medium. In that case, the Hamiltonian describing neutrino flavour evolution in an electrically neutral and unpolarised medium is given by the Hamiltonian in vacuum --- denoted by $H_0$ --- and the contribution from the matter potentials, both standard and non-standard --- which we denote $V_{\text{matt}}$ and $V_{\text{NSI}}$ respectively. Namely, it reads
\begin{align}
&H = H_0 + V_{\text{matt}} + V_{\text{NSI}}\, ,
\label{eq:ch5-ham-nsi}
\end{align}
with
\begin{align}
&  H_0 = U\frac{1}{2E}\begin{pmatrix} 0 & 0 & 0 \\ 0 & \Delta m^2_{21} & 0 \\ 0 &  0 & \Delta m^2_{31}\end{pmatrix} U^\dagger\, ,\end{align}
and
\begin{align}V_{\text{matt}} + V_{\text{NSI}} = \sqrt{2}G_{\rm F} N_e\left[\begin{pmatrix} 1 & 0 & 0 \\ 0 & 0 & 0\\ 0 & 0 & 0\end{pmatrix} + \begin{pmatrix} \varepsilon^{V}_{ee} & \varepsilon^{V}_{e\mu} & \varepsilon^{V}_{e\tau} \\ \varepsilon^{V*}_{e\mu} & \varepsilon^{V}_{\mu\mu} & \varepsilon^{V}_{\mu\tau} \\ \varepsilon^{V*}_{e\tau} & \varepsilon^{V*}_{\mu\tau} & \varepsilon^{V}_{\tau\tau}\end{pmatrix} \right] \, .
\end{align} 
Here, we have defined the effective NSI parameters,
\begin{align}
\varepsilon^{V}_{\alpha\beta} =  \varepsilon^{eV}_{\alpha\beta} + \varepsilon^{uV}_{\alpha\beta} \frac{N_u}{N_e} +\varepsilon^{dV}_{\alpha\beta}\frac{N_d}{N_e}\, ,
\label{eq:ch5-eps-def}
\end{align}
which include the contributions from these non-standard interactions with the matter fields: electrons, $u$-quarks and $d$-quarks. For simplicity, from this moment on and unless otherwise specified, we drop the superindices $V$ referring to the vector projection of the interaction.
 
%%%%%%%%%%%%%%%%%%%%%%%%%%%%%%%%%%%%%%%%%%%%%%%
\subsection{Effective two-neutrino approach}
\label{sec:ch5-eff-2nu}
%%%%%%%%%%%%%%%%%%%%%%%%%%%%%%%%%%%%%%%%%%%%%%%
In the \textit{primed basis} defined in Equation \ref{eq:ch4-primed-basis}, the evolution equation with vanishing NSI parameters was shown in Equation \ref{eq:ch4-vac_evol}. In that scenario, the evolution of solar neutrinos inside the Sun and through the Earth satisfies the condition
\begin{align}
\sqrt{2}G_FN_e \lesssim \Delta m^2_{21}/2E \ll \Delta m^2_{31}/2E\, .
\end{align} 
In addition, since the reactor mixing angle is very small, the evolution of one of the eigenstates --- namely $\nu'_\tau$ --- decouples from the other two. Then, the electron survival probability is given by
\begin{align}
P_{ee} = \cos^4\theta_{13}P^{2\nu}_{ee} + \sin^4\theta_{13} \, ,
\label{eq:ch5-sol-surv}
\end{align} 
where $P^{2\nu}_{ee}$ denotes the electron survival probability in the effective \mbox{two-neutrino} approach --- which corresponds to solving only the evolution of $\nu'_e$ and $\nu'_\mu$.

In the presence of NSIs, since the parameters $\varepsilon^{V}_{\alpha\beta}$ are, at maximum $\mathcal{O}(1)$, the overall picture does not differ significantly. The three-neutrino Hamiltonian in the \textit{primed basis} is
\begin{align}
i\frac{\text{d}}{\text{d}t}\begin{pmatrix}\nu'_{eL} \\ \nu'_{\mu L} \\ \nu'_{\tau L}\end{pmatrix} &= \left[\begin{pmatrix}2 \delta s^2_{12} + c^2_{13}V_e & 2 \delta s_{12}c_{12} & s_{13}c_{13}V_e \\ 2 \delta s_{12}c_{12} & 2\delta c^2_{12} & 0 \\ c_{13}s_{13}V_e & 0 & 2\Delta + s^2_{13}V_e  \end{pmatrix} \right. \nonumber \\
& \left. +\sqrt{2}G_F N_e\begin{pmatrix}c^2_{13} + \tilde{\varepsilon}_{ee} & \tilde{\varepsilon}_{e\mu} & c_{13}s_{13} +\tilde{\varepsilon}_{e\tau} \\ \tilde{\varepsilon}^*_{e\mu} & \tilde{\varepsilon}_{\mu\mu} & \tilde{\varepsilon}_{\mu\tau} \\ c_{13}s_{13} + \tilde{\varepsilon}^*_{e\tau} & \tilde{\varepsilon}^*_{\mu\tau} & s^2_{13}+\tilde{\varepsilon}_{\tau\tau}
\end{pmatrix} \right]
\begin{pmatrix}\nu'_{eL} \\ \nu'_{\mu L} \\ \nu'_{\tau L}\end{pmatrix} \, ,
\label{eq:ch5-ham-nsi-primed}
\end{align}
where we have defined the NSI parameters in the \textit{primed basis},
\begin{align}
\tilde{\varepsilon}_{ee} &= c^2_{13}\varepsilon_{ee} + s^2_{13}\left[s^2_{23}\varepsilon_{\mu\mu} + c^2_{23}\varepsilon_{\tau\tau} + 2c_{23}s_{23}\,\text{Re}(\varepsilon_{\mu\tau})\right]    \nonumber \\ & \hspace{0.5cm}- 2s_{13}c_{13}\,\text{Re}\left[(s_{23}\varepsilon_{e\mu} + c_{23}\varepsilon_{e\tau})e^{i\delta_\text{CP}}\right] \, ,
\\
\tilde{\varepsilon}_{\mu\mu} &= c^2_{23}\varepsilon_{\mu\mu} + s^2_{23}\varepsilon_{\tau\tau} - s_{23}c_{23} \,\text{Re}(\varepsilon_{\mu\tau})\, ,
\\
\tilde{\varepsilon}_{\tau \tau} &= s^2_{13}\varepsilon_{ee} + c^2_{13}\left[s^2_{23}\varepsilon_{\mu\mu} + c^2_{23}\varepsilon_{\tau\tau} + 2c_{23}s_{23}\,\text{Re}(\varepsilon_{\mu\tau})\right] \nonumber \\& \hspace{0.5cm} + 2s_{13}c_{13}\,\text{Re}\left[(s_{23}\varepsilon_{e\mu} + c_{23}\varepsilon_{e\tau})e^{i\delta_\text{CP}}\right] \, ,
\\
\tilde{\varepsilon}_{e\mu} &= c_{13}\left[c_{23}\varepsilon_{e\mu}- s_{23}\varepsilon_{e\tau}\right] \nonumber \\ & \hspace{0.5cm} - s_{13}e^{-i \delta_\text{CP}}\left[c_{23}s_{23}(\varepsilon_{\mu\mu} - \varepsilon_{\tau\tau}) + c^2_{23}\varepsilon^*_{\mu\tau}-s^2_{23}\varepsilon_{\mu\tau}\right]\, ,
\\
\tilde{\varepsilon}_{e\tau} &= c_{13}s_{13}\left[ \varepsilon_{ee} +s^2_{23}\varepsilon_{\mu\mu} + c^2_{23}\varepsilon_{\tau\tau} + 2c_{23}s_{23}\,\text{Re}(\varepsilon_{\mu\tau})\right] \nonumber\\& \hspace{0.5cm} + c^2_{13}e^{i\delta_\text{CP}}(s_{23}\varepsilon_{e\mu} + c_{23}\varepsilon_{e\tau}) - s^2_{13}e^{-i\delta_\text{CP}}(s_{23}\varepsilon^*_{e\mu} + c_{23}\varepsilon^*_{e\tau})\, ,
\\
\tilde{\varepsilon}_{\mu\tau} &= s_{13}\left[c_{23}\varepsilon^*_{e\mu} - s_{23}\varepsilon^*_{e\tau}\right]\nonumber \\ & \hspace{0.5cm}+c_{13}e^{i\delta_\text{CP}}\left[c_{23}s_{23}(\varepsilon_{\mu\mu} - \varepsilon_{\tau\tau})+c^2_{23}\varepsilon_{\mu\tau}-s^2_{23}\varepsilon^*_{\mu\tau}\right]\, .
\label{eq:ch5-deq-nsi}
\end{align}
Some of the NSI parameters are strongly constrained, mainly those involving muon neutrino \cite{Farzan:2017xzy}. Hence, for simplicity, we neglect all $\varepsilon_{\alpha\mu}$. Then, the evolution of $\nu'_{\tau L}$ essential decouples and the system is well-described by the effective 2-flavour Hamiltonian in the \textit{primed basis}
\begin{equation}
H^{2\nu} = \begin{pmatrix}2\delta s^2_{12} & 2\delta s_{12}c_{12}\\ 2\delta s_{12}c_{12} & 2\delta c^2_{12} \end{pmatrix} + \sqrt{2}G_FN_e \left[\begin{pmatrix}c^2_{13} & 0 \\ 0 & 0 \end{pmatrix} + \begin{pmatrix} 0 & \varepsilon \\ \varepsilon^* & \varepsilon'\end{pmatrix}\right]  \, ,
\label{eq:ch5-eff-2nu-nsi}
\end{equation}
where we have introduced two effective NSI coefficients following the usual notation in the literature,
\begin{align}
\varepsilon' \equiv \tilde{\varepsilon}_{e\mu} \quad \quad \text{and} \quad \quad
\varepsilon \equiv \tilde{\varepsilon}_{\mu\mu} - \tilde{\varepsilon}_{ee}\, .
\label{eq:ch5-eff-nsiparam}
\end{align}

The effective two-neutrino approach is also valid for the description of \mbox{medium-baseline} reactor antineutrino experiments when the energy resolution is such that it is not possible to resolve the subleading effect arising from the interference between $\Delta m^2_{31}$ and $\Delta m^2_{32}$. This description is also valid for long-baseline reactor experiments since they are not sensitive to these mass splittings. 

%%%%%%%%%%%%%%%%%%%%%%%%%%%%%%%%%%%%%%%%%%%%%%%%%%%%%%%%%%%%%%%%%%%%%%%%%%%%
\subsection{Generalised mass ordering degeneracy and the LMA-D solution}
\label{sec:ch5-degen-teo}
%%%%%%%%%%%%%%%%%%%%%%%%%%%%%%%%%%%%%%%%%%%%%%%%%%%%%%%%%%%%%%%%%%%%%%%%%%%%%
One can see that in the Hamiltonian in vacuum, as in Equation \ref{eq:ch5-ham-nsi}, the simultaneous transformations
\begin{align}
&\theta_{12} \to \pi/2 - \theta_{12}\, ,\nonumber\\ &\Delta m^2_{31} \to -\Delta m^2_{31} + \Delta m^2_{21} = - \Delta m^2_{32} \, ,\nonumber \\
&\text{and} \quad \delta_\text{CP} \to \pi - \delta_\text{CP}\, ,
\label{eq:ch5-deg-cond-vac}
\end{align}
lead to the transformation of the Hamiltonian $H_{vac} \to -H^*_{vac}$. This means that these three simultaneous transformations would not alter neutrino propagation. In matter, the degeneracy is broken and as a consequence, solar neutrinos allow to determine $\sin^2 \theta_{12} < 0.5$. Nevertheless, in the presence of NSIs, the degeneracy is restored when the transformations above --- see Equations \ref{eq:ch5-deg-cond-vac} --- are performed, together with the following ones:
\begin{align}
\varepsilon_{ee} \to -\varepsilon_{ee} -2 \quad \text{and} \quad \varepsilon_{\alpha\beta} \to -\varepsilon^*_{\alpha\beta} \quad \text{with } (\alpha, \, \beta \neq e)\, . 
\label{eq:ch5-deg-cond-nsi}
\end{align}
Note that, in the two-neutrino effective approach, the degeneracy corresponds to
\begin{align}
\theta_{12}\to \pi/2 -\theta_{12}\, , \quad \varepsilon' \to -\varepsilon' +2c^2_{13} \quad \text{and} \quad \varepsilon \to -\varepsilon^*\, .
\label{eq:ch5-deg-cond-nsi-2nu}
\end{align}
This degeneracy can not be broken unless large values of NSIs are excluded experimentally by other observables or the mass ordering is determined by means insensitive to the hypothetical presence of NSIs. Otherwise, for large values of the NSI parameter $\varepsilon_{ee}$, there exists a solution such that $\sin^2 \theta_{12} > 0.5 $, often referred to as the LMA-D solution --- Large Mixing Angle -Dark solution. Consequently, stringent bounds from scattering experiments can help explore this scenario of large NSIs.

%%%%%%%%%%%%%%%%%%%%%%%%%%%%%%%%%%%%%%%%%%%%%%%%%%%%%%%%%
%%%%%%%%%%%%%%%%%%%%%%%%%%%%%%%%%%%%%%%%%%%%%%%%%%%%%%%%%
\section{Future experiments: Hyper-Kamiokande}
\label{sec:ch5-HK}
%%%%%%%%%%%%%%%%%%%%%%%%%%%%%%%%%%%%%%%%%%%%%%%%%%%%%%%%%
%%%%%%%%%%%%%%%%%%%%%%%%%%%%%%%%%%%%%%%%%%%%%%%%%%%%%%%%%
Hyper-Kamiokande is a next-generation water Cherenkov neutrino observatory in Japan, conceived as the heir of Kamiokande and Super-Kamiokande's successful research program. This multi-purpose detector has a wide physics program which includes the study of accelerator, atmospheric and solar neutrinos. The most conservative proposal consists of a water tank with a fiducial volume of 187 kt --- that is 8.3 times greater than its predecessor Super-Kamiokande. However, there exists the possibility of locating a second tank in South Korea, so that the volume will be duplicated.

%%%%%%%%%%%%%%%%%%%%%%%%%%%%%%%%%%%%%%%
\subsection{Simulation and analysis}
\label{sec:ch5-HK-simu}
%%%%%%%%%%%%%%%%%%%%%%%%%%%%%%%%%%%%%%%
In our analysis, we will address the impact of the experimental configuration in the simultaneous determination of the oscillation parameters and NSI parameters. To do so, we consider the scenario in which only one tank will be built and consider an optimistic energy threshold of 3.5 MeV, as achieved in Super-Kamiokande Run IV, and a more realistic one of 5 MeV. We refer to these configurations as Configuration A and B respectively and we assume a running time of 10 years. The interest in this comparison relies on the fact that a smaller energy threshold allows studying part of the transition region which is very sensitive to neutrino propagation in the Sun, and hence, to neutrino interactions. We consider a third option in which we assume a second tank will be built in Korea and run for 3 years, with a threshold of 5 MeV. In this case, the configuration is characterised by having larger statistics and, therefore, being more sensitive to the day-night asymmetry induced due to neutrino propagation across the Earth. In~\reftab{ch5-confsHK} we summarise the three experimental configurations considered.

\begin{table*}
\renewcommand*{\arraystretch}{1.2}
	\centering
	\begin{tabular}{lccc}
	\toprule[0.25ex]
	    Configuration & Low energy threshold (MeV) & Running time (years) \\ \midrule
	     A (optimistic) & 3.5 &  10 \\ 
	     B (conservative) & 5 & 10 \\ 
	     C (2 tanks) & 5  & 10 + 3 \\ 
	    \bottomrule[0.25ex]
    \end{tabular}
    \caption{
        \labtab{ch5-confsHK}
        Summary of the main characteristics of the three possible configurations studied for Hyper-Kamiokande. The fiducial volume of each tank is 187 kt.
    }
\end{table*}

Hyper-Kamiokande will be sensitive to solar neutrinos through electron elastic scattering, $\nu_x + e^- \rightarrow \nu_x + e^-$. For this process, we take the cross-section from~\cite{Bahcall:1995mm} and we estimate the response of the detector using a Gaussian function with the same energy resolution as in Super-Kamiokande Run IV, which is given by
\begin{align}
    \sigma (T) = -0.0839 \text{MeV} + 0.349 \text{MeV}^{1/2}\sqrt{E_e} + 0.0397 E_e\, ,
    \label{eq:ch5-sigmaSKIV}
\end{align}
where $E_e = T + m_e$. Here, $m_e$ is the mass of the electron and $T$ is the electron recoil kinetic energy expressed in MeV.

The expected number of events in the detector, in the $i$-th energy bin, for neutrinos from the solar chain $j = \lbrace ^8\text{B}, \, hep\rbrace$ and in  the zenith angle bin $k = \lbrace$day (D), night (N)$\rbrace$, is
\begin{align}
\mathcal{R}_{i,j,k}^{\text{ osc}}= \mathcal{A} \int \text{d}E_\nu  \phi_j(E_\nu) \times \left[ \sigma^i_{\text{eff}, e} \times P_{ee}^{j,k} +
\sigma^i_{\text{eff}, x} \times (1- P_{ee}^{j,k}) \right]\, .
\label{eq:ch5-roscHK}
\end{align} 
The number of events depends on the flux, $\phi_j (E_\nu)$,  of $^8$B~\cite{Winter:2004kf} and $hep$ solar neutrinos~\cite{Bahcall:1997eg}. It also depends on the electron survival probability for each of these two chains and for day and night zenith-angle bins, $P^{j,k}_{ee}$.\footnote{Note that since the calculation of the survival probability requires the averaging over the production region, it is different for each chain.}  In addition, we have introduced the effective cross-section, which results from the convolution of the detection cross-section with the response of the detector, for the reconstructed energy bin $i$. Finally, $\mathcal{A}$ includes all the prefactors such as the exposure time, efficiency and the number of targets in the detector.

Let us define the rate of events normalised to the expected rate in the absence of oscillations,
\begin{align}
    \mathcal{R}_{i,j,k} = \frac{\mathcal{R}_{i,j,k}^{\text{ osc}}}{\mathcal{R}_{i,^8\text{B}, k}^{\text{ unosc}}+\mathcal{R}_{i,hep, k}^{\text{ unosc}}} \, ,
\label{eq:ch5-HK-rate}
\end{align}
where $\mathcal{R}_{i,^8\text{B}, k}^{\text{ unosc}}$ is defined as $\mathcal{R}_{i,j,k}^{\text{ osc}}$ in Equation~\ref{eq:ch5-roscHK}, but assuming no oscillations, i.e setting the electron survival probability to one. 

Following the analysis from~\cite{Nakano:2016uws}, one can define a $\chi^2$ function which includes spectral and angular information. In our case, it reads:
\begin{align}
    \chi^2 = \sum_{k = D,N}\sum_{i = 1}^{i = 23} \frac{\left(\mathcal{D}_{i,k} - \mathcal{B}_{i,k}  - \mathcal{H}_{i,k}\right)^2}{(\sigma^{i,k}_{stat})^2 + (\sigma^i_{uncorr})^2} 
    &+ \left(\frac{\alpha}{\sigma_{\alpha}}\right)^2 + \left(\frac{\beta}{\sigma_{\beta}}\right)^2 \nonumber \\ &+ \epsilon^2_{^8\text{B}}+ \epsilon^2_{scale} + \epsilon^2_{resol}\, .
\label{eq:ch5-HK-chi2}
\end{align}
The complete function depends on the \textit{observed number of events}, $\mathcal{D}_{i,k}$, in each of the 23 energy bins --- denoted with the index $i$ --- both for day and night zenith-angle bins --- i.e.  $k \in \lbrace D, N \rbrace$. The \textit{observed number of events} is generated as mock data assuming the best-fit values from~\cite{deSalas:2020pgw} --- see \reftab{ch5-oscparams}. We have defined the theoretically estimated number of events from the $^8\text{B}$ chain as  
\begin{align}
&\mathcal{B}_{i,j}  = (1 + \alpha + \epsilon_{^8\text{B}} \, \sigma ^{i,k}_{^8\text{B}} + \epsilon_{scale}\, \sigma^{i,k}_{scale} + \epsilon_{resol}\, \sigma^{i,k}_{resol})\,  \mathcal{R}_{i, ^8 B, k} \, .
\label{eq:ch5-HK-events-1}
\end{align}
It includes the effect of the energy-correlated systematics due to flux shape uncertainty, $\sigma^{i,k}_{^8\text{B}}$, the energy resolution $\sigma^{i,k}_{resol}$ of the detector, and its energy scale $\sigma^{i,k}_{scale}$.  The energy-correlated uncertainties here introduced are weighted by three corresponding pull parameters ($\epsilon_{^8\text{B}}$, $\epsilon_{scale}$ and $\epsilon_{resol}$). Similarly, we have defined the theoretically estimated number of events from the $hep$ chain
\begin{align}
&\mathcal{H}_{i,k}  = (1 + \beta)\,  \mathcal{R}_{i, hep, k} \, ,
    \label{eq:ch5-HK-events-2}
\end{align}
where we do not include the contributions of these uncertainties, since this flux already provides a subdominant contribution to the analysis. We include two additional pull parameters, $\alpha$ and $\beta$, to account for uncertainties in the fluxes of $^8\text{B}$ and $hep$ neutrinos, respectively, and the corresponding penalty terms. Regarding these uncertainties on the total flux, the one on the $^8$B flux, $\sigma_\alpha = 0.04$, is taken from the NC measurement carried out by the SNO collaboration~\cite{SNO:2009uok}, while the one for $hep$ neutrinos, $\sigma_\beta = 200\%$, is determined based on observations.\footnote{Note that the error from theoretical predictions is smaller, $\sigma_\beta = 30\%$.}

\begin{table}
\renewcommand*{\arraystretch}{1.2}
	\centering
	\begin{tabular}{ccc}
	\toprule[0.25ex]
	     \multicolumn{3}{c}{Neutrino oscillation parameters} \\ \midrule
	    $\sin^2 \theta_{12} = 0.32$ & $\sin^2\theta_{13} = 0.022$ & $\sin^2\theta_{23} = 0.574$ \\ 
	    $\Delta m^2_{21} = 7.5 \times 10 ^{-5}\, \text{eV}^2$ & $\Delta m^2_{31} = 2.55\times 10^{-3}\,\text{eV}^2$ & $\Delta_{\text{CP}} = 1.2 \pi $ \\ 
	    \bottomrule[0.25ex]
    \end{tabular}
    \caption{Choice of best-fit values for the oscillation parameters for the analyses presented in this chapter.\labtab{ch5-oscparams}}
\end{table}

Finally, our $\chi^2$ function also accounts for the statistical and energy-uncorrelated uncertainties. \footnote{Note that we do not indicate the dependence of $\mathcal{B}_{i,j}$, $\mathcal{H}_{i,j}$, $\sigma^{i,k}_{^8\text{B}}$, $\sigma^{i,k}_{scale}$ and $\sigma^{i,k}_{resol}$ on the oscillation parameters explicitly.} Energy-uncorrelated systematics were assumed to be equal to Super-Kamiokande Run IV, as in~\cite{Nakano:2016uws}.
The statistical systematics were scaled from those of Super-Kamiokande, assuming that the number of events follows a Poissonian distribution. Then, the standard deviation will be $\sigma^{i,k}_{\text{stat}} = \sqrt{N^i_{\text{events}}}$ for each bin $i$. A longer running time and a larger volume,
\begin{align}
T_{\text{HK}} >T_{\text{SK}} \quad \text{and} \quad V_{\text{HK}} > V_{\text{SK}}\, ,
\end{align}
results in a reduction of the statistical error,
\begin{align}
\frac{\sigma^{i,k}_{stat, HK}}{N^{i,k}_{\text{events, SK}}} = \frac{\sigma^{i,k}_{\text{stat, SK}}}{N^{i,k}_{\text{events, SK}}} \sqrt{\frac{T_{SK}}{T_{HK}} \frac{V_{SK}}{V_{HK}}}\, .
\label{eq:ch5-HK-syst}
\end{align}

In~\reffig{fig:ch5-stdHK}, we depict the expected sensitivity of Hyper-Kamiokande to the oscillation parameters $\theta_{12}$ and $\Delta m^2_{21}$ in the absence of new physics and assuming Configuration A from \reftab{ch5-confsHK}.

%%%%%%%%%%%%%%%%%%%%%%%%%%%%%%%%%%%%%%%%%%%%%%%%%%%%%%%
\subsection{NSIs on solar neutrinos}
\label{sec:ch5-HK-nsi}
%%%%%%%%%%%%%%%%%%%%%%%%%%%%%%%%%%%%%%%%%%%%%%%%%%%%%%%
Even with the inclusion of a new interaction framework, the evolution of neutrinos inside the Sun can be described accurately in the adiabatic approximation. Nonetheless, NSIs alter the mixing in the production region, which translates into a change in the position and shape of the transition region in the energy profile. Neglecting matter effects on $\sin^2 \theta_{13}$, which are known to be small, the survival probability during the day is:
\begin{align}
P^{D}_{ee,\odot} = \cos ^4\theta_{13} \left[\cos^2\theta_{13} \cos^2 \tilde{\theta}_{12} + \sin^2\theta_{12} \sin^2\tilde{\theta}_{12}\right] + \sin^4 \theta_{13}\, ,
\label{eq:ch5-HK-pday}
\end{align}
where we have defined the mixing at the production point in the Sun as
\begin{align}
\cos{2\tilde{\theta}_{12}} = \frac{\Delta m^2_{21}\cos 2\theta_{12}+ 2 \sqrt{2}G_F E N^0_e \left(\varepsilon ' - 1\right)}{\Delta \tilde{m}^2_{21}}\, ,
\label{eq:ch5-angle-nsi}
\end{align}
and where the effective mass splitting is
\begin{align}
\begin{aligned}
\left[\Delta \tilde{m}^2_{21}\right]^2 =  \left[\Delta m^2_{21}\cos 2\theta_{12} + 2 \sqrt{2}G_F E N^0_e\left(\varepsilon ' -1\right) \right]^2 \\+ \left[ \Delta m^ 2_{21} \sin 2 \theta_{12} + 4\sqrt{2} G_F E N^0_e \varepsilon \right] ^2 \, ,
\end{aligned}
\label{eq:ch5-mass-nsi}
\end{align}
In the above expressions, $N^0_e$ refers to the electron number density in the production region and the parameters $\varepsilon$ and $\varepsilon'$ are defined in Equation \ref{eq:ch5-eff-nsiparam}.

\begin{figure}
\centering
\includegraphics[width=0.75\textwidth]{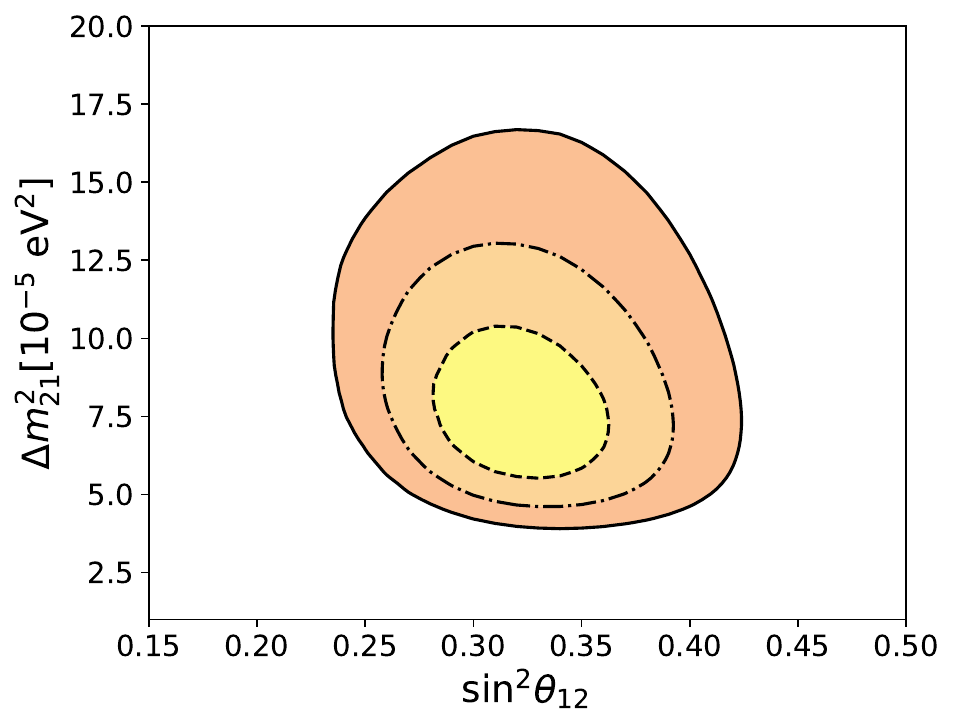}
\caption{Expected sensitivity to the solar neutrino oscillation parameters, $\theta_{12}$ and $\Delta m^2_{21}$, in Hyper-Kamiokande in the absence of NSIs. The 1$\sigma$, 2$\sigma$ and 3$\sigma$ confidence levels are indicated by the dashed, dot-dashed and solid lines, respectively. We assume Configuration A from~\reftab{ch5-confsHK}. \labfig{fig:ch5-stdHK}}
\end{figure}

When solar neutrinos travel through the Earth before reaching the detector, one should take into account that, despite arriving from the Sun as an incoherent sum of mass eigenstates, they will undergo flavour oscillations as they traverse the Earth. This gives rise to a zenith-angle dependence in the oscillation probability, which is generally referred to as the Day-Night asymmetry --- denoted by $A_{D/N}$. Although it is possible to gain some understanding of this observable from analytical studies~\cite{Akhmedov:2004rq}, it is often computed numerically, since it requires solving the evolution of the system in a varying matter potential. 

Hence, one expects that significant distortions with respect to the standard picture arise in the presence of neutrino non-standard interactions.

\textbf{Numerical results}

For the experimental configurations we are considering here, only one side of the neutrino spectrum is accessible --- i.e. energies above a certain threshold. This means that, although the large number of statistics expected would allow the differentiation between the spectra for day and night, the transition region and the low-energy side of the neutrino spectrum will not be measurable and Hyper-Kamiokande will have to rely on previous measurements from other experiments.

At this point, it is important to clarify that we will restrict ourselves to the case of NC-NSIs with $d$-type quarks. In that scenario, the cross-section of the detection channel in Hyper-Kamiokande is not altered. Since the predicted solar neutrino flux results from charged-current interactions, NC-NSIs do not modify the predicted flux. We will present our results in terms of the NSI parameters $\varepsilon_d$ and $\varepsilon_d'$, defined as 
\begin{align}
\varepsilon_d = \frac{N_e}{N_d} \varepsilon \quad \text{and} \quad \varepsilon_d = \frac{N_e}{N_d} \varepsilon\, .
\end{align}

In the right panel of~\reffig{fig:ch5-oneatatimeHK}, we show the projected sensitivity of \mbox{Hyper-Kamiokande} to solar oscillation parameters in the presence of a non-zero NSI coefficient $\varepsilon_d'$. In this figure, we have assumed the most optimistic experimental configuration --- which we refer to as Configuration A in \reftab{ch5-confsHK}. It can be seen that non-universal NSIs would worsen the determination of the mass splitting by more than an order of magnitude. Moreover, it should be noted that a solution in the second octant arises for very large values of $\varepsilon_d'$. This solution corresponds to the LMA-D solution discussed in Section~\ref{sec:ch5-degen-teo}~\cite{Miranda:2004nb}. Both features are expected from the arguments presented in previous sections.

\begin{figure*}
\centering
\includegraphics{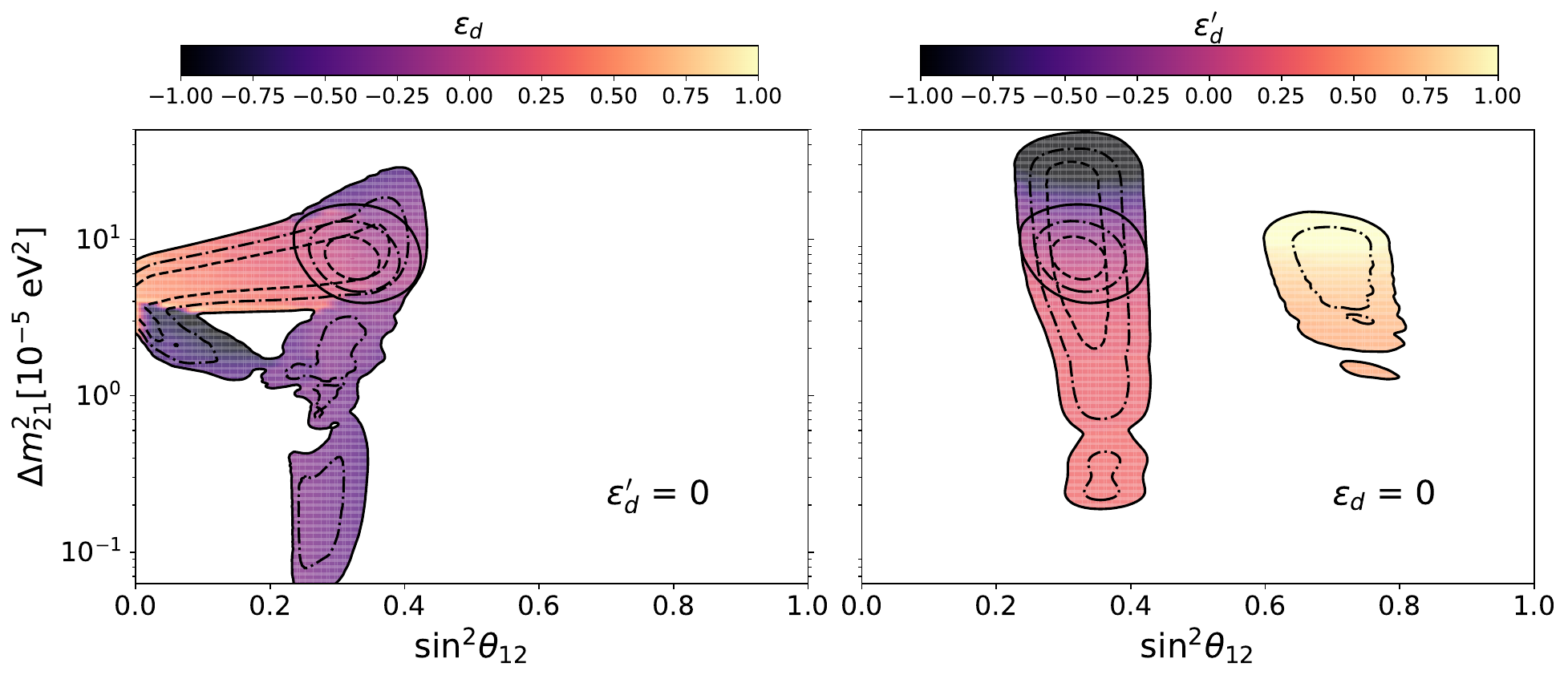}
\caption{
Effect of the effective NSI parameters on the sensitivity of Hyper-Kamiokande, varying one at a time. Left panel shows the 1$\sigma$, 2$\sigma$ and 3$\sigma$ C.L. in the $\sin^2 \theta_{12}- \Delta m^2_{21}$ plane when varying $\varepsilon_d$ between -1 and 1. The same confidence levels are drawn in the right panel for the case of $\varepsilon_d = 0$ and $\varepsilon _d$ allowed to vary within the same range. The colour map indicates the best-fit value of the effective NSI parameters. Unfilled contours correspond to the same confidence levels expected for Hyper-Kamiokande in the absence of NSIs. \labfig{fig:ch5-oneatatimeHK}}
\end{figure*}

Regarding flavour-changing NSI, there is a strong degeneracy between the effective parameter $\varepsilon_d$ and the oscillation parameters, $\sin^2\theta_{12}$ and $\Delta m^2_{21}$. This is shown in the left panel of~\reffig{fig:ch5-oneatatimeHK}, where one can see how the allowed parameter space in this plane is significantly enlarged with respect to the standard LMA solution in the absence of NSIs~\cite{deSalas:2020pgw}.

Likewise, these degeneracies increase significantly once both $\varepsilon_d$ and $\varepsilon_d'$ are considered simultaneously. In order to break these degeneracies, the inclusion of other datasets is crucial, as Hyper-Kamiokande cannot resolve them by measuring the high-energy range of solar neutrinos alone. In fact, though small differences in this energy range are expected in the presence of NSIs, an experimental configuration aimed at maximising statistics --- i.e. one involving two tanks --- would not be able to set significant constraints on NSI parameters on its own. Similarly, lowering the energy threshold to \mbox{3.5 MeV} would not help to resolve the degeneracies of $\varepsilon_d$ and $\varepsilon_d'$ with the oscillation parameters.
This can be seen in~\reffig{fig:ch5-HK-confs}, where we compare the three experimental set-ups considered --- see Table~\ref{tab:ch5-confsHK} --- and no significant difference is found. Some slight improvement can be seen in the low-threshold configuration --- i.e. Configuration A. Nevertheless, this happens mainly in regions that will be excluded later on after combining the results with those from other experiments. A more detailed discussion on the impact of each different configuration on the combined analysis with JUNO is presented in Section~\ref{sec:ch5-configurations}.

\begin{figure*}
\centering
\includegraphics{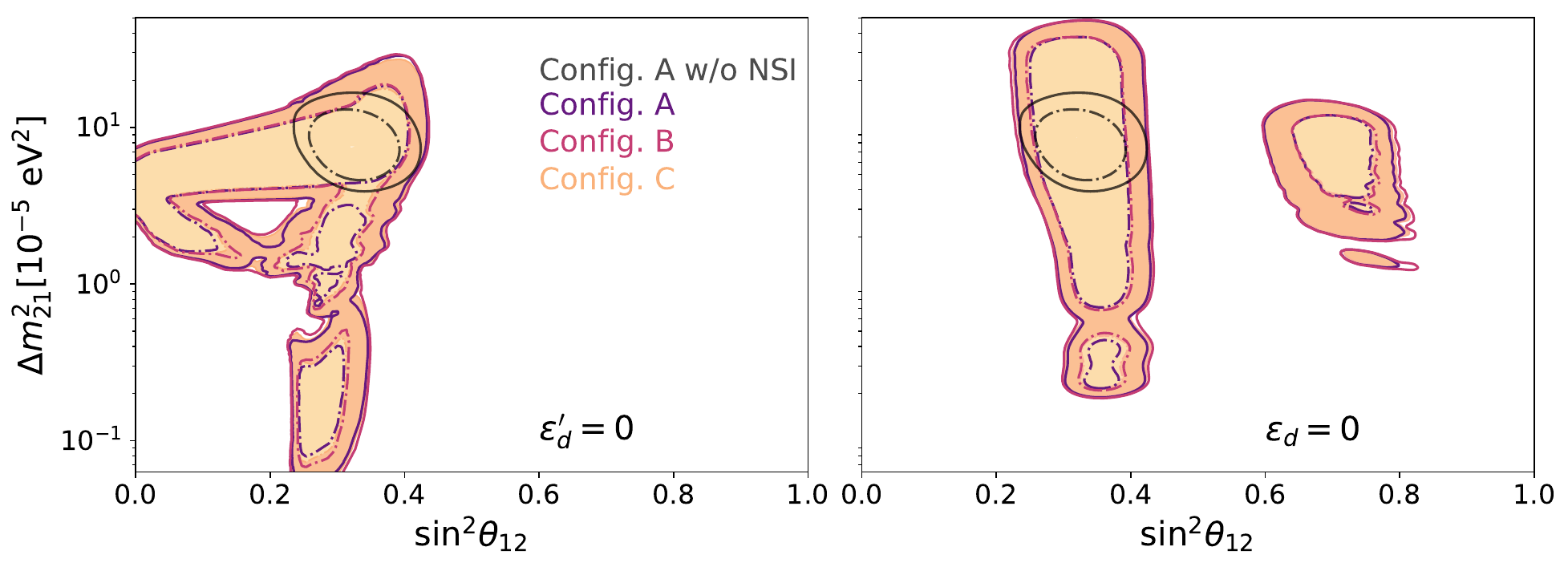}
\caption{Impact of effective NSI parameters on the sensitivity of Hyper-Kamiokande, varying one at a time, for each of the three experimental configurations Table~\ref{tab:ch5-confsHK}. Contours correspond to 2$\sigma$ and 3$\sigma$ C.L. In the left panel, $\varepsilon_d'$ is fixed to zero; in the right panel, $\varepsilon_d = 0$ is considered. Unfilled contours correspond to the same confidence levels expected for Hyper-Kamiokande in the absence of NSIs for Configuration A. \labfig{fig:ch5-HK-confs}}
\end{figure*}

%%%%%%%%%%%%%%%%%%%%%%%%%%%%%%%%%%%%%%%%%%%%%%%%
%%%%%%%%%%%%%%%%%%%%%%%%%%%%%%%%%%%%%%%%%%%%%%%%
\section{Future experiments: JUNO}
\label{sec:ch5-JUNO}
%%%%%%%%%%%%%%%%%%%%%%%%%%%%%%%%%%%%%%%%%%%%%%%%
%%%%%%%%%%%%%%%%%%%%%%%%%%%%%%%%%%%%%%%%%%%%%%%%
The Jiangmen Underground Neutrino Observatory (JUNO) is a \mbox{multi-purpose} neutrino experiment, whose programme includes studies on solar, atmospheric, supernova and reactor neutrino, as well as indirect dark matter searches, among others~\cite{JUNO:2015zny}. Its main scientific goal is to perform a direct measurement of the neutrino mass hierarchy and its location has been optimised consequently. 

%%%%%%%%%%%%%%%%%%%%%%%%%%%%%%%%%%%%%%%%%
\subsection{ Simulation and analysis}
\label{sec:ch5-JUNO-simu}
%%%%%%%%%%%%%%%%%%%%%%%%%%%%%%%%%%%%%%%%%
The neutrino detector is a liquid scintillator detector with a fiducial mass of 20 kton and it will be sensitive to electron antineutrino disappearance via inverse beta decay (IBD). The flux of reactor antineutrinos is mainly due to the Yangjiang and Taishan Nuclear Power Plants, although contributions from the Daya Bay and Huizhou reactors are also expected. The baseline and thermal power of each core are summarised in~\reftab{ch5-JUNObaselines}.

The energy resolution expected at JUNO is $3\,\% /\sqrt{E(\text{MeV})} $. This will allow an accurate measurement of the solar oscillation parameters $\theta_{12}$ and $\Delta m^2_{21}$, as well as a determination of the mass ordering~\cite{JUNO:2015zny,JUNO:2022mxj}.

\begin{table}
    \centering
    \renewcommand*{\arraystretch}{1.2}
    \caption{Baselines and thermal power of the cores contributing to the expected antineutrino flux at the JUNO detector. The shorthand notation refers to Yangjiang (YJ) and Taishan(TS) different cores (C), as well as the Daya Bay (DYB) and Huizhou (HZ) complexes~\cite{JUNO:2021vlw}.}
    \begin{tabular}{lcccccc}
    \toprule[0.25ex]
    Cores & YJ-C1 & YJ-C2 & YJ-C3 & YJ-C4 & YJ-C5 & YJ-C6 \\ \midrule
    Power (GW) & 2.9 & 2.9 & 2.9 & 2.9 & 2.9 & 2.9 \\ 
    Baseline(km) & 52.74 & 52.82 & 52.41 & 52.49 & 52.11 & 52.19\\			\bottomrule[0.25ex]
 & & & & & & \\
\toprule[0.25ex]
    Cores & TS-C1 & TS-C2 & \multicolumn{2}{c}{DYB} & \multicolumn{2}{c}{HZ} \\ \midrule
    Power (GW) & 4.6 & 4.6 & \multicolumn{2}{c}{17.4} & \multicolumn{2}{c}{17.4} \\ 
    Baseline(km) & 52.77 & 52.64 & \multicolumn{2}{c}{215} & \multicolumn{2}{c}{265}  \\ 
    \bottomrule[0.25ex]
    \end{tabular}
    \labtab{ch5-JUNObaselines}
\end{table}

Let us define the \textit{observed number of events} in the $i$-th energy bin, as $N_i$. We simulate them as mock data with the best fit in \reftab{ch5-oscparams} \cite{deSalas:2020pgw}. We denote the predicted number of events in the $i$-th energy bin due to the $j$-th reactor core by $T_{ij}$. Inspired by the oscillation analyses in~\cite{JUNO:2015zny,IceCube-Gen2:2019fet}, we consider 200 equal-size bins for the incoming neutrino energy ranging from 1.8\,MeV to 8.0\,MeV. 

For our analysis, we define the following $\chi^2$ function:
\begin{align}
&\chi^2 = \sum_{i = 1}^{200} \frac{\left[N_i - \sum_{j = 1}^{12} (1 + \xi _a)(1+ \xi_{r,j})(1+\xi_{s,i})T_{ij}\right]^2}{N_i (1+ \sigma_d N_i)}  \nonumber \\ & \hspace{3cm}+ \sum_{j=1}^{10} \left(\frac{\xi_{r,j}}{\sigma_r}\right)^2 + \left( \frac{\xi_a}{\sigma_a}\right)^2 + \sum_{i=1}^{200} \left(\frac{\xi_{s,i}}{\sigma_{s}}\right)\, .
\label{eq:ch5-JUNO-chi2}
\end{align}

Here, the systematic uncertainties are accounted for by introducing a total of 211 nuisance parameters. We have included an absolute uncertainty on the reactor flux, $\sigma_a=2\,\%$, an uncertainty related to each reactor, $\sigma_{r} = 0.08\,\%$, and an uncertainty on the shape of the spectrum, $\sigma_{s}=1\,\%$. The nuisance parameter associated with the reactor flux uncertainties are denoted by $\xi_{a}$ and the pull parameters corresponding to the reactor power uncertainty are $\xi_{r,j}$, with $j \in \lbrace 1, 10\rbrace$. Similarly, the pull parameters included for spectral error in the $i$-th bin are $\xi_{s, i}$, with $i \in \lbrace 1,200 \rbrace$. Likewise, an uncorrelated uncertainty from the detector, $\sigma_{d} = 1\,\%$, is also included. 

In our calculations, we implement the IBD cross-section as in \cite{Vogel:1999zy}, the energy spectra from \cite{Mueller:2011nm}, and the reactor fission fractions from \cite{Zhan:2008id}. Event computation and the minimisation of our $\chi^2$ function were performed using \texttt{GLoBES} (General Long Baseline Experiment Simulator) \cite{Huber:2004ka, Huber:2007ji}.

\begin{figure}
\centering
\includegraphics[width=0.72\textwidth]{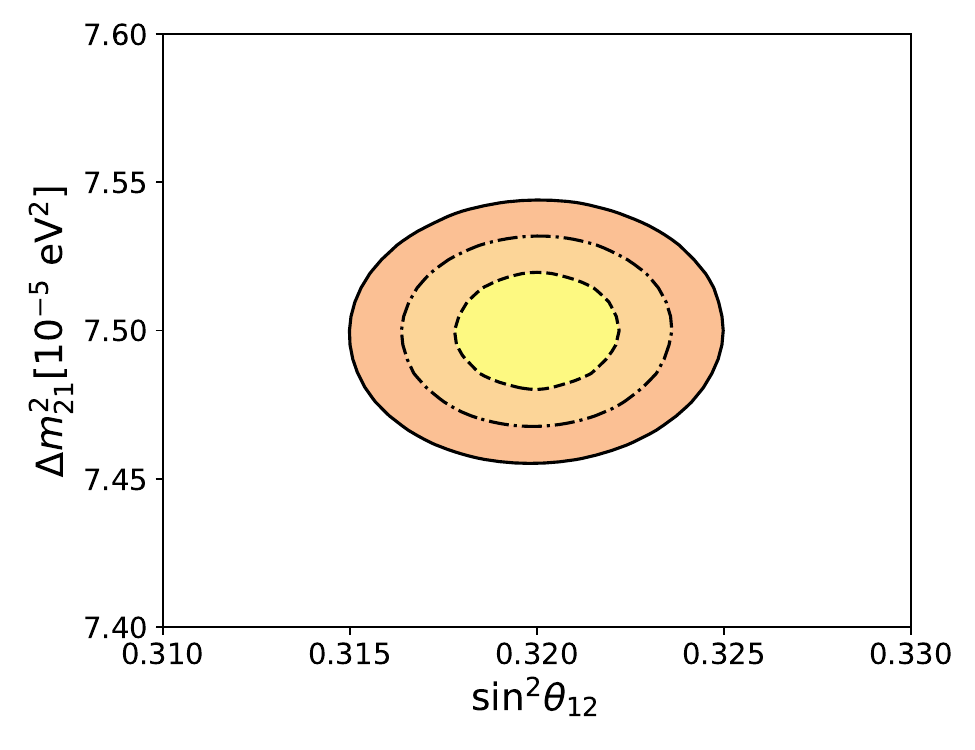}
\caption{\labfig{fig:ch5-stdJUNO} JUNO's expected sensitivity to the oscillation parameters $\theta_{12}$ and $\Delta m^2_{21}$ in the absence of NSIs, where the 1$\sigma$, 2$\sigma$ and 3$\sigma$ C.L. contours are shown as before.}
\end{figure}

%%%%%%%%%%%%%%%%%%%%%%%%%%%%%%%%%%%%%%%%%%%%%%%%%%%%%%%%%%%%%%%%%%%
\subsection{NSIs on medium-baseline reactor experiments}
\label{sec:ch5-JUNO-nsi}
%%%%%%%%%%%%%%%%%%%%%%%%%%%%%%%%%%%%%%%%%%%%%%%%%%%%%%%%%%%%%%%%%%%

If matter effects and NSIs are not considered, the survival probability in medium-baseline reactor experiments is given by:
\begin{align}
&P^{\text{MBR}}_{\overline{\nu}_e \to \overline{\nu}_e} = 1- \cos^4\theta_{13}\sin^2 2\theta_{12}\sin^2\left(\frac{\Delta m^2_{21}}{4E}\right) \nonumber \\ & \, - \sin^2 2\theta_{13}\left[\cos ^2\theta_{12}\sin ^2 \left(\frac{\Delta m^2_{31}}{4E}\right) + \sin^2 \theta_{12}\sin^2 \left(\frac{\Delta m^2_{32}}{4E}\right)\right] \, .
\label{eq:ch5-JUNOp}
\end{align}
Non-standard neutrino interactions with matter will have a similar impact on oscillation parameters in the solar sector as those discussed in Equation~\ref{eq:ch5-angle-nsi} and Equation~\ref{eq:ch5-mass-nsi} for solar neutrinos, with the key difference that, since reactors emit electron antineutrinos, the matter and NSI terms in the Hamiltonian will have an opposite sign to their counterparts in the case of neutrinos. For completeness, \reffig{fig:ch5-stdJUNO} shows the expected sensitivity of JUNO to $\sin^2\theta_{12}$ and $\Delta m^2_{21}$ in the absence of NSIs, after marginalising over $\theta_{13}$ and $\Delta m^{2}_{31}$ for normal ordering. The other two oscillation parameters influencing the survival probability expected at JUNO --- the reactor mixing angle $\theta_{13}$ and the atmospheric mass splitting $\Delta m^2_{31}$ --- are not significantly affected by matter effects or NSIs according to current constraints~\cite{JUNO:2015zny}. Hence, one would expect JUNO to be capable of providing an accurate measurement of these oscillation parameters even in the presence of NSIs.

\textbf{Numerical results}

In our analysis, we limit ourselves to the study of two NSI parameters simultaneously, $\varepsilon_{ee}^{d}$ and $\varepsilon_{e\tau}^{d}$. The motivation behind this choice is twofold: firstly, these two parameters are among the least constrained~\cite{Farzan:2017xzy} and, secondly, they can be easily mapped onto the two effective parameters ($\varepsilon$ and $\varepsilon '$) used to describe NSIs in solar neutrinos.

Moreover, we will assume all NSI coefficients to be real, so that
\begin{flalign}
\varepsilon_d ' = \sin 2\theta_{13}\cos \theta_{23} \cos\delta_{\text{CP}}\,\varepsilon_{e\tau}^{d} -\cos ^2 \theta_{13}\, \varepsilon_{ee}^{d}\,,
\label{eq:ch5-JUNO-epsprime}
\end{flalign}
and
\begin{flalign}
\varepsilon_d  = -\cos \theta_{13} \sin \theta_{23}\,\varepsilon_{e\tau}^{d}\,.
\label{eq:ch5-JUNO-eps}
\end{flalign}

\begin{figure*}
\centering
\includegraphics{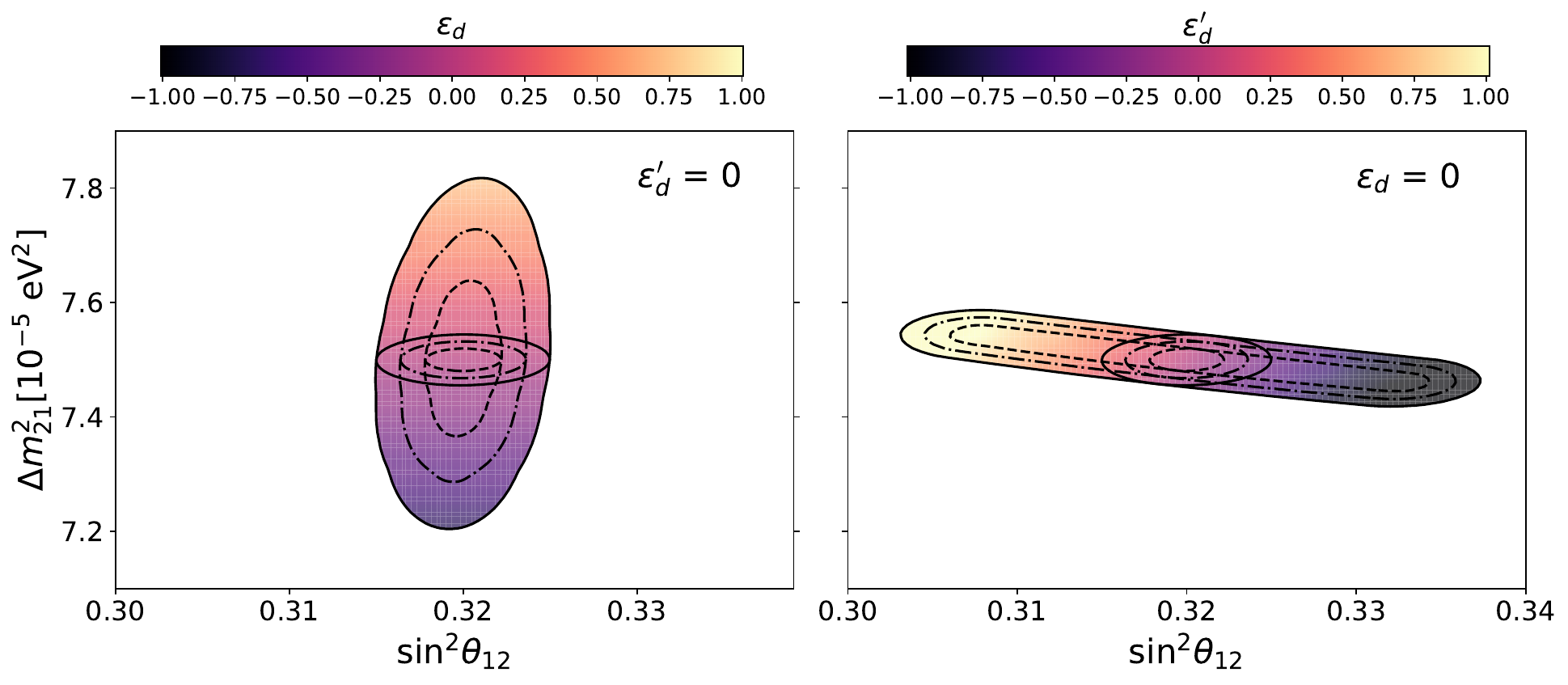}
\caption{Impact of the effective NSI parameters on the sensitivity of JUNO, varying one at a time. Left panel shows the 1$\sigma$, 2$\sigma$ and 3$\sigma$ C.L. in the $\sin^2 \theta_{12}- \Delta m^2_{21}$ plane when varying $\varepsilon_d$ between -1 and 1. The same confidence levels are drawn in the right panel for the case of $\varepsilon_d = 0$ and $\varepsilon_d'$ allowed to vary. The colour map indicates the best-fit value of the effective NSI parameters. The confidence levels expected for JUNO when NSIs are not included in the analysis are indicated by unfilled contours. \labfig{fig:ch5-JUNOoneatatime}}
\end{figure*}

In a medium-baseline reactor experiment aiming to measure the oscillation parameters $\theta_{12}$ and $\Delta m^2_{21}$ with high precision, matter effects have been shown to play an important role~\cite{Khan:2019doq, Li:2016txk}. In particular, matter effects produce an approximately $ 0.2\,\%$ shift in the effective mass splitting and a $1.2 \,\%$ shift in the value of effective $\sin^2 \theta_{12}$ with respect to the values which would be obtained if matter effects were not considered~\cite{Khan:2019doq}. Since JUNO aims to measure these two oscillation parameters with a precision of $\sim 0.5-0.7\,\%$, matter effects are very relevant. 

Likewise, the existence of non-standard interactions, even if smaller than standard matter effects, could greatly affect the precision goals of this experiment. In~\reffig{fig:ch5-JUNOoneatatime}, it can be seen how the sensitivity to the solar oscillation parameters is affected if the effective NSI couplings $\varepsilon_d$ and $\varepsilon_d'$ are included in the analysis and allowed to vary between -1 and 1. For both panels, the absence of NSIs was assumed as the true hypothesis, with the best-fit values for the oscillation parameters taken from Table~\ref{tab:ch5-oscparams}, while the test hypothesis consisted of $\varepsilon_d\neq0$ in the left panel and $\varepsilon_d'\neq0$ in the right panel.

It can be seen that a non-zero $\varepsilon_d$ results mainly in a shift in the effective mass splitting, whereas the main impact of a non-zero $\varepsilon_d'$  would be a distortion in the reconstructed value of the solar mixing angle.

%%%%%%%%%%%%%%%%%%%%%%%%%%%%%%%%%%%%%%%%%%%%%%%%%%%%%%%%%%%%%%%%%
%%%%%%%%%%%%%%%%%%%%%%%%%%%%%%%%%%%%%%%%%%%%%%%%%%%%%%%%%%%%%%%%%
\section{Projected sensitivity from complementarities}
\label{sec:ch5-results}
%%%%%%%%%%%%%%%%%%%%%%%%%%%%%%%%%%%%%%%%%%%%%%%%%%%%%%%%%%%%%%%%%
%%%%%%%%%%%%%%%%%%%%%%%%%%%%%%%%%%%%%%%%%%%%%%%%%%%%%%%%%%%%%%%%%

%%%%%%%%%%%%%%%%%%%%%%%%%%%%%%%%%%%%%%
\subsection{Combining JUNO and Hyper-Kamiokande}
\label{sec:ch5-combined}
%%%%%%%%%%%%%%%%%%%%%%%%%%%%%%%%%%%%%%

The impact of non-standard interactions on solar neutrinos and long and medium-baseline reactor experiments is significantly different. Therefore, this fact can be employed to constrain these interactions. If absent, JUNO would dominate the determination of the parameters $\sin^2\theta_{12}$ and $\Delta m^2_{21}$. In fact, the main contribution of solar neutrinos would be to break the degeneracy between the two octants. Nonetheless, if one considers additional neutrino interactions, the picture changes radically.

As shown in the previous section, JUNO on its own would provide a robust determination of the oscillation parameters. This results from the fact that matter effects --- as well as non-standard matter effects due to NSIs --- barely alter neutrino propagation in this context. Given the accuracy aimed by this experiment, even these small effects would spoil its precision goals. On the contrary, solar neutrinos are very sensitive to any new physics affecting propagation and, as such, they can deliver powerful tests of neutrino interactions during propagation as long as the oscillation picture is well-established. This complementarity motivates the combination of both experiments as a way to get stronger bounds on non-standard interactions and ensure a precise measurement of the oscillation parameters. 

\begin{figure*}[t!]
\centering
\includegraphics[width=0.365\paperwidth]{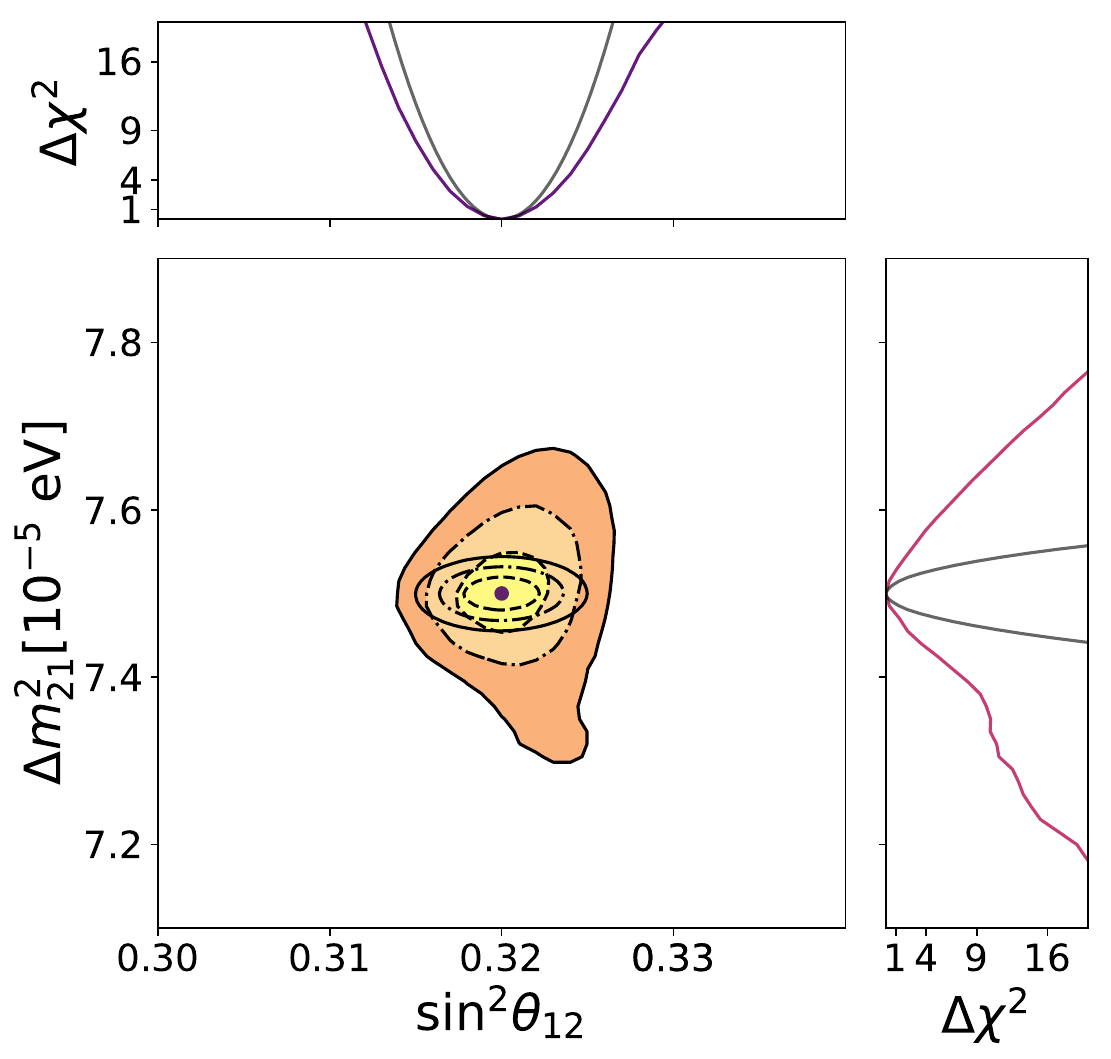}
\hspace{0.02\textwidth}
\includegraphics[width=0.365\paperwidth]{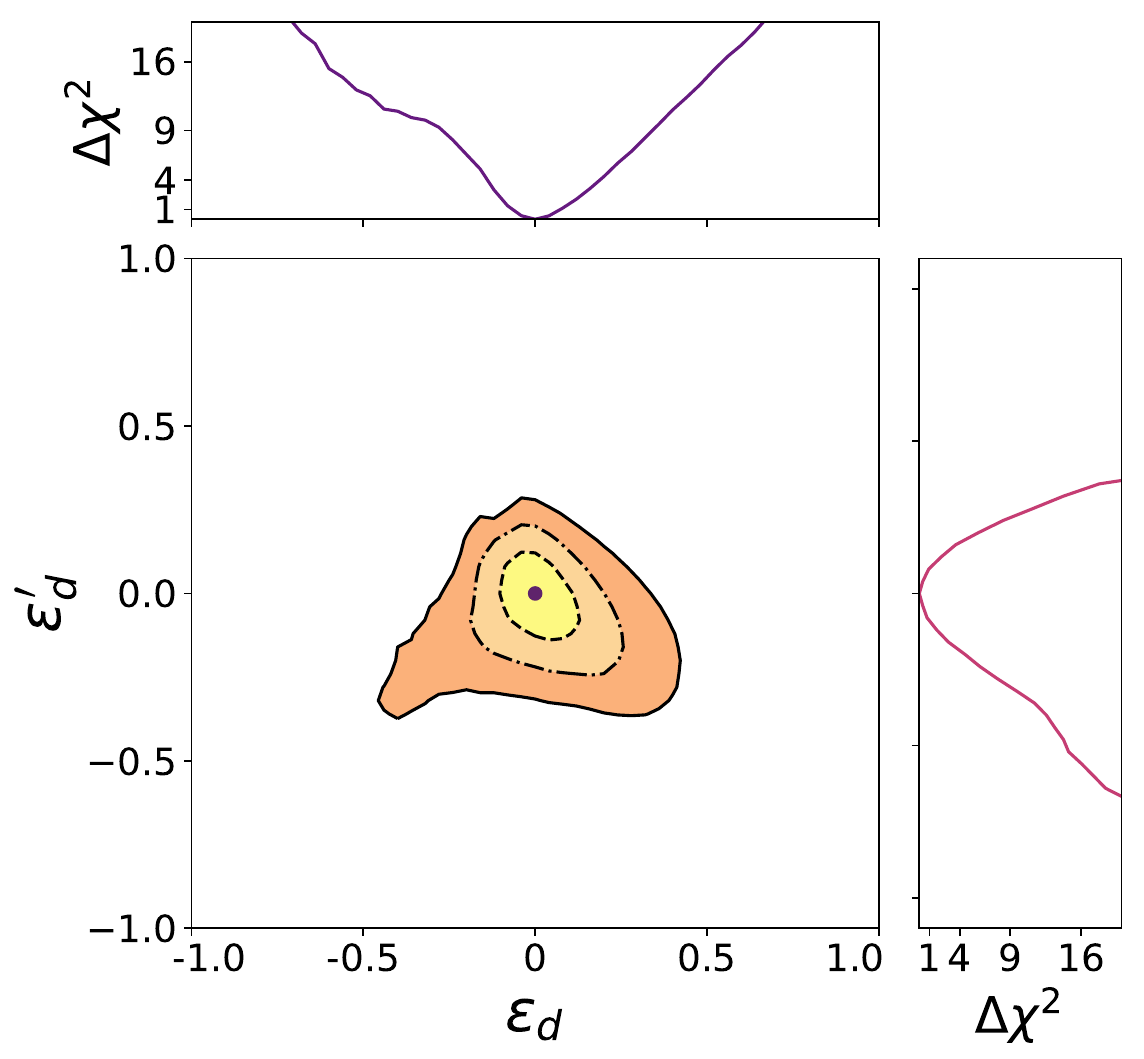}
\caption{Two-dimensional projections onto the $\sin^2 \theta_{12}$ - $\Delta m^2_{21}$ plane --- in the left panel --- and the $\varepsilon$-$\varepsilon'$ plane --- in the right panel --- of the expected sensitivity from a combined analysis of Hyper-Kamiokande and JUNO. Contours correspond to 1$\sigma$, 2$\sigma$ and 3$\sigma$ allowed regions --- indicated with dashed, dot-dashed and solid lines respectively. One-dimensional projections are shown for completeness. In the left panel, the corresponding confidence levels expected in the absence of NSIs are again shown using unfilled contours.\labfig{fig:ch5-comb1stoct}}
\end{figure*}

In \reffig{fig:ch5-comb1stoct}, we present the combined sensitivity of JUNO and \mbox{Hyper-Kamiokande} using the most optimistic configuration --- referred to as Configuration A in \reftab{ch5-confsHK}. In the left panel of this figure, one can see that, from a combination of both experiments, it will be possible to reach an impressive level of precision in both parameters of the solar sector. Notably, the allowed regions for the solar mixing and the solar mass splitting at 90\%~C.L. would be
\begin{gather}
0.318 \, <\,  \sin^2\theta_{12}\,  < \,  0.322 \nonumber \\
7.48 \times 10 ^{-5} \text{eV}^2 \,<\, \Delta m^2_{21} \,<\, 7.52 \times 10^{-5} \text{eV}^2 \,.
\label{eq:ch5-osc-results}
\end{gather}
The projected sensitivity to $\sin^2 \theta_{12}$ is very close to what JUNO alone would obtain if NSIs were not considered. This means that JUNO will reach its precision goals for $\sin^2\theta_{12}$ independently of NSIs. However, the sensitivity to the solar mass splitting will be significantly degraded if one allows for NSIs, while still presenting a remarkable improvement with respect to its current level.

The right panel of \reffig{fig:ch5-comb1stoct} shows the projected sensitivity to NSI parameters after combining JUNO and Hyper-Kamiokande. At 90$\%$~C.L., the allowed regions for NSI parameters would read
\begin{align}
&-0.153 < \varepsilon_d' <\,  0.135 \, ,\nonumber \\
&-0.113 < \varepsilon_d <\,  0.144 \, ,
\label{eq:ch5-limitsNSI}
\end{align}
where these limits have been obtained allowing for one non-zero parameter at a time. Note that large values of $\varepsilon_d'$ are excluded. This is because we are assuming the same mass ordering for the true values and the ones being tested. In this case, the LMA-D solution --- which is only possible for large $\varepsilon_d'$ and different orderings for each set of values --- does not arise. This approach is justified as long as the mass ordering is determined independently of NSIs. In the next subsection, we relax this constraint and study the case in which the mass orderings are allowed to be different.

\begin{figure*}[t!]
\centering
\includegraphics[width=0.76\paperwidth]{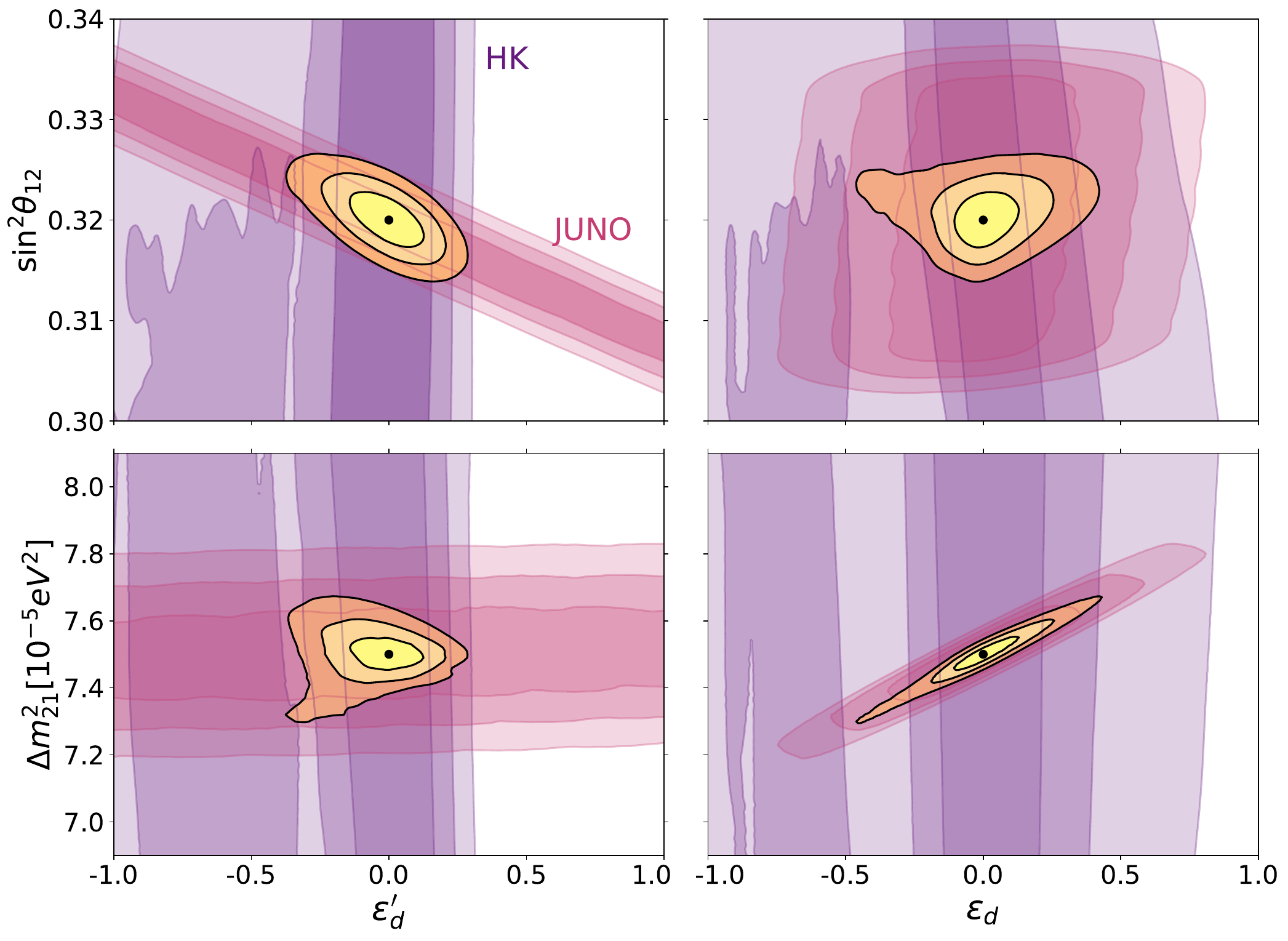}
\caption{Two-dimensional projections onto the $\varepsilon_d'$ - $\sin^2 \theta_{12}$ plane in the top left panel, $\varepsilon_d$ - $\sin^2 \theta_{12}$ plane in the top right panel, $\varepsilon_d'$ - $\Delta m^2_{21}$ plane in the bottom left panel and $\varepsilon_d$ - $\Delta m^2_{21}$ plane in the bottom right panel. They result from a combined analysis of Hyper-Kamiokande (HK) and JUNO. Contours correspond to 1$\sigma$, 2$\sigma$ and 3$\sigma$ C.L. The allowed regions from HK and JUNO individually are shown in purple and pink, respectively. \labfig{fig:ch5-othercomb}}
\end{figure*}

In \reffig{fig:ch5-othercomb}, one can see the remaining two-dimensional projections. In this figure, the individual constraints from Hyper-Kamiokande (HK) and JUNO are shown to illustrate that the sensitivity to oscillation and NSI parameters arises from the combination of both experiments. As discussed previously, the sensitivity of JUNO to the oscillation parameters is not greatly affected when NSIs are included in the analysis. Conversely, non-standard interactions would induce large deviations and a significant loss of accuracy in the measurement of neutrino oscillation parameters by Hyper-Kamiokande. 

%%%%%%%%%%%%%%%%%%%%%%%%%%%%%%%%%%%%%%%%%
\subsection{About the LMA-D solution}
\label{sec:ch5-lmad}
%%%%%%%%%%%%%%%%%%%%%%%%%%%%%%%%%%%%%%%%%

Up until this point, we have assumed that the true mass ordering was known from an NSI-independent probe --- or at least that there existed a strong preference in favour of one ordering which had been determined from an analysis accounting for NSIs. In this way, we were only considering one of the two possible mass orderings. If this assumption is lifted, one expects a second region of parameter space to become allowed. This is the so-called LMA-D solution and it is a consequence of the generalised mass degeneracy as explained in Section \ref{sec:ch5-degen-teo}.
The allowed regions for the oscillation parameters and the NSI parameters are shown in the left and right panels of~\reffig{fig:ch5-comb_2ndoct} respectively.
\begin{figure*}
\centering
\includegraphics[width=0.365\paperwidth]{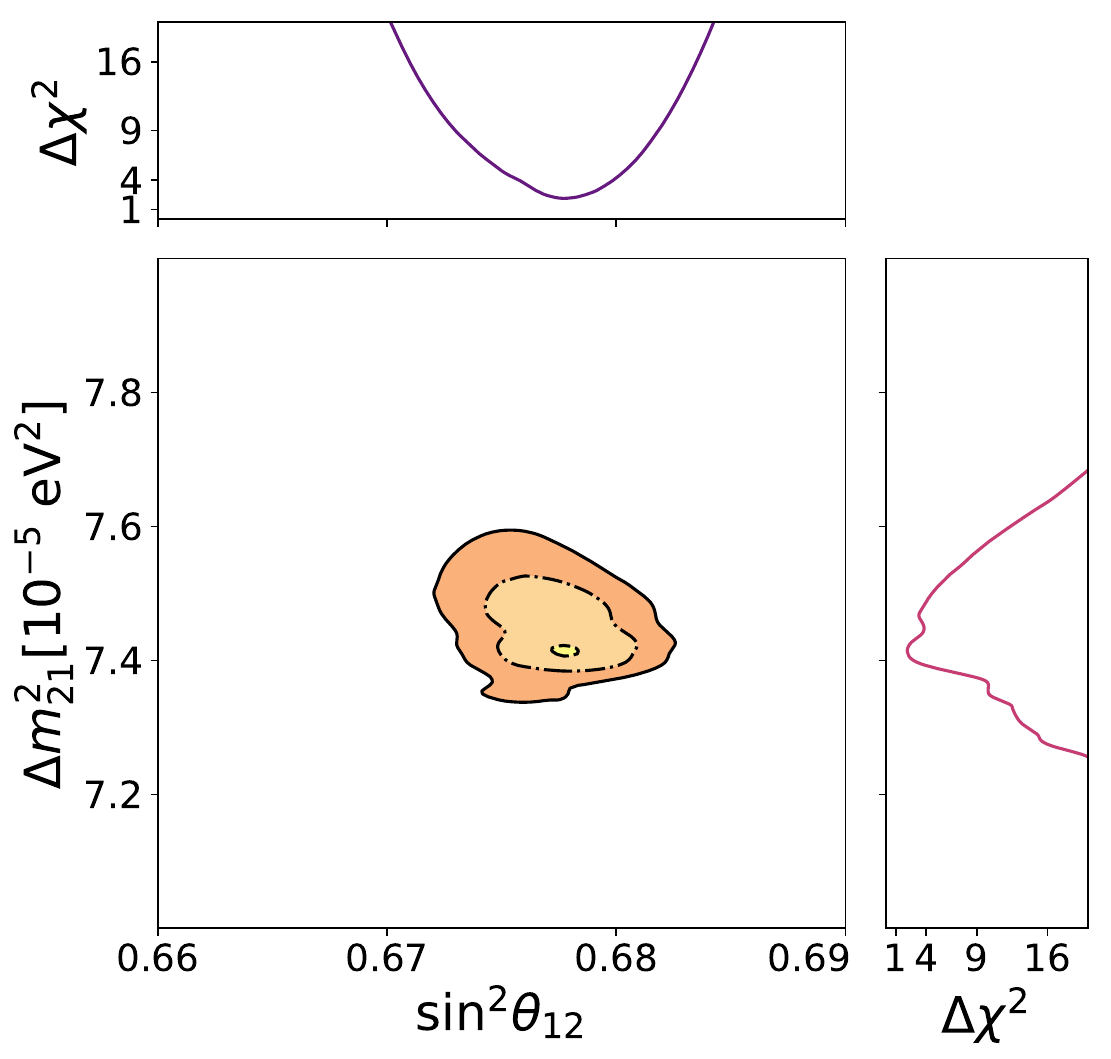}
\hspace{0.02\textwidth}
\includegraphics[width=0.365\paperwidth]{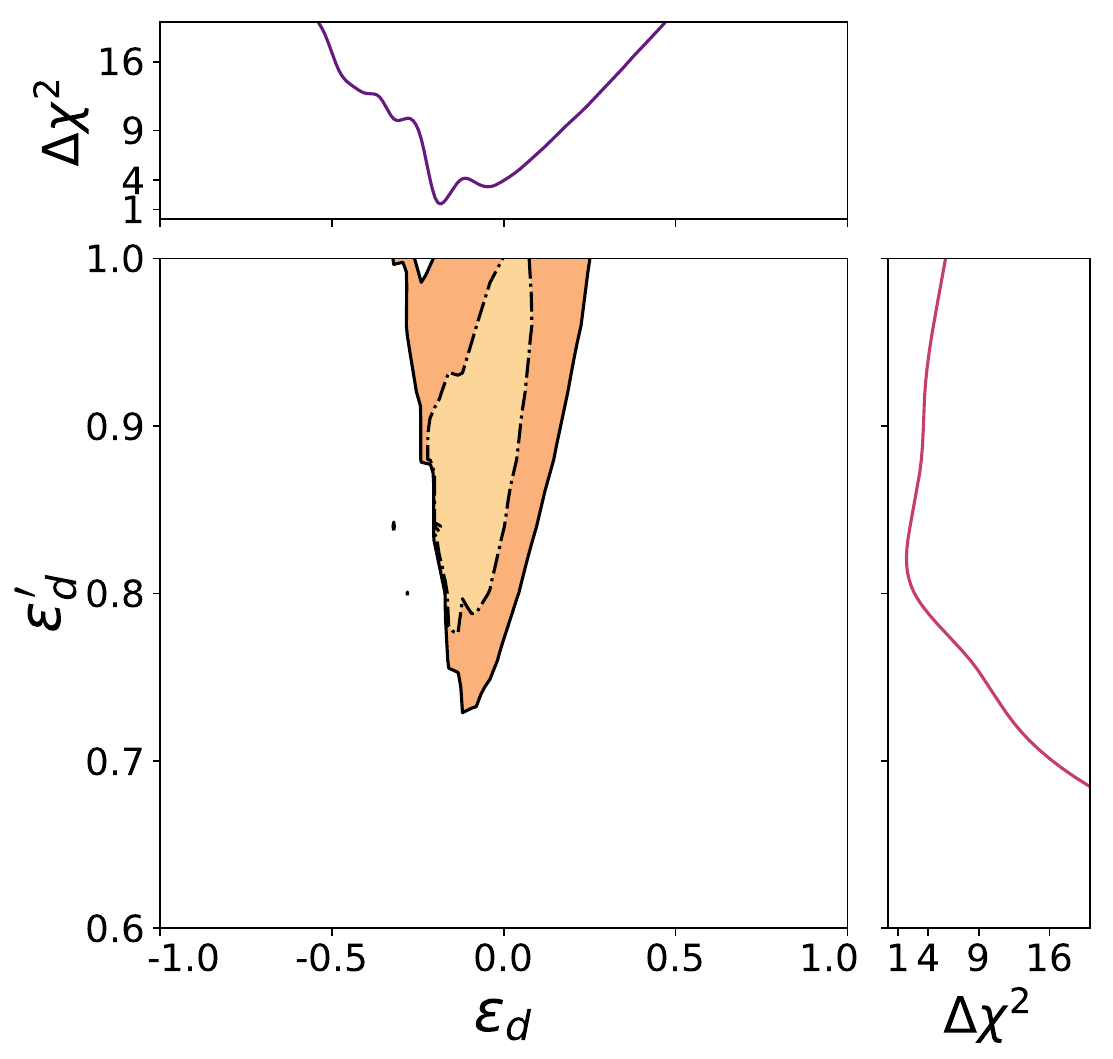}
\caption{Two-dimensional projections onto the $\sin^2 \theta_{12}$ - $\Delta m^2_{21}$ plane --- in the left panel --- and the $\varepsilon_d$ - $\varepsilon_d'$ plane --- in the right panel ---  of the expected sensitivity from a combined analysis of Hyper-Kamiokande and JUNO for the LMA-D solution. Contours correspond to 1$\sigma$, 2$\sigma$, and 3$\sigma$ allowed regions ---  indicated with dashed, dot-dashed and solid --- with respect to the best fit in the first octant from Table \ref{tab:ch5-oscparams} and assuming normal ordering. One-dimensional projections are shown for completeness.\labfig{fig:ch5-comb_2ndoct}}
\end{figure*}

In the analysis of solar neutrinos, the dependence on $\Delta m^2_{31}$ and its sign is negligible. Thus, our analysis of Hyper-Kamiokande remains identical to the one in the previous subsection, except for the fact that we also explore values of $\sin^2\theta_{12} > 0.5$ in order to cover the \mbox{LMA-D} solution.
For JUNO, however, the picture depends significantly on $\Delta m^2_{31}$. Our \textit{mock data} had been generated assuming no NSIs, normal ordering and a solar mixing angle in the first octant --- i.e $\sin ^2\theta_{12} < 0.5$. Then, we performed our analysis and fixed the value of $\Delta m^2_{31}$ to its best-fit point under normal mass ordering we were systematically excluding the LMA-D solution. Therefore, the appropriate procedure to explore this degenerate solution is to allow $\Delta m^2_{31}$ to take both positive and negative values, thus accounting for both hierarchies.

\begin{figure*}[t!]
\centering
\includegraphics[width=0.76\paperwidth]{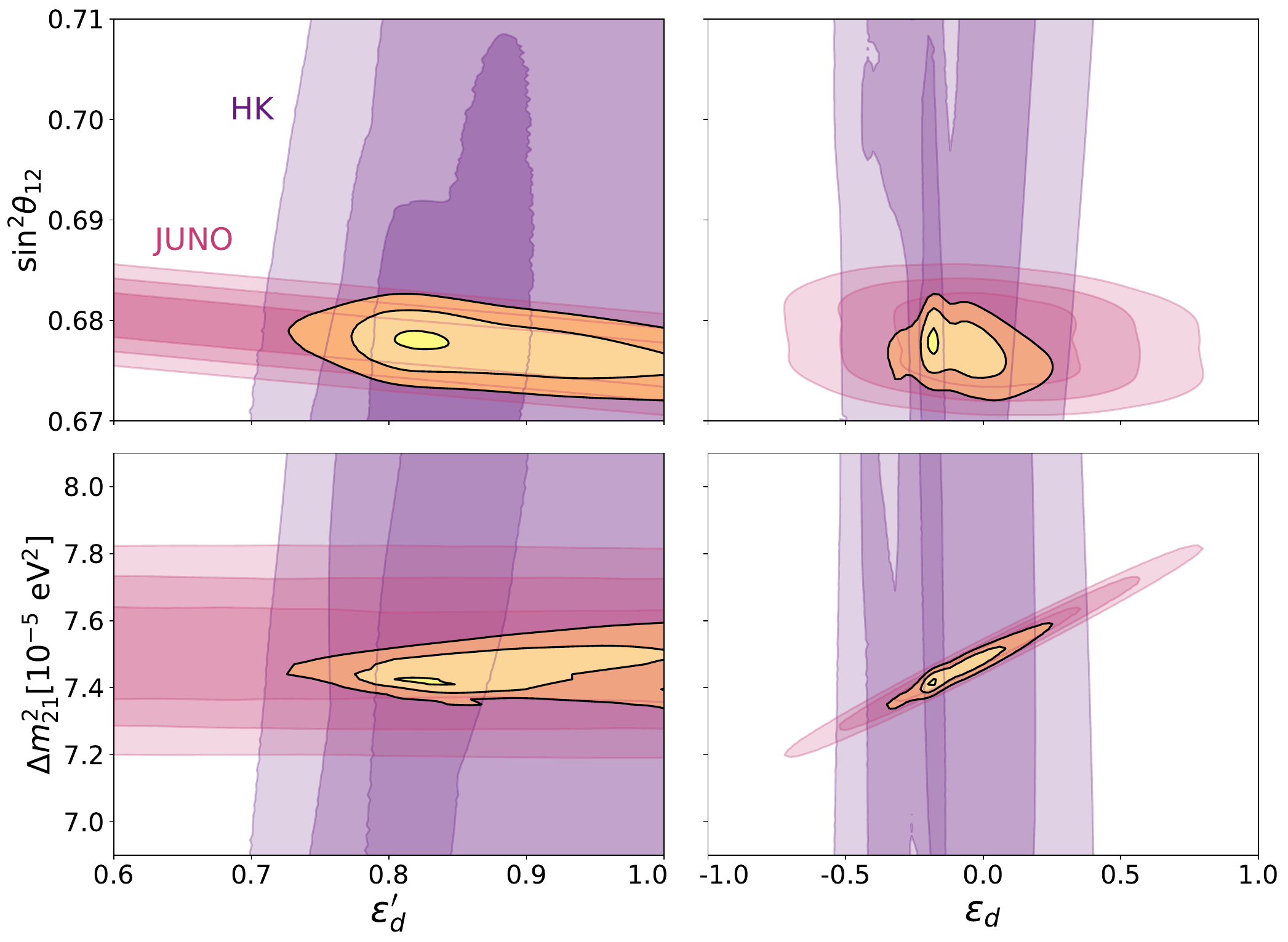}
\caption{Two-dimensional projections onto the $\varepsilon_d'$ - $\sin^2 \theta_{12}$ plane (top left), $\varepsilon_d$ - $\sin^2 \theta_{12}$ plane (top right), $\varepsilon_d'$ - $\Delta m^2_{21}$ plane (bottom left) and $\varepsilon_d$ - $\Delta m^2_{21}$ plane (bottom right), from a combined analysis of Hyper-Kamiokande (HK) and JUNO, for the LMA-D solution. Contours correspond to 1$\sigma$, 2$\sigma$ and 3$\sigma$ C.L. The allowed regions from HK and JUNO individually are shown in purple and pink, respectively.  \labfig{fig:ch5-other2ndoct}}
\end{figure*}

In this case, a second solution arises in the second octant, that is, for \mbox{$\sin^2 \theta_{12} > 0.5$.} The best fit for this degenerate solution is slightly disfavoured with respect to the one in the first octant since $\Delta \chi^2$ is larger than zero for this solution. This is because the number density of $d$-quarks is different in the Earth's crust and core and along the neutrino trajectory in the Sun. Therefore, different values of $\varepsilon_d'$ can account for the generalised mass ordering degeneracy in each medium. Nevertheless, in a combined analysis of JUNO and Hyper-Kamiokande, the LMA-D solution would not be completely discounted. Notice that this solution also arises for a slightly smaller $\Delta m^2_{21}$, as seen in \reffig{fig:ch5-comb_2ndoct}. 

After marginalising over the solar neutrino oscillation parameters, one can obtain the sensitivity to the effective NSI parameters $\varepsilon_d$ and $\varepsilon_d'$, as shown in the right panel of~\reffig{fig:ch5-comb_2ndoct}. We find that the LMA-D region requires very large values of $\varepsilon_d'$, as well as a non-zero $\varepsilon_d$, which is in agreement with previous works~\cite{Miranda:2004nb,Escrihuela:2009up}. 
Despite being very sizeable non-standard interactions, these two neutrino oscillation datasets alone would not be able to exclude them. However, scattering data and results from coherent elastic neutrino-nucleus scattering  experiments~\cite{COHERENT:2017ipa} are a powerful complementary probe for this scenario~\cite{Coloma:2017ncl,Esteban:2018ppq,Coloma:2019mbs}. In fact, the combination of current solar neutrino data with results from the COHERENT experiment excludes the LMA-D solution at more than 3$\sigma$ in models with NSIs involving only a single quark flavour~\cite{Coloma:2017ncl,COHERENT:2021xmm}. However, this constraint is relaxed when allowing for non-zero NSI with both $u$ and $d$-type quarks simultaneously. Then, so far, it is not possible to exclude the LMA-D solution in these scenarios~\cite{Esteban:2018ppq, Coloma:2019mbs}.

Finally, and for completeness, \reffig{fig:ch5-other2ndoct} depicts the remaining two-dimensional projections for the LMA-D solution. Once again, the sensitivity obtained for JUNO and Hyper-Kamiokande individually is shown together with the resulting sensitivity from a combined analysis and illustrates how both experiments complement each other. 

%%%%%%%%%%%%%%%%%%%%%%%%%%%%%%%%%%%%%%%%%%%
\subsection{The experimental details of Hyper-Kamiokande}
\label{sec:ch5-configurations}
%%%%%%%%%%%%%%%%%%%%%%%%%%%%%%%%%%%%%%%%%%%

In a previous section, we discussed the impact of three possible experimental configurations of Hyper-Kamiokande on its determination of solar oscillation parameters in the presence of NSIs --- see \reffig{fig:ch5-HK-confs}.
It is also interesting to examine whether our final results depend significantly on the exact configuration of the Hyper-Kamiokande detector. The sensitivity profiles of each of the four parameters under consideration --- $sin^2\theta_{12}$, $\Delta m^2_{21}$, $\varepsilon_d$, and $\varepsilon_d'$ --- are presented in \reffig{fig:ch5-chi2_profiles}.
\begin{figure*}[t!]
\centering
\includegraphics[width=0.76\paperwidth]{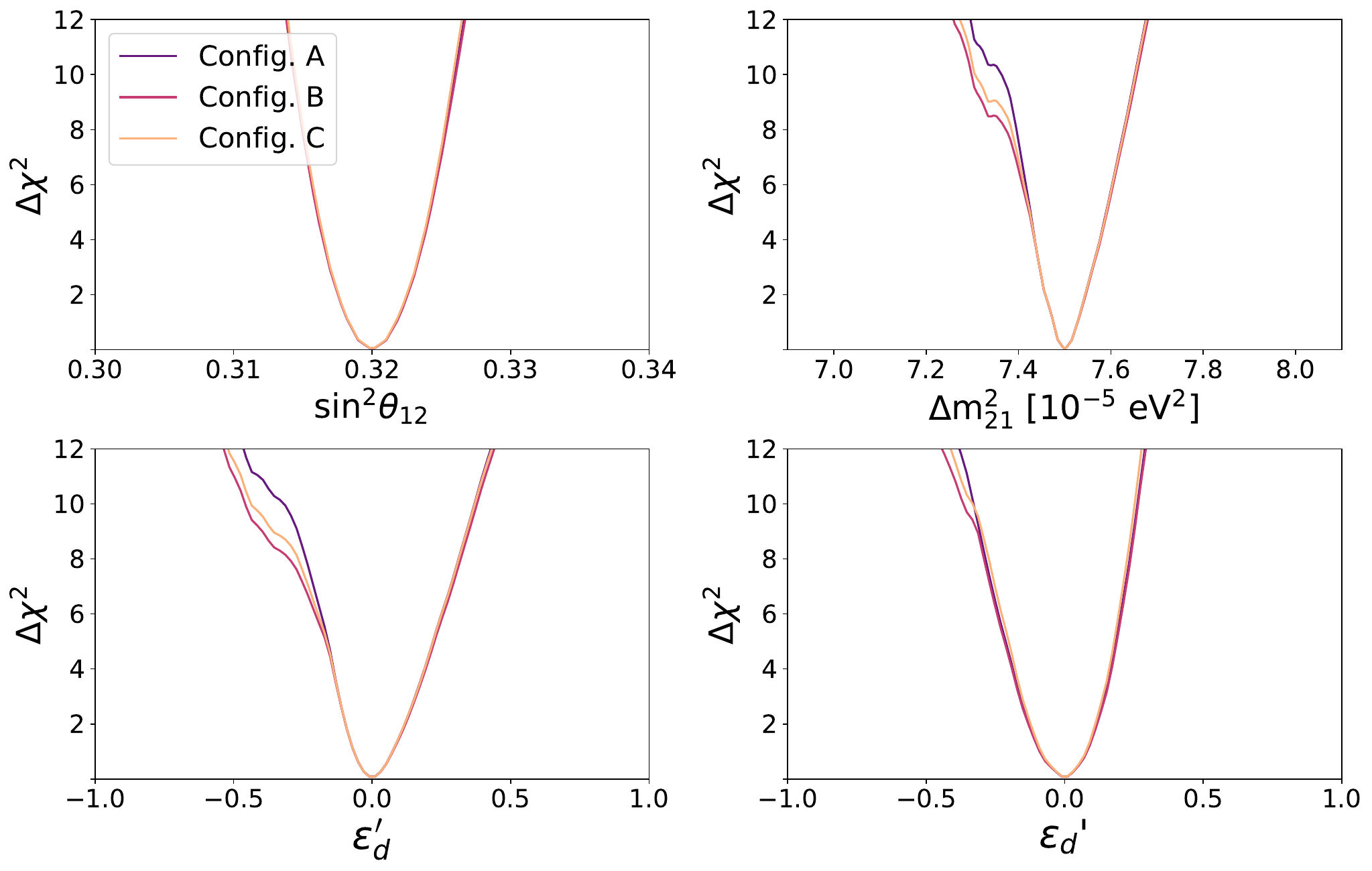}
\caption{
Sensitivity to neutrino oscillation and NSI parameters for the three different potential configurations of the Hyper-Kamiokande detector considered in this work. \labfig{fig:ch5-chi2_profiles}}
\end{figure*}
One can see that the choice of one experimental setup over another would not have a significant impact on the determination of the oscillation parameters $\sin^2 \theta_{12}$ and $\Delta m^2_{21}$, since the accuracy is dominated by JUNO.
The only noticeable distinction arises at the $\sim 2-3\,\sigma$ level for the low-$\Delta m^2_{21}$ and low-$\varepsilon_d$ sides of the profiles. These regions correspond to the lobules which appear in the two-dimensional regions presented in Section \ref{sec:ch5-combined}. Thus, the differences between the three configurations listed in Table \ref{tab:ch5-confsHK} are not relevant in a joint analysis with JUNO.

%%%%%%%%%%%%%%%%%%%%%%%%%%%%%%%%%%%%%%%%%%
%%%%%%%%%%%%%%%%%%%%%%%%%%%%%%%%%%%%%%%%%%
\section{Concluding remarks}
\label{sec:ch5-conclusion}
%%%%%%%%%%%%%%%%%%%%%%%%%%%%%%%%%%%%%%%%%%
%%%%%%%%%%%%%%%%%%%%%%%%%%%%%%%%%%%%%%%%%%

In the past, the complementarity between solar and reactor neutrinos has been proven to be extremely powerful for the determination of the oscillation parameters of the solar sector and also for constraining non-standard neutrino interactions. This chapter addresses the expected improvements from a combined analysis of the future neutrino experiments JUNO and Hyper-Kamiokande, focusing on NSIs with $d$-type quarks. 

Including non-standard interactions in the analysis of JUNO data would degrade significantly its sensitivity to the solar oscillation parameters $\sin^2\theta_{12}$ and $\Delta m^2_{21}$. Nonetheless, a combined analysis with Hyper-Kamiokande would allow a subpercent precision measurement of $\sin^2\theta_{12}$ and $\Delta m^2_{21}$ at 90\% confidence level. We have shown that these results do not depend strongly on the exact experimental setup used for Hyper-Kamiokande. In particular, we have considered different configurations aiming to reach a lower energy threshold and collect larger statistics. These results rely strongly on a more precise determination of the oscillation parameters at JUNO and therefore, the experimental details of Hyper-Kamiokande are not very influential. Nevertheless, the day-night asymmetry and the upturn in the solar neutrino spectrum remain key observables for verifying our understanding of the solar neutrino picture.

In Equation \ref{eq:ch5-limitsNSI}, we provide the expected limits, excluding the LMA-D solution. Those constraints are comparable to other sensitivity studies for future neutrino experiments~\cite{Liao:2017awz, Bakhti:2020fde}. In all cases, next-generation experiments will improve the current bounds from combined analyses of solar and KamLAND data~\cite{Miranda:2004nb,Escrihuela:2009up}, as expected. 

In principle, oscillation experiments could also constrain the LMA-D solution since they are affected by matter effects in different ways. We have explored this possibility based on the projected sensitivities for Hyper-Kamiokande and JUNO. Although it will not be possible to exclude this solution only with these two experiments, only a very small region of parameter space would still be allowed at 1 $\sigma$. 

However, in order to truly exclude large values of $\varepsilon_d'$ so that this assumption is supported by data, external input from independent probes would be needed. Global fits to neutrino oscillation and coherent elastic neutrino-nucleus scattering data ~\cite{Esteban:2018ppq,Coloma:2019mbs} exploit the complementarities between these experimental probes. Along that lines, including data from future experiments like Hyper-Kamiokande and JUNO will improve the already stringent bounds and provide additional information on the nature of neutrino interactions.

%\setchapterpreamble[u]{\margintoc}
\chapter{Testing CPT symmetry with neutrino oscillations}
%\addcontentsline{toc}{chapter}{Testing CPT with neutrino oscillations} 
\labch{ch6-cpt}

Global fits to neutrino data strongly rely on the fact that masses and mixings are the same for neutrinos and antineutrinos in vacuum. This assumption is a direct consequence of CPT conservation in local relativistic quantum field theories, the theoretical framework on which we rely for describing nature. However, in the context of non-local theories, the difference between the masses of particles and antiparticles is an observable of CPT violation.

This chapter discusses how neutrinos provide a powerful test of this fundamental symmetry through the measurements performed in oscillation experiments. In Section~\ref{sec:ch6-cpt}, we introduce CPT symmetry and describe in which context it could be broken. Next, Section~\ref{sec:ch6-cpt-osc} discusses the test of CPT invariance that can be performed with oscillation data and summarises the existing results. The prospects from the future solar sector --- in particular, from DUNE, JUNO and Hyper-Kamiokande --- are discussed in Section~\ref{sec:ch6-solar}. Finally, Section~\ref{sec:ch6-othercpt} provides a discussion of other existing tests of CPT invariance to contextualise the results here presented and some parting thoughts are outlined in Section~\ref{sec:ch6-thoughts}.

\section{CPT symmetry}
\label{sec:ch6-cpt}
Symmetries are a key ingredient in our description of nature. Among them all, CPT symmetry plays an essential role in our understanding of particle physics and its invariance serves as a guideline in model building. That is the reason motivating the extensive set of experimental analyses aiming to test it~\cite{Widmann:2021krf}.

The CPT theorem~\cite{Jost:1957zz,Streater:1989vi} guarantees that CPT is conserved in any local relativistic quantum theory preserving Lorentz invariance and formulated on flat spacetimes. CPT invariance implies that, if one performs simultaneously the operations of parity transformation, charge conjugation and time reversal, the measurable properties of the system are not modified.

The CPT operator is anti-unitary and connects the S-matrix of a process with the S-matrix of the inverse process --- i.e. the result of replacing particles with their antiparticles and reverting all spin components. At this point, one should remember that CPT conjugate processes do not necessarily occur with the same probability. For instance, the decay branching ratios of particles and antiparticles are not necessarily the same. Nonetheless, the sum of all of the decay rates --- the lifetime --- is the same for a particle and its antiparticle.

If CPT is broken, then it means that --- at least --- one of the underlying assumptions does not hold. These assumptions are Lorentz invariance, locality and hermicity of the Hamiltonian. CPT violation associated with the breakdown of Lorentz invariance has been extensively studied in the literature due to its connection with theories of quantum gravity~\cite{Addazi:2021xuf}. In this context, the Standard Model Extension~\cite{Colladay:1996iz,Colladay:1998fq} provides a useful framework to address this topic. It consists of an effective field theory in which the effective Lagrangian respects the symmetries of the Standard Model but it is not Lorentz invariant. Alternatively, one can build a non-local --- yet causal and Lorentz invariant --- theory, in which CPT is not a symmetry~\cite{Barenboim:2002tz,Chaichian:2011fc,Chaichian:2012ga}.

As one can see, since the origin of a hypothetical CPT non-conservation is deeply rooted in the fundamental pillars of our theories, it is interesting to test it using elementary particles, like neutrinos~\cite{Barenboim:2022rqu}. Nonetheless, other limits based on non-elementary particles exist~\cite{Cheng:2022omt}.

\section{Testing CPT with neutrino oscillations}
\label{sec:ch6-cpt-osc}
Neutrinos, as neutral elementary particles, are an ideal system to test CPT symmetry.  The first reason is purely theoretical. We know neutrino masses prove that there is physics beyond the Standard Model. In our attempt to explain the origin of these masses, the existence of a new --- high or very high --- scale is invoked. If that scale lies somewhere near the scale at which gravity becomes non-local or Lorentz invariance breaks, then neutrinos provide a unique opportunity to explore such high scales through the imprint left at low energies. The second reason is that the field of neutrino physics is blooming experimentally. Current and next-generation experiments aim to reach unprecedented precision and test a wide variety of new physics scenarios.

Notice that testing the predictions of CPT conservation is not the same as setting constraints on CPT violation since, in a way, the observables which are sensitive to CPT non-conservation depend on the underlying model. For instance, when originating from Lorentz invariance violation, the \mbox{CPT-breaking} effects would manifest as altered dispersion \mbox{relations~\cite{Kostelecky:2003cr,Barenboim:2018ctx,Barenboim:2019hso}.} These modifications in neutrino propagation can be explored in neutrino oscillation experiments and would manifest as additional energy and baseline dependences in the oscillation probabilities. Conversely, CPT violation from non-locality can result in different masses for neutrinos and antineutrinos. From this moment on, we will consider this second scenario.

In the neutrino sector, one could think of testing CPT invariance from the endpoint of beta decay. However, since no direct measurement has been achieved yet, this is not a viable possibility for the moment. However, it is possible to set limits on CPT violation from the difference in the mass splittings responsible for flavour oscillations, $\Delta m^2_{ij}$, for neutrinos and antineutrinos~\cite{Super-Kamiokande:2011dgc,MINOS:2013utc,Ohlsson:2014cha,T2K:2017krm,Barenboim:2017ewj}. Along these lines, the most recent bounds on CPT violation from the neutrino sector at 3$\sigma$ level were reported in~\cite{Tortola:2020ncu} --- using the same datasets as in~\cite{deSalas:2020pgw}--- and read
\begin{align}
&|\Delta m^2_{21} - \Delta \overline{m}^2_{21}| <  4.7\times 10^{-5}\, \text{eV}^2\, , \nonumber\\
&|\Delta m^2_{31} - \Delta \overline{m}^2_{31}| < 2.5\times 10^{-4}\, \text{eV}^2\, , \nonumber\\
&|\sin^2\theta_{12} - \sin^2\bar{\theta}_{12}| < 0.14\, , \nonumber \\
&|\sin^2\theta_{13} - \sin^2\bar{\theta}_{13}| < 0.029\, ,\nonumber \\
&|\sin^2\theta_{23} - \sin^2\bar{\theta}_{23}| < 0.19
\label{eq:ch6-current-cpt}
\end{align}
One can see that the solar sector is already giving the strongest bound on CPT violation from the solar mass splitting. Actually, when including the most recent preliminary results from Super-Kamiokande \cite{yasuhiro_nakajima_2020_4134680,yusuke_koshio_2022_6695966}, the bounds read \cite{Barenboim:2023krl}
\begin{align}
&|\Delta m^2_{21} - \Delta \overline{m}^2_{21}| <  3.7\times 10^{-5}\, \text{eV}^2\, , \nonumber\\
&|\sin^2\theta_{12} - \sin^2\bar{\theta}_{12}| < 0.187
\label{eq:ch6-current-cpt-2}
\end{align}
at 3$\sigma$. As one can see, the change in the limits is small. Notice that the bound on $|\sin^2\theta_{12} - \sin^2\bar{\theta}_{12}|$ is weaker now. The reason is that the current \mbox{best-fit} value for the solar mixing angle, $\sin^2\theta_{12}=0.306$, is in slightly worse accord with the KamLAND best-fit point, $\sin^2\theta_{12}=0.316$, than the previous solar best-fit value --- which used to be $\sin^2\theta_{12}=0.320$. Regarding the limit on $|\Delta m^2_{21} - \Delta \overline{m}^2_{21}|$, it becomes more stringent since the current agreement between the measurement of the solar mass splitting has improved. For further discussion on the topic, see~\cite{Barenboim:2023krl}.

\section{Prospects for CPT limits from the solar sector}
\label{sec:ch6-solar}
\subsection{Reactor antineutrinos at JUNO}
As introduced previously in Section~\ref{sec:ch5-JUNO}, the JUNO experiment will determine the parameters of the solar sector, $\sin^2\theta_{12}$ and $\Delta m^2_{21}$ with unprecedented accuracy using reactor antineutrinos. Here, we will denote those parameters as $\sin^2 \bar{\theta}_{21}$ and $\Delta \overline{m}^2_{21}$. Our simulation of the experiments follows the considerations we discussed in the previous chapter.

\subsection{Solar neutrinos at Hyper-Kamiokande}
The capability of Hyper-Kamiokande to detect solar neutrinos was also introduced in the previous chapter --- in Section~\ref{sec:ch5-HK-simu}. For the discussion that occupies us here, we will consider two experimental configurations. The first one corresponds to having one single tank and a 5 MeV energy threshold --- see Configuration B in~\reftab{ch5-confsHK}. This conservative configuration considers the same efficiency and energy resolution as in Super-Kamiokande IV. We also simplify slightly our $\chi^2$ function by considering only a systematic error in the energy resolution and uncorrelated systematics.\footnote{Note that in Equation \ref{eq:ch5-HK-events-1} we included also a spectral error in the flux and an error in the energy scale.} We adopt this simplification since we know those are the dominating systematics in the analysis. We also consider a 30\% error in the normalisation of the \textit{hep} flux --- whereas in the previous chapter, we considered a 200\%. Since \textit{hep} neutrinos barely contribute to the overall flux, this change has a negligible impact on our analysis. 

Next, we consider an improved version of Hyper-Kamiokande resulting from the upgrade of the photomultiplier detectors~\cite{Hyper-Kamiokande:2018ofw}. Mainly, we assume a factor two improvement of the efficiency and a reduction of the energy resolution in a factor 2 with respect to Super-Kamiokande IV~\cite{Nishimura:2020eyq}. Additionally, we consider a 4.5 MeV energy threshold and we also reduce to half the magnitude of the energy-uncorrelated systematics in the analysis. These improvements seem reasonable in light of preliminary sensitivity studies~\cite{Yano:2021usb}. When discussing our results, we will refer to this analysis as optimal. Notice that an improvement in the determination of the oscillation parameters would translate into more stringent limits on CPT violation. A comparison of the sensitivity of both analyses is presented in \reffig{fig:ch6-newHK}. For the simulation, we assume the best-fit values of the oscillation parameters according to the most recent preliminary results from the Super-Kamiokande Collaboration~\cite{yusuke_koshio_2022_6695966}. 

\begin{figure}
\includegraphics[width = 0.6\textwidth]{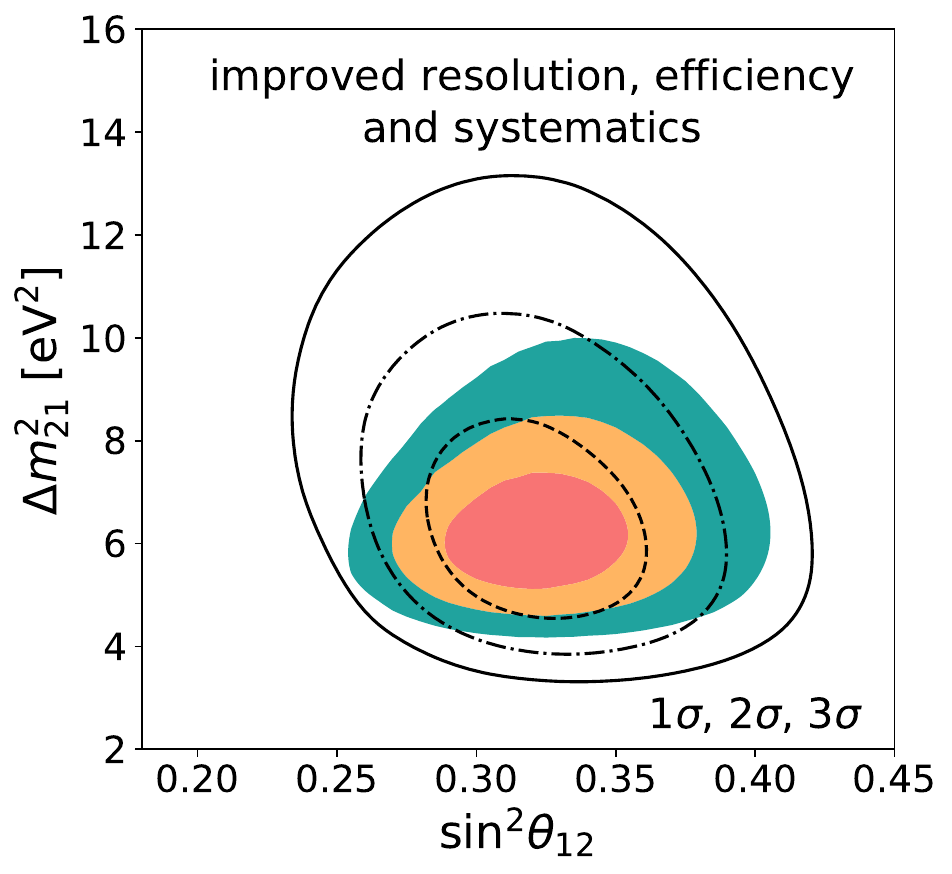}
\caption{Comparison between the 1$\sigma$, 2$\sigma$ and 3$\sigma$ allowed regions for Hyper-Kamiokande in the $\sin^2\theta_{12}$ - $\Delta m^2_{21}$ plane for the conservative and the optimal analysis --- in black lines and with colour regions, respectively. The best-fit values of the oscillation parameters are assumed to be $\sin^2\theta_{12} = 0.32$ and $\Delta m^2_{21} = 6.1 \times 10^{-5}\, \text{eV}^2$.\labfig{fig:ch6-newHK}}
\end{figure}

\subsection{Solar neutrinos at DUNE}
The capabilities of the next-generation multi-purpose neutrino experiment DUNE include the measurement of MeV neutrinos, among which are solar neutrinos. Using the charged-current reaction $\nu_e + ^{ 40}{\rm Ar} \rightarrow e^- + ^{ 40}{\rm K}$, DUNE will be sensitive to electron neutrinos from the Sun, with energies above 9 MeV.\footnote{A lower energy threshold could be possible if a significant background reduction is achieved.} 
Our sensitivity analysis considers 10 years of data and the projected full size of DUNE's far detector consisting of 40 kT of liquid argon. We consider an energy resolution~\cite{DUNE:2020ypp,Castiglioni:2020tsu}
\begin{align}
\frac{\sigma(E)}{E} = 0.2\, ,
\end{align} 
and for the cross-section, we consider the baseline configuration implemented in \texttt{SNOwGLoBES}~\cite{snowglobes}. We also include backgrounds from $^{222}$Rn and neutron capture~\cite{Pershey2020} with a 10\% uncertainty each and we take into account the efficiency linearly increasing from 30\% at 9 MeV to 60\% at 21 MeV~\cite{Ilic2020}. Likewise, the uncertainty in the flux normalisation for $^8$B and $hep$ neutrinos is the same as in the analysis carried out for Hyper-Kamiokande.

\begin{figure*}
\includegraphics[width = 0.74\paperwidth]{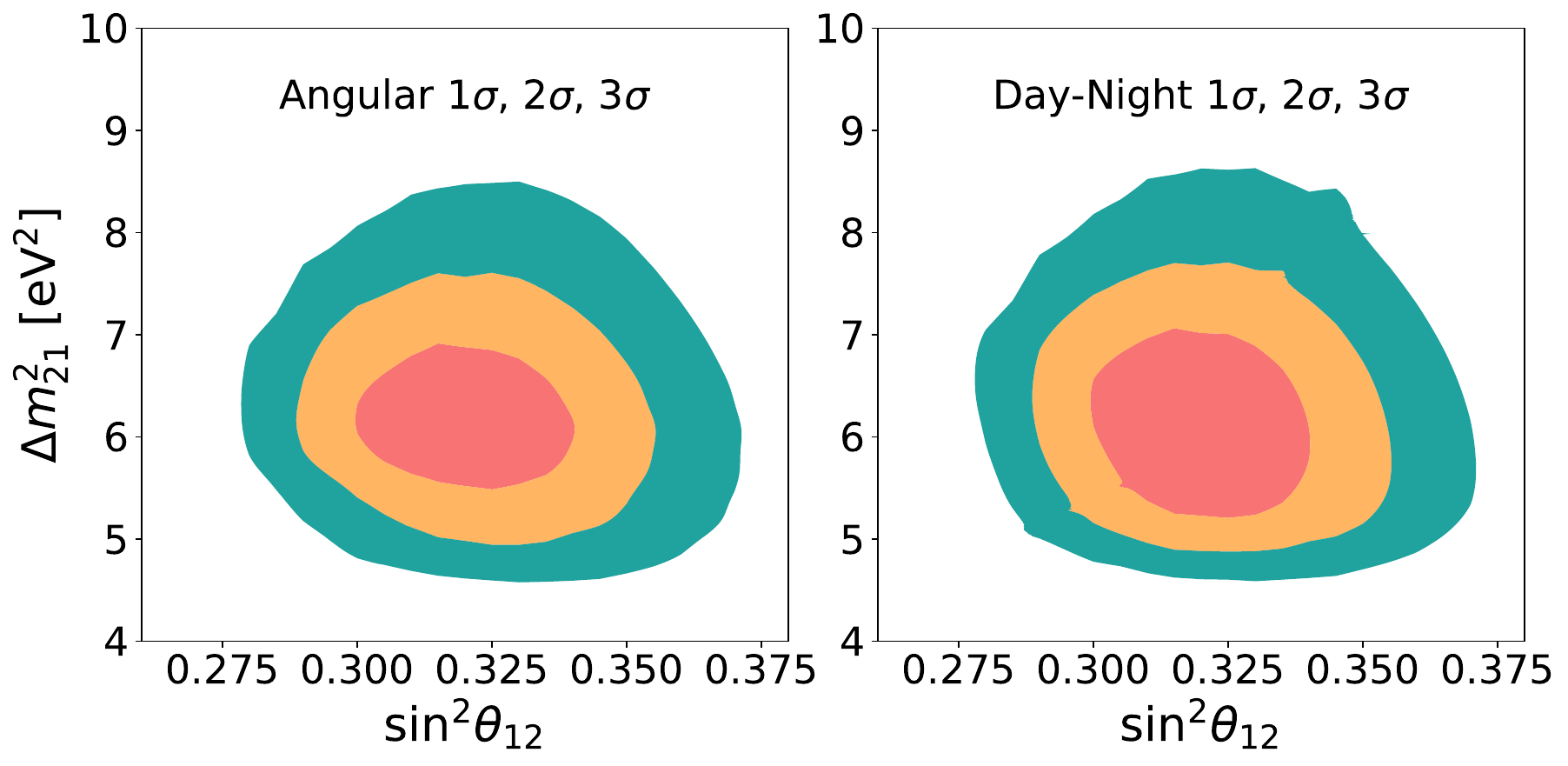}
\caption{Allowed regions at 1$\sigma$, 2$\sigma$ and 3$\sigma$ in the $\sin^2\theta_{21}$ - $\Delta m^2_{21}$ plane for DUNE. The best-fit value corresponds to $\sin^2\theta_{12} = 0.32$ and $\Delta m^2_{21} = 6.1 \times 10^{-5}\,\text{eV}^2$. The left panel shows the spectral analysis with angular binning and the right panel is the result of a day-night spectral fit. \labfig{fig:ch6-dn-angular}}
\end{figure*}

We consider two different approaches to the treatment of angular information. The first one uses only the spectral information for two angular bins, labelled as day and night, while in the second one, we separate the angular information for the night events into 10 bins of the same width. \reffig{fig:ch6-dn-angular} shows the results of our analysis where we have generated our mock data assuming the true values $\Delta m^2_{21} = 6.1 \times 10^{-5} \text{ eV}^2$ and $\sin^2\theta_{12} = 0.32\,$. One can see that there is a small improvement in the determination of the mass splitting when the angular binning is included in the analysis --- left panel --- with respect to the case in which only a day-night spectral fit is performed --- right panel. In light of this result, from this point on, we will only report the results obtained when angular binning for night events is incorporated and refer to it as conservative.

Next, we discuss the role of some potential improvements that could significantly boost DUNE's sensitivity to solar neutrinos. We studied the changes that would result from a reduction of the neutron background to a 10\% of its nominal value~\cite{Zhu:2018rwc,Capozzi:2018dat,Borkum:2023dsu}. We also considered the possibility of achieving perfect efficiency for energies above 9 MeV~\cite{Ankowski:2016lab,Moller:2018kpn,DUNE:2020zfm}. Finally, we also look at the possibility of improving the energy resolution to $\sigma (E)/E = 0.1$. In~\reffig{fig:ch6-improve}, we show how the allowed regions change if these improvements in the detector were achieved --- denoted by coloured regions. For comparison, we show in black lines the same contours for the conservative configuration. It is clear to see that, whereas the energy resolution does not change the expected sensitivity, a reduction of the neutron background level or an enhancement of the efficiency will allow a much more accurate determination of $\Delta m^2_{21}$. The bottom right panel of this figure shows the best-case scenario in which the three improvements are simultaneously achieved. The reason why the improvement is so significant is two folded. A reduction of the neutron background improves the signal-to-background ratio in the energy bins from 9 MeV to 11 MeV approximately. In addition, an overall increase in efficiency results in a reduced statistical error. Even if these proposals might seem a bit far-fetched, they illustrate what would be the optimal results that can be expected and indicate in which directions the improvement would be more significant. Hence, we will refer to this analysis as optimal.

\begin{figure*}[t!]
\includegraphics[width = 0.74\paperwidth]{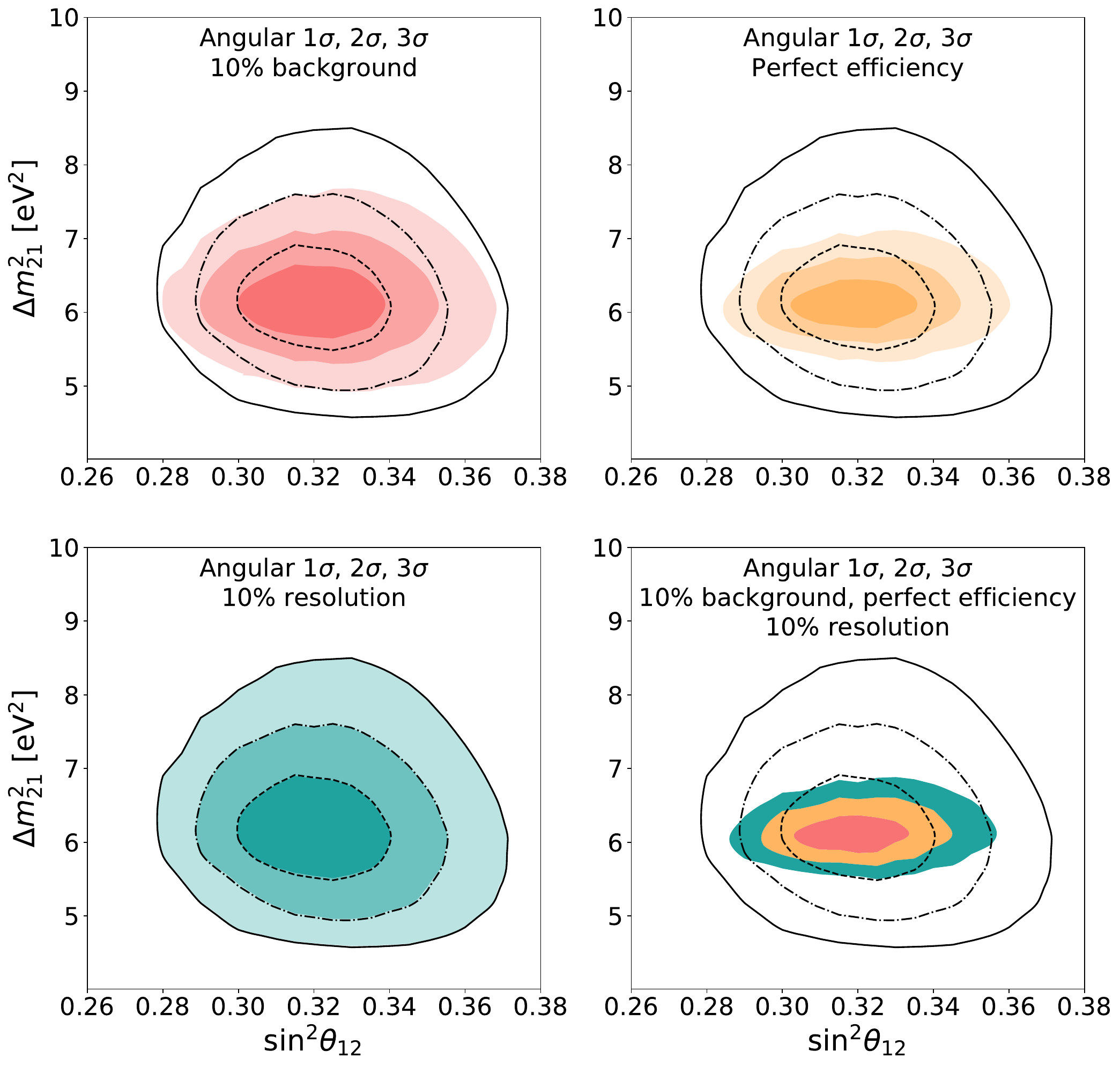}
\caption{Allowed 1$\sigma$, 2$\sigma$ and 3$\sigma$ regions in the $\sin^2\theta_{12}$ - $\Delta m^2_{21}$ plane. Black lines correspond to the conservative analysis. Coloured regions are the contours for the improvements in neutron background, efficiency, and resolution --- in the top left, top right and bottom left panels respectively. The bottom right panel corresponds to the comparison when the three improvements are considered simultaneously in an optimal scenario.\labfig{fig:ch6-improve}}
\end{figure*}

\subsection{CPT limits from the solar sector}
In the future, JUNO will provide a subpercent determination of both $\sin^2\bar{\theta}_{12}$ and $\Delta \overline{m}^2_{21}$ using antineutrinos. Parallely, a combined analysis of solar neutrinos at DUNE and Hyper-Kamiokande, exploiting the complementarities in the detection channels employed, will also reduce the uncertainty in the determination of $\sin^2\theta_{12}$ and $\Delta m^2_{21}$. Moreover, ongoing studies show that the accuracy achieved could be better than what was initially considered~\cite{Capozzi:2018dat}. Then, it is expected that the next-generation experiments JUNO, Hyper-Kamionkande and DUNE will significantly improve the current bounds on CPT violation shown in Equations~\ref{eq:ch6-current-cpt}. 

Let us first consider the possibility that the best-fit values for neutrinos and antineutrinos are
\begin{align}
&\sin^2\theta_{12} = 0.32  \quad \text{and} \quad \Delta m^2_{21} = 6.10 \times 10^{-5}\, \text{eV}^2\,, \nonumber \\
&\sin^2\bar{\theta}_{12} = 0.32 \quad \text{and} \quad \Delta \bar{m}^2_{21} = 7.53 \times 10 ^{-5}\, \text{eV}^2\, .
\label{eq:ch6-bf}
\end{align}
The choice of an equal mixing angle is motivated by the good agreement between KamLAND and solar neutrinos in this measurement. For the mass splitting, $\Delta m^2_{21}$ is chosen as the most recent best-fit point reported by Super-Kamiokande \cite{yasuhiro_nakajima_2020_4134680}, whereas we choose for $\Delta \overline{m}^2_{21}$ the best fit from KamLAND \cite{deSalas:2020pgw}.

\begin{figure*}
\includegraphics[width = 0.74\paperwidth]{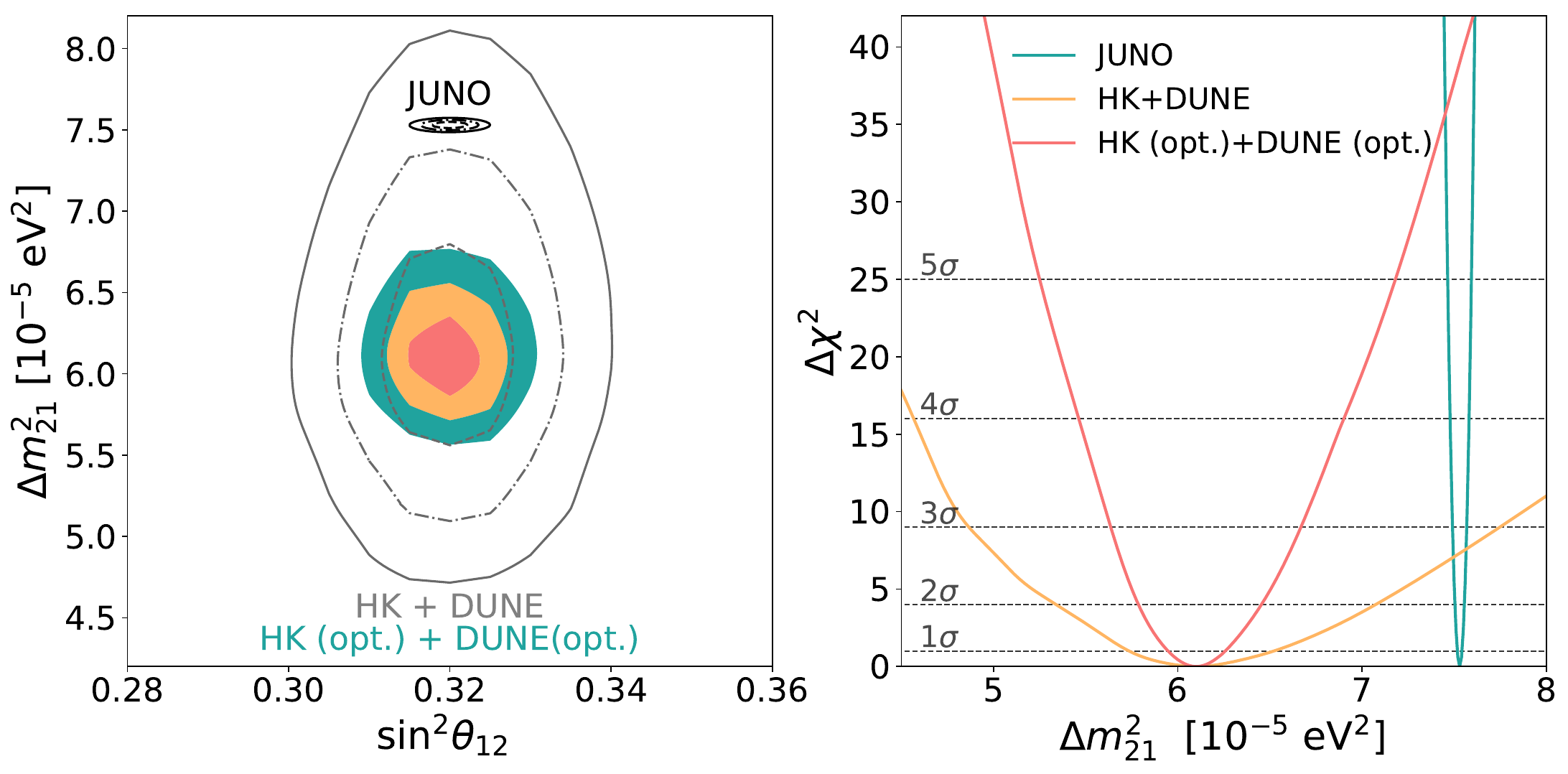}
\caption{Left panels shows the 1$\sigma$, 2$\sigma$ and 3$\sigma$ contours in the $\sin^2\theta_{12}$ - $\Delta m^2_{21}$. Results are shown in black for JUNO, and in grey for the combination of Hyper-Kamiokande (HK) and DUNE using its conservative configuration. Colour regions correspond to the results expected from the combination of the optimal versions of DUNE and Hyper-Kamiokande discussed in the text. The right panel describes the $\Delta\chi^2$ profiles for the mass splittings for JUNO and the two possible combinations of Hyper-Kamiokande and DUNE. Best-fit values for the oscillation parameters were chosen according to Equation~\ref{eq:ch6-bf}.\labfig{fig:ch6-cpt-61}}
\end{figure*}

The allowed regions at 1$\sigma$, 2$\sigma$ and 3$\sigma$ from JUNO are shown in black in the left panel of~\reffig{fig:ch6-cpt-61}. We also show the result from the combined analysis of Hyper-Kamiokande and DUNE. In grey, we depict the results for the conservative configuration in DUNE and Hyper-Kamiokande and with coloured contours we indicate the allowed regions when the improvements mentioned above are considered in both experiments. Notice, from the comparison with results in \reffig{fig:ch6-newHK} and \reffig{fig:ch6-improve}, that the combination improves significantly the measurement of the solar mixing angle. The reason is that the degeneracy between the normalisation of the $^8$B flux and $\sin^2 \theta_{12}$ is broken when using two different detection channels. Likewise, depending on the exact efficiency and background reduction achieved in the DUNE experiment and the reach of the improvements in \mbox{Hyper-Kamiokande,} the best-fit value of $\Delta m^2_{21}$ from JUNO could be excluded at more than 5$\sigma$. This can be seen in the right panel of~\reffig{fig:ch6-cpt-61}, where we show the one-dimensional $\chi^2$ profiles as a function of the mass splitting for one degree of freedom --- after profiling over the solar mixing angle. 

\begin{figure*}
\includegraphics[width = 0.74\paperwidth]{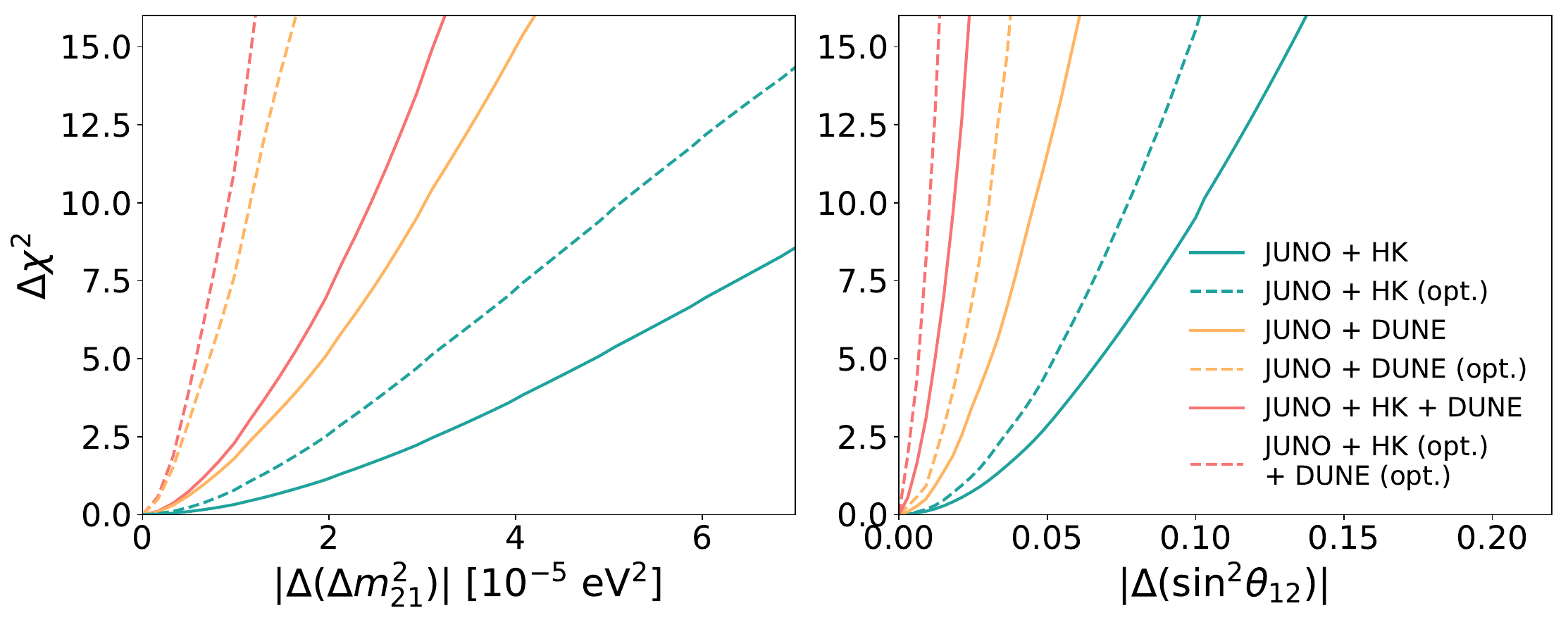}
\caption{The left and right panels show the $\Delta \chi^2$ profiles for the differences $|\Delta (\Delta m^2_{21})|$ and $|\Delta (\sin^2\theta_{12})|$ respectively, as defined in Equation~\ref{eq:cpt-diffs}. Results are shown for JUNO and Hyper-Kamiokande, JUNO and DUNE, and the combination of the three of them --- in green, orange and red, respectively. Dashed lines indicate that the optimal configuration of DUNE and Hyper-Kamiokande were considered, instead of the conservative one. \labfig{fig:ch6-cpt-limits}}
\end{figure*}

Alternatively, one can consider the case in which, future measurements both from neutrino and antineutrinos would perfectly agree in the best-fit value of both parameters. It is interesting to see how the limits on the differences,
\begin{align}
&|\Delta (\Delta m^2_{21})| = |\Delta m^2_{21} - \Delta \overline{m}^2_{21}|\, ,\nonumber\\
&|\Delta(\sin^2\theta_{12})| = |\sin^2\theta_{12} - \sin^2\bar{\theta}_{12}|\, ,
\label{eq:cpt-diffs}
\end{align}
would look like. Let us assume that the best-fit value of the oscillation parameters for neutrinos and antineutrinos is
\begin{gather}
\sin^2\bar{\theta}_{12} = 0.32 \quad \text{and} \quad \Delta \bar{m}^2_{21} = 7.53 \times 10 ^{-5}\, \text{eV}^2\, .
\end{gather}
In~\reffig{fig:ch6-cpt-limits}, we depict the $\chi^2 $ profiles resulting from comparing JUNO and different mock datasets from DUNE and \mbox{Hyper-Kamiokande} --- considering the conservative and the optimal configurations --- along with the corresponding combinations. From this analysis, one can extract the expected bounds on CPT violation in the neutrino sector, which are summarised in \reftab{ch6-sum-cpt}. 

\begin{table}
\renewcommand*{\arraystretch}{1.2}
\centering
\begin{tabular}{lccc}
\toprule[0.25ex]
& \multicolumn{3}{c}{$|\Delta m^2_{21} - \Delta \overline{m}^2_{21}|$ [10$^{-5}$ eV$^2$]} \\[2.4mm]
Experiment & 90\% C.L. & 99\% C.L. & 3$\sigma$ \\ \midrule
JUNO + HK & 3.3 & 5.9 & 7.3 \\
JUNO + HK (opt.) & 2.0 & 3.8 & 4.7 \\
JUNO + DUNE & 1.3 & 2.3 & 2.8 \\
JUNO + DUNE (opt.) & 0.5 & 0.9 & 1.1 \\
JUNO + HK + DUNE & 1.1 & 1.9 & 2.3 \\
JUNO + HK (opt.) + DUNE (opt.) & 0.4 & 0.7 & 0.8 \\
\bottomrule[0.25ex]
 & & & \\
\toprule[0.25ex]
& \multicolumn{3}{c}{$|\sin^2 \theta_{12} - \sin^2 \bar{\theta}_{12}|$}  \\[2.4mm]
Experiment & 90\% C.L. & 99\% C.L. & 3$\sigma$ \\ \midrule
JUNO + HK & 0.049 & 0.080 & 0.097 \\
JUNO + DUNE &   0.022 & 	0.037 & 0.043\\
JUNO + DUNE (opt.) & 0.015 	& 0.024 &  0.029\\
JUNO + HK + DUNE &  0.008 &	0.014 & 0.019\\
JUNO + HK + DUNE (opt.) & 0.007 & 0.012 &  0.014\\
\bottomrule[0.25ex]
\end{tabular}
\caption{Projected sensitivity to the difference in the solar mass splitting and solar mixing angle measured from neutrinos and antineutrinos at 90\%, \mbox{99\% C.L.} and 3$\sigma$ for different sets of solar neutrino data. Best fit values were chosen to be $\Delta m^2_{21} = 7.53 \times 10^{-5}\, \text{eV}^2$ and $\sin^2\theta_{12} = 0.32$. }
\label{tab:ch6-sum-cpt}
\end{table}

Let us remark that the successful establishment of the three-neutrino oscillation picture has strongly relied on combining datasets --- including combining data from neutrinos and antineutrinos. However, as neutrino physics is entering a precision era, it is possible to reach great precision in the measurement of the oscillation parameters without analysing jointly neutrino and antineutrino data. This provides a very useful test of CPT symmetry. Here we have shown that, in the future, JUNO, Hyper-Kamiokande and DUNE will improve the bound from the solar sector on CPT violation in a factor $\sim 2 - 6$ for $|\Delta (\Delta m^2_{21})|$ and in a factor $\sim 10-14$ for $|\Delta (\sin^2\theta_{12})|$. The exact improvement on $|\Delta (\Delta m^2_{21})|$ will strongly depend on experimental features of the DUNE experiment such as its efficiency in the MeV range and the background levels eventually achieved in the detectors and on the technical improvements in Hyper-Kamiokande with respect to its predecessor Super-Kamiokande.

\section{Other tests on CPT symmetry}
\label{sec:ch6-othercpt}
Historically, tests of CPT have been associated with the system of neutral kaons. The bound from the difference between the masses at 90\% C.L., as quoted by the Particle Data Group PDG~\cite{ParticleDataGroup:2022pth}, reads
\begin{align}
2\frac{m_{\text{K}^0} - m_{\overline{\text{K}}^0}}{m_{\text{K}^0} +  m_{\overline{\text{K}}^0}} < 6 \times 10 ^{-19} \, .
\label{eq:ch6-kaon}
\end{align}

In light of this result, it is clear that such relative precision is unachievable in neutrino experiments. However, two important points must be made at this point. The first one, which was already mentioned previously, is the fact that testing CPT is not the same as putting bounds on CPT violation. Signatures of the breakdown of CPT are not unique but model-dependent. For instance, the difference between the masses of particles and antiparticles is a CPT-violating observable but it is not necessarily the one in which the breaking manifests. Moreover, it is not guaranteed that the hypothetical breaking of CPT would equally affect all sectors of the Standard Model. Hence, complementary tests are necessary. The second aspect to remark regarding the choice of the neutral kaon system as a reference is that kaons are not elementary particles. Besides that, the reference scale in the denominator of Equation \ref{eq:ch6-kaon}  --- the average mass of the neutral kaons --- is rather arbitrary and not connected to any model of CPT violation. Actually, this limit can be seen as a test of CPT in QCD instead. Let us consider a model of CPT violation in which particles and antiparticles would have different masses. Then, for neutral kaons the parameter that enters the Lagrangian is the square of the mass. In that case, the limit would read
\begin{align}
|m^2_{\text{K}^0} - m^2_{\bar{\text{K}}^0}| < 0.28 \,\text{eV}^2\,.
\end{align}
In this context, the relevance of the limits from the neutrino sector becomes evident.

Other tests of CPT exist from the comparison of the decay widths of the neutral kaons, the difference in the charge and masses between electrons and positrons, and the difference between the charge-to-mass ratio of protons and antiprotons. Besides that, precision measurement of the mass of hydrogen and antihydrogen and some other relevant spectroscopic measurements on hydrogen and antihydrogen --- two-photon 1s to 2s transition, the Lamb shift 2s-2p and the ground-state hyperfine structure --- also provide interesting tests of CPT symmetry. Some of these limits are summarised in~\reftab{ch6-review-cpt}. 

\begin{table}
\renewcommand*{\arraystretch}{1.2}
\centering
\begin{tabular}{ccc}
\toprule[0.25ex]
System  & Limit & Reference \\\midrule
Neutral kaons & $2\frac{m_{\text{K}^0} - m_{\overline{\text{K}}^0}}{m_{\text{K}^0} +  m_{\overline{\text{K}}^0}} < 0.6 \times 10 ^{-18} $ at 90\% C.L. & \cite{ParticleDataGroup:2022pth} \\[3.2mm]
 & $2\frac{\Gamma_{\text{K}^0} - \Gamma_{\overline{\text{K}}^0}}{\Gamma_{\text{K}^0} +  \Gamma_{\overline{\text{K}}^0}} = (0.8\pm 0.8) \times 10 ^{-17} $ & \cite{ParticleDataGroup:2022pth}\\[3.2mm]
$e^- \,  - \, e^+$ & $2\frac{m_{e^+} - m_{e^-}}{m_{e^+} +  m_{e^-}} < 8 \times 10 ^{-19} $ at 90\% C.L. & \cite{ParticleDataGroup:2022pth} \\[3.2mm]
$p\, - \, \bar{p}$ & $\frac{|q_p|/m_p}{|q_{\bar{p}}|/m_{\bar{p}}} - 1 = (0.3\pm 1.6) \times 10^{-16}$ & \cite{BASE:2022yvh} \\[1.6mm]
\bottomrule[0.25ex]
\end{tabular}
\caption{Summary of some CPT tests using neutral kaons, electrons and protons. }
\label{tab:ch6-review-cpt}
\end{table}

\section{Parting thoughts}
\label{sec:ch6-thoughts}
In our quest to gain a better understanding of nature, the origin of neutrino masses is not the only enthralling puzzle. The validity of the underlying assumptions employed in our description of nature in terms of local relativistic quantum field theories is another matter of great interest. The precision measurements that are being achieved in neutrino physics could also shed light on this topic. In particular, regarding tests of CPT invariance, next-generation neutrino experiments JUNO, DUNE, and Hyper-Kamiokande could provide fascinating results. These tests, when compared with other existing results in the literature, might seem not competitive. However, one should remember that the hypothetical breaking of CPT might not manifest in the same way in all the sectors and that, depending on the origin of the breaking, the signatures might differ. Hence, any piece of information is useful to get a better understanding of the overall picture. A particularly motivating aspect for CPT tests in the neutrino sector is the fact that they already provide evidence of physics beyond the Standard Model. If the new physics scale required to explain their masses is very high --- as predicted in many scenarios --- then, it might be close to the scale of CPT breaking. Thus, neutrinos will be providing a window to the study of new physics.

As a last remark, notice that the bounds here presented rely on the \mbox{non-observation} of a difference between the values of the oscillation parameters as measured using neutrinos and antineutrinos. In this analysis, matter effects from Standard Model interactions are taken into account. However, as it was shown in the previous chapter, non-standard interactions with matter fields can induce a shift between the measured and the true value of these oscillation parameters --- see Equations \ref{eq:ch5-angle-nsi} and \ref{eq:ch5-mass-nsi}. Hence, there is a significant similarity between the signatures from NSIs and a more fundamental violation of CPT, coming, for instance, from Lorentz invariance violation~\cite{Diaz:2015dxa,Barenboim:2018lpo,MartinezMirave:2022lhj,Barenboim:2023krl}. The inclusion of additional datasets --- in particular the inclusion of data from neutrino scattering, which is also sensitive to neutrino non-standard interactions --- can disentangle both effects.

\chapter{Non-standard interactions and neutrino decoupling}
%\addcontentsline{toc}{chapter}{Non-standard interactions and neutrino decoupling} 
\labch{ch7-nsicosmo}
Primordial neutrinos have played a very relevant role in the evolution of the universe. They contribute to the total energy density and pressure of the universe and hence are partially responsible for its expansion rate. Moreover, they were also key ingredients in the nucleosynthesis of primordial elements, often referred to as Big Bang Nucleosynthesis. Likewise, they left their imprint in the cosmic microwave background and prevented the growth of small-scale structures due to their relativistic character. From several cosmological observations such as the matter power spectrum, the anisotropies in the CMB or the relative abundances of elements, one can infer neutrino properties such as their masses --- see \refch{ch3-fit}. 

This chapter addresses the role of neutrino interactions in the process of decoupling. In particular, it focuses on how non-standard neutrino interactions with electrons alter the contribution of neutrinos to the cosmological radiation energy density. In Section \ref{sec:ch7-decoupling}, we introduce the standard picture of neutrino decoupling. Next, Section \ref{sec:ch7-nsie} reviews the status of non-standard interactions (NSIs) with electrons and the existing limits. Then, Section \ref{sec:ch7-nsi-decoupling} presents the role of NSIs in neutrino decoupling and the results obtained. Finally, Section \ref{sec:ch7-conclude} provides some concluding comments on our findings.

%%%%%%%%%%%%%%%%%%%%%%%%%%%%%%%%%%%%%%
%%%%%%%%%%%%%%%%%%%%%%%%%%%%%%%%%%%%%%
\section{Neutrino decoupling}
\label{sec:ch7-decoupling}
%%%%%%%%%%%%%%%%%%%%%%%%%%%%%%%%%%%%%%
%%%%%%%%%%%%%%%%%%%%%%%%%%%%%%%%%%%%%%

%%%%%%%%%%%%%%%%%%%%%%%%%%%%%%%%%%%%%%%%%%%%%%%%%%%%%%%%%%%%%%%%
\subsection{Neutrino decoupling in the instantaneous limit}
%%%%%%%%%%%%%%%%%%%%%%%%%%%%%%%%%%%%%%%%%%%%%%%%%%%%%%%%%%%%%%%%

As a first approximation, one can state that cosmological species are in thermodynamical equilibrium if there is at least one interaction per Hubble time --- in other words, as long as the expansion of the universe does not prevent the species in the fluid from continuing to interact. One can estimate the temperature of neutrino decoupling as the temperature at which the interaction rate of neutrinos with the other species in the cosmic plasma, $\Gamma$, equals the Hubble rate, $H$.

The interaction rate can be estimated as the thermal average of the cross section times the number density and the particle velocity --- i.e. $\Gamma = n\langle v \sigma \rangle$. At temperatures well below the mass of the gauge bosons, $W$ and $Z$, neutrino interaction cross sections are
\begin{align}
\sigma \sim G^2_F T^2\, .
\label{eq:ch7-xsec}
\end{align}
Since the number density for relativistic species is proportional to $T^3$, then the interaction rate is
\begin{align}
\Gamma \sim G^2_F T^5\, .
\label{eq:ch7-gamma}
\end{align}
Regarding the Hubble rate, during radiation domination, it evolves with temperature as
\begin{align}
H(T)^2 = \frac{8\pi G}{3} g_* \frac{ \pi^2}{30} T^4 \sim g_* \frac{T^4}{m_{Pl}^2} \, ,
\end{align}
and it depends on the effective degrees of freedom $g_*$ --- defined in Equation~\ref{eq:ch1-edof}.
Consequently, and according to the argument that neutrinos stopped being in thermal contact with the plasma --- which at that time consisted of electrons, positrons and photons --- when $\Gamma \sim H$,
\begin{align*}
\frac{\Gamma}{H} \sim \frac{G^2_F T^5 m_{Pl}}{\sqrt{g_*} T^2} \sim \left(\frac{T}{g^{1/6}_* \text{MeV}}  \right)^3\, .
\end{align*}
the temperature of neutrino decoupling is estimated to be $\mathcal{O}$(MeV). From that moment on, their distribution evolved as a frozen Fermi-Dirac distribution with a temperature that decreased as the universe expanded,
\begin{align}
T_\nu = T_{\nu, \text{ decoupling}}\frac{a_{\text{dcoupling}}}{a}  \, .
\end{align}

As the universe expanded, the cosmic plasma cooled down. Once the temperature dropped below the electron mass, the process of electron-positron pair production stopped occurring. Consequently, only electron-positron annihilations happened, resulting in an entropy transfer to the photon bath and an increase of their temperature in a sort of reheating. Since at this point neutrinos had already decoupled from the plasma, no entropy was transferred to them. This leads to a difference between the temperature of photons and neutrinos.

In a universe expanding adiabatically, the entropy per comoving volume is conserved. Let us denote the situation right after neutrino decoupling by subindex 1 and the situation after electron-positron annihilation by subindex 2. Then, the condition of entropy conservation in a comoving volume reads
\begin{align}
s(a_1)a_1^3 =s(a_2)a_2^3\, ,
\label{eq:ch7-entropy}
\end{align}
where the entropy of the fluid of electrons, positrons and photons is 
\begin{align}
s_{e^\pm, \, \gamma}(a_i)a^3_i \propto \left(\frac{T_{i, \, \gamma}}{T_{i, \, \nu}}\right)^3 g_{e^\pm, \, \gamma} (T_\gamma)\, .
\end{align}
Right after neutrino decoupling, the temperature of neutrino and photons were equal, $T_\gamma = T_\nu$, and larger than the electron mass, $T_\gamma \gg m_e$,
\begin{align}
g_{e^\pm, \, \gamma} = 2 + 4\times\frac{7}{8} = \frac{11}{2}
\end{align}
whereas after electron-positron annihilation,$T_ \gamma \ll m_e$ and hence the effective degrees of freedom $g_{e^\pm, \gamma} = 2$. Then, requiring that the entropy per comoving volume is conserved, one finds the relation between the temperature of relic neutrinos and photons,
\begin{align}
\frac{T_\nu}{T_\gamma} = \left(\frac{4}{11}\right)^{1/3}\,.
\end{align}

The energy density of all the species that contribute as radiation to the energy budget of the universe is generally parametrised in terms of the \textit{effective number of relativistic species} --- also dubbed \textit{effective number of neutrinos}, $N_{\text{eff}}$. At temperatures below the electron mass, the cosmological radiation energy density only contains contributions from photons and neutrinos
\begin{align}
\rho_{\text{rad}} = \rho_\gamma + \rho_\nu &= \rho_\gamma \left[ 1 + \frac{7}{8}\left(\frac{T_\nu}{T_\gamma}\right)^4 N_{\text{eff}}\right]\nonumber\\ &= \rho_\gamma \left[ 1 + \frac{7}{8}\left(\frac{4}{11}\right)^{4/3} N_{\text{eff}}\right]\, .
\end{align}

%%%%%%%%%%%%%%%%%%%%%%%%%%%%%%%%%%%%%%%%%%%%%%%%%%%%%%%%%%%%%%%%
\subsection{Neutrino decoupling beyond the instantaneous limit}
%%%%%%%%%%%%%%%%%%%%%%%%%%%%%%%%%%%%%%%%%%%%%%%%%%%%%%%%%%%%%%%%
The calculation assuming that both neutrino decoupling and electron-positron annihilations are instantaneous processes allows us to have a relatively accurate understanding of the picture. Nevertheless, a precise calculation of the value $N_{\text{eff}}$ needs to incorporate several effects such as distortions in the distribution functions from collisional processes, finite temperature Quantum Electrodynamics (QED) effects, the differences between the three neutrino flavours and the role of flavour oscillations.

To account for flavour oscillations, we study the evolution of the neutrino density matrix,
\begin{align}
\varrho(t,p) = \begin{pmatrix}
\varrho_{ee} & \varrho_{e\mu} & \varrho_{e\tau} \\ \varrho_{\mu e} & \varrho_{\mu \mu} & \varrho_{\mu \tau}  \\ \varrho_{\tau e} & \varrho_{\tau \mu} & \varrho_{\tau \tau} 
\end{pmatrix} = \begin{pmatrix}
f_{\nu_e} & a_1 + i a_2 & b_1 + ib_2 \\ a_1 -i a_2 & f_{\nu_\mu} & c_1  + i c_2 \\ b_1 - i b_2 & c_1 - i c_2 & f_{\nu_\tau} 
\end{pmatrix}\, .
\label{eq:ch7-densitymatrix}
\end{align}
Note that in the expression above, the diagonal terms denote the occupation number of each flavour, $f_{\nu_e}$, $f_{\nu_\mu}$, and $f_{\nu_\tau}$. In addition, we have parameterised the off-diagonal terms of the density matrix with 6 real parameters  --- $a_i$, $b_i$, and $c_i$, with $i = 1,2$. These terms are non-zero in the presence of mixing and account for the coherence of the system~\cite{deSalas:2016ztq}.

In our calculations, we neglect a potential asymmetry between neutrinos and antineutrinos and hence, we compute the evolution of the density matrix for neutrinos and assume that it is the same one for antineutrinos --- i.e. $\varrho = \bar{\varrho}$.

To compute the evolution of the density matrix, we define the following set of comoving variables,
\begin{align}
x = m_e a\, , \quad y = p a\, , \quad \text{and} \quad z = a T_\gamma\, ,
\label{eq:ch7-variables}
\end{align}
which are convenient for numerical calculations. Note that $x$ is a measure of the cosmic expansion, $y$ is the comoving momentum, and $z$ is related to the enhancement of the photon temperature with respect to the neutrino temperature, which evolves as $a^{-1}$. Then, the evolution of the neutrino density matrix is given by the system of Boltzmann equations,
\begin{align}
i \frac{\text{d}\varrho(t,p)}{\text{d}t} = [\mathbb{H}, \varrho(t,p)] +i \mathcal{I}(\varrho(t,p))\, ,
\end{align}
where the square brackets denote the commutator --- so that $[\mathbb{H}, \varrho] = \mathbb{H}\varrho - \varrho\mathbb{H}$ --- and the collisional integrals $\mathcal{I}(\varrho(t,p))$ encode all the scattering of the system involving electrons and positrons. For massive neutrinos propagating in an expanding universe consisting of a gas of charged leptons and ultrarelativistic neutrinos, the evolution equation reads
\begin{align}
&i \frac{\text{d}\varrho(t,p)}{\text{d}t} = \left[ \frac{\mathbb{M}_F}{2p} -\frac{2\sqrt{2}G_{\text{F}} p}{3m^2_W}\left(\mathbb{E}_l + \mathbb{P}_l + \frac{4}{3}\mathbb{E}_\nu\right), \varrho(t,p)\right] \nonumber\\ & \hspace{7cm}+ \mathcal{I}(\varrho(t,p))\, .
\end{align}
The expression above is normally presented in terms of the comoving variables introduced above so that it reads
\begin{align}
&i \frac{\text{d}\varrho(t,p)}{\text{d}t} = \nonumber \\ &\hspace{0.5cm}\sqrt{\frac{3 m^2_{Pl}}{8\pi\rho}} \left\lbrace \frac{x^2}{m^3_e}  \left[ \frac{\mathbb{M}_F}{2y} -\frac{2\sqrt{2}G_{\text{F}} y m^6_e}{3m^2_W x^6} \right. \right. \left.\left(\mathbb{E}_l + \mathbb{P}_l + \frac{4}{3}\mathbb{E}_\nu\right), \varrho(t,p)\right]  \nonumber \\ &\hspace{7cm}\left. + i\frac{m^3_e}{x^4}\mathcal{I}(\varrho(t,p)) \right\rbrace\, .
\end{align}
Here, $m_{Pl}$ is the Planck mass, $m_W$ and $m_Z$ denote the mass of the gauge bosons $W^\pm$ and $Z$, $G_{\text{F}}$ is the Fermi constant, and $\rho$ is the total energy density of the universe. The first term in the commutator is related to neutrino oscillations in vacuum,
\begin{align}
\mathbb{M}_F = U \begin{pmatrix}
0 & 0 & 0 \\ 0 & \Delta m^2_{21} & 0 \\ 0 & 0 & \Delta m^2_{31} 
\end{pmatrix} U ^\dagger\, ,
\end{align}
which depends on the mass splittings $\Delta m^2_{21}$ and $\Delta m^2_{31}$ and on the lepton mixing matrix $U$. Note that the second term in the commutator accounts for neutrino interactions in matter and neutrino self-interactions. In this expression, $\mathbb{E}_l$ and $\mathbb{P}_l$ denote the energy density and pressure of charged leptons in the medium and $\mathbb{E}_\nu$ is the energy density of neutrinos.

At temperatures of $\mathcal{O}$(MeV), when neutrino decoupling takes place, the contributions of muon and tau lepton are Boltzmann suppressed. Thus, we only consider the energy density and pressure of electrons and positrons assuming an ideal gas, i.e.
\begin{align}
\mathbb{E}_l = \text{diag} (\rho_e, \, 0,\, 0) \quad \text{and} \quad \mathbb{P}_l = \text{diag} (P_e, \, 0,\, 0) \, ,
\end{align}
where
\begin{align}
\rho_e = \frac{1}{\pi^2}\int \text{d}y y^2 \sqrt{y^2 + a^2 m_e^2} f_{\nu_e}(y)\, ,  \\
P_e = \frac{1}{3\pi^2}\int \text{d}y y^2 \frac{y^2}{\sqrt{y^2 + a^2 m_e^2}} f_{\nu_e}(y)\, .
\end{align}
The refractive term arising for neutrino-neutrino interactions in the medium is the equivalent of the term presented above when, instead of charged leptons, one considers a gas of ultrarelativistic neutrinos, 
\begin{align}
\mathbb{E}_\nu = \frac{1}{\pi^2}\int \text{d}y y^3 \varrho(y)\, .
\end{align}

Regarding the collisional integrals, they can be separated into two sets: the ones involving two neutrinos and two electrons distributed between the initial and final states and neutrino-neutrino scattering.

Following the notation and procedure for integral reduction in \cite{Bennett:2020zkv}, the neutrino-electron collision integral is separated in the terms accounting for scattering and annihilations,
\begin{align}
    \mathcal{I}_{\nu  e}[\varrho (x,y)] = \frac{G^2_F}{(2\pi)^3y^2}\lbrace I_{\nu e}^\text{ scatt} [\varrho(x,y)] + I_{\nu e}^{ann}[\varrho(x,y)]\rbrace\, ,
\end{align}
using the comoving variables defined in Equation \ref{eq:ch7-variables}. 

The term accounting for scattering reads
\begin{align}
    &I_{\nu e}^\text{ scatt} [\varrho(x,y)] = \nonumber \\ &\hspace{0.5cm} \int \text{d}y_2 \text{d}y_3 \hspace{-7pt} \sum_{a,b=L,R} \hspace{-7pt} A_{ab}(x,y,y_2,y_3)  F_\text{sc}^{ab}\left(\varrho^{(1)}, f_e^{(2)}, \varrho^{(3)}, f_e^{(4)}\right)\, ,
\end{align}
whereas the one related to annihilations is
\begin{align}
    &I_{\nu e}^\text{ann} [\varrho(x,y)] =  \nonumber \\ &\hspace{0.5cm}\int \text{d}y_2 \text{d}y_3\hspace{-7pt} \sum_{a,b=L,R}\hspace{-7pt} B_{ab}(x,y,y_2,y_3) F^\text{ann}_{ab}\left(\varrho^{(1)}, \varrho^{(2)}, f_e^{(3)}, f_e^{(4)}\right)\, .
\end{align}

The exact dependence on the comoving variables of the scattering kernels, $A_{ab}$ and $B_{ab}$, can be found in~\cite{Bennett:2020zkv}. Their expressions do not depend on the basis in which the evolution of the density matrix is computed nor on the strength and structure of the interactions.\footnote{Note that, in this chapter, we work in the flavour basis whereas, in the next chapter, we do so in the mass basis.}

The fact that we are working in the flavour basis only manifests in the \mbox{phase-space} factors $F^\text{ann}_{ab}$ and $F^\text{scatt}_{ab}$, which depend on the matrices $G_L$ and $G_R$ defined as,
\begin{align}
G_L = \text{diag} (\tilde{g}_L,\, g_L, \, g_L)
\label{eq:ch7-gl}
\end{align}
and
\begin{align}
G_R = \text{diag}(g_R, \, g_R, \, g_R) \, ,
\label{eq:ch7-gr}
\end{align}
where $g_L = \sin^2\theta_W - 1/2 $, $\tilde{g}_L = 1 + g_L$, $g_R = \sin^2 \theta_W$ and $\theta_W$ is the weak mixing angle. 

The expressions of the factors accounting for the phase-space of scattering and annihilation processes read
\begin{align}
&F^\text{scatt}_{ab}\left(\varrho^{(1)}, f_e^{(2)}, \varrho^{(3)}, f_e^{(4)}\right) \nonumber\\
& \hspace{0.2cm}= f_e^{(4)}(1-f_e^{(2)})\left[G_a\varrho^{(3)}G_b(1-\varrho^{(1)})+(1-\varrho^{(1)})G_b\varrho^{(3)}G_a\right]
\nonumber\\
&\hspace{0.2cm}-
f_e^{(2)}(1-f_e^{(4)})\left[\varrho^{(1)}G_b(1-\varrho^{(3)})G_a+G_a(1-\varrho^{(3)})G_b\varrho^{(1)}\right]\, ,
\label{eq:ch7-F_ab_sc}
\end{align}
and
\begin{align}
&F^\text{ann}_{ab}\left(\varrho^{(1)}, \varrho^{(2)}, f_e^{(3)}, f_e^{(4)}\right)
\nonumber\\
&\hspace{0.2cm}=\, f_e^{(3)}f_e^{(4)}\left[G_a(1-\varrho^{(2)})G_b(1-\varrho^{(1)})+(1-\varrho^{(1)})G_b(1-\varrho^{(2)})G_a\right]
\nonumber\\
&\hspace{0.2cm}-
(1-f_e^{(3)})(1-f_e^{(4)})\left[G_a\varrho^{(2)}G_b\varrho^{(1)}+\varrho^{(1)}G_b\varrho^{(2)}G_a\right],
\label{eq:ch7-F_ab_ann}
\end{align}

where $a,b = L,R$. These expressions depend both on the electron momentum distribution function, $f_e$, and on the density matrix $\varrho^{(i)}= \varrho(y_i)$, where $y_i$ is the comoving momentum of particle $i$. 

As for neutrino-neutrino interactions, we will compute them as in \cite{Bennett:2020zkv}. No further discussion on their contribution is given since these processes do not play a significant role in the scenarios that we consider.

At present, the effective number of neutrinos is constrained to be $N_{\text{eff}} = 2.99^{+0.34}_{-0.33}$ at 95\%~C.L. as determined from measurements of the cosmic microwave background anisotropies by Planck, together with other cosmological data \cite{Planck:2018vyg}. This measurement is in good agreement with the most accurate theoretical calculations, which find a value of $N_{\text{eff}} = 3.0440 \pm 0.0002 $~\cite{Akita:2020szl,Froustey:2020mcq,Bennett:2020zkv}.

%%%%%%%%%%%%%%%%%%%%%%%%%%%%%%%%%%%%%%%%%%%%%%%%%%%%%%
%%%%%%%%%%%%%%%%%%%%%%%%%%%%%%%%%%%%%%%%%%%%%%%%%%%%%%
\section{Non-standard interactions with electrons}
\label{sec:ch7-nsie}
%%%%%%%%%%%%%%%%%%%%%%%%%%%%%%%%%%%%%%%%%%%%%%%%%%%%%%
%%%%%%%%%%%%%%%%%%%%%%%%%%%%%%%%%%%%%%%%%%%%%%%%%%%%%%
Current limits on neutrino non-standard interactions with electrons result from the detailed study of neutrino oscillations and scattering measurements in terrestrial experiments. Besides, several observables from the Large \mbox{Electron-Positron} Collider (LEP) provide complementary information.

Following the notation introduced in the previous chapter, we define \mbox{neutral-current} \mbox{non-standard} neutrino interactions with electrons through the effective Lagrangian in Equation \ref{eq:ch5-NCNSI}. However, for simplicity, we will denote the NSI parameters by $\varepsilon^X_{\alpha\beta}$, where it is implicit that we are only considering NSIs with electrons.

Neutrino non-standard interactions with electrons impact neutrino oscillations since they modify the effective potential arising from coherent elastic forward scattering. The non-observation of deviations from the standard oscillation picture is therefore translated into limits on the NSI parameters. It is known that neutrino propagation is sensitive solely to the vector component of the interaction. Nonetheless, NSIs may also modify the cross-section of the detection processes involved in the experiment --- depending on the type of NSI considered and the detection channel employed. In that case, neutrino oscillation experiments are sensitive to both sets of chiral NSI parameters, $\varepsilon_{\alpha \beta}^{L}$ and $\varepsilon_{\alpha\beta}^{R}$. For instance, the parameter space involving non-universal and/or flavour-changing NSIs can be explored using data from solar experiments and \mbox{KamLAND~\cite{Miranda:2004nb,Bolanos:2008km,Escrihuela:2010zz,Agarwalla:2012wf, Khan:2017oxw,Coloma:2022umy}.} Alternatively, long-baseline and atmospheric experiments --- which are also sensitive to NSIs --- set constraints in the vector parameters $|\varepsilon_{\tau \tau}^{V} - \varepsilon_{\mu \mu }^{V}|$ and $|\varepsilon_{\mu \tau}^{V}|$ \cite{Gonzalez-Garcia:2011vlg, Salvado:2016uqu, Demidov:2019okm}. 

\begin{table*}
\centering
\renewcommand{\arraystretch}{1.2}
\begin{tabular}{cc}
\toprule[0.25ex]
Parameter and $90\%$ C.L range & Experimental probe \\ \midrule
  -0.021  $ <  \varepsilon^L_{ee} < $  0.052 & Neutrino oscillations \cite{Bolanos:2008km} \\[2.4mm]
-0.07  $< \varepsilon^R_{ee} <$  0.08   & Scattering \cite{TEXONO:2010tnr} \\
-0.23 $ <\varepsilon^R_{ee} <$ 0.07 & Neutrino oscillations \cite{Coloma:2022umy} \\[2.4mm]
-0.03 $ < \varepsilon^L_{\mu\mu}$, $\varepsilon^R_{\mu\mu} < $ 0.03 & Scattering and accelerator data \cite{Barranco:2007ej}\\ [2.4mm]
-0.12  $< \varepsilon^L_{\tau\tau} <$ 0.06  & Neutrino oscillations \cite{Bolanos:2008km} \\ [2.4mm]
-0.98  $< \varepsilon^R_{\tau\tau} <$  0.23 &  Neutrino oscillation \cite{Bolanos:2008km, Agarwalla:2012wf} \\
-0.25  $< \varepsilon^R_{\tau\tau} <$  0.43  &Scattering and accelerator data \cite{Bolanos:2008km}\\ 
\midrule
  -0.13  $ <  \varepsilon^L_{e\mu}, \varepsilon^R_{e\mu} < $  0.13 & Scattering and accelerator data \cite{Barranco:2007ej}\\ [2.4mm]
 -0.33  $< \varepsilon^L_{e\tau} <$  0.33   & Scattering and accelerator data \cite{Barranco:2007ej} \\ [2.4mm]
-0.28 $ < \varepsilon^R_{e\tau} < $ -0.05 $\&$ 0.05 $ < \varepsilon^R_{e\tau} < $ 0.28 &  Scattering and accelerator data \cite{Barranco:2007ej}\\ 
-0.19 $ < \varepsilon^R_{e\tau} < $ 0.19 & Scattering \cite{TEXONO:2010tnr} \\ [2.4mm]
-0.10  $< \varepsilon^L_{\mu\tau}, \varepsilon^R_{\mu\tau} <$ 0.10  & Scattering and accelerator data \cite{Barranco:2007ej}\\  \bottomrule[0.25ex]
\end{tabular}
\caption{Current bounds on non-universal and flavour-changing NSIs with electrons at 90$\%$ C.L. for 1 degree of freedom.\label{tab:ch7-nsielimits}}
\end{table*}

Accurate measurements of the cross-section of purely leptonic processes, like neutrino-electron scattering, allow us to constrain the NSI parameters too. Reactor experiments like TEXONO \cite{TEXONO:2010tnr}, MUNU \cite{MUNU:2003peb}, Irvine \cite{Reines:1976pv}, and Rovno \cite{Derbin:1993wy} studied electron antineutrino scattering on electrons,
\begin{align}
\bar{\nu}_e \,  + \, e^- \, \longrightarrow \, \bar{\nu}_e \, + \, e^-
\end{align} and the LSND Collaboration performed measurements of electron neutrino scattering on electrons \cite{Davidson:2003ha,LSND:2001akn} 
\begin{align}
\nu_e \, +\,e^{\pm}\, \longrightarrow \, \nu_e\, + \, e^{\pm}\,.
\end{align} 
The combination of both experiments can set constraints on the NSI parameters involved: $\varepsilon^L_{ee}$, $\varepsilon^R_{ee}$, $\varepsilon^L_{e\mu}$, $\varepsilon^R_{e\mu}$, $\varepsilon^L_{e\tau}$ and $\varepsilon^R_{e\tau}$ \cite{Barranco:2005ps, Khan:2016uon}. Moreover, precise measurements of muon neutrino scattering on electrons 
\begin{align}
\bar{\nu}_\mu \, + \, e^- \, \longrightarrow \, \bar{\nu}_\mu \, + \, e^- \quad \text{and} \quad \nu_\mu \, + \, e^{\pm} \, \longrightarrow \, \nu_\mu\, + \, e^{\pm}\, ,
\end{align}
by the CHARM Collaboration set very stringent constraints on $\varepsilon^L_{\mu \mu}$, $\varepsilon^R_{\mu \mu}$, $\varepsilon^L_{e\mu}$, $\varepsilon^R_{e\mu}$, $\varepsilon^L_{\mu\tau}$ and $\varepsilon^R_{\mu\tau}$~\cite{CHARM-II:1994dzw}. Likewise, the measurement of the forward-backward asymmetry in the process 
\begin{align}
e^+ \, + e^- \, \longrightarrow \, e^+\, + \, e^-\, 
\end{align} 
at LEP breaks a degeneracy existing in the limits on non-universal NSIs from scattering experiments. On top of that, one can also study the reaction 
\begin{align}
e^{+}\, + \, e^{- }\, \longrightarrow \, \nu \, + \,  \overline{\nu} \, + \, \gamma 
\end{align}
which is mediated by the $Z$ and $W$ bosons in the Standard Model. Further constraints on neutrino-electron NSIs can be placed from additional coherent contributions in the presence of NSIs, which modify the expected number of events \cite{Barranco:2007ej}. This process has been studied by the four collaborations operating at LEP.

Table \ref{tab:ch7-nsielimits} summarises the existing limits on neutrino NSIs with electrons, $\varepsilon^L_{\alpha\beta}$ and $\varepsilon^R_{\alpha\beta}$. These limits are derived assuming that only one parameter differs from zero at a time. When the possible correlation between parameters is considered, the allowed parameter space is enlarged.

%%%%%%%%%%%%%%%%%%%%%%%%%%%%%%%%%%%%%%%%%%%%%%%%%%%%%%%%%%%%%%%%%%%%%%%%%
%%%%%%%%%%%%%%%%%%%%%%%%%%%%%%%%%%%%%%%%%%%%%%%%%%%%%%%%%%%%%%%%%%%%%%%%%
\section{The role of non-standard interactions in neutrino decoupling}
\label{sec:ch7-nsi-decoupling}
%%%%%%%%%%%%%%%%%%%%%%%%%%%%%%%%%%%%%%%%%%%%%%%%%%%%%%%%%%%%%%%%%%%%%%%%%
%%%%%%%%%%%%%%%%%%%%%%%%%%%%%%%%%%%%%%%%%%%%%%%%%%%%%%%%%%%%%%%%%%%%%%%%%
Neutrino non-standard interactions with electrons alter the picture of neutrino decoupling in two different ways. Firstly, they modify the interaction rate of the processes keeping neutrinos in thermal contact with the cosmic plasma. NSIs modify the cross-section of neutrino scattering on electrons and positrons and also the ones for pair productions and annihilations. Secondly, they also modify the evolution of the system since they alter neutrino flavour oscillations, similar to the role matter effects play in terrestrial experiments. These modifications can advance or delay the decoupling and induce a shift in the predicted value of $N_{\text{eff}}$.

%%%%%%%%%%%%%%%%%%%%%%%%%%%%%%%%%%%%%%%
\subsection{Implementation of NSIs with electrons in the calculation of $\mathbf{N_{\text{eff}}}$}
%%%%%%%%%%%%%%%%%%%%%%%%%%%%%%%%%%%%%%%
In the presence of non-zero NSIs with electrons, the matrices $G_L$ and $G_R$ that were previously defined in Equations \ref{eq:ch7-gl} and \ref{eq:ch7-gr} get additional contributions, namely, 
\begin{align}
G_L = \begin{pmatrix}
\tilde{g}_L + \varepsilon^L_{ee} & \varepsilon^L_{e\mu} & \varepsilon^L_{e\tau}\\
\varepsilon^{L*}_{e\mu} & g_L + \varepsilon^L_{\mu\mu} & \varepsilon^L_{\mu \tau} \\
\varepsilon^{L*}_{e\tau} & \varepsilon^{L*}_{\mu\tau} & g_L + \varepsilon^L_{\tau\tau}
\end{pmatrix}\, ,
\label{eq:ch7-gLR_nsi-1}
\\
G_R = \begin{pmatrix}
\tilde{g}_R + \varepsilon^R_{ee} & \varepsilon^R_{e\mu} & \varepsilon^R_{e\tau}\\
\varepsilon^{R*}_{e\mu} & g_R + \varepsilon^R_{\mu\mu} & \varepsilon^R_{\mu \tau} \\
\varepsilon^{R*}_{e\tau} & \varepsilon^{R*}_{\mu\tau} & g_R + \varepsilon^R_{\tau\tau}
\end{pmatrix}\, ,
\label{eq:ch7-gLR_nsi-2}
\end{align}
and therefore the collisional integrals are modified. As a consequence, the presence of NSIs modifies the \mbox{phase-space} factors of collision integrals.

The interactions between neutrinos and electrons are proportional to the Standard Model coefficients $g_L^2$, $g_R^2$ and a mixed term $g_Lg_R$ --- or $\tilde{g}^2_L$, $g^2_R$ and $\tilde{g}_Lg_R$ in the case of electrons. In the presence of NSIs, the values of these coefficients are effectively shifted, notably
\begin{eqnarray}
g_L^{2}
&\longrightarrow&
\left(g_{L} + \varepsilon^{L}_{\alpha\alpha}\right)^2 + \sum_{\beta \neq \alpha} |\varepsilon^{L}_{\alpha \beta}|^2
\,,
\label{eq:ch7-gLsqshift}
\\
g_R^{2}
&\longrightarrow& \left(g_{R} + \varepsilon^{R}_{\alpha\alpha}\right)^2 + \sum_{\beta \neq \alpha} |\varepsilon^{R}_{\alpha \beta}|^2
\,,
\label{eq:ch7-gRsqshift}
\\
g_Lg_R
&\longrightarrow&
\left(g_L + \varepsilon^L_{\alpha\alpha} \right)\left(g_R + \varepsilon^R_{\alpha\alpha}\right) + \sum_{\beta \neq \alpha} |\varepsilon^L_{\alpha\beta}||\varepsilon^R_{ \alpha\beta}|
\,.
\label{eq:ch7-gLgRshift}
\end{eqnarray}
One can see that for $\varepsilon^X_{\alpha \beta} = 0$ and $\varepsilon^X_{\alpha\alpha} = - g_X$, the energy transfer between neutrinos and the bath of electrons and positrons is minimised. As a consequence, we expect a minimum value of $N_{\text{eff}}$ whenever at least one of these conditions is satisfied.

Apart from the collisional terms in the Boltzmann equations, the terms in the Hamiltonian accounting for coherent elastic neutrino forward scattering with electrons needs to be modified too. Let us define the matrix
\begin{align}
\mathbb{E}_{\text{NSI}} = \begin{pmatrix} 1 + \varepsilon^V_{ee} & \varepsilon^V_{e\mu} & \varepsilon^V_{e\tau}\\
\varepsilon^{V*}_{e\mu} & \varepsilon^V_{\mu\mu} & \varepsilon^V_{\mu \tau} \\
\varepsilon^{V*}_{e\tau} & \varepsilon^{V*}_{\mu\tau} & \varepsilon^V_{\tau\tau}
\end{pmatrix}\, ,
\end{align}
so that the matter potentials become
\begin{align}
\mathbb{E}_e = \rho_e \, \mathbb{E}_{\text{NSI}} \quad \text{and} \quad
\mathbb{E}_e = P_e \,  \mathbb{E}_{\text{NSI}}\, .
\label{eq:ch7-matterNSI}
\end{align}
One can see that both terms include the standard contribution --- the factor one --- plus the additional term arising from NSIs. Notice that the effect of non-standard interactions in the matter potentials can be null if $\varepsilon^V_{\alpha\beta}$ are zero.

%%%%%%%%%%%%%%%%%%%%%%%%%%%%%%%%%%%%%%%
\subsection{Results}
%%%%%%%%%%%%%%%%%%%%%%%%%%%%%%%%%%%%%%%

In order to compute the evolution of the neutrino density matrix and the value of $N_{\text{eff}}$, we rely on the publicly available code \texttt{FortEPiaNO} (Fortran-Evolved Primordial Neutrino Oscillations) \cite{Gariazzo:2019gyi}.\footnote{\href{https://bitbucket.org/ahep_cosmo/fortepiano_public}{https://bitbucket.org/ahep\_cosmo/fortepiano\_public}} We adopt the numerical setting that ensures a determination of $N_{\text{eff}}$ with a numerical error $\leq 5\times 10^{-4}$ in the standard scenario. Such precision exceeds the one required to study the impact of non-standard interactions.  Specifically, we set an absolute and relative numerical precision for the differential equation solver of $10^{-6}$ or less, and an initial temperature given by $x_{in}$ = 0.01. In addition, we use a Gauss-Laguerre spacing for the neutrino momenta with $N_y = 30$ and \mbox{$y_{max}$ = 20}. Likewise, we compute the full collisional integrals --- without damping terms --- for neutrino–electron and neutrino–neutrino interactions.

\begin{table}
	\renewcommand{\arraystretch}{1.2}
    \centering
    \begin{tabular}{ccccc}
    \toprule[0.25ex]
         $\varepsilon_{ee}^L$ & $\varepsilon_{\tau \tau}^L$ & $N_{\rm eff}^{\text{no muons}}$ & $N_{ \rm eff}^{\text{with muons}}$ & $N_{\rm eff}^{\text{no muons}}$ - $N_{ \rm eff}^{\text{with muons}}$
         \\ \midrule
         0 & 0 & 3.04364 & 3.04358 & 6 $\times 10 ^{-5}$ \\ [2.4mm]
         0.2 & -0.3 & 3.05714 & 3.05710 & $4\times10^{-5}$\\ [2.4mm]
         -0.3 & 0.2 & 3.03199 & 3.03188 & $1.1\times10^{-4}$ \\ \bottomrule[0.25ex]
    \end{tabular}
    \caption{
    \label{tab:ch7-muons}
    Comparison between the value of $N_{\rm eff}$ obtained for two sets of NSI parameters with and without including the contribution of muons, $N_{\rm eff}^{\text{with muons}}$ and $N_{\rm eff}^{\text{no muons}}$, respectively.
    The difference is well below the experimental uncertainty and at the level of the numerical precision.
    }
\end{table}

In our calculations, we neglect the contribution of muons for two reasons. Firstly, we only consider NSIs between neutrinos and electrons, which means that interactions with muons remain unchanged. Secondly, at the time of neutrino decoupling the muon density is already Boltzmann-suppressed. Muons may have a small effect on $N_{\text{eff}}$\ in scenarios in which the thermalisation process starts much before neutrino decoupling~\cite{Gariazzo:2019gyi}. We checked numerically if these considerations are justified. The results --- summarised in \reftab{ch7-muons} --- showed that not including muons introduces a variation on $N_{\text{eff}}$\  slightly larger than $10^{-4}$ yet below the numerical precision aimed here.

Neutrino non-standard interactions with electrons change the collision terms involving neutrinos and electrons and they also alter neutrino oscillations via matter effects. \reftab{ch7-contributions} shows the total and the separate effect of including NSIs in the two different terms --- the Hamiltonian and the collisional integrals. One can see that the effect provided by altering neutrino oscillations is very small. This is in accord with the results of \cite{Bennett:2020zkv}, where the authors showed that the effect on $N_{\text{eff}}$\ from neutrino oscillations is at most of $\sim10^{-4}$ within the allowed range of mixing parameters \cite{deSalas:2020pgw}. For the problem at hand, we have estimated the numerical uncertainties to be around $10^{-4}$. Hence, the effect of NSIs on oscillations could be safely neglected --- although for consistency, we do include it in our calculations. Therefore, the impact that NSIs have on $N_{\text{eff}}$\ comes almost entirely from their effect on the collision terms. The variations induces are $\mathcal{O}(10^{-2}$), according to \reftab{ch7-contributions}. These aspects will be discussed in detail in what follows. 

\begin{table}
	\renewcommand{\arraystretch}{1.2}
    \centering
    \begin{tabular}{cccccc}
    \toprule[0.25ex]
         $\varepsilon_{ee}^L$ & $\varepsilon_{\tau \tau}^L$ & $N_{\rm eff}$ & $N_{ \rm eff}$ - $N_{ \rm eff}^{\text{no NSI}}$ & $N_{ \rm eff}^{\text{osc}}$ - $N_{ \rm eff}^{\text{no NSI}}$ &  $N_{ \rm eff}^{\text{coll}}$ - $N_{ \rm eff}^{\text{no NSI}}$  \\ \midrule
        0.2 & -0.3 & 3.05714 & 1.4 $\times 10 ^{-2}$ &  -7 $\times 10 ^{-5}$ &   1.4 $\times 10 ^{-2}$\\ [2.4mm]
         -0.3 & 0.2 & 3.03199 & -1.2 $\times 10 ^{-2}$ &  3$\times 10^{-5}$ & -1.2 $\times 10 ^{-2}$ \\ \bottomrule[0.25ex]
    \end{tabular}
    \caption{
    \label{tab:ch7-contributions}
    Comparison between the value of $N_{\rm eff}$ obtained for two sets of NSI parameters considering only its impact on oscillations through Equation~\ref{eq:ch7-matterNSI} --- denoted by $N_{ \rm eff}^{\text{osc}}$ --- or in the collisional integrals through the $G^X$ matrices in Equations~\ref{eq:ch7-gLR_nsi-1} and~\ref{eq:ch7-gLR_nsi-1} --- referred to as $N_{ \rm eff}^{\text{coll}}$. The deviation from the value of $N_{\text{eff}}$\ expected in the absence of NSIs under the same assumptions is presented as a reference, where $N_{\rm eff}^{\text{no NSI}} = 3.04364$. Muons are not included.
    }
\end{table}

\begin{figure*}
    \centering
    \includegraphics[width = 0.375\paperwidth]{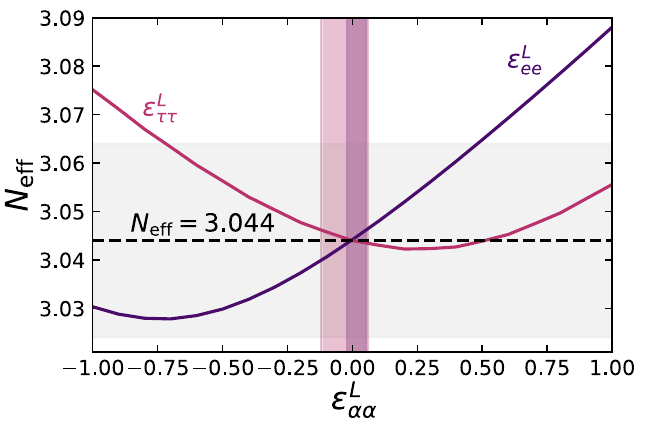}
    \includegraphics[width = 0.375\paperwidth]{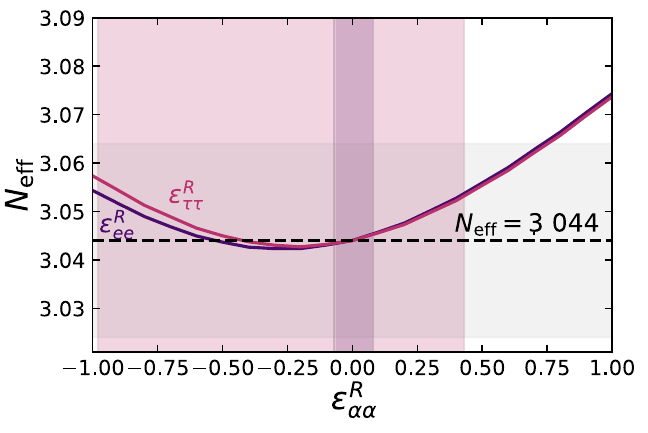}
    \caption{
    \labfig{fig:ch7-aaoneatatime}
    Values of $N_{\rm eff}$ in the presence of non-zero $\varepsilon^L_{\alpha \alpha}$ and $\varepsilon^R_{\alpha \alpha}$ --- in the left and right panel respectively --- for $\alpha = \lbrace e , \tau \rbrace$. The shaded vertical regions correspond to the 90$\% $ C.L. bounds presented in \reftab{ch7-nsielimits}. The value of $N_{\rm eff} = 3.044$ expected in the absence of NSIs is indicated by a dashed line, together with a shaded region corresponding to $\pm 0.02$, which is a reference value for the expected $1 \sigma$ uncertainty from future cosmological observations~\cite{CMB-S4:2016ple}.}
\end{figure*}

\textbf{Non-universal NSIs}

In \reffig{fig:ch7-aaoneatatime}, we show the values of $N_{\text{eff}}$ predicted in the presence of non-zero non-universal NSIs, namely for $\varepsilon^L_{\alpha\alpha}$ or $\varepsilon^R_{\alpha\alpha}$, with $\alpha=e,\tau$.
In the left panel, one can see that for negative values of $\varepsilon^L_{ee}$ --- indicated by a purple line --- the value of $N_{\text{eff}}$ decreases with respect to the Standard Model prediction, since the strength of the coupling is reduced.
One can see that there is a minimum for
\begin{align}
\varepsilon^L_{ee} = -\tilde g_L = -\frac{1}{2} - \sin^2\theta_{W}\, .
\end{align}
Concerning the parameter $\varepsilon^L_{\tau\tau}$ --- which corresponds to the pink line --- the minimum is found at
\begin{align}
\varepsilon^L_{\tau\tau} = - g_L = \frac{1}{2} - \sin^2 \theta_W\,.
\end{align} 
Notice that the impact of this parameter on $N_\text{eff}$ is mild.
The trend observed for the right-handed chiral couplings --- depicted in the right panel --- is different. For both $\varepsilon^R_{ee}$ and $\varepsilon^R_{\tau\tau}$, the minimum value of $N_{\text{eff}}$ is located close to 
\begin{align}
\varepsilon^R_{\alpha\alpha} = - g_R = -\sin ^2 \theta_{W}\,.
\end{align}
The small difference between the effect of the two parameters $\varepsilon^R_{ee}$ and $\varepsilon^R_{\tau\tau}$ at large negative values is a consequence of the terms proportional to $g_L g_R$ in the Standard Model --- since the left coupling of neutrinos to electrons or taus is different.
Notice that the minima in $N_{\text{eff}}$ are found where they were expected from the argument of the shift in the coupling. This reinforces the argument that the exact value of the oscillation parameters is not playing a relevant role.

From the \reffig{fig:ch7-aaoneatatime}, one can see that only large negative values of $\varepsilon^L_{ee}$ would lead to an effective number of neutrinos considerably smaller than $3.044$. It is important to note that NSIs are among the scenarios that can lead to a value of $N_{\text{eff}}$ smaller than the Standard Model prediction. However, such large values of $\varepsilon^L_{ee}$ are strongly in tension with other limits from terrestrial experiments. 
\begin{figure*}
    \centering
    \includegraphics[width = 0.375\paperwidth]{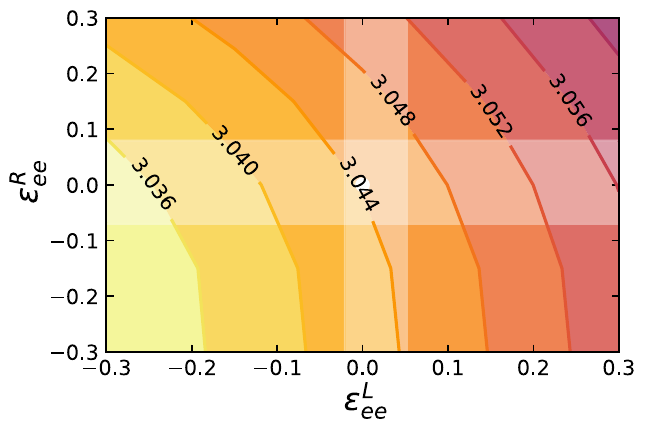}
    \includegraphics[width = 0.375\paperwidth]{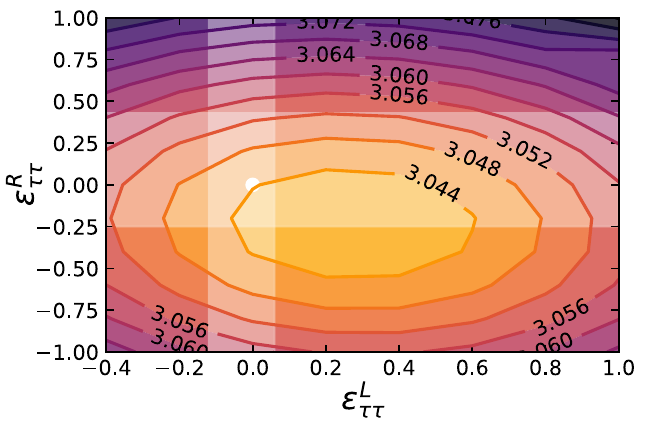}
    \includegraphics[width = 0.375\paperwidth]{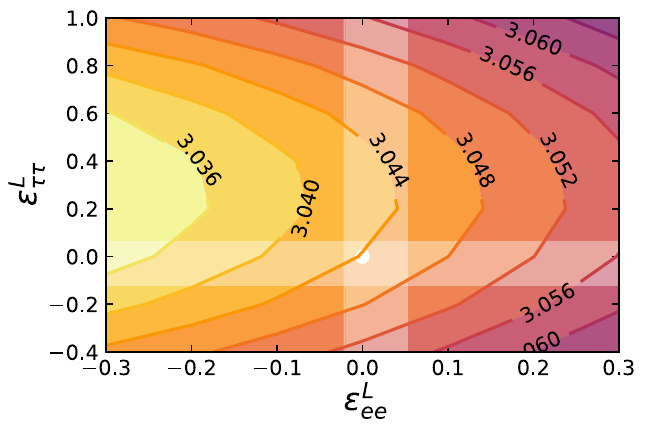}
     \includegraphics[width = 0.375\paperwidth]{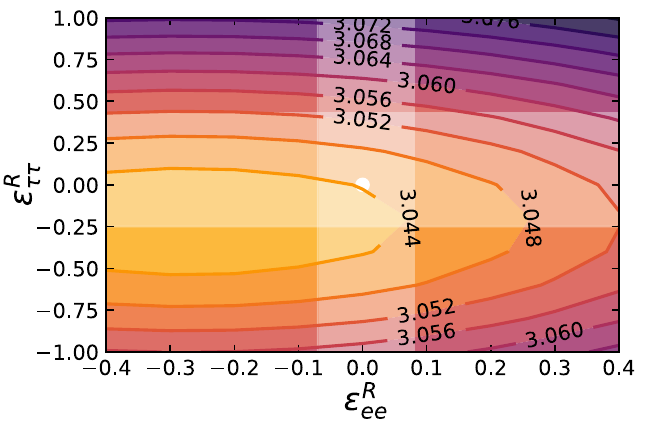}
    \caption{Values of $N_{\text{eff}}$\ when two NSI parameters are varied simultaneously: \mbox{$\varepsilon^L_{ee}$- $\varepsilon^R_{ee}$}, \mbox{$\varepsilon^L_{\tau \tau}$- $\varepsilon^R_{\tau \tau}$}, \mbox{$\varepsilon^L_{ee}$- $\varepsilon^L_{\tau \tau}$}, and \mbox{$\varepsilon^R_{ee}$- $\varepsilon^R_{\tau \tau}$} --- in the top left, top right, bottom left and bottom right panels, respectively. White-shaded regions correspond to the 90\%~C.L. experimental bounds obtained varying one parameter at a time, extracted from \cite{Farzan:2017xzy}.}\labfig{fig:ch7-2dnu}    
\end{figure*}

It is also interesting to see the interplay between the NSI parameters once two of them are allowed to vary simultaneously. \reffig{fig:ch7-2dnu} shows the variation of $N_{\text{eff}}$ induced by the simultaneous presence of two non-zero NSI couplings.
Notice that the iso-$N_{\text{eff}}$ contours presented in the plots are ellipses, as one can predict from the shift in the couplings entering the collisions --- see Equations~\ref{eq:ch7-gLsqshift} to \ref{eq:ch7-gLgRshift}. This confirms again that the contribution of NSIs on collisions dominates over the effect on oscillations. 
See for instance that, along the line for which $\varepsilon^L_{\alpha \alpha} = - \varepsilon^R_{\alpha \alpha}$, the vectorial component of the NSIs cancels out, leaving oscillations unchanged. In that case, the change in the value of $N_{\text{eff}}$, therefore, is exclusively due to the axial coupling entering the collision term. 

From \reffig{fig:ch7-aaoneatatime} and \reffig{fig:ch7-2dnu}, we can also see that next-generation cosmological observations --- whose determination of $N_{\text{eff}}$ could reach an uncertainty of $\sigma(N_{\text{eff}})=0.02$ \cite{CMB-S4:2016ple} --- will be able to set limits on non-universal NSIs of the same order of magnitude of those from current laboratory experiments. Remember, however, that constraints from cosmology are indirect and, therefore, much more model-dependent than laboratory results.

\textbf{Flavour-changing NSIs}

Non-standard interactions leading to flavour-changing processes result in higher values of $N_{\text{eff}}$. Such interactions are not present in the Standard Model. Hence, non-zero $\varepsilon^X_{\alpha\beta}$ increase the interaction rates between neutrinos and electrons and positrons --- see from Equation~\ref{eq:ch7-gLsqshift} to Equation~\ref{eq:ch7-gLgRshift}.Also, the effect on the collisions is independent of the sign of $\varepsilon^X_{\alpha\beta}$ and, hence, one expects the contours to be symmetric with respect to $\varepsilon^X_{\alpha\beta} = 0$. In \reffig{fig:ch7-etoneatatime}, we show how $N_{\text{eff}}$ depends on $\varepsilon^L_{e\tau}$ and $\varepsilon^R_{e\tau}$. This figure verifies our understanding of the role of flavour-changing NSIs.

As it happened for the non-universal NSI parameters, we can notice that future CMB constraints --- denoted by the horizontal grey band --- will test flavour-changing NSIs. Although weaker than current laboratory limits, they provide a relevant and independent confirmation of the existing limits

Similarly, one can repeat the calculations for other parameters. The results are not reported since the dependence of $N_{\text{eff}}$\ on these NSI couplings is practically indistinguishable from the effect of the corresponding $\varepsilon^L_{e\tau}$ or $\varepsilon^R_{e\tau}$ parameters.

\begin{figure*}
    \centering
    \includegraphics[width = 0.375\paperwidth]{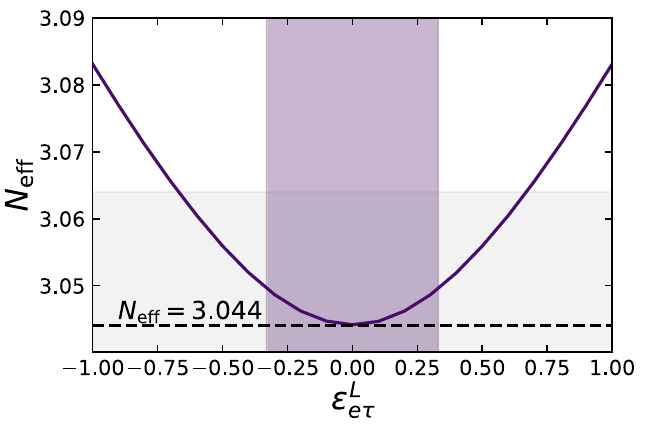}
    \includegraphics[width = 0.375\paperwidth]{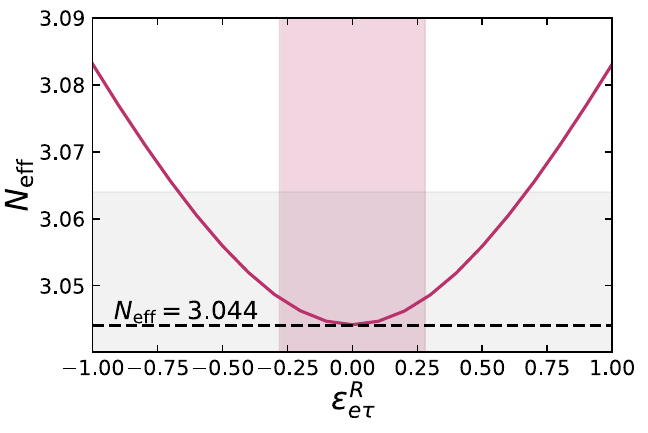}
    \caption{
    \labfig{fig:ch7-etoneatatime}
    Same as \reffig{fig:ch7-aaoneatatime} for the flavour-changing NSI parameters $\varepsilon^L_{e\tau}$ and $\varepsilon^R_{e\tau}$ in the left and right panels respectively.}
\end{figure*}

\begin{figure*}
    \centering
    \includegraphics[width = 0.375\paperwidth]{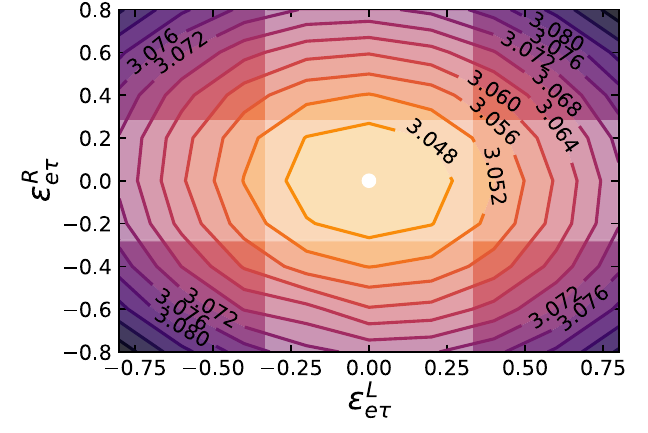}
    \includegraphics[width = 0.375\paperwidth]{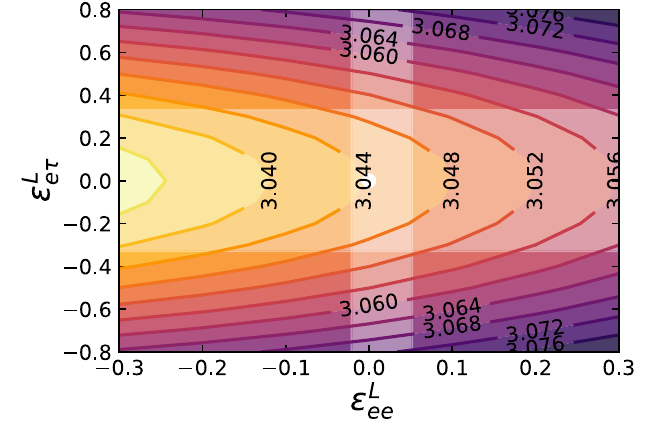}
    \includegraphics[width = 0.375\paperwidth]{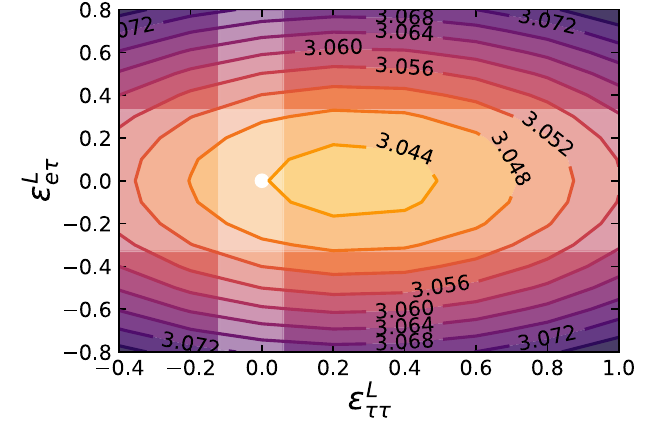}
        \includegraphics[width = 0.375\paperwidth]{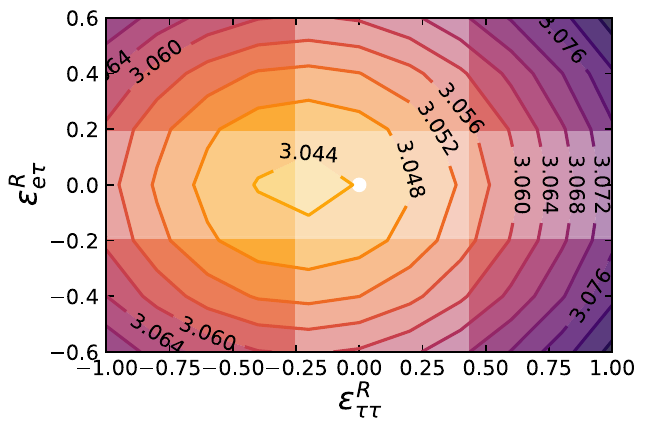}
    \caption{Same as \reffig{fig:ch7-2dnu} for four possible combinations of two NSI parameters: $\varepsilon^L_{e\tau}$- $\varepsilon^R_{e\tau}$, $\varepsilon^L_{ee}$- $\varepsilon^L_{e \tau}$, $\varepsilon^L_{\tau\tau}$- $\varepsilon^L_{e\tau}$ and $\varepsilon^R_{\tau\tau}$- $\varepsilon^R_{e\tau}$ --- in the top left, top right, bottom left and bottom right panels respectively. Again, white-shaded regions correspond to the 90\% CL experimental bounds obtained varying one parameter at a time, extracted from \cite{Farzan:2017xzy}.\labfig{fig:ch7-2dfc}}
\end{figure*}

In \reffig{fig:ch7-2dfc}, we display how the interplay between two different NSI couplings affects the prediction for the effective number of neutrinos.
Moreover, one can also vary three NSI coefficients at the same time, as shown in \reffig{fig:ch7-Neff3eps} %\footnote{An interactive version of \reffig{fig:ch7-Neff3eps} in HTML format, that can be rotated, is available at \url{https://www.astroparticles.es/NSI_Neff/GL11_GR11_GL13.html} --- in the left panel --- or \url{https://www.astroparticles.es/NSI_Neff/GL33_GR33_GL13.html} --- in the right panel.}.
The ellipses and ellipsoids are well understood based on the argument that NSI parameters induce a shift in the couplings between neutrinos and electrons. These figures demonstrate that complementary information can be gained from cosmological observations and that a multi-variable analysis is feasible.

Notice that, when several NSIs depart from zero, values of $N_{\text{eff}}$ significantly larger than 3.044 could be reached. For instance, for $\varepsilon^L_{\tau \tau} = -0.60$, $\varepsilon^R_{\tau \tau} = -0.36$, $\varepsilon^L_{e\tau} = 0.132$, $\varepsilon^R_{e\tau} = -0.80$, $\varepsilon^L_{\mu \tau}$ = $\varepsilon^R_{\mu\tau}$ = 0.52, one obtains $N_{\text{eff}}$ = 3.10. Such a large deviation from the standard prediction in the absence of NSIs could be tested by future determinations of $N_{\text{eff}}$ with a significance around 2--3$\sigma$. The values of the parameters needed to reach such a significant deviation from $N_{\text{eff}} = 3.044$ are excluded by terrestrial experiments with a very high confidence level. Nonetheless, one should keep in mind that constraints are often derived by varying one parameter at a time, without taking into account the degeneracies between the various parameters. 

Note that cosmology provides access to combinations that are unlikely to be probed on Earth, namely those involving tau neutrinos in the initial state. At a terrestrial experiment, one would need, for instance, a beam of tau neutrinos in order to study $\nu_\tau$ scattering on electrons. However, the expected constraints are not very competitive: they would rather be useful to provide an independent probe of the validity of terrestrial constraints than to provide strong bounds on the parameters themselves.

\begin{figure*}
    \centering
    \includegraphics[width = 0.375\paperwidth]{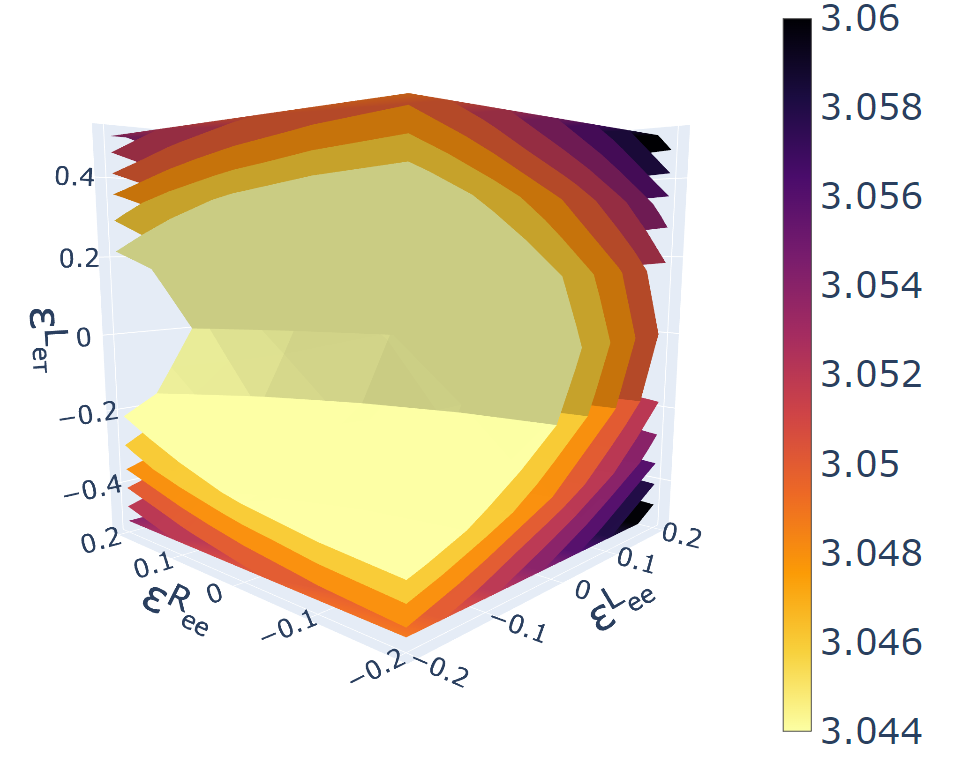}
    \includegraphics[width = 0.375\paperwidth]{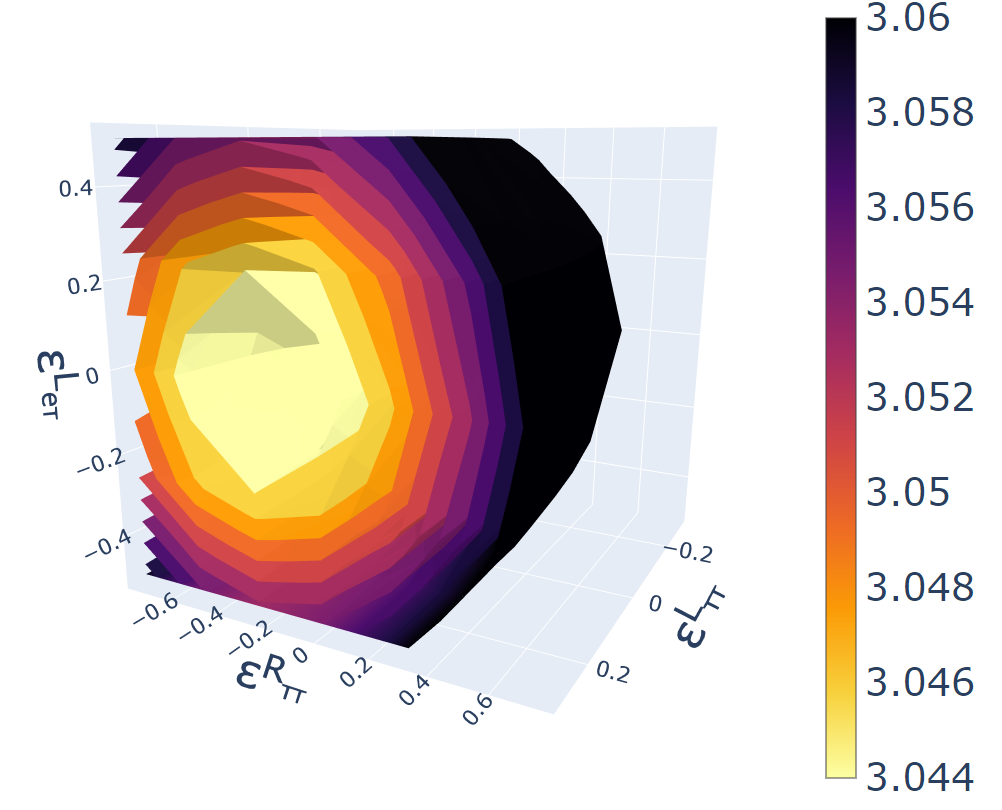}
    \caption{ Isosurfaces of $N_{\text{eff}}$ obtained from the simultaneous variation of three NSI couplings.    \labfig{fig:ch7-Neff3eps}}

\end{figure*}

%%%%%%%%%%%%%%%%%%%%%%%%%%%%%%%%
%%%%%%%%%%%%%%%%%%%%%%%%%%%%%%%%
\section{Concluding remarks}
\label{sec:ch7-conclude}
%%%%%%%%%%%%%%%%%%%%%%%%%%%%%%%%
%%%%%%%%%%%%%%%%%%%%%%%%%%%%%%%%
In this chapter, we have addressed how neutrino non-standard interactions would alter the process of decoupling from the primaeval plasma. We have studied the evolution of neutrinos in the MeV epoch of the universe using an extended version of the \texttt{FortEPiaNO} code, adapted to account for non-zero NSIs. In this scenario, two main differences exist with regard to the standard picture. Firstly, flavour oscillations --- which take place in a dense medium --- are altered due to NSIs. Secondly, these additional interactions also change the scattering, annihilation and pair-production processes keeping neutrinos in thermal equilibrium with the rest of the plasma. Although the latter is the dominant effect, we included both aspects in our calculations.

For the first time, we performed accurate calculations of the effective number of neutrinos, $N_{\rm eff}$. We did so for broad ranges of the NSI parameters and allowed for several of them to be non-zero at a time.
From these calculations, it stems that non-universal NSIs alter the energy transfer between neutrinos and electrons, advancing or delaying the process of decoupling, and shifting the predicted value of $N_{\text{eff}}$. Conversely, flavour-changing NSIs enhance the interaction rate between neutrinos and the rest of the plasma and would manifest as a larger value of $N_{\text{eff}}$. The interplay between different non-zero NSI couplings can be well understood from the modifications that they induction in the collisional integrals of the Boltzmann equations.

It is also interesting to contextualise these findings in light of the forecasted sensitivity from forthcoming CMB observations. Generally, one does not expect future limits on NSIs from cosmological data to be competitive with current bounds from terrestrial experiments. However, a determination of $N_{\text{eff}}$ with an error of $\sigma(N_{\text{eff}})=0.02-0.03$  \cite{CMB-S4:2016ple} could constrain non-universal NSIs --- for instance\ $\varepsilon^R_{\tau\tau}$ --- to the same level as current laboratory experiments, or could bound some particular combinations of several NSI parameters. Anyhow, cosmological observations provide a complementary --- and independent --- test of neutrino non-standard interactions with electrons.

\chapter{Non-unitary three-neutrino mixing in cosmology}
%\addcontentsline{toc}{chapter}{Non-unitary three-neutrino mixing in cosmology} 
\labch{ch8-nucosmo}

A common feature in neutrino mass generation mechanisms is the mediation of heavy neutral leptons (HNLs). In such models, the lepton mixing matrix has to be extended to account for the mixing between light neutrinos and heavy states. In canonical see-saw models \cite{Minkowski:1977sc,Gell-Mann:1979vob,Yanagida:1979as,Schechter:1980gr,Magg:1980ut,Cheng:1980qt,Mohapatra:1980yp,Foot:1988aq}, the smallness of neutrino masses would indicate a very high mass scale for the heavy neutral leptons and a negligible admixture between heavy and light states. Then, the three-neutrino mixing matrix departs negligibly from unitarity \cite{Antusch:2006vwa,Abada:2007ux,Grimus:2000vj}. However,  other realisations of the see-saw mechanism, such as the linear~\cite{Akhmedov:1995ip,Malinsky:2005bi} and inverse seesaw~\cite{Mohapatra:1980yp,Wyler:1982dd,Mohapatra:1986bd,Gonzalez-Garcia:1988okv,Forero:2011pc}, can render significant departures from unitarity for the three-neutrino mixing matrix \cite{Mohapatra:1986bd, Akhmedov:1995ip,Akhmedov:1995vm,Malinsky:2005bi,Malinsky:2009df, Malinsky:2009gw}. Within this framework, two scenarios can be contemplated: \mbox{(i) not-so-heavy} neutral leptons, with masses such that they can participate in neutrino oscillations at the energies probed in different experiments and (ii) canonical heavy neutral leptons, kinematically inaccessible in oscillation experiments. The former scenario is being scrutinised due to its potential to explain several short-baseline anomalies in neutrino experiments \cite{Gariazzo:2013gua,Kopp:2013vaa,Gariazzo:2017fdh,Dentler:2018sju,Giunti:2019aiy,Giunti:2019fcj,Boser:2019rta}. Nonetheless, when discussing the non-unitarity of the three-neutrino mixing matrix in this chapter, only the latter case is considered.

The content is organised as follows. \refsec{ch8-nu} introduces the notation employed to parametrise a non-unitary three-neutrino mixing matrix. Next, \refsec{ch8-nuincosmo} is devoted to the discussion of the implications of non-unitarity in cosmology and in \refsec{ch8-nubounds}, the current limits and projected sensitivities on this scenario are presented. Finally, to contextualise the results, in \refsec{ch8-nudiscuss} we outline the synergies with the formalism of NSIs and we compare the results with the existing bounds on non-unitarity from laboratory experiments. Finally, we draw our conclusions in Section \ref{sec:ch8-conclusion}.

%%%%%%%%%%%%%%%%%%%%%%%%%%%%%%%%%%%%%%%%%%%%%%%%%%%%%%
%%%%%%%%%%%%%%%%%%%%%%%%%%%%%%%%%%%%%%%%%%%%%%%%%%%%%% 
\section{Non-unitary three-neutrino mixing}
\labsec{ch8-nu}
%%%%%%%%%%%%%%%%%%%%%%%%%%%%%%%%%%%%%%%%%%%%%%%%%%%%%%
%%%%%%%%%%%%%%%%%%%%%%%%%%%%%%%%%%%%%%%%%%%%%%%%%%%%%%
Let $\upnu_L$ denote the $n$-vector representing the neutral lepton states with definite mass, including both light and heavy ones. In this basis, the charged-current and neutral-current Lagrangians --- for which we adopt the short-hand notation CC and NC respectively --- read
\begin{align}
    \mathcal{L}_ {\rm CC} = \frac{g}{\sqrt{2}} (W^-)_\mu \bar{e}_L\gamma^\mu \mathrm{K} \upnu_L + {\rm h.c.}\, , 
\end{align}
and
\begin{align}\mathcal{L}_{\rm NC} = \frac{g}{2 \cos \theta_W} Z_\mu \bar{\upnu}_L \gamma^\mu (\mathrm{K}^\dagger \mathrm{K}) \upnu_L +{\rm h.c. }\, ,
\end{align}
where $\theta_W$ is the weak mixing angle and $g$ is the coupling of the interactions. We have also introduced the matrix $\mathrm{K}$, which is the $3\times n$ matrix relating the three active flavour eigenstates and the $n$ mass eigenstates. Note that this matrix also includes the rotation required to define the charged-leptons mass basis. At low energies, one can rewrite the CC and NC interactions in terms of four-fermion interactions, namely
\begin{align}
    \mathcal{L}_{\rm CC} =-2\sqrt{2}G_{\rm F}\sum_{i,j}(\mathrm{K}^\dagger)_{i e}\mathrm{K}_{e j}\left(\bar{\nu}_i \gamma^\mu P_L \nu_j\right)\left(\bar{e}\gamma_\mu P_L e \right)~,
    \label{eq:ch8-nucc}
\end{align}
and 
\begin{align}
    \mathcal{L}_{\rm NC} = -2\sqrt{2}G_{\rm F}\sum_{X = L, R}g_X\sum_{i,j}\left(\mathrm{K}^\dagger \mathrm{K} \right)_{ij}\left(\bar{\nu}_i \gamma^\mu P_L \nu_j\right)\left(\bar{e}\gamma_\mu P_X e \right) \, ,
    \label{eq:ch8-nunc}
\end{align}
Notice that in the case of Equation \ref{eq:ch8-nucc} we have performed a Fierz rearrangement. The chiral projectors $P_{L,\,R}= (1 \mp \gamma_5)/2$ are denoted by $P_X$ with  $X=\lbrace L,\,R\rbrace$ and $G_\text{F}$ is the Fermi constant. The Lagrangians are also expressed in terms of the couplings $g_L = -1/2 + \sin^2\theta_W$ and $g_R = \sin^2\theta_W$.

Note that, when computing the amplitude of physical processes involving real neutrinos, the sum over mass eigenstates is limited to the largest mass eigenstate kinematically accessible. In what remains, we assume that only three mass eigenstates lie below the energy scale of interest here and thus are kinematically accessible. Then, the sums in Equations \ref{eq:ch8-nucc} and \ref{eq:ch8-nunc} only include the three lightest mass eigenstates. Thus, let us separate the matrix $\mathrm{K}$~in two blocks,
\begin{align}
    \mathrm{K} = \begin{pmatrix}
        N & S 
    \end{pmatrix}\, ,
\end{align}
where $N$ describes the mixing among the three lightest states and $S$ accounts for the mixing between the three lightest states and the $n-3$ heavy ones. Then, only the $3\times3$ submatrix $N$ is involved in the calculations discussed in this chapter.

Notice that the unitarity of the full $n\times n$ lepton mixing matrix does not imply that the submatrix $N$ is unitary too. Following previous results in the literature, we adopt the following parametrisation for the non-unitarity $N$ matrix in terms of a triangular matrix \cite{Xing:2007zj,Xing:2011ur,Escrihuela:2015wra,Han:2021qum}
\begin{equation}
    N = \begin{pmatrix}
    \alpha_{11} & 0 & 0 \\ \alpha_{21} & \alpha_{22} & 0 \\ \alpha_{31} & \alpha_{32} & \alpha_{33} 
    \end{pmatrix} U\, .
    \label{eq:ch8-parametrisation}
\end{equation}
Here, $U$ denotes the usual unitary $3\times3$ mixing matrix. The diagonal parameters, $\alpha_{ii}$, are real whereas the off-diagonal ones, $\alpha_{ij} ~ (i \neq j)$, are complex. All these $\alpha_{ij}$ parameters can be related to the full $n\times n$ unitary lepton mixing matrix, $U_{n \times n}$ as follows.

Let us adopt the following notation: $c_{ij} = \cos \theta_{ij}$, \mbox{$\eta_{ij} = \sin \theta_{ij}\, e^{- \phi_{ij}}$} and \mbox{$\bar{\eta}_{ij} = -\sin \theta_{ij} \,e ^{i\phi_{ij}}$}. Then, one can define a complex rotation matrix as
\begin{align}
    \omega_{ij} = \begin{pmatrix}
    1& 0 & & \cdots &0& \cdots& & &0\\
    0& 1 & & & & & & & \vdots \\
    \vdots & & c_{ij} & \cdots & 0 & \cdots & \eta_{ij} & & \\
    & & \vdots & \ddots & &  &\vdots & \\
    & & 0 & &1& & 0 & &\\
    & & \vdots & & \ddots& &\vdots & &\\
    & & \bar{\eta}_{ij} &\cdots&0& \cdots& c_{ij}&  &\vdots \\
    \vdots & & & & & & & 1 & 0 \\
    0 & & & \cdots & 0 & \cdots& & 0 & 1\end{pmatrix} \, .
\end{align}

With these definitions, we express the full $n\times n$ lepton mixing matrix as
\begin{align}
    &U_{n\times n} = \nonumber\\ & \hspace{0.7cm}\omega_{n-1 \, n}\, \omega_{n-2 \, n} ...\, \omega_{1\, n} \omega_{n-2 \, n-1}\, \omega_{n-3\, n-1} ...\, \omega_{1\, n-1} ...\, \omega_{23}\,\omega_{13}\,\omega_{12}\, .
\end{align}

One can see that the diagonal parameters $\alpha_{ii}$ in Equation~\ref{eq:ch8-parametrisation} are real and expressed as~\cite{Escrihuela:2015wra,Forero:2021azc}
\begin{align}
    \alpha_{ii} = c_{i\,n}\, c_{i\, n-1}...\, c_{i\, 4} \, .
\end{align}
whereas the off-diagonal ones are 
\begingroup
\allowdisplaybreaks
\begin{align}
&\alpha_{21} = \, c_{2\,n}\,c_{2\, n-1} ...\, c_{25}\,\eta_{24}\,\bar{\eta}_{14} + c_{2\,n}...\, c_{26}\,\eta_{25}\,\bar{\eta}_{15}\,c_{14} \nonumber \\&  \hspace{4.4cm} + ...+\eta_{2\,n}\,\bar{\eta}_{1n}\,c_{1\,n-1}...\,c_{14} \, ,\\
&\alpha_{31} =\, c_{3\,n\,}c_{3\, n-1}...\, c_{35}\,\eta_{34}\,c_{24}\,\bar{\eta}_{14} +c_{3\,n}...\,c_{36}\,\eta_{35}\,c_{25}\,\bar{\eta}_{15}\,c_{14} \nonumber\\ & \hspace{4.4cm} +...+\eta_{3\,n}\,c_{2\,n}\,\bar{\eta}_{1\, n}\,c_{1\, n-1}...\, c_{14}\, ,\\
&\alpha_{32} = \,c_{3\,n}\,c_{3\,n-1}...\, c_{35}\,\eta_{34}\,\bar{\eta}_{24}+ c_{3\,n}...\,c_{36}\,\eta_{35}\,\bar{\eta}_{25}\,c_{24} \nonumber \\ & \hspace{4.4cm}+ ... +\eta_{3\, n}\,\bar{\eta}_{2\,n}\, c_{2\, n-1} ...\, c_{24}\, .
\end{align}
\endgroup
It is now clear that the off-diagonal parameters are complex and provide additional sources of CP violation --- related to the additional phases needed to parameterise an $n\times n$ lepton mixing matrix.

The triangular inequality 
\begin{equation}
    |\alpha_{ij}| \leq \sqrt{(1- \alpha^2_{ii})(1-\alpha^2_{jj})}\, ,
\end{equation}
relates the diagonal and off-diagonal parameters \cite{Escrihuela:2016ube}. Additionally, from the unitarity of the full $n \times n$ mixing matrix~\cite{Forero:2021azc}, the $\alpha_{ij}$ parameters satisfy the following conditions:
\begin{equation}
    \alpha^2_{11}|\alpha_{21}|^2 \leq (1 -\alpha^2_{11})(1-\alpha^2_{22} - |\alpha_{21}|^2) \, , \label{eq:ch8-NUcond1}
\end{equation}
\begin{equation}
    \alpha^2_{11} |\alpha_{31}|^2 \leq (1 -\alpha^2_{11})(1-\alpha^2_{33} - |\alpha_{31}|^2-|\alpha_{32}|^2 ) \, ,
    \label{eq:ch8-NUcond2}
\end{equation}
\begin{equation}
     |\alpha_{22}\alpha_{32} + \alpha^*_{21}\alpha_{31} |^2 \leq (1 -\alpha^2_{22}-|\alpha_{21}|^2)(1-\alpha^2_{33} - |\alpha_{31}|^2 -|\alpha_{32}|^2) \,.
    \label{eq:ch8-NUcond3}
\end{equation}

%%%%%%%%%%%%%%%%%%%%%%%%%%%%%%%%%%%%%%%%%%%%%%%%%%%%%%
%%%%%%%%%%%%%%%%%%%%%%%%%%%%%%%%%%%%%%%%%%%%%%%%%%%%%%
\section{Implications in cosmology}
\labsec{ch8-nuincosmo}
%%%%%%%%%%%%%%%%%%%%%%%%%%%%%%%%%%%%%%%%%%%%%%%%%%%%%%
%%%%%%%%%%%%%%%%%%%%%%%%%%%%%%%%%%%%%%%%%%%%%%%%%%%%%%

%%%%%%%%%%%%%%%%%%%%%%%%%%%%%%%%%%%%%%%%%%%%%%%%%%%%%%
\subsection{Non-unitarity in neutrino decoupling}
%%%%%%%%%%%%%%%%%%%%%%%%%%%%%%%%%%%%%%%%%%%%%%%%%%%%%%

Let us focus on a scenario in which HNLs can not be produced at temperatures of $\mathcal{O}$(MeV) and, therefore, their distribution is Boltzmann-suppressed. As a consequence, one only needs to study the evolution of the three lightest mass eigenstates. Previous studies of neutrino decoupling study the evolution in terms of flavour states \cite{Froustey:2020mcq,Bennett:2020zkv}. However, a correct treatment of non-unitarity requires working in the mass basis, as in \cite{Akita:2020szl}. The reason is that the truncation in the sum over accessible states is only well-defined for mass eigenstates.

Let us adopt the same comoving variables we defined in Equation~\ref{eq:ch7-variables} --- $x = m_e \,a$, $y = p\,a$, $z = T_\gamma\,a$ --- in terms of the electron mass $m_e$, the scale factor $a$, the neutrino momentum $p$, and the photon temperature $T_\gamma$ \cite{Gariazzo:2019gyi,Bennett:2020zkv}. With these variables, the evolution of the density matrix $\varrho$ for three neutrinos in the mass basis reads \cite{Akita:2020szl}
\begin{align}
\label{eq:ch8-drho_dx_nxn}
&\frac{{\rm d}\varrho(y)}{{\rm d}x} = \sqrt{\frac{3 m^2_{\rm Pl}}{8\pi\rho}}
\left\{-i \frac{x^2}{m_e^3} \left[\frac{\mathbb{M}}{2y} - \frac{2\sqrt{2}G_{\rm F} y m_e^6}{x^6} \left(\frac{\mathbb{E}_\ell+\mathbb{P}_\ell}{m_W^2}\right), \varrho \right]\right. \nonumber \\& \hspace{6cm} \left.+\frac{m_e^3}{x^4}\mathcal{I(\varrho)} \right\}\,.
\end{align}
Here, we have introduced the Planck mass, $m_{\rm Pl}$, the total energy density of the universe, $\rho$, the Fermi constant, $G_\text{F}$, and the mass of the W boson, $m_W$. We have also defined the diagonal neutrino mass matrix 
\begin{align}
\mathbb{M} = \begin{pmatrix}
0 & 0 & 0 \\ 0 & \Delta m^2_{21} & 0 \\ 0 & 0 & \Delta m^2_{31}\, .
\end{pmatrix}
\end{align}
It is convenient to define the following matrices
\begin{align}
    \label{eq:ch8-YL}
    \left(Y_L\right)_{ij} &= g_L \left(N^\dagger N\right)_{ij} + (N^\dagger)_{i e} N_{ej}\, , \\
    \label{eq:ch8-YR}
    \left(Y_R\right)_{ij} &= g_R \left(N^\dagger N\right)_{ij}\, ,
\end{align}
where the indices $i$ and $j$ run from 1 to 3. Since only the vector component of the interactions contribute to the effective potentials in the Hamiltonian, the charged-lepton energy density and pressure --- $\mathbb{E}_{\ell}$ and $\mathbb{P}_{\ell}$ respectively --- are proportional to the matrix
\begin{equation}
    \mathbb{E}_{\rm NU} = Y_L  - Y_R \, .
\end{equation}
For the temperatures of the plasma that concerns us, the contributions of muon and tau leptons are Boltzmann-suppressed and, therefore, only electrons contribute to the energy density and pressure of charged-leptons, notably
\begin{align}
     \mathbb{E}_{\ell} = \rho_e \, \mathbb{E}_{\rm NU}~, & & \mathbb{P}_{\ell} = P_e\, \mathbb{E}_{\rm NU}   \simeq  \rho_e \, \mathbb{E}_{\rm NU}/3~.
\end{align}
Here, we have expressed the contributions in terms of the energy density and pressure of electrons and positrons --- $\rho_e$ and $P_e$ respectively.

Finally, let us analyse how the term $\mathcal{I(\varrho)}$ in Equation \ref{eq:ch8-drho_dx_nxn} -- which encodes the collision integrals --- changes when the three neutrino mixing is non-unitary. 
The terms accounting for neutrino-electron interactions are proportional to $G_{\rm F}^2$ and the factors related to phase-space considerations read
%\begingroup
%\allowdisplaybreaks
\begin{align}
F^\text{scatt}_{ab}&\left(\varrho^{(1)}, f_e^{(2)}, \varrho^{(3)}, f_e^{(4)}\right) \nonumber\\
=\, & f_e^{(4)}(1-f_e^{(2)})\left[Y_a\varrho^{(3)}Y_b(1-\varrho^{(1)})+(1-\varrho^{(1)})Y_b\varrho^{(3)}Y_a\right]
\nonumber\\
&-
f_e^{(2)}(1-f_e^{(4)})\left[\varrho^{(1)}Y_b(1-\varrho^{(3)})Y_a+Y_a(1-\varrho^{(3)})Y_b\varrho^{(1)}\right]\, ,
\label{eq:ch8-F_ab_sc}\\
F^\text{ann}_{ab}&\left(\varrho^{(1)}, \varrho^{(2)}, f_e^{(3)}, f_e^{(4)}\right)
\nonumber\\
=\, & f_e^{(3)}f_e^{(4)}\left[Y_a(1-\varrho^{(2)})Y_b(1-\varrho^{(1)})+(1-\varrho^{(1)})Y_b(1-\varrho^{(2)})Y_a\right]
\nonumber\\
&-
(1-f_e^{(3)})(1-f_e^{(4)})\left[Y_a\varrho^{(2)}Y_b\varrho^{(1)}+\varrho^{(1)}Y_b\varrho^{(2)}Y_a\right]\, ,
\label{eq:ch8-F_ab_ann}
\end{align}
%\endgroup
with $a,b = L,R$. Notice that they are analogous to the definitions made in the flavour basis in Equations \ref{eq:ch7-F_ab_sc} and \ref{eq:ch7-F_ab_ann}. They depend on the electron momentum distribution, $f_e$, and on the density matrix $\varrho^{(i)}= \varrho(y_i)$ --- being $y_i$ the comoving momentum of particle $i$. However, in this case, they depend on the matrices $Y_L$ and $Y_R$, as defined in Equation~\ref{eq:ch8-YL} and Equation~\ref{eq:ch8-YR}.

A non-unitary three-neutrino mixing would also modify neutrino self-interactions. The low-energy Lagrangian for neutrino-neutrino interactions is given by
\begin{align}
&\mathcal{L}_{\nu SI} = \nonumber\\ & \quad -2 \sqrt{2}G_{\rm F} \sum_{i,j,k,m} \left(N^\dagger N\right)_{ij}\left(N^\dagger N\right)_{mk}(\bar{\nu}_i \gamma^\mu P_L \nu_j)(\bar{\nu}_k \gamma_\mu P_L \nu_m)\, .
\end{align}
Such interactions among neutrinos are included in the formalism as a refractive term in the Hamiltonian. It has been shown that, in the absence of an initial asymmetry between neutrinos and antineutrinos, this term only gives a subleading contribution to $N_{\text{eff}}$. We assume there is not such an asymmetry~\cite{Castorina:2012md} and ignore the refractive term in our calculations.

Notice that, up until this point, we have discussed the impact on non-unitarity in terms of the Fermi constant $G_{\rm F}$, which is a fundamental constant in nature. However, in the presence of a non-unitary three-neutrino mixing matrix, the quantity that is experimentally accessible is not that fundamental constant --- it depends on the mixing between light neutrinos and HNLs~\cite{Escrihuela:2015wra}. The effective parameter measured in the case of beta decay is
\begin{equation}
    G^\beta_{\rm F} = G_{\rm F} \sqrt{(NN^ \dagger)_{ee}} = G_{\rm F} \alpha_{11}~,
\end{equation}
whereas for muon decay it is
\begin{equation}
    G^\mu_{\rm F}=G_{\rm F} \sqrt{(NN^\dagger)_{ee} (N N^\dagger)_{\mu\mu}} = G_{\rm{F}} \sqrt{\alpha^2_{11}(\alpha^2_{22}+ |\alpha_{21}|^2)} \, .
    \label{eq:ch8-GFmu}
\end{equation}

We will take the value determined from muon decay --- namely \mbox{$G^\mu_{\rm F} = 1.1663787(6)\times 10^{-5} \, {\rm GeV}^{-2}$} \cite{Tiesinga:2021myr} --- since it has achieved a higher precision.

%%%%%%%%%%%%%%%%%%%%%%%%%%%%%%%%%%%%%%%%%%%%%%%%%%%%%%
\subsection{Other implications of non-unitarity in cosmology}
%%%%%%%%%%%%%%%%%%%%%%%%%%%%%%%%%%%%%%%%%%%%%%%%%%%%%%
The distortions in the distributions of neutrinos after neutrino decoupling could also alter Big Bang Nucleosynthesis. In addition, the rate of inverse beta decay, which is known to play a relevant role in fixing the \mbox{proton-to-neutron} ratio, depends on $G^\beta_{\text{F}}$. In the presence of non-unitarity, theoretical computations would also need to account for the shift between the value of $G_{\text{F}}$ and the experimentally accessible one.

Regarding neutrino self-interactions, their role in cosmology has been extensively studied and proposed as a solution to the Hubble tension --- see~\cite{DiValentino:2021izs} and references therein. However, the self-interactions invoked in those scenarios are several orders of magnitude larger than the ones due to the non-unitarity of the three-neutrino mixing.

%%%%%%%%%%%%%%%%%%%%%%%%%%%%%%%%%%%%%%%%%%%%%%%%%%%%%%
%%%%%%%%%%%%%%%%%%%%%%%%%%%%%%%%%%%%%%%%%%%%%%%%%%%%%%
\section{Current limits and future constraints from neutrino decoupling}
\labsec{ch8-nubounds}
%%%%%%%%%%%%%%%%%%%%%%%%%%%%%%%%%%%%%%%%%%%%%%%%%%%%%%
%%%%%%%%%%%%%%%%%%%%%%%%%%%%%%%%%%%%%%%%%%%%%%%%%%%%%%
We derive our numerical results using again the publicly available code  \texttt{FortEPiaNO} (Fortran Evolved Primordial Neutrino Oscillations) \cite{Gariazzo:2019gyi,Bennett:2020zkv}. In this case, it was adapted in order to work in the mass basis and to include the effects of non-unitarity that were discussed previously. The evolution is computed from $x \simeq 0.010 -0.015$ to $x = 35$ and using a momentum grid with 50 nodes between $y = 0.01$ and $y = 20$. With these settings, an absolute precision below $10^{-3}$ on the value of $N_{\text{eff}}$ is reached. This is around one order of magnitude smaller than the estimated sensitivity from future surveys.

%%%%%%%%%%%%%%%%%%%%%%%%%%%%%%%%%%%
\subsection{The role of diagonal parameters $\mathbf{\alpha_{ii}}$}
%%%%%%%%%%%%%%%%%%%%%%%%%%%%%%%%%%%

According to the parametrisation in Equation \ref{eq:ch8-parametrisation}, non-unitarity could manifest as a deviation from unity in the diagonal parameters $\alpha_{ii}$. We vary each of them separately to gain a better understanding of their individual role. Later, we also study the interplay between them.

The left panel of \reffig{fig:ch8-one_at_a_time} illustrates how $N_{\text{eff}}$ depends on each of the $\alpha_{ii}$  parameters. The top right panel depicts the value of $N_{\text{eff}}$ when non-unitarity effects are included only in collisional integrals --- and the Hamiltonian is left as in the standard picture. For comparison, in the bottom right panel, we show the values of $N_{\text{eff}}$ when only the Hamiltonian is modified --- through~$G_{\text{F}}$ --- and the collisional terms are left unchanged. 

\begin{figure*}
    \centering
    \includegraphics[width =0.72\paperwidth]{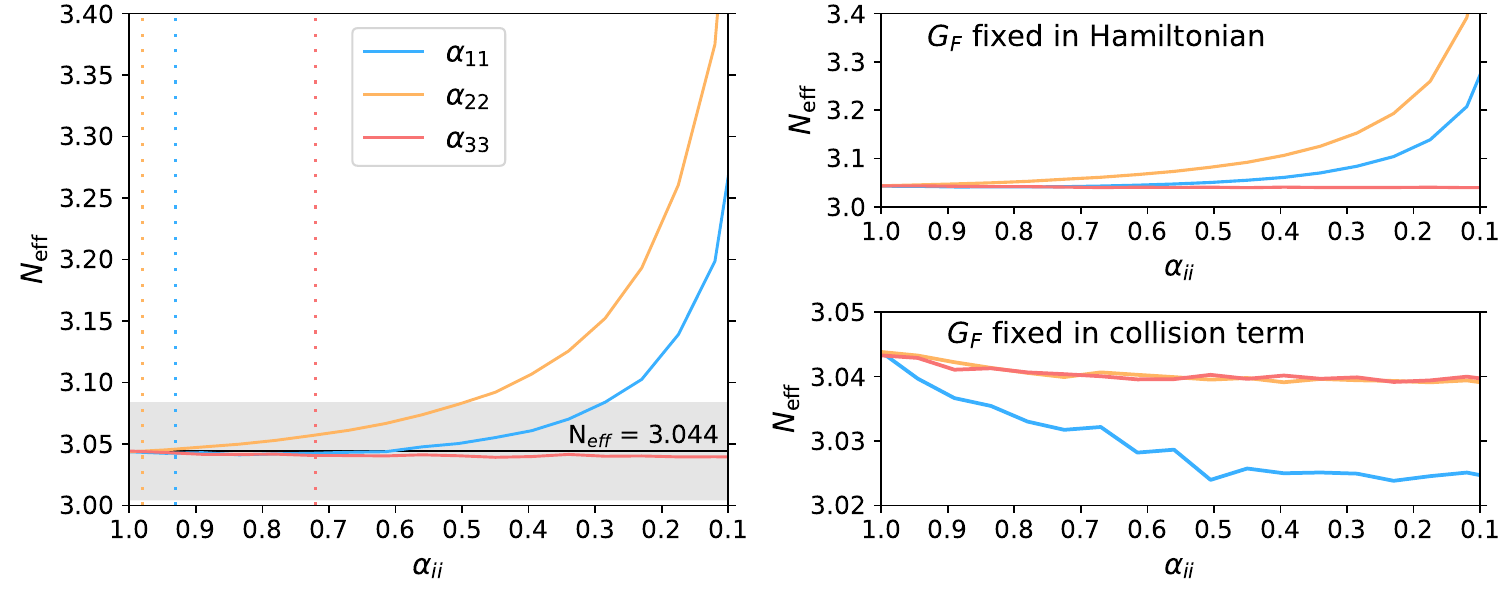}
    \caption{Values of $N_{\text{eff}}$ as a function of the diagonal non-unitarity parameters $\alpha_{ii}$, when only one of them is allowed to be non-zero.  For comparison, the left panel shows the exact calculation, where the vertical dotted lines indicate the existing limits from neutrino oscillations --- see \reftab{ch8-bounds}. The grey shaded band corresponds to the expected 95$\%$ projected sensitivity of CMB-S4 \cite{CMB-S4:2016ple}. The upper and lower right panels display the cases when the dependence of $G_{\rm F}$ on the $\alpha_{ii}$ parameters is considered only in the collision integrals or only in the Hamiltonian, respectively. \labfig{fig:ch8-one_at_a_time}}
\end{figure*}

Three key aspects should be noticed about these results. Firstly, $N_{\text{eff}}$ is barely sensitive to $\alpha_{33}$. Indeed, the main impact of non-unitarity on neutrino decoupling arises from the mismatch between the true and the observed value of $G_{\rm F}$. Since the value of $G_{\rm F}$ does not depend on $\alpha_{33}$, $N_{\text{eff}}$ is almost insensitive to it.
Secondly, as one can see from the right panel of \reffig{fig:ch8-one_at_a_time}, the collision terms dominate the dependence of $N_{\text{eff}}$ on the non-unitarity parameters. The changes that non-unitarity parameters induce in the effective $G_{\rm F}$ delay the decoupling and result in a larger value of $N_{\text{eff}}$. We have also shown in the previous chapter that changes in the collisional integrals are generally more relevant than matter effects. Hence, although non-unitarity also modifies neutrino oscillations, this effect has a much smaller impact.
Finally, notice that the results in \reffig{fig:ch8-one_at_a_time} are not equal for $\alpha_{11}$ and $\alpha_{22}$. One could initially think that they should, since $G^\mu_{\text{F}}$ has the same dependence on both of them --- see Equation \ref{eq:ch8-GFmu} for $\alpha_{21} = 0$. Nonetheless, these parameters also modify the matrices $Y_L$ and $Y_R$ although in different ways. Developing a physical intuition to justify these results is not straightforward when working on the mass basis. For this reason, in Subsection~\ref{subsec:ch8-nsi-nu}, we map this scenario to non-standard interactions. This analogy is valid for small departures from unitarity and helps gain a better understanding of the differences between the role of $\alpha_{11}$ and $\alpha_{22}$.

At present, the number of effective neutrinos is known to be $N_{\text{eff}}=2.99^{+0.34}_{-0.33}$ at 95$\%$~C.L. \cite{Planck:2018vyg}.  For this result, one can already constrain the non-unitarity of the three-neutrino mixing matrix, namely
\begin{align}
\centering
    & \alpha_{11} > 0.07 \, ,& & \alpha_{22}> 0.15 \, ,
\end{align}
both at $95\%$~C.L. Similarly, it is possible to study the sensitivity to the scenario in light of the forecasted CMB observations. Let us assume that $N_{\text{eff}}$ will be determined with an uncertainty $\sigma(N_{\text{eff}}) = 0.02$, as expected from CMB-S4 \cite{CMB-S4:2016ple}. Let us also assume a Gaussian posterior distribution for this parameter. We define the following $\chi^2$ test as
\begin{align}
    \chi^2 = \frac{[(N_{\rm eff})_0 - N_{\rm eff}]^2}{\sigma^2(N_{\text{eff}})} \, ,
    \label{eq:ch8-chi2}
\end{align}
where $(N_{\text{eff}})_0$ is the observed value of $N_{\text{eff}}$ which we assume that matches the standard theoretical prediction --- that is $(N_{\text{eff}})_0 = 3.044$. Then, future CMB measurements will significantly improve the existing limits from Planck, namely
\begin{align}
\centering
    & \alpha_{11} > 0.29 \,,& & \alpha_{22}> 0.50\, ,
\end{align}
at 95\% C.L.

\begin{figure*}[t!]
    \includegraphics[width = 0.37\paperwidth]{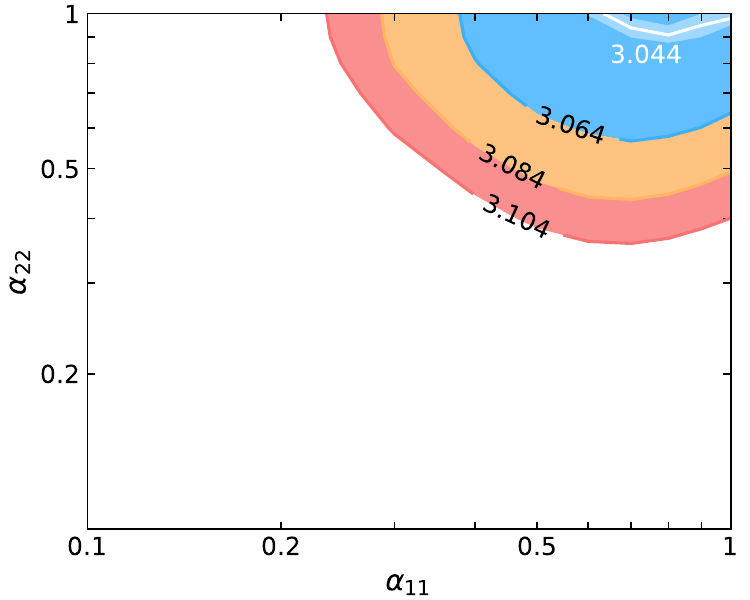} \raggedright
    \\
    \includegraphics[width = 0.37\paperwidth]{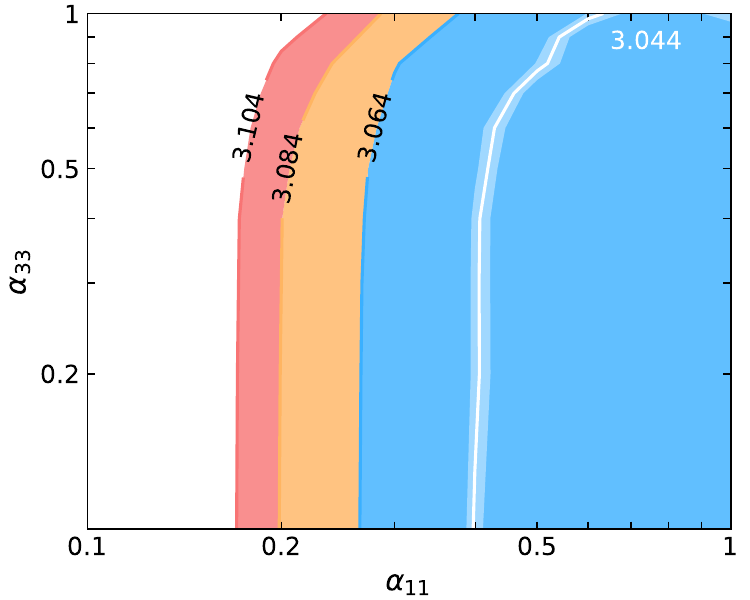}
    \includegraphics[width = 0.37\paperwidth]{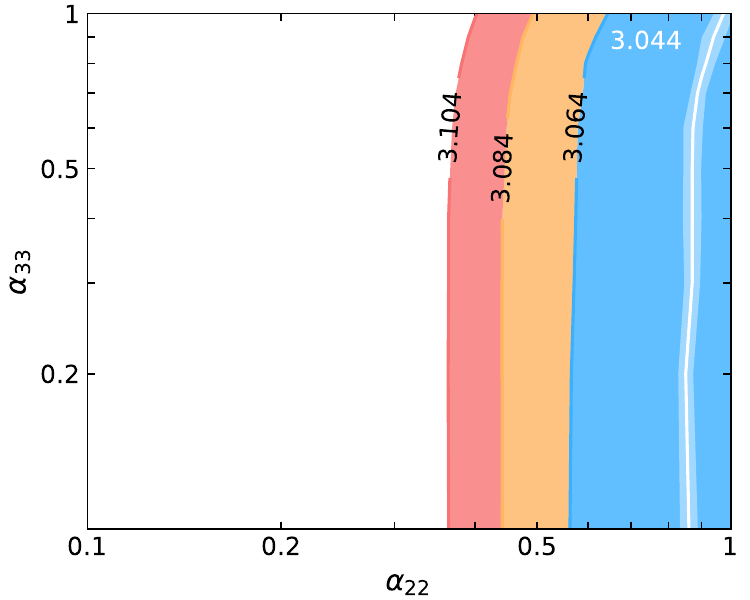}
    \caption{The three panels show the 1$\sigma$, 2$\sigma$ and 3$\sigma$ allowed regions expected from the measurement of $N_{\text{eff}}$ when two of the diagonal parameters are allowed to vary simultaneously. The top panel shows the $\alpha_{11} - \alpha_{22}$ plane. The bottom left and bottom right panels correspond to the $\alpha_{11} - \alpha_{33}$ and $\alpha_{22} - \alpha_{33}$ planes, respectively. The isocontour of $N_{\text{eff}}$ =$(N_{\rm eff})_0$=3.044  is shown in white, together with a shaded band indicating the numerical uncertainty --- which we consider to be $(N_{\rm eff})_0\pm 0.001$.\labfig{fig:ch8-nu_two_at_a_time}}
\end{figure*}

Next, we perform a more exhaustive study of the correlations between parameters. From the $\chi^2$ test definition in Equation \ref{eq:ch8-chi2}, one can also derive allowed regions when two or more parameters are varied simultaneously in the analysis. In \reffig{fig:ch8-nu_two_at_a_time}, we show the expected 1$\sigma$, 2$\sigma$ and 3$\sigma$ confidence levels in the two-dimensional planes defined when two of the diagonal parameters are allowed to depart from unitarity. The white line shows the region of parameter space where the standard value is recovered --- and the shaded banded corresponds to the region that is compatible given the estimated error from our choice of numerical settings for the calculation.

The top panel displays the allowed regions in the $\alpha_{11}-\alpha_{22}$ plane given the expected sensitivity of CMB-S4. These regions, although far from being the most restrictive limits, provide independent constraints on the non-unitarity diagonal parameters. The lower panels show the expected sensitivity in the $\alpha_{11}-\alpha_{33}$ and $\alpha_{22}-\alpha_{33}$ planes, which have been derived following the same procedure. As it was discussed in the light of the results shown in \reffig{fig:ch8-one_at_a_time}, even if $\alpha_{33}$ deviates significantly from unity, the predicted value $N_{\text{eff}}$ does not deviate remarkably. In particular, for $\alpha_{33}< 0.5$, the value of $N_{\text{eff}}$ is almost independent of $\alpha_{33}$. The reason is that this parameter only affects the interaction rates of processes involving tau neutrinos. These are known to play a minor role in the decoupling process. Conversely, having $\alpha_{11}$ and $\alpha_{22}$ different from one generally leads to a larger value of $N_{\text{eff}}$ as a consequence of their effect on $G_{\rm F}$. 

%%%%%%%%%%%%%%%%%%%%%%%%%
\subsection{Including off-diagonal parameters}
%%%%%%%%%%%%%%%%%%%%%%%%%%
Non-zero off-diagonal non-unitarity parameters --- $\alpha_{ij} > 0$ with $i < j$ --- modify the structure of neutrino interactions with the electrons and positrons in the plasma. Likewise, $\alpha_{21}$ changes the relation between the true value of the fundamental constant $G_{\text{F}}$ and the value accessible at experiments measuring muon decay. Notice that, in order for $\alpha_{ij}$ to differ from zero, $\alpha_{ii}$ and $\alpha_{jj}$ have to be different from one.

Here, we extend the previous analysis to include one non-zero off-diagonal parameter besides the two diagonal ones. For simplicity, we limit the study to real values of the parameters. This choice is motivated by the results in~\cite{Froustey:2021azz}, where it was shown that the CP phase does not alter the predicted $N_{\text{eff}}$ in the case of unitary mixing. As a consequence, only minor changes are expected from additional phases in the scenario that occupies us.

\reffig{fig:ch8-3at_a_time} shows the allowed regions in the  $\alpha_{11}-\alpha_{22}$ two-dimensional plane after profiling over the $\alpha_{21}$ parameter. The values of $\alpha_{21}$ are required to satisfy the conditions in Equations \ref{eq:ch8-NUcond1}, \ref{eq:ch8-NUcond2} and \ref{eq:ch8-NUcond3}. Comparing with the upper left panel of \reffig{fig:ch8-nu_two_at_a_time}, one can see that a non-zero $\alpha_{21}$ enlarges the parameter space allowed in this plane. 
Due to the effect on $G_{\text{F}}$, a non-zero $\alpha_{21}$ can partially compensate the larger value of $N_{\text{eff}}$ expected when $\alpha_{11}$ and $\alpha_{22}$ are different from unity.
Conversely, when the same calculation is repeated for non-zero values of the off-diagonal parameters $\alpha_{31}$ and $\alpha_{32}$, nearly imperceptible changes are found in the planes $\alpha_{11}-\alpha_{33}$ and $\alpha_{22}-\alpha_{33}$, respectively. This means that non-zero values of $\alpha_{31}$ and $\alpha_{32}$ increase the predicted value of $N_{\text{eff}}$ instead of reducing it. Therefore, after profiling over the off-diagonal parameters, the contours in the $\alpha_{11}-\alpha_{33}$ and $\alpha_{22}-\alpha_{33}$ planes remain almost unaltered. For this reason, we do not display these results.

\begin{figure}
    \centering
    \includegraphics[width=0.37\paperwidth]{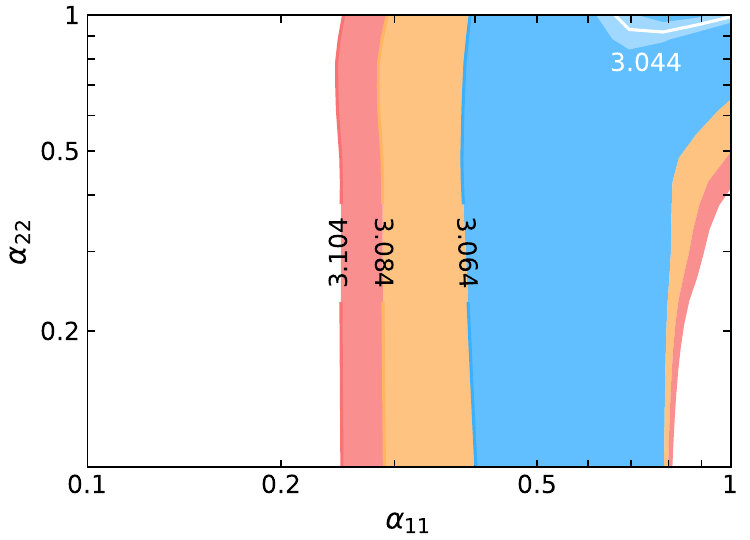}
    \caption{Expected allowed regions at 1$\sigma$, 2$\sigma$ and 3$\sigma$ significance from the measurement of $N_{\text{eff}}$ in the $\alpha_{11}-\alpha_{22}$ plane, when $\alpha_{21}$ is allowed to vary while respecting the unitarity condition and after profiling over $\alpha_{21}$.\labfig{fig:ch8-3at_a_time}}
\end{figure}

%%%%%%%%%%%%%%%%%%%%%%%%%%%%%%%%%%%%%%%%%%%%%%%%%%%%%%
%%%%%%%%%%%%%%%%%%%%%%%%%%%%%%%%%%%%%%%%%%%%%%%%%%%%%%
\section{Discussion}
\labsec{ch8-nudiscuss}
%%%%%%%%%%%%%%%%%%%%%%%%%%%%%%%%%%%%%%%%%%%%%%%%%%%%%%
%%%%%%%%%%%%%%%%%%%%%%%%%%%%%%%%%%%%%%%%%%%%%%%%%%%%%%

%%%%%%%%%%%%%%%%%%%%%%%%%%%%%%%%%%%%%%%%%%%%%%%%%%%%%%
\subsection{Synergies with non-standard interactions}
\label{subsec:ch8-nsi-nu}
%%%%%%%%%%%%%%%%%%%%%%%%%%%%%%%%%%%%%%%%%%%%%%%%%%%%%%
If the departures from unitarity of the three-neutrino mixing matrix are small, it is possible to write the low energy effective Lagrangian including the contributions from Equation~\ref{eq:ch8-nucc} and Equation~\ref{eq:ch8-nunc} in terms of the active neutrino states, as
\begin{align}
&\mathcal{L}_{\text{NU}} = 
     -2\sqrt{2}G_{F} \sum_{\alpha , \beta}\left(\bar{\nu}_\alpha \gamma^\mu P_L \nu_\beta\right)\left[ g_R\left[ N N^\dagger\right]^2_{\alpha \beta}\left(\bar{e}\gamma_\mu P_R e \right) \right. \nonumber \\ & \hspace{2cm}  \left.+\left((N N^\dagger)_{\alpha e}( N N^\dagger)_{e \beta} +g_L \left[ N N^\dagger\right]^2_{\alpha \beta}\right)\left(\bar{e}\gamma_\mu P_L e \right)\right]\,.
    \label{eq:ch8-nu-lag}
\end{align}

Then, it is possible to map this scenario of small deviations from unitarity to the one of neutrino non-standard interactions. To do that, one also needs to rewrite $G_{\text{F}}$ in terms of the quantity measured in muon decays --- in \mbox{Equation \ref{eq:ch8-GFmu}.} Let us define the Lagrangian including Standard Model interactions and neutral-current NSIs between neutrinos and electrons as
\begin{align}
&\mathcal{L}_{\text{SM+NSIe}}= - 2\sqrt{2}\,G_{\rm F}  \bigg[ \left(\overline{\nu}_e \gamma^\mu P_{L} e \right) (\overline{e}\gamma_\mu P_{L} \nu_e) \nonumber\\ & \hspace{3.5cm} +  \sum_{X, \alpha} g_{X}\left(\overline{\nu}_\alpha \gamma^\mu P_L \nu_\alpha \right) (\overline{e}\gamma_\mu P_{X} e)\nonumber \\ & \hspace{3.5cm} + \sum_{X, \alpha, \beta} \varepsilon_{\alpha \beta}^{eX} \left(\overline{\nu}_\alpha \gamma^\mu P_L \nu_\beta \right) (\overline{e}\gamma_\mu P_{X} e)\bigg],
\label{eq:ch8-nsie}
\end{align}
so that the effective NSI parameters read
\begin{align}
    \varepsilon ^{eL}_{\alpha \beta} = -(\delta_{\beta e}\delta_{\alpha e} +g_L\delta_{\alpha\beta}) + \frac{(N N^\dagger)_{\alpha e}( N N^\dagger)_{e \beta} + g_L\left[ N  N^\dagger\right]^2_{\alpha \beta}}{\sqrt{\alpha^2_{11} (\alpha^2_{22} + |\alpha_{21}|^2)}}~,
\end{align}
and
\begin{align}
    \varepsilon ^{eR}_{\alpha \beta} = -g_R\delta_{\alpha\beta }  + g_R\frac{\left[ N  N^\dagger\right]^2_{\alpha \beta}}{\sqrt{\alpha^2_{11} (\alpha^2_{22} + |\alpha_{21}|^2)}} \, .
\end{align}

Note that we only show the effective NSI parameters related to interactions with electrons since those are the only ones relevant for the neutrino decoupling. However, identical effective NSIs would also be expected between neutrinos and all charged leptons.

A better understanding of the results in the previous section can be achieved after computing the matrix elements $(N N^\dagger)_{\alpha \beta}$ ~\cite{Escrihuela:2016ube} at leading order. \reftab{ch8-nsi} provides a summary of the mapping between the diagonal non-unitarity parameters and the NSI ones. It is known that weaker interactions between neutrinos and the cosmic plasma would have resulted in an earlier decoupling. On the contrary, stronger interactions would have kept neutrinos in thermal contact with the plasma for a longer time and as a result, they would have received more energy from the electron-positron pair annihilation.

\begin{table*}
\renewcommand*{\arraystretch}{1.2}
    \centering
    \begin{tabular}{cll}
         \toprule[0.25ex] NU Parameter  &  \multicolumn{2}{l}{Effective NSI} \\ \midrule 
         $\alpha_{11} < $  1 & $\varepsilon^{eL}_{\alpha\beta } $ = &  $ ( 1+g_L)(-1+ \alpha_{11}^3)\delta_{\beta e}\delta_{\alpha e}$ \\ & & $+ g_L \left(\frac{1}{\alpha_{11}} -1\right)(\delta_ {\alpha \mu}\delta_{\beta \mu} +\delta_ {\alpha \tau}\delta_{\beta \tau})$   \\[2.4mm]
         & $\varepsilon^{eR}_{\alpha\beta } = $ &  $ -g_R(1 -\alpha^3_{11})\delta_{\alpha e}\delta_{\beta e} + g_R\left(\frac{1}{\alpha_{11} }-1 \right)(\delta_ {\alpha \mu}\delta_{\beta \mu} +\delta_ {\alpha \tau}\delta_{\beta \tau})$
          \\[2.4mm]
         \midrule
         $\alpha_{22} < $  1 & $\varepsilon^{eL}_{\alpha\beta } $ = &  $ -(1 +g_L)\left( 1 - \frac{1}{\alpha_{22}}\right)\delta_{\alpha e} \delta_{\beta e} $ \\ & & $-g_L(1-\frac{1}{\alpha_{22}})\delta_{\alpha\tau} \delta_{\beta \tau}- g_L(1 -{\alpha^3_{22}}) \delta_ {\alpha \mu}\delta_{\beta \mu}$   \\[2.4mm]
         & $\varepsilon^{eR}_{\alpha\beta } = $ &  $ -g_R\left(1 - \frac{1}{\alpha_{22}}\right)(\delta_{\alpha e}\delta_{\beta e} + \delta_{\alpha \tau }\delta_{\beta \tau} )- g_R(1-\alpha^3_{22}) \delta_ {\alpha \mu}\delta_{\beta \mu} $
        \\[2.4mm]
         \midrule
         $\alpha_{33} < $  1 & $\varepsilon^{eL}_{\alpha\beta}$ = &  $g_L(-1 + \alpha_{33}^4)\delta_{\alpha\tau}\delta_{\tau\beta}$  \\[2.4mm]
         & $\varepsilon^{eR}_{\alpha\beta }  = $ & $ g_R(-1 + \alpha_{33}^4)\delta_{\alpha\tau}\delta_{\tau\beta}$
         \\
         \bottomrule[0.25ex]
    \end{tabular}
    \caption{\label{tab:ch8-nsi} Mapping between the diagonal non-unitarity (NU) parameters and effective NSI with electrons.}

\end{table*}

With the results on \reftab{ch8-nsi}, one can better understand the results in \reffig{fig:ch8-one_at_a_time}. In particular, it is possible to see that, depending on the values of $\alpha_{11}$ and $\alpha_{22}$, the decoupling process is shifted to earlier or later times.
Regarding $\alpha_{33}$, it is clear that a value smaller than unity reduces the strength of the effective neutral-current interactions and therefore the decoupling would occur earlier.

Although useful, the validity of this approach is strictly limited to small deviations from unitarity. In any other case, it is necessary to work in the mass basis to include only the states that are kinematically accessible. A more detailed discussion on non-unitarity, light sterile neutrinos and non-standard interactions can be found in~\cite{Blennow:2016jkn}.

%%%%%%%%%%%%%%%%%%%%%%%%%%%%%%%%%%%%%%%%%%%%%%%%%%%%%%
\subsection{Other constraints on non-unitarity}
\label{subsec:ch8-otherlimits}
%%%%%%%%%%%%%%%%%%%%%%%%%%%%%%%%%%%%%%%%%%%%%%%%%%%%%%
The unitarity of the three-neutrino mixing matrix is also constrained by several flavour observables and precision measurements at terrestrial experiments. In this subsection, we provide an overview of the most restrictive bounds from references \cite{Antusch:2014woa,Antusch:2016brq,Escrihuela:2016ube,Escrihuela:2019mot, Fernandez-Martinez:2016lgt} to contextualise our results.

At tree level, the Fermi constant and the weak mixing angle are related, notably
\begin{align}
    \sin ^2\theta_W \cos^2 \theta_W = \frac{\alpha_{\text{EM}}\pi}{\sqrt{2}G_{\rm F} \,m^2_Z} = \frac{\alpha_{\text{EM}} \pi}{\sqrt{2}G^\mu_F \,m^2_Z}\sqrt{(N N^\dagger)_{ee}(NN^\dagger)_{\mu\mu}}\, ,
\end{align}
being $\alpha_{\text{EM}}$ the fine structure constant. Regarding the mass of the W boson, at tree level, it can be related to the Fermi constant,
\begin{align}
    m^2_W = \frac{\alpha_{\text{EM}} \pi}{\sqrt{2}G_{\rm F} \sin^2\theta_W}\,.
\end{align}
In both cases, requiring the agreement between the measurements of the weak mixing angle and the $m_W$ mass and the value inferred from $G_{\rm F}$, one can set limits on the unitarity of the three-neutrino mixing matrix.

Previously, we discussed the differences between the measurements of $G_{\rm F}$ derived from beta and muon decays when a non-unitary three-neutrino mixing is considered. Along these lines, it is possible to recast the limits on the unitarity of the Cabibbo-Kobayashi-Maskawa matrix into limits on the non-unitarity of the three-neutrino mixing matrix as follows:
\begin{align}
    \sum_{i=1}^3 |V_{ui}|^2 = \left(\frac{G^\beta_{\rm F}}{G^\mu_{\rm F}}\right)^2 = \frac{1}{\alpha_{22}^2 + |\alpha_{21}|^2}\, .
\end{align}

In the Standard Model, couplings between leptons and gauge bosons are flavour-independent and semileptonic decay rates too --- before introducing quantum corrections. This is no longer true in scenarios where non-unitarity is present. Then, the measurement of pion and kaon decays can restrict deviations from unitarity. Besides that, one can test universality in the leptonic sector and W-boson decays and relate them to the non-unitarity of the three-neutrino mixing matrix --- see for instance
\begin{align}
    R ^W_{\alpha \beta} = \frac{\Gamma(W\rightarrow l_\alpha  \bar{\nu}_\alpha ) }{\Gamma(W\rightarrow  l_\beta \bar{\nu}_\beta)} = \sqrt{\frac{(N N^\dagger)_{\alpha\alpha}}{(N N^\dagger)_{\beta\beta}}}\, .
\end{align}

Combining these and other probes it is possible to set stringent constraints to departures from unitarity~\cite{Antusch:2016brq,Fernandez-Martinez:2016lgt,Escrihuela:2016ube,Escrihuela:2019mot}. We have already commented on how non-unitarity would also alter neutrino oscillations. From the non-observation of deviations in the three-neutrino picture, and using data from both near and far detectors,  the most up-to-date limits were derived in~\cite{Forero:2021azc}. \reftab{ch8-bounds} summarises the main results.

\begin{table}
\renewcommand{\arraystretch}{1.2}
\centering
\begin{tabular}{lccc}
\toprule[0.25ex]
Diagonal NU parameters & $\alpha_{11}$ & $\alpha_{22}$ & $\alpha_{33}$ \\ 
\midrule
3$\sigma$ bounds & $>  0.93$ & $>  0.98$ & $>  0.72$ \\
\bottomrule[0.25ex]
& & & \\
\toprule[0.25ex]
Off-diagonal NU parameters & $|\alpha_{21}|$ & $|\alpha_{31}|$ & $|\alpha_{32}|$  \\ 
\midrule
3$\sigma$ bounds   & $< 0.025$ & $< 0.075$ & $<  0.02$ \\
\bottomrule[0.25ex]
\end{tabular}
\caption{\label{tab:ch8-bounds}
Current constraints on the non-unitary parameters from neutrino oscillation data \cite{Forero:2021azc}.}

\end{table}

Table \ref{tab:ch8-limitsleptons} includes the constraints obtained in \cite{Escrihuela:2016ube} after combining data from neutrinos and charged leptons. One can see that the limits from neutrino oscillations  --- reported in \reftab{ch8-bounds}~\cite{Forero:2021azc} --- are not really competitive yet they provide an independent cross-check. 
\begin{table}
\renewcommand*{\arraystretch}{1.2}
    \begin{tabular}{lccc}
    \toprule[0.25ex]
        Diagonal NU parameters & $\alpha_{11}$ &  $\alpha_{22}$ &  $\alpha_{33}$ \\ \midrule
          90$\%$ C.L. bounds & $>$ 0.9961 & $>$ 0.9990 & $>$ 0.9973  \\ \bottomrule[0.25ex]
	& & & \\    
    \toprule[0.25ex]
        Off-diagonal NU parameters &  $|\alpha_{21}|$ &  $|\alpha_{31}|$ &  $|\alpha_{31}|$   \\ \midrule
          90$\%$ C.L. bounds & $<$ 2.6$\cdot 10 ^{-3}$ & $<$ 5.0$\cdot 10 ^{-3}$ & $<$ 2.4$\cdot 10 ^{-3}$ \\ \bottomrule[0.25ex]
    \end{tabular}
    \caption{Combined limits on the non-unitarity parameters $\alpha_{ij}$ from charged lepton and neutrino measurements at 90\% C.L.  --- considering 6 degrees of freedom --- from \cite{Escrihuela:2016ube}.}
    \label{tab:ch8-limitsleptons}
\end{table}

%%%%%%%%%%%%%%%%%%%%%%%%%%%%%%%%%
%%%%%%%%%%%%%%%%%%%%%%%%%%%%%%%%%
\section{Concluding remarks}
\label{sec:ch8-conclusion}
%%%%%%%%%%%%%%%%%%%%%%%%%%%%%%
%%%%%%%%%%%%%%%%%%%%%%%%%%%%%%%%

In this chapter, we have analysed the impact of a non-unitary three-neutrino mixing matrix in the process of neutrino decoupling. The discovery of \mbox{non-unitarity} would be a clear low-energy manifestation of the presence of heavy neutral leptons. Since it alters the strength and structure of neutral-current and charged-current interactions, it also affects how neutrinos decouple from the cosmic plasma in the early Universe.

We have addressed the deviations from the standard prediction for $N_{\text{eff}}$ in the presence of a non-unitary mixing. Likewise, using the parametrisation in Equation \ref{eq:ch8-parametrisation}, we have derived the first limits from cosmology in the parameters $\alpha_{11}$ and $\alpha_{22}$. Non-zero values of diagonal parameters $\alpha_{11}$ and $\alpha_{22}$ would delay neutrino decoupling, resulting in a larger value of $N_{\text{eff}}$. From current cosmological observations, we have derived the following bounds at 95\% C.L.:
\begin{align}
\centering
    & \alpha_{11} > 0.07 \, ,& & \alpha_{22}> 0.15 \, .
\end{align}
Besides that, we have studied the sensitivity expected from forthcoming cosmological observations. When considering the future CMB-S4 mission combined with large-scale structure observations, we find that the limits will be improved to be  $\alpha_{11} > 0.29$ and $\alpha_{22}> 0.50$, both at~$95\%$~C.L. Regarding the third non-unitarity diagonal parameter, $\alpha_{33}$, its role in neutrino decoupling can be safely neglected since it would barely alter the overall picture. We have addressed the interplay between several non-zero parameters. Interestingly, we find that a non-zero value $\alpha_{21}$ could relax the limits in the $\alpha_{11}-\alpha_{22}$ plane considerably. 

This chapter shows how future cosmological measurements will provide \mbox{high-redshift} independent constraints on the non-unitarity of the three-neutrino mixing matrix, which are complementary to laboratory-based searches. Hence, they could falsify neutrino mass mechanisms predicting large departures from unitarity. As a final remark, we should emphasise that, in spite of being considerably weaker than terrestrial limits, these constraints from cosmological observation are independent and serve as a consistency check of our understanding of the neutrino decoupling process. In that sense, and given that the non-unitarity of the three neutrino mixing matrix is being tested at a different epoch in the universe, they should be regarded as supportive searches.

\pagelayout{wide} % No margins
\addpart{Neutrino connections to dark matter}
\pagelayout{margin} % Restore margins
\chapter{Neutrinos from primordial black hole evaporation}
%\addcontentsline{toc}{chapter}{Neutrinos from primordial black hole evaporation} 
\labch{ch9-nupbh}

Primordial black holes (PBHs) could conform --- at least --- part of the dark matter in the universe. This dark matter candidate has attracted a lot of attention from the scientific community due to the many different ways in which their existence could be proved. Among them, one finds Hawking radiation \cite{Hawking:1974rv}, which would consist of the emission of particles at the event horizon of a  black hole due to quantum effects in curved spacetime. In particular, for relatively light PBHs, with masses between $10^{15} \, -\, 10^{17}$g, the expected neutrino emission from Hawking radiation would fall within the reach of future neutrino observatories. 

In this chapter, we address the sensitivity of such experiments to neutrinos from the evaporation of PBHs under several different assumptions such as the PBH spin or the mass distribution of the hypothetical PBH population. In \refsec{ch9-hawking}, we present the concept of Hawking radiation and the computational tools employed in the analysis. Next, in \refsec{ch9-nuflux}, we discuss the different components and spectral features of the expected neutrino flux. Finally, the sensitivity of future neutrino observatories DUNE and THEIA is presented in \refsec{ch9-observatories} and the results are discussed in \refsec{ch9-discussion}.

\section{Hawking evaporation in a nutshell}
\labsec{ch9-hawking}

\subsection{Evaporation of Schwarzschild and Kerr Black Holes}
Schwarzschild black holes are spherically symmetric and their evaporation can be predicted solely in terms of their mass, $M_{\text{BH}}$. Their horizons emit elementary particles as blackbodies with a temperature ~\cite{Page:1976df,MacGibbon:1990zk,MacGibbon:1991tj,MacGibbon:2007yq}
\begin{equation}
    T = \frac{1}{8\pi M_{\text{BH}}} \indent ,
\end{equation}
where we have chosen units such that the gravitational constant $G=1$. The spectrum of particles of type $i$ emitted per unit of time and energy is
\begin{equation}
     \frac{\text{d}^2 \mathcal{N}_i}{\text{d}E \text{d}t} = \sum \frac{1}{2\pi}\frac{\Gamma_i (E,M_{\text{BH}}, s , m_0)}{e^{E/T} \pm 1} (2l+1)\,,
\end{equation}
where the sum is over the total multiplicity of the particles --- colour, helicity and angular momentum --- and $m_0$ and $s$ are the mass and spin of the emitted particles. The global $(2l+1)$ factor comes from the fact that for a spherically symmetric --- Schwarzschild  --- black hole, all the projections of the angular momentum $m \in \lbrace -l,\, l \rbrace$ contribute equally. The function $\Gamma_i$ is the so-called greybody factor and it describes the probability that a spherical wave representing an elementary particle produced at the horizon of the black hole escapes to infinity. 

Kerr black holes are axially symmetric, uncharged rotating black holes and their rotation can be due to the formation mechanism, accretion or mergers. In addition to the mass, they are described by their spin per unit mass \mbox{$a =J_{\text{BH}}/M _{\text{BH}}\in \lbrace 0, M_{\text{BH}} \rbrace$,} where $J_{\text{BH}}$ is the angular momentum of the black hole. Alternatively, one can define the reduced spin parameter, $a_{*} = a/(GM_{\text{BH}})$. Then, the temperature of a rotating black hole is given by~\cite{Page:1976ki,}
\begin{equation}
    T = \frac{1}{4M_{\text{BH}}\pi} \frac{\sqrt{1 -a_{*}^2}}{1 + \sqrt{1 -a_{*}^2}}\, .
\end{equation}

In this case, the emission of particles of type $i$ per unit of time and energy is
\begin{align}
     \frac{\text{d}^2 \mathcal{N}_i}{\text{d}E \text{d}t} = \sum \frac{1}{2\pi}\frac{\Gamma_i (E,M_{\text{BH}},a_{*}, m_0, s)}{e^{\Tilde{E}/T} - (-1)^{2s}}\,.
\end{align}
Here, we sum again over the degrees of freedom of the emitted particles. We have also defined $\Tilde{E}$ as the effective energy of the emitted particles, including the black hole rotational velocity,
\begin{align}
\Tilde{E} = E - m \frac{a_{*}}{2M_{\text{BH}}( 1 + \sqrt{1-a^2_{*}})}\, .
\end{align}

For a given population of black holes, the rate of emission of particles per unit of time and energy is 
\begin{equation}
    \frac{\text{d} ^2 N_i}{\text{d}t \text{d}E} = \int _{M^{\text{min}}_{\text{BH}}} ^{M^{\text{max}}_{\text{BH}}} \int _{a^{\text{min}}_{*}}^{a^{\text{max}}_{*}} \frac{\text{d}^2 \mathcal{N}_i}{\text{d}E \text{d}t} \frac{\text{d}^2n}{\text{d}M_{\text{BH}} \text{d}a_{*}} \text{d}M_{\text{BH}} \text{d}a_{*}\,,
\end{equation}
where $\text{d}^2 n/\text{d}M_{\text{BH}}\text{d}a_{*}$ is the mass and spin distribution of the population of black holes.

Since some of the particles emitted are not stable or only exist when forming hadrons, the expected flux of a population of black holes does not depend only on this primary emission.  Therefore, one expects a secondary component of the flux resulting from particle decay and hadronisation. The emission rate of secondary particles $j$ per units of time and energy is
\begin{equation}
    \frac{\text{d}^2 N_j}{\text{d}t\text{d}E} = \int \sum_i \frac{\text{d}^2N_i}{\text{d}t\text{d}E'} \frac{\text{d}\mathcal{N}^i_j (E, E')}{\text{d}E} \text{d}E'\,.
\end{equation}
Here, $\text{d}\mathcal{N}^i_j (E, E')/\text{d}E$ is the spectrum of particles of type $j$ with energy E resulting from the decay of particles of type $i$ with energy $E'$ and from hadronisation processes. 

\subsection{Black Hole evolution}
Hawking evaporation induces mass and angular momentum losses in black holes, 
\begin{align}
&\frac{\text{d}M_{\text{BH}}}{\text{d}t} = - \frac{f(M_{\text{BH}}, \, a_{*})}{M_{\text{BH}}^2}\, , \\
&\frac{\text{d}J_{\text{BH}}}{\text{d}t} = - \frac{a_{*}g(M_{\text{BH}},\, a_{*})}{M_{\text{BH}}}.
\end{align}
The Page factor $f(M_{\text{BH}}, a_{*})$ is related to the number of quantum degrees of freedom that a black hole of mass of $M_{\text{BH}}$ can emit and the impact of this emission on the mass loss. It is given by
\begin{align}
    f(M_{\text{BH}},\, a_{*}) &= -M_{\text{BH}}^2 \frac{\text{d}M_{\text{BH}}}{\text{d}t} \nonumber\\ &= M_{\text{BH}}^2 \int_{0}^{\infty} \frac{E}{2\pi} \sum_i \sum \frac{\Gamma_i (E,M_{\text{BH}}, a_{*}, m_0, s)}{e^{E/T} - (-1)^{2s}} \text{d}E.
\end{align}
The second Page factor is
\begin{align}
    g(M_{\text{BH}}, a_{*}) & = - \frac{M_{\text{BH}}}{a_{*}} \frac{\text{d}J_{ \text{BH}}}{\text{d}t} \nonumber \\ &= - \frac{M_{\text{BH}}}{a_{*}}\int_{0}^{\infty} \sum_{i} \sum \frac{m}{2\pi}\frac{\Gamma_i (E,M_{\text{BH}}, a_{*}, m_0, s)}{e^{E'/T} - (-1)^{2s}} \text{d}E \, .
\end{align}
It accounts for the angular momentum loss occurring in a rotating --- axially symmetric --- black hole due to the enhancement in the emission of particles with high angular momentum, which effectively extract angular momentum from the system.

\subsection{\texttt{BlackHawk}}
We rely on the publicly available code \texttt{BlackHawk} \cite{Arbey:2019mbc,Arbey:2021mbl} to compute the emission of elementary particles from the evaporation of a black hole, given its mass and angular momentum. The secondary emission of particles is computed using \texttt{PYTHIA} \cite{Sjostrand:2014zea} .\footnote{This option is implemented as part of \texttt{BlackHawk}. Other choices for computing the hadronisation, like \texttt{HERWIG} \cite{Bellm:2019zci}, exist too.} With \texttt{BlackHawk}, we also calculate the flux emitted as a function of time, taking into account the evolution of the mass and angular momentum of the black hole. By default, neutrinos are treated as massless Majorana fermions in the code. We comment on the difference between Majorana and Dirac neutrinos in the text. Regarding neutrino masses, they are of no relevance for the energy range that we are interested in and would only play a role for emission at lower energies, when~$T \sim m_\nu$~\cite{Lunardini:2019zob}. 

\section{Neutrinos from primordial black hole evaporation}
\labsec{ch9-nuflux}

The neutrino spectrum from the evaporation of a distribution of primordial black holes (PBHs) has two components. The first one is composed of neutrinos that are emitted directly from the PBH. The second component results from the hadronisation and subsequent decays of the unstable particles radiated. We refer to these two contributions as \textit{primary} and \textit{secondary}, respectively. Likewise, one expects contributions from primordial black holes in the galactic halo and outside of it. We denote these two contributions as \textit{galactic} and \textit{extragalactic}, respectively, 
\begin{equation}
 \frac{\text{d}\phi^\nu_{\rm tot}} {\text{d}E}= \frac{\text{d}\phi^\nu_{\rm gal}} {dE} +
 \frac{\text{d}\phi^\nu_{\rm exg}} {\text{d}E}.
 \label{eq:ch9-totaldiffflux}
 \end{equation}

Let us consider a monochromatic distribution of non-rotating primordial black holes, notably
\begin{align}
 \frac{\text{d}^2 n_{\text{PBH}}}{\text{d}M'_{\text{PBH}}\text{d}a'_{*}} =\delta (M'_{\text{PBH}} - M_{\text{PBH}})\delta (a'_{*})\, .
\end{align} 
Then, the galactic contribution is
\begin{equation}
 \frac{\text{d}\phi^\nu_{\rm gal}} {\text{d}E}=  \frac{f_{\rm PBH}}{M_{\rm{PBH}}}   \frac{\text{d}^2N}{\text{d}E \text{d}t} \int \frac{1}{4 \pi} \text{d} \Omega \int \rho_{\rm MW}\,[r(\ell,\psi)]\,\text{d}\ell \,,
 \label{eq:ch9-dFdEgal}
 \end{equation}
where $f_{\rm PBH}$ is the fraction of dark matter in the form of primordial black holes. The dark matter profile of the Milky Way (MW) halo --- $\rho_{\rm MW}\,[r(\ell,\psi)]$ --- is expressed in terms of the galactocentric distance,
\begin{align}
r(\ell,\psi) = \sqrt{d_{\odot}^2 + \ell^2 - 2\ell d_\odot \rm cos(\psi)}\, ,
\end{align} 
which depends on the galactocentric distance of the Sun, $d_\odot$, and the line of sight, $\ell$. It also depends on the angle of view, $\psi$, defined by the line of sight and the differential solid angle considered, $\text{d} \Omega$. Throughout our analysis, we assume that the density profile of the Milky Way halo follows the \mbox{Navarro-Frenk-White} (NFW) parametrisation~\cite{Navarro:1995iw},
\begin{align}
\rho_{\text{MW}} (r) = \frac{\rho_0}{\frac{r}{r_s}\left(1 + \frac{r}{r_s}\right)^2}\, ,
\end{align} 
with scale radius $r_s = 20$ kpc and normalisation $\rho_0 = 0.4~ \rm GeV/cm^3$. Had we considered a cored dark matter density profile instead --- for instance an isothermal profile with scale radius $r_s = 1.5$ kpc --- the galactic contribution would be approximately a factor 2 larger and our final results would change accordingly. Notice that, when computing the galactic contribution, we are neglecting the redshift of the spectrum.

For the extragalactic contribution, we assume that, at sufficiently large scales, dark matter follows an isotropic distribution. Then, the flux of neutrinos would be the full-sky integrated redshifted emission from all epochs~\cite{Carr:2009jm,Arbey:2019vqx,Dasgupta:2019cae},
\begin{equation}
\frac{d\phi^\nu_{\rm exg}} {dE}= 
%\frac{\Delta \Omega}{4\pi} 
\frac{f_{\rm{PBH}}\,\bar{\rho}_{\rm DM}}{M_{\rm{PBH}}}  \int_{t_{\rm min}}^{t_{\rm max}} dt [1+z(t)]\, \frac{d^2N}{dE_0 dt}\Bigr\rvert_{E_0 = [1+z(t)]E}\,.
\label{eq:ch9-dFdEexg}
\end{equation}
We take the average dark matter density at present to be \mbox{$\bar{\rho}_{\rm DM} = 2.35\times 10^{-30}$ g cm$^{-3}$} as from~\cite{Planck:2018vyg}. Regarding the integral limits, we consider $t_{\rm min} = 1 s$ and  $t_{\rm max}$ to be the minimum value between the primordial black hole lifetime and the age of the Universe. \footnote{Note that, according to~\cite{Dasgupta:2019cae}, changing the lower limit of the integral has essentially no impact on the results.} The neutrino energy at the source $E_0$ is related to the neutrino energy in the observer's frame via the redshift parameter $z(t)$. The remaining cosmological parameters are fixed to the latest Planck results~\cite{Planck:2018vyg}. 

\reffig{fig:ch9-totaldFdE} shows the total $\nu_e$ flux from PBH evaporation expected on Earth, including the primary and secondary components of the galactic and extragalactic contributions. The dark blue lines correspond to a primordial black hole with a mass $M_{\rm{PBH}} = 1\times 10^{15}$ g, whereas the fluxes from PBHs with masses  $5\times 10^{15}$ g and $1\times 10^{16}$ g are indicated in blue and cyan, respectively. For each mass, the results are shown for three different spins --- $a_*=0$,  $a_*=0.5$, and $a_*=0.9$ --- with solid, dashed and large-dashed lines. For illustrative purposes, we assume the totality of the dark matter of the universe exists in the form of primordial black holes, i.e.~$f_{\rm PBH} =1$. In all cases, we have assumed a monochromatic distribution in mass and spin, i.e
\begin{align}
\frac{\text{d}^2 n}{\text{d}M'_{\text{PBH}}\text{d}a'_*} = \delta (M'_{\text{PBH}}- M_{\text{PBH}})\, \delta (a'_* - a_*)\, .
\label{eq:ch9-monochrom}
\end{align}
One can see that for larger PBH masses the emitted flux in the region of interest is reduced and shifted towards smaller neutrino energies. As shown in~\reffig{fig:ch9-totaldFdE}, for PBHs with a large spin parameter, the flux features a smooth contribution and several characteristic peaks. The dominant --- smooth --- contribution to the flux stems from particles emitted with no additional angular momentum aside from their intrinsic spin, whereas the peaks are related to the emission of particles with angular momentum. For a detailed discussion, see for instance~\cite{Page:1976df,Page:1976ki}. 

\begin{figure}
\centering
\includegraphics[width=0.66\textwidth]{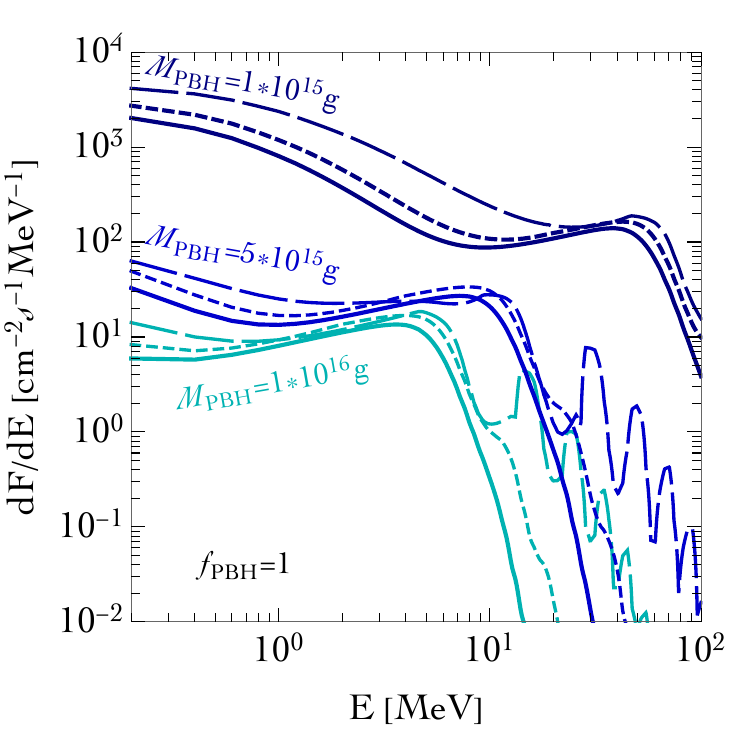}
\caption{Total electron neutrino flux from PBHs, for three different PBH masses  --- $M_{\rm{PBH}} = 1\times 10^{15}$ g  in dark blue, $5\times 10^{15}$ g in blue and $1\times 10^{16}$ g in cyan --- and spins --- solid lines for $a_*=0$, dashed lines for $a_*=0.5$ and large-dashed lines for $a_*=0.9$. We assume $f_{\text{PBH}}= 1 $ and a monochromatic distribution of PBHs, as in Equation \ref{eq:ch9-monochrom} .\labfig{fig:ch9-totaldFdE} }
\end{figure}

So far, we have considered a monochromatic mass distribution. However, formation mechanisms generally predict extended mass distributions for PBHs. The constraints derived on the mass and fraction of primordial black holes in the universe depend on their underlying distribution~\cite{Carr:2017jsz,Kuhnel:2017pwq,Bellomo:2017zsr,Arbey:2019vqx,Dasgupta:2019cae}. In order to study the impact of the mass distribution in our limits, we consider a more realistic distribution, in particular, a log-normal distribution~\cite{Carr:2017jsz} for the comoving number density, namely
\begin{equation}
\frac{\text{d}^2n_{\rm PBH}}{\text{d}M'_{\rm PBH}\text{d}a_{*}}= \frac{1}{\sqrt{2\pi}\sigma M_{\rm PBH }}\, \exp\left[- \dfrac{ {\rm ln} \left(M_{\rm PBH}/M_c\right)^2}{2 \sigma^2} \right]\delta(a_{*})\,,
\label{eq:ch9-log-normal-mass-distribution}
\end{equation} 
where $M_c$ is the critical mass --- related to the maximum of the distribution --- and $\sigma$ is its width. The distribution is normalised to unity when the lower and upper integration limits tend to 0 and infinity, respectively. Note that this condition can not be exactly met numerically. Nonetheless, we adapt our sampling of PBH masses over more than two orders of magnitude to ensure the validity of the result.

In this case, the galactic and extragalactic contributions to the neutrino flux read
\begin{align}
 \frac{\text{d}\phi^\nu_{\rm gal}} {\text{d}E} = \frac{f_{\rm PBH}}{\overline{M}_{\rm{PBH}}} \int \frac{1}{4 \pi} \text{d} \Omega & \int \rho_{\rm MW}\,[r(\ell,\psi)]\,\text{d}\ell \nonumber \\ &\times \int_{M^{\rm min}_{\rm PBH}}^{M^{\rm max}_{\rm PBH}} \text{d} M_{{\rm PBH}}\frac{\text{d}n_{\rm{PBH}}}{\text{d}M_{\rm{PBH}}}\, \frac{\text{d}^2\mathcal{N}}{\text{d}E \text{d}t} \,,
 \label{eq:ch9-dFdEgal-lognorm}
 \end{align}
 and
\begin{align}
\frac{\text{d}\phi^\nu_{\rm exg}} {\text{d}E}= &
\frac{f_{\rm{PBH}}\,\bar{\rho}_{\rm DM}}{\overline{M}_{\rm{PBH}}}  \int_{t_{\rm min}}^{t_{\rm max}} \text{d}t [1+z(t)]\, \nonumber \\ &\times  \int_{M_{\rm min}}^{M_{\rm max}} \text{d}M_{{\rm PBH}} \frac{\text{d}\mathcal{N}_{\rm{PBH}}}{\text{d}M_{\rm{PBH}}} \frac{\text{d}^2N}{\text{d}E_0 \text{d}t}\Bigr\rvert_{E_0 = [1+z(t)]E}\,,
\label{eq:ch9-dFdEexg-lognorm}
\end{align}
where $\overline{M}_{\rm PBH}$ denotes the mean mass of the distribution in Equation~\ref{eq:ch9-log-normal-mass-distribution}. In \reffig{fig:ch9-lognorm}, we show the expected neutrino fluxes from evaporating PBHs with a log-normal mass distribution, assuming different values of the standard deviation, and comparing them with the expectations from a monochromatic mass distribution. One can see that when a wider mass distribution is considered, the characteristic peak of the primary galactic contribution is smoothed down and, consequently, the flux becomes larger at higher energies.
 
\begin{figure}
    \centering
    \includegraphics[width = 0.66\textwidth]{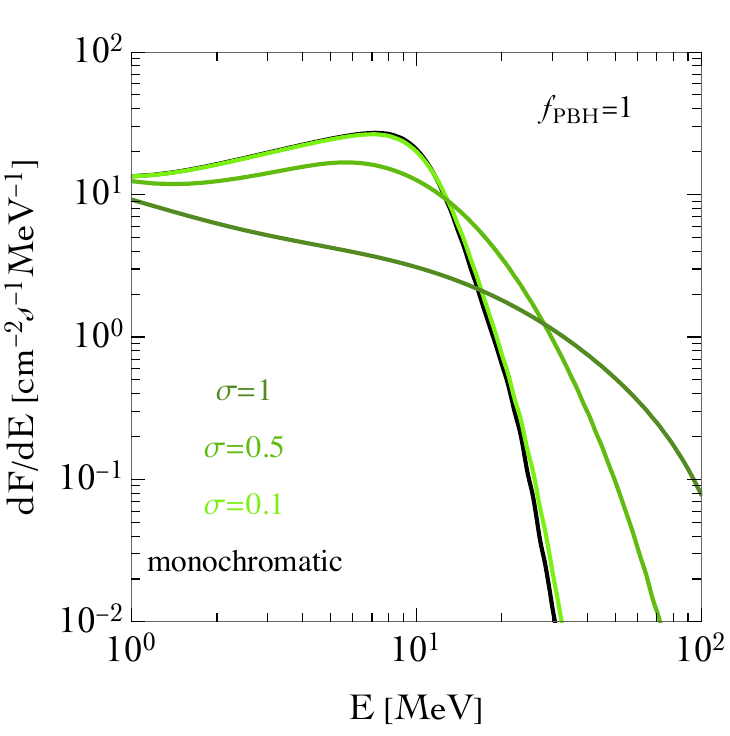}
    \caption{\labfig{fig:ch9-lognorm}Comparison between the total neutrino flux expected from PBH evaporation, for a monochromatic mass distribution with $M_{\rm PBH} = 5 \times 10^{15}$ g  --- in black --- and a log-normal distribution with critical mass \mbox{$M_c = 5 \times 10 ^{15}$ g} and $\sigma = \lbrace 0.1, 0.5, 1 \rbrace$ --- in light, medium and dark green respectively. The fraction of dark matter consisting of PBHs is fixed to $f_{\rm PBH} = 1$ and  $a_*=0$.}
\end{figure}

As we have mentioned, the emission of particles depends on the number of degrees of freedom of the elementary particles. Hence, a slightly larger differential flux would be expected in the case of Dirac neutrinos~\cite{Lunardini:2019zob}. In~\reffig{fig:ch9-vDiracvsMaj}, we show the total differential flux of electron neutrinos from a monochromatic population of PBHs with masses $M_{\rm PBH}=10^{15}$ g and fixing $f_{\rm PBH}=5.5 \times 10^{-4}$ --- which corresponds to the current upper limit from Super-Kamiokande~\cite{Dasgupta:2019cae}. The results for Majorana neutrinos correspond to the solid blue line, while the cyan, dashed curve refers to Dirac neutrinos. As one can see, for Dirac neutrinos --- which have double the number of degrees of freedom --- the flux around the peak is approximately a factor 2 large~\cite{Lunardini:2019zob}. Nevertheless, notice that a neutrino detector would never observe a difference due to the nature of neutrinos, since in the case of Dirac neutrinos, half of the emitted states would be inert.

\begin{figure}
\centering
\includegraphics[width=0.66\textwidth]{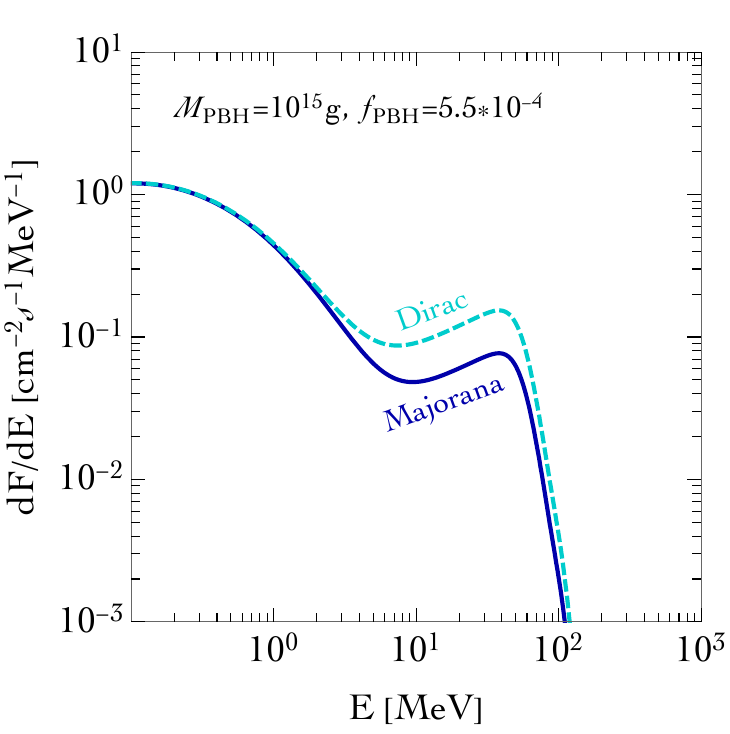}
\caption{Total differential electron neutrino flux from PBHs of mass \mbox{$M_{\rm PBH}=10^{15}$ g} and $a_*=0$, and assuming $f_{\rm PBH}=5.5 \times 10^{-4}$ . The blue, solid line is for Majorana neutrinos, while the cyan, dashed curve refers to Dirac neutrinos.\labfig{fig:ch9-vDiracvsMaj}}
\end{figure}

As a last remark, let us mention that our analysis does not include neutrino flavour mixing. We consider that the primary component of the flux -- which arises directly from evaporation --- consists of mass eigenstates. Moreover, it has been shown that the effect of their masses is negligible for the energy range of interest~\cite{Lunardini:2019zob}. Consequently, flavour conversions would only be relevant for the secondary component of the flux. This contribution to the total flux stems from the decay of heavier particles. Hence, neutrinos are produced as flavour eigenstates. Including the effects of neutrino mixing, the secondary flux expected on Earth reads
\begin{equation}
    \label{eqn:ch9-osc}
    \left. \frac{d\phi^{\nu_e}_{\rm sec}}{dE} \right\vert_{\rm detector} = \sum_{\alpha } P (\nu_\alpha \rightarrow \nu_e)   \frac{d\phi^{\nu_\alpha}}{dE} \approx \sum_{\alpha, j} |U_{\alpha j}|^2 |U_{ej}|^2  \frac{d\phi^{\nu_\alpha}}{dE},
\end{equation}
where the Greek and Latin indices refer to flavour and mass eigenstates, respectively. Given the values of the oscillation parameters in \cite{deSalas:2020pgw}, including neutrino mixing would reduce the total differential flux by less than $\sim 2$\%. In addition, this reduction will be mainly present at the lowest end of the energy range accessible at neutrino observatories. Therefore, for this work, we neglect the role of flavour mixing. However, neutrino mixing changes considerably --- approximately in a $\mathcal{O} (10\%)$ --- the expected flux of electron neutrinos for energies much smaller than the ones at which the spectrum of the galactic contribution peaks.

%%%%%%%%%%%%%%%%%%%%%%%%%%%%%%%%%%%%%%%%%%%%%%%%%%%%%%%%%%%%%%%%
%%%%%%%%%%%%%%%%%%%%%%%%%%%%%%%%%%%%%%%%%%%%%%%%%%%%%%%%%%%%%%%%
\section{Sensitivity of future neutrino observatories: DUNE and THEIA}
\labsec{ch9-observatories}
%%%%%%%%%%%%%%%%%%%%%%%%%%%%%%%%%%%%%%%%%%%%%%%%%%%%%%%%%%%%%%%%
%%%%%%%%%%%%%%%%%%%%%%%%%%%%%%%%%%%%%%%%%%%%%%%%%%%%%%%%%%%%%%%%
The expected flux of MeV neutrinos can be sizeable in magnitude if it comes from the evaporation of primordial black holes with masses \mbox{$M_{\rm PBH}\sim 10^{15} -10^{16}$ g,} as shown in \reffig{fig:ch9-totaldFdE}. Even if those PBHs do not form the totality of the dark matter of the universe, small fluxes of MeV neutrinos --- reduced proportionally to the value of $f_{\text{PBH}}$ --- may still be detected at future neutrino experiments. 
Heavier, asteroid-mass PBHs, are more interesting as dark matter candidates since they are currently unconstrained. Nonetheless, due to the larger masses, they would produce a weaker flux of neutrinos with sub-MeV energies. As a consequence, their detection is not realistic since the flux of solar and reactor neutrinos at the same energies would act as an irreducible --- and much larger --- background for these searches.  

In this section, we focus on the capabilities of DUNE --- a liquid argon time-projection chamber --- and THEIA --- a proposed water-based liquid scintillator neutrino detector --- to test the existence of primordial black holes. Other current experiments --- such as Super-Kamiokande~\cite{Beacom:2003nk, Kibayashi:2009ih} --- and \mbox{next-generation} neutrino observatories --- like JUNO~\cite{Wang:2020uvi,JUNO:2021vlw}, \mbox{Hyper-Kamiokande~\cite{Hyper-Kamiokande:2018ofw},} and coherent elastic neutrino-nucleus scattering detectors --- can also search for a flux of neutrinos originated from PBHs.  

%%%%%%%%%%%%%%%%%%%%%%%%%%%%%%%%%
\subsection{The DUNE experiment}
%%%%%%%%%%%%%%%%%%%%%%%%%%%%%%%%%

For the far detector of the DUNE experiment, we consider four tanks of liquid argon with a total fiducial mass of 40 kton~\cite{DUNE:2020ypp}. Although several independent detection channels are available in such a detector, we consider only the interactions $\nu_e + ^{40}$Ar $\rightarrow e^- + ^{40}K^*$ and focus on its potential to detect electron neutrinos. Then, the predicted $\nu_e$ event rate is computed using {SNOwGLoBES}~\cite{snowglobes} --- a numerical tool which is partially based on \texttt{GLoBES}~\cite{Huber:2004ka,Huber:2007ji}. We assume an energy resolution~\cite{snowglobes,ICARUS:2003zvt} 
\begin{equation}
    \left(\frac{\sigma}{E}\right)^{2} = \left(\frac{0.11}{\sqrt{E~({\rm MeV})}}\right)^{2}~+~(0.02)^{2} \,. 
\end{equation}
For simplicity, we also assume perfect efficiency in the detection.

However, the detection of neutrinos from primordial black hole evaporation is rather challenging given the numerous backgrounds at the energy range of interest. On the one hand, solar neutrinos and atmospheric neutrinos constitute the dominant and irreducible backgrounds for these searches. Therefore, we restrict our analysis to energies between 16 MeV and 100 MeV, where the corresponding fluxes are relatively small. Actually, in that energy window, we can neglect the flux of solar neutrinos. Concerning atmospheric neutrinos, we rescale the prediction for the atmospheric neutrino flux calculated with a FLUKA simulation for the Gran Sasso laboratory in the range of energies between 100 MeV and 300 MeV~\cite{Battistoni:2005pd}.\footnote{The latitude of the Gran Sasso laboratory and the one of the Sanford Underground Research Facility (SURF) are similar so that one should not worry about the dependence of the flux on the latitude. The flux predicted with FLUKA is then rescaled to the HKKM atmospheric neutrino flux \cite{Honda:2015fha} for the same energies and at the right location --- the Homestake mine. The estimated uncertainty of this flux prediction is $\mathcal{O}$(35$\%$)~\cite{Battistoni:2005pd, Guo:2018sno,Sawatzki:2020mpb}.} On the other hand, the diffuse supernova neutrino background (DSNB) would also be a relevant background for this search. Although yet unobserved, the DUNE experiment is expected to measure electron neutrinos from the \mbox{DSNB~\cite{Moller:2018kpn,DUNE:2020ypp}.} We calculate the expected flux following~\cite{DeGouvea:2020ang} and we will consider an uncertainty of $\mathcal{O}$(35\%) --- mainly from the uncertainties in the star formation rate.

In \reffig{fig:ch9-events_spectra-DUNE}, we show the predicted number of events in DUNE in 0.5~MeV energy bins and for 40~ktons and 10~years of exposure. We considered different masses, spins and dark matter fractions, as indicated in the labels of the plot. Unless stated differently, we assume a monochromatic mass distribution --- indicated by \textit{mc} in the labels of the figure. Notice that, as we discussed previously, larger PBH masses produce a smaller neutrino flux, which also shifts towards lower energies. 

\begin{figure}
\centering
\includegraphics[width=0.87\textwidth]{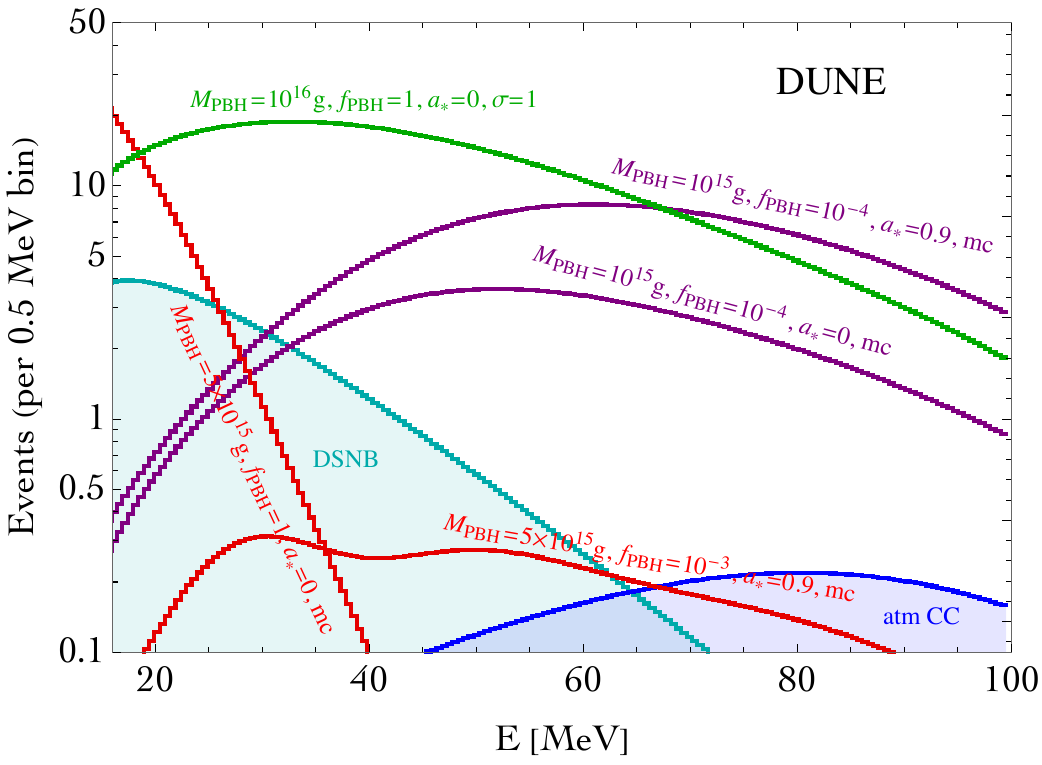}
\caption{Number of events as a function of the neutrino energy at DUNE. We assume the main detection channel $\nu_e + ^{40}$Ar $\rightarrow e^- + ^{40}K^*$, 40 kton of fiducial volume for the far detector, 10 years of operation and $0.5$ MeV energy bins. Different PBH masses and spins, as well as DM fractions, are considered as described in the legends. Expected background events from DSNB and charged-current (CC) atmospheric neutrino interactions are shown for comparison.\labfig{fig:ch9-events_spectra-DUNE}}
\end{figure}

For our analysis, we define the following $\chi^2$ function,
\begin{align}
  \label{eq:ch9-chiSqDUNE}
  &\chi_{\rm DUNE}^2 \nonumber \\ &=\sum_{i}\frac{\left(N^\text{\rm PBH}_i
    +(1+\alpha)N^\text{\rm atmCC}_i
    +(1+\beta)N^\text{\rm DSNB}_i-N^\text{\rm atmCC}_i-N^\text{\rm DSNB}_i\right)^2}{N^\text{\rm PBH}_i+N^\text{\rm atmCC}_i+N^\text{\rm DSNB}_i} 
  \nonumber \\ & \hspace{6.5cm}+ \left(\frac{\alpha}{\sigma_\alpha}\right)^2
  + \left(\frac{\beta}{\sigma_\beta}\right)^2\ ,
\end{align}
where $N^\text{\rm PBH}_i$ is the predicted number of neutrino events from PBH evaporation in the energy bin $i$, $N^\text{\rm DSNB}_i$ refers to the backgrounds events from the DSNB and $N^\text{\rm atmCC}_i$ denotes the events from the atmospheric neutrino charged-current background in the same energy bin. We introduce two nuisance parameters --- $\alpha$ and $\beta$ --- which account for uncertainties on the backgrounds. As we discussed previously, the associated uncertainties are $\sigma_\alpha=0.35$ and \mbox{$\sigma_\beta=0.35$.}

\reffig{fig:ch9-DUNEsens} displays the sensitivity of the DUNE experiment to electron neutrinos from primordial black hole evaporation. In the left panel, we show the 95$\%$ C.L. exclusion region in the two-dimensional plane determined by PBH abundance $f_{\rm PBH}$ and mass $M_{\rm PBH}$ for a monochromatic mass distribution. The results for different PBH spins are shown in orange, red and dark red for $a_*=0$, $a_*=0.5$ and $a_*=0.9$ respectively. For comparison, we display the upper limit from Super-Kamiokande~\cite{Dasgupta:2019cae} and from other data as in~\cite{PBHbounds, bradley_j_kavanagh_2019_3538999}. These constraints come mainly from the extragalactic $\gamma-$ray background~\cite{Carr:2009jm,Ballesteros:2019exr,Arbey:2019vqx,Carr:2020gox}, CMB and MeV extragalactic $\gamma$-ray background~\cite{Clark:2016nst} and soft $\gamma$-ray observations from COMPTEL~\cite{Coogan:2020tuf}. The sensitivity of JUNO for monochromatic non-rotating population of PBHs at 90\% C.L. is also included~\cite{Wang:2020uvi}. One can see that DUNE can improve the existing limits from Super-Kamiokande. For $a_*$ = 0, at masses below $\sim 3 \times 10^{15}$ g, the expected sensitivity of DUNE is also considerably better than for JUNO. This is due to the large statistics expected. Conversely, note that for larger masses, the sensitivity of JUNO will be slightly better. Whereas solar neutrinos do not allow to perform these searches at DUNE to energies below 16 MeV, JUNO --- which is sensitive to electron antineutrinos --- could extend the search to energies around 10-12 MeV. As a consequence, JUNO will be more sensitive to larger PBH masses.

\begin{figure*}
\centering
\includegraphics[width=0.37\paperwidth]{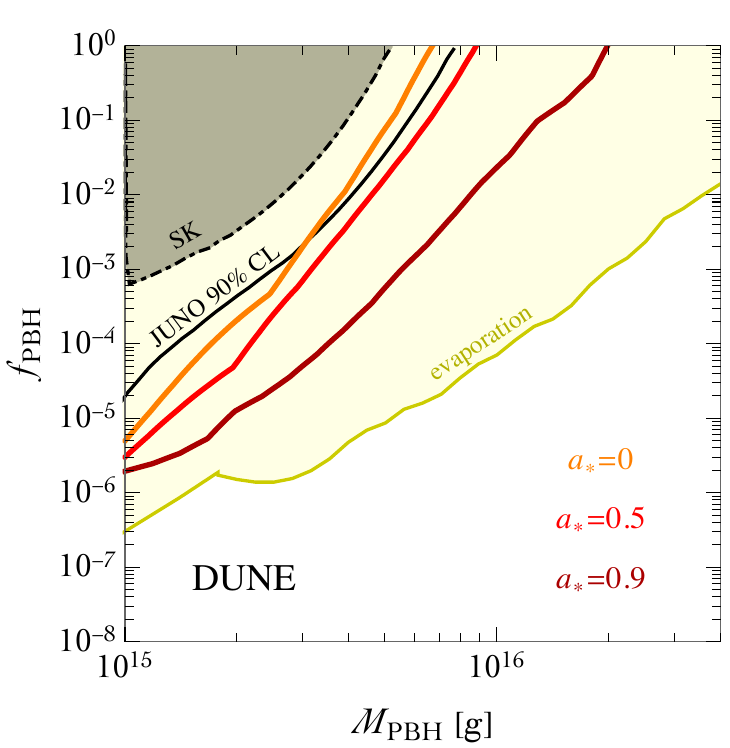}
\includegraphics[width=0.37\paperwidth]{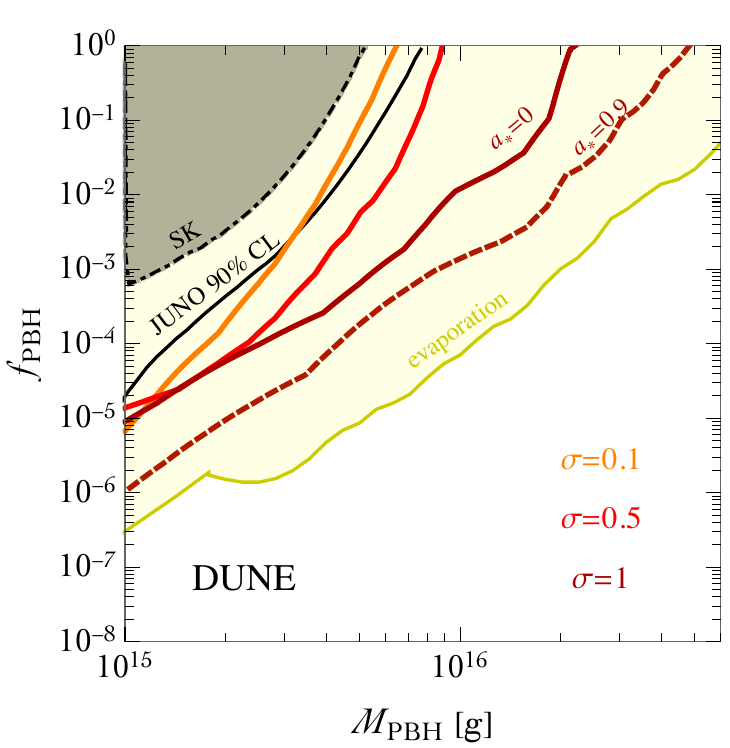}
\caption{\labfig{fig:ch9-DUNEsens} Expected 95$\%$ C.L. sensitivities on the fraction of DM in the form of PBHs, $f_{\rm PBH}$, as a function of $M_{\rm PBH}$ at DUNE. The left panel assumes a monochromatic mass distribution of PBHs and three different spins. The right panel is for a log-normal PBH mass distribution with different widths. In this case, PBHs are assumed to be non-rotating except for the dark-red dashed contour which assumes PBHs with $a_*=0.9$.  For comparison, current Super-Kamiokande~\cite{Dasgupta:2019cae} and evaporation \cite{Clark:2016nst,Coogan:2020tuf,PBHbounds, bradley_j_kavanagh_2019_3538999} bounds, together with the 90\% C.L. expected sensitivity at JUNO~\cite{Wang:2020uvi}, are shown.} 
\end{figure*}

The right panel of \reffig{fig:ch9-DUNEsens} displays the sensitivity of DUNE to an extended PBH mass distribution. We consider a distribution as in Equation \ref{eq:ch9-log-normal-mass-distribution}, with different widths. The results correspond to non-rotating PBHs, except if a non-zero spin parameter is explicitly indicated. For a narrow log-normal distribution --- when $\sigma = 0.1$ --- the sensitivity is similar to the one expected from a monochromatic distribution with $M_{\rm PBH}=M_{\rm c}$. As we explore wider distributions, the projected limits become less stringent for low PBH masses. However, at the same time, it becomes possible to constrain heavier PBHs. This effect is the result of the broadening of the flux and the smoothing of the peak in the primary galactic contribution, as shown in~\reffig{fig:ch9-lognorm}. The most stringent bound is expected for a rather extreme scenario in which the mass distribution is broad and PBHs have extremely large initial angular momentum --- the curve corresponding to $\sigma=1$ and $a_*=0.9$.

\subsection{THEIA}

Water-based liquid scintillator (WbLS) detector technology aims to simultaneously detect scintillation photons and Cherenkov light~\cite{Bignell:2015oqa, Caravaca:2020lfs}. This can be achieved by adding ultrapure water to an organic scintillator to obtain a transparent medium which --- if adequately instrumented --- could detect both forms of light. THEIA is a proposal for a future neutrino detector based on this technology \cite{Theia:2019non}. Since the exact configuration is yet undecided, we will explore two options: a 25~kton tank and a 100~kton one. Inverse-beta decay (IBD) is the dominant reaction channel in a THEIA-like detector~\cite{Theia:2019non}. In our analysis, we focus on this detection channel and implement the cross-section of the process following \cite{Strumia:2003zx}. We consider an energy resolution of $7\% / \sqrt{E \text{ (MeV)}}$ and assume that the fiducial volumes in both configurations will be 20~kton for the smaller of the tanks and 80~kton for the larger one --- which correspond to $1.55 \times 10^{33}$  and $6.2 \times 10^{33}$ targets respectively.

As in the case of DUNE, backgrounds also play a relevant role in THEIA and would limit its sensitivity. In addition to the background of \mbox{charged-current} (CC) atmospheric neutrino interactions at energies above 30 MeV and the DSNB --- which would act as a background in this scenario --- reactor antineutrinos constitute an irreducible background for energies below \mbox{$\sim$ 10~MeV.} To exclude them, we limit our analysis to the energy window between 10~MeV and 100~MeV. In a dedicated analysis for THEIA, it was shown that neutral current (NC) interactions of high-energy atmospheric neutrinos can be significantly reduced. We consider the reduced backgrounds resulting from this study~\cite{Sawatzki:2020mpb} and implement an 80\%~detection efficiency in our analysis. 

In \reffig{fig:ch9-events_spectra-THEIA}, we show the event rate expected in THEIA for different spectra from PBH evaporation. We assume IBD as the detection channel, a fiducial volume of 80 kton and 10 years of exposure. The backgrounds are also shown in the figure. Notice that there is an additional background with respect to DUNE --- see~\reffig{fig:ch9-events_spectra-DUNE}. It corresponds to atmospheric neutrinos interacting via neutral currents, which can mimic an IBD event. The differences in the cross-section and energy resolution lead to different event spectra in each detector. For instance, THEIA would be capable of resolving the characteristic peaks of a PBH population with large spin --- $a_* = 0.9$ --- while, in DUNE, this feature of the spectrum would not be resolved.

Regarding our $\chi^2$ function for THEIA, it is similar to the one presented in Equation~\ref{eq:ch9-chiSqDUNE} but it includes an additional pull parameter which accounts for the normalisation of the background of atmospheric neutrons-- with an uncertainty $\sigma_\gamma = 0.3$~\cite{KamLAND:2011bnd,Cheng:2020oko} --- so that it reads
\begin{align}
  \label{eq:chiSqTHEIA}
  \chi_{\rm THEIA}^2 = \sum_{i}\frac{\left(N^\text{\rm PBH}_i
    +\alpha N^\text{\rm atmCC}_i
    +\beta N^\text{\rm DSNB}_i 
    +\gamma N^\text{\rm atmNC}_i \right)^2}{N^\text{\rm PBH}_i+N^\text{\rm atmCC}_i+N^\text{\rm DSNB}_i +N^\text{\rm atmNC}_i} \nonumber \\
    + \left(\frac{\alpha}{\sigma_\alpha}\right)^2
  + \left(\frac{\beta}{\sigma_\beta}\right)^2
  + \left(\frac{\gamma}{\sigma_\gamma}\right)^2\ .
\end{align}

\begin{figure}
\centering
\includegraphics[width=0.87\textwidth]{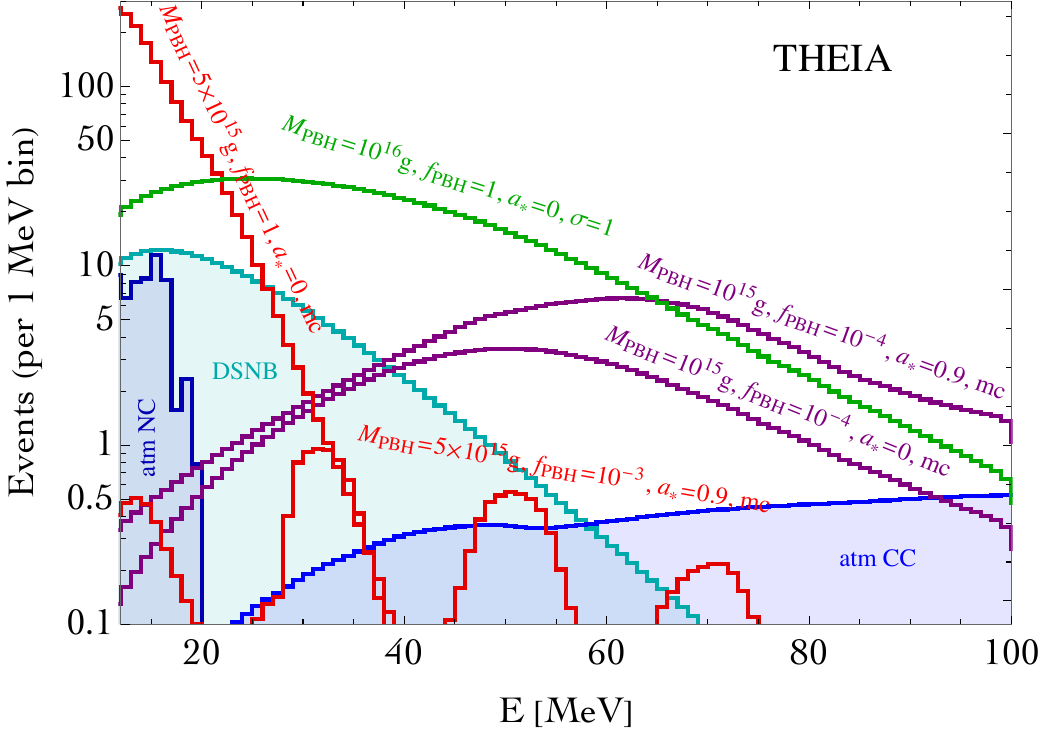}
\caption{Number of events as a function of the neutrino energy at THEIA, assuming  IBD is the detection channel, a detector with a fiducial volume of 80~kton, 10 years of operation and $1$ MeV energy bins. Different PBH masses and spins, as well as DM fractions, are considered as described in the legends. Expected background events from DSNB, charged-current (CC) and \mbox{neutral-current} (NC) atmospheric neutrino interactions are also shown for comparison.\labfig{fig:ch9-events_spectra-THEIA}}
\end{figure}

The future THEIA detector will provide complementary limits to those from DUNE since it will be sensitive to the electron antineutrino component of the flux from primordial black hole evaporation. The expected sensitivity to spherically symmetric and axially symmetric  PBHs is depicted in the left panels of \reffig{fig:ch9-THEIAsens}. For low PBH masses, the expected constraints are comparable to those from JUNO~\cite{Wang:2020uvi}. Despite relying on IBD for the detection of electron antineutrinos, one would expect THEIA to have better statistics due to the larger volume. Note, however, that our prediction for the galactic component is a factor 2 smaller than the one reported in the sensitivity study for JUNO~\cite{Wang:2020uvi}. For larger PBH masses, the antineutrino flux peaks at low energies around 10 - 20 MeV.  The potential of WbLS detectors to reduce the backgrounds in that energy region enhances their sensitivity to PBHs of masses larger than $\sim 8 \times 10^{15}$ g with respect to the sensitivity of a conventional liquid scintillator. In the right panels of~\reffig{fig:ch9-THEIAsens}, we display the results of repeating the calculation using a non-monochromatic distribution. For narrow distribution, the departures from the results of a monochromatic one are barely distinguishable --- see for instance the case of $\sigma = 0.1$. In contrast, as we discussed for DUNE, for a broader mass distribution, the expected constraints for PBHs with masses close to $10^{15}$ g are weakened, whereas the sensitivity to masses around $10^{16}$ g is improved. Finally, one can compare the upper and lower panels in \reffig{fig:ch9-THEIAsens}, which correspond to the two possible configurations of THEIA --- 25 kton and 100 kton tanks respectively. One can see that, if they were located at a similar latitude, increasing the size of the detector would lead to a factor 2 improvement in the limits on $f_{\rm PBH}$.

\begin{figure*}
\centering
\includegraphics[width=0.37\paperwidth]{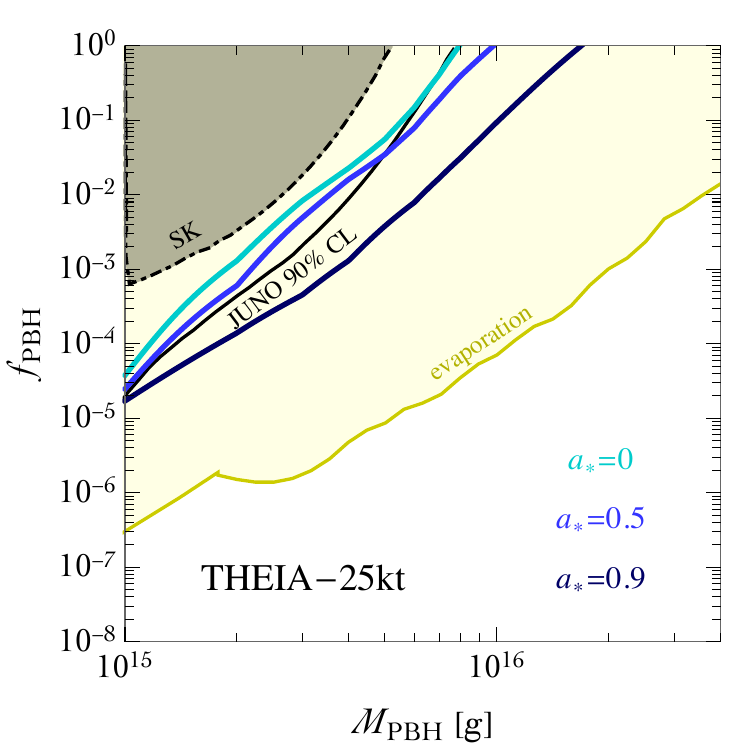}
\includegraphics[width=0.37\paperwidth]{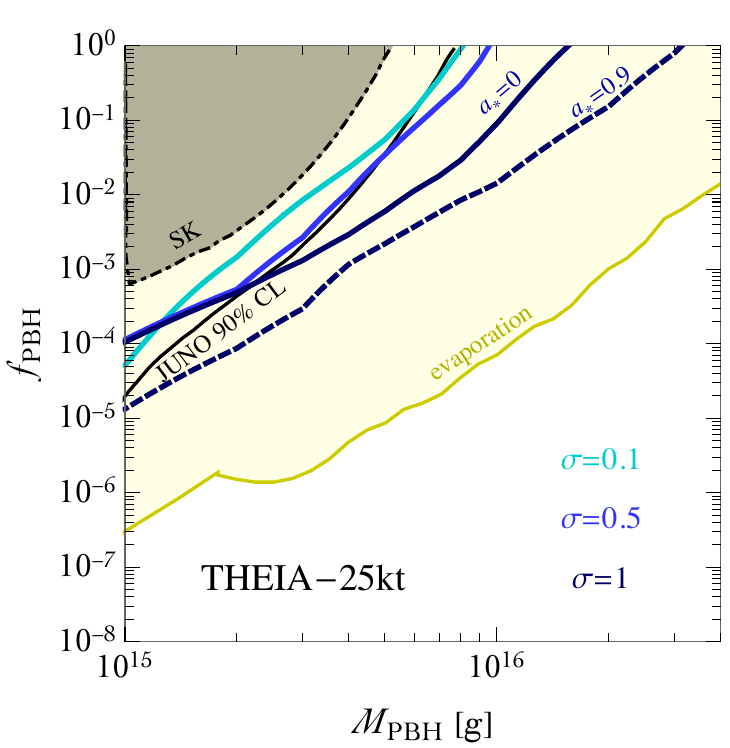}\\
\includegraphics[width=0.37\paperwidth]{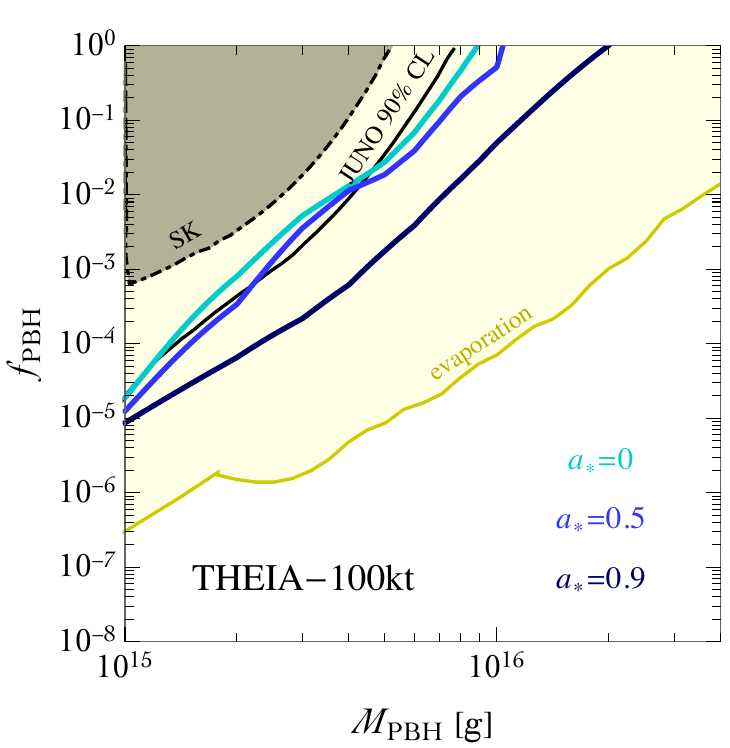}
\includegraphics[width=0.37\paperwidth]{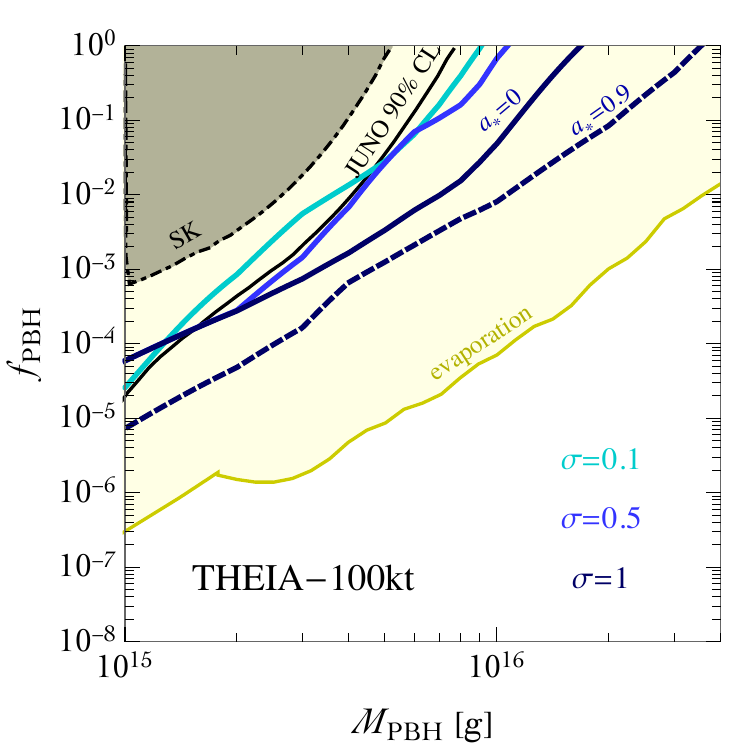}
\caption{Expected 95$\%$ C.L. sensitivities to the two-dimensional plane$f_{\rm PBH}$- $M_{\rm PBH}$ at THEIA. Upper and lower panels consider a fiducial volume of 20~kton and 80~kton respectively.  The left panels assume a monochromatic mass distribution of PBHs and three different spins. The right panels are for a \mbox{log-norma}l PBH mass distribution with different widths. PBHs are assumed to be \mbox{non-rotating} except for the \mbox{dark-blue} dashed contour which considers PBHs with $a_*=0.9$.  Current limits from \mbox{Super-Kamiokande~\cite{Dasgupta:2019cae}} and from evaporation \cite{Clark:2016nst,Coogan:2020tuf,PBHbounds, bradley_j_kavanagh_2019_3538999} are shown for comparison, together with the 90\% C.L. expected sensitivity at JUNO~\cite{Wang:2020uvi}.\labfig{fig:ch9-THEIAsens}}
\end{figure*}

%%%%%%%%%%%%%%%%%%%%%%%%%%%%%
%%%%%%%%%%%%%%%%%%%%%%%%%%%%%
\section{Discussion}
\labsec{ch9-discussion}
%%%%%%%%%%%%%%%%%%%%%%%%%%%%%
%%%%%%%%%%%%%%%%%%%%%%%%%%%%%

Primordial black holes with masses in the range $10^{15}-10^{17}$ g are viable and interesting dark matter candidates. Such --- relatively --- light PBHs may evaporate and produce a sizeable flux of MeV neutrinos. This chapter focuses on the prospects for detecting such flux in the next-generation experiment DUNE and in a water-based liquid scintillator like the proposed detector THEIA. 
We have studied the impact of the mass and spin of the primordial black hole population on the expected neutrino flux. We find that, the larger the PBH mass, the lower the energy of the expected flux is. In addition, the overall magnitude of the flux is also suppressed with respect to the expectations for lighter PBHs. Regarding rotating PBHs, we show that an enhancement of the emitted flux of neutrinos and the appearance of the characteristic structure of peaks are expected. Likewise, we have analysed the role of the mass distribution of PBHs by considering monochromatic and extended PBH mass distributions. For a broad mass distribution, we have shown that the spectrum is smoothed and it extends up to larger neutrino energies than in the case of monochromatic distributions.

Our projected sensitivities show that both DUNE and THEIA would improve significantly the existing bounds on the primordial black hole abundances in the mass range between $10^{15}-10^{16}$ g. Depending on the mass distribution and spin of the PBH population, masses of $\sim 5 \times 10^{16}$ g could also be probed.
 
Our studies also indicate that the bounds from these next-generation neutrino facilities would be weaker than the ones already existing from other cosmic messengers, in particular, from photons. However, they would provide a consistency check of those limits and of the underlying assumptions made when deriving those constraints. Moreover, DUNE and THEIA would complement each other since they are sensitive to electron neutrinos and antineutrinos respectively. 

As a closing comment, let us remark that the analyses here proposed involved the study of neutrinos in the same energy range as the diffuse supernova neutrino background. Hence, the study of this scenario --- the quest for signatures of primordial black hole dark matter --- can be easily integrated into the physics program of neutrino observatories with access to astrophysical fluxes of neutrinos in the MeV range. 

\chapter{Neutrinos and ultralight scalar dark matter}
%\addcontentsline{toc}{chapter}{Neutrinos and ultralight dark matter} 
\labch{ch10-uldm}
Ultralight scalars are phenomenologically rich dark matter candidates and could explain some small-scale cosmological discrepancies between simulations and observations. Such fields can be produced non-thermally in the early universe and are well described by a classical-number field. Motivated by the feebly interacting nature of dark matter and neutrinos, we propose a portal connecting both species. If such a connection existed, a plethora of testable experimental signatures could be observed in neutrino experiments. 

In Sections \ref{sec:ch10-uldm-intro} and \ref{sec:ch10-uldm-nu} of this chapter, we motivate the interest in ultralight scalar dark matter candidates and how they could couple to neutrinos. Then, in Sections \ref{sec:ch10-uldm-osc} and \ref{sec:ch10-DUNE}, we explore the signatures predicted in oscillation experiments, taking DUNE as an example. Next, Section \ref{sec:ch10-uldm-beta} discusses the impact in beta decay experiments. Finally, Section \ref{sec:ch10-uldm-critical} wraps up the chapter with a critical discussion on the scenario proposed, other existing limits, and future avenues.

\section{Ultralight scalars as dark matter candidates}
\label{sec:ch10-uldm-intro}

Ultralight scalars with masses well below the eV scale could compose the dark matter of the universe. Being produced coherently via misalignment, such dark matter candidate can be represented by a Glauber-Sudarshan state \cite{Davidson:2014hfa,Marsh:2015xka,Dvali:2017ruz,Veltmaat:2020zco}, namely
\begin{align}
\ket{\Phi} = \frac{1}{N} \exp \left[ \int \frac{\text{d}^3k}{(2\pi)^3} \widetilde{\Phi}(\vec{k})\hat{a}^\dagger _k  \right] \ket{0}\, ,
\end{align}
where $N$ is a normalisation factor. 

Then, the expectation value of the field operator in this coherent state,
\begin{align}
\bra{\Phi}\hat{\Phi}(x)\ket{\Phi} =\int \frac{\text{d}^3 k}{(2\pi)^3}\frac{1}{\sqrt{2E_k}}\widetilde{\Phi}(\vec{k})e^{-i\vec{k}\vec{x}} \equiv \Phi(x)\, ,
\end{align}
obeys the classical equation of motion  
\begin{align}
(g_{\mu\nu}\partial^\mu \partial_\nu + m_\phi^2)\Phi(x) = 0
\end{align}
where we are using the metric
\begin{align}
\text{d}s^2 = \text{d}t^2 - a^2 \delta_{ij}\text{d}x_i\text{d}x_j \, .
\end{align}
Thus, it can be parametrised as a classical field, namely
\begin{align}
\Phi(x) = \phi\sin(m_\phi t - \vec{k}\cdot\vec{x})\, ,
\end{align}
where $\phi$ denotes the amplitude of the modulating field. 

Moreover, the local value of the ultralight dark matter field can be expressed as
\begin{equation}
    \Phi(\vec{x},t)\simeq\frac{\sqrt{2\rho_\phi}}{m_\phi}\sin\left[m_\phi( t-\vec{v}_\phi\cdot \vec{x}) \right],
\end{equation}
where the local density $\rho_\phi$ should not exceed the local DM density \mbox{$\rho_{\rm DM}\sim0.4~{\rm GeV}/{\rm cm}^3$}~\cite{deSalas:2019rdi}, $m_\phi$ is the mass of the scalar and $v_\phi \sim \mathcal{O}(10^{-3}$c) is the virial velocity. Note that the space-dependent phase $m_\phi\vec{v}_\phi\cdot\vec{x}$ of the local field value is much smaller than $m_\phi t$ and, therefore, it will be neglected henceforth. 

Furthermore, notice that the mass of $\Phi$ defines the modulation period via
\begin{equation}
   \tau_\phi\equiv \frac{2\pi\hbar}{m_\phi}=0.41 \left(\frac{10^{-14}~{\rm eV}}{m_\phi}\right)~{\rm s}\, .
\end{equation} 

The de-Broglie wavelength of an ultralight scalar with velocity $v_\phi$ is given by
\begin{align}
\lambda_{\text{dB}} = \frac{1}{m_\phi v_\phi} \simeq 600\, \text{pc} \left( \frac{10^{-22}\text{eV}}{m_\phi}\right)\left(\frac{10^{-3}\text{c}}{v_\phi}\right)\, .
\end{align}

This means that, for a Milky Way-like galaxy, with typical virial velocities $v_\phi \sim 600$ km/s, the de-Broglie wavelength is of order $\mathcal{O}$(0.1 kpc). Notice that the de-Broglie wavelength can reach the size of galaxies depending on the virial velocity and the mass of the ultralight scalar. 

In a given volume of $\lambda_{\text{dB}}^3$, the occupation number in the Milky Way is
\begin{align}
\mathcal{N} \sim n \lambda^3_{\text{dB}} \sim \frac{\rho_{\text{DM, MW}}}{m_\phi}\lambda_{\text{dB}}^3 \sim 10^{94}\left(\frac{10^{-22}\text{eV}}{m_\phi}\right)^4\, ,
\end{align}
where $n$ is the number density. Being so large, it justifies the approximation that the ultralight scalar field behaves like a classical field, since quantum fluctuations in the expectation value  $\bra{\Phi}\hat{\Phi}\ket{\Phi}$ are of order $\mathcal{O}(\mathcal{N}^{-1})$.

Hence, the overall picture is the following: in the outskirts of galaxies or outside them --- for distances $d > \lambda_{\text{dB}}$ --- ultralight dark matter behaves very similarly to standard cold dark matter, whereas for smaller scales --- when $d < \lambda_{dB}$ --- a condensate core is expected to form in the inner parts of the galaxy. Due to these properties, ultralight scalar fields are invoked to address several small-scale discrepancies between observations and calculations, mainly \textit{cusp-vs-core}, \textit{missing satellite} and \textit{too-big-to-fail} problems.
\begin{itemize}
\item Cosmological simulations considering only dark-matter predict that the Navarro-Frenk-White profile approximately describes all halo masses. In this profile, the density increases towards the centre and forms a \textit{cusp}. However, measurements of rotation curves in bright dwarf galaxies are in better agreement with constant density profiles in their centre \cite{KuziodeNaray:2007qi,Flores:1994gz,Moore:1994yx,McGaugh:1998tq,deBlok:2008wp}. This is known as the \textit{cusp-vs-core puzzle}.
 
\item The \textit{missing-satellite problem} consists of a discrepancy between the number of dwarf galaxies we observe and the number of halos with low mass predicted in $\Lambda$CDM \cite{Klypin:1999uc,Moore:1999nt,Ferreira:2020fam}.

\item It has also been pointed out that, when simulating subhalos in Milky Way-like systems, those are found to have higher central densities than what is observed in any Milky Way satellite~\cite{Boylan-Kolchin:2011qkt,Boylan-Kolchin:2011lmk}. Simulated subhalos seem to be too massive for star formation to occur and hence, they should not host any observable satellite galaxy. This is known as the \textit{too-big-to-fail problem}.
\end{itemize}  In the three cases, ultralight dark matter could provide a solution to these puzzles \cite{Hui:2021tkt}. For a complete review of several aspects of ultralight dark, including different descriptions, phenomenological implications, and connections with observations, see \cite{Ferreira:2020fam}.

Given the need to explain the dark matter of the universe and the fact that ultralight candidates are phenomenologically motivated to address the \mbox{small-scale} puzzles discussed before, in this chapter we focus on the signatures expected from neutrinophilic ultralight scalars.

\section{Neutrinophilic ultralight dark matter}
\label{sec:ch10-uldm-nu}

From a coupling to an ultralight scalar, neutrinos would get a contribution to their masses, in analogy to the Higgs mechanism. As a result of classical dark matter field modulations, the generated mass term for the neutrinos would become time-varying. A simple way to realise this scenario is to couple a classical scalar field ($\Phi$) to neutrinos via the operator $\bar{\nu}\nu \Phi$, which generates a time-varying neutrino mass term. In a UV-complete model, this can be achieved by directly coupling $\Phi$ to the lepton doublet $l_L \, \equiv \, (\nu,\, e)_L$, as in the type-II seesaw \cite{Magg:1980ut,Schechter:1980gr,Cheng:1980qt}. However, there exist stringent constraints because of the accurate determination of electron properties. Particularly, a time-varying neutrino mass would also induce mass variations in charged leptons. In the case of the electron, this phenomenon is strongly constrained by the stability of atomic clocks~\cite{Luo:2011cf}.

Another way to couple light neutrinos to the scalar field is via the mixing with gauge singlets, such as a right-handed sterile neutrino $N$. It is also natural to have a considerable Yukawa coupling between the sterile component and the scalar field. \footnote{Note that this field $N$ is not necessarily involved in the generation of neutrino masses.} To explore such a scenario, we consider the following Lagrangian:
\begin{align}
-\mathcal{L} \supset y_D \overline{l_L} \tilde{h}N \, + \, \frac{1}{2}(m_N + g\Phi)\overline{N^c} N \, &+ \, \frac{\kappa}{2} \left(\overline{l^c_L} \tilde{h}^*\right)\left( \tilde{h}^\dagger l_L \right)\, \nonumber \\ & + \, \text{h.c.} \, + \, ... \, .
\label{eq:ch10-lagrangian-phinu}
\end{align}
Here, $\Phi$ represents the ultralight dark matter field and $h$ is the Standard Model Higgs. The active Majorana neutrino mass term  --- $m_\nu \equiv \kappa v^2/2$ --- and the Dirac mass --- $m_D = y_D v/\sqrt{2}$ --- will be generated after Higgs takes the vacuum expectation value  $\langle \tilde{h}\rangle \equiv v/\sqrt{2} = 174\, \text{GeV}$. The Majorana mass term of the sterile neutrino is denoted by $m_N$. On top of that, the Yukawa interaction $g\Phi \overline{N^c} N$ with a coupling constant $g$ generates a time-varying mass term to the sterile neutrino. 

In general, the Majorana mass terms $m_\nu$ and $m_N$ can be forbidden by imposing additional symmetries --- such as lepton number conservation.\footnote{Then the remaining Lagrangian will look similar to the singlet Majoron model.} In the most economical case, one can even generate small neutrino masses, with a minimal interaction form $y_D\overline{l_L} \tilde{h} N + g\phi \overline{N^c}N$, but the number of sterile neutrinos should be extended to at least two in order to explain the oscillation data. Note that even though we assume Majorana neutrinos in this work, the generalization to Dirac neutrinos is not difficult. An alternative scenario is the one in which right-handed neutrinos have no vacuum mass ($m_N$ = 0) but get a tiny Majorana mass from their coupling to the scalar field. This can give rise to ultralight scalar-induced Quasi-Dirac neutrinos~\cite{Dev:2022bae}.

From the Lagrangian in Equation \ref{eq:ch10-lagrangian-phinu}, the equation of motion of the scalar is
\begin{align}
\ddot{\Phi} + 3 H \dot{\Phi} + m^2_\phi \Phi = \frac{g}{2} \left(\overline{N^c}N + \overline{N}N^c\right)\, ,
\end{align}
with $H$ being the Hubble expansion rate and the dot denoting the derivative with respect to proper time. The Hubble dilution term $3H\dot{\Phi}$ is relevant when considering the evolution of the scalar over cosmological time scales in the expanding universe. However, it can be neglected when addressing the behaviour at present and in our local galaxy. In the absence of the source term on the right-hand side, the scalar field evolves freely in the universe after production. 

\textbf{Neutrino masses and mixing in the presence of an ultralight scalar.}

We derive the effective neutrino masses and mixing in the presence of the scalar potential $g\Phi$ in a two-flavour framework --- where $\nu_{\rm a}$ and $N$ are the active and sterile states respectively. In the absence of a coupling between sterile states and the ultralight scalar, the diagonalisation of the mass matrix leads to a vacuum mixing $U$, and two masses $m_1$ and $m_4$ --- which dominantly mix with active and sterile neutrinos, respectively.

The addition of the scalar potential leads to a mass matrix which in the flavour basis, $(\nu_{\rm a}, N^{\rm c})$, reads
\begin{align}
\widetilde{M}_{\nu} =  U^{\dagger}
\left( \begin{matrix}
m_1 & 0 \\
0 & m_4  
\end{matrix} \right)  U^* 
+ \left( \begin{matrix}
0 & 0 \\
0 & g \Phi
\end{matrix} \right)
 =  \widetilde{U}^{\dagger}
\left( \begin{matrix}
\widetilde{m}_1 & 0 \\
0 & \widetilde{m}_4 
\end{matrix} \right)  \widetilde{U}^*  \,.
\end{align}

The effective neutrino masses and mixing in the presence of ultralight scalar have the following expressions
\begin{align} \label{eq:ch10-wtm1}
&2\widetilde{m}_1  =   m_1 + m_4  + g\Phi  \nonumber \\ & \hspace{1.8cm} - \sqrt{ (m_4 -m_1)^2+(g \Phi)^2  + 2 (m_4 -m_1) g\Phi \cos{2 \theta_{14}} }\, , \\ \label{eq:ch10-wtm4}
&2\widetilde{m}_4  =  m_1 + m_4  + g\Phi  \nonumber \\ & \hspace{1.8cm} + \sqrt{ (m_4 -m_1)^2+(g\Phi)^2 +\, 2 (m_4 -m_1) g\Phi\cos{2 \theta_{14}}}\,, \\ \label{eq:ch10-wtth14}
&\tan{2\widetilde{\theta}_{14}} = \frac{(m_4 -m_1) \sin{2\theta_{14}}}{(m_4 -m_1) \cos{2\theta_{14}} + g\Phi} \,.
\end{align}

In the limit when $|g\Phi| \ll m_4  \cos{2\theta_{14}}$, one can do a series of approximations and get
\begin{align}
&\widetilde{m}_1 \simeq   m_1 + \sin^2{\theta_{14}} \cdot  g \Phi \,, \\
\label{eq:ch10-wtm1-2}
&\widetilde{m}_4 \simeq m_4  + \cos^2{\theta_{14}} \cdot  g \Phi \;, \\ 
&\tan{2\widetilde{\theta}_{14}}  \simeq  \tan{2{\theta}_{14}} \,,
\end{align}
which usually applies to scenarios where sterile neutrinos are heavy. 

In the case of $g \Phi \gg m_4 >0~{\rm eV}$, we have the approximation
\begin{align}\label{eq:ch10-wtm1_3}
&\widetilde{m}_1 \simeq \frac{m_1+m_4  - (m_4 -m_1)\cos{2\theta_{14}}}{2} - \frac{(m_4 -m_1)^2 \sin^2{2\theta_{14}}}{4 g \Phi} \,, \\
&\widetilde{m}_4 \simeq \frac{m_1+m_4 +(m_4 -m_1)\cos{2\theta_{14}}}{2}+ g\Phi \,, \\
&\tan{2\widetilde{\theta}_{14}} \simeq  \frac{(m_4 -m_1) \sin{2\theta_{14}}}{ g\Phi}\,,
\end{align}
whereas, in the case of $g \Phi \ll -m_4 < 0~{\rm eV}$, we instead have 
\begin{align}\label{eq:ch10-wtm1_4}
&\widetilde{m}_1 \simeq - \frac{m_1+m_4 +(m_4 -m_1)\cos{2\theta_{14}}}{2}+ |g\Phi|   \,, \\
&\widetilde{m}_4 \simeq \frac{m_1+m_4  - (m_4 -m_1)\cos{2\theta_{14}}}{2} - \frac{(m_4 -m_1)^2 \sin^2{2\theta_{14}}}{4 g \Phi}  \,, \\
&\tan{2\widetilde{\theta}_{14}} \simeq  \frac{(m_4 -m_1) \sin{2\theta_{14}}}{ g\Phi}\,.
\end{align}

\textbf{Ultralight scalars coupling to heavy sterile neutrinos}

From Equation \ref{eq:ch10-wtm1-2}, in the limit that the sterile neutrino is very heavy compared to the scalar potential --- i.e. $m_4 \gg g \phi$ --- the light neutrinos will receive an effective mass in addition to the original vacuum one,
\begin{eqnarray}\label{eq:mtphi}
\widetilde{m}_{i}(t)  & \approx & 
m_i +  \sin^2\theta_{14}\,  g  {\phi} \sin{m_{\phi}t} \,.
\end{eqnarray}
The active-sterile mixing angle $\theta_{14}$ can simply be absorbed by redefining $y \equiv \sin^2{\theta_{14}} \cdot g $ such that the neutrino mass correction reads as $y \phi  \sin{m_\phi t}$. In such a case, it is technically indistinguishable whether active neutrinos couple directly to the scalar field or by mixing with the sterile neutrino. 

In this case, the neutrino mass matrix will therefore receive a contribution from $\Phi$ given by
\begin{equation}
   \delta m_\nu =  y\frac{\sqrt{2\rho_\phi}}{m_\phi}\sin(m_\phi t) \, .
\end{equation}
Without a flavour model, the Yukawa couplings $y$ can have any structure in flavour space, and thus the modulations of neutrino masses and mixing can bear any correlation.

For a given neutrino oscillation experimental setup, there are three characteristic time scales: the neutrino time of flight \begin{align}
\tau_\nu = L/c = 3.4 \left(\frac{L}{1000~{\rm km}}\right)~{\rm msec}\, ,
\end{align}
the time between two detected events $\tau_{\rm evt}$ --- which is the inverse of the ratio of events --- and the running time of the experiment $\tau_{\rm exp}$. It is therefore possible to identify three different regimes depending on how large the modulation period of the ultralight scalar is compared to the characteristic time scales of each particular oscillation experiment:
\begin{itemize}
   \item \textbf{Time modulation regime ($\mathbf{\tau_{\rm evt}\lesssim\tau_\phi \lesssim\tau_{\rm exp}}$).} When the modulation period of $\Phi$ is similar in order of magnitude to the experiment running time, one may observe a variation of the signal with time. In that case, large statistics and high event rates would be crucial features for neutrino experiments to be sensitive to modulation periods significantly smaller than the data-taking lifetime of the experiments.
   
   \item \textbf{Averaged Distorted Neutrino Oscillations ($\mathbf{\tau_\nu\ll\tau_\phi\ll\tau_{\rm exp}}$).} Even when the rate of change of neutrino oscillation parameters is too fast to be observed as a modulating signal, the time average oscillation probability may be distorted by such effects and deviate from the standard three-neutrino scenario. For a modulating mixing angle, the effect can be approximately mapped onto standard oscillation with an inferred value of the mixing angle different from its true value. Nevertheless, the averaging effect for a modulating mass splitting is a non-trivial distortion of the neutrino oscillation probability. This non-trivial averaging produces a smearing effect similar to the one expected from the energy resolution of the detector~\cite{Krnjaic:2017zlz}. This regime covers a large range of scalar masses and can be easily searched for in oscillation experiments.
   
	\item \textbf{Dynamically Distorted Neutrino Oscillations ($\mathbf{\tau_\phi\sim\tau_\nu}$).} As the modulating period of $\Phi$ gets closer to the neutrino time of flight, the changes in oscillation parameters need to be treated at the Hamiltonian level and can be modelled by a modified matter effect~\cite{Brdar:2017kbt}. This matter potential is time-dependent and, therefore, it changes as the neutrino propagates towards the detector. When the variation of the matter potential is too slow compared to the neutrino time of flight, dynamically Distorted Neutrino Oscillations recover the averaged case. In the opposite limit, when the variations are too fast compared to the neutrino time of flight, they cannot be observed and thus, the scenario can not be distinguished from the standard oscillation picture.   
\end{itemize}

These different regimes are discussed in more detail in Sections~\ref{sec:ch10-uldm-osc}~and~\ref{sec:ch10-DUNE}. For $\beta$-decay experiments, only two of these regimes are relevant and will be discussed in Section \ref{sec:ch10-uldm-beta}. The first one corresponds to the case in which the modulation period is of the order of magnitude of the data-taking campaigns. In that case, the value of $m^2_\beta$ inferred from data would be modulating with time. The second regime of interest corresponds to the case in which $\mathbf{\tau_\nu\ll\tau_\phi\ll\tau_{\rm exp}}$, which would manifest as non-trivial distortions in the end-spectrum of beta decays. Notice that these two regimes are analogous to the time modulation regime and the averaged Distorted Neutrino Oscillations discussed previously.

\section{Phenomenology in oscillation experiments}
\label{sec:ch10-uldm-osc}
Let us now look at the phenomenological signatures expected in neutrino oscillation experiments for the three different regimes presented previously, using the DUNE experimental setup as a case study. For simplicity, we will consider a single parameter modulating at a time. The modulation of mixing angles is assumed to be of the form
\begin{equation}\label{eq:ch10-angles}
   \theta_{ij}(t)=\theta_{ij}+\eta\sin(m_\phi t),
\end{equation}
where $\theta_{ij}$ represents the undistorted value of the mixing angle and $\eta$ is related to the amplitude of modulation of $\Phi$.\footnote{The connection between this parameter and physical quantities depends on the exact realisation of the model.}
On the other hand, for $\eta \ll 1$, mass splittings modulate like 
\begin{equation}\label{eq:ch10-dmsq}
   \Delta m^2_{ij}(t) \equiv m_i^2(t)-m_j^2(t) \simeq \Delta m^2_{ij}\left[1 + 2\eta \sin(m_\phi t)\right],
\end{equation}
where, similarly to above, $\Delta m^2_{ij}$ represents the undistorted value of the mass splitting. Here, we also quantify the amplitude of the time modulation by an effective parameter $\eta$.\footnote{Note that the parameter $\eta$ is not, in general, the same one for each modulating mass splitting and mixing angle, since it depends on the flavour structure of the coupling and hence, it depends on the underlying UV completion of the model.} All across the text, we assume normal mass ordering and the best-fit values of oscillation parameters from \cite{deSalas:2017kay}, namely $\Delta m^2_{31}=2.5\times 10^{-3}$~eV$^2$ and $\sin^2\theta_{23}=0.55$.

\subsection{Time modulation \label{sec:ch10-time-modulation}}
The phenomenology in the time modulation regime is very intuitive: mixing angles or mass splittings are modulating with time. Thus, the oscillation probability also acquires a dependence on time, namely
\begin{equation}
   P_{\alpha\beta}\equiv P(\nu_\alpha\to\nu_\beta, t) = P\left(\nu_\alpha\to\nu_\beta, \{ \theta_{ij}(t), \Delta m^2_{ij}(t)\}\right)\, ,
\end{equation}
and one would expect a time dependence in the signal measured at experiments.

In the case of modulating angles, the oscillation probability for muon-neutrino disappearance in vacuum, in a simplified two-neutrino framework, reads
\begin{align}
   P_{\mu\mu}^{\rm angle} &\simeq 1-\sin^2\left(2\theta(t)\right)\sin^2\left(\frac{\Delta m^2 L}{4E}\right)\nonumber \\ &=1-\sin^2\left(2\theta+2\eta\sin(m_\phi t)\right)\sin^2\left(\frac{\Delta m^2 L}{4E}\right),
\end{align}
where $L$ is the baseline of the experiment and $E$ is the neutrino energy.
Notice that the oscillation probability displays a time modulation via the $\sin(m_\phi t)$ term.

\begin{figure*}
    \centering
    \includegraphics[width= 0.78\paperwidth]{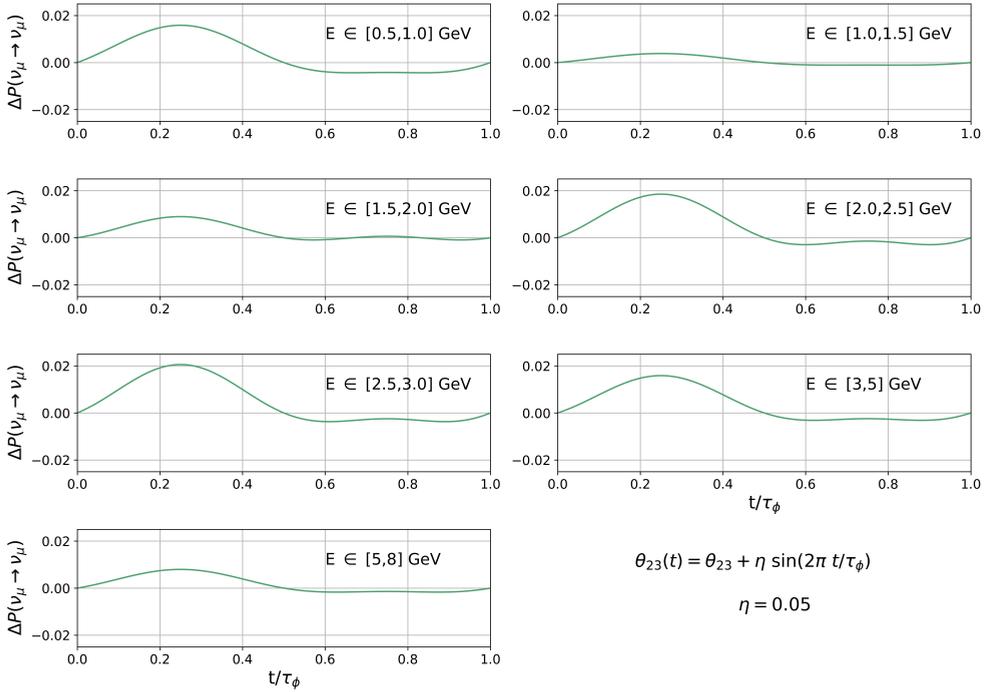}
    \caption{The difference in muon-neutrino disappearance probability \mbox{$\Delta P(t)\equiv P_{\mu\mu}(t)-P_{\mu\mu}(0)$} at DUNE for a modulating $\theta_{23}$, assuming $\eta=0.05$, and for several energy bins. We have assumed normal mass ordering and the best-fit values of oscillation parameters from \cite{deSalas:2017kay}.\labfig{fig:ch10-timemod-th}}
\end{figure*}

In \reffig{fig:ch10-timemod-th}, we show the change in the disappearance probability \mbox{$\Delta P(t) \equiv P_{\mu\mu}(t)-P_{\mu\mu}(0)$} for different energy bins at DUNE, considering a modulating~$\theta_{23}$. For illustrative purposes, we show the effect for $\eta=0.05$. 

One should note three aspects here. In the first place, the modulation phases across different energies are fully correlated. Secondly, the change in the instantaneous mixing angle is approximately
\begin{equation}
  \sin^2[2\theta+2\eta\sin(m_\phi t)]\simeq \sin^2(2\theta) + 2\sin(4\theta)\eta \sin (m_\phi t),
\end{equation}
and thus, it depends on $\eta$ and the mixing angle itself. Finally, when $\sin(4\theta)$ is near zero, we observe a shrinking of the modulation amplitude. This shrinking --- which is sometimes referred to as a Jacobian suppression --- will be discussed later in the text in other contexts too.

The shapes in \reffig{fig:ch10-timemod-th} are due to the fact that the changes in the probability for a modulating mixing angle depend on the value of $\theta_{23}$, which in this case is in the upper octant yet close to maximal mixing. Alternatively, for a modulation of $\theta_{12}$, which could be studied in the JUNO experiment~\cite{JUNO:2015zny}, the change in oscillation probability would have been more symmetric.

If instead, one considered that the mass splitting modulates on time, the simplified time-dependent oscillation probability would read 
\begin{align}
   P_{\mu\mu}^{\rm mass} &\simeq 1-\sin^2(2\theta)\sin^2\left[\frac{\Delta m^2(t) L}{4E}\right]\nonumber \\ &=1-\sin^2(2\theta)\sin^2\left\{\left(\frac{\Delta m^2 L}{4E}\right) \left[1+2\eta\sin(m_\phi t)\right]\right\}.
\end{align}
Here, what modulates is not the amplitude of the oscillation probability, but rather the position of the maxima and minima of the oscillation probability. In \reffig{fig:ch10-timemod-dm}, we show again the change in muon-neutrino disappearance oscillation probability $\Delta P(t) \equiv P_{\mu\mu}(t)-P_{\mu\mu}(0)$ at DUNE, but now for a modulating $\Delta m^2_{31}$ assuming $\eta=0.05$ for several energy bins. 

In this case, the main distinctive features are the correlations and \mbox{anti-correlations} across multiple energies, which result from the shift in the maxima and minima of the oscillation probability for a fixed baseline.

\begin{figure*}
    \centering
    \includegraphics[width = 0.78\paperwidth]{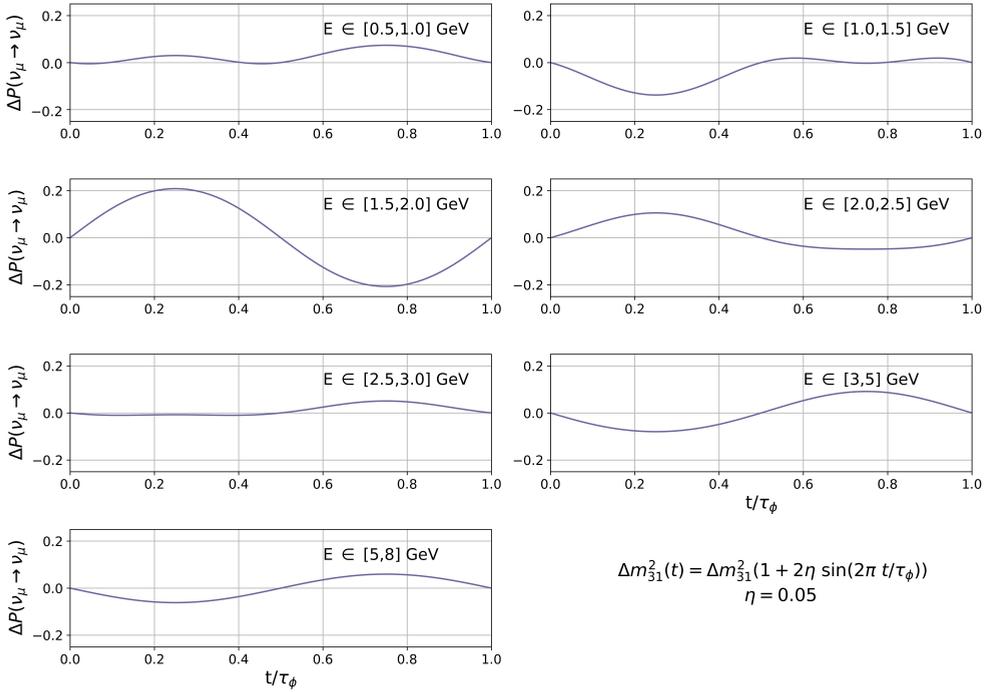}
    \caption{The difference in muon-neutrino disappearance oscillation probability $\Delta P(t)\equiv P_{\mu\mu}(t)-P_{\mu\mu}(0)$  at DUNE for a modulating $\Delta m^2_{31}$ assuming $\eta=0.05$ for several energy bins. We have assumed normal mass ordering and the best-fit values of oscillation parameters from \cite{deSalas:2017kay}.  \labfig{fig:ch10-timemod-dm}}
\end{figure*}

A search for time-dependent frequencies in the data would seem suitable to probe this scenario, for modulations of either mixing angles or mass splittings. In the next section, we illustrate how one can perform an analysis of a set of time-dependent mock data --- using DUNE as an example --- employing the Lomb-Scargle method to estimate the experimental sensitivity.

\subsection{Averaged Distorted Neutrino Oscillations}\label{subsec:average-dino}
The second regime we consider corresponds to the case in which the modulations of mixing angles or mass splittings are too fast to be observed, but an averaging effect on oscillation probabilities remains. Besides that, the modulation period has to be much larger than the neutrino time of flight too. We refer to this effect as Distorted Neutrino Oscillations --- or DiNOs, for short. Then, the average oscillation probability would be given by~\cite{Krnjaic:2017zlz}
\begin{align}
   \langle P_{\alpha\beta}\rangle=\frac{1}{\tau_\phi}\int_0^{\tau_\phi} dt \, P_{\alpha\beta}(t),
\end{align}
where $\tau_\phi$ denotes the modulation period of the ultralight scalar field.

Let us first focus on the averaging of modulating angles. 
The averaging of the mixing would be~\cite{Krnjaic:2017zlz}
\begin{align}
   \frac{1}{\tau_\phi}\int_0^{\tau_\phi} &\text{d}t \, \sin^2\left[2\theta+2\eta\sin(m_\phi t)\right] \nonumber \\ &= \frac{1}{2}\left[1-J_0(4\eta)\cos(4\theta)\right]\simeq \sin^2(2\theta)(1-4\eta^2)+2\eta^2,
\end{align}
where $J_0$ is a Bessel function of the first kind.  In the last step, we have expanded the result to second order in $\eta$. In this way, one can see that this effect simply maps into standard oscillation probability with a different mixing angle. The effect of averaging is pushing apparent mixing angles away from zero or maximal mixing. Hence, the experimental sensitivity to averaged DiNOs effects on mixing angles depends not only on the precision with which the experiment can determine the mixing angles but also on the value of the measured angle itself. 

Regarding the averaging effect for time-varying mass splittings, the disappearance channel --- in vacuum and in a two-neutrino framework --- reads
\begin{align}
   \langle P^{\rm mass}_{\alpha\beta}\rangle &= \frac{1}{\tau_\phi}\int_0^{\tau_\phi} \text{d}t \left\{ 1-\sin^2(2\theta) \sin^2\left[\left(\frac{\Delta m^2 L}{4E}\right) \left[1+2\eta\sin(m_\phi t)\right]\right]\right\}\nonumber \\
        & \simeq 1-\sin^2(2\theta)\left\{\sin^2\left(\frac{\Delta m^2 L}{4E}\right)\right. \nonumber \\ & \hspace{2.5cm}\left.+2\eta^2 \left(\frac{\Delta m^2 L}{4E}\right)^2\cos\left(\frac{\Delta m^2 L}{2E}\right)\right\}\, ,
\label{eq:ch10-average-prob}
\end{align}
where the last expression was expanded to order $(\eta \Delta m^2 L/4E)^2$. This is typically fine for the first oscillation minimum and $\eta<0.05$. 
As shown in \reffig{fig:ch10-av-dm}, for oscillation parameters from~\cite{deSalas:2017kay} and different values of $\eta$, the effect of mass splitting averaging is a smearing in the oscillation probability, similar to the effect of a finite energy resolution. Therefore, one would expect experiments like DUNE, KamLAND and JUNO to be ideal to probe such scenarios.

\begin{figure}
    \centering
    \includegraphics[width = 0.8\textwidth]{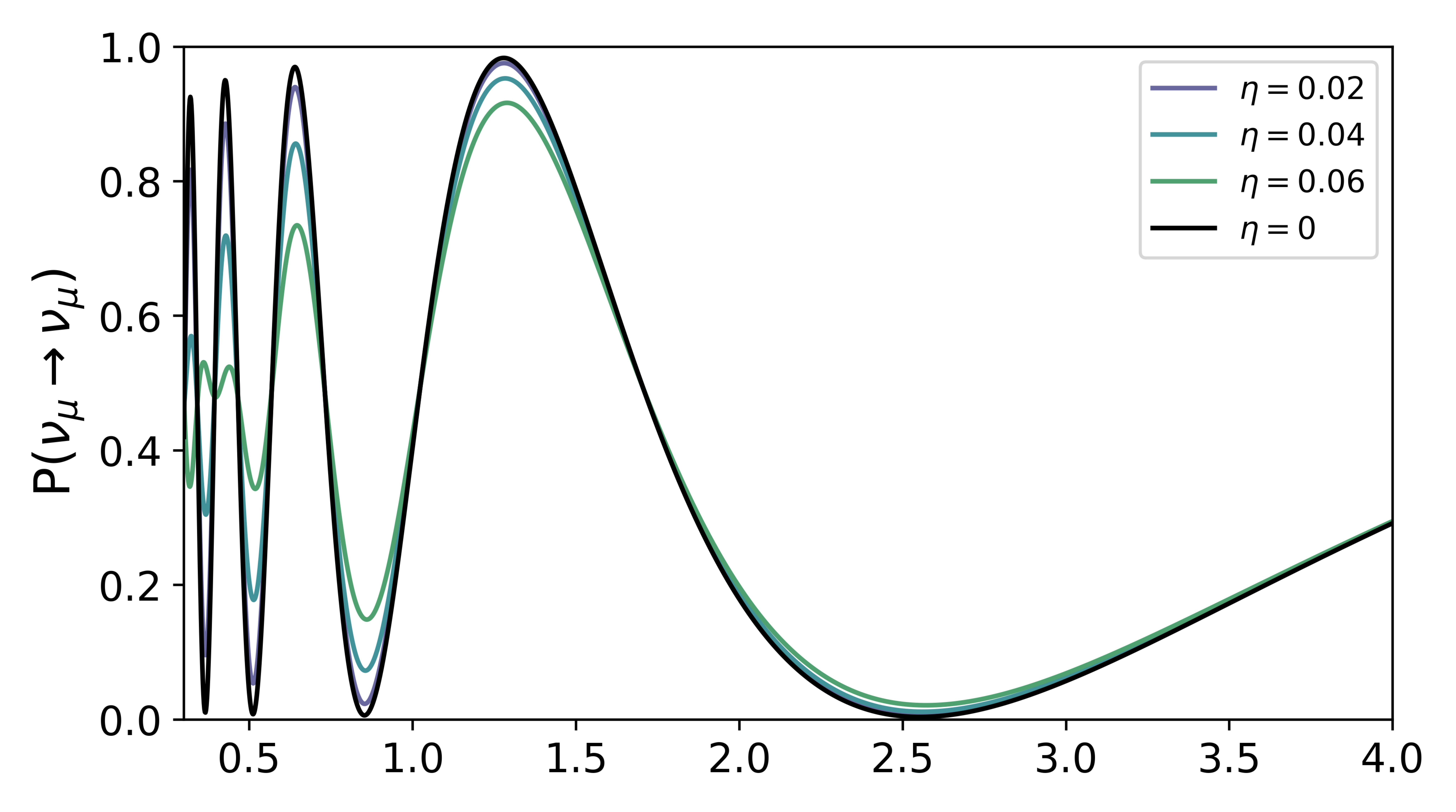}
    \caption{Averaged $\nu_\mu$ disappearance probability at DUNE for a modulating $\Delta m^2_{31}$ for three different values of $\eta$ and in the absence of DiNOs ($\eta$= 0). \labfig{fig:ch10-av-dm}} 
\end{figure}

\subsection{Dynamically Distorted Neutrino Oscillations \label{sec:ch10-dynamical-dinos}}
The last regime is the one in which the effect of modulation of the neutrino mass matrix in oscillations needs to be taken into account at the Hamiltonian level. The matter potential induced by the ultralight scalar field can be written as~\cite{Brdar:2017kbt} 
\begin{equation}
  V_\phi(t) = \frac{1}{2E}\left[(Y m_\nu + m_\nu Y)\Phi(t)+Y^2\Phi^2(t)\right],
  \label{eq:ch10-effpot}
\end{equation}
where we have defined the coupling matrix $Y\equiv m_\nu g/m_N$. Then, the Hamiltonian that drives the evolution of the system is
\begin{equation}
  H(t) = H_{0} + V_{\rm matt}+ V_\phi(t),
\end{equation}
where $H_{0}$ is the vacuum Hamiltonian in the mass basis and $V_{\rm matt}$ is the usual matter potential in the same mass basis. Since $\Phi$ evolves in time, the full Hamiltonian also depends on time. We refer to this scenario as dynamically distorted neutrino oscillations.

We propose a simplification of this treatment which is valid when considering a modulation effect either solely in the mixing angles or solely in the mass splittings. We also neglect the terms that would arise from the term in Equation \ref{eq:ch10-effpot}, since it would be suppressed. \footnote{Notice that $Y ^2$ is proportional to $g^2/m^2_N$ and we are working in the limit of large $m_N$.} If the modulation of $\Phi$ affects solely the mass splittings, we can work in the mass basis, denoted by a \textit{mass} subscript, and write
\begin{align}
&H_{\rm mass}(t)\equiv \nonumber\\ & \hspace{0.6cm} \frac{1}{2E}
\left(\begin{array}{ccc}
 0 & 0  &0   \\
 0 & \Delta m^2_{21}(t)  & 0  \\
 0 & 0  & \Delta m^2_{31}(t)
\end{array}\right)
   +U^\dagger \left(\begin{array}{ccc}
 V_e & 0  & 0  \\
 0 & 0  & 0  \\
 0 & 0  & 0
\end{array}\right)U\, ,
\end{align}
where the time-dependent mass splittings are given in \ref{eq:ch10-dmsq}. 

The evolution of this state has to be solved numerically and leads to an \textit{instantaneous} oscillation probability from flavour $\alpha$ to $\beta$, namely, $P_{\alpha\beta}(t_0,L)$. Numerically, in the adiabatic approximation, the propagation of the neutrino from the source to the detector can be implemented by dividing the path into layers. The oscillation probability from flavour $\alpha$ to flavour $\beta$ is given by
\begin{equation}
    P_{\alpha\beta} (t_0, L) = \Bigg|\bra{\nu_\alpha} U \left\{\prod_{n=1}^N \exp\left[i H_{\text{mass}}(t_n) \Delta L\right]\right\}U^\dagger\ket{\nu_\beta}\Bigg|^2 \, ,
\end{equation}
for a modulating mass-splitting. Here, $\ket{\nu_{\alpha}}$ is a flavour state and \mbox{$t_n\equiv t_0+n\Delta L$.} 

Throughout the duration of the experiment, all possible initial phases will be scanned randomly. 
Therefore, the observed oscillation probability is the time average of $P_{\alpha \beta}(t_0,L)$, namely
\begin{equation}\label{eq:ch10-dyn-obs-prob}
  \langle P_{\alpha\beta}(L)\rangle = \frac{1}{\tau_\phi}\int_0^{\tau_\phi}\text{d}t_0 P_{\alpha \beta}(t_0,L),
\end{equation}
where again, $\tau_\phi=2\pi/m_\phi$ is the period of oscillation of $\Phi$.

When $ L/c \ll \tau_\phi$, the variation of oscillation parameters during the journey is negligible and the only non-standard effect is the different initial phases \mbox{$\varphi=2\pi t_i/\tau_\phi$.} In this limit, the time-averaged oscillation probability in Equation~\ref{eq:ch10-dyn-obs-prob} reduces to the formula for averaged oscillations in Equation~\ref{eq:ch10-average-prob}. It is also easy to see that, in the limit when $L/c \gg \tau_\phi$, the rapid oscillation in the masses from layer to layer leads to cancellations in the time-varying part such that we retain only the standard oscillation formula, as expected.

For the case of modulating mixing angle, it is more convenient to write the Hamiltonian in the flavour basis, denoted by the subscript \textit{fl}. Let us write the full Hamiltonian as
\begin{align}
&H_{\rm fl}(t)\equiv\nonumber\\ &\hspace{0.6cm} \frac{1}{2E}U(t)
\begin{pmatrix}
 0 & 0  &0   \\
 0 & \Delta m^2_{21}  & 0  \\
 0 & 0  & \Delta m^2_{31}
\end{pmatrix}U^\dagger(t) + \left(\begin{array}{ccc}
 V_e & 0  & 0  \\
 0 & 0  & 0  \\
 0 & 0  & 0
\end{array}\right),
\end{align}
where the lepton mixing matrix is now changing in time.
The oscillation probability can be calculated with
\begin{equation}
    P_{\alpha \beta} (t_0, L) = |\bra{\nu_\alpha} \left\{\prod_{n=1}^N \exp\left[i H_{\rm fl}(t_n) \Delta L\right]\right\}\ket{\nu_\beta}|^2\, ,
\end{equation}
for a modulating mixing angle. Note that the observed oscillation probability is again given by the average over $t_0$, as in Equation \ref{eq:ch10-dyn-obs-prob}.

We can see the effect of dynamically Distorted Neutrino Oscillations in \reffig{fig:ch10-dynprob} for modulations of $\Delta m^2_{31}$ and $\theta_{23}$  --- in the left and right panels respectively --- for a modulation amplitude of $\eta=0.1$ and $L=1300$~km. In the left panel, it is clear to see that as $\tau_\phi/\tau_\nu$ gets larger, the effect shrinks, as the changes in the matter effect become too fast to affect neutrino oscillations. Besides, as $\tau_\phi/\tau_\nu$ gets smaller, the effect of dynamically Distorted Neutrino Oscillations tends to the one of averaged DiNOs. For mixing angle modulations --- in the right panel --- the Jacobian effect discussed previously in Subsection \ref{sec:ch10-time-modulation} --- i.e. the fact that the amplitude of the modulation is approximately proportional to $\sin(4\theta_{23}) $ --- suppresses the impact of modulating $\theta_{23}$ in DUNE. 

Notice also that there is a small displacement of the minima and maxima of oscillations. This is simply because the effective mass splitting measured in long baseline muon neutrino disappearance is not exactly $\Delta m^2_{31}$, but rather a function of the mass splittings and mixing angles~\cite{Nunokawa:2005nx}. In this plot, we have chosen a fixed value of $\Delta m^2_{31}$ to obtain the curves. In a realistic analysis, this displacement would simply be mapped into a different value of the atmospheric mass splitting and therefore, there would be a correlation between the $\Delta m^2_{31}$ and the new parameters from this scenario, $m_\phi$ and $\eta$. 

\begin{figure*}
    \includegraphics[width = 0.37\paperwidth]{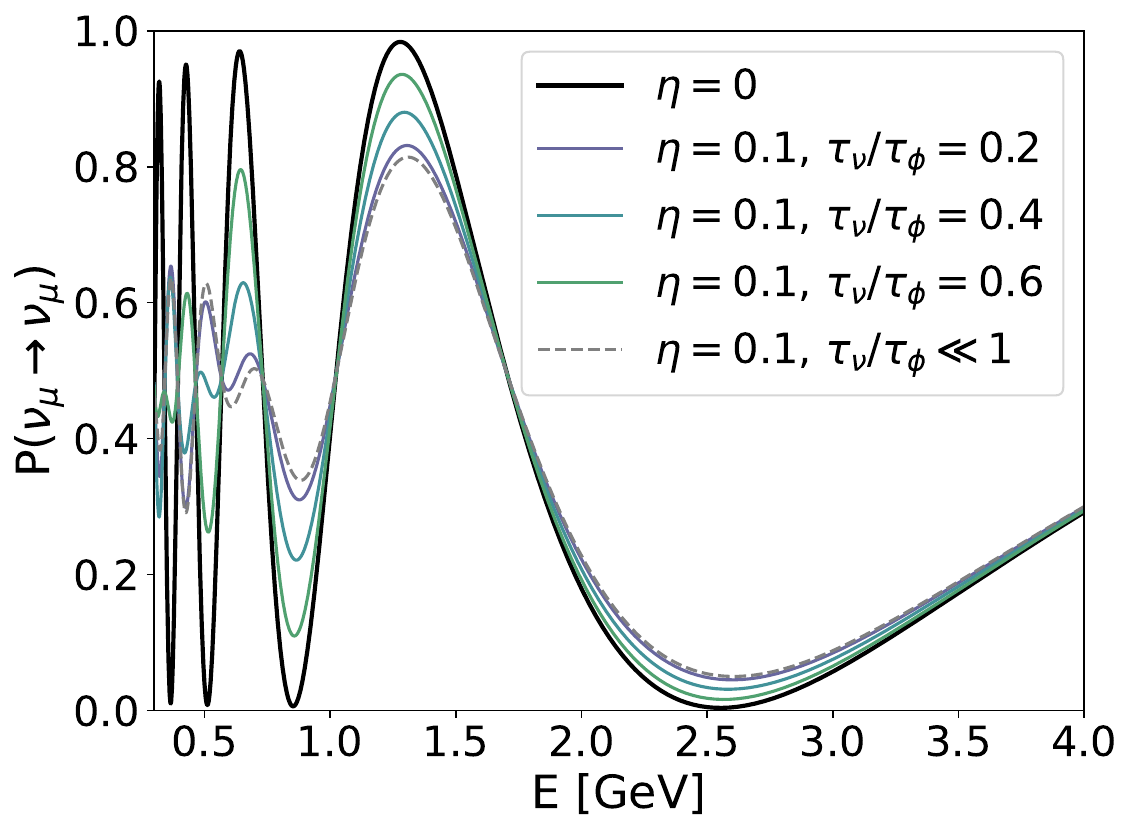}~
    \includegraphics[width = 0.37\paperwidth]{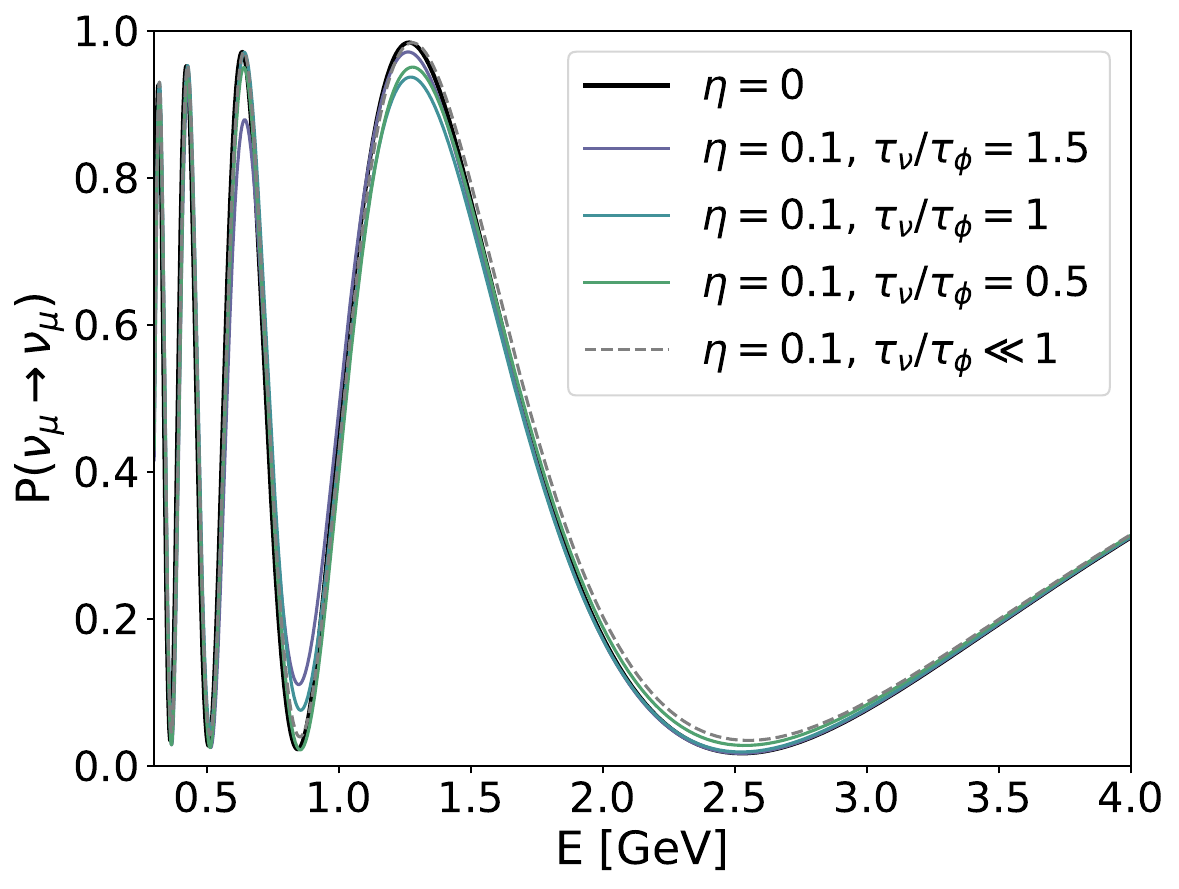}
    \caption{The muon neutrino disappearance probability at DUNE for a modulating $\Delta m^2_{31}$ and $\theta_{23}$ in the dynamical regime --- in the left and right panels respectively --- for different values of the ratio between the period of the modulation, $\tau_\phi$, and the neutrino time of flight, $\tau_{\nu}$. \labfig{fig:ch10-dynprob}}
\end{figure*}

\section{Case study: expected sensitivity at DUNE}
\label{sec:ch10-DUNE}
In this section, we will perform a case study of how the oscillation experiments can probe the ultralight scalars that we introduced in the previous sections. We will analyze the sensitivity of DUNE to the aforementioned three regimes. For all analyses performed here, we follow the simulation of \cite{DUNE:2016ymp} using \texttt{GLoBES}~\cite{Huber:2004ka, Huber:2007ji}. We have assumed the DUNE experiment to have a 1.07~MW beam, four far detectors with a total mass of 40~kton, and a total run time of 7 years, equally divided between neutrino and antineutrino mode. We take the matter density to be constant $\rho=2.8~{\rm g}/{\rm cm}^{3}$. We will focus on the disappearance channels --- for neutrinos and antineutrinos --- since they have the largest statistics and hence are the most suitable to study the phenomenology of neutrinophilic ultralight scalar.

In all cases, we generate our mock data assuming normal ordering and take the best-fit values of the oscillation parameters from~\cite{deSalas:2017kay} as the true ones.

\subsection{Time modulation phenomenology at DUNE}
The signature of the time modulation regime is the presence of a periodic signal in the neutrino spectrum at the far detector. As can be seen in Equations \ref{eq:ch10-angles} and \ref{eq:ch10-dmsq}, the period of modulation is given by the mass of the scalar field, and thus a positive signal of time modulation at DUNE would provide a measurement of the mass of this field.

In general, the search for periodic signals in data sets can be performed using the Lomb-Scargle periodogram~\cite{Lomb:1976wy,Scargle:1982bw}, which is an extension of the classical periodogram for unevenly separated data --- for a pedagogical review, see~\cite{VanderPlas_2018}. Such searches are not new in the context of neutrino physics. For instance, many analyses of periodicities in the solar neutrino flux have been performed in SNO~\cite{SNO:2005ftm,Ranucci:2006rz} and SuperKamiokande~\cite{Super-Kamiokande:2003snd,Ranucci:2005ep}, as well as for other time-varying signals in the context of Lorentz and CPT violation in Daya Bay~\cite{DayaBay:2018fsh}. 

Let us define the Lomb-Scargle (LS) power for a frequency $\omega$ as 
\begin{align}
    P_{\text{LS}}(\omega) &=  
\frac{1}{2}\Bigg\{ \bigg( \sum_{n} g_n \cos [2\pi \omega (t_n -\tau)]\bigg)^2 \bigg/ \sum _n \cos^2 [2\pi \omega (t_n -\tau)] \nonumber \\ & +
\bigg( \sum_{n} g_n \sin [2\pi \omega (t_n -\tau)]\bigg)^2  \bigg/ \sum _n \sin^2 [2\pi \omega (t_n -\tau)]   \Bigg\}
\end{align}   
where $g_n = g(t_n)$ is the signal at the time of the measurement $t_n$ and the parameter $\tau$ is defined by solving 
\begin{align}
\tan (4\pi \omega \tau) = \sum_n \sin(4 \pi \omega t_n) / \cos(4\pi \omega t_n)\, .
\end{align}
As for a classical periodogram, the LS periodogram --- which results from representing the magnitude-squared Lomb-Scargle power as a function of the frequency --- will display a pattern of peaks. The size of the peak at a frequency $\omega_0$ can be related to the significance with which one could state that data shows a periodic behaviour with frequency $\omega_0$.
One can quantify the significance of a certain LS power with the False Alarm Probability Test (FAP), which is a measure of how likely it is that a data set with no periodic signal would give rise to a spurious peak --- form background noise --- of the same magnitude. We estimate the FAP following the Baluev approach~\cite{Baluev:2007su}.

In our analysis, we consider different forms of binning in energy to enhance the sensitivity to this scenario. In an experimental setup, different choices of the time bins can be explored \textit{a posteriori}, ensuring that the whole accessible parameter space is covered. In the case of a positive modulation signal, the correlations and anticorrelations discussed in \reffig{fig:ch10-timemod-dm} can be exploited to further constrain the model.

\begin{figure*}
    \centering
    \includegraphics[width = 0.74\paperwidth]{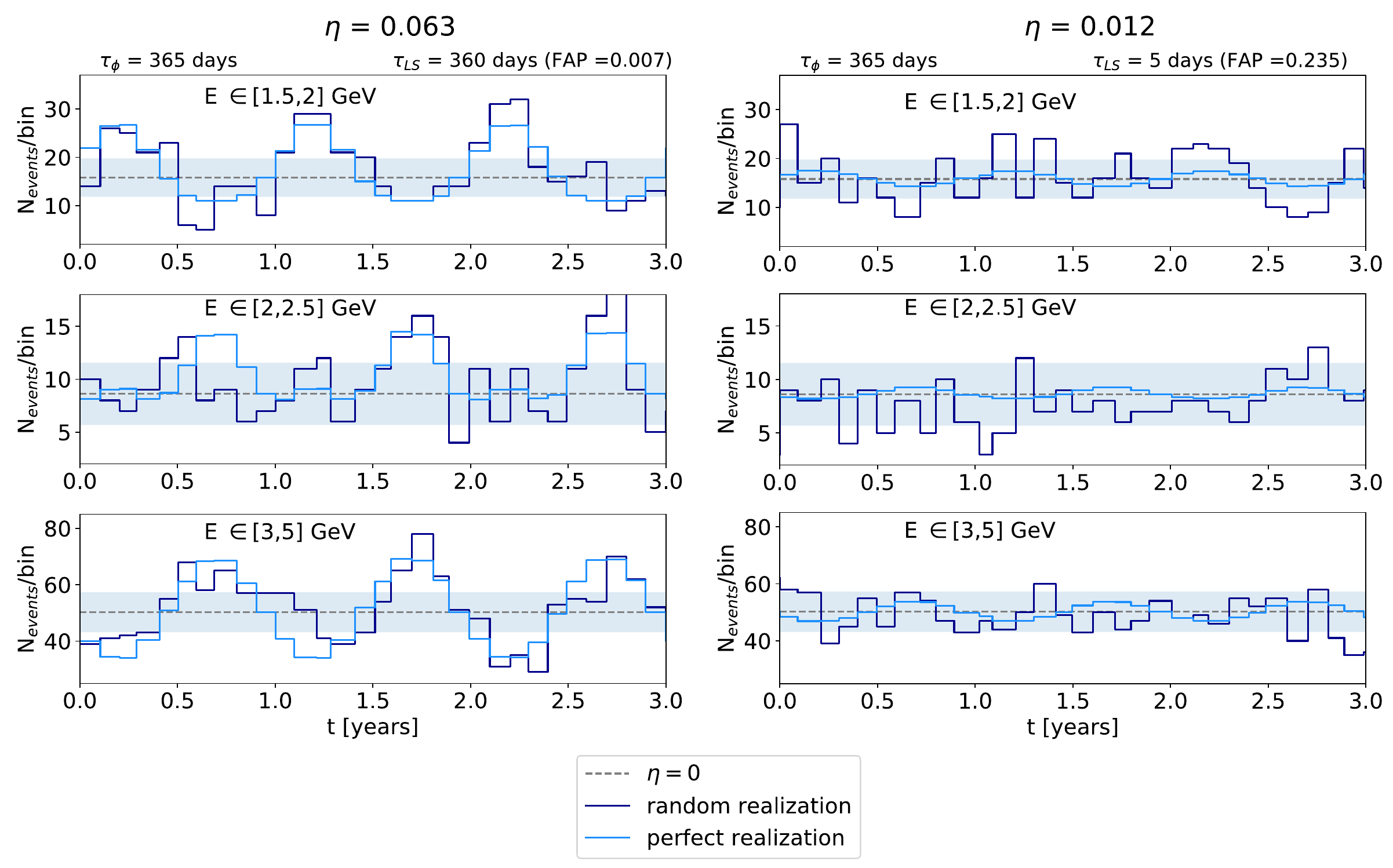}
    \caption{Expected number of events in different time bins at three different energy bins for a particular realisation of DUNE. Gray dashed lines represent the number of events expected with $\eta = 0$ and the shadowed region corresponds to the $1\sigma$ statistical uncertainty. Light blue lines are the expected signal with non-zero $\eta$ and $m_\phi$, while dark blue lines represent a random realisation of the signal which includes statistical fluctuations. The left panel corresponds to a set of parameters $\eta$ and  $m_\phi$ in the 90$\%$ CL region above shown. The right panel corresponds to a point in the $\eta-m_\phi$ plane for which the frequency would not be identified correctly with the LS method. \labfig{fig:ch10-allowed}}
\end{figure*}

For illustrative purposes, we present in \reffig{fig:ch10-allowed} two realisations of the DUNE experiment with a modulating mass splitting. Two modulation amplitudes are shown, namely $\eta=0.063$ and $\eta=0.012$. As we will see later, DUNE is expected to be sensitive to the larger one but not to the smaller one. The expected number of events is presented for three different energy bins assuming 3 years running time in the neutrino mode. The original period of the data ($\tau_\phi$) and the one determined with the LS method ($\tau_{LS}$) are presented together with the FAP score.

In \reffig{fig:ch10-LS}, we show the DUNE sensitivity to the modulations of the mass splitting $\Delta m^2_{31}(t)$. We present region in the two-dimensional plane $m_\phi$ - $\eta$ for which the Lomb-Scargle method would identify a frequency such that one would be able to state \textit{there is a 90$\%$ probability that the periodic signal found in data is not due to random noise}. Notice that this statement is different from affirming that there is a periodic signal with a given frequency in the data set.

\begin{figure}
    \centering
    \includegraphics[width = 0.86\textwidth]{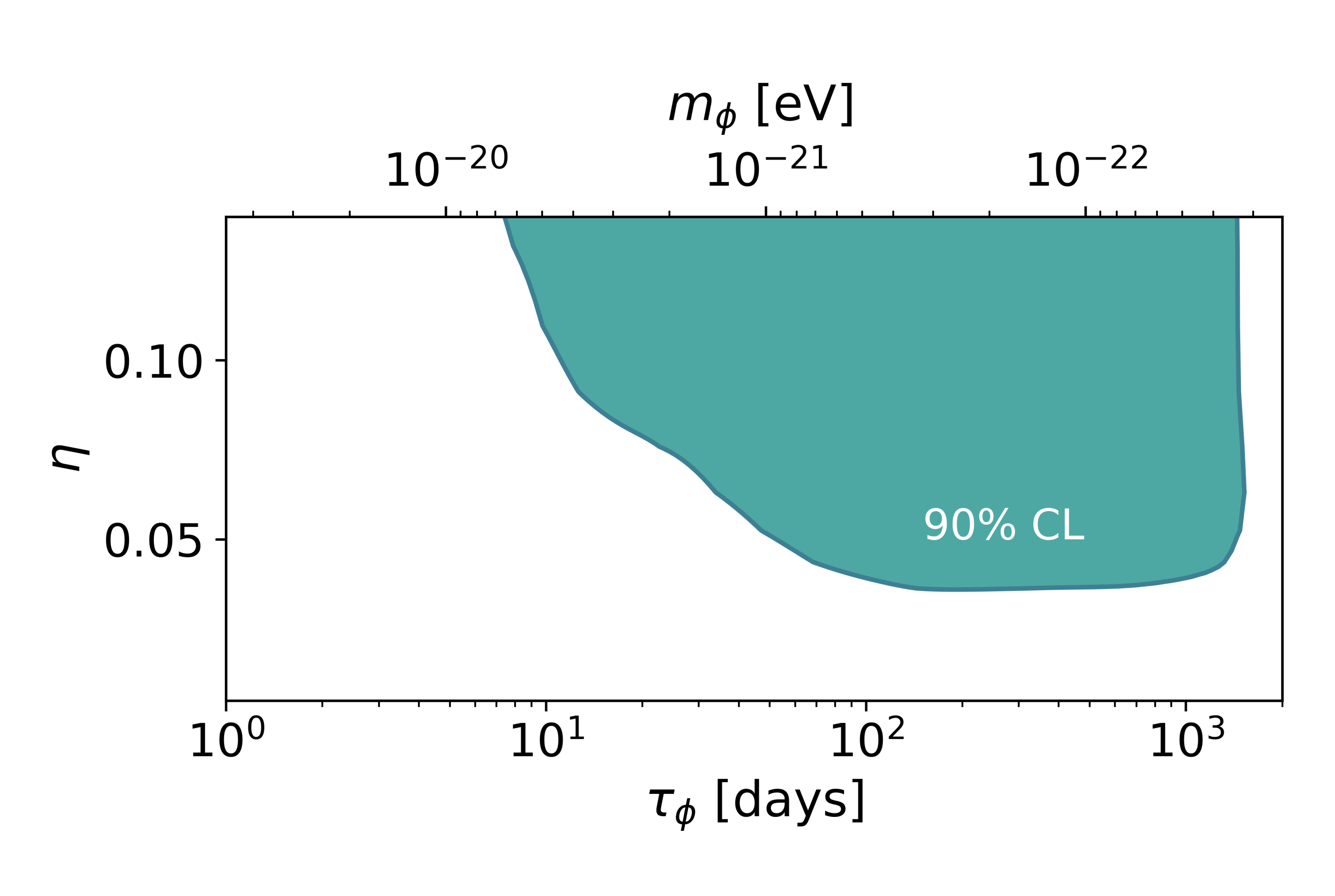}
    \vspace{-0.5cm}
    \caption{Region in the modulation amplitude in the $\tau_\phi$ - $\eta$ plane for which the analysis performed using the Lomb-Scargle periodogram would be capable to state at a 90$\%$ C.L. that the periodic signal found in 7 years of DUNE data is not spurious. The upper x-axis presents the corresponding range for the mass of the ultralight scalar, $m_\phi$. \labfig{fig:ch10-LS}}
\end{figure}

The smallest period --- the largest frequency --- to which DUNE would be sensitive corresponds to a modulation period of about a few days. To be sensitive to larger frequencies smaller time bins are required. This would lead to fewer events per bin and larger statistical fluctuations. Consequently, the sensitivity to the amplitude decreases when looking for shorter periods. The largest period is determined by the running time of the experiment and is expected to be a couple of years.  This means that DUNE would be sensitive to the range of scalar masses  $5\times 10^{-23}\lesssim m_\phi \lesssim 10^{-20}$~eV. Regarding the amplitude of the modulation, we estimate DUNE would be sensitive to values of $\eta \gtrsim 0.4$.

DUNE would not be ideal to search for modulations of the mixing angle $\theta_{23}$. The reason is that since the modulation is approximately proportional to $\sin(4\theta_{23})$ and this mixing angle is close to $\pi/4$, the effect would be strongly suppressed. Therefore, we consider that the search for a modulation of the mass splitting would be a more promising way of detecting the presence of ultralight bosonic dark matter. Nevertheless, the same Lomb-Scargle technique described above could be used to probe the modulations of mixing angles too.

As a last remark, let us comment that this analysis is not free from systematic uncertainties. For instance, deadtime intervals would constitute an important source of systematic errors when analysing time modulations. In particular, quasi-periodic deadtimes due to scheduled breaks of runs or maintenance and calibration operations would manifest as peaks in the Lomb-Scargle power spectrum. However, there exist ways to deal with these systematics related to periodicities in the process of data-taking~\cite{VanderPlas_2018}. If incorporated into the physics program of any experimental collaboration, a specific and detailed analysis of other sources of periodic o quasi-periodic sources of uncertainties would be needed.

\subsection{Averaged Distorted Neutrino Oscillations at DUNE}
The second regime we will study in detail in DUNE is the case of averaged Distorted Neutrino Oscillations. As discussed in Subsection~\ref{subsec:average-dino}, if $\Delta m^2 _{31}$  varies in time, the maxima and minima in the oscillation probability are displaced periodically. For a very fast modulation, such displacement manifests as a non-trivial averaging and has to be carefully studied in order to disentangle it from the distortion caused by the finite energy resolution~\cite{Krnjaic:2017zlz}. Consequently, the searches here presented would benefit from improvements in the energy resolution, as the one proposed in \cite{Friedland:2018vry}. In this analysis, we include systematics as from~\cite{DUNE:2016ymp} and we compute the average oscillation probability in Equation \ref{eq:ch10-average-prob} numerically. 

\begin{figure*}
    \centering
    \includegraphics[width = 0.46\paperwidth]{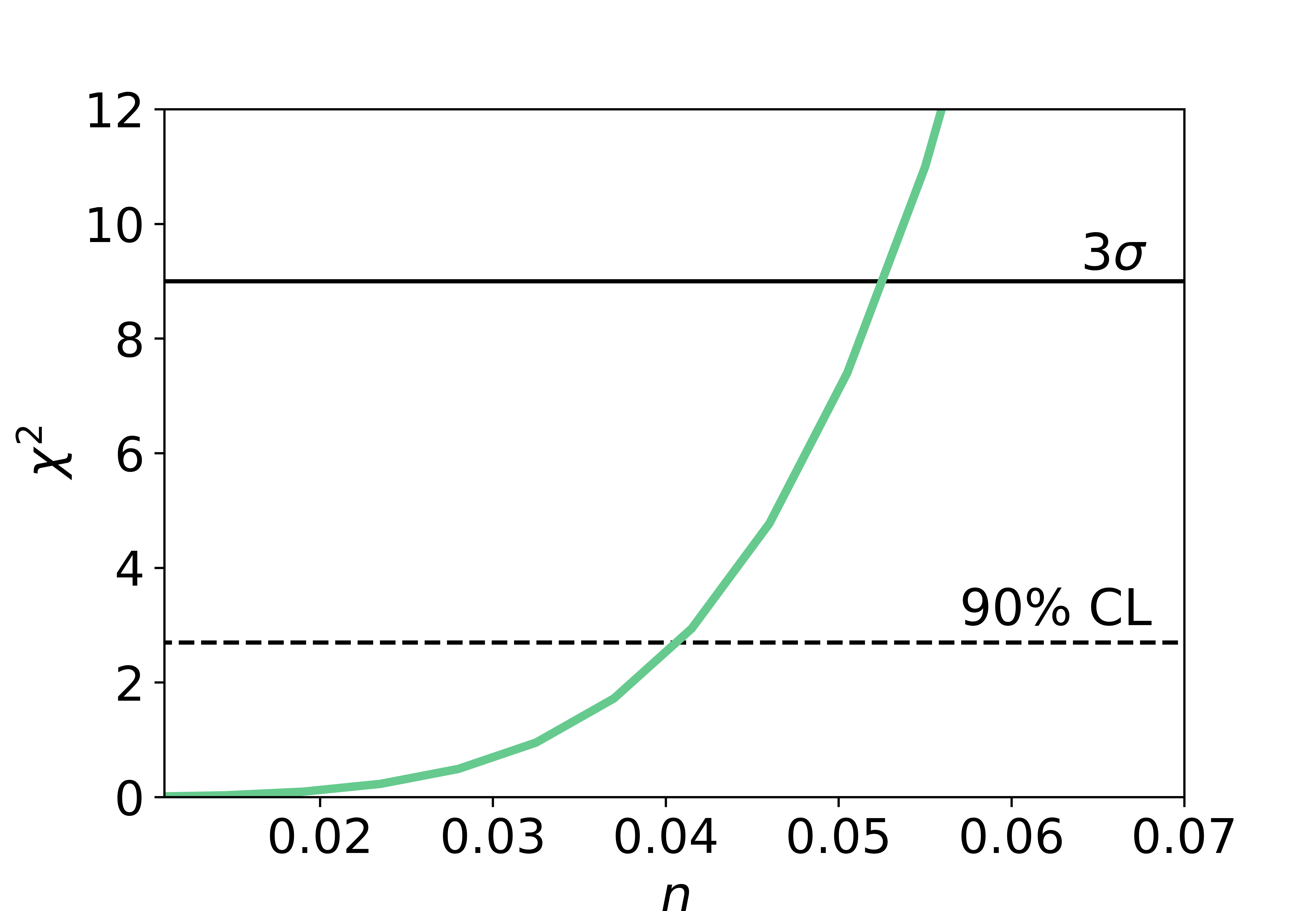}~    
    \includegraphics[width = 0.32\paperwidth]{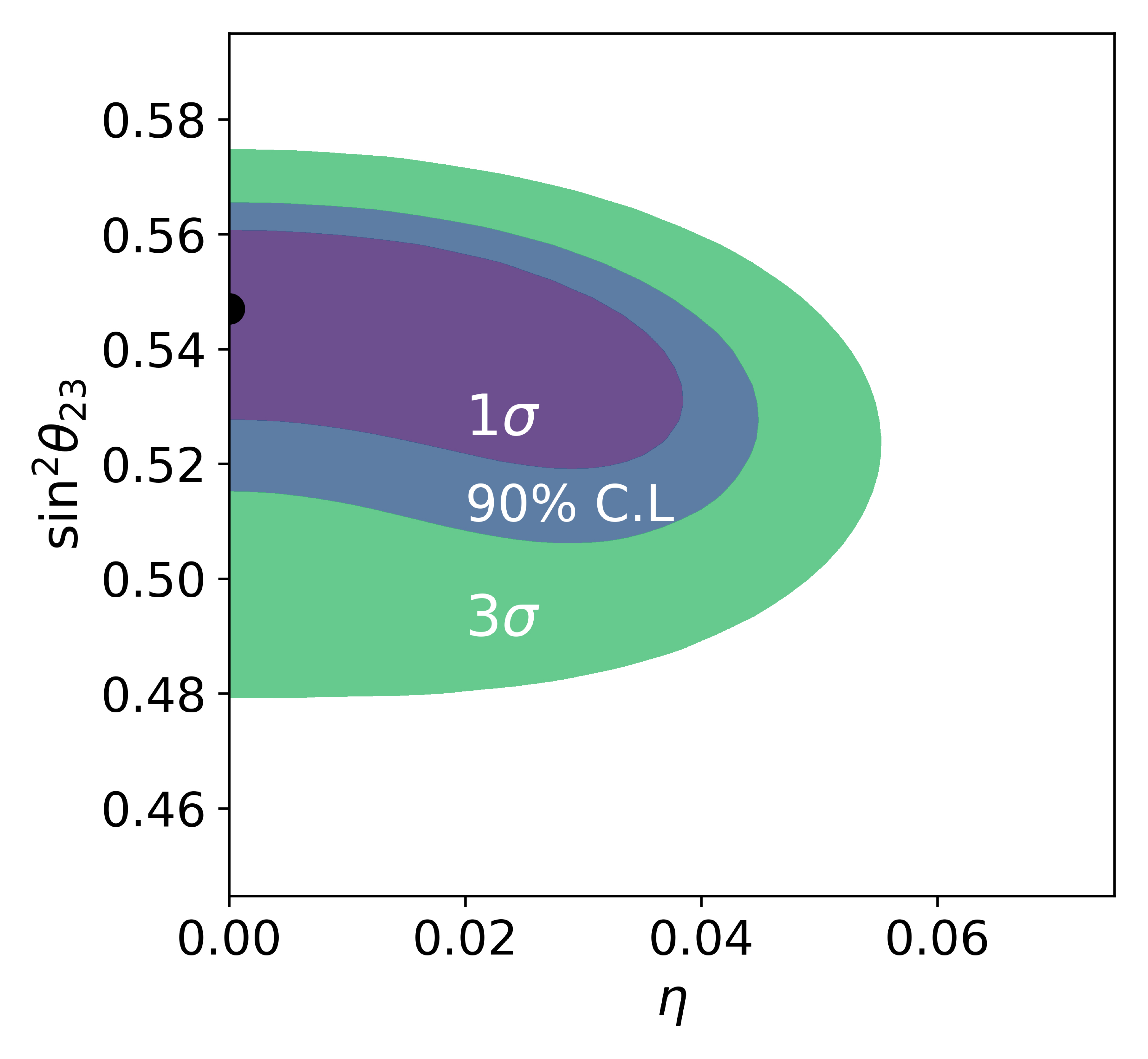}
    \caption{$\chi^2$ profile as a function of $\eta$ and confidence levels to discriminate with respect to the standard three-neutrino picture --- in the left panel --- and allowed region in the $\eta$ -  $\sin^2\theta_{23}$ plane in DUNE ---in the right panel. Best-fit values are marked by a black circle and standard oscillations were assumed as the true hypothesis.\labfig{fig:ch10-chi-av}}
\end{figure*}

In \reffig{fig:ch10-chi-av}, we display the sensitivity of DUNE to a fast modulation of the mass splitting. This regime corresponds to modulation periods between a year and tens of milliseconds --- which translate into a very large range of masses: $2\times 10^{-23} \lesssim m_\phi\lesssim 3\times 10^{-14}$~eV. However, notice that if a signal of averaged DiNOs is observed at DUNE, one would not be able to determine the exact mass of $\Phi$ straightforwardly, as its effect would have been averaged out. From the left panel of \reffig{fig:ch10-chi-av}, one can see that the expected value of $\eta$ at 90\%C.L. corresponds to modulation amplitudes of order 4\%. As we mentioned before, improvements in energy resolution would translate into better sensitivity to this effect.

Finally, it is of interest to see if distorted neutrino oscillations could affect our determination of the oscillation parameters. In the right panel of \reffig{fig:ch10-chi-av}, we display the allowed region by DUNE in the plane $\eta$ - $\sin^22\theta_{23}$ for this regime. Notice that there is some mild degeneracy between $\theta_{23}$ and the modulation amplitude for small values of $\eta$.

\subsection{Dynamically Distorted Neutrino Oscillations}
\begin{figure}[t]
    \centering
    \includegraphics[width = 0.86\textwidth]{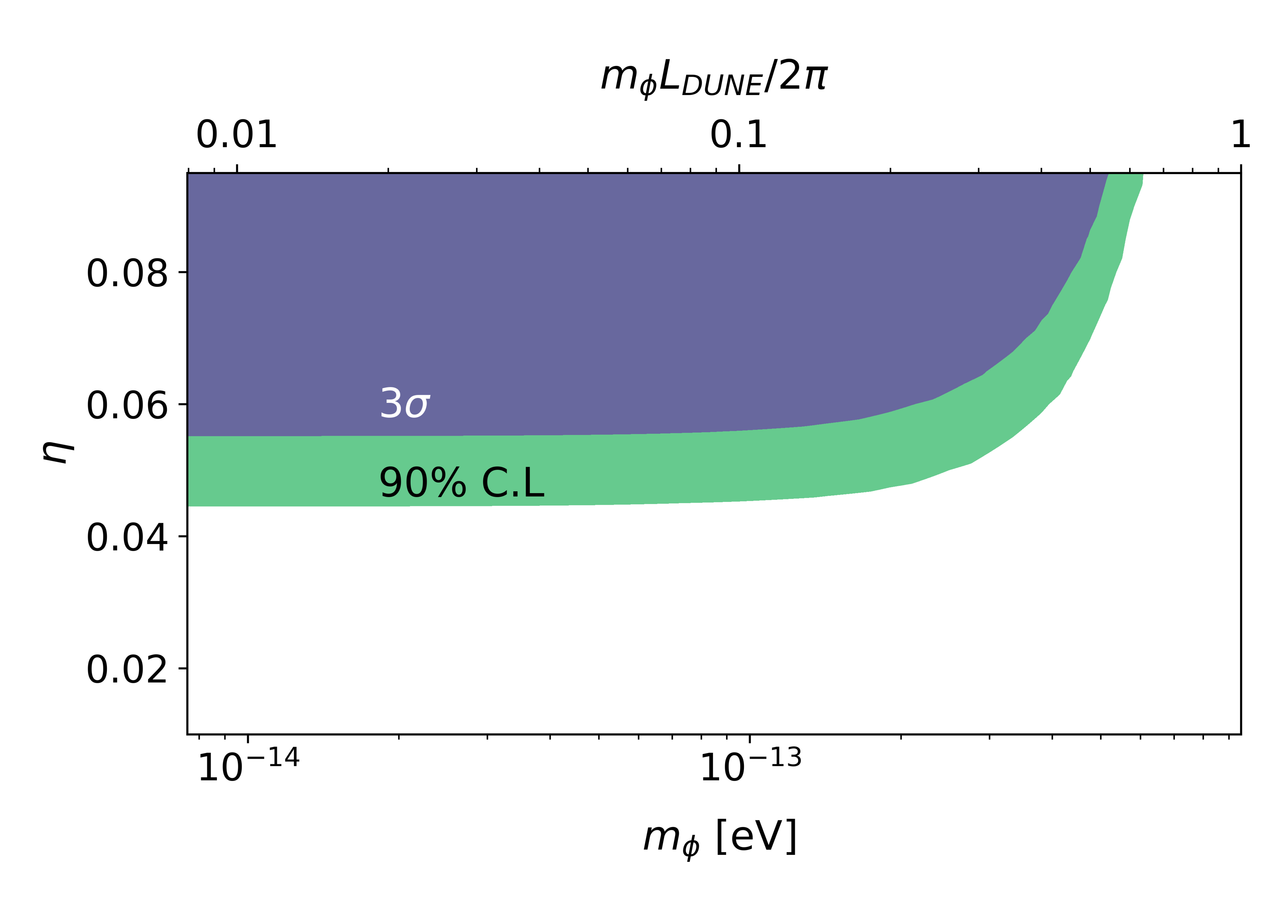}
    \caption{DUNE sensitivity in a fast modulation of the $\Delta m^2_{31}$ mass splitting in the $\eta$ - $m_\phi$ plane at 90$\%$ C.L and $3\sigma$ --- in green and blue respectively. The upper x-axis presents the corresponding range for the dimensionless quantity $m_\phi L_{\rm DUNE} / 2\pi$.\labfig{fig:ch10-dyn-mass}}
\end{figure}
The last regime corresponds to that of dynamically distorted neutrino oscillations --- when the time of flight of the neutrino is comparable to the magnitude of the modulation period. We evaluate the DUNE sensitivity to the dynamical DiNO regime for the case of mass splitting modulations and present the results in \reffig{fig:ch10-dyn-mass}. As expected, when the modulation period is much smaller than the neutrino time-of-flight --- of about 4.3~msec --- the experimental sensitivity degrades as the oscillation probability tends to the standard one. Conversely, when the modulation period is sufficiently large, we recover the sensitivity shown in \reffig{fig:ch10-chi-av} for averaged distorted neutrino oscillations.

\subsection{Combined sensitivity at DUNE}
Finally, in \reffig{fig:ch10-all} we present DUNE's sensitivity to ultralight scalar dark matter incorporating the three searches proposed in this paper. Remarkably, a neutrino oscillation experiment can probe scalar masses ranging about 10 orders of magnitude. One can see that the transition between the dynamically distorted scenario and the averaged one is smooth. Likewise, when time modulation can be directly determined --- the region label as \textit{LS} --- the expected sensitivity is better. While its sensitivity spans a very large region in parameter space, DUNE would only be able to measure the mass of $\Phi$ by the determination of the modulation frequency, which is only possible in the \textit{LS} region. Elsewhere, even if distorted neutrino oscillations are observed, the determination of the ultralight scalar mass might not be feasible.

\begin{figure}[t]
    \centering
    \includegraphics[width = 0.9\textwidth]{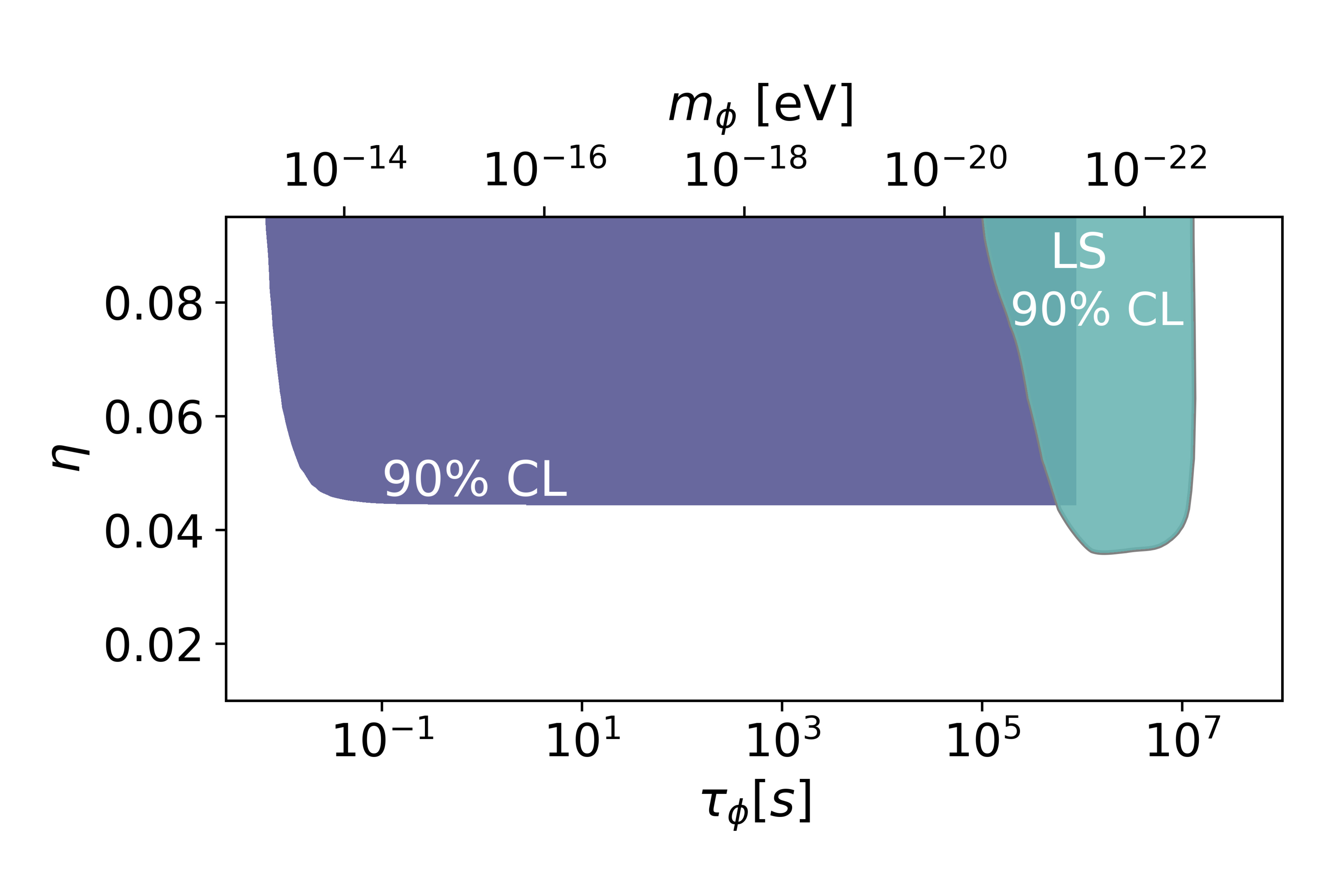}
    \caption{DUNE sensitivity to ultralight scalars via modulation of the atmospheric mass splitting $\Delta m^2_{31}$ at 90\% C.L. The amplitude of modulation is $\eta$ --- as from Equation~\ref{eq:ch10-dmsq} --- and the modulation period and mass of the scalar field is denoted by $\tau_\phi$ and $m_\phi$, respectively.\labfig{fig:ch10-all}}
\end{figure}

\section{Impact in beta-decay measurements}
\label{sec:ch10-uldm-beta}
In the standard three-neutrino scenario, beta-decay experiments --- like Mainz~\cite{Kraus:2012he}, Troisk~\cite{Belesev:2012hx,Belesev:2013cba} and KATRIN~\cite{KATRIN:2019yun, KATRIN:2021uub} --- measure the effective neutrino mass
\begin{align} 
m_{\beta} \equiv  \sqrt{\sum_{i=1}^3 |{U}_{ei}|^2 m^{2}_{i}}\, ,
\end{align}
which receives incoherent contributions from three generations of neutrinos. By analysing the electron spectrum from tritium beta decays, the KATRIN Collaboration has set stringent constraints on the effective neutrino mass ${m}_{\beta} < 0.8~{\rm eV}$ at 90\% C.L. from the combination of the first and second data-taking campaigns (KNM1 + KNM2)~\cite{KATRIN:2021uub}.

A coupling between ultralight scalars to sterile neutrinos can also leave potentially observable imprints in $\beta$-decay experiments, depending on the mass of the sterile neutrino and the mass and amplitude of the scalar field~$\Phi$. 

\textbf{Ultralight scalar coupling to a heavy sterile neutrino}

When the sterile neutrino is heavy, larger than $g\phi$, and decoupled from the energy scale of $\beta$-decay experiments, the decay spectrum is affected due to the mixing of sterile neutrinos with the active ones. To show the effect of a time-varying scalar on the beta-decay spectrum, we consider a simplified modification to the effective neutrino mass, in the limit $g\phi < m_4$, as
\begin{align}
\widetilde{m}_{\beta}(\phi)  \approx 
{m}_{\beta} + y {\phi} \sin{m_{\phi}t} \; ,
\label{eq:ch10-heavylimit}
\end{align}
where $y \equiv \sin^2{\theta}_{14} \cdot g $ represents the effective neutrino coupling suppressed by the active-sterile mixing angle.
For Equation~\ref{eq:ch10-heavylimit}, a uniform mixing of the sterile neutrino to three generations of neutrinos has been assumed, such that $\widetilde{U}_{ei}(\phi) = {U}_{ei}$ and $\widetilde{m}_i(\phi) = m_i+ y\Phi$ hold --- for $i=1,\, 2,\, 3$ thereby leading to the simplified relation in Equation~\ref{eq:ch10-heavylimit}. In a general scenario, a more complex dependence on the scalar field is expected. However, for the accuracy of current beta-decay experiments, the observable effect can be ascribed to a single mass parameter $\widetilde{m}_{\beta}$. Hence, we assume that different coupling patterns do not affect significantly the overall magnitude of modifications.

The $\beta$-decay spectrum with the effective neutrino mass $\widetilde{m}_{\beta}(\phi) $ can be parameterised as
\begin{align}\label{eq:betaExact}
\frac{\text{d}R_{\beta}(E_{e},\widetilde{m}_{\beta}, K_f)}{\text{d}K_f} =  \frac{G^2_{\rm F}}{2\pi^3}|V_{\rm ud}|^2  & \left(g^{2}_{\rm V} + 3  g^{2}_{\rm A}\right)  \frac{m_{^3 {\rm  He}}}{m_{^3 {\rm H}}}  F(Z, E_{e}) \nonumber \\ &\times E_e \sqrt{E^2_{e} - m^2_e}  H(E_{e}, \widetilde{m}_{\beta}, K_f)   \, ,
\end{align}
with the spectral function
\begin{align}
&H(E_{e}, \widetilde{m}_{\beta}, K_f)=\nonumber \\ &\hspace{2cm}
\left(K_{\rm end,0} - K_{e}- K_f\right) \sqrt{(K_{\rm end,0}-K_{e}-K_f)^2 - \widetilde{m}^{2}_{\beta} } \, .
\end{align}
Here, $G_{\rm F}$ is the Fermi coupling constant, $V_{\rm ud}$ is the weak mixing matrix element, $g_{\rm V}=1$ and $g_{\rm A}=1.247$ are the vector and axial-vector weak coupling constants of tritium respectively. \footnote{The step function, which results from requiring a positive neutrino energy, is omitted but should be understood.} The function $F(Z,E_{e})$ is the ordinary Fermi function describing the spectral distortion in the atomic Coulomb potential, 
\begin{equation}
    F(E_e, Z) = 2(1+ \gamma)(2\,  R\,  p_e )^{-2(1+\gamma)} \frac{|\Gamma(\gamma + i \xi)|^2}{\Gamma(2\gamma +1)^2} \,
\end{equation}
where we have defined the quantities
\begin{equation}
    \gamma = (1 -\alpha^2 _{\rm EM}Z^2)^ {1/2} \quad \text{ and } \quad \xi = \frac{\alpha_{\rm EM}ZE_e}{p_e} \, .
\end{equation}
In these expressions, $\alpha_{\rm EM}$ denotes the fine-structure constant, $p_e$ is the electron momentum and we take $R = 2.8840 \times 10^{-3} m_e ^{-1}$~\cite{Ludl:2016ane,Mertens:2014nha}. Throughout this work, we use $E_{e}$ and $K_{e}$ to distinguish the total and kinematic electron energies, while $K_{\rm end,0}$ denotes the electron endpoint energy in the massless neutrino limit.

Likewise, one needs to account for the fact that the final state can have some different energy depending on the final atomic state --- which we denote with the subscript $f$. The reason is that not only is the atomic ground state populated, but a fraction $P_f$ of the decay also ends in states with excitation energy $K_f$ \cite{Saenz:2000dul},
\begin{align}
    R_\beta (E_e, \widetilde{m}_\beta) = \int \text{d}K_f P_f(K_f)\frac{\text{d}R_\beta(E_e, \widetilde{m}_\beta, K_f)}{\text{d}K_f}  \, .
\end{align}
Our analysis is based on a $\chi^2$ function comparing the experimental count rate for different average values of the retarding energy, $\langle qU \rangle$, with the one expected for different values of the parameters of the model,
\begin{align}
R\left(\langle qU \rangle_i \right)= N_T A_s \int \text{d}E_e R_\beta (E_e, \widetilde{m}_\beta)f(E_e - qU) + R_{\text{bg}}\, ,
\end{align}
following the procedure detailed in~\cite{KATRIN:2019yun}. Here, the calculation of the expected count rate, $R\left(\langle qU \rangle_i \right)$, depends on the calculated number of tritium atoms in the source multiplied with the accepted solid angle of the \mbox{setup, $N_T$}. It also depends on the response function of the detector, $f(E_e -qU)$ --- which relates the electron energy and the applied retarding potential. In the analysis, we include two free parameters: the background rate, $R_{\text{bg}}$, and the signal amplitude, $A_s$. We also consider the statistical and systematic uncertainties reported by the KATRIN Collaboration in \cite{KATRIN:2019yun}.

The ultralight scalar may manifest itself as a modulation effect in the spectrum of the decay. However, this would require careful planning of the duration of the scans over the spectrum. Alternatively, one can look for the distortion effect by averaging over the modulation of the ultralight scalar.

\begin{figure*}
	\centering
	\includegraphics[width=0.32\paperwidth]{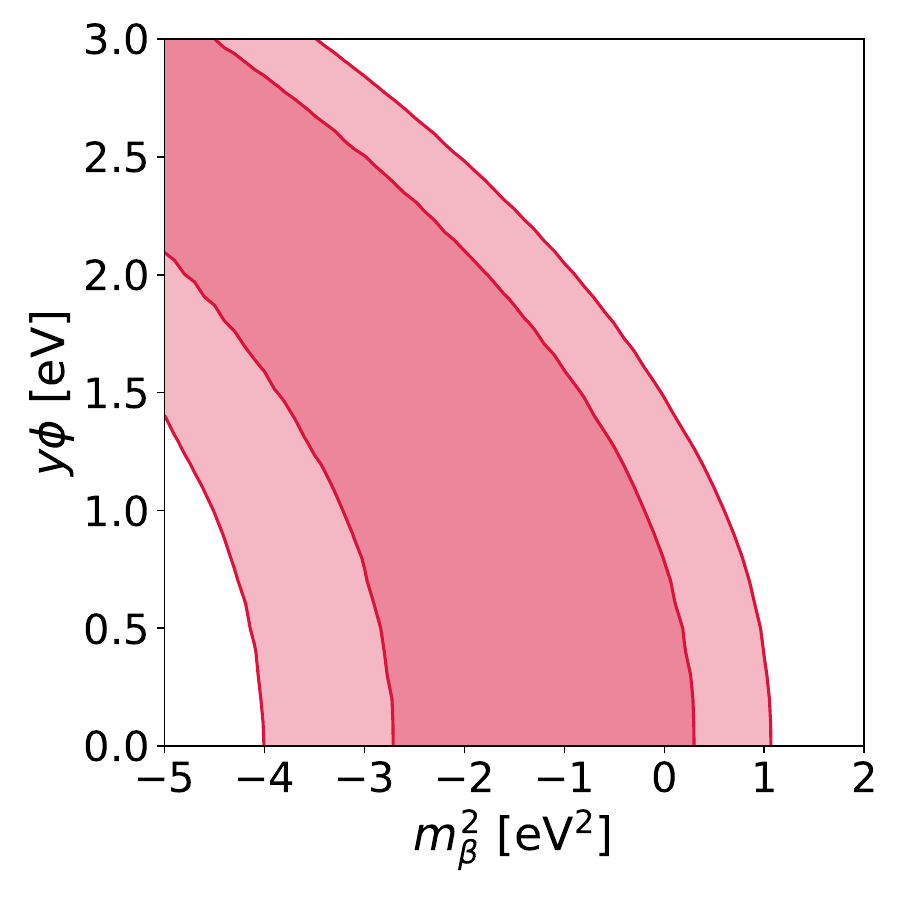}
	\includegraphics[width=0.4\paperwidth]{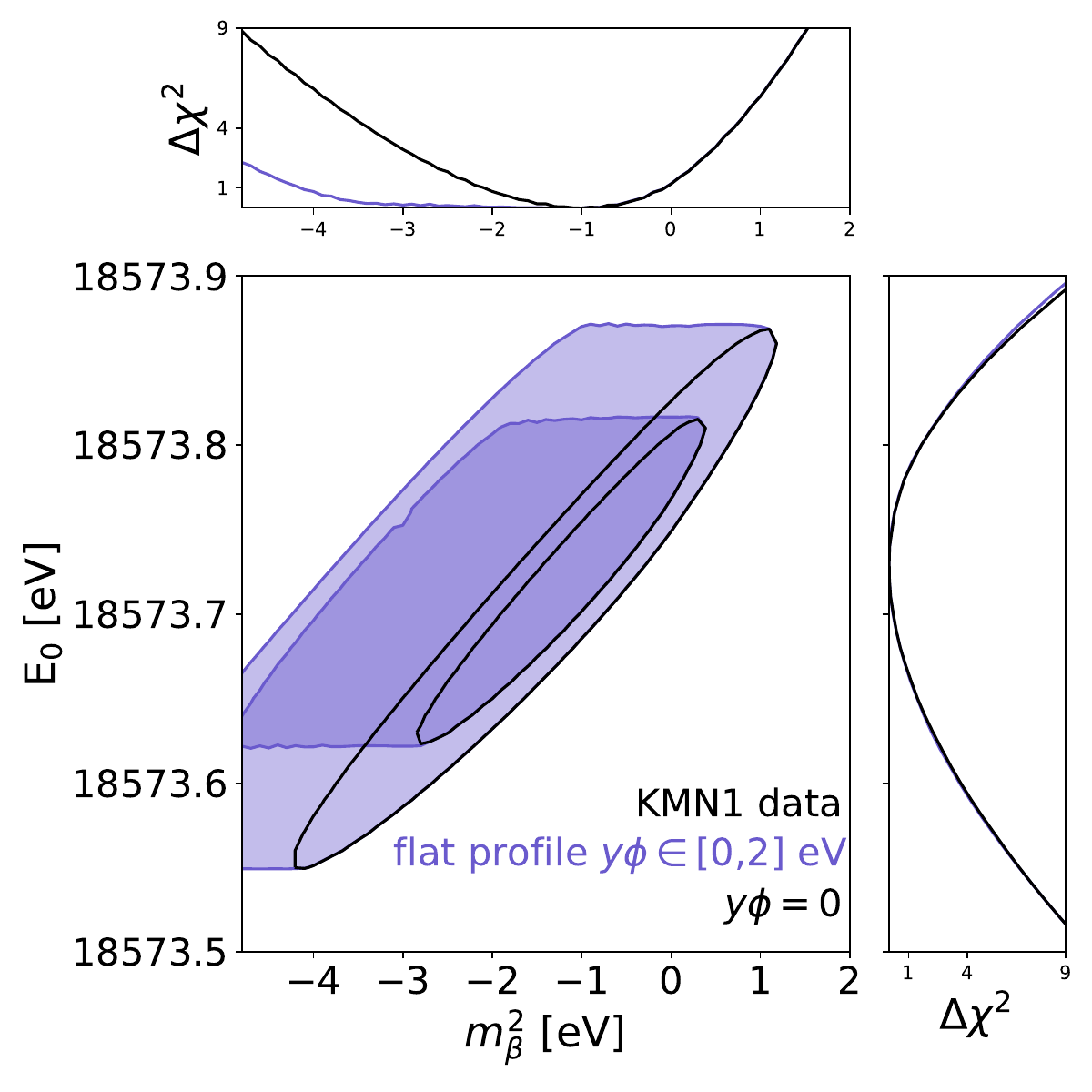}
	\caption{The left panel depicts the allowed regions in the ${m}^2_\beta - y\phi$ plane at 68\% and 95\% C.L. using data from KATRIN's first campaign (KNM1) and in the limit in which $m_4$ is very heavy and $\widetilde{m}^2_\beta$ exhibits a time modulation as in Equation~\ref{eq:ch10-heavylimit}. The right panel shows the corresponding allowed regions in the ${m}^2_\beta - E_0$ plane at the same confidence levels under the assumption of a flat profile for $y\phi$ between $[0,2]$ eV. \labfig{fig:ch10-gphi_mbeta}}
\end{figure*}

Let us investigate analytically how the averaged scalar field would modify the beta spectrum. For the current KATRIN sensitivity, it is a good approximation to keep up to the first order of the perturbative expansions on $\widetilde{m}^{2}_{\beta}$  ---unless $y\phi$ is very large --- in the spectral function, namely,
\begin{align}
H(E_{e}, \widetilde{m}_{\beta}) \propto (K_{\rm end,0}-K_{e}) - \frac{\widetilde{m}^{2}_{\beta}}{2 (K_{\rm end,0}-K_{e})}\, .
\end{align}
The square of effective neutrino mass as in Equation \ref{eq:ch10-heavylimit} averaged over one modulation period is
\begin{align}
\left\langle \widetilde{m}^{2}_{\beta} \right\rangle = m^2_{\beta} + \frac{(y \phi )^2}{2} \, .
\label{eq:ch10-effective_mbeta}
\end{align}
Hence, in such a case the presence of the scalar field directly adds a constant term to the square of effective neutrino mass. For the sensitivity of the first KATRIN campaign, there is a degeneracy between the averaged scalar effect and a usual neutrino mass, but it leads to large neutrino-mass cosmology~\cite{Alvey:2021xmq}.
This degeneracy is non-linear and obvious from \reffig{fig:ch10-gphi_mbeta}, where we have fitted the parameter space of $m^2_{\beta}$ and $y\phi$ using the KATRIN data from the first campaign (KMN1). Following the analysis strategy from the KATRIN Collaboration, we allow $m^{2}_{\beta}$ to become negative during the fit and also leave the end-point energy, $E_0$ as a free parameter. From \reffig{fig:ch10-gphi_mbeta} and Equation \ref{eq:ch10-effective_mbeta}, one can see that in this scenario, the effective neutrino mass measured, $\langle \widetilde{m}^{2}_{\beta} \rangle$, is always larger than the true $m^2 _{\beta}$ and hence, the upper limits derived when assuming $y \phi = 0$ are conservative. From the right panel of \reffig{fig:ch10-gphi_mbeta}, one can also see that the determination of the end-point energy, $E_0$, is not compromised in this scenario.

\textbf{Ultralight scalar coupling to a light sterile neutrino}

Up to this point, we have limited the discussion to the case in which the sterile neutrino is heavy. However, when the sterile neutrino is light enough, an additional emission channel of beta decay will be open. In the (3+1)-neutrino scenario, we can split the contribution from the three active neutrinos and the sterile one as~\cite{KATRIN:2020dpx}

\begin{equation}
R^{(3+1)\nu}_{\beta}(E_{e}) = \left(1- \left|\widetilde{U}_{e4}(\Phi) \right|^2 \right)	R_{\beta}(E_{e},\widetilde{m}_{\beta}) + \left|\widetilde{U}_{e4}(\Phi) \right|^2 R_{\beta}(E_{e},\widetilde{m}_4) \;,
\end{equation}

with $|\widetilde{U}_{e4}|= \sin{\widetilde{\theta}}$. The $\Phi(t)$-dependent mixing angle $\widetilde{\theta}$, the masses $\widetilde{m}_{\beta}$, and $\widetilde{m}_4$ can be calculated from Equations \ref{eq:ch10-wtm1}, \ref{eq:ch10-wtm4} and \ref{eq:ch10-wtth14}, respectively. For simplicity, we assume that the $\Phi(t)$ dependence in $\widetilde{m}_{\beta}$ is approximately that of $\widetilde{m}_1$. This is well justified in light of the current KATRIN sensitivity to the effective neutrino mass. 

\begin{figure*}
	\centering
	\includegraphics[width=0.74\paperwidth]{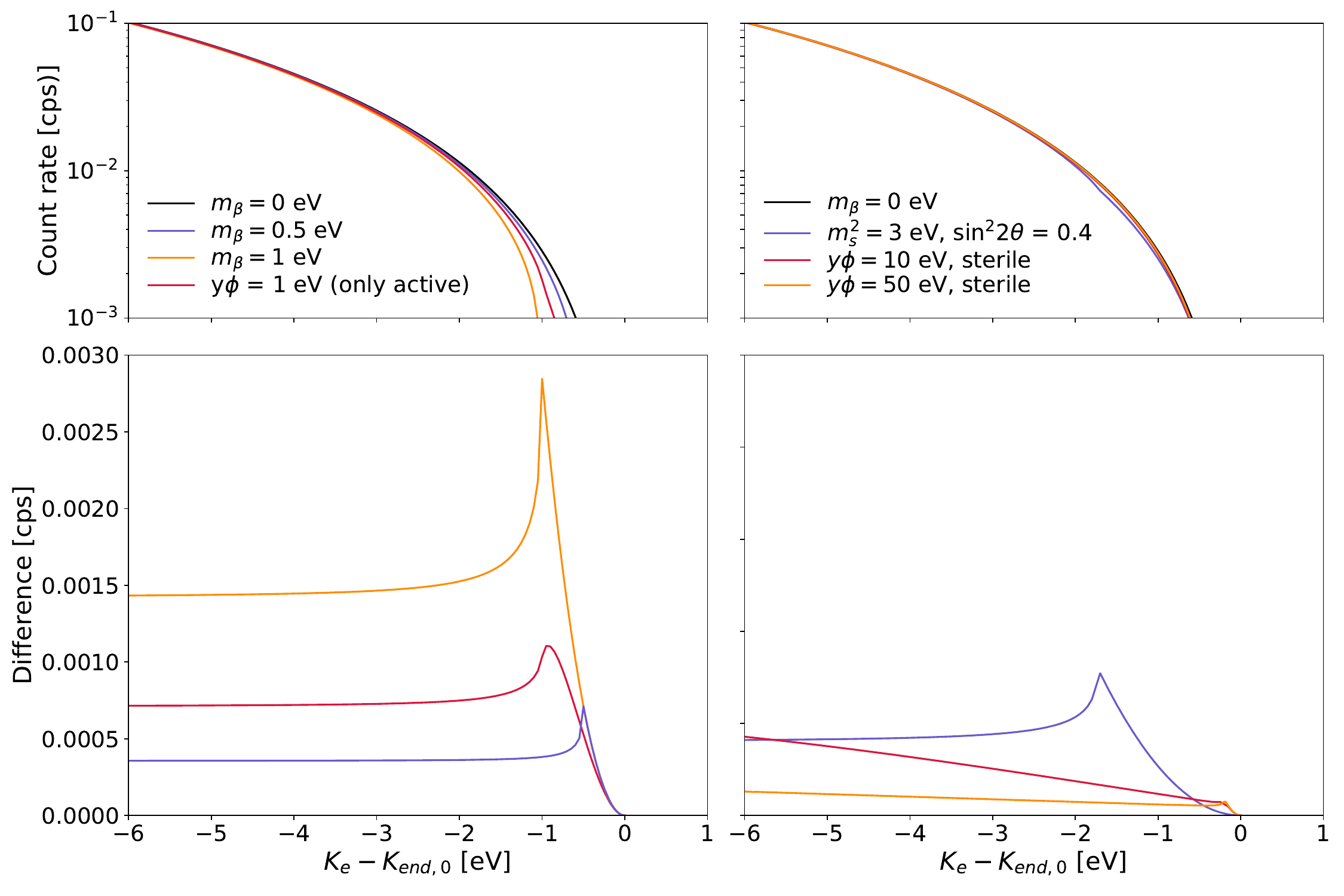}
	\caption{Beta-decay spectrum for various scenarios including the standard beta decays with $m_{\beta} = \{0, 0.5, 1\}~{\rm eV}$ -- -curves from right to left. The heavy sterile neutrino case as in \ref{eq:ch10-heavylimit} with an effective potential $y \phi = 1~{\rm eV}$ is shown by the orange curve in the left panel. 
The right panel shows a benchmark choice of a light sterile neutrino with $m^2_4  = 3~{\rm eV}^2$ and $\sin^2{2\theta} = 0.4$, and the addition of scalar potentials $g\phi = 10~{\rm eV}$ and $50~{\rm eV}$ --- in purple, red, and orange respectively. Note that we assume $m_{\beta} = 0$ for all the cases with non-zero scalar potentials. \labfig{fig:ch10-spectrum}}
\end{figure*}

For a general light sterile neutrino, one has to integrate numerically the exact spectral function over the modulation period. To provide a deeper comprehension of the experimental signatures, we show in the upper panel of \reffig{fig:ch10-spectrum} the $\beta$-decay spectra for different scenarios --- assuming a KATRIN-like experimental configuration. The lower panel depicts the difference of rates between the various scenarios and the standard one --- when $m_{\beta} = 0~{\rm eV}$.

A finite neutrino mass will induce a constant shift to the beta-decay spectrum away from the endpoint and a kink. This is illustrated in the left panel of \reffig{fig:ch10-spectrum} for three different values of $m_{\beta} = \{0,0.5,1\}~{\rm eV}$.
However, the different kinks are still unresolvable given the sensitivity of the current generation of $\beta$-decay experiments.  

A light sterile neutrino will induce an additional kink in the decay spectrum. Such spectral feature --- which is expected for energies around \mbox{$K_e - K_{\rm end,0} \lesssim m_4$} --- can be well resolved provided that the kink is away from the endpoint. An example of such a kink is shown by the purple curve in the right panel of \reffig{fig:ch10-spectrum}. Note that the magnitude of the distortion is related to the size of the mixing $|U_{e4}|^2$.  From these features, beta-decay experiments can set limits to the sterile neutrino mass and mixing \cite{Giunti:2019fcj,KATRIN:2022ith}.
In the right panel of \reffig{fig:ch10-spectrum}, we also illustrate how adding large scalar potentials to the sterile neutrino will smooth these distortions, mainly because the mixing will be suppressed by the potential $g\Phi$. Consequently, this scenario allows to open up the sterile-neutrino parameter space in the context of short-baseline anomalies. %For comparison, we also display the expected spectrum when the sterile neutrino is heavy. We have shown that the effect of the scalar is degenerate with the mass term --- see \reffig{fig:ch10-gphi_mbeta}.

In \reffig{fig:ch10-sterile-yphi}, we show the allowed region in the plane determined by the mass of the sterile neutrino $m^2_s$ and the mixing $|U_{e4}|^2$ for different values of $y\phi$. For simplicity, we have assumed $m^2_\beta = 0$ eV$^2$. Notice that for increasing values of $y\phi$, larger values of the mixing become allowed for masses below 10 eV. Extending the analysis to allow for a non-zero $m^2_\beta$ would smooth the displayed curves. As a related issue, one can see that for non-zero $y\phi$, maximal mixing is not allowed for sterile neutrino masses of $\mathcal{O}$(50 - 100 eV). That region of parameter space corresponds to the limit in which sterile neutrinos are integrated out and only the time modulation affecting active neutrinos can manifest in beta-decay experiments. In that case, there is a degeneracy
between $y\phi$ and $m^2_\beta$ --- see discussion around Equation \ref{eq:ch10-effective_mbeta}. Hence, it is likely that in a more general analysis including $m^2_\beta$ as a free parameter, the constraint in that region will be relaxed. 

\begin{figure}
\includegraphics[width=0.7\textwidth]{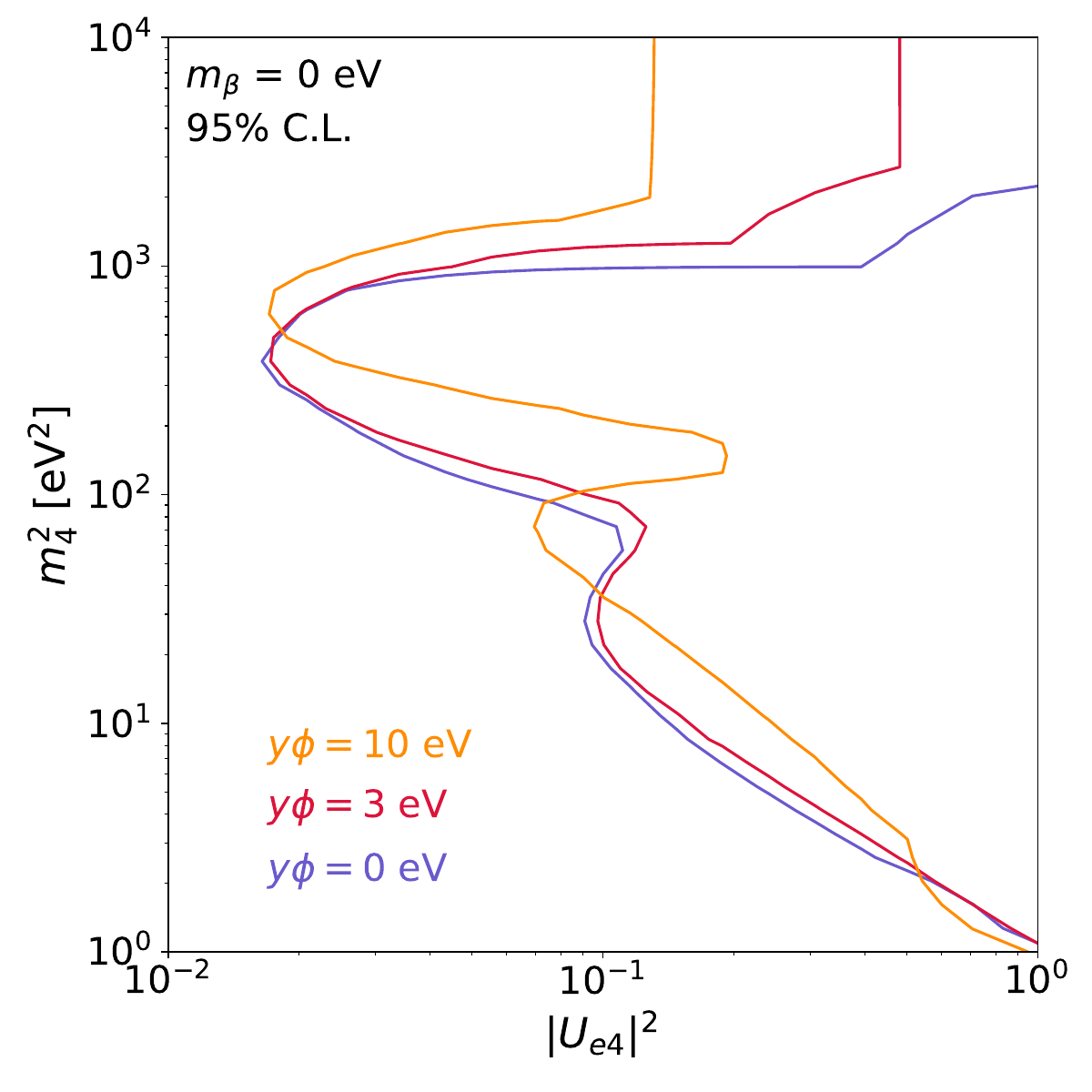}
\caption{Allowed regions at $95\%$ C.L. in the $m^2_s - |U_{e4}|^2$ plane for \mbox{2 degrees} of freedom and different values of $y\phi = \lbrace 0 \, \text{eV}, 3\,  \text{eV, } 10 \text{eV}\rbrace$ for KMN1. \labfig{fig:ch10-sterile-yphi}}
\end{figure}

\section{Critical discussion}
\label{sec:ch10-uldm-critical}
In this chapter, we have discussed several experimental signatures of neutrinophilic ultralight dark matter scalars. We have shown that such a scenario can be realised via a sterile neutrino portal. If a singlet fermion couples to an ultralight scalar, a rich phenomenology is expected in laboratory experiments for a sizeable active-sterile neutrino mixing. However, other realisations have been discussed in the literature.

In the case of oscillation experiments, and depending on the mass of the scalar, such a coupling to ultralight dark matter would result in a time-varying probability, a distortion in the oscillation probability or an additional effective potential. Taking DUNE as an example, we have addressed the sensitivity to these three regimes and commented on the analysis techniques required for such searches. We have also discussed the degeneracies with the standard oscillation parameters.

Regarding $\beta$-decay, a time modulation of neutrino masses would also lead to a non-trivial distortion of the electron --- or positron --- spectrum. Depending on the mass of the sterile, one can differentiate between two situations. 

\begin{itemize}
\item For heavy sterile neutrinos, and given the current experimental sensitivity at KATRIN, there is a degeneracy between the distortion effect and the mass parameter $m_\beta$. Moreover, in this scenario, the effective parameter $\widetilde{m}^2_\beta$ is always larger than the true parameter $m^2_\beta$. This result could be of potential relevance in case KATRIN would determine a value of the neutrino mass scale incompatible with cosmological observations.

\item In the case of light sterile neutrinos and large scalar potentials, \mbox{$y\phi > m_s$,} neutrinophilic ultralight dark matter could enlarge the parameter space in the $m^2_s$ - $|U_{e4}|^2$ plane. In particular, in this scenario, the constraints from $\beta$-decays on the size of active-sterile mixing are relaxed for sterile neutrino masses in the range $m^2_s \sim \mathcal{O}$(1 - 10 eV).
\end{itemize}
Generally, this scenario faces severe constraints from cosmology. According to \cite{Huang:2022wmz}, it is possible to make this scenario compatible with observations from Big Bang Nucleosynthesis and neutrino decoupling. The reason is that, the ultralight scalar redshifts as
\begin{align}
\phi(z) \propto (1 + z)^{3/2} \, ,
\end{align}
and so do neutrino masses as long as $y\phi < m_s$. However, once the size of $y\phi$ overcomes the mass of the sterile neutrino to which it is coupled, the effective active-sterile mixing is suppressed. This fact prevents sterile neutrinos from being produced and thermalised in the early universe. Other consequences in the history of the universe such as the impact on large-scale structure formation and the cosmic microwave background have also been explored in the literature~\cite{Huang:2021kam, Huang:2022wmz}. 

Likewise, searches for neutrinoless double beta decay~\cite{Huang:2021kam} and black hole superradiance~\cite{Brito:2015oca,Davoudiasl:2019nlo} also set constraints in neutrino portals to ultralight dark matter or even in the existence of ultralight scalars.

Although the phenomenology here outlined is very rich and interesting, if regarded with criticism, it is complicated to provide a cosmology-friendly framework that simultaneously presents testable signatures in laboratory experiments. The proposal here discussed, involving one sterile neutrino, provides a first insight into the avenues that could successfully achieve this goal. 

There exist alternative proposals featuring ultralight degrees of freedom. Particularly, neutrinophilic ultralight vectors are known to leave imprints in neutrino experiments which are similar to the ones here discussed. The interest in both ultralight dark matter radicates in the fact that they could explain the small-scale puzzles arising from discrepancies between observations and simulations, and in the rich phenomenology expected when coupled to the Standard Model content.

\pagelayout{wide} % No margins
\addpart{Concluding remarks}
\pagelayout{margin} % Restore margins
\chapter*{Conclusions and final thoughts}
\addcontentsline{toc}{chapter}{Conclusions and final thoughts} 
\label{finale}

Over the last decades, neutrinos have taught us many relevant lessons, the most important one being that the Standard Model of particle physics is an extraordinarily accurate yet incomplete theory. It took more than 20 years to detect for the first time these elusive particles after they were first --- and desperately --- postulated by Pauli. Their weakly interacting nature has constantly challenged our technology and understanding of nature. However, the reward for all the years of patiently waiting and the great generational efforts devoted to neutrino physics is beyond doubt: it is now well-established that neutrinos are massive particles --- at least two of them. Likewise, we also know neutrinos play a relevant role in astrophysical environments such as the Sun or supernova explosions, as well as in the evolution of the universe. 

The lessons left to learn from neutrinos are still numerous -- probably many more than we could think of. Yet, despite our ignorance, the roadmap for exploring the neutrino sector has three well-delineated directions: the study of human-made neutrinos in the laboratory, neutrinos of astrophysical origin and cosmological neutrinos.

Laboratory experiments allow the study of neutrinos in detail under particular conditions, in a relatively wide range of baselines and for different energies. They also probe neutrinos from diverse sources and use various experimental techniques, which also give a handle on the physics behind the production and detection processes. Our current understanding of these experiments and the physics involved are mature enough to say that we are entering the precision era. Moreover, there is an ongoing worldwide effort to build a series of next-generation multi-purpose experiments covering many possible phenomena related to neutrino masses and mixing, other neutrino properties and possible connections with dark matter. However, laboratory experiments have a main limitation: one can not study neutrinos under extreme conditions. For instance, it is not possible to investigate their propagation in large matter densities, in media with high neutrino number densities, under the effects of strong magnetic fields or at extremely high energies.

Cosmological observations are reaching unprecedented sensitivity too. Consequently, the study of neutrino properties through their role in several relevant moments in the history of the universe is becoming a reality. These probes shed light on neutrino masses, mixings and interactions through the impact neutrinos have had on the expansion rate of the universe, the production of light elements and the growth of small-scale structures. In cosmology, neutrinos are studied as an ensemble, with the limitations that the initial conditions are given, the lack of reproducibility, and the fact that all probes so far are indirect and do not involve the direct detection of neutrinos. Nonetheless, the strength of neutrino cosmology lies in the fact that they allow us to test the largest scales ever achievable in terms of time and distance.

The study of astrophysical neutrinos has been of great relevance in previous stages of neutrino physics. In particular, the observation of neutrinos from SN1987A opened the door to studying neutrinos produced in extremely violent events. Besides that,  solar neutrinos triggered interest in neutrino flavour conversions. In recent years, neutrino astronomy and the study of \mbox{high-energy} neutrinos produced in astrophysical environments is a reemerging field. In this case, astrophysical sources are both a blessing and a curse for neutrino physics. Because of the variety of sources, large distances and extreme conditions achieved in astrophysical objects, neutrino astrophysics allow us to prove our fundamental understanding of these particles. Nonetheless, our lack of understanding of the sources is at the same time a major limiting factor. Thus, the leading direction to explore in the near future consists in exploiting complementarities with other cosmic messengers and synergies with laboratory experiments and cosmological probes to push our understanding of the many unanswered questions related to neutrinos. 

This thesis can be thought of as a logbook exploring the different complementarities between laboratory experiments, astrophysical searches, and cosmological observations for the determinations of neutrino properties. It starts from the main fact that we know about neutrinos: at least two have non-zero masses. As a result, we expect neutrino flavour conversions --- a quantum phenomenon that takes place at macroscopical scales --- to occur. Combining data from several experiments studying neutrinos and antineutrinos from different sources and at various baselines and energies, one can learn about neutrino masses and mixing. At this point, other experimental probes such as beta decay measurements can constrain the neutrino mass scale. Other indirect or circumstantial probes like neutrinoless double beta decay searches and cosmological observables can also boost our understanding of the three-neutrino picture. The results from the global fit to experimental data presented in~\cite{deSalas:2020pgw} serve as a guideline in the quest for answers to questions. \textit{What is the mechanism responsible for neutrino masses?} \textit{Why are the mass scales of neutrinos and other fermions of the Standard Model several orders of magnitude apart?} \textit{Why is the mixing in the quark and leptonic sector so different?} \textit{Are neutrino Dirac or Majorana particles?} \textit{Is it possible to connect neutrinos to the origin of the matter-antimatter asymmetry of the universe?}

The next question we can ask ourselves is whether there is more new physics in the neutrino sector related to the origin of neutrino masses. The reason is that many extensions of the Standard Model accounting for neutrino masses predict additional neutrino properties. For instance, the fact that neutrinos are massive also implies that they have non-zero electromagnetic properties, such as magnetic dipole moments. Depending on the mechanism behind neutrino masses, one could expect sizeable interaction between neutrinos and electromagnetic fields and electromagnetic interactions between neutrinos and other charged particles. Non-standard interactions between neutrinos and electrons or quarks are also a usual prediction in many neutrino mass models. In some of these extensions, heavy neutral leptons are also a common ingredient. In that case, the three-neutrino mixing matrix is not necessarily unitary, providing the full lepton mixing matrix is. This non-zero admixture between active neutrinos and heavy neutral leptons has potentially detectable signatures too. These neutrino properties (electromagnetic properties, \mbox{non-standard} interactions and mixing with heavy neutral leptons) can also be studied and constrained in laboratory experiments and from cosmology. 

This thesis emphasises the synergies between different probes and experiments through particular examples. For example, for Majorana neutrinos, sizeable non-zero magnetic moments could give rise to a potentially detectable flux of solar antineutrinos \cite{Akhmedov:2022txm}. Likewise, a joint analysis of data from Hyper-Kamiokande and JUNO will allow the determination of the oscillation parameters of the solar sector with unprecedented accuracy while simultaneously constraining non-standard interactions between neutrinos and quarks \cite{Martinez-Mirave:2021cvh}. Besides that, the solar sector will also provide an accurate test of CPT symmetry by comparing the measurements using neutrinos and antineutrinos. From cosmology, non-standard interactions between neutrinos and electrons can be constrained through their impact in the process of neutrino decoupling too~\cite{deSalas:2021aeh}. The last example concerns the limits on a non-unitary three-neutrino mixing from the measurement of $N_{\text{eff}}$~\cite{Gariazzo:2022evs}.

On a different page, the second most relevant evidence that the Standard Model does not provide a complete description of the universe is the lack of a dark matter candidate. From the hope that a connection between dark matter and the Standard Model particle content exists, the community is scrutinising many scenarios with ongoing efforts both on the theoretical and experimental side. Unfortunately, this quest remains --- so far --- unsuccessful. Then, we have reasons to believe that dark matter might be very feebly interacting.

Motivated by the fact that neutrinos also share this weakly interactive nature, this thesis explores two distinct scenarios featuring both species. In particular, we focus on dark matter candidates sitting in the extremes of the allowed mass range: primordial black holes and ultralight scalars. In the first case, we study the sensitivity of future neutrino observatories DUNE and THEIA to the potentially observable flux of neutrinos from primordial black hole evaporation \cite{DeRomeri:2021xgy}. In the second case, we address the signatures in oscillation experiments \cite{Dev:2020kgz} and in the $\beta$-decay spectral measurements \cite{Huang:2022wmz} in the presence of a coupling between neutrinos and ultralight scalar dark matter.

I personally believe two main messages stem from this thesis. The first one is that exploiting complementarities between experimental probes is still a relevant approach to shed light on the many unknowns of neutrino physics. The second message is that forthcoming cosmological observation, next-generation terrestrial experiments, and the blooming of particle astrophysics offer numberless learning opportunities. Therefore, we shall be ready for the theoretical and phenomenological challenges accompanying the immense amount of data available --- and to come --- so that we learn the most out of it. 

\chapter*{Summary of the thesis}
\addcontentsline{toc}{chapter}{Summary of the thesis} 
\label{ch12-summary}
The field of neutrino physics has been evolving rapidly in the last decades due to significant advances in the technical and technological domains. As a consequence, we are reaching an era of precision measurements, both in scattering and oscillation experiments, which are boosting our understanding of neutrino interactions and properties. Furthermore, current and future cosmological observations are allowing the study of the role of neutrinos in the evolution of the universe.

This thesis discusses recent advances in the field. It is structured in three main parts devoted to neutrino masses and mixing, other neutrino properties beyond masses and mixing, and connections between neutrinos and dark matter respectively.

\vspace{1cm}
\textbf{FUNDAMENTAL COSMOLOGY AND THE PHYSICS OF MASSIVE NEUTRINOS}

The first part of this thesis consists of three chapters discussing fundamental aspects of cosmology, neutrino masses and our current knowledge regarding neutrino flavour oscillations. The first chapter consists of an introduction to the basics of the cosmological model $\Lambda$CDM. It discusses the expansion of the universe, its matter content, and some fundamental concepts of thermodynamics, and it also provides an outline of the thermal history of the universe, highlighting the contribution of neutrinos to it. Besides that, it includes an overview of the observational evidence of the existence of dark matter and a summary of some --- of the many --- dark matter candidates. It is of relevance for the chapters discussing the determination of neutrino properties from cosmological observables and the third part of this thesis  --- dedicated to connections between neutrinos and dark matter. 

The second chapter introduces some topics of neutrino physics that are of relevance for the research included in this thesis, in particular, in connection with the topic of neutrino masses. Firstly, we review some of the minimal neutrino mass mechanisms. Next, we discuss neutrino oscillations in vacuum and matter --- which provided the evidence for neutrino non-zero masses. Later, we describe other direct and indirect probes of neutrino masses, namely beta decays, neutrinoless \mbox{double-beta} decays and several cosmological observables. We conclude the introduction by delineating some other phenomenological consequences of neutrino masses --- as well as some of the predicted signatures from different neutrino mass mechanisms --- that are studied in this thesis or related to the research conducted during this period.

\refch{ch3-fit} is devoted to a combined analysis of neutrino data from oscillation experiments. Such joint analyses exploit the complementarities between datasets to maximise the sensitivity to the mass splittings\footnote{The mass splittings are defined as the mass squared differences $\Delta m^2_{ij} = m^2_i - m^2_j$.} and mixing parameters describing the phenomenon of flavour conversion in the \mbox{three-neutrino} picture. This chapter is specially focused on the status in 2020 ~\cite{deSalas:2020pgw}, which we summarise in \reftab{ch12-sum-2020}. Several experimental collaborations have presented more recent results and revisited previous analyses. Although these are not included in the results of this thesis, we describe the expected implications in the determination of the oscillation parameters and the overall picture --- which are not expected to be of importance. 

\begin{table}
\renewcommand*{\arraystretch}{1.2}
\centering
\begin{tabular}{lccc}
\toprule[0.25ex]
parameter & best fit $\pm$ $1\sigma$ & \hphantom{x} 2$\sigma$ range \hphantom{x} & \hphantom{x} 3$\sigma$ range \hphantom{x}
\\ \midrule 
$\Delta m^2_{21} [10^{-5}$eV$^2$]  &  $7.50^{+0.22}_{-0.20}$  &  7.12--7.93  &  6.94--8.14  \\[2.4mm]
$|\Delta m^2_{31}| [10^{-3}$eV$^2$] (NO)  &  $2.55^{+0.02}_{-0.03}$  &  2.49--2.60  &  2.47--2.63  \\
$|\Delta m^2_{31}| [10^{-3}$eV$^2$] (IO)  &  $2.45^{+0.02}_{-0.03}$  &  2.39--2.50  &  2.37--2.53  \\[2.4mm]
$\sin^2\theta_{12} / 10^{-1}$         &  $3.18\pm0.16$  &  2.86--3.52  &  2.71--3.69  \\[2.4mm]

$\sin^2\theta_{23} / 10^{-1}$       (NO)  &  $5.74\pm0.14$  &  5.41--5.99  &  4.34--6.10  \\
$\sin^2\theta_{23} / 10^{-1}$       (IO)  &  $5.78^{+0.10}_{-0.17}$  &  5.41--5.98  &  4.33--6.08  \\[2.4mm]

$\sin^2\theta_{13} / 10^{-2}$       (NO)  &  $2.200^{+0.069}_{-0.062}$  &  2.069--2.337  &  2.000--2.405  \\
$\sin^2\theta_{13} / 10^{-2}$       (IO)  &  $2.225^{+0.064}_{-0.070}$  &  2.086--2.356  &  2.018--2.424  \\[2.4mm]

$\delta/\pi$                        (NO)  &  $1.08^{+0.13}_{-0.12}$  &  0.84--1.42  &  0.71--1.99  \\
$\delta/\pi$                        (IO)  &  $1.58^{+0.15}_{-0.16}$  &  1.26--1.85  &  1.11--1.96  \\
\bottomrule[0.25ex]
\end{tabular}
\caption{\label{tab:ch12-sum-2020}
Best-fit values and 1$\sigma$, 2$\sigma$, and 3$\sigma$ ranges for normal and inverted ordering from the global fit presented in \refch{ch3-fit}.}

\end{table}

As shown in \reftab{ch12-sum-2020}, at present, four of the oscillation parameters are determined with great accuracy: $\theta_{12}$, $\theta_{13}$, $\Delta m^2_{21}$ and $|\Delta m^2_{31}|$. However, three important questions remain unanswered.
\begin{itemize}
\item The first one is related to the value of $\theta_{23}$ --- in particular, whether it is larger, smaller or equal to $\pi/4$. This fact is of relevance for flavour models and neutrino mass mechanisms trying to relate the mass scales and mixing patterns among Standard Model fermions.
\item The second question is the value of the Dirac CP-phase, $\delta_{\text{CP}}$. Depending on its value, CP symmetry would be conserved --- or not --- in the neutrino sector.
\item The third aspect is the mass hierarchy --- whether the lightest neutrino mass eigenstate is the one with a larger fraction of $\nu_e$ flavour eigenstate.
\end{itemize}
The answers to these questions are targeted by present and next-generation experiments.

Additional information on the mass ordering and the absolute neutrino mass scale --- which is not accessible in oscillation experiments --- can be obtained from beta-decay spectral measurements, neutrinoless \mbox{double-beta} decay --- if neutrinos were Majorana fermions --- and cosmological observables. \refch{ch3-fit} also discusses the role of these probes in our current understanding of the neutrino sector, based on available data by 2020. From these, together with the information from neutrino oscillations, we know that currently there is a preference for normal mass ordering, it is yet far from being conclusive. In this chapter, we also comment on the absolute neutrino mass scale and the current upper limits on the mass of the lightest neutrino.

\vspace{1cm}
\textbf{NEUTRINO PROPERTIES BEYOND MASSES AND MIXING}

The observation of flavour oscillations evidences that neutrinos are massive particles and, therefore, that the Standard Model provides an incomplete description of nature. The proposals to extend the existing theory to account for neutrino masses often come together with a variety of predictions of additional neutrino properties. The second part of the thesis presents research published --- or in the process of peer review --- on such neutrino properties.

It is known that Dirac and Majorana fermions can have non-zero electromagnetic properties. Despite being neutral fermions, such properties are expected from quantum loop effects. However, the details of the predictions depend strongly on the underlying realisation of neutrino masses. In any case, since neutrinos are massive, they can interact electromagnetically with other charged particles or directly with external electromagnetic fields. Yet, whether the consequences will ever be observed or not strongly depends on the mass-generation mechanism.

If neutrinos had a non-zero magnetic dipole moment, they could interact with external magnetic fields and undergo a flip in chirality. For Dirac neutrinos, such interaction would turn active states into inert ones --- magnetic dipole interactions turn left-handed neutrinos and right-handed antineutrinos into right-handed neutrinos and left-handed antineutrinos, respectively. Conversely, for Majorana neutrinos, magnetic dipole interactions transform neutrinos into antineutrinos of a different flavour. The phenomenon of simultaneous flipping chirality and undergoing a flavour transformation is known as \textit{spin-flavour precession.} The topic of \refch{ch4-magnetic} is the study of spin-flavour precession of solar neutrinos, under the assumption that neutrinos are Majorana particles. In that case, from the interaction of solar neutrinos with the magnetic field in the Sun, one would expect a flux of solar antineutrinos. Although small, such flux is potentially detectable in searches for electron antineutrino fluxes of astrophysical origin at experiments like KamLAND, Super-Kamiokande, Borexino and SNO.

In this fourth chapter, we derive an analytical expression for the expected electron antineutrino appearance probability of solar neutrinos from spin-flavour precession ~\cite{Akhmedov:2022txm}. The calculation is performed in the three-neutrino picture and includes recent information on Standard Solar Models. Additionally, it considers the role of a possible magnetic field twist. This analytical expression is of interest for experimental collaborations since it can be easily implemented, it is considerably faster than full numerical calculations, and it is reliable. In this chapter, we check its validity against exact calculations and discuss the role of several assumptions like the magnetic field profile of the Sun. Additionally, we show that if all magnetic moments were of the same order of magnitude, the expected electron antineutrino flux would be proportional to $|\mu_{12} B_\perp (r_0)|^2$ --- being $\mu_{12}$ one of the transition magnetic moments in the mass basis and $B_\perp (r_0)$ the magnetic field in the neutrino production region. This means that, if this phenomenon were ever observed, it will provide entangled information on neutrino magnetic moments and the magnetic field inside the Sun.

Solar neutrinos and, in general, the solar sector --- which also includes medium- and long-baseline reactor experiments sensitive to $\Delta m^2_{21}$ and $\theta_{12}$ --- can also prove the existence of other neutrino properties beyond masses, flavour mixing and electromagnetic properties. For instance, in \refch{ch5-nsisolar}, the projected sensitivity of these experiments to non-standard neutrino interactions (NSIs) is discussed. The rise of new interactions between neutrinos and other Standard Model fermions is commonly predicted in extensions of the Standard Model accounting for neutrino masses. This chapter describes how the complementarities between solar neutrino observatories and \mbox{medium-baseline} reactor experiments can be used to maximise the sensitivity to \mbox{non-standard} interactions with quarks while ensuring a precise determination of the oscillation parameters $\Delta m^2_{21}$ and $\theta_{12}$~\cite{Martinez-Mirave:2021cvh}. 

We consider the medium-baseline reactor experiment JUNO, which will determine $\Delta m^2_{21}$, $\Delta m^2_{31}$, $\sin^2\theta_{12}$ and $\sin^2\theta_{13}$ with unprecedented accuracy. Nevertheless, this experiment is not very sensitive to any effect altering neutrino propagation. We also address the sensitivity of the next-generation water-Cherenkov detector Hyper-Kamiokande, which will precisely measure the high-energy end of the solar neutrino spectrum. Neutrino propagation in the Sun is very sensitive to matter effects and hence, to possible non-standard interactions. However, its determination of the oscillation parameters $\sin^2\theta_{12}$ and $\Delta m^2_{21}$  will not be competitive with the one from JUNO. A combined analysis will allow determining the oscillation parameters of the solar sector with subpercent precision at 90\% C.L.,
\begin{align}
0.318 < &\sin^2\theta_{12} < 0.322 \, , \nonumber \\
7.48 \times 10 ^{-5} \text{eV}^2 < &\Delta m^2_{21} < 7.52 \times 10^{-5} \text{eV}^2 \, ,
\label{eq:ch12-nsiosc-results}
\end{align}
and to exclude NSIs larger than $\sim 15\%$ the strength of Standard-Model interactions, namely
\begin{align}
&-0.153 < \varepsilon_d' < 0.135\, , \nonumber \\
&-0.113 < \varepsilon_d < 0.144 \, ,
\label{eq:ch12-nsiosc-results2}
\end{align}
simultaneously. These results are specific for NSIs with $d$-quarks.  Analogous qualitative results would be found for non-standard interactions with \mbox{$u$-quarks.} An estimate of the limits for $u$-quarks can be obtained from the average ratio between the number density of both quark types in the Sun. 

We point out that a second solution -- the so-called LMA-D solution --- which corresponds to large non-standard interactions and a value of $\sin^2\theta_{21} > 0.5$ can not be ruled out exclusively by a combined analysis of these two experiments --- JUNO and Hyper-Kamiokande. This means that additional input would be needed, for instance from scattering data. Finally, we explore the impact of the experimental configuration of Hyper-Kamiokande on these results. Particularly, we allow for a reasonably lower energy threshold or the addition of a second tank --- and hence duplicating the target volume --- would barely affect the results of a joint analysis. 

The solar sector can also provide a fundamental test of CPT symmetry. Our current description of nature is based on local relativistic quantum field theories, which are CPT invariant. However, the breaking of the pillars of this framework --- mainly Lorentz invariance and locality --- is expected in some high-energy theories. Then, CPT violation resulting from non-locality could manifest as differences between the masses of particles and antiparticles. Along these lines, \refch{ch6-cpt} discusses how the future neutrino solar sector will improve significantly the existing limits on CPT violation from a precise measurement of the mass splitting $\Delta m^2_{21}$ using neutrinos and antineutrinos~\cite{Barenboim:2023krl}. The limits from the solar mixing angle will also improve with respect to current data. In this chapter, we considered JUNO and Hyper-Kamiokande as in \refch{ch5-nsisolar}. Besides that, we also address the capabilities of the DUNE experiment to measure solar neutrinos using electron neutrino scattering on argon. Under conservative assumptions, a joint analysis of the three experiments would result in the following bounds on the differences in the solar parameters at 3$\sigma$:
\begin{align}
&|\Delta m^2_{21} - \Delta \overline{m}^2_{21}| < 2.3 \times 10^{-5}\, \text{eV}^2\, , \\
&|\sin^2 \theta_{12} - \sin^2 \bar{\theta}_{12}| <  0.019\, .
\end{align}

We comment on the importance of some experimental aspects of the low-energy physics program of the DUNE experiment and the relevance of reducing the neutron backgrounds and maximising the detection efficiency. Likewise, improvements in the energy resolution and a reduction of the systematics in \mbox{Hyper-Kamiokande} would also boost the sensitivity to CPT violation. Moreover, we show that a significant improvement along those lines could particularly improve the bound on $|\Delta m^2_{21} - \Delta \overline{m}^2_{21}|$. Actually, in an optimal scenario, the limits would be further improved and would read
\begin{align}
&|\Delta m^2_{21} - \Delta \overline{m}^2_{21}| < 0.8 \times 10^{-5}\, \text{eV}^2\, , \\
&|\sin^2 \theta_{12} - \sin^2 \bar{\theta}_{12}| <  0.014\, ,
\end{align}
at 3$\sigma$. Finally, we also remark that neutrinos, as elementary particles, are ideal to test fundamental theoretical aspects of our description of nature.

The following two chapters of the manuscript --- which are also the last two chapters devoted to neutrino properties beyond masses and mixing --- focus on the process of neutrino decoupling and the lessons that could be learned from it in the years to come. \refch{ch7-nsicosmo} addresses the impact of non-standard interactions between neutrinos and electrons in the process of neutrino decoupling and how one can set constraints on them from a precise determination of the effective number of neutrinos --- also known as~$N_{\text{eff}}$. New interactions between neutrinos and electrons would alter the processes keeping neutrinos in thermal equilibrium with the primaeval plasma. Consequently, the process of neutrino decoupling would be advanced or delayed and, hence, the cosmological radiation density of the universe would be altered. \mbox{Non-observation} of departures from the standard picture will allow constraining non-standard interactions ~\cite{deSalas:2021aeh}. 

We show that the effect of \textit{non-universal} and \textit{flavour-changing} NSIs is mainly due to the effective shift they induce in the Standard Model coefficients $g^2_L$, $g^2_R$ and $g_Lg_R$. NSIs cancel out or enhance the interactions between neutrinos and the electrons and positrons in the cosmic plasma, altering the moment of decoupling and the cosmological radiation density. Likewise, we show that studying several non-zero NSI parameters is currently feasible and we provide an example in which cosmological bounds might be competitive with those from terrestrial experiments.

The second scenario considered in the context of neutrino cosmology is the \mbox{non-unitarity} of the three-neutrino mixing matrix. Heavy neutral leptons are commonly invoked as additional degrees of freedom in neutrino mass models. The admixture of such heavy neutral leptons with the light neutrino states would be parametrised in terms of a unitary lepton mixing matrix. However, this does not imply that the $3\times 3$ submatrix accounting for the mixing between light states is unitary. In \refch{ch8-nucosmo}, we discuss how the existence of heavy neutral leptons and a non-unitary three-neutrino mixing matrix would affect neutrino cosmology, and in particular, neutrino decoupling ~\cite{Gariazzo:2022evs}. To do so, we show the importance of understanding and describing the process in the mass basis --- as compared with the usual calculation which employs the flavour basis. For the first time, using the determination of $N_{\text{eff}}$ from Planck 2018 data, we set limits on two non-unitary parameters, namely
\begin{align}
\centering
    & \alpha_{11} > 0.07 \, , \\
    & \alpha_{22}> 0.15 \, ,
\end{align}
at $95\%$~C.L.

Besides that, we show how the projected sensitivities of forthcoming cosmological observations --- which could reach a sensitivity $\sigma(N_{\text{eff}}) = 0.02$ --- would improve those bounds. Likewise, we discuss the interplay between different parameters and discuss the limits in terms of non-standard interactions, under the assumption that deviations from unitarity are small.

The limits on neutrino non-standard interactions with electrons and the \mbox{non-unitarity} of the three-neutrino mixing matrix are generally far from the existing constraints from laboratory experiments. Even in light of the sensitivity of forthcoming cosmological observations, terrestrial bounds are still much more stringent. Notwithstanding, they are a valuable test of our understanding of the role of neutrinos in cosmology and at the same time provide independent constraints on neutrino properties.

\vspace{1cm}
\textbf{NEUTRINO CONNECTIONS TO DARK MATTER}

Dark matter has been observed indirectly via its gravitational effects. However, within the Standard Model, no candidate could account for a form of matter with the required properties. This fact gives additional motivation to extend the Standard Model of particle physics. The third part of the thesis is devoted to two scenarios involving a connection between neutrinos and dark matter. Particularly, very heavy and extremely light dark matter candidates --- primordial black holes and ultralight scalars --- are scrutinised.

In \refch{ch9-nupbh}, we discuss the future constraints that the next-generation observatories THEIA and DUNE will set on the abundance of primordial black holes with masses $M_{\text{PBH}} \sim 10^{15}\, - \, 10^{16}\,$ g ~\cite{DeRomeri:2021xgy}. A flux of MeV neutrinos is expected from Hawking evaporation of a population of primordial black holes in that mass range. The observation --- or non-observation --- of a flux of the predicted characteristics can be employed to constrain the fraction of the dark matter that could be present in the form of primordial black holes.  We study the sensitivity of DUNE to the electron neutrino component of the flux while with THEIA one would have access to the electron antineutrino one. We analyse the dependence of our results on the underlying assumption regarding the mass distribution of the primordial black hole population. We also address the changes in the flux from a population of rotating black holes. In that case, the emission of neutrinos is enhanced at the high-energy end of the spectrum and it would be possible to detect primordial black holes with larger masses.

The existing bounds from neutrino emission from evaporation --- from \mbox{Super-Kamiokande} --- and the projected sensitivity of next-generation experiments --- JUNO, THEIA and DUNE --- are far from the current limits from gamma-ray observations and CMB. Notice, however, that the limits are based on a phenomenon --- Hawking radiation --- that has been theoretically proposed yet never observed. Hence, it is interesting to cross-check our limits to ensure that our understanding of the physics case is correct. Moreover, astrophysics-related uncertainties are present in these analyses and therefore it is interesting to validate the results with a different cosmic messenger.

\refch{ch10-uldm} is devoted to the signatures from neutrinophilic ultralight scalars. Such dark matter candidates have gained interest since they can reconcile discrepancies between cosmological simulations and observations at small scales. Produced via misalignment in the early universe, their large occupation number allows treating ultralight scalar as classical --- oscillating --- fields. A hypothetical coupling between neutrinos would result in an effective mass term that shows a time dependence, where the modulation is inversely proportional to the mass of the ultralight scalar. In this chapter, we show how such an effective coupling between active neutrinos and the dark matter candidate can be realised via a coupling to sterile neutrinos --- not necessarily related to the mass-generation mechanism. 

If the sterile neutrino was heavy, one would get an effective coupling between active neutrinos and the ultralight scalar field. We study the phenomenology expected in a neutrino oscillation experiment~\cite{Dev:2020kgz} --- using DUNE as an example --- and in $\beta$ decay experiments~\cite{Huang:2022wmz} --- like KATRIN. In oscillation experiments and for masses $m_\phi \sim 10^{-20} - 10^{-22}$ eV, the modulation period of the neutrino mass splittings would manifest as a time-varying signal. If the amplitude of the modulation were large enough, this effect could be studied using the Lomb-Scargle periodogram and a determination of $m_\phi$ would be possible. For larger masses, the modulation would be too fast to be observed but its integrated effect would distort the oscillation probabilities in a non-trivial way. Hence, searches for these distortions in the oscillated neutrino spectrum could also shed light on this scenario. These searches are limited to masses $m_\phi \sim 10^{-14}$ eV, which correspond to modulation periods as small as the neutrino time-of-flight. Regarding measurements of $\beta$-decay experiments, we show that for a fast modulation, the spectrum is also distorted. Given the current experimental sensitivity, the effect can not be disentangled from the determination of the effective mass \mbox{parameter $m_\beta$.}

In the context of the sterile neutrino - ultralight scalar portal, and for sterile neutrinos with masses in the eV range, we show that in this scenario, the existing limits from KATRIN in the mass and mixing of the light sterile neutrino could be significantly relaxed. 

As a final comment, one should be aware that this scenario in general faces strong constraints from cosmology. Further --- and more detailed ---research on the possibilities to evade these limits is needed.

\vspace{1cm}

Aside from the specific results and conclusions presented in each chapter, this thesis aims to convey two main general ideas. The first one is the importance of continuing to understand and exploit the synergies and complementarities between experimental probes. The second one is that, at present and in the near future, the available data from laboratory experiments, cosmological observation and astrophysical neutrinos will keep on questioning and, hence, improving our understanding of neutrino physics --- and its plausible connection to dark matter. Creativity, inventiveness and the ability to put together all the available pieces of information --- regardless of their origin --- have proven so far --- and will prove to be in the future --- a successful approach to circumventing the challenges arising from the elusive character of neutrinos. 

\chapter*{Resum de la tesi}
\addcontentsline{toc}{chapter}{Resum de la tesi} 
\label{ch13-resum}
\vspace{1cm}

Els neutrins són partícules elementals i peces fonamentals del Model Estàndard de la Física de partícules. En 1930 foren postulats per Wolfgang Pauli com a explicació per a l'espectre continu observat en les desintegracions beta. En tractar-se de partícules que interaccionen molt feblement, poden travessar enormes quantitats de matèria sense deixar traça. Per eixe motiu, a vegades són denominades partícules 'fantasma'. És precisament pel seu caràcter poc interactiu que no va ser fins a 1956 quan Frederick Reines i Clyde Cowan detectaren per primera vegada els antineutrins electrònics. Anys més tard, Leon Lederman, Mel Schwartz i Jack Steinberger descobriren un segon tipus de neutrí: el neutrí muònic. Aquest descobriment va permetre entendre que els neutrins s'integren en el Model Estàndard com a doblets i que es produeixen acompanyats d'un altre leptó carregat. Els tres investigadors reberen el Premi Nobel de Física de 1988 per aquesta gesta.

En els anys posteriors va tindre lloc la detecció de neutrins solars i neutrins de l'explosió supernova SN1987A. L'acadèmia sueca de la ciència va atorgar a Raymond Davis i a Masatoshi Koshiba el Premi Nobel de Física 2002 pels seus treballs pioners en l'estudi de neutrins còsmics. A més a més, en aquest període també s'observaren per primera vegada els neutrins atmosfèrics, produïts quan els rajos còsmics col·lideixen amb altres partícules en les capes més altes de l'atmosfera. Finalment, l'any 2000 la col·laboració experimental DONUT va detectar la tercera generació de neutrins, el neutrí del taó. Aquesta troballa suposà una victòria per al Model Estàndard, que prediu que les partícules que constitueixen la matèria s'estructuren en tres rèpliques amb propietats similars, que reben el nom de famílies o generacions.

Així i tot, una sorpresa esperava a la comunitat científica. A principis d'aquest segle es va mesurar per primera vegada un fenomen quàntic anomenat oscil·lacions de sabor, pel qual Takaaki Kajita i Arthur B. McDonald van ser guardonats amb el Premi Nobel de Física 2015. Aquest fenomen únicament és possible si almenys dos dels neutrins tenen massa. Tanmateix, el Model Estàndard prediu que els neutrins no tenen massa. Per tant, malgrat que el Model Estàndard proporciona una descripció extremadament precisa de les partícules fonamentals i les seues interaccions, gràcies als neutrins sabem que és una teoria incompleta.

\vspace{1cm}
\textbf{OBJECTIUS}

Arran d'aquests esdeveniments, aquesta tesi se centra en l'estudi de les conseqüències fenomenològiques de les teories i models que intenten explicar l'origen de la massa dels neutrins i en la recerca de física més enllà del Model Estàndard. Els ràpids desenvolupaments tecnològics dels darrers anys han permés a la física de neutrins aconseguir una certa maduresa de forma relativament ràpida. En l'actualitat, ens trobem en una època on és possible fer mesures de precisió en experiments d'oscil·lacions i de dispersió. Així mateix, l'estudi de neutrins d'origen astrofísic i la cosmologia també ens han permés arribar a una millor comprensió del paper dels neutrins en l'evolució de l'univers i d'alguns dels seus fenòmens més fascinants.

Aquesta tesi inclou resultats de la recerca realitzada durant el període de doctorat en relació amb l'estudi dels neutrins, gastant dades actuals i anticipant la precisió que es podrà obtenir en les pròximes dècades. El contingut s'ha estructurat en tres parts. La primera d'elles correspon a les implicacions fenomenològiques de les masses dels neutrins i la mescla de sabors, així com a la mesura amb precisió dels paràmetres que les descriuen en aquest escenari. La segona se centra en l'estudi d'altres propietats dels neutrins que són predites de forma general per extensions del Model Estàndard on s'explica l'origen de la massa dels neutrins. Finalment, la tercera part està dedicada a la possible connexió entre els neutrins i una altra evidència de física més enllà del Model Estàndard: la matèria fosca. Mitjançant els seus efectes gravitacionals, sabem que més d'un 20\% del contingut de l'univers és una forma de matèria desconeguda i, per tant, no descrita pel Model Estàndard. Els capítols de la tercera part de la tesi presenten dos escenaris en els quals els neutrins podrien ajudar-nos a entendre la natura d'aquesta forma de matèria.

\vspace{1cm}
\textbf{METODOLOGIA}

El treball presentat en aquesta tesi combina una gran varietat d'aspectes tècnics de natura teòrica. En primer lloc, la descripció dels neutrins, de la seua fenomenologia i del seu rol en l'evolució de l'univers, necessita la mecànica quàntica, aspectes de teoria quàntica de camps, teoria de grups i relativitat general.

Les prediccions fenomenològiques que deriven de les diferents propietats dels neutrins estudiades es realitzen analíticament o mitjançant càlculs numèrics. En general aquests requereixen resoldre equacions diferencials o sistemes d'equacions diferencials acoblades. Com s'il·lustra en els diversos capítols, en la majoria de les situacions el càlcul exacte ha de dur-se a terme numèricament. No obstant això, es pot aconseguir una millor comprensió dels resultats amb aproximacions i estudis analítics. Els càlculs numèrics que s'inclouen en aquesta tesi s'han portat a terme amb ferramentes d'accés públic o bé amb codis desenvolupats específicament per a aquests treballs. Entre elles es troben \texttt{GLoBES} (General Long Baseline Experiment Simulator) \cite{Huber:2007ji,Huber:2004ka}, \texttt{BlackHawk} \cite{Arbey:2019mbc,Arbey:2021mbl} o \texttt{FortEPiaNO} (FORTran Evolved PrimordIAl Neutrino Oscillations)~\cite{Gariazzo:2019gyi}, entre altres. En ocasions, s'han desenvolupat de manera individual peces de \textit{software} específic per a la realització de les tasques requerides.  Pel que fa a la comparació de les prediccions teòriques amb els resultats reportats per les col·laboracions experimentals, les anàlisis estadístiques són a vegades freqüentistes i en altres casos, bayesianes. De fet, l'anàlisi estadística és de gran rellevància en aquesta tesi ja que permet quantificar la significància dels resultats obtinguts en els diferents estudis.

En conclusió, en tractar-se d'una tesi en fenomenologia de física de partícules que cobreix una temàtica d'una amplada significativa, la metodologia emprada és molt variada i s'adapta a les necessitats i particularitats dels diferents estudis que la componen. Els detalls tècnics són descrits en cadascun dels capítols, particularitzant per a la problemàtica abordada en cada cas.

\vspace{1cm}
\textbf{FONAMENTS DE COSMOLOGIA I LA FÍSICA DELS NEUTRINS AMB MASSA}

El primer capítol de la tesi consisteix en una introducció a les bases del Model Estàndard de la cosmologia, també conegut com a model $\Lambda$CDM. En ell, s'explica com la radiació, la matèria, els neutrins i l'energia fosca contribueixen a l'expansió de l'univers. A continuació, s'introdueixen conceptes fonamentals de la termodinàmica i la mecànica estadística necessaris per a entendre alguns dels càlculs presentats en aquesta tesi. Seguidament, es presenta un resum de la història de l'univers, focalitzant la discussió en les etapes i processos on els neutrins jugaren un paper rellevant. El final del capítol està dedicat a la matèria fosca, en particular a les evidències observacionals de la seua existència. Aquestes provenen tant del camp de l'astrofísica com de la cosmologia. Aquesta darrera discussió també inclou una recapitulació d'algunes propostes teòriques sobre la natura de la matèria fosca en relació amb la física de partícules.

A continuació, la tesi inclou un segon capítol introductori que versa sobre la massa dels neutrins i la fenomenologia associada. En primer lloc, es presenten de forma breu els principals mecanismes postulats per a explicar les masses dels neutrins. En segon lloc, es descriu el fenomen de les oscil·lacions de sabor en el buit i en matèria, establint el formalisme i la notació que es farà servir al llarg de la tesi. Posteriorment, s'introdueix la importància d'altres mesures com la determinació de l'espectre de les desintegracions beta, la cerca de senyals de desintegracions beta dobles sense neutrins i les observacions cosmològiques, per tal de determinar l'escala absoluta de la massa dels neutrins i la seua ordenació. El capítol conclou delineant les principals prediccions obtingudes en els diferents models de massa i que podrien detectar-se en experiments actuals i futurs, així com a través de mesures indirectes d'observables cosmològics.

El capítol \refchval{ch3-fit} està dedicat a les oscil·lacions de sabor. Aquest fenomen es pot descriure en funció de dues diferències de masses al quadrat, que denotem $\Delta m^2_{ij} = m^2_i - m^2_j$, tres angles de mescla, $\theta_{ij}$ i una fase relacionada amb la simetria sota conjugació de càrrega i paritat (CP), $\delta_{\text{CP}}$. En ell, es presenta l'anàlisi elaborada combinant les dades dels experiments d'oscil·lacions de neutrins públicament disponibles en 2020 \cite{deSalas:2020pgw}. L'objecte d'aquest treball és explotar els aspectes complementaris dels diferents experiments amb el fi de maximitzar la precisió amb la qual es determinen els paràmetres d'oscil·lacions. Posteriorment a la publicació d'aquesta anàlisi, algunes de les col·laboracions experimentals han mostrat noves mesures a partir de més dades o millorant aspectes tècnics. Malgrat que la determinació dels paràmetres d'oscil·lacions presentada en aquesta tesi no es troba completament actualitzada, el capítol conté discussions sobre les implicacions dels resultats experimentals no inclosos. Tanmateix, els canvis deguts a estes actualitzacions no són significatius.

Actualment, s'ha aconseguit una gran precisió en la mesura de quatre dels paràmetres d'oscil·lació, en particular de $\theta_{12}$, $\theta_{13}$, $\Delta m^2_{21}$ i $|\Delta m^2_{31}|$. Així i tot, encara queden tres preguntes de gran rellevància per respondre.
\begin{itemize}
\item La primera està relacionada amb el valor de l'angle $\theta_{23}$, que podria ser major, menor o igual a $\pi/4$. Aquest fet és de gran importància per a les teories de sabor i els models que intenten explicar els patrons i relacions entre els diferents fermions del Model Estàndard.
\item La segona qüestió és el valor de la fase $\delta_{\text{CP}}$. Depenent del seu valor, la simetria CP no es conservaria en el sector dels neutrins. Aquest fet proporcionaria una possible explicació a l'asimetria entre matèria i antimatèria de l'univers.
\item El tercer aspecte d'interés és l'ordenació de la massa dels neutrins. La seua determinació permetria excloure alguns models teòrics i a més, en ser determinada a través de mesures independents, serviria per a posar a prova la nostra comprensió de la fenomenologia dels neutrins.
\end{itemize}
La següent generació d'experiments de neutrins té com a objectiu abordar aquestes tres incògnites.

Addicionalment, l'estudi de les desintegracions beta i les observacions cosmològiques permetrien, en un futur, determinar la jerarquia dels neutrins i l'escala absoluta de massa, que no és accessible als experiments d'oscil·lacions. Si els neutrins foren fermions de Majorana, l'observació de desintegracions beta doble sense neutrins també proporcionaria informació sobre l'escala de massa i la jerarquia. En aquest tercer capítol es discuteix l'estatus experimental d'aquestes cerques l'any 2020 i com contribueixen a la nostra descripció dels neutrins. Per exemple, les dades mostren una preferència per l'ordenació normal de les masses, on el neutrí més lleuger és el que conté la fracció més gran de sabor electrònic. No obstant això, aquest resultat encara no és conclusiu i podria canviar en les anàlisis futures.

\vspace{1cm}
\textbf{ALTRES PROPIETATS DELS NEUTRINS MÉS ENLLÀ DE LES MASSES I LA MESCLA}

Per tal d'explicar les oscil·lacions de sabor, que evidencien la necessitat de nova física, és necessari estendre el Model Estàndard. Les propostes per fer-ho sovint inclouen noves interaccions i afegeixen noves partícules al contingut del Model Estàndard. Com a resultat, en molts d'aquests escenaris els neutrins adquireixen no sòls massa, sinó també altres propietats. La segona part de la tesi està dedicada a la recerca feta en el marc d'aquestes noves propietats. Els aspectes abordats inclouen com es manifestarien experimentalment i les seues implicacions en l'evolució de l'univers.

Pel fet de ser partícules amb massa i spin 1/2, els neutrins podrien ser fermions de Dirac o Majorana. En qualsevol dels dos casos, a conseqüència de l'efecte dels bucles quàntics, també adquireixen propietats electromagnètiques. Per tant, els neutrins amb massa poden interaccionar electromagnèticament amb altres partícules carregades i amb camps electromagnètics externs. Les indicacions de l'existència de neutrins amb propietats electromagnètiques no nul·les poden manifestar-se en una enorme varietat d'experiments. Com que els diferents models teòrics prediuen propietats electromagnètiques de diferent magnitud, l'observació o absència de senyals degudes a aquestes propietats serveix per a validar o refutar models teòrics.

El quart capítol de la tesi està dedicat a l'estudi del moment magnètic dipolar dels neutrins. Aquesta propietat permetria als neutrins interaccionar amb un camp magnètic extern i experimentar un canvi en la seua quiralitat. Per als neutrins de Dirac, això suposaria la transformació d'estats actius, que interaccionen amb la matèria, en estats inerts. És a dir, el moment magnètic dipolar transforma neutrins levògirs en neutrins dextrògirs i antineutrins dextrògirs en antineutrins levògirs. Al contrari, per a neutrins de Majorana, aquest canvi resulta en la transformació de neutrins en antineutrins i viceversa. En aquest context, el canvi simultani de quiralitat i sabor es denomina \textit{precessió espín-sabor}. Aquest fenomen s'estudia en profunditat en el capítol \refchval{ch4-magnetic} per als neutrins solars i assumint que els neutrins són fermions de Majorana. D'eixa forma, la interacció dels neutrins solars amb el camp magnètic del Sol donaria lloc a un flux d'antineutrins solars. Aquest flux, encara que fora menut, podria detectar-se en experiments que estudien antineutrins electrònics d'origen astrofísic, com KamLAND, Super-Kamiokande, Borexino i SNO.

Fent servir tècniques d'anàlisi matemàtica, en aquest capítol s'estudia de manera analítica el procés de precessió espín-sabor amb el fi d'obtenir una expressió per a la probabilitat d'aparició d'antineutrins electrònics del Sol \cite{Akhmedov:2022txm}. El càlcul es realitza considerant que existeixen tres famílies de neutrins i inclou informació actualitzada sobre dos Models Estàndard Solars. A més, s'estudia com afecta la torsió del camp magnètic als resultats. Aquest tipus d'expressió analítica és molt rellevant per a les col·laboracions experimentals, ja que pot ser fàcilment integrada en les anàlisis i permet estalviar una considerable quantitat de temps en càlculs numèrics. Com a part de la discussió dels resultats, s'estudia l'impacte que tenen les distintes aproximacions i consideracions fetes comparant la predicció analítica amb un càlcul exacte. Entre els resultats de més impacte destaca el fet que, si tots els moments magnètics dipolars foren similars en magnitud, aleshores el flux d'antineutrins electrònics que s'espera del Sol dependria del quadrat del producte del moment magnètic $\mu_{12}$ i del camp magnètic en el punt on es produeixen els neutrins $B_\perp(r_0)$. Per tant, si en un futur s'aconseguira aquesta mesura, tindríem simultàniament informació sobre el moment magnètic dipolar dels neutrins i el camp magnètic a l'interior del Sol. A més a més, és confirmaria que els neutrins són fermions de Majorana.

Històricament, la contribució dels neutrins solars a l'estudi de les propietats dels neutrins ha sigut remarcable. De fet, resultats encara més rellevants han sigut obtinguts estudiant el sector solar, que inclou també els experiments amb neutrins de reactor sensibles a $\Delta m^2_{21}$ i $\theta_{12}$. En el capítol \refchval{ch5-nsisolar}, es presenten les prediccions per a la sensibilitat dels experiments futurs a l'existència d'interaccions no estàndard entre neutrins i altres partícules. Aquestes noves interaccions, generalment referides com a NSIs per les sigles en anglés de \textit{non-standard interaccions}, apareixen sovint en models que expliquen la massa dels neutrins. Aquest capítol descriu com es complementen els experiments que estudien neutrins del Sol i alguns dels experiments que estudien els antineutrins produïts en reactors. La combinació d'aquests tipus d'experiments permet determinar amb gran precisió els paràmetres d'oscil·lació $\Delta m^2_{21}$ and $\theta_{12}$ i fitar l'existència d'interaccions no predites al Model Estàndard entre neutrins i quarks \cite{Martinez-Mirave:2021cvh}.

Per a l'anàlisi, considerem el futur experiment de reactor JUNO, situat a una distància mitjana de 50 kilòmetres de les plantes nuclears on es produeixen els antineutrins electrònics. Entre els seus objectius científics destaca la mesura simultània de quatre paràmetres d'oscil·lació, $\Delta m^2_{21}$, $\Delta m^2_{31}$, $\sin^2\theta_{12}$ i $\sin^2\theta_{13}$. Tanmateix, per les seues característiques, aquest experiment no serà quasi sensible a possibles senyals de nova física degudes a canvis en la propagació dels neutrinos. D'altra banda, pel que fa als neutrins solars, el nostre càlcul pren com a exemple l'experiment de pròxima generació Hyper-Kamiokande. Aquest consistirà en un detector de radiació Cherenkov en aigua que mesurara els neutrins solars de major energia. Per a aquestes energies, les interaccions dels neutrins quan es propaguen en el Sol són fonamentals i, per tant, s'espera que aquest experiment siga molt sensible a l'existència d'interaccions no estàndard entre els neutrins i els components de la matèria. En canvi, no s'espera la seua mesura dels paràmetres d'oscil·lacions siga competitiva amb la precisió esperada en JUNO.

En aquest cinqué capítol, mostrem com una anàlisi conjunta dels dos experiments permetria aconseguir un error menor de l'1\% en la determinació dels paràmetres d'oscil·lacions amb un nivell de confiança del 90\%, és a dir,
\begin{align}
0.318 < &\sin^2\theta_{12} < 0.322\, , \nonumber \\
7.48 \times 10 ^{-5} \text{eV}^2 < &\Delta m^2_{21} < 7.52 \times 10^{-5} \text{eV}^2 \, .
\label{eq:ch13-nsiosc-results}
\end{align}
Al mateix temps, aquesta anàlisi podria excloure l'existència d'interaccions no estàndard entre neutrins i quarks de tipus $d$ amb una intensitat major al 15\% de la magnitud de les interaccions predites al Model Estàndard. En particular, amb la notació usualment empleada en aquesta tesi, els límits serien
\begin{align}
&-0.153 < \varepsilon_d' < 0.135\, , \nonumber \\
&-0.113 < \varepsilon_d < 0.144 \, .
\label{eq:ch13-nsiosc-results2}
\end{align}

A més a més, en aquest capítol es mostra que la solució corresponent a interaccions no estàndard de gran magnitud i amb $\sin^2\theta_{21} > 0.5$, formalment coneguda com a solució LMA-D, no podrà ser exclosa únicament amb les dades de JUNO i Hyper-Kamiokande. Per tant, es requerirà informació addicional provinent, per exemple, d'experiments de dispersió elàstica coherent neutrí-nucli.

En últim lloc, el capítol inclou un estudi sobre l'impacte d'alguns detalls experimentals del detector Hyper-Kamiokande, com l'energia mínima detectable o la possibilitat de construir un segon detector. Les prediccions mostren que cap d'aquests aspectes canviarien significativament les conclusions de l'estudi. Tots els resultats corresponen a noves interaccions de caràcter vectorial entre neutrins i quarks de tipus $d$. No obstant això, per a interaccions no estàndard amb quarks de tipus $u$ s'esperarien resultats qualitativament anàlegs.

La física que es pot estudiar amb el sector solar dels neutrins també compren els tests de la simetria CPT. Actualment, la nostra descripció de la natura se sustenta en l'ús de teories de camps quàntics locals i relativistes. Aquestes són invariants sota transformacions simultànies de paritat, conjugació de càrrega i inversió temporal, és a dir, són invariants CPT. Malgrat això, d'acord amb algunes teories de física d'altes energies, s'esperaria que alguns dels pilars fonamentals d'aquesta descripció no foren inqüestionables. En eixe cas, i en particular si les propietats de localitat i invariància Lorentz no es mantingueren, la simetria CPT no seria necessàriament respectada en la natura. Aquest escenari es podria manifestar, per exemple, com a una diferència entre les masses de les partícules i les corresponents antipartícules.

Partint d'aquest concepte, el capítol \refchval{ch6-cpt} argumenta que els futurs experiments del sector solar podrien millorar significativament les fites que existeixen per a aquesta hipòtesi, principalment a conseqüència de la precisió esperada en la mesura de $\Delta m^2_{21}$ en experiments independents de neutrins i antineutrins. Addicionalment, els límits obtinguts de la mesura de l'angle de mescla solar $\theta_{12}$ també esdevindrien més restrictius~\cite{Barenboim:2023krl}.

L'anàlisi presentada en aquest capítol considera els experiments JUNO i \mbox{Hyper-Kamiokande} com en el cas anterior. A més, s'afegeix el detector DUNE, que podria mesurar la interacció dels neutrins solars en argó líquid. S'estima que, com a resultat de fer una combinació realista tenint en compte les capacitats dels tres experiments, es podrien fitar amb gran precisió les diferències entre els paràmetres dels neutrins i antineutrins,
\begin{align}
&|\Delta m^2_{21} - \Delta \overline{m}^2_{21}| < 2.3 \times 10^{-5}\, \text{eV}^2 \, \\
&|\sin^2 \theta_{12} - \sin^2 \bar{\theta}_{12}| < 0.019\, ,
\end{align}
amb una significança de 3$\sigma$.

El paper de l'experiment DUNE s'analitza amb detall en aquest capítol, estudiant com una reducció del fons de neutrons o la millora de l'eficiència de detecció tindrien un impacte positiu en el programa de física de baixes energies de l'experiment. Una millor resolució d'energia i una reducció dels errors sistemàtics en Hyper-Kamiokande també afectaria positivament als límits de violació de CPT.  En el context dels tests de la simetria CPT, el limit en $|\Delta m^2_{21} - \Delta \overline{m}^2_{21}|$ podria tornar-se encara més restrictiu. En el cas més favorable, les fites a 3$\sigma$ serien
\begin{align}
&|\Delta m^2_{21} - \Delta \overline{m}^2_{21}| < 0.8 \times 10^{-5}\, \text{eV}^2\, \\
&|\sin^2 \theta_{12} - \sin^2 \bar{\theta}_{12}| < 0.014\, .
\end{align}
Per acabar, cal remarcar que els neutrins són partícules elementals i, per tant, constitueixen el sistema ideal per a provar la validesa d'aspectes teòrics fonamentals de la nostra descripció de la natura.

Els darrers dos capítols d'aquesta part de la tesi presenten estudis de propietats de neutrins a partir d'observacions cosmològiques, centrant-se en el procés de desacoblament dels neutrins de la resta del plasma còsmic. La primera situació analitzada és l'efecte d'interaccions no estàndard, en aquest cas entre neutrins i electrons i positrons, en el desacoblament dels neutrins. El paràmetre $N_{\text{eff}}$, sovint denominat \textit{nombre efectiu de neutrins}, caracteritza la contribució dels neutrins a la densitat d'energia de l'univers en forma de radiació. En cas que existiren noves interaccions entre els neutrins i els electrons i positrons, canviaria l'instant de la història de l'univers en què les dues espècies deixaren d'estar en equilibri tèrmic. En conseqüència, la densitat d'energia en forma de radiació canviaria i això alteraria la mesura de $N_{\text{eff}}$. La no observació de desviacions en el valor d'aquest paràmetre respecte de la predicció teòrica estàndard, que no considera cap efecte de nova física, serveix per a fitar l'existència d'aquestes interaccions no estàndard \cite{deSalas:2021aeh}.

L'efecte dels diferents tipus d'interaccions no estàndard, les anomenades interaccions \textit{no universals} i de \textit{canvi de sabor}, es pot entendre pel canvi efectiu que produeixen en els coeficients del Model Estàndard dels quals depenen la magnitud de les interaccions. Les noves interaccions poden cancel·lar o potenciar la possibilitat d'interacció dels neutrins amb els electrons i positrons del plasma primordial. D'aquesta forma, el desacoblament es veu modificat i si aquest canvi és gran, seria observable per exemple al fons còsmic de microones. Aquest capítol \refchval{ch7-nsicosmo} permet extraure dues idees principals. D'una banda, és possible fer una anàlisi sistemàtica dels paràmetres de nova física en cosmologia, en concret en el cas d'interaccions no estàndard. D'altra banda, encara que els límits que s'esperen de les observacions de final d'aquesta dècada no són tan forts com els obtinguts al laboratori, aquests permetrien confirmar la validesa dels resultats i consolidar la comprensió física del procés estudiat.

El segon escenari que s'estudia en el context de la cosmologia de neutrins és la no unitarietat de la matriu de mescla per a tres neutrins. Aquest efecte resultaria de la presència de leptons neutres pesats, en acord amb les prediccions de diversos mecanismes per a dotar als neutrins de massa. La mescla entre els neutrins lleugers i els leptons neutres pesats en aquests escenaris és unitària. Tanmateix, la submatriu de dimensió 3 que descriu la mescla dels neutrins lleugers no ho seria necessàriament. El capítol \refchval{ch8-nucosmo} descriu l'impacte que tindrien els leptons neutres pesats i la no unitarietat en la mescla dels neutrins lleugers en l'evolució de l'univers \cite{Gariazzo:2022evs}. Igual que al capítol anterior, la discussió i els càlculs se centren al procés del desacoblament dels neutrins. Per tal de fer-ho, cal comprendre la importància d'abordar el problema en la base de masses, en lloc de la base de sabor empleada sovint en aquests càlculs. La raó és que el problema únicament està bé descrit en aquesta base. Posteriorment, i fent servir els resultats de $N_{\text{eff}}$ del satèl·lit Planck de 2018, s'obtenen límits en dos dels paràmetres emprats per a descriure aquest escenari,
\begin{align}
\centering
& \alpha_{11} > 0.07 \, , \\
& \alpha_{22}> 0.15 \, ,
\end{align}
amb un nivell de confiança del 95\%.

De manera similar, és possible calcular les fites esperades en vistes dels objectius de precisió fixats per a futures observacions. El capítol també explora les degeneracions entre els diferents paràmetres i com una descripció en termes d'interaccions no estàndard pot ajudar a visualitzar les implicacions fenomenològiques quan les desviacions respecte d'una matriu unitària són menudes.

És important emfatitzar que, malgrat que els límits derivables de l'estudi de la cosmologia dels neutrins no sempre són realment competitius amb les fites aconseguides en experiments de laboratori, el seu valor radica en el fet que es deriven de forma independent i serveixen com a prova del nostre coneixement de diversos aspectes de la cosmologia.

\vspace{1cm}
\textbf{CONNEXIONS ENTRE ELS NEUTRINS I LA MATÈRIA FOSCA}

La segona evidència de la necessitat de física més enllà del Model Estàndard és l'existència de matèria fosca. Aquesta ha sigut observada de manera indirecta a través dels seus efectes gravitacionals. Tanmateix, la detecció directa o indirecta mitjançant altres mètodes continua resistint-se. D'aquests fets hem aprés que la matèria fosca no interacciona o ho fa molt feblement amb la resta de partícules que conformen la matèria ordinària. En un principi, els neutrins foren candidats a matèria fosca. Malgrat això, pel fet de tindre una massa molt menuda i comportar-se de manera relativista al llarg de la major part de la història es va descartar aquesta hipòtesi.

Actualment, existeix una varietat de propostes teòriques i fenomenològiques per a explicar la natura de la matèria fosca i aconseguir una prova experimental addicional, directa o indirecta, de la seua existència. En la tercera i última part d'aquesta tesi s'exploren dos escenaris que connecten la matèria fosca i els neutrins. La motivació darrere d'aquestes propostes és la feblesa amb què interactuen ambdues espècies. En particular, s'estudien dos candidats a matèria fosca amb masses que disten en desenes d'ordres de magnitud: els forats negres primordials i els escalars ultralleugers.

El capítol \refchval{ch9-nupbh} presenta les fites a l'abundància de forats negres primordials a partir de la no observació d'un flux de neutrins amb energies de l'ordre del megaelectronvolt als experiments DUNE i THEIA ~\cite{DeRomeri:2021xgy}. Per a forats negres primordials amb masses $M_{\text{PBH}} \sim 10^{15}\, - \, 10^{16}\,$ g, l'evaporació Hawking produiria neutrins i antineutrins que podrien ser detectats en futurs observatoris de neutrins. Aquest flux dependria de quina és la fracció de matèria fosca formada per forats negres primordials. Aquest estudi analitza la capacitat de l'experiment DUNE per a detectar neutrins electrònics d'aquestes característiques i com es complementaria amb la detecció d'antineutrins electrònics en el detector THEIA.

L'anàlisi discuteix els canvis en les prediccions en funció de la distribució en massa de la població de forats negres primordials considerada. També s'analitza el canvi en l'espectre que s'esperaria si els forats negres primordials estiguessen rotant. En aquest cas, s'espera una emissió de neutrins en la regió de més energia de l'espectre. Aquest fet, permetria la detecció de forats negres primordials amb masses més grans.

Els límits de l'experiment Super-Kamiokande sobre l'emissió de neutrins de l'evaporació de forats negres primordials, i les prediccions per a les fites dels futurs experiments de neutrins, JUNO, THEIA i DUNE, es troben lluny de la sensibilitat aconseguida a partir de l'observació de rajos gamma i el fons còsmic de microones. Aquestes fites estan basades en el fenomen de la radiació Hawking, que ha sigut predit teòricament, però mai ha sigut observat. Per tant, és important validar els límits amb mètodes alternatius i independents per a confirmar la nostra comprensió de la física és correcta. A més, aquests estudis presenten una sèrie d'incerteses relacionades amb conceptes d'astrofísica. Com a resultat, comprovar la validesa de les hipòtesis amb dos missatgers còsmics, fotons i neutrins, és de gran rellevància.

En darrer lloc, el capítol \refchval{ch10-uldm} està dedicat a l'estudi de connexions entre els neutrins i escalars ultralleugers. Aquests són un dels candidats a matèria fosca que més interés han despertat en els darrers anys en la comunitat científica, ja que permetrien resoldre discrepàncies entre simulacions cosmològiques i observacions a escales menudes. Aquests camps escalars es podrien haver produït en l'univers primigeni i es comportarien com a camps clàssics oscil·lants. Si existira un acoblament entre els neutrins i la matèria fosca ultralleugera, aleshores els neutrins adquiririen una massa que oscil·laria amb el temps. El període d'aquesta variació estaria relacionat amb la massa de l'escalar. Aquest capítol presenta un escenari en el qual l'acoblament efectiu entre els neutrins lleugers i l'escalar resulta d'un acoblament amb neutrins estèrils, és a dir, que són singlets del Model Estàndard. No obstant, aquests neutrins estèrils no necessàriament estarien relacionats amb el mecanisme que dona massa als neutrins lleugers.

Si aquests fermions estèrils foren pesats, els neutrins actius tindrien un acoblament efectiu directe amb els escalars lleugers. L'anàlisi que es presenta inclou la fenomenologia que s'esperaria d'aquest escenari en un experiment d'oscil·lacions~\cite{Dev:2020kgz}, com DUNE, i en la mesura de l'espectre de les desintegracions beta~\cite{Huang:2019tdh}, com a l'experiment KATRIN. En els experiments d'oscil·lacions, si la massa de l'escalar és $m_\phi \sim 10^{-20} - 10^{-22}$ eV, l'efecte es manifestaria com una variació del senyal detectat amb el temps. Per a modulacions d'amplada significativa, la massa d'aquest candidat a matèria fosca es podria determinar amb el mètode del periodograma de Lomb-Scargle. Per a masses majors, la modulació de la massa seria massa ràpida per a ser detectada directament, però deixaria una empremta en forma d'una distorsió en la probabilitat d'oscil·lació. No obstant això, aquestes cerques estarien limitades a masses $m_\phi \sim 10^{-14}$ eV. Pel que fa a les mesures de la desintegració beta, la forma espectral també es veuria distorsionada. D'acord amb la sensibilitat actual dels experiments, aquest efecte no es podria diferenciar perquè estaria degenerat amb el valor del paràmetre de massa efectiva $m_\beta$.

En el context del model introduït en aquest capítol, on existeix un acoblament entre un neutrí estèril i l'escalar ultralleuger, per a masses del neutrí estèril de l'ordre de l'electronvolt, els límits de l'experiment KATRIN a la presència de neutrins estèrils lleugers es tornarien significativament més febles. Tanmateix, cal tindre consciència del fet que, en aquest tipus de models, existeixen fites prou restrictives i que, per tant, es necessitaria un estudi més detallat.

\vspace{1cm}

\textbf{CONCLUSIONS}

En aquesta tesi queden recollits els principals resultats de la recerca feta en el context de la física dels neutrins i l'estudi de les seues propietats amb experiments de laboratori i a través del seu paper en el cosmos. Aquestes anàlisis combinen conceptes de diferents branques de la física amb mètodes de càlcul analític i numèric amb el fi de maximitzar l'aprenentatge sobre la natura d'aquestes partícules i de la matèria fosca.

A més dels resultats específics presentats en cadascun dels capítols, aquesta tesi té com a objectiu transmetre dues idees generals. La primera és la importància d'utilitzar els trets complementaris d'experiments i mesures distintes com a estratègia per a l'estudi dels neutrins. Aquesta metodologia ha provat ser molt exitosa en el passat i els estudis de sensibilitat ací presentats confirmen que ho continuarà sent. La segona idea que es projecta en aquesta tesi és la necessitat de connectar els estudis de física de neutrins realitzats en els laboratoris, a partir d'observacions cosmològiques i de fonts astrofísiques. En un camp ple d'incògnites, la multidisciplinarietat i l'enginy són necessaris per a superar els reptes i obstacles que es presenten en la física de neutrins i en la cerca de la matèria fosca.

\appendix % From here onwards, chapters are numbered with letters, as is the appendix convention

\pagelayout{wide} % No margins
\addpart{Appendix}
\pagelayout{margin} % Restore margins

\chapter{Other limits on neutrino magnetic moments}
\label{app1-magnetic}

This appendix provides an updated overview of the existing limits on neutrino magnetic moments from a variety of experimental searches. Special attention is paid to the relations between the quantities that are experimentally accessible and the neutrino magnetic moments --- which are truly fundamental quantities.  
These relations often depend on the Dirac or Majorana nature of neutrinos, the mixing parameters and the flavour of the incoming neutrinos. Moreover, sometimes they even depend on specific experimental details, which determine whether flavour conversions occurred between neutrino production and detection.

The four sections of the appendix are dedicated to limits from astrophysics, cosmology, scattering experiments, and colliders, respectively.

\section{Astrophysical limits}
\labsec{magn_limits}

\textbf{Plasmon decay and related processes in astrophysical environments}

Photons in a non-relativistic plasma --- often referred to as plasmons and denoted by $\gamma^*$ --- have a dispersion relation $\omega^2_\gamma - \textbf{k}^2 _\gamma = \omega^ 2_P$, where $(\omega_\gamma, \textbf{k})$ is the plasmon four-momentum and $\omega_P$ is the plasma frequency. Whenever $\omega_P > 2m_\nu$, plasmon decay to a neutrino-antineutrino pair,
\begin{align}\gamma^* \,  \longrightarrow\, \, \nu \, + \, \bar{\nu}\, ,
\end{align} 
becomes kinematically allowed \cite{Raffelt:1987yb,Raffelt:1996wa}. The plasmon decay rate is proportional to an effective magnetic moment defined as \cite{Giunti:2014ixa}
\begin{equation}
    \mu^2_{\nu, \text{PLASMON}} = \sum_{i,j} |\mu_{ij}|^2 \, ,
    \label{eqn:a1-plasmon_mu}
\end{equation}
where $\mu_{ij}$ are the neutrino magnetic moments in the mass basis.

These decays provide an additional mechanism of energy loss in stellar environments which would alter the luminosity of stars. From this fact, it is possible to constrain the effective neutrino magnetic moment, as shown in \reftab{a1-mu_astro}. In the case of red giants, plasmon decay would be an additional source of cooling and helium ignition would be delayed. Then, the tip of the red-giant branch would be brighter than predicted by stellar \mbox{models \cite{Raffelt:1992pi,Capozzi:2020cbu,Franz:2023gic}.} Likewise, there exist bounds from searches of changes in the frequency of pulsating white dwarfs --- in particular from those whose spectrum only has helium absorption lines \cite{Corsico:2014mpa}. 

In the presence of non-zero neutrino magnetic moments, plasmon decay is not the only source of additional cooling. Processes like 
\begin{align}
&\gamma\, + e^- \, \longrightarrow \, e^-\,  + \, \bar{\nu}\, +\, \nu\, , \nonumber \\
&e^+\, + \,  e^- \, \longrightarrow \, \bar{\nu}\, + \, \nu\, ,\nonumber\\
&e^- \, +  \, (Ze) \, \longrightarrow \, (Ze)\, + \, e^- \, + \, \bar{\nu}\, + \, \nu \, ,
\end{align} would induce significant changes in the evolution of stars with masses between 7$M_\odot$ and 18 $M_\odot$ \cite{Heger:2008er}. These considerations allow to set constraints on the same parameter combination as plasmon decay and hence, this bound is also included in  \reftab{a1-mu_astro}.

\begin{table*}
\renewcommand*{\arraystretch}{1.2}
\centering
\begin{tabular}{ccc}
\toprule[0.25ex]
Limit at 95\% C.L.& Reference & Method  \\ \midrule  $\mu_{\nu\text{ PLASMON}} < 1.2 \times 10^{-12} \mu_B $ & \cite{Capozzi:2020cbu} & Tip of red-giant branch \\ 
$\mu_{\nu\text{ PLASMON}} < 6.0 \times 10^{-12} \mu_B $ & \cite{Franz:2023gic} & Tip of red-giant branch \\ 
$\mu_{\nu\text{ PLASMON}} < 1.0 \times 10^{-11} \mu_B $ & \cite{Corsico:2014mpa}  & Pulsating white dwarfs \\
$\mu_{\nu\text{ PLASMON}} < 2.2 \times 10^{-12} \mu_B $ & \cite{Diaz:2019kim} & Luminosity \\ 
$\mu_{\nu\text{ PLASMON}} < 2.2 \times 10^{-12} \mu_B $  & \cite{Arceo-Diaz:2015pva} & Luminosity \\
\midrule
$\mu_{\nu, \text{ PLASMON}} < (2-4) \times 10^{-11} \mu_B$ & \cite{Heger:2008er} & Stellar evolution\\
\bottomrule[0.25ex]
\end{tabular}
\caption{\label{tab:a1-mu_astro}Limits on effective neutrino magnetic moments from plasmon decays and other related processes in stars.} 

\end{table*}

\textbf{Limits from SN1987A}

For Dirac neutrinos, non-zero magnetic moments would induce the conversion of part of the flux of left-handed neutrinos into --- practically inert --- right-handed ones. Electromagnetic scattering of neutrinos on electrons and protons,
\begin{align}
\nu_{L} \, + \, e^- \, \longrightarrow \, \nu_R\, + \,e^- \quad \text{and} \quad \nu_L \, + \, p \, \longrightarrow \, \nu_R \, + \, p \, ,
\end{align}  
would produce $\nu_R$ states which would escape the supernova environment. If a large number of neutrinos were converted into sterile states, the resulting reduction of the supernova luminosity would not be consistent with the observations from SN1987A. The limit based on this argument is \cite{Barbieri:1988nh}
\begin{equation}
    \mu_\nu \leq (0.1-1)\times10^{-12} \mu_B\,.
\end{equation}

Detailed analyses of plasmon-mediated neutrino scattering with electrons and protons in the plasma set limits on Dirac neutrino magnetic moments to be~\cite{Ayala:1999xn, Kuznetsov:2009we, Kuznetsov:2009zm}
\begin{equation}
    \mu_\nu < (1.1 -2.7)\times 10^{-12} \mu_B \, ,
\end{equation}
for different realistic models of the supernova core. The limit is set on a flavour- and time-averaged combination of magnetic moments for Dirac neutrinos. 

It is complicated to translate these limits directly into limits on fundamental magnetic moments. The reason is that, in the calculation, the contribution of each different neutrino flavour varies together with their energies. Besides, the argument relies on the --- very strong --- assumption that neutrino propagation in the supernova environment is well-understood. 

\textbf{Conversion of $\mathbf{\nu_e}$ from supernova neutronisation burst into $\mathbf{\bar{\nu}_e}$}

Similarly to the scenario explored in \refch{ch4-magnetic}, spin-flavour precession could also occur in the environment of a core-collapse supernova. In particular, a simultaneous chirality flip and flavour conversion can take place in the presence of strong magnetic fields. In the case of a supernova and due to the large number densities, it could also be resonantly enhanced~\cite{Akhmedov:1992ea,Akhmedov:2003fu,Ando:2003is,Jana:2022tsa}. 

The neutronisation burst is the first phase of the supernova explosion and its observation could be used to study neutrino magnetic moments. It is characterised by a prompt emission of electron neutrinos. Hence, the appearance of electron antineutrinos would be a clear signature of \mbox{spin-flavour} precession. The electron antineutrino appearance probability would depend on the product of the effective magnetic moment $\mu_{\nu, \text{SN}}$ and the strength of the magnetic field at the resonance of spin-flavour precession, $B_0$. The actual value of $\mu_\nu$ depends on the mass hierarchy and it can be most simply expressed in the \textit{primed basis} defined in Equation \ref{eq:ch4-primed-basis} --- which can also be related to the neutrino magnetic moments in the mass basis. Then, for normal ordering,
\begin{align}
\mu_{\nu, \text{SN}} = \mu'_{e\mu'}=\mu_{12}c_{13} e^{-i\lambda_2}+(\mu_{13}s_{12}-\mu_{23}c_{12}
e^{-i\lambda_2})s_{13} e^{i(\delta_\text{ CP} - \lambda_3)}\,,
\end{align}
whereas for inverted ordering
\begin{align}
\mu_{\nu, \text{SN}} = \mu'_{e\tau'}=(\mu_{13}c_{12}+\mu_{23}s_{12}e^{-i\lambda_2})e^{-i\lambda_3}\,.
\end{align}

Next-generation neutrino experiments such as Hyper-Kamiokande could search for a fraction of electron antineutrinos from a supernova neutronisation burst \cite{Jana:2022tsa}.
\newpage
\textbf{Other limits from astrophysical sources}

Many of the existing limits and projected sensitivities to neutrino magnetic moments from the study of astrophysical sources strongly depend on the underlying assumptions. For example, for Dirac neutrinos, the chirality flip resulting from magnetic dipole moment interactions can modify the flavour ratio of high-energy neutrino fluxes from distant sources~\cite{Kopp:2022cug}. Note, however, that the predictions strongly rely on the magnetic field profile considered in the astrophysical source and along the propagation. Another example for Majorana neutrinos is the resonant enhancement of spin-flavour precession that would occur in the presence of twisting magnetic fields~\cite{Jana:2023ufy}. Such phenomenon could be searched for but it is limited by our very restricted understanding of the magnetic field in the source.

Hence, whereas a positive signal could be observed for neutrino magnetic moments as small as $\mathcal{O}(10^{-17}\mu_B)$ if these scenarios were realised in nature, the absence of a discovery would not straightforwardly correspond to an equivalently strong limit on the magnitude of neutrino magnetic moments.

\section{Cosmological limits}

Non-zero neutrino magnetic moments could have also altered the evolution of the universe. For instance, additional neutrino electromagnetic interactions would have kept neutrinos in thermal contact with the cosmic plasma for a longer time, even until the epoch of electron-positron annihilation. The impact of such delay in neutrino decoupling on the production of deuterium in Big Bang Nucleosynthesis is addressed in \cite{Morgan:1981psa}.
For Majorana neutrinos, the constraints from the changes in neutrino decoupling temperatures and the abundances of primordial elements constrain the magnetic moments in the flavour basis to be of order $\mathcal{O}(10^{-10} \mu_B)$ \cite{Vassh:2015yza}.
There are similar studies considering Dirac neutrinos instead. In that case, the interactions with the cosmic plasma result in a population of right-handed neutrinos -- and \mbox{left-handed} antineutrinos. From their contribution to the radiation energy density and the modification of the decoupling picture, the limits on the magnetic moment are $\mu_\nu \sim 5 \times 10^{-12} \mu_B$~\cite{Carenza:2022ngg,Grohs:2023xwa}.

\section{Scattering limits}

Photon-mediated interactions resulting from non-zero neutrino magnetic moments would alter the cross-section of several scattering processes such as elastic neutrino-electron scattering or coherent elastic neutrino-nucleus scattering (CE$\nu$NS). Electromagnetic dipole interactions flip neutrino chirality whereas Standard Model weak interactions preserve it. Then, both contributions add up incoherently to the cross-section of the process under study. 

The effective neutrino magnetic moment in an scattering experiment \mbox{is~\cite{Grimus:2000tq,Grimus:2002vb}}
\begin{equation}
    \mu^2_{\nu_\alpha} = \nu^\dagger_{L}\, (\upmu^\dagger \upmu) \, \nu_{L} + 
\nu^\dagger_{R}\, (\upmu\upmu^ \dagger) \, \nu_{R} \, .
    \label{eqn:a1-effective_mu}
\end{equation}
Here, $\nu_{L}$ and $\nu_{R}$ denote the vectors of the amplitudes of the incoming left- and right-handed neutrinos, respectively, and $\upmu$ is the matrix of neutrino magnetic moments. This expression is basis-independent and it is valid for both Dirac and Majorana neutrinos --- as long as the neutrino in the final states is not detected in the scattering process \cite{Akhmedov:2022txm}.

%%%%%%%%%%%%%%%%%%%%%%%%%%%%%%%%%%%%%%%%%%%%%%%%%%%%%%%%%%
\textbf{Limits from short-baseline experiments}
%%%%%%%%%%%%%%%%%%%%%%%%%%%%%%%%%%%%%%%%%%%%%%%%%%%%%%%%%%

In short-baseline scattering experiments, the distance between the neutrino source and the detector is, in general, much shorter than the distance needed for flavour oscillations to develop. Then, the effective magnetic moment is
\begin{align}
    \mu^2_{\nu_\alpha{\rm SB}} = (\upmu^\dagger \, \upmu )_{\alpha \alpha} 
= \mu^2_{\bar{\nu}_\alpha{\rm SB}}\,.
\end{align}

In terms of the neutrino magnetic moments in the flavour basis, it reads
\begin{align}
    \mu^2_{\nu_\alpha{\rm SB}} = \sum _\beta |\mu_{\beta\alpha}|^2 \,, 
\label{eqn:a1-sb_flavour_mu}
\end{align}
for Dirac or Majorana neutrinos --- with the subtlety that for Majorana neutrinos $\mu_{\alpha\alpha} = 0$. 

\begin{table*}[t!]
\renewcommand*{\arraystretch}{1.2}
\centering
\begin{tabular}{cccc}
\toprule[0.25ex]
Experiment  & Limit at 90\% C.L. & Reference & Method   \\ \midrule
        LAMPF & $\mu_{\nu_e} < 1.08\times 10^{-9} \mu_B $ & 
        \cite{Krakauer:1990cd} & Accelerator $\nu_e \, + \,e^-$ \\
        LSND & $\mu_{\nu_e} < 1.1\times 10^{-9} \mu_B $  & 
        \cite{LSND:2001akn} & Accelerator $\nu_e \, + \,e^-$ \\ \midrule
       
        Krasnoyarsk & $\mu_{\nu_e} < 1.4\times 10^{-10} \mu_B$ & 
        \cite{Aleshin:2008zz} & Reactor $\bar{\nu}_e \, + \,e^-$\\
        ROVNO & $\mu_{\nu_e} < 1.9\times 10^{-10} \mu_B$ & 
        \cite{Derbin:1993wy} & Reactor $\bar{\nu}_e \, + \,e^-$\\
        MUNU & $\mu_{\nu_e} < 9\times 10^{-11} \mu_B$ & 
        \cite{MUNU:2005xnz} & Reactor $\bar{\nu}_e \, + \,e^-$\\
        TEXONO & $\mu_{\nu_e} < 7.4\times 10^{-11} \mu_B$ & 
        \cite{TEXONO:2006xds} & Reactor $\bar{\nu}_e \, + \,e^-$\\
        GEMMA & $\mu_{\nu_e} < 2.9\times 10^{-11} \mu_B$ & 
        \cite{Beda:2012zz} & Reactor $\bar{\nu}_e \, + \,e^-$\\ \midrule
        
CONUS & $\mu_{\nu_e} < 7.5 \times 10^{-11} \mu_B$ &  
        \cite{CONUS:2022qbb}& Reactor CE$\nu$NS \\
        Dresden-II & $\mu_{\nu_e} < 2.2 \times 10^{-10} \mu_B$ & \cite{Coloma:2022avw,AtzoriCorona:2022qrf} & Reactor CE$\nu$NS \\ \midrule       
        LAMPF & $\mu_{\nu_\mu} < 7.4\times 10^{-10} \mu_B$ & 
        \cite{Krakauer:1990cd} & Accelerator $\nu_\mu \, + \,e^-$ \\
        BNL-E-0734 & $\mu_{\nu_\mu} < 8.5\times 10^{-10} \mu_B$ & 
        \cite{Ahrens:1990fp} & Accelerator $\nu_\mu \, + \,e^-$ \\
        LSND & $\mu_{\nu_\mu} < 6.8 \times 10^{-10} \mu_B$& 
        \cite{LSND:2001akn} & Accelerator $\nu_\mu \, + \,e^-$ \\ \midrule
        DONUT & $\mu_{\nu_\tau} < 3.9 \times 10^{-7} \mu_B$ & 
        \cite{DONUT:2001zvi} & Accelerator $\nu_\tau \, + \,e^-$ \\
\bottomrule[0.25ex]
\end{tabular}
\caption{\label{tab:a1-sb_limits} Current experimental limits on the effective neutrino magnetic moments from short-baseline reactor and accelerator experiments, using elastic neutrino or antineutrino scattering on electrons or coherent elastic \mbox{neutrino-nucleus} scattering (CE$\nu$NS).}
\end{table*}

Alternatively, one can express the effective neutrino magnetic moment in terms of the fundamental neutrino magnetic moments in the mass basis. Particularly, for Dirac neutrinos it is given by
\begin{align}
\mu^ 2_{\nu_\alpha{\rm SB}} =& |U_{\alpha1}|^2\left(|\mu_{11}|^2 + 
|\mu_{21}|^2+ |\mu_{31}|^2\right) + |U_{\alpha2}|^2\left(|\mu_{12}|^2 + 
|\mu_{22}|^2+ |\mu_{32}|^2\right)\nonumber \\ &+|U_{\alpha3}|^2
\left(|\mu_{13}|^2 + |\mu_{23}|^2+ |\mu_{33}|^2\right)\nonumber \\ & +2\text{ Re}\lbrace 
U_{\alpha1}U^*_{\alpha2} \left(\mu^*_{11} \mu_{12} + \mu_{21}^* \mu_{22} + 
\mu^*_{31} \mu_{32}\right) \rbrace \nonumber \\ &+ 
2\text{ Re}\lbrace U_{\alpha1}U^*_{\alpha3} \left(\mu^*_{11} \mu_{13} + 
\mu^*_{21} \mu_{23} + \mu^*_{31} \mu_{33}\right)\nonumber \\ & + U_{\alpha2}U^*_{\alpha3} 
\left(\mu^*_{12} \mu_{13} + \mu^*_{22} \mu_{23} + \mu^*_{32} \mu_{33}\right) 
\rbrace \,,
\end{align}
whereas for Majorana neutrinos, it is
\begin{align}
    \mu^ 2_{\nu_\alpha{\rm SB}} =& |U_{\alpha1}|^2\left(|\mu_{12}|^2+ 
    |\mu_{13}|^2\right) + |U_{\alpha2}|^2\left(|\mu_{12}|^2 + 
    |\mu_{23}|^2\right) \nonumber \\ & +|U_{\alpha3}|^2\left(|\mu_{13}|^2 + 
    |\mu_{23}|^2\right)  +2\text{ Re}\lbrace 
    U^*_{\alpha1}U_{\alpha2} \mu_{13} \mu^*_{23}\rbrace \nonumber \\ & - 2\text{ Re}\lbrace 
    U^*_{\alpha1}U_{\alpha3} \mu_{12} \mu^*_{23} \rbrace + 2\text{ Re}
    \lbrace U^*_{\alpha2}U_{\alpha3}\mu_{12} \mu^*_{13}  \rbrace \, .
\end{align}

\reftab{a1-sb_limits} reports the existing limits in the three effective neutrino magnetic moments from short-baseline accelerator and reactor experiments.

%%%%%%%%%%%%%%%%%%%%%%%%%%%%%%%%%%%%%%%%%
\textbf{Limits from solar-neutrino scattering}
%%%%%%%%%%%%%%%%%%%%%%%%%%%%%%%%%%%%%%%%%

Solar neutrino scattering can constrain neutrino electromagnetic properties not only from the search for spin-flavour precession --- as in \refch{ch4-magnetic} --- but also from the study of the scattering of solar neutrinos on electrons. The main difference in this case with respect to the limits from short-baseline experiments is that, for solar neutrinos, flavour conversions are of great relevance.

The expression for the effective magnetic moment accessible in solar neutrino experiments depends only on the left-chirality amplitudes \mbox{$\nu_L$ = ($\nu_{eL} \; \nu_{\mu L} \; \nu_{\tau L})^T$,} which can be obtained in the standard three-flavour picture. We find that the effective magnetic moment probed is
\begin{align}
    \mu^2_{\nu{\rm SOLAR}}  =& \, \nu^\dagger _L (\upmu^\dagger \upmu)\nu_L 
     \nonumber \\ 
     = & \, (|\mu_{11}|^2 + |\mu_{21}|^2 +|\mu_{31}|^2) |\nu_{1L}|^2 + 
      (|\mu_{12}|^2 + |\mu_{22}|^2 +|\mu_{32}|^2) |\nu_{2L}|^2 \nonumber \\
    & + (|\mu_{13}|^2 + |\mu_{23}|^2 +|\mu_{33}|^2) |\nu_{3L}|^2 \nonumber \\ 
      & +2\rm{Re}\lbrace (\mu^*_{11}\mu_{12} + \mu^*_{21} \mu_{22} + 
      \mu^*_{31}\mu_{32}) (\nu_{1L}\nu^*_{2L})\rbrace \nonumber \\ & + 
      2 \rm{Re}\lbrace (\mu^*_{11}\mu_{13} + \mu^*_{21} \mu_{23} + 
      \mu^*_{13}\mu_{33}) (\nu_{1L}\nu^*_{3L})\rbrace \nonumber \\ 
   &+2 \rm{Re}\lbrace (\mu^*_{12}\mu_{13} + \mu^*_{22} \mu_{23} + 
     \mu^*_{32}\mu_{33}) (\nu_{2L}\nu^*_{3L})\rbrace \, .
\label{eq:a1-munuSol1}
\end{align}

Notice that the agreement with experimental data requires that, if existing, the right-chiral amplitudes have to be much smaller than the left-chiral ones. Since the coherence of different neutrino mass eigenstates is lost when travelling to Earth, the terms $\nu^*_{iL} \nu_{jL}$ average to zero for $i \neq j$~\cite{Dighe:1999id, Grimus:2002vb}. Under the approximation of adiabatic flavour conversion in the Sun, we find that 
\begin{align}
    |\nu_{1L}|^2 
 = c^2_{13} \cos^2\tilde{\theta}\,, 
\quad
        |\nu_{2L}|^2 = 
c^2_{13} \sin ^2 \Tilde{\theta} 
\quad \text{ and } 
    \quad     |\nu_{3L}|^2 
 = s^2_{13} \, ,
\end{align}
where the mixing angle $\tilde{\theta}(r)$ was defined in Equation~\ref{eq:ch4-tildetheta} and the averaging over the coordinate of the neutrino production point in the Sun is implied. Then, one can find that the effective neutrino magnetic moment measured from the scattering of Dirac solar neutrinos is
\begin{align}
    \mu^2_{\nu{\rm SOLAR}} = & \, (|\mu_{11}|^2 + |\mu_{21}|^2 +|\mu_{31}|^2)
    c^2_{13} \cos ^2 \tilde{\theta}  \nonumber \\  & + (|\mu_{12}|^2 + 
    |\mu_{22}|^2 +|\mu_{32}|^2)c^2_{13} \sin ^2 \tilde{\theta} \nonumber \\ 
    & + (|\mu_{13}|^2 + |\mu_{23}|^2 +|\mu_{33}|^2)s^2_{13} \, ,
\label{eq:a1-muDir}
\end{align}
while for Majorana neutrinos
\begin{align}
    \mu^2_{\nu{\rm SOLAR}} = &\,  |\mu_{12}|^2 c^2_{13} + |\mu_{13}|^2(c^2_{13} 
    \cos^2 \Tilde{\theta} + s^2_{13}) & \nonumber \\   & + 
    |\mu_{23}|^2(c^2_{13} \sin^2 \Tilde{\theta} + s^2_{13}) \, .
\label{eq:a1-muMaj}
\end{align}
For neutrino energies $E\lesssim 1$ MeV, matter effects barely affect neutrino evolution in the Sun and $\Tilde{\theta} \simeq \theta_{12}$, whereas for $E\gtrsim 5-7$ MeV, the effective mixing is $\tilde{\theta} \simeq \pi/2$. Then, in both limits, the effective neutrino magnetic moments can be directly found from Equation \ref{eq:a1-muDir} and Equation \ref{eq:a1-muMaj}~\cite{Miranda:2020kwy}.  Generally, to extract consistent limits on neutrino magnetic moments from the scattering of solar neutrinos, one has to account for the energy dependence of $\tilde{\theta}$ in detail. 
\reftab{a1-mu_solar} collects the existing limits from the solar experiments Borexino and Super-Kamiokande, and from the dark-matter experiments LUX-ZEPLIN \cite{AtzoriCorona:2022jeb,LZ:2022ufs} and XENONnT \cite{XENON:2022ltv}. 

\begin{table*}
\renewcommand*{\arraystretch}{1.2}
\centering
\begin{tabular}{cccc}
\toprule[0.25ex]
Experiment  & Limit at 90\% C.L. & Reference & Energy range \\ \midrule
Borexino & $\mu_{\nu{\rm SOLAR}} < 2.8 \times 10^{-11} \mu_B $ & 
\cite{Borexino:2017fbd,Coloma:2022umy} & 0.19 MeV - 2.93 MeV \\
Super-Kamiokande & $\mu_{\nu{\rm SOLAR}} < 1.1 \times 10^{-10} 
\mu_B $  & \cite{Super-Kamiokande:2004wqk} & 5 MeV - 20 MeV \\
LUX-ZEPLIN & $\mu_{\nu{\rm SOLAR}} < 6.2 \times 10^{-12} \mu_B $  & \cite{AtzoriCorona:2022jeb} &   E $\leq$ 2 MeV \\ 
XENONnT & $\mu_{\nu{\rm SOLAR}} < 6.3 \times 10^{-12} \mu_B $  & \cite{XENON:2022ltv} &  E $\leq$ 1 MeV \\
\bottomrule[0.25ex]
\end{tabular}
\caption{ \label{tab:a1-mu_solar} Limits on the effective neutrino magnetic moment from the elastic scattering of solar neutrinos on electrons.} 
\end{table*}

Note that a combined analysis of the available data from scattering experiments allows deriving stringer limits on neutrino magnetic moments --- see for instance \cite{Canas:2015yoa,DeRomeri:2022twg}. Such analyses can also shed some light on the so-called \textit{blind spots} in the neutrino parameter space \cite{Canas:2016kfy,AristizabalSierra:2021fuc}.

\section{Collider limits}
There are limits on neutrino magnetic moments also from collider searches, although these bounds are among the weakest. For instance, the process 
\begin{align}
e^+\, + \, e^- \, \longrightarrow \, \bar{\nu}\, + \, \nu \, +\, \gamma
\end{align} 
would be enhanced in the presence of non-zero neutrino magnetic moments~\cite{Grotch:1988ac}. Off the resonance, the dominant mechanism for this process to occur is bremsstrahlung from the electron or positron in the initial state. Alternatively, one can search for anomalous production of energetic single photons in electron-positron annihilation at the $Z$ resonance \cite{Gould:1994gq,L3:1997exg}. In that case, the dominant mechanism for the production of single-photon events via the neutrino magnetic moment interaction is the radiation of a photon from the final-state neutrino or anti-neutrino.  
These searches are sensitive to the same combination of magnetic moments as in plasmon decay --- see Equation~\ref{eqn:a1-plasmon_mu} --- and the best constraints, which come mainly from LEP, are of the order \mbox{of $10^{-6} \mu_B$~\cite{Gould:1994gq,L3:1997exg}.}

Other processes sensitive to neutrino magnetic moments, like pion decay
\begin{align}
\pi^0 \, \longrightarrow \gamma \, + \, \nu \, + \bar{\nu}\, ,
\end{align} 
can also be explored. So far the obtained limits are of the same order of magnitude as those from LEP \cite{Grasso:1991qy,Grasso:1993kn}. Again, in this case, the combination of magnetic moments constrained is the same as that in plasmon decay.

%----------------------------------------------------------------------------------------
%\cleardoubleevenpage
\blankpage
\blankpage
\backmatter % Denotes the end of the main document content
\setchapterstyle{plain} % Output plain chapters from this point onwards
%----------------------------------------------------------------------------------------
%	BIBLIOGRAPHY
%----------------------------------------------------------------------------------------

% The bibliography needs to be compiled with biber using your LaTeX editor, or on the command line with 'biber main' from the template directory
%\defbibnote{bibnote}{} % Prepend this text to the bibliography
\printbibliography[heading=bibintoc, title=Bibliography]
%\bibliographystyle{kaorefs} % We choose the "plain" reference style
%\bibliography{bibliography}
\blankpage
\blankpage
\end{document}